\documentclass{report}

\usepackage[english]{babel}
\usepackage[a4paper, left=3cm, right=3cm, top=3cm, bottom=3cm]{geometry}
\usepackage[T1]{fontenc}
\usepackage[utf8]{inputenc}
\usepackage{amsmath}
\usepackage{tikz}
\usepackage{graphicx}
\usepackage{amssymb}
\usepackage{tabularx}
\usepackage{fancyhdr}
\usetikzlibrary{matrix}
\usepackage[font=small,labelfont=bf]{caption}
\usepackage[all]{xy}
\newcommand{\tikzmark}[2]{
    \tikz[overlay,remember picture,baseline] 
    \node[anchor=base] (#1) {$#2$};
}
\usetikzlibrary{decorations.text}

\usepackage{csquotes}
\usepackage[backend=bibtex, natbib=true, maxcitenames=2, maxbibnames=100, dashed=false, citestyle=authoryear, style=authoryear]{biblatex}
\DeclareNameAlias{sortname}{family-given}
\renewbibmacro{in:}{}
\DeclareFieldFormat[article]{title}{#1}
\DeclareFieldFormat{pages}{#1}
\renewbibmacro*{journal+issuetitle}{%
  \usebibmacro{journal}%
  \setunit*{\addspace}%
  \iffieldundef{series}
    {}
    {\newunit
     \printfield{series}%
     \setunit{\addspace}}%
  \printfield{volume}%
  \iffieldundef{number}
     {}
      {\mkbibparens{\printfield{number}}}%
  \setunit{\addcomma\space}%
  \printfield{eid}%
  \setunit{\addspace}%
  \usebibmacro{issue+date}%
  \setunit{\addcolon\space}%
  \usebibmacro{issue}%
  \newunit}

\bibliography{mybib.bib}

\usepackage{listings}
\usepackage{dcolumn}
\usepackage{multirow} 
\usepackage{rotating}
\usepackage[justification=centering]{caption}
\usepackage{url} 
\usepackage{xcolor}
\usepackage{pgfsys}
\usepackage{pgfplots}
\pgfplotsset{every axis/.append style={
    axis x line=middle,    
    axis y line=middle,    
    axis line style={->},  
    xlabel={$t$},          
    },
    cmhplot/.style={color=black,mark=none,line width=1pt,->},
    soldot/.style={color=black, only marks,mark=*},
    holdot/.style={color=black,fill=white,only marks,mark=*},
}
\tikzset{>=stealth}
\definecolor{mygray}{rgb}{0.5,0.5,0.5}
\definecolor{commentgreen}{RGB}{2,112,10}
\usepackage{listings}
\usepackage{dcolumn}
\usepackage{multirow} 
\usepackage{rotating}
\usepackage[justification=centering]{caption}
\usepackage{url} 
\usepackage{xcolor}
\usepackage{pgfsys}
\usepackage{pgfplots}
\pgfplotsset{every axis/.append style={
    axis x line=middle,    
    axis y line=middle,    
    axis line style={->},  
    xlabel={$t$},          
    },
    cmhplot/.style={color=black,mark=none,line width=1pt,->},
    soldot/.style={color=black, only marks,mark=*},
    holdot/.style={color=black,fill=white,only marks,mark=*},
}
\tikzset{>=stealth}

\pgfplotsset{every axis/.append style={
    axis x line=middle,    
    axis y line=middle,    
    axis line style={->},  
    xlabel={$t$},          
    },
    cmhplot/.style={color=black,mark=none,line width=1pt,->},
    soldot/.style={color=black, only marks,mark=*},
    holdot/.style={color=black,fill=white,only marks,mark=*},
}
\tikzset{>=stealth}
\definecolor{mygray}{rgb}{0.5,0.5,0.5}
\definecolor{commentgreen}{RGB}{2,112,10}
\usepackage{keyval}
\usepackage{bigfoot}
\usepackage{suffix}
\usetikzlibrary{arrows,shapes,positioning}
\newcolumntype{Y}{>{\centering\arraybackslash}X}
\usetikzlibrary{shapes,arrows,chains}
\usepackage{stmaryrd}
\usepackage{bbm}
\newcommand{\rem}[1]{}
\usetikzlibrary{patterns}
\usepackage{framed}
\usetikzlibrary{arrows.meta,arrows}
\usetikzlibrary{decorations.pathreplacing}
\usepackage{color}
\usepackage{bbm}
\usetikzlibrary{automata, positioning}
\usepackage{amssymb}
\newtheorem{definition}{Definition}[section]

\newtheorem{theorem}{Theorem}[section]

\usetikzlibrary{decorations.pathreplacing,positioning, arrows.meta}
\usepackage[nohyperlinks]{acronym}
\usepackage{bm}
\usepackage{array}
\newcolumntype{P}[1]{>{\centering\arraybackslash}p{#1}}
\usepackage{lscape}
\usepackage{subfig}

\usepackage{float}

\begin{document}
\setlength{\parindent}{0em}

\begin{titlepage}
\centering

\begin{minipage}[t]{0.75\linewidth}
\centering
\includegraphics[width=0.9\linewidth]{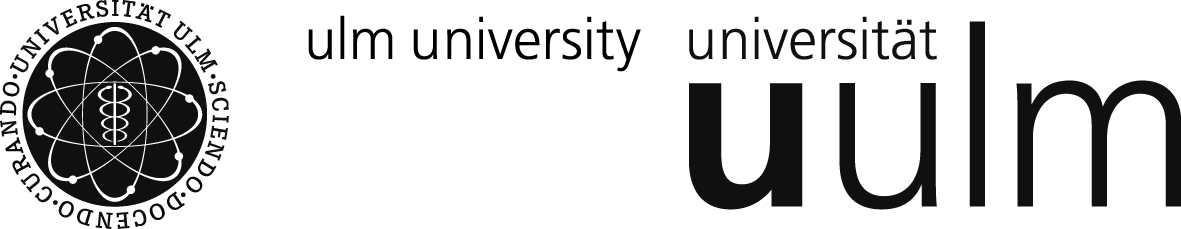}
\end{minipage}
\hfill
\begin{minipage}[t]{0.25\linewidth}
\centering
\includegraphics[width=1.3\linewidth]{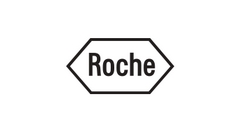}
\end{minipage}

\vspace{0.7cm}
{\Large Ulm University}\\
{\Large Faculty of Mathematics and Economics}

\vspace{2.0 cm}
{\scshape \LARGE Master Thesis \par}
\vspace{1.5 cm}
{\huge\bfseries Comparison of Time-to-First-Event \newline and Recurrent Event Methods in \newline Multiple Sclerosis Trials \par}

\vspace{2.0cm}

{\Large  by \par}
\vspace{0.3cm}
{\Large \bfseries Alexandra Bühler \par}

\vspace{2.5cm}
{\Large \bfseries \emph{Supervisors:} \par}
\smallskip
{\Large Prof. Dr. Jan Beyersmann, University of Ulm, Institute of Statistics \par}
{\Large Dr. Marcel Wolbers, F. Hoffmann-La Roche, Biostatistics Basel \par}
{\Large Dr. Fabian Model, F. Hoffmann-La Roche, Biostatistics Basel \par}
{\Large Dr. Qing Wang, F. Hoffmann-La Roche, Biostatistics Basel \par}

\vspace{2.0cm}
{\Large Ulm, 18 September 2019 \par}

\end{titlepage}

\section*{Acknowledgements} 
\thispagestyle{empty}
\hspace*{2cm}

I would like to thank

\begin{itemize}
\item Prof. Dr. Jan Beyersmann for his invaluable and constant support throughout my entire studies at the University of Ulm. His research enthusiasm for complex event history analysis has strongly influenced my interests in a positive sense. In the process of thesis writing, his inspiring guidance and expert advice helped me all the time. 
\item my external supervisors Dr. Qing Wang, Dr. Fabian Model and Dr. Marcel Wolbers for giving me the opportunity to work on this relevant and fascinating topic and for their excellent efforts. Their helpful suggestions and constructive comments contributed greatly to my thesis. 
\item Dr. Tobias Bluhmki for sharing his research experience in this topic and for his support with the high-performance computing cluster. 
\item everyone else who contributed to this thesis or supported me during my studies. 
\end{itemize}

\begin{verbatim}

\end{verbatim}


\pagenumbering{Roman}
\setcounter{page}{2}
\tableofcontents 
\newpage
\addcontentsline{toc}{chapter}{List of Figures}
\listoffigures 
\newpage
\addcontentsline{toc}{chapter}{List of Tables}
\listoftables
\newpage 

\addcontentsline{toc}{chapter}{List of Abbreviations}
\chapter*{List of Abbreviations}
\begin{acronym}[Bash]
\acro{AG}{Andersen Gill}
\acro{ARR}{Annualized Relapse Rate}
\acro{BMI}{Body Mass Index}
\acro{CDF}{Cumulative Density Function}
\acro{CDP}{Confirmed Disability Progression}
\acro{CDP12}{12-Week Confirmed Disability Progression} 
\acro{CDPW12}{imputed 12-Week Confirmed Disability Progression}
\acro{cf}{compare}
\acro{CI}{Confidence Interval}
\acro{CMF}{Cumulative Mean Function}
\acro{CP}{Counting Process}
\acro{EDSS}{Expanded Disability Status Scale}
\acro{e.g.}{exempli gratia (for example)}
\acro{Eq.}{Equation}
\acro{et al.}{et alia (and others)}
\acro{FDCE}{First Demyelinating Clinical Episode}
\acro{FS}{Functional Systems}
\acro{GT}{Gap Time}
\acro{IDP}{Initial disability/disease progression}
\acro{i.e.}{id est (that is, in other words)}
\acro{IFN}{Interferon beta-1a treatment group}
\acro{iid}{independent and identically distributed}
\acro{IPCW}{Inverse Probability of Censoring Weighting}
\acro{HR}{Hazard Ratio}
\acro{RR}{Rate Ratio}
\acro{KM}{Kaplan Meier}
\acro{LWA}{Lee Wei Amato}
\acro{LWYY}{Lin Wei Yang Ying}
\acro{MC}{Monte Carlo}
\acro{MRI}{Magnetic Resonance Imaging}
\acro{MS}{Multiple Sclerosis}
\acro{MSE}{Mean Squared Error}
\acro{NB}{Negative Binomial}
\acro{OCR}{Ocrelizumab treatment group}
\acro{PLA}{Placebo treatment group}
\acro{PDF}{Probability Density Function}
\acro{PWP}{Prentice Williams Peterson}
\acro{PPMS}{Primary Progressive Multiple Sclerosis}
\acro{RCT}{Randomized Clinical Trial}
\acro{RIS}{Radiologically Isolated Syndrome}
\acro{RRMS}{Relapsing-Remitting Multiple Sclerosis}
\acro{ROW}{Rest Of the World}
\acro{SD}{Standard Deviation}
\acro{SE}{Standard Error}
\acro{T25FW}{Timed 25-Foot Walk}
\acro{WLW}{Wei Lin Weissfeld}
\acro{9HPT}{9-Hole Peg Test}
\end{acronym}

\chapter{Introduction}
\label{Intro}
\pagestyle{fancy}
\fancyhead[R]{\thepage}
\fancyfoot[]{}
\pagenumbering{arabic}

\begin{figure}[H]
\centering
\resizebox{0.75\textwidth}{!}{
\includegraphics[keepaspectratio]{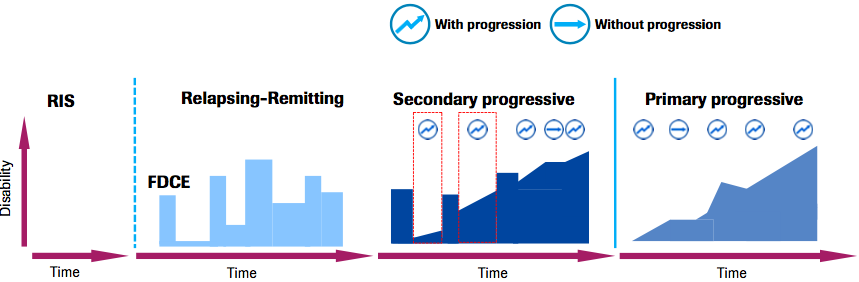}}
\caption[Forms of MS - RRMS, SPMS and PPMS]{Forms of MS - RRMS, SPMS and PPMS}
\label{FormsMS}
\end{figure}

Currently, more than $2.3$ million people are affected by multiple sclerosis (MS), for which no known cure has been detected yet. MS is a chronic, inflammatory and degenerative demyelinating disease of the human central nervous system that manifests itself through neurological deficits caused by damage to the brain, spinal cord and optic nerves. Resulting symptoms of MS include weakness, spasticity, gait and coordination imbalances, sensory dysfunction, visual loss, fatigue and cognitive impairment. While MS is a very heterogeneous disease, it is differentiated between three major disease courses varying in occurrence and timing of relapses relative to disease onset and disability progression: \textit{relapsing-remitting}, \textit{secondary progressive} and \textit{primary progressive} MS \parencite{Lublin2014, NMSS2019}. Figure $\ref{FormsMS}$ depicts potential disease activities that may occur in the different subtypes of MS. 
\newline \newline
\textbf{Relapsing-remitting MS (RRMS):} \\
RRMS is the most frequent MS type, affecting around $85 \%$ of the MS patients at diagnosis. The relapsing-remitting form is characterized by clearly defined relapses of new or increasing neurological symptoms that vary in time and subside with either partial or complete recovery (remission), and no disease progression between attacks. During remissions, the symptoms may disappear without causing any change in disability level or may persist, leading to an increased disability level. In most RRMS patients, the disease advances to a secondary progressive form after many years. 
\newline \newline
\textbf{Secondary progressive MS (SPMS):} \\
SPMS is characterized by an initial relapsing-remitting disease course followed by progression, where progression refers to as the accumulation of clinical disability independent of relapse activity over time. In SPMS, complete recovery from relapses in the relapsing-remitting part is unlikely.
\newline \newline
\textbf{Primary progressive MS (PPMS):} \\
PPMS is a relatively rare form of MS, accounting for approximately $15 \%$ of all cases of MS. The primary progressive form of MS is phenotypically characterized by a progressive, disabiling course of the disease from onset of symptoms, typically without distinct relapses or periods of remission. PPMS differs from the relapsing-remitting form in that symptoms steadily get worse over time rather than appearing as unpredictable and sudden relapses. 

In all forms of MS, suppression of disease activity and disability progression as early as possible is an important goal for treating MS. The phase III OPERA trials showed that Roche's MS drug OCREVUS (substance: ocrelizumab) is more effective than the standard treatment REBIF (substance: interferon beta-1a) in both reducing the number of relapses and delaying worsening of disability in patients with RRMS, including active SPMS with relapses \parencite{opera}. Based on the phase III ORATORIO trial, OCREVUS also provided significant benefits against disease progression among patients with early PPMS versus placebo \parencite{oratorio}. Although OCREVUS has been approved for all relapsing MS forms and early PPMS, there is still a need for demonstrating its efficacy in a broader spectrum of progressive MS (including PPMS and SPMS) as well as in underrepresented populations, such as PPMS patients who are more advanced in their disease level \parencite{NMSS2018, Roche2018}. This calls for further clinical trials in MS. As claimed by \citet{Pardini2019}, efficacy evaluation of therapies for such progressive MS forms requires innovative clinical trial designs, since the existing ones have either some limitations or have been primarily designed for relapsing populations. In recent MS research, the focus of clinical MS trials is moving towards clinical trials aimed at progressive patients to fully address the medical need for PPMS and SPMS. 
\newline \newline
Confirmed disability progression (CDP) measured on the Expanded Disability Status Scale (EDSS) is the most common outcome measure in progressive MS trials, evaluated as time-to-first-event endpoint \parencite{Ebers2008}. Since disability accumulates slowly in some patients, randomized clinical trials (RCT) usually require large sample sizes and long-term follow-up to assess relevant treatment effects with adequate statistical power. This makes RCTs time-consuming and expensive \parencite{Zhang2019, Manouchehri2019}. However, especially in progressive forms of MS, a substantial proportion of patients may experience repeated CDP events. Compared to conventional time-to-first-event analyses, where only the first CDP is included, recurrent event analyses incorporate all CDP events and could therefore improve statistical power. Additionally, recurrent event methods have also been expected to better characterize a patient's disease burden, leading to improved statistical precision and clinical interpretation of treatment effect measures \parencite{Claggett2018}. Due to the fact that progressive MS patients may progress several times during follow-up, a clinical trial based on a recurrent CDP endpoint seems to be a more appropriate design for PPMS and SPMS studies, as compared to designs featuring the first CDP only. In contrast, recurrent CDP analyses in RRMS have not been supposed to show any advantages over time-to-first-event analyses, as most RRMS patients experience disability progression at best once. 
\newline \newline 
The main objective of this thesis is to evaluate the benefit of recurrent event over time-to-first-event analyses in randomized PPMS and RRMS trials with regard to statistical properties. \newline 
While standard methods of survival analysis (e.g., Kaplan Meier estimator, Cox model, log-rank test) can be used to evaluate the treatment effect on the time to the first CDP, there exists a broad variety of recurrent event methods, classified as either conditional (Andersen-Gill, Prentice-Williams-Peterson, ...) or marginal (Wei-Lin-Weissfeld, Lin-Wei-Yang-Ying, ...). Due to conditioning on previous events, it has been argued that conditional intensity-based models are not optimal for the analysis of recurrent events in RCTs. Instead, marginal models can provide the treatment effect estimate with a clear causal interpretation \parencite{Cook2007}. Although the focus of this thesis is on marginal models, conditional models are also reported to give an overview of the whole recurrent event methodology and to emphasize the reason why marginal models are preferred in RCTs. The methods are illustrated using data from the ORATORIO and OPERA trials to examine treatment effects on recurrent CDP events and on the first CDP only. 
\newline 
In two simulation studies, analyses of the time to the first CDP are compared with recurrent event analyses including negative binomial, Andersen-Gill and Lin-Wei-Yang-Ying models. The first simulation study is generic and recurrent event data is simulated according to a mixed non-homogeneous Poisson process. The second simulation study is MS-specific, where longitudinal measurements of the ordinal EDSS scale are simulated using a homogeneous multistate model and recurrent CDP events are derived based on the resulting EDSS scores. Simulation parameters are chosen to mimic a typical trial population in PPMS and include scenarios with heterogeneity. Based on the simulation results, recommendations for the choice of an appropriate endpoint and analysis method of progressive MS trials with disability progression as primary outcome are made.
\newpage
This thesis is structured as follows. Chapter $2$ gives detailed insights into the derivation of repeated CDP events from longitudinal EDSS measurements. First, the standard definition of a first CDP event, as commonly used in clinical MS trials, is introduced. Based on this time-to-first-event endpoint, a new definition of recurrent CDP events is proposed, since a recurrent event analysis of MS progression has not been considered so far. Time-to-first-event methods are shortly summarized in Chapter $3$. In Chapter $4$, main characteristics of recurrent event data are briefly reviewed before discussing conditional and marginal recurrent event models in more detail. It further gives an overview on recurrent events in RCTs. The models suggested for the analysis of recurrent events in RCTs are applied to data from the ORATORIO and OPERA trials in PPMS and RRMS in Chapter $5$. In order to advance the clinical understanding of MS progression and to investigate potential risk factors on repeated disability progression, results from multivariate intensity-based and rate-based models are also presented. While Chapter $6$ motivates and describes the design of the two simulation studies, Chapter $7$ presents the corresponding simulation results. Finally, concluding remarks as well as an outlook on further research are provided in Chapter $8$. 

\chapter{CDP endpoint}
Prevention of or at least slowing down disability progression is a major goal of disease-modifying therapies for MS. In general, disability defines the loss of abilities resulting from (irreversible) damage to the central nervous system. In order to assess therapeutic effects of different drugs in MS patients, time to the onset of the first CDP is a widely used and well-established endpoint in RCTs, where CDP is expressed on the EDSS scale developed by John Kurtzke in 1983. As motivated in Chapter $\ref{Intro}$, recurrent event endpoints may be more suitable from a clinical and statistical perspective rather than endpoints based on the time to the first event only. \newline 
This chapter focuses on defining recurrent CDP events in both RRMS and PPMS. First, Section $\ref{SectionEDSS}$ gives a brief overview of the EDSS framework. Section $\ref{DefCDPOnset}$ extends the classical MS trial definition of a first CDP event to treat recurrent CDP events. Additionally, it also outlines the rationale behind using a roving rather than the standard fixed reference system for CDP derivation. In Section $\ref{DefCDPConfir}$, a 'new' endpoint definition based on the time to the confirmation of the $j^{th}$ CDP is proposed, differing from the classical concept with respect to the timing of events. Section $\ref{DefOverview}$ summarizes the different CDP definitions considered in this work.

\section{Expanded Disability Status Scale}
\label{SectionEDSS}
\begin{figure}[H]
\centering
\scalebox{0.12}{
\includegraphics[keepaspectratio]{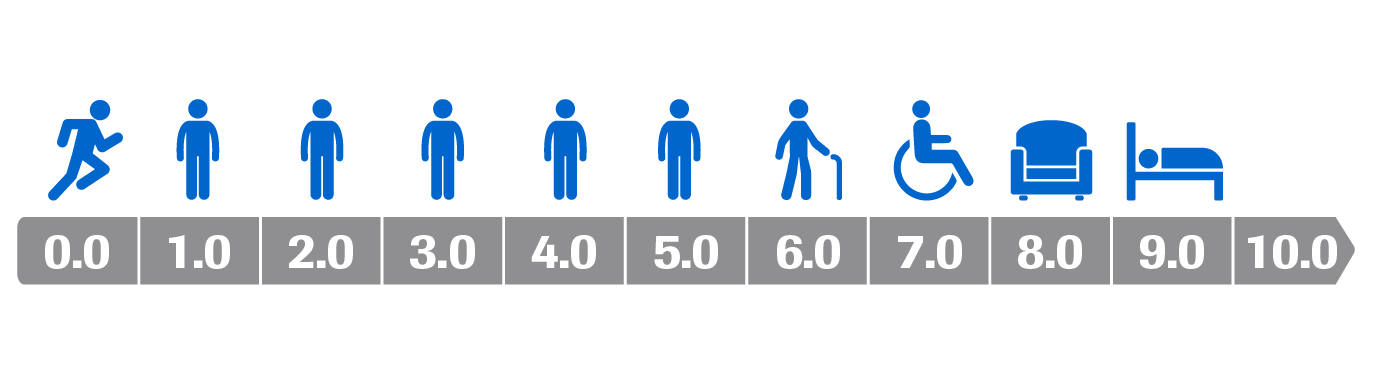}
}
\vspace*{-5mm}
\caption[Kurtzke Expanded Disability Status Scale]{Kurtzke EDSS scale}
\label{FigureEDSS}
\end{figure}

The EDSS measures the degree of physical disability based on a neurological exam of seven functional systems (FS) throughout the body (pyramidal, cerebellar, brain stem, sensory, bowel and bladder, visual, cerebral plus 'other') and a patient's walking ability. The FS are rated on a scale of $0$ to $5$ or $6$, except for the 'other' category which determines other neurological findings related to MS and is dichotomous (0 = none, 1 = present). The walking ability is assessed using the ambulation score which ranges from $0$ to $12$. Based on the FS scores and some other information (e.g., ambulation and use of medical assistive devices), EDSS scores are calculated according to the rules defined by \citet{Kurtzke1983}. In the end, the EDSS is an ordinal score ranging from $0.0$ to $10.0$ in half-point increments (only $0.5$ is not defined), with higher scores indicating worse disability. Figure $\ref{FigureEDSS}$ graphically illustrates the clinical meaning of some specific EDSS scores. For instance, a score of $0.0$ indicates normal neurological examination, a score of $2.0$ signifies minimal disability, a score of $4.0$ corresponds to relatively severe disability, a score of $6.0$ indicates that the patient requires assistance to walk such as a crane and a score of $7.0$ is associated with restriction to a wheelchair. As extracted from Figure $\ref{FigureEDSS}$, EDSS scores from $1.0$ to $5.0$ refer to MS patients who are able to walk without any aid, whereas EDSS scores equal to or higher than $5.5$ are defined by the impairment to walking. It is important to recognize that a one-point increase from $2.0$ to $3.0$ is not as severe as from $8.0$ to $9.0$. Changes at the lower or middle part of the scale describe more subtle changes in disability than at the upper part of the scale.

\section{Time to onset of CDP}
\label{DefCDPOnset}
In clinical MS trials, patients are seen approximately every $12$ weeks at predefined scheduled study visits to monitor changes in their EDSS scores over time. Apart from examinations at scheduled visits, individual patients do also have additional EDSS assessments at unscheduled study visits (e.g., during a MS relapse), withdrawal-from-treatment or end-of-study visits. Based on these longitudinal EDSS measurements, the time to the onset of the first CDP or, in a more general sense, the time to the onset of the $j^{th}$ CDP can be derived. The standard definition of the first CDP, as used in the OPERA and ORATORIO trials, will be described first.  

\subsection{Standard definition of the first CDP event}
\label{DefStandardCDP}
Disability progression is defined as an increase in the EDSS score of $\geq 1.0$ point from the baseline EDSS score if the baseline EDSS score is $\leq 5.5$, or an increase of $\geq 0.5$ points if the baseline EDSS score is $> 5.5$. The baseline EDSS score is the average score of the EDSS assessments at screening and 'day 1' study visit. The time to the onset of the first 12-week CDP (CDP12) is defined as the time from baseline to the onset of the first disability progression that is confirmed at the next regularly scheduled study visit $\geq 12$ weeks ($\geq 84$ days) after the initial disability progression (IDP). Baseline for the time to onset of CDP12 is the date of randomization \parencite{oratorio, opera}. \newline
According to this definition, a CDP event consists of two parts: \textit{initial disability progression} and \textit{confirmation of disability progression}. Roughly speaking, a confirmed IDP is a CDP. MS patients with IDP but without confirmation of initial disease progression are usually considered as progression-free, showing importance of the confirmation part. The IDP must happen during the double-blind treatment period and can occur at both scheduled or unscheduled study visits after randomization, whereas confirmation of disability progression must occur at a regularly scheduled visit that is $\geq 12$ weeks after the IDP. The confirmation visit can occur during the double-blind treatment period or even after the double-blind treatment period (e.g., open label extension phase or safety follow up phase). If there are EDSS assessments at unscheduled or non-confirmatory scheduled study visits between the IDP and the confirmation of disability progression, the corresponding EDSS scores must be at least as high as the minimum change required for progression. For example, a patient with a baseline score of $4.0$ must have EDSS scores of $\geq 5.0$ at all study visits between the visit with IDP and the scheduled visit to confirm the $12$-week CDP. Thus, non-confirmatory EDSS assessments between IDP and confirmation of IDP need to fulfill the requirements for progression as well. Otherwise, the IDP is not confirmed. EDSS assessments within $30$ days after a protocol-defined relapse cannot be used for confirmation of disability progression. \newline 
Patients who do not have an IDP by the end of the double-blind treatment period, time of early discontinuation or at time of loss to follow-up are censored at the date of their last EDSS assessment. Patients on treatment with no confirmation after an IDP are also censored at the date of their last EDSS assessment. However, \citet{Ebers2008} provided evidence of higher CDP12 confirmation rates in progressive versus relapsing MS, with confirmation rates in PPMS patients of approximately $80 \%$. PPMS patients with IDP have an increased risk of CDP compared to other patients without an initial event of neurological worsening. Therefore, PPMS patients who have an IDP and then withdraw from treatment early with no confirmatory EDSS assessment are not censored, as this would introduce bias, but are imputed as having a CDP12 event at time of withdrawal. These events only occur in PPMS trials and are often referred to as 'imputed' CDPW12 events. Table $\ref{TabelDefCens}$ summarizes the different censoring mechanisms following an IDP.
\begin{table}[b]
\centering
\scalebox{0.65}{ 
\begin{tabular}{|l|c|l|}
\hline
\multicolumn{1}{|c|}{\textbf{\begin{tabular}[c]{@{}c@{}}Availability of EDSS assessments \\ for patients with IDP\end{tabular}}}                                 & \multicolumn{2}{l|}{\textbf{\begin{tabular}[c]{@{}l@{}}Recorded disability progression status\\ $$\end{tabular}}} \\ \hline
\begin{tabular}[c]{@{}l@{}}Confirmatory EDSS assessment on treatment or \\ after withdrawal from treatment\end{tabular}                                         & \multicolumn{2}{c|}{CDP}                                                                                          \\ \hline
\begin{tabular}[c]{@{}l@{}}No confirmatory EDSS assessment or patients \\ on treatment at the time of study closure\end{tabular}                                 & \multicolumn{2}{c|}{censored at last EDSS assessment}                                                             \\ \hline
\begin{tabular}[c]{@{}l@{}}Discontinuation from treatment and loss to follow-up,\\ i.e., no available EDSS assessment at $\geq 84$ days\\ after IDP\end{tabular} & \multicolumn{2}{c|}{\begin{tabular}[c]{@{}c@{}}$$\\ 'imputed' CDP (= CDPW12)\\ $$\end{tabular}}                      \\ \hline
\end{tabular}}
\caption{Censoring algorithms of patients after initial disability progression}
\label{TabelDefCens}
\end{table}

\subsection{Definition of recurrent CDP events}
\label{DefStandardRecurrentCDP}
The standard definition of the time to the onset of the first CDP will be now reformulated to capture recurrent progression events. More generally, a $j^{th}$ disability progression is defined as an increase in the EDSS score of $\geq 1.0$ or $\geq 0.5$ points from the $j^{th}$ reference EDSS score, where one-step disability progression is used for reference EDSS scores $\leq 5.5$ and half-step disability progression is applied to reference EDSS scores $> 5.5$, $\ j=1,2,...,$ \parencite{Weinshenker1996}. The reference score for the $j^{th}$ disability progression is the EDSS value associated with the $(j-1)^{th}$ IDP of the previous event. Following each identified CDP, the reference disability level must be readjusted. The baseline EDSS value serves as the reference EDSS score for the first progression event, in which case the newly proposed definition of recurrent CDP events coincides with the classical first event definition. There may be other clinically meaningful definitions for the reference EDSS score but, in this work, the $j^{th}$ reference score is restricted to the EDSS score recorded at the study visit with the $(j-1)^{th}$ IDP. As in the classical time-to-first-event setting, the time to the onset of the $j^{th}$ CDP12 is then defined as the time from baseline to the onset of the $j^{th}$ disability progression that is confirmed at the next regularly scheduled study visit $\geq 12$ weeks ($\geq 84$ days) after the $j^{th}$ IDP, where baseline corresponds to the date of randomization. \newline 
All other rules (e.g., IDP at unscheduled or scheduled visits, confirmation only at scheduled visits, visits between IDP and confirmation must also meet the requirements for progression, ...) and censoring algorithms after an IDP can be extracted from Section $\ref{DefStandardCDP}$, as the same concepts apply to recurrent CDP events. 
\newline \newline 
Derivation of recurrent CDP12 events from longitudinal EDSS data and the corresponding event times is exemplified in Figure $\ref{FigDefinition}$ for $3$ different scenarios. Example $(a)$ shows a typical PPMS EDSS profile from which $3$ CDP12 events can be determined, with readjustment of the reference EDSS score at the patient's week $24$, week $60$ and week $84$ visit. As illustrated in panel $(a)$, the reference EDSS score for the first CDP12 is the baseline value of $3.0$, the reference EDSS score for the second CDP12 is $4.0$ (= EDSS at first IDP) and the reference EDSS value for the third CDP12 is $5.0$ (= EDSS at second IDP). The event date corresponds to the date of IDP, if disability progression is confirmed. Example $(b)$ is slightly different from $(a)$ in the sense that confirmation of the second IDP and registration of a third CDP12 event happen exactly at the same study visit. In example $(c)$, IDPs at week $36$ and week $60$ can not be confirmed because the EDSS scores at potential confirmation visits do not fulfill the requirements for progression. 

\subsection{Roving reference system}
\label{sectionRoving}
So far, a fixed reference system has been considered to deduce recurrent progression events from longitudinal EDSS data. Using a fixed system, the reference EDSS score for a particular CDP is kept at a fixed disability level and reference measurements vary only across different events. On the other hand, \citet{Kappos2018} demonstrated in their publication that a roving reference system is more efficient in detecting progression events (unrelated to relapses) in RRMS patients. Their findings are limited to time-to-first-event analyses in RRMS though. Based on the fundamental ideas of \citet{Kappos2018}, this work intends to evaluate the use of a roving reference system to derive repeated progression events in both RRMS and PPMS populations within the scope of a sensitivity analysis. A roving reference system resets the original reference score after a $\geq 12$ -week or even $\geq 24$- week confirmation of a new score. To be more precise, a recorded EDSS score is qualified for a 'new' reference if this EDSS score is lower than the current EDSS reference score and can be confirmed by the same EDSS score $12$- or $24$- weeks later. The $12$- week or $24$- week confirmation of the new score aims at detecting the 'true' reference level in the absence of natural variation in the EDSS assignments. Using a roving reference system, reference measurements may vary within and across different progression events. \newline 
In order to clarify the difference between a fixed and roving reference system, Figure $\ref{FigRovingFixed}$ illustrates a hypothetical EDSS profile with an initial decrease in the EDSS score after randomization. Starting with a baseline EDSS score of $5.0$, the patient's EDSS score decreases to $4.0$ at the week $12$ study visit and keeps stable thereafter until week $48$. From week $48$ to study closure, the patient seems to experience disability progression, as an increasing trend in EDSS scores can be observed. 

\newpage
\definecolor{darkpowderblue}{rgb}{0.0, 0.2, 0.6}
\begin{figure}[H]
\centering
\subfloat[Example 1]{ 
\resizebox{\textwidth}{!}{
\begin{tikzpicture}[decoration=brace]
	\draw [color=gray!50]  [step=5mm] (0,0) grid (15.5,7.0);
	\draw[->,thick] (0,0) -- (15.5,0) node[right] {$\textnormal{time}$};
	\draw[->,thick] (0,0) -- (0,7.0) node[above] {$\textnormal{EDSS score}$};
	\draw (0.5,-.2) -- (0.5,0) node[below=4pt] {$\textnormal{B}$};
	\draw (2.0,-.2) -- (2.0,0) node[below=4pt] {$\textnormal{W12}$};
	\draw (3.5,-.2) -- (3.5,0) node[below=4pt] {$\textnormal{W24}$};
	\draw (5.0,-.2) -- (5.0,0) node[below=4pt] {$\textnormal{W36}$};
	\draw (6.5,-.2) -- (6.5,0) node[below=4pt] {$\textnormal{W48}$};
	\draw (8.0,-.2) -- (8.0,0) node[below=4pt] {$\textnormal{W60}$};
	\draw (9.5,-.2) -- (9.5,0) node[below=4pt] {$\textnormal{W72}$};
	\draw (11.0,-.2) -- (11.0,0) node[below=4pt] {$\textnormal{W84}$};
	\draw (12.5,-.2) -- (12.5,0) node[below=4pt] {$\textnormal{W96}$};
	\draw (14.0,-.2) -- (14.0,0) node[below=4pt] {$\textnormal{...}$};
	\draw (-.1,1.0) -- (.1,1.0) node[left=4pt] {$\scriptstyle 3.0$};
	\draw (-.1,1.5) -- (.1,1.5) node[left=4pt] {$\scriptstyle 3.5$};
	\draw (-.1,2.0) -- (.1,2.0) node[left=4pt] {$\scriptstyle 4.0$};
	\draw (-.1,2.5) -- (.1,2.5) node[left=4pt] {$\scriptstyle 4.5$};
	\draw (-.1,3.0) -- (.1,3.0) node[left=4pt] {$\scriptstyle 5.0$};
	\draw (-.1,3.5) -- (.1,3.5) node[left=4pt] {$\scriptstyle 5.5$};
	\draw (-.1,4.0) -- (.1,4.0) node[left=4pt] {$\scriptstyle 6.0$};
	\draw (-.1,4.5) -- (.1,4.5) node[left=4pt] {$\scriptstyle 6.5$};
	\draw (-.1,5.0) -- (.1,5.0) node[left=4pt] {$\scriptstyle 7.0$};
  \node[outer sep=0pt,circle, fill=black,inner sep=1.5pt] (E0) at (0.5, 1.0) {};
  \node[outer sep=0pt,circle, fill=black,inner sep=1.5pt] (E1) at (2.0, 1.0) {};
  \node[outer sep=0pt,circle, fill=black,inner sep=1.5pt] (E2) at (3.5, 2.0) {};
  \coordinate[label=right: $\textnormal{IDP}_{1}$]() at (3.0,2.2);
  \coordinate[label=right: $\textnormal{= CDP}_{1}$]() at (2.7,2.5);
  \node[outer sep=0pt,circle, fill=black,inner sep=1.5pt] (E3) at (5.0, 2.0) {};
  \coordinate[label=right: $\textnormal{C}_{1}$]() at (4.5,2.3);
  \node[outer sep=0pt,circle, fill=black,inner sep=1.5pt] (E4) at (6.5, 2.0) {};
  \node[outer sep=0pt,circle, fill=black,inner sep=1.5pt] (E5) at (8.0, 3.0) {};
  \coordinate[label=right: ${\textnormal{IDP}_{2} = \textnormal{CDP}_{2}}$]() at (8.0,2.8);
  \node[outer sep=0pt,circle, fill=black,inner sep=1.5pt] (E6) at (9.5, 3.5) {};
  \coordinate[label=right: $\textnormal{C}_{2}$]() at (9.5,3.3);
  \node[outer sep=0pt,circle, fill=black,inner sep=1.5pt] (E7) at (11.0, 4.5) {};
  \coordinate[label=right: $\textnormal{IDP}_{3}$]() at (10.5,4.7);
  \coordinate[label=right: $\textnormal{= CDP}_{3}$]() at (10.7,4.3);
  \node[outer sep=0pt,circle, fill=black,inner sep=1.5pt] (E8) at (12.5, 4.5) {};
  \coordinate[label=right: $\textnormal{C}_{3}$]() at (12.0,4.8);
  \node[outer sep=0pt,circle, fill=black,inner sep=1.5pt] (E9) at (14.0, 4.0) {};
  \draw [color=black, dashed, line width=0.25mm] plot [dashed] coordinates
	{(0.5,1.0) (2.0,1.0) (3.5, 2.0) (5.0, 2.0) (6.5, 2.0) (8.0, 3.0) (9.5, 3.5) (11.0, 4.5) (12.5, 4.5) (14.0, 4.0)} node[right]{};
	\draw[dotted, color=black](0.5,1.0) -- (3.5,1.0);
	\coordinate[label=right: $\textit{Ref}_{1}$]() at (3.5,1.0);
	\draw[dotted, color=black](3.5,2.0) -- (8.0,2.0);
	\coordinate[label=right: $\textit{Ref}_{2}$]() at (8.0,2.0);
	\draw[dotted, color=black](8.0,3.0) -- (11.0,3.0);
	\coordinate[label=right: $\textit{Ref}_{3}$]() at (11.0,3.0);
	\draw[dotted, color=black](11.0,4.5) -- (14.5,4.5);
	\coordinate[label=right: $\textit{Ref}_{4}$]() at (14.5,4.5);
	\draw[->, dotted](3.5,6.5) -- (3.5,5.5);
	\coordinate[label=above: $\textit{readjustment}$]() at (3.5,6.5);
	\draw[->, dotted](8.0,6.5) -- (8.0,5.5);
	\coordinate[label=above: $\textit{readjustment}$]() at (8.0,6.5);
	\draw[->, dotted](11.0,6.5) -- (11.0,5.5);
	\coordinate[label=above: $\textit{readjustment}$]() at (11.0,6.5);
	\draw[decorate, yshift=7ex, color=darkpowderblue, thick] (0.5,1.7) -- node[above=0.4ex] {$\boldsymbol{=T_{1}}$} (3.5,1.7);
	\draw[decorate, yshift=7ex, color=darkpowderblue, thick] (0.5,2.5) -- node[above=0.4ex] {$\boldsymbol{=T_{2}}$} (8.0,2.5);
	\draw[decorate, yshift=7ex, color=darkpowderblue, thick] (0.5,4.0) -- node[above=0.4ex] {$\boldsymbol{=T_{3}}$} (11,4.0);
\end{tikzpicture}
}
}

\subfloat[Example 2]{ 
\resizebox{\textwidth}{!}{
\begin{tikzpicture}[decoration=brace]
	\draw [color=gray!50]  [step=5mm] (0,0) grid (15.5,7.0);
	\draw[->,thick] (0,0) -- (15.5,0) node[right] {$\textnormal{time}$};
	\draw[->,thick] (0,0) -- (0,7.0) node[above] {$\textnormal{EDSS score}$};
	\draw (0.5,-.2) -- (0.5,0) node[below=4pt] {$\textnormal{B}$};
	\draw (2.0,-.2) -- (2.0,0) node[below=4pt] {$\textnormal{W12}$};
	\draw (3.5,-.2) -- (3.5,0) node[below=4pt] {$\textnormal{W24}$};
	\draw (5.0,-.2) -- (5.0,0) node[below=4pt] {$\textnormal{W36}$};
	\draw (6.5,-.2) -- (6.5,0) node[below=4pt] {$\textnormal{W48}$};
	\draw (8.0,-.2) -- (8.0,0) node[below=4pt] {$\textnormal{W60}$};
	\draw (9.5,-.2) -- (9.5,0) node[below=4pt] {$\textnormal{W72}$};
	\draw (11.0,-.2) -- (11.0,0) node[below=4pt] {$\textnormal{W84}$};
	\draw (12.5,-.2) -- (12.5,0) node[below=4pt] {$\textnormal{W96}$};
	\draw (14.0,-.2) -- (14.0,0) node[below=4pt] {$\textnormal{...}$};
	\draw (-.1,1.0) -- (.1,1.0) node[left=4pt] {$\scriptstyle 3.0$};
	\draw (-.1,1.5) -- (.1,1.5) node[left=4pt] {$\scriptstyle 3.5$};
	\draw (-.1,2.0) -- (.1,2.0) node[left=4pt] {$\scriptstyle 4.0$};
	\draw (-.1,2.5) -- (.1,2.5) node[left=4pt] {$\scriptstyle 4.5$};
	\draw (-.1,3.0) -- (.1,3.0) node[left=4pt] {$\scriptstyle 5.0$};
	\draw (-.1,3.5) -- (.1,3.5) node[left=4pt] {$\scriptstyle 5.5$};
	\draw (-.1,4.0) -- (.1,4.0) node[left=4pt] {$\scriptstyle 6.0$};
	\draw (-.1,4.5) -- (.1,4.5) node[left=4pt] {$\scriptstyle 6.5$};
	\draw (-.1,5.0) -- (.1,5.0) node[left=4pt] {$\scriptstyle 7.0$};
  \node[outer sep=0pt,circle, fill=black,inner sep=1.5pt] (E0) at (0.5, 1.0) {};
  \node[outer sep=0pt,circle, fill=black,inner sep=1.5pt] (E1) at (2.0, 1.0) {};
  \node[outer sep=0pt,circle, fill=black,inner sep=1.5pt] (E2) at (3.5, 2.0) {};
  \coordinate[label=right: $\textnormal{IDP}_{1}$]() at (3.0,2.2);
  \coordinate[label=right: $\textnormal{= CDP}_{1}$]() at (2.7,2.5);
  \node[outer sep=0pt,circle, fill=black,inner sep=1.5pt] (E3) at (5.0, 2.0) {};
  \coordinate[label=right: $\textnormal{C}_{1}$]() at (4.5,2.3);
  \node[outer sep=0pt,circle, fill=black,inner sep=1.5pt] (E4) at (6.5, 2.0) {};
  \node[outer sep=0pt,circle, fill=black,inner sep=1.5pt] (E5) at (8.0, 3.0) {};
  \coordinate[label=right: $\textnormal{IDP}_{2}$]() at (7.0,3.1);
  \coordinate[label=right: $\textnormal{= CDP}_{2}$]() at (6.8,3.4);
  \node[outer sep=0pt,circle, fill=black,inner sep=1.5pt] (E6) at (9.5, 4.0) {};
  \coordinate[label=right: $\textnormal{IDP}_{3}$]() at (9.15,4.2);
  \coordinate[label=right: $\textnormal{= CDP}_{3}$]() at (8.95,4.5);
  \coordinate[label=right: $\textnormal{C}_{2}$]() at (9.5,3.75);
  \node[outer sep=0pt,circle, fill=black,inner sep=1.5pt] (E9) at (11.0, 4.0) {};
  \coordinate[label=right: $\textnormal{C}_{3}$]() at (10.5,4.3);
  \node[outer sep=0pt,circle, fill=black,inner sep=1.5pt] (E8) at (12.5, 4.0) {};
  \node[outer sep=0pt,circle, fill=black,inner sep=1.5pt] (E9) at (14.0, 4.0) {};
  \draw [color=black, dashed, line width=0.25mm] plot [dashed] coordinates
	{(0.5,1.0) (2.0,1.0) (3.5, 2.0) (5.0, 2.0) (6.5, 2.0) (8.0, 3.0) (9.5, 4.0) (11.0, 4.0) (12.5, 4.0) (14.0, 4.0)} node[right]{};
	\draw[dotted, color=black](0.5,1.0) -- (3.5,1.0);
	\coordinate[label=right: $\textit{Ref}_{1}$]() at (3.5,1.0);
	\draw[dotted, color=black](3.5,2.0) -- (8.0,2.0);
	\coordinate[label=right: $\textit{Ref}_{2}$]() at (8.0,2.0);
	\draw[dotted, color=black](8.0,3.0) -- (9.5,3.0);
	\coordinate[label=right: $\textit{Ref}_{3}$]() at (9.5,3.0);
	\draw[dotted, color=black](9.5, 4.0) -- (14.5,4.0);
	\coordinate[label=right: $\textit{Ref}_{4}$]() at (14.5,4.0);
	\draw[->, dotted](3.5,6.5) -- (3.5,5.5);
	\coordinate[label=above: $\textit{readjustment}$]() at (3.5,6.5);
	\draw[->, dotted](8.0,6.5) -- (8.0,5.5);
	\coordinate[label=above: $\textit{readjustment}$]() at (8.8,6.5);
	\draw[->, dotted](9.5,6.5) -- (9.5,5.5);
	\draw[decorate, yshift=7ex, color=darkpowderblue, thick] (0.5,1.7) -- node[above=0.4ex] {$\boldsymbol{=T_{1}}$} (3.5,1.7);
	\draw[decorate, yshift=7ex, color=darkpowderblue, thick] (0.5,2.5) -- node[above=0.4ex] {$\boldsymbol{=T_{2}}$} (8.0,2.5);
	\draw[decorate, yshift=7ex, color=darkpowderblue, thick] (0.5,3.7) -- node[above=0.4ex] {$\boldsymbol{=T_{3}}$} (9.5,3.7);
\end{tikzpicture}
}
}

\subfloat[Example 3]{ 
\resizebox{\textwidth}{!}{
\begin{tikzpicture}[decoration=brace]
	\draw [color=gray!50]  [step=5mm] (0,0) grid (15.5,7.0);
	\draw[->,thick] (0,0) -- (15.5,0) node[right] {$\textnormal{time}$};
	\draw[->,thick] (0,0) -- (0,7.0) node[above] {$\textnormal{EDSS score}$};
	\draw (0.5,-.2) -- (0.5,0) node[below=4pt] {$\textnormal{B}$};
	\draw (2.0,-.2) -- (2.0,0) node[below=4pt] {$\textnormal{W12}$};
	\draw (3.5,-.2) -- (3.5,0) node[below=4pt] {$\textnormal{W24}$};
	\draw (5.0,-.2) -- (5.0,0) node[below=4pt] {$\textnormal{W36}$};
	\draw (6.5,-.2) -- (6.5,0) node[below=4pt] {$\textnormal{W48}$};
	\draw (8.0,-.2) -- (8.0,0) node[below=4pt] {$\textnormal{W60}$};
	\draw (9.5,-.2) -- (9.5,0) node[below=4pt] {$\textnormal{W72}$};
	\draw (11.0,-.2) -- (11.0,0) node[below=4pt] {$\textnormal{W84}$};
	\draw (12.5,-.2) -- (12.5,0) node[below=4pt] {$\textnormal{W96}$};
	\draw (14.0,-.2) -- (14.0,0) node[below=4pt] {$\textnormal{...}$};
	\draw (-.1,1.0) -- (.1,1.0) node[left=4pt] {$\scriptstyle 3.0$};
	\draw (-.1,1.5) -- (.1,1.5) node[left=4pt] {$\scriptstyle 3.5$};
	\draw (-.1,2.0) -- (.1,2.0) node[left=4pt] {$\scriptstyle 4.0$};
	\draw (-.1,2.5) -- (.1,2.5) node[left=4pt] {$\scriptstyle 4.5$};
	\draw (-.1,3.0) -- (.1,3.0) node[left=4pt] {$\scriptstyle 5.0$};
	\draw (-.1,3.5) -- (.1,3.5) node[left=4pt] {$\scriptstyle 5.5$};
	\draw (-.1,4.0) -- (.1,4.0) node[left=4pt] {$\scriptstyle 6.0$};
	\draw (-.1,4.5) -- (.1,4.5) node[left=4pt] {$\scriptstyle 6.5$};
	\draw (-.1,5.0) -- (.1,5.0) node[left=4pt] {$\scriptstyle 7.0$};
  \node[outer sep=0pt,circle, fill=black,inner sep=1.5pt] (E0) at (0.5, 4.0) {};
  \node[outer sep=0pt,circle, fill=black,inner sep=1.5pt] (E1) at (2.0, 2.5) {};
  \node[outer sep=0pt,circle, fill=black,inner sep=1.5pt] (E2) at (3.5, 4.0) {};
  \node[outer sep=0pt,circle, fill=black,inner sep=1.5pt] (E3) at (5.0, 4.5) {};
  \node[outer sep=0pt,circle, fill=black,inner sep=1.5pt] (E4) at (6.5, 4.0) {};
  \node[outer sep=0pt,circle, fill=black,inner sep=1.5pt] (E5) at (8.0, 4.5) {};
  \node[outer sep=0pt,circle, fill=black,inner sep=1.5pt] (E6) at (9.5, 4.0) {};
  \node[outer sep=0pt,circle, fill=black,inner sep=1.5pt] (E9) at (11.0, 4.0) {};
  \node[outer sep=0pt,circle, fill=black,inner sep=1.5pt] (E8) at (12.5, 4.5) {};
  \coordinate[label=right: $\textnormal{IDP}_{1}$]() at (12.0,4.7);
  \coordinate[label=right: $\textnormal{= CDP}_{1}$]() at (11.8,5.0);
  \node[outer sep=0pt,circle, fill=black,inner sep=1.5pt] (E9) at (14.0, 4.5) {};
  \coordinate[label=right: $\textnormal{C}_{1}$]() at (13.5,4.8);
  \draw [color=black, dashed, line width=0.25mm] plot [dashed] coordinates
	{(0.5, 4.0) (2.0,2.5) (3.5, 4.0) (5.0, 4.5) (6.5, 4.0) (8.0, 4.5) (9.5, 4.0) (11.0, 4.0) (12.5, 4.5) (14.0, 4.5)} node[right]{};
	\draw[dotted, color=black](0.5,4.0) -- (12.5,4.0);
	\coordinate[label=right: $\textit{Ref}_{1}$]() at (12.5,4.0);
	\draw[dotted, color=black](12.5,4.5) -- (15.0,4.5);
	\coordinate[label=right: $\textit{Ref}_{2}$]() at (15.0,4.5);
	\draw[->, dotted](14.0,6.5) -- (14.0,5.5);
	\coordinate[label=above: $\textit{readjustment}$]() at (14.0,6.5);
	\draw[decorate, yshift=7ex, color=darkpowderblue, thick] (0.5,4.2) -- node[above=0.4ex] {$\boldsymbol{=T_{1}}$} (12.5,4.2);
\end{tikzpicture}
}
}
\caption[Derivation of time-to-onset-of-CDP12 endpoint from EDSS measurements]{Derivation of time-to-onset-of-CDP12 endpoint from EDSS measurements ($\text{IDP}_{j}$ = $j^{th}$ initial disability progression, $\text{C}_{j}$ = confirmation of $IDP_{j}$, $\text{CDP}_{j}$ = $j^{th}$ confirmed disability progression (event), $\text{Ref}_{j}$ = reference EDSS score for $j^{th}$ CDP, $T_{j}$ = time to onset of the $j^{th}$ CDP)}
\label{FigDefinition}
\end{figure}
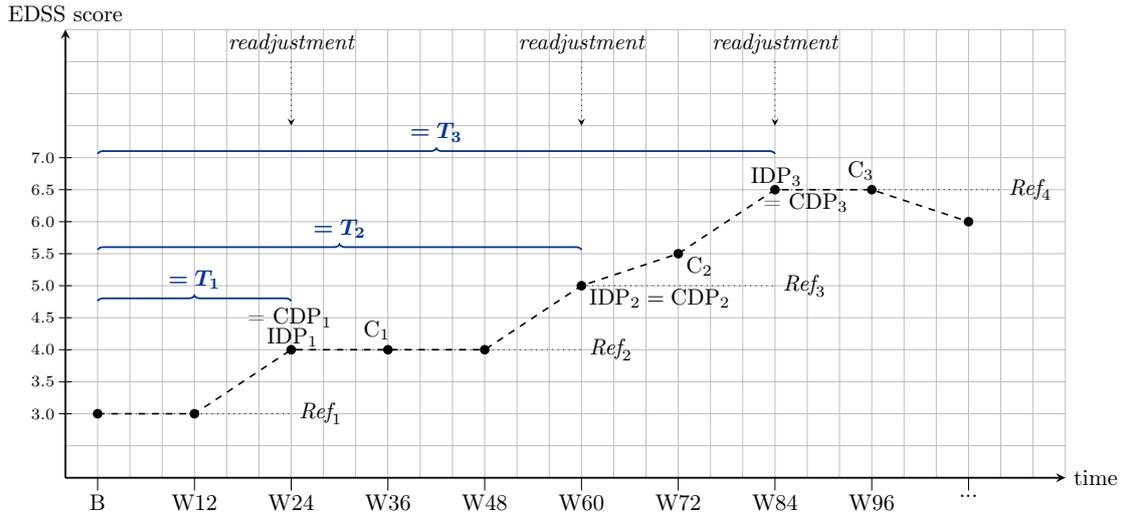
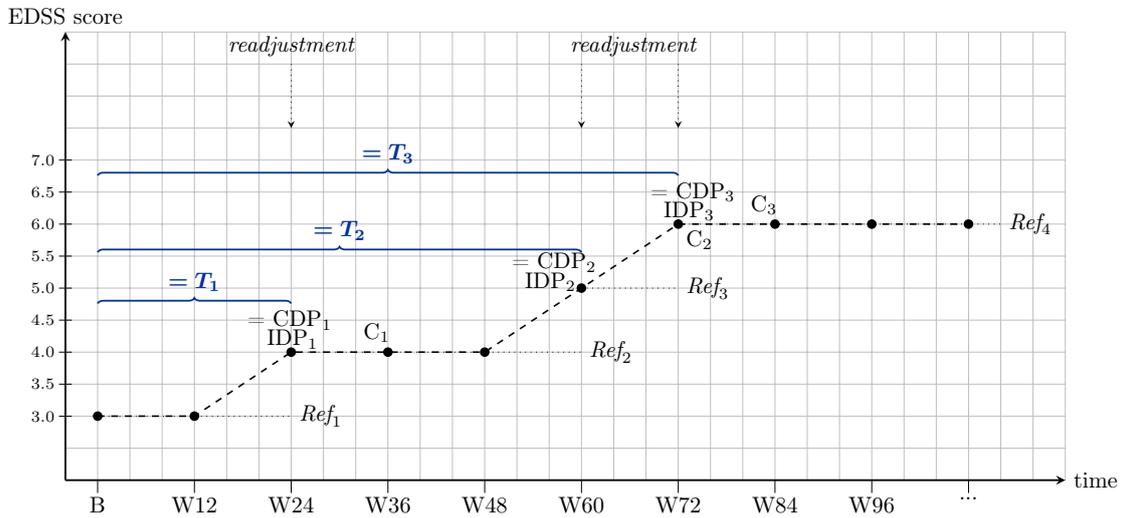
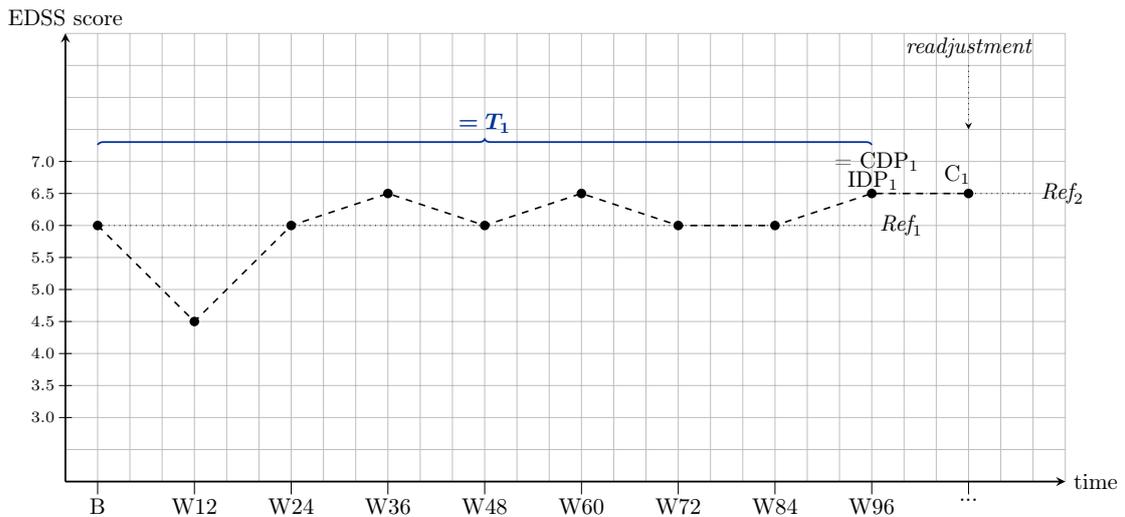

The EDSS profile considered in Figure $\ref{FigRovingFixed}$ is typical for both PPMS and RRMS patients enrolled into clinical trials. The reasons for disability improvement after randomization are different for PPMS and RRMS patients. The study population for the OPERA trials included RRMS patients who were supposed to have clinical disease activity. Although inclusion criteria for the OPERA trials request neurological stability for $\geq 30$ days prior to baseline (i.e., no relapse has been observed), RRMS patients often experience a decrease in the EDSS score during the first $6$ months under observation due to a prolonged recovery from relapse. This phenomenon is also known as EDSS score regression \parencite{Kappos2018}. In PPMS patients, relapses are very rare and initial improvement in disability after randomization may be explained by time-dependent natural variability in patients' disease status or misclassification of the baseline EDSS score by the investigator. With regard to Figure $\ref{FigRovingFixed}$, the EDSS score $4.0$ observed at the week $12$ and week $48$ study visits would better reflect the current disability level and would consequently serve as a more meaningful reference measurement for progression events. Such an initial decrease in the EDSS score may be associated with a reduced detection rate or even non-detection of 'true' progression events, as patients first need to progress back to the reference measurement and then beyond it to record disability progression. Non-detection versus detection of progression events under a fixed and roving reference system is illustrated in Figure $\ref{FigRovingFixed}$. 
\begin{figure}[H]
\centering
\subfloat[Fixed reference system]{ 
\resizebox{\textwidth}{!}{
\begin{tikzpicture}
	\draw [color=gray!50]  [step=5mm] (0,0) grid (15.5,5.5);
	\draw[->,thick] (0,0) -- (15.5,0) node[right] {$\textnormal{time}$};
	\draw[->,thick] (0,0) -- (0,5.5) node[above] {$\textnormal{EDSS score}$};
	\draw (0.5,-.2) -- (0.5,0) node[below=4pt] {$\textnormal{B}$};
	\draw (2.0,-.2) -- (2.0,0) node[below=4pt] {$\textnormal{W12}$};
	\draw (3.5,-.2) -- (3.5,0) node[below=4pt] {$\textnormal{W24}$};
	\draw (5.0,-.2) -- (5.0,0) node[below=4pt] {$\textnormal{W36}$};
	\draw (6.5,-.2) -- (6.5,0) node[below=4pt] {$\textnormal{W48}$};
	\draw (8.0,-.2) -- (8.0,0) node[below=4pt] {$\textnormal{W60}$};
	\draw (9.5,-.2) -- (9.5,0) node[below=4pt] {$\textnormal{W72}$};
	\draw (11.0,-.2) -- (11.0,0) node[below=4pt] {$\textnormal{W84}$};
	\draw (12.5,-.2) -- (12.5,0) node[below=4pt] {$\textnormal{W96}$};
	\draw (14.0,-.2) -- (14.0,0) node[below=4pt] {$\textnormal{...}$};
	\draw (-.1,1.0) -- (.1,1.0) node[left=4pt] {$\scriptstyle 3.0$};
	\draw (-.1,1.5) -- (.1,1.5) node[left=4pt] {$\scriptstyle 3.5$};
	\draw (-.1,2.0) -- (.1,2.0) node[left=4pt] {$\scriptstyle 4.0$};
	\draw (-.1,2.5) -- (.1,2.5) node[left=4pt] {$\scriptstyle 4.5$};
	\draw (-.1,3.0) -- (.1,3.0) node[left=4pt] {$\scriptstyle 5.0$};
	\draw (-.1,3.5) -- (.1,3.5) node[left=4pt] {$\scriptstyle 5.5$};
	\draw (-.1,4.0) -- (.1,4.0) node[left=4pt] {$\scriptstyle 6.0$};
	\draw (-.1,4.5) -- (.1,4.5) node[left=4pt] {$\scriptstyle 6.5$};
	\draw (-.1,5.0) -- (.1,5.0) node[left=4pt] {$\scriptstyle 7.0$};
  \node[outer sep=0pt,circle, fill=black,inner sep=1.5pt] (E0) at (0.5, 3.0) {};
  \node[outer sep=0pt,circle, fill=black,inner sep=1.5pt] (E1) at (2.0, 2.0) {};
  \node[outer sep=0pt,circle, fill=black,inner sep=1.5pt] (E2) at (3.5, 2.0) {};
  \node[outer sep=0pt,circle, fill=black,inner sep=1.5pt] (E3) at (5.0, 2.0) {};
  \node[outer sep=0pt,circle, fill=black,inner sep=1.5pt] (E4) at (6.5, 2.0) {};
  \node[outer sep=0pt,circle, fill=black,inner sep=1.5pt] (E5) at (8.0, 3.0) {};
  \node[outer sep=0pt,circle, fill=black,inner sep=1.5pt] (E6) at (9.5, 3.0) {};
  \node[outer sep=0pt,circle, fill=black,inner sep=1.5pt] (E9) at (11.0, 3.5) {};
  \node[outer sep=0pt,circle, fill=black,inner sep=1.5pt] (E8) at (12.5, 3.5) {};
  \node[outer sep=0pt,circle, fill=black,inner sep=1.5pt] (E9) at (14.0, 3.5) {};
  \draw [color=black, dashed, line width=0.25mm] plot [dashed] coordinates
	{(0.5,3.0) (2.0,2.0) (3.5, 2.0) (5.0, 2.0) (6.5, 2.0) (8.0, 3.0) (9.5, 3.0) (11.0, 3.5) (12.5, 3.5) (14.0, 3.5)} node[right]{};
	\draw[dotted, color=black](0.5,3.0) -- (14.0,3.0);
	\coordinate[label=right: $\textit{Ref}_{1}$]() at (14.0,3.0);
\end{tikzpicture}
}
}

\subfloat[Roving reference system]{ 
\resizebox{\textwidth}{!}{
\begin{tikzpicture}[decoration=brace]
	\draw [color=gray!50]  [step=5mm] (0,0) grid (15.5,6.5);
	\draw[->,thick] (0,0) -- (15.5,0) node[right] {$\textnormal{time}$};
	\draw[->,thick] (0,0) -- (0,6.5) node[above] {$\textnormal{EDSS score}$};
	\draw (0.5,-.2) -- (0.5,0) node[below=4pt] {$\textnormal{B}$};
	\draw (2.0,-.2) -- (2.0,0) node[below=4pt] {$\textnormal{W12}$};
	\draw (3.5,-.2) -- (3.5,0) node[below=4pt] {$\textnormal{W24}$};
	\draw (5.0,-.2) -- (5.0,0) node[below=4pt] {$\textnormal{W36}$};
	\draw (6.5,-.2) -- (6.5,0) node[below=4pt] {$\textnormal{W48}$};
	\draw (8.0,-.2) -- (8.0,0) node[below=4pt] {$\textnormal{W60}$};
	\draw (9.5,-.2) -- (9.5,0) node[below=4pt] {$\textnormal{W72}$};
	\draw (11.0,-.2) -- (11.0,0) node[below=4pt] {$\textnormal{W84}$};
	\draw (12.5,-.2) -- (12.5,0) node[below=4pt] {$\textnormal{W96}$};
	\draw (14.0,-.2) -- (14.0,0) node[below=4pt] {$\textnormal{...}$};
	\draw (-.1,1.0) -- (.1,1.0) node[left=4pt] {$\scriptstyle 3.0$};
	\draw (-.1,1.5) -- (.1,1.5) node[left=4pt] {$\scriptstyle 3.5$};
	\draw (-.1,2.0) -- (.1,2.0) node[left=4pt] {$\scriptstyle 4.0$};
	\draw (-.1,2.5) -- (.1,2.5) node[left=4pt] {$\scriptstyle 4.5$};
	\draw (-.1,3.0) -- (.1,3.0) node[left=4pt] {$\scriptstyle 5.0$};
	\draw (-.1,3.5) -- (.1,3.5) node[left=4pt] {$\scriptstyle 5.5$};
	\draw (-.1,4.0) -- (.1,4.0) node[left=4pt] {$\scriptstyle 6.0$};
	\draw (-.1,4.5) -- (.1,4.5) node[left=4pt] {$\scriptstyle 6.5$};
	\draw (-.1,5.0) -- (.1,5.0) node[left=4pt] {$\scriptstyle 7.0$};
  \node[outer sep=0pt,circle, fill=black,inner sep=1.5pt] (E0) at (0.5, 3.0) {};
  \node[outer sep=0pt,circle, fill=black,inner sep=1.5pt] (E1) at (2.0, 2.0) {};
  \node[outer sep=0pt,circle, fill=black,inner sep=1.5pt] (E2) at (3.5, 2.0) {};
  \node[outer sep=0pt,circle, fill=black,inner sep=1.5pt] (E3) at (5.0, 2.0) {};
  \node[outer sep=0pt,circle, fill=black,inner sep=1.5pt] (E4) at (6.5, 2.0) {};
  \node[outer sep=0pt,circle, fill=black,inner sep=1.5pt] (E5) at (8.0, 3.0) {};
  \coordinate[label=right: $\textnormal{IDP}_{1}$]() at (8.0,2.8);
  \coordinate[label=right: ${= \textnormal{CDP}_{1}}$]() at (7.6,3.3);
  \node[outer sep=0pt,circle, fill=black,inner sep=1.5pt] (E6) at (9.5, 3.0) {};
  \coordinate[label=right: $\textnormal{C}_{1}$]() at (9.5,2.8);
  \node[outer sep=0pt,circle, fill=black,inner sep=1.5pt] (E9) at (11.0, 3.5) {};
  \node[outer sep=0pt,circle, fill=black,inner sep=1.5pt] (E8) at (12.5, 3.5) {};
  \node[outer sep=0pt,circle, fill=black,inner sep=1.5pt] (E9) at (14.0, 3.5) {};
  \draw [color=black, dashed, line width=0.25mm] plot [dashed] coordinates
	{(0.5,3.0) (2.0,2.0) (3.5, 2.0) (5.0, 2.0) (6.5, 2.0) (8.0, 3.0) (9.5, 3.0) (11.0, 3.5) (12.5, 3.5) (14.0, 3.5)} node[right]{};
	\draw[dotted, color=black](0.5,3.0) -- (3.5,3.0);
	\coordinate[label=right: $\textit{Ref}_{1,1}$]() at (3.5,3.0);
	\draw[dotted, color=black](3.5,2.0) -- (9.5,2.0);
	\coordinate[label=right: $\textit{Ref}_{1,2}$]() at (9.5,2.0);
	\draw[dotted, color=black](8.0,3.0) -- (14.0,3.0);
	\coordinate[label=right: $\textit{Ref}_{2}$]() at (14.0,3.0);
	\draw[->, dotted](3.5,6.0) -- (3.5,5.0);
	\coordinate[label=above: $\textit{readjustment of reference}$]() at (3.5,6.0);
	\draw[->, dotted](9.5,6.0) -- (9.5,5.0);
	\coordinate[label=above: $\textit{readjustment}$]() at (9.5,6.0);
	\draw[decorate, yshift=7ex, color=darkpowderblue, thick] (0.5,2.5) -- node[above=0.4ex] {$\boldsymbol{=T_{1}}$} (8.0,2.5);
\end{tikzpicture}
}
}
\caption[Fixed versus roving reference system based on time-to-onset-of-CDP analyses]{Fixed versus roving reference system based on time-to-onset-of-CDP analyses ($\text{IDP}_{j}$ = $j^{th}$ initial disability progression, $\text{C}_{j}$ = confirmation of $\text{IDP}_{j}$ , $\text{CDP}_{j}$ = $j^{th}$ confirmed disability progression, $\text{Ref}_{j}$ = original reference EDSS score for $j^{th}$ CDP, $\text{Ref}_{j,k}$ = $k^{th}$ readjusted reference EDSS score for $j^{th}$ CDP, $T_{j}$ = time to onset of the $j^{th}$ CDP)}
\label{FigRovingFixed}
\end{figure}
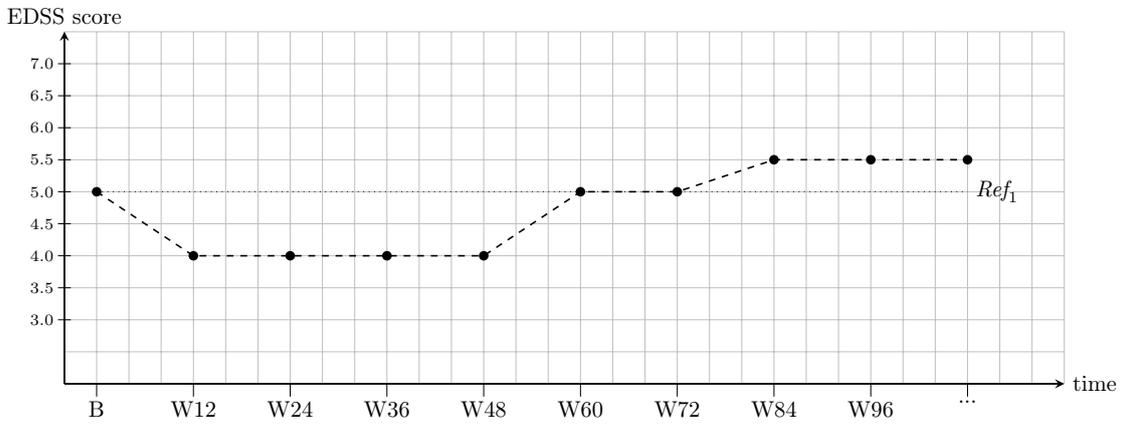
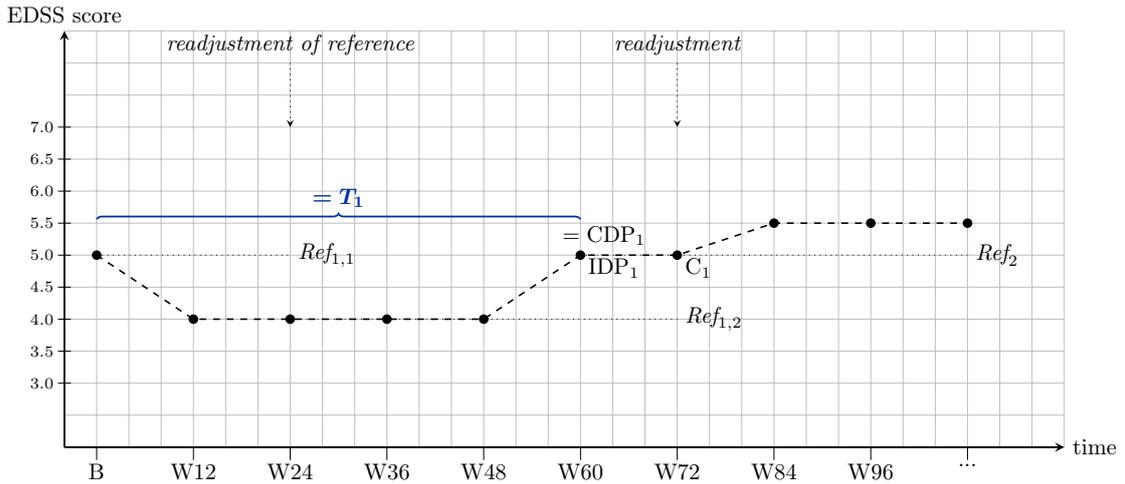
\newpage 
In the fixed approach $(a)$, the baseline EDSS value $5.0$ is chosen as fixed reference score for a potential progression event so that an EDSS score of at least $6.0$ at two consecutive study visits separated in time by at least $12$ weeks is required to observe disability progression. Following this definition, the progression event assumed to exist for this hypothetical patient would not be registered. In the roving system approach $(b)$, the initial reference value $5.0$ is replaced by a new score of $4.0$ after $24$ weeks because this new score fulfills both conditions required for resetting the original reference score. From week $24$ onwards, disability progression is referred to as achieving an EDSS level of at least $5.0$ followed by a $12$-week confirmation period (CDP12). A progression event would then be observed at the week $60$ study visit. In summary, disability progression would be captured using a roving system but would not be accounted for using a fixed reference system. \newline
For PPMS patients, under the no misclassification of EDSS condition, the roving reference approach would be almost the same as the fixed approach, considering that reference also needs to be confirmed to account for variability in EDSS. 

\subsection{Limitations}
Regardless of whether a fixed or roving reference system is used, there are two shortcomings of the time-to-onset-of-CDP endpoint. As already described previously, confirmed disability progression in MS patients is characterized by IDP and confirmation of IDP. Confirmation of disability progression is essentially required to robustify the endpoint against variability in EDSS assessments and, thus, it reduces the probability of capturing progression events that may subsequently revert. Only if disability progression can be confirmed, CDP is justified. Although a CDP event is not completely approved until the time of confirmation, it is reasonable, from a clinical perspective, to define the event time as time to the onset of disability progression. However, from a statistical point of view, this definition induces the so-called look-ahead bias, as a CDP12 event at time $t$ does not only rely on information prior to time $t$ (= past) but also on EDSS assessments behind time $t$ (= future) because of confirmation. \newline
Further, it is important to realize that the derivation process for the $j^{th}$ CDP12 event overlaps with the process for the $(j-1)^{th}$ CDP12 event, inducing dependency between the processes. For instance, in Figure $\ref{FigDefinition}$ $(b)$, the first CDP12 event is derived based on EDSS assessments from baseline to week $36$ and the second CDP12 event relies on EDSS information from week $24$ to week $72$, with an overlapping period $[\textnormal{week} \ 24, \textnormal{week} \ 36]$. This bias may be accounted for by robust variance estimation in some recurrent event methods. \newline
Despite these limitations, time to the onset of confirmed disability progression is the most widely used and accepted outcome measure in clinical MS trials. 
\section{Time to confirmation of CDP}
\label{DefCDPConfir}
In order to avoid the limitations of time-to-onset-of-CDP analyses and to evaluate the look-ahead bias in simulation studies, a slightly modified endpoint definition based on the time to confirmation of CDP is proposed. Both the $j^{th}$ disability progression and the $j^{th}$ reference EDSS score for recurrent events are specified as before in Section $\ref{DefStandardRecurrentCDP}$. The time to the confirmation of the $j^{th}$ 12-week CDP is defined as the time from baseline to the confirmation of the $j^{th}$ disability progression. \newline 
The main difference between the time to onset of CDP and the time to confirmation of CDP is given in Table $\ref{DiffOnsetConfir}$. While for the former approach the event date is the date of IDP (if confirmed), the event date in the latter approach reflects the date of confirmation of IDP. 

\begin{table}[H]
\centering
\scalebox{0.90}{ 
\begin{tabular}{l|cc}
\textbf{Endpoint}                    & \textbf{Starting date} & \textbf{Event date}        \\ \hline
\textbf{Time to onset of CDP}        & date of randomization  & date of IDP,  if confirmed \\
\textbf{Time to confirmation of CDP} & date of randomization  & date of confirmation      
\end{tabular}}
\caption{Difference between time-to-event endpoints}
\label{DiffOnsetConfir}
\end{table}

\newpage

\newpage
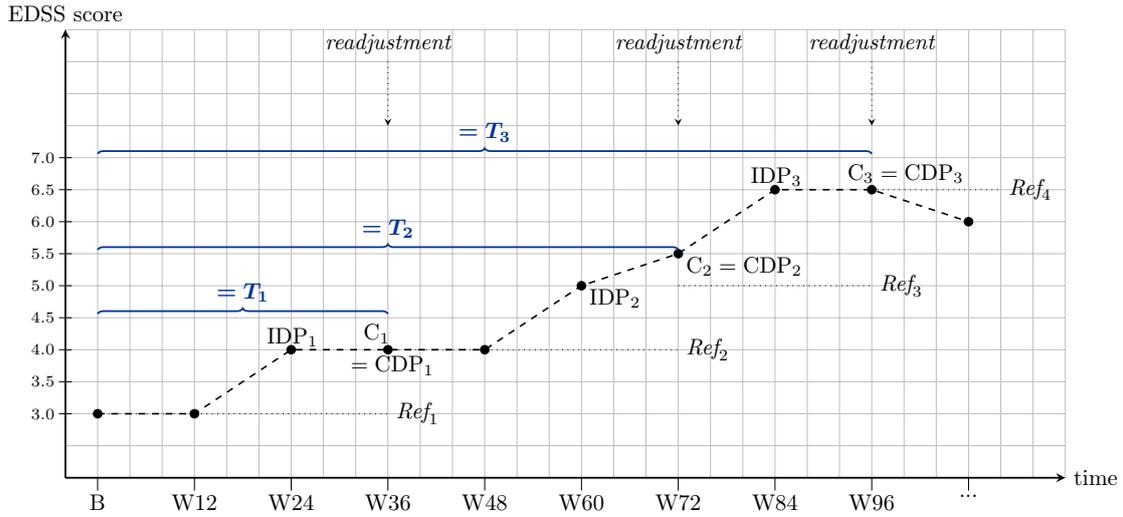
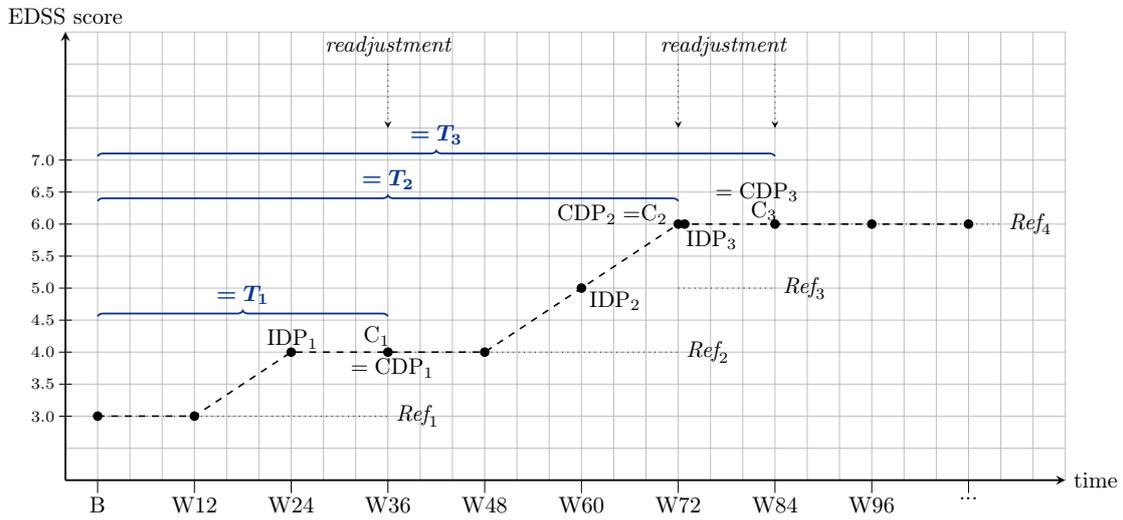
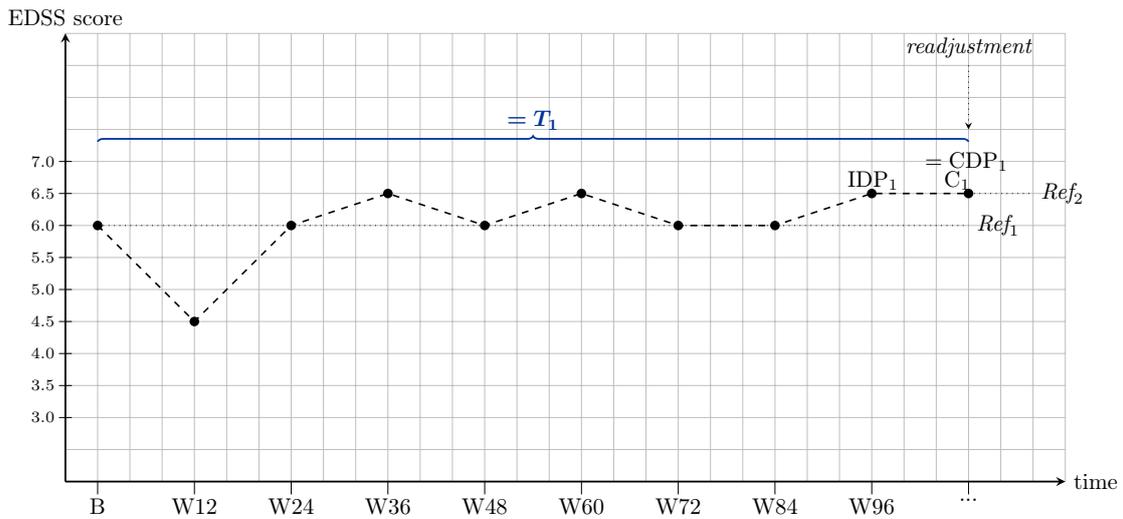
\begin{figure}[H]
\centering
\subfloat[Example 1]{ 
\resizebox{\textwidth}{!}{
\begin{tikzpicture}[decoration=brace]
	\draw [color=gray!50]  [step=5mm] (0,0) grid (15.5,7.0);
	\draw[->,thick] (0,0) -- (15.5,0) node[right] {$\textnormal{time}$};
	\draw[->,thick] (0,0) -- (0,7.0) node[above] {$\textnormal{EDSS score}$};
	\draw (0.5,-.2) -- (0.5,0) node[below=4pt] {$\textnormal{B}$};
	\draw (2.0,-.2) -- (2.0,0) node[below=4pt] {$\textnormal{W12}$};
	\draw (3.5,-.2) -- (3.5,0) node[below=4pt] {$\textnormal{W24}$};
	\draw (5.0,-.2) -- (5.0,0) node[below=4pt] {$\textnormal{W36}$};
	\draw (6.5,-.2) -- (6.5,0) node[below=4pt] {$\textnormal{W48}$};
	\draw (8.0,-.2) -- (8.0,0) node[below=4pt] {$\textnormal{W60}$};
	\draw (9.5,-.2) -- (9.5,0) node[below=4pt] {$\textnormal{W72}$};
	\draw (11.0,-.2) -- (11.0,0) node[below=4pt] {$\textnormal{W84}$};
	\draw (12.5,-.2) -- (12.5,0) node[below=4pt] {$\textnormal{W96}$};
	\draw (14.0,-.2) -- (14.0,0) node[below=4pt] {$\textnormal{...}$};
	\draw (-.1,1.0) -- (.1,1.0) node[left=4pt] {$\scriptstyle 3.0$};
	\draw (-.1,1.5) -- (.1,1.5) node[left=4pt] {$\scriptstyle 3.5$};
	\draw (-.1,2.0) -- (.1,2.0) node[left=4pt] {$\scriptstyle 4.0$};
	\draw (-.1,2.5) -- (.1,2.5) node[left=4pt] {$\scriptstyle 4.5$};
	\draw (-.1,3.0) -- (.1,3.0) node[left=4pt] {$\scriptstyle 5.0$};
	\draw (-.1,3.5) -- (.1,3.5) node[left=4pt] {$\scriptstyle 5.5$};
	\draw (-.1,4.0) -- (.1,4.0) node[left=4pt] {$\scriptstyle 6.0$};
	\draw (-.1,4.5) -- (.1,4.5) node[left=4pt] {$\scriptstyle 6.5$};
	\draw (-.1,5.0) -- (.1,5.0) node[left=4pt] {$\scriptstyle 7.0$};
  \node[outer sep=0pt,circle, fill=black,inner sep=1.5pt] (E0) at (0.5, 1.0) {};
  \node[outer sep=0pt,circle, fill=black,inner sep=1.5pt] (E1) at (2.0, 1.0) {};
  \node[outer sep=0pt,circle, fill=black,inner sep=1.5pt] (E2) at (3.5, 2.0) {};
  \coordinate[label=right: $\textnormal{IDP}_{1}$]() at (3.0,2.2);
  \node[outer sep=0pt,circle, fill=black,inner sep=1.5pt] (E3) at (5.0, 2.0) {};
  \coordinate[label=right: ${= \textnormal{CDP}_{1}}$]() at (4.3,1.75);
  \coordinate[label=right: $\textnormal{C}_{1}$]() at (4.5,2.25);
  \node[outer sep=0pt,circle, fill=black,inner sep=1.5pt] (E4) at (6.5, 2.0) {};
  \node[outer sep=0pt,circle, fill=black,inner sep=1.5pt] (E5) at (8.0, 3.0) {};
  \coordinate[label=right: $\textnormal{IDP}_{2}$]() at (8.0,2.8);
  \node[outer sep=0pt,circle, fill=black,inner sep=1.5pt] (E6) at (9.5, 3.5) {};
  \coordinate[label=right: ${\textnormal{C}_{2} = \textnormal{CDP}_{2}}$]() at (9.5,3.3);
  \node[outer sep=0pt,circle, fill=black,inner sep=1.5pt] (E7) at (11.0, 4.5) {};
  \coordinate[label=right: $\textnormal{IDP}_{3}$]() at (10.5,4.7);
  \node[outer sep=0pt,circle, fill=black,inner sep=1.5pt] (E8) at (12.5, 4.5) {};
  \coordinate[label=right: ${\textnormal{C}_{3} = \textnormal{CDP}_{3}}$]() at (12.0,4.75);
  \node[outer sep=0pt,circle, fill=black,inner sep=1.5pt] (E9) at (14.0, 4.0) {};
  \draw [color=black, dashed, line width=0.25mm] plot [dashed] coordinates
	{(0.5,1.0) (2.0,1.0) (3.5, 2.0) (5.0, 2.0) (6.5, 2.0) (8.0, 3.0) (9.5, 3.5) (11.0, 4.5) (12.5, 4.5) (14.0, 4.0)} node[right]{};
	\draw[dotted, color=black](0.5,1.0) -- (5.0,1.0);
	\coordinate[label=right: $\textit{Ref}_{1}$]() at (5.0,1.0);
	\draw[dotted, color=black](5.0,2.0) -- (9.5,2.0);
	\coordinate[label=right: $\textit{Ref}_{2}$]() at (9.5,2.0);
	\coordinate[label=right: $\textit{Ref}_{3}$]() at (12.5,3.0);
	\draw[dotted, color=black](9.5,3.0) -- (12.5,3.0);
	\draw[dotted, color=black](12.5,4.5) -- (14.5,4.5);
	\coordinate[label=right: $\textit{Ref}_{4}$]() at (14.5,4.5);
	\draw[->, dotted](5.0,6.5) -- (5.0,5.5);
	\coordinate[label=above: $\textit{readjustment}$]() at (5.0,6.5);
	\draw[->, dotted](9.5,6.5) -- (9.5,5.5);
	\coordinate[label=above: $\textit{readjustment}$]() at (9.5,6.5);
	\draw[->, dotted](12.5,6.5) -- (12.5,5.5);
	\coordinate[label=above: $\textit{readjustment}$]() at (12.5,6.5);
	\draw[decorate, yshift=7ex, color=darkpowderblue, thick] (0.5,1.5) -- node[above=0.4ex] {$\boldsymbol{=T_{1}}$} (5.0,1.5);
	\draw[decorate, yshift=7ex, color=darkpowderblue, thick] (0.5,2.5) -- node[above=0.4ex] {$\boldsymbol{=T_{2}}$} (9.5,2.5);
	\draw[decorate, yshift=7ex, color=darkpowderblue, thick] (0.5,4.0) -- node[above=0.4ex] {$\boldsymbol{=T_{3}}$} (12.5,4.0);
\end{tikzpicture}
}
}

\subfloat[Example 2]{ 
\resizebox{\textwidth}{!}{
\begin{tikzpicture}[decoration=brace]
	\draw [color=gray!50]  [step=5mm] (0,0) grid (15.5,7.0);
	\draw[->,thick] (0,0) -- (15.5,0) node[right] {$\textnormal{time}$};
	\draw[->,thick] (0,0) -- (0,7.0) node[above] {$\textnormal{EDSS score}$};
	\draw (0.5,-.2) -- (0.5,0) node[below=4pt] {$\textnormal{B}$};
	\draw (2.0,-.2) -- (2.0,0) node[below=4pt] {$\textnormal{W12}$};
	\draw (3.5,-.2) -- (3.5,0) node[below=4pt] {$\textnormal{W24}$};
	\draw (5.0,-.2) -- (5.0,0) node[below=4pt] {$\textnormal{W36}$};
	\draw (6.5,-.2) -- (6.5,0) node[below=4pt] {$\textnormal{W48}$};
	\draw (8.0,-.2) -- (8.0,0) node[below=4pt] {$\textnormal{W60}$};
	\draw (9.5,-.2) -- (9.5,0) node[below=4pt] {$\textnormal{W72}$};
	\draw (11.0,-.2) -- (11.0,0) node[below=4pt] {$\textnormal{W84}$};
	\draw (12.5,-.2) -- (12.5,0) node[below=4pt] {$\textnormal{W96}$};
	\draw (14.0,-.2) -- (14.0,0) node[below=4pt] {$\textnormal{...}$};
	\draw (-.1,1.0) -- (.1,1.0) node[left=4pt] {$\scriptstyle 3.0$};
	\draw (-.1,1.5) -- (.1,1.5) node[left=4pt] {$\scriptstyle 3.5$};
	\draw (-.1,2.0) -- (.1,2.0) node[left=4pt] {$\scriptstyle 4.0$};
	\draw (-.1,2.5) -- (.1,2.5) node[left=4pt] {$\scriptstyle 4.5$};
	\draw (-.1,3.0) -- (.1,3.0) node[left=4pt] {$\scriptstyle 5.0$};
	\draw (-.1,3.5) -- (.1,3.5) node[left=4pt] {$\scriptstyle 5.5$};
	\draw (-.1,4.0) -- (.1,4.0) node[left=4pt] {$\scriptstyle 6.0$};
	\draw (-.1,4.5) -- (.1,4.5) node[left=4pt] {$\scriptstyle 6.5$};
	\draw (-.1,5.0) -- (.1,5.0) node[left=4pt] {$\scriptstyle 7.0$};
  \node[outer sep=0pt,circle, fill=black,inner sep=1.5pt] (E0) at (0.5, 1.0) {};
  \node[outer sep=0pt,circle, fill=black,inner sep=1.5pt] (E1) at (2.0, 1.0) {};
  \node[outer sep=0pt,circle, fill=black,inner sep=1.5pt] (E2) at (3.5, 2.0) {};
  \coordinate[label=right: $\textnormal{IDP}_{1}$]() at (3.0,2.2);
  \node[outer sep=0pt,circle, fill=black,inner sep=1.5pt] (E3) at (5.0, 2.0) {};
  \coordinate[label=right: ${= \textnormal{CDP}_{1}}$]() at (4.3,1.75);
  \coordinate[label=right: $\textnormal{C}_{1}$]() at (4.5,2.25);
  \node[outer sep=0pt,circle, fill=black,inner sep=1.5pt] (E4) at (6.5, 2.0) {};
  \node[outer sep=0pt,circle, fill=black,inner sep=1.5pt] (E5) at (8.0, 3.0) {};
  \coordinate[label=right: $\textnormal{IDP}_{2}$]() at (8.0,2.8);
  \node[outer sep=0pt,circle, fill=black,inner sep=1.5pt] (E6) at (9.5, 4.0) {};
  \coordinate[label=right: ${\textnormal{C}_{2}}$]() at (8.8,4.15);
  \coordinate[label=right: ${\textnormal{CDP}_{2} = }$]() at (7.5,4.15);
  \node[outer sep=0pt,circle, fill=black,inner sep=1.5pt] (E7) at (9.6, 4.0) {};
  \coordinate[label=right: $\textnormal{IDP}_{3}$]() at (9.5,3.75);
  \node[outer sep=0pt,circle, fill=black,inner sep=1.5pt] (E9) at (11.0, 4.0) {};
  \coordinate[label=right: $\textnormal{C}_{3}$]() at (10.5,4.2);
  \coordinate[label=right: ${=\textnormal{CDP}_{3}}$]() at (9.95,4.5);
  \node[outer sep=0pt,circle, fill=black,inner sep=1.5pt] (E8) at (12.5, 4.0) {};
  \node[outer sep=0pt,circle, fill=black,inner sep=1.5pt] (E9) at (14.0, 4.0) {};
  \draw [color=black, dashed, line width=0.25mm] plot [dashed] coordinates
	{(0.5,1.0) (2.0,1.0) (3.5, 2.0) (5.0, 2.0) (6.5, 2.0) (8.0, 3.0) (9.5, 4.0) (11.0, 4.0) (12.5, 4.0) (14.0, 4.0)} node[right]{};
	\draw[dotted, color=black](0.5,1.0) -- (5.0,1.0);
	\coordinate[label=right: $\textit{Ref}_{1}$]() at (5.0,1.0);
	\draw[dotted, color=black](5.0,2.0) -- (9.5,2.0);
	\coordinate[label=right: $\textit{Ref}_{2}$]() at (9.5,2.0);
	\coordinate[label=right: $\textit{Ref}_{3}$]() at (11.0,3.0);
	\draw[dotted, color=black](9.5,3.0) -- (11.0,3.0);
	\draw[dotted, color=black](11.0, 4.0) -- (14.5,4.0);
	\coordinate[label=right: $\textit{Ref}_{4}$]() at (14.5,4.0);
	\draw[->, dotted](5.0,6.5) -- (5.0,5.5);
	\coordinate[label=above: $\textit{readjustment}$]() at (5.0,6.5);
	\draw[->, dotted](9.5,6.5) -- (9.5,5.5);
	\coordinate[label=above: $\textit{readjustment}$]() at (10.2,6.5);
	\draw[->, dotted](11.0,6.5) -- (11.0,5.5);
	\draw[decorate, yshift=7ex, color=darkpowderblue, thick] (0.5,1.5) -- node[above=0.4ex] {$\boldsymbol{=T_{1}}$} (5.0,1.5);
	\draw[decorate, yshift=7ex, color=darkpowderblue, thick] (0.5,3.3) -- node[above=0.4ex] {$\boldsymbol{=T_{2}}$} (9.5,3.3);
	\draw[decorate, yshift=7ex, color=darkpowderblue, thick] (0.5,4.0) -- node[above=0.4ex] {$\boldsymbol{=T_{3}}$} (11,4.0);
\end{tikzpicture}
}
}

\subfloat[Example 3]{ 
\resizebox{\textwidth}{!}{
\begin{tikzpicture}[decoration=brace]
	\draw [color=gray!50]  [step=5mm] (0,0) grid (15.5,7.0);
	\draw[->,thick] (0,0) -- (15.5,0) node[right] {$\textnormal{time}$};
	\draw[->,thick] (0,0) -- (0,7.0) node[above] {$\textnormal{EDSS score}$};
	\draw (0.5,-.2) -- (0.5,0) node[below=4pt] {$\textnormal{B}$};
	\draw (2.0,-.2) -- (2.0,0) node[below=4pt] {$\textnormal{W12}$};
	\draw (3.5,-.2) -- (3.5,0) node[below=4pt] {$\textnormal{W24}$};
	\draw (5.0,-.2) -- (5.0,0) node[below=4pt] {$\textnormal{W36}$};
	\draw (6.5,-.2) -- (6.5,0) node[below=4pt] {$\textnormal{W48}$};
	\draw (8.0,-.2) -- (8.0,0) node[below=4pt] {$\textnormal{W60}$};
	\draw (9.5,-.2) -- (9.5,0) node[below=4pt] {$\textnormal{W72}$};
	\draw (11.0,-.2) -- (11.0,0) node[below=4pt] {$\textnormal{W84}$};
	\draw (12.5,-.2) -- (12.5,0) node[below=4pt] {$\textnormal{W96}$};
	\draw (14.0,-.2) -- (14.0,0) node[below=4pt] {$\textnormal{...}$};
	\draw (-.1,1.0) -- (.1,1.0) node[left=4pt] {$\scriptstyle 3.0$};
	\draw (-.1,1.5) -- (.1,1.5) node[left=4pt] {$\scriptstyle 3.5$};
	\draw (-.1,2.0) -- (.1,2.0) node[left=4pt] {$\scriptstyle 4.0$};
	\draw (-.1,2.5) -- (.1,2.5) node[left=4pt] {$\scriptstyle 4.5$};
	\draw (-.1,3.0) -- (.1,3.0) node[left=4pt] {$\scriptstyle 5.0$};
	\draw (-.1,3.5) -- (.1,3.5) node[left=4pt] {$\scriptstyle 5.5$};
	\draw (-.1,4.0) -- (.1,4.0) node[left=4pt] {$\scriptstyle 6.0$};
	\draw (-.1,4.5) -- (.1,4.5) node[left=4pt] {$\scriptstyle 6.5$};
	\draw (-.1,5.0) -- (.1,5.0) node[left=4pt] {$\scriptstyle 7.0$};
  \node[outer sep=0pt,circle, fill=black,inner sep=1.5pt] (E0) at (0.5, 4.0) {};
  \node[outer sep=0pt,circle, fill=black,inner sep=1.5pt] (E1) at (2.0, 2.5) {};
  \node[outer sep=0pt,circle, fill=black,inner sep=1.5pt] (E2) at (3.5, 4.0) {};
  \node[outer sep=0pt,circle, fill=black,inner sep=1.5pt] (E3) at (5.0, 4.5) {};
  \node[outer sep=0pt,circle, fill=black,inner sep=1.5pt] (E4) at (6.5, 4.0) {};
  \node[outer sep=0pt,circle, fill=black,inner sep=1.5pt] (E5) at (8.0, 4.5) {};
  \node[outer sep=0pt,circle, fill=black,inner sep=1.5pt] (E6) at (9.5, 4.0) {};
  \node[outer sep=0pt,circle, fill=black,inner sep=1.5pt] (E9) at (11.0, 4.0) {};
  \node[outer sep=0pt,circle, fill=black,inner sep=1.5pt] (E8) at (12.5, 4.5) {};
  \coordinate[label=right: $\textnormal{IDP}_{1}$]() at (12.0,4.7);
  \node[outer sep=0pt,circle, fill=black,inner sep=1.5pt] (E9) at (14.0, 4.5) {};
  \coordinate[label=right: ${= \textnormal{CDP}_{1}}$]() at (13.2,5.0);
  \coordinate[label=right: $\textnormal{C}_{1}$]() at (13.5,4.7);
  \draw [color=black, dashed, line width=0.25mm] plot [dashed] coordinates
	{(0.5, 4.0) (2.0,2.5) (3.5, 4.0) (5.0, 4.5) (6.5, 4.0) (8.0, 4.5) (9.5, 4.0) (11.0, 4.0) (12.5, 4.5) (14.0, 4.5)} node[right]{};
	\draw[dotted, color=black](0.5,4.0) -- (14.0,4.0);
	\coordinate[label=right: $\textit{Ref}_{1}$]() at (14.0,4.0);
	\draw[dotted, color=black](14.0,4.5) -- (15.0,4.5);
	\coordinate[label=right: $\textit{Ref}_{2}$]() at (15.0,4.5);
	\draw[->, dotted](14.0,6.5) -- (14.0,5.5);
	\coordinate[label=above: $\textit{readjustment}$]() at (14.0,6.5);
	\draw[decorate, yshift=7ex, color=darkpowderblue, thick] (0.5,4.25) -- node[above=0.4ex] {$\boldsymbol{=T_{1}}$} (14.0,4.25);
\end{tikzpicture}
}
}
\caption[Derivation of time-to-confirmation-of-CDP12 endpoint from EDSS measurements]{Derivation of time-to-confirmation-of-CDP12 endpoint from EDSS measurements ($\text{IDP}_{j}$ = $j^{th}$ initial disability progression, $\text{C}_{j}$ = confirmation of $\text{IDP}_{j}$, $\text{CDP}_{j}$ = $j^{th}$ confirmed disability progression (event), $\text{Ref}_{j}$ = reference EDSS score for $j^{th}$ CDP, $T_{j}$ = time to confirmation of the $j^{th}$ CDP)}
\label{FigDefinition1}
\end{figure}

Figure $\ref{FigDefinition1}$ depicts the same hypothetical EDSS profiles as in Figure $\ref{FigDefinition}$ but illustrates the derivation of repeated CDP12 events based on the time-to-confirmation definition. When comparing Figure $\ref{FigDefinition}$ $(a)$ with Figure $\ref{FigDefinition1}$ $(a)$, it can be seen that the times to confirmation of CDP are longer for at least $12$ weeks. Example $(b)$ demonstrates the case in which a patient experiences an IDP shortly after a recorded confirmed disability progression. For this patient, a second CDP12 event is recorded at week 72 with corresponding IDP at week $60$, followed by readjustment of the  reference EDSS score at (week $72)+$. As mentioned before, the reference EDSS score for a new progression event is equal to the EDSS score associated with the IDP from the previous event, which implies that, from week $72$ onwards, progression events are derived based on a reference EDSS score of $5.0$. Consequently, an increase in EDSS score of $1.0$ point from a reference score $5.0$ is observed at (week $72)+$, leading to an IDP at week $72$ and a third CDP $12$ weeks later. In particular, example $(b)$ clearly shows that the derivation process for the $j^{th}$ CDP12 event only uses information from the past (e.g., EDSS score at IDP) and does not start before the $(j-1)^{th}$ event process has been completed $\Longrightarrow$ non-overlapping property and absence of look-head bias. 

\section{Overview of CDP definitions}
\label{DefOverview}
In total, repeated CDP12 events can be derived using different combinations of the following criteria: 
\renewcommand{\labelenumi}{\roman{enumi})}
\begin{enumerate}
\item Readjustment of reference disability level after each event: reference $=$ baseline EDSS score (only for first event) or reference $=$ EDSS score associated with IDP of previous event
\item Reference system: fixed or roving reference value 
\item Magnitude of EDSS change: increase in EDSS by $\geq 1$ point if reference EDSS is $\leq 5.5$, or increase in EDSS by $\geq 0.5$ point if
reference EDSS is $>5.5$ 
\item Confirmation of disability progression at two or more consecutive study visits separated in time by a minimum of $12$ weeks
\item Time-to-event endpoint: time to onset of CDP or time to confirmation of CDP. 
\end{enumerate}

\chapter{Time-to-first-event methods}
In time-to-first-event analyses, not all individuals under study experience the event of interest by the end of the observation period so that the actual event times for some individuals are unknown. Due to those incomplete (censored) observations, time-to-first-event analyses require special statistical techniques based on hazards. This chapter briefly outlines the basic concepts of appropriate statistical methods for analyzing time-to-first-event data, including the Nelson Aalen estimator, Kaplan Meier estimator, log-rank test and the Cox proportional hazards model. \newline 
In general, time-to-first-event processes can at best be modelled through counting process theory and intensity functions \parencite{Andersen1993, Aalen2008, Beyersmann2012}. A detailed discussion of recurrent event methods within the counting process framework is given in Chapter $4$, with the time-to-first-event setting included as a special case. For this reason, this chapter only summarizes the main ideas of time-to-first-event methods, leaving out mathematical details. In Section $\ref{Chapter31}$, main characteristics of time-to-first-event data are briefly described. While Section $\ref{Chapter32}$ introduces the Kaplan Meier and Nelson Aalen estimators, Section $\ref{Chapter33}$ focuses on the Cox proportional hazards model. 

\section{Characteristics of time-to-first-event data}
\label{Chapter31}
In time-to-first-event analyses, individuals are followed from time origin until the first occurrence of an event of interest or until the end of study. For instance, time origin can be birth, diagnosis of disease, randomization in a RCT, start of a specific intervention or admission to hospital. Examples for the event of interest may be a patient's death, relapse, progression or disease onset. With regard to MS disease, clinical trials in PPMS or RRMS patients focus on the time from randomization to the occurrence of the first confirmed disability progression. Individuals who have not experienced an event during follow-up are said to be right-censored, as only a minimum event time can be observed. 

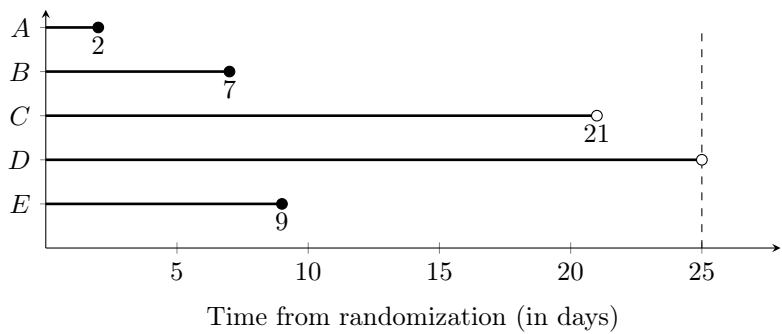
\begin{figure}[H]
\centering
\begin{tikzpicture}
    \begin{axis}[height=0.2\textheight, width=0.75\textwidth, xmin=0.0, xmax=28, ymin=0.0, ymax=2.7, xtick={5, 10, 15, 20, 25}, xticklabels={5, 10, 15, 20, 25}, ytick={0.5, 1, 1.5, 2.0, 2.5},
                 yticklabels={$E$, $D$, $C$, $B$, $A$}, xlabel={Time from randomization (in days)}, x label style={at={(axis description cs:0.5,-0.2)},anchor=north}]
        \addplot[cmhplot,-,domain=0:21]{1.5};
        \addplot[cmhplot,-,domain=0:7]{2.0};
        \addplot[cmhplot,-,domain=0:2]{2.5};
        \addplot[cmhplot,-,domain=0:25]{1.0};
        \addplot[cmhplot,-,domain=0:9]{0.5};
        \addplot[holdot]coordinates{(21,1.5)};
        \addplot[soldot]coordinates{(7,2.0)};
        \addplot[soldot]coordinates{(2,2.5)};
        \addplot[holdot]coordinates{(25,1.0)};
        \addplot[soldot]coordinates{(9,0.5)};
        \addplot[dashed]coordinates {(25,0) (25,2.5)};
        \node [below] at (axis cs: 21, 1.5) {21};
        \node [below] at (axis cs: 7, 2.0) {7};
        \node [below] at (axis cs: 9, 0.5) {9};
        \node [below] at (axis cs: 2, 2.5) {2};
\end{axis}
\end{tikzpicture}
\caption[Illustration of time-to-first-event data]{Illustration of time-to-first-event data ($\bullet$ = event, $\circ$ = censoring, dashed vertical line is end of follow-up)}
\label{GraphicTTFE}
\end{figure}

\newpage
Figure $\ref{GraphicTTFE}$ illustrates time-to-first-event data on $5$ individuals from a hypothetical study, in which each individual is observed over a fixed time period of $25$ days to assess whether a specific event occurs or not. Observations are given in study time scale. While individuals $A$, $B$ and $E$ experience an event at day $2$, $7$ and $9$, individuals $C$ and $D$ have not experienced an event by study closure, so their observations are right-censored. Specifically, for individual $D$, censoring is caused by the administrative end of the observation period, whereas individual $E$ is lost to follow-up due to reasons unrelated to the event process. 

\section{Multistate model for the first event setting} 
\label{Chapter32}
Figure $\ref{MultiStateModelSurvival}$ depicts the simplest multistate model reflecting time-to-first-event analyses, with only two states $0$ and $1$ \parencite{Beyersmann2012}. In Figure $\ref{MultiStateModelSurvival}$, states of the multistate model are represented by boxes and transitions between the states are shown by arrows. State $0$ may be interpreted as being 'event-free' and state $1$ as having experienced an event.  Each individual enters the initial state $0$ at time origin $t=0$ and stays there until the occurrence of the event of interest. This means, at a random time $T$, individuals make transitions into the absorbing state $1$. State $1$ is said to be absorbing because transitions out ouf state $1$ do not exist and individuals can therefore not move out of this state. The failure or event time $T$ (= time to occurrence of the first event) is defined as the smallest time at which the multistate process is not in the initial state $0$ anymore. More specifically, the absolutely continuous random variable $T \in [0, \infty)$ can be formalized as
\begin{align}
\nonumber T := \inf \{ t \geq 0: \ (\textnormal{state at} \ t) \neq 0 \}.
\end{align}
The relationship between the multistate process and the event time $T$ is presented in Figure $\ref{RelationshipTMP}$, for an arbitrary individual with event time $T=t_{1}$. The individual is in state $0$ for all times $t \in [0, t_{1})$ and in state 1 for all times $t \geq t_{1}$. Since the state occupied at $T$ is $1$, the sample paths of the multistate process are right-continuous. 

\begin{figure}[H]
\begin{minipage}[t]{\textwidth}
\centering
\begin{minipage}[b]{0.48\textwidth}
\centering
\begin{tikzpicture}[node distance=1.9cm]
  \tikzset{node style/.style={state, fill=gray!20!white, rectangle}}
        \node[node style,]               (I)   {0};
        \node[node style, right=of I]   (II)  {1};
    \draw[>=latex,
          auto=left,
          every loop]
         (I)   edge node {$\tiny \alpha(t)$} (II); 
\end{tikzpicture}
\caption{Multistate representation of a time-to-first-event process}
\label{MultiStateModelSurvival}
\end{minipage}%
\hfill
\begin{minipage}[t]{0.48\textwidth}
\centering 
\begin{tikzpicture}
    \begin{axis}[height=0.2\textheight, width=0.75\textwidth, xmin=0, xmax=22, ymin=0.0, ymax=4.5, xtick={6}, xticklabels={$T=t_{1}$}, ytick={1, 2}, 
    yticklabels={0, 1}, xlabel={Time since study start}, ylabel={Multistate process}, y label style={at={(axis description cs:-0.1,.5)},rotate=90,anchor=south}, 
    x label style={at={(axis description cs:0.5,-0.2)},anchor=north}]
        \addplot[cmhplot,-,domain=0:6]{1};
        \addplot[cmhplot,-,domain=6:22]{2};
        \addplot[holdot]coordinates{(6,1)};
        \addplot[soldot]coordinates{(6,2)};
    \end{axis}
\end{tikzpicture}
\caption[Relationship between the multistate process and event time]{Multistate process and event time $T$  ($\bullet$ = included, $\circ$ = not included)}
\label{RelationshipTMP}
\end{minipage}
\end{minipage}
\end{figure}

The statistical analysis of $T$ relies on the hazard function $\alpha(t)$ defined as 
\begin{align}
\label{hazardTTE} \alpha(t) &:= \lim \limits_{\Delta t \rightarrow 0}{\dfrac{P(T \in [t, t + \Delta t) \ | \ T \geq t)}{\Delta t}}  \\
\nonumber \\
\nonumber \Longleftrightarrow: \ \ \alpha(t)dt &= P(T \in [t, t + dt) \ | \ T \geq t) \\
\nonumber &= P(0 \longrightarrow 1 \ \textnormal{transition between} \ t \ \textnormal{and} \ t + dt \ | \ \textnormal{state} \ 0 \ \textnormal{at} \ t-) \\
\nonumber &= P(\textnormal{state} \ 1 \ \textnormal{at} \ t \ | \ \textnormal{state} \ 0 \ \textnormal{at} \ t-)
\end{align}
The hazard rate $(\ref{hazardTTE})$ can be any non-negative function and specifies the conditional probability that an event is observed within the next very small time interval $[t, t+dt)$ given that the event has not happened before time $t$. The corresponding cumulative hazard $A(t)$ is defined via 
\begin{align}
\nonumber A(t) &= \int_{0}^{t} \alpha(u) du. 
\end{align}
As already explained above, time-to-first-event data is characterized by incomplete observations in the sense that for some individuals the event of interest has not happened during follow-up. In order to model right-censored time-to-first-event data, let $C \in [0, \infty)$ be a right-censoring time assumed to be independent of the event time $T$. This is often referred to as the random censorship model, i.e., $T \perp C$. Then, the observation is given by
\begin{align}
\nonumber \bigl( \min(T, C), \ \mathbbm{1}(T \leq C) \bigr)  &= \left\{ \begin{array}{ll} (T, \ 1), & T \leq C \ \ \ \ (observed) \\
         (C, \ 0), & T > C \ \ \ \ (right-censored) \\
         \end{array} \right.  . 
\end{align}
The event indicator $\delta := \mathbbm{1}(T \leq C) \in \{0 , 1 \}$ indicates whether $\min(T, C)$ equals the actual event time $T$ or the right-censoring time $C$. Under the random censorship model and the assumption that either $T$ or $C$ happens in $[t, t+dt)$, it can be shown that the hazard function is undisturbed by censoring \parencite{Beyersmann2012}. That is, 
\begin{align}
\nonumber \alpha(t)dt &= P(T \in [t, t + dt) \ | \ T \geq t) \\
\label{hazard1} &= P(T \in [t, t + dt), \ T \leq C \ | \ T \geq t, \ C \geq t). \\
\nonumber \\
\label{hazard2} \widehat{{\alpha(t)dt}} &= \dfrac{\textnormal{no. of individuals observed to have an event at} \ t}{\textnormal{no. of individuals at-risk at} \ t-}
\end{align}
Thus, the probability that an event occurs in $[t, t+dt)$ given both $T \geq t$ and $C \geq t$ is the same as in the absence of censoring. Individuals with $T \geq t$ and $C \geq t$ are considered to be at-risk for an event at time $t$. Now, Eq. $(\ref{hazard1})$ specifies the conditional probability that an observed event happens in the next very small time interval $[t, t+dt)$ given the fact that neither event nor censoring have happened before time $t$. As seen from Eq. $(\ref{hazard2})$, $\alpha(t)dt$ can be consequently estimated from censored time-to-first-event data. Informally, the nonparametric Nelson Aalen estimator of the cumulative hazard function $A(t)$ is then given by
\begin{align}
\label{NAestimator} \hat{A}(t) &= \sum_{u \leq t} \dfrac{\textnormal{no. of individuals observed to have an event at} \ u}{\textnormal{no. of individuals at-risk at} \ u-}, 
\end{align}
where the sum goes over all unique observed event times $u$, with $u \leq t$. Eq. $(\ref{NAestimator})$ implies further that $\hat{A}(t)$ is an increasing right-continuous step function with jumps at the observed event times $u$. \citet{Aalen2008} used counting process formulation to give a more formal derivation of the Nelson Aalen estimator and its statistical properties. \newline 
If $T$ denotes the time to the first event, the underlying survival function $S(t)$ with 
\begin{align}
\label{KMestimator} S(t) = P(\textnormal{state} \ 0 \ \textnormal{at} \ t) = P(T > t) = \exp \bigl(- \int_{0}^{t} \alpha(u) du \bigr) = \exp(- A(t))
\end{align} 
gives the unconditional probability that the event of interest has not happened by time $t$. The survival curve is a function that is equal to $1$ for $t=0$, i.e., $S(0)=1$, and declines over time. Moreover, the survival function given in Eq. $(\ref{KMestimator})$ can be estimated by the so-called Kaplan Meier (KM) estimator
\begin{align}
\nonumber \hat{S}(t) = \prod_{u \leq t} (1- \Delta \hat{A}(u)), 
\end{align}
where the product is over all unique event times $u$, $u \leq t$, and $\Delta \hat{A}(u)$ is the increment of the Nelson Aalen estimator $\hat{A}$ at time $u$. 

\newpage
\section{Cox proportional hazards model}
\label{Chapter33}
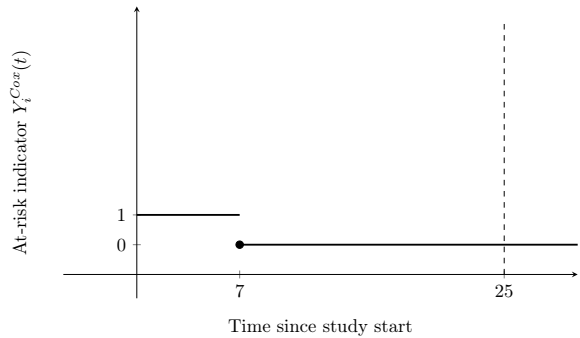
\begin{figure}[H]
\centering
\scalebox{0.7}{
\begin{tikzpicture}
    \begin{axis}[height=0.3\textheight, width=0.75\textwidth, xmin=-5.0, xmax=30, ymin=-0.8, ymax=9, xtick={7, 25}, xticklabels={7, 25}, ytick={1, 2}, yticklabels={0,1}, xlabel={Time since study start}, ylabel={At-risk indicator $Y_{i}^{Cox}(t)$},y label style={at={(axis description cs:-0.05,.5)},rotate=90,anchor=south},
    x label style={at={(axis description cs:0.5,-0.05)},anchor=north}]
        \addplot[cmhplot,-,domain=7:30]{1};
        \addplot[cmhplot,-,domain=0:7]{2};
        \addplot[soldot]coordinates{(7,1)};
        \addplot[dashed]coordinates {(25,0) (25,8.5)};
    \end{axis}
\end{tikzpicture}}
\caption[At-risk indicator under the Cox proportional hazards model]{At-risk indicator under the Cox model for individual $B$ from hypothetical example}
\label{AtRiskCox}
\end{figure}

The Cox proportional hazards model \parencite{Cox1972} is a regression model commonly used for evaluating the association between the event times of individuals and multiple covariates. The hazard function for the Cox model has the following form: 
\begin{align}
\label{CoxModel} \alpha(t \ | \ Z_{i}) = \alpha_{0}(t) \exp( \beta^\intercal Z_{i}(t)), 
\end{align}
where $\alpha_{0}(t)$ is an unspecified baseline hazard, $Z_{i}(t)$ a $q$-dimensional covariate vector and $\beta \in \mathbb{R}^{q}$ is a $q$-dimensional vector of regression coefficients. The exponential form of the relative risk function $\exp( \beta^\intercal Z_{i}(t))$ specifies the relationship between the covariates and the hazard function. It can be seen from Eq. $(\ref{CoxModel})$ that covariates are assumed to have a multiplicative effect on the hazard rate. The baseline hazard $\alpha_{0}(t)$ corresponds to $Z_{i}(t)=(0, ..., 0)^\intercal$ for all times $t$, with $A_{0}(t) = \int_{0}^{t} \alpha_{0}(u) du < \infty$. The Cox model is said to be semiparametric, as it involves both a nonparametric part (= baseline hazard) and a parametric part (= relative risk function). The underlying at-risk process $Y^{Cox}_{i}(t) = \mathbbm{1}(T_{i} \geq t, C_{i} \geq t) =  \mathbbm{1}(\min(T_{i},C_{i}) \geq t)$ for individual $B$ from the hypothetical study (cf. Figure $\ref{GraphicTTFE}$) is illustrated in Figure $\ref{AtRiskCox}$.

Let $z_{i}(t)$ and $z_{\tilde{i}}(t)$ be the covariate vectors for individuals $i$ and $\tilde{i}$, with $i \neq \tilde{i}$ and $i, \tilde{i} = 1,2,...,n$ . The ratio of the hazard rates $\alpha(t \ | \ z_{i})$ and $\alpha(t \ | \ z_{\tilde{i}})$ is
\begin{align}
\label{ratio} \dfrac{\alpha(t \ | \ z_{\tilde{i}})}{\alpha(t \ | \ z_{i})} = \dfrac{\alpha_{0}(t) \exp( \beta^\intercal z_{\tilde{i}}(t))}{\alpha_{0}(t) \exp( \beta^\intercal z_{i}(t))} = \dfrac{\exp( \beta^\intercal z_{\tilde{i}}(t))}{\exp( \beta^\intercal z_{i}(t))}.
\end{align}
If all covariates are fixed and time-independent, the ratio given in Eq. $(\ref{ratio})$ is constant over time and the corresponding Cox model $(\ref{CoxModel})$ is a proportional hazards model. \newline
Further, it is assumed that all components of $z_{i}(t)$ and $z_{\tilde{i}}(t)$ are identical, except for the $k^{th}$ component, where $z_{\tilde{i}k}(t) = z_{ik}(t) + 1$ and $k \in \{1, 2, ..., q \}$. Then, the hazard ratio $(\ref{ratio})$ reduces to
\begin{align}
\nonumber \dfrac{\alpha(t \ | \ z_{\tilde{i}})}{\alpha(t \ | \ z_{i})} = \exp \bigl( \beta^\intercal (z_{\tilde{i}}(t)- z_{i}(t)) \bigr) = \exp(\beta_{k}), 
\end{align}
i.e., the effect of a one-unit increase in the $k^{th}$ covariate when all other covariates are kept the same and irrespective of the baseline hazard. 
\newline \newline 
\textbf{Inference for $\boldsymbol{\beta}$ and large sample theory} \newline 
Because of the nonparametric baseline hazard, standard maximum likelihood arguments can not be used to estimate the regression coefficient $\beta \in \mathbb{R}^{q}$. Instead, \citet{Cox1972} derived a partial likelihood function for the estimation of $\beta$ in the Cox model and proved asymptotic properties of the proposed estimator $\hat{\beta}$ using counting process and martingale theory. \citet{Andersen1982} discussed partial likelihood estimation in a more general model featuring recurrent events, with the Cox model included as a special case. Inference for $\beta$ in this general model will be extensively described in Section $\ref{IntensityAGmodel}$ and includes the Cox model, with only $Y_{i}^{AG}(t)$ replaced by $Y_{i}^{Cox}(t)$. Thus, it is referred to Section $\ref{IntensityAGmodel}$ for further details on partial likelihood estimation in the Cox model. 
\chapter{Recurrent event methods}
\label{RETheory}
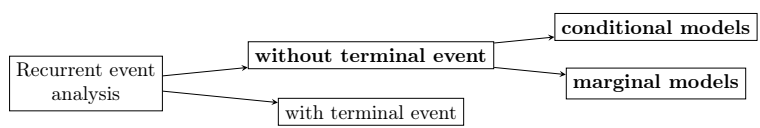
\begin{figure}[H]
\centering
\scalebox{0.75}{
\begin{tikzpicture}[->, grow=right,level 1/.style={sibling distance=10mm, level distance=5cm},level 2/.style={sibling distance=10mm, level distance=5cm}, 
every node/.style = {shape=rectangle, draw, align=center, top color=white, bottom color=white}]]
  \node {Recurrent event \\ analysis}
    child { node {with terminal event}} 
    child { node {\textbf{without terminal event}}
      child { node {\textbf{marginal models}} }
      child { node {\textbf{conditional models}} }
    };
\end{tikzpicture}}
\caption[Classification of recurrent event methodology]{Classification of recurrent event methodology}
\label{IntroRecurrentEvents}
\end{figure}

Recurrent events refer to the repeated occurrence of the same type of event over time for the same individual. There has been considerable progress in methodology for analyzing recurrent events in the past few decades \parencite{Cook2007, Andersen1993, Kalbfleisch2002, Therneau2000}. These advances have been mainly motivated by biomedical studies in which individuals are subject to experience repeated events. For instance, patients with chronic heart failure may be admitted to hospital multiple times \parencite{Rogers2014}, patients with asthma may have repeated attacks \parencite{Duchateau2003} and cancer patients may develope recurrent tumor metastases \parencite{Rondeau2010}. Other examples of recurrent events include infections, myocardial infarctions, epileptic seizures and disease relapses. A broad range of models have been developed to analyze recurrent event data: Poisson model, negative binomial (NB) model, Andersen-Gill (AG) model, Prentice-Williams-Peterson (PWP) model, Wei-Lin-Weissfeld (WLW) model, Lee-Wei-Amato (LWA) model, Lin-Wei-Yang-Ying (LWYY) model or frailty models. \newline
In many settings, e.g., when analyzing hospitalizations in chronic heart failure, recurrent event processes are often permanently terminated by a patient's death. The underlying disease is associated with both recurrent complications and high mortality. Patients who experience such an early terminated event are more likely to have fewer events than patients who experience the terminal event later during follow-up. Inversely, repeated occurrence of recurrent events may also increase the risk for the terminal event. As a consequence, using inappropriate methods that ignore terminal events may lead to biased results. Methods for the analysis of recurrent events in the presence of terminal events have been considered by \citet{Gosh2000}, \citet{Gosh2002}, \citet{Miloslavsky2004} and \citet{CharlesNelson2019}. In MS trials, the process of recurrent progression events hasn't been observed to be stopped by a terminal event, e.g., death from MS. Since this work is motivated by repeated progression events in RRMS and PPMS, methodology without consideration of terminal events is of major interest. \newline 
As illustrated in Figure $\ref{IntroRecurrentEvents}$, there are two major approaches for the analysis of recurrent events: \textit{conditional} and \textit{marginal models}. Conditional models are intensity-based methods that attempt to fully specify the entire recurrent event process by modelling the past through internal time-varying covariates (e.g., AG model and PWP model) or random effects (e.g., frailty models). In marginal models, the dependence structure between successive events may remain unspecified and the focus is essentially on marginal parameters (e.g., expected number of events in $[0,t]$, rate functions or the marginal distribution of times to the first, second, ... event). Models that belong to this class are the WLW model, LWA model and the LWYY model. While conditional models are appealing when the purpose is to understand the disease process and to identify risk factors, marginal models have been proposed for the analysis of recurrent events in RCTs, where treatment effect estimates should yield a clear causal interpretation. 

This chapter aims at describing the existing recurrent event methodology in the absence of terminal events and at highlighting the differences between conditional and marginal approaches. Before that, the characteristics of recurrent event data and the respective methods are described in Section $\ref{CharaRecurrentEvents}$, followed by an introduction into counting process theory for recurrent events in Section $\ref{SectionSP}$. Conditional models are discussed in Section $\ref{condmodel}$, while marginal models are presented in Section $\ref{margMoDels}$. Finally, recommendations for choosing appropriate methods for analyzing recurrent events in RCTs are provided in Section $\ref{RCTrecurrentevents}$. 

\section{Characteristics of recurrent event data and methods}
\label{CharaRecurrentEvents}
\begin{figure}[H]
\centering
    \subfloat[Recurrent event data]{
    \scalebox{0.87}{
\begin{tikzpicture}
    \begin{axis}[height=0.2\textheight, width=0.75\textwidth, xmin=0.0, xmax=28, ymin=0.0, ymax=2.7, xtick={5, 10, 15, 20, 25}, xticklabels={5, 10, 15, 20, 25}, ytick={0.5, 1, 1.5, 2.0, 2.5},
                 yticklabels={$E$, $D$, $C$, $B$, $A$}, xlabel={Time from randomization (in days)}, x label style={at={(axis description cs:0.5,-0.2)},anchor=north}]
        \addplot[cmhplot,-,domain=0:21]{1.5};
        \addplot[cmhplot,-,domain=0:25]{2.0};
        \addplot[cmhplot,-,domain=0:13]{2.5};
        \addplot[cmhplot,-,domain=0:25]{1.0};
        \addplot[cmhplot,-,domain=0:25]{0.5};
        \addplot[holdot]coordinates{(21,1.5)};
        \addplot[soldot]coordinates{(7,2.0)};
        \addplot[soldot]coordinates{(11,2.0)};
        \addplot[soldot]coordinates{(16,2.0)};
        \addplot[holdot]coordinates{(25,2.0)};
        \addplot[soldot]coordinates{(2,2.5)};
        \addplot[soldot]coordinates{(5,2.5)};
        \addplot[holdot]coordinates{(13,2.5)};
        \addplot[holdot]coordinates{(25,1.0)};
        \addplot[holdot]coordinates{(25,0.5)};
        \addplot[soldot]coordinates{(9,0.5)};
        \addplot[dashed]coordinates {(25,0) (25,2.5)};
        \node [below] at (axis cs: 21, 1.5) {21};
        \node [below] at (axis cs: 7, 2.0) {7};
        \node [below] at (axis cs: 11, 2.0) {11};
        \node [below] at (axis cs: 9, 0.5) {9};
        \node [below] at (axis cs: 16, 2.0) {16};
        \node [below] at (axis cs: 2, 2.5) {2};
        \node [below] at (axis cs: 5, 2.5) {5};
        \node [below] at (axis cs: 13, 2.5) {13};
\end{axis}
\end{tikzpicture}
    }}
\hfil
    \subfloat[Time-to-first-event data]{
    \scalebox{0.87}{
\begin{tikzpicture}
    \begin{axis}[height=0.2\textheight, width=0.75\textwidth, xmin=0.0, xmax=28, ymin=0.0, ymax=2.7, xtick={5, 10, 15, 20, 25}, xticklabels={5, 10, 15, 20, 25}, ytick={0.5, 1, 1.5, 2.0, 2.5},
                 yticklabels={$E$, $D$, $C$, $B$, $A$}, xlabel={Time from randomization (in days)}, x label style={at={(axis description cs:0.5,-0.2)},anchor=north}]
        \addplot[cmhplot,-,domain=0:21]{1.5};
        \addplot[cmhplot,-,domain=0:7]{2.0};
        \addplot[cmhplot,-,domain=0:2]{2.5};
        \addplot[cmhplot,-,domain=0:25]{1.0};
        \addplot[cmhplot,-,domain=0:9]{0.5};
        \addplot[holdot]coordinates{(21,1.5)};
        \addplot[soldot]coordinates{(7,2.0)};
        \addplot[soldot]coordinates{(2,2.5)};
        \addplot[holdot]coordinates{(25,1.0)};
        \addplot[soldot]coordinates{(9,0.5)};
        \addplot[dashed]coordinates {(25,0) (25,2.5)};
        \node [below] at (axis cs: 21, 1.5) {21};
        \node [below] at (axis cs: 7, 2.0) {7};
        \node [below] at (axis cs: 9, 0.5) {9};
        \node [below] at (axis cs: 2, 2.5) {2};
\end{axis}
\end{tikzpicture}
}}
\caption[Illustration of time-to-first-event and recurrent event data]{Illustration of time-to-first-event and recurrent event data ($\bullet$ = event, $\circ$ = censoring, dashed vertical line is end of follow-up)}
\label{GraphicRE}
\end{figure}
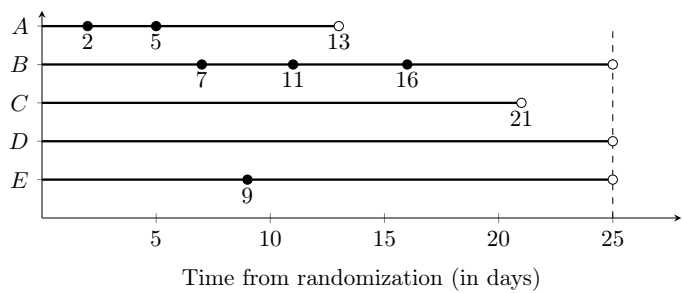
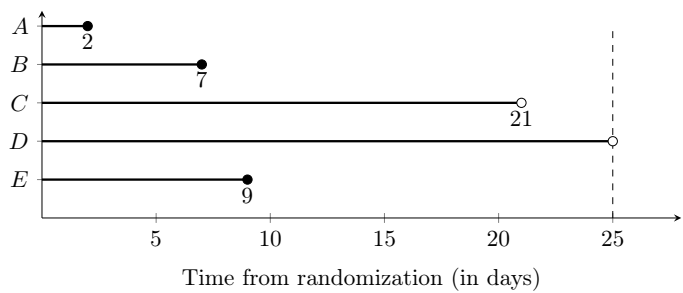

Stochastic processes that generate events repeatedly over time are known as recurrent event processes and the corresponding data produced by such processes are referred to as recurrent event data. \newline
Figure $\ref{GraphicRE}$ displays a hypothetical example of recurrent event and time-to-first-event data for $5$ individuals, where events are depicted by filled circles and censoring is marked by unfilled circles. Time is measured from randomization (= time origin) and maximum length of follow-up is $25$ days for each individual. With regard to recurrent events (a), three individuals are administratively right-censored at the end of follow-up ($B$, $D$, $E$) and the other two individuals ($A$ and $C$) withdraw from study earlier for reasons unrelated to the recurrent event process. While individual $A$ experiences two events at times $2$ and $5$ days and drops out of study at day $13$, individual $C$ is right-censored after $21$ days without monitoring an event. No event has also be observed for individual $D$ who is censored at the end of follow-up. Individual $B$ is observed to have $3$ events at day $7$, $11$ and $16$, followed by administrative right-censoring at day $25$. For individual $E$, the event of interest has occurred once and administrative censoring takes place at day $25$. 
In terms of time-to-first-event data (b), individuals $A$, $B$ and $E$ experience an event, whereas individuals $C$ and $D$ are event-free and, thus, they are right-censored at day $21$ and $25$. Comparing panel (a) with panel (b), only individuals who have not experienced an event during follow-up provide the same information in both time-to-first-event and recurrent event analyses. For all other individuals, recurrent event processes provide richer information on the underlying disease than single event processes. 

\newpage 
Specific to recurrent event data without terminating event and unlike the case of time-to-first-event data, all individuals under study are eventually right-censored. If repeated events are considered, the observation process is ongoing after a first, second, .... event and is theoretically never finished. Termination of the observation process happens only due to end of study or a study participant withdrawing earlier from the trial. However, the event process may continue beyond the right-censoring time but it is not possible to observe it. Such a termination needs to be distinguished from termination due to another type of event (e.g., death), which completely terminates the process under study. In time-to-first-event approaches, observation of the process is stopped upon occurrence of an event and only event-free patients are subject to right-censoring at the end of follow-up.
\subsubsection*{Components of recurrent event methods}
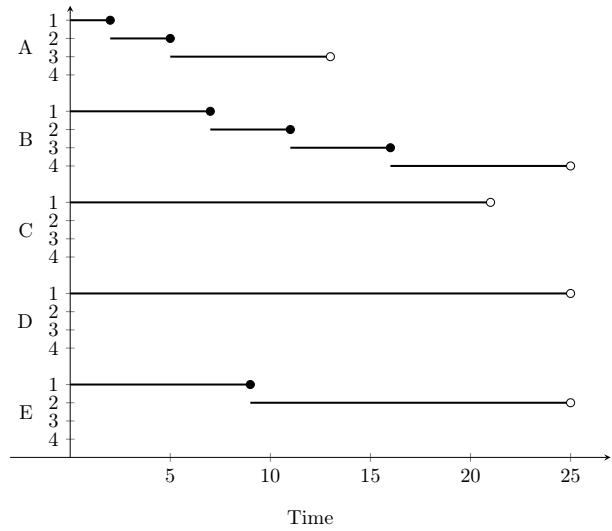
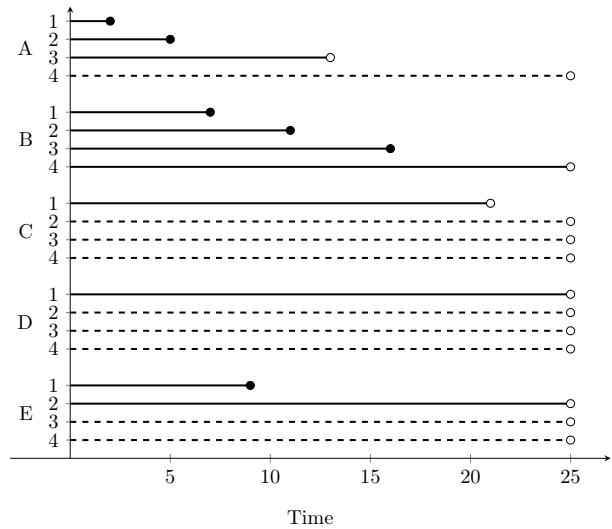
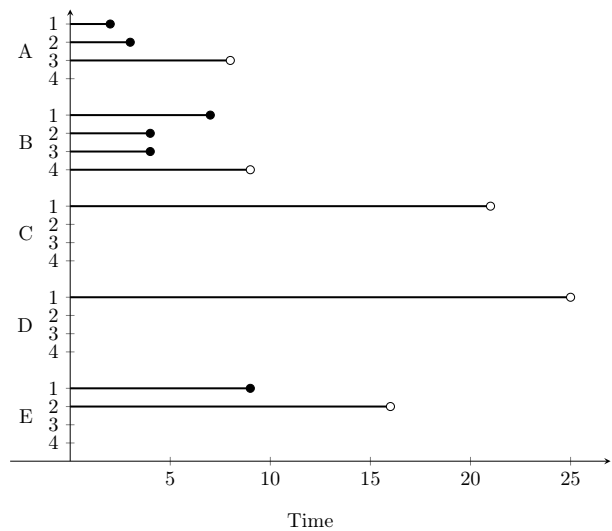
\begin{figure}[H]
\centering
\subfloat[\textbf{Counting process}]{
\resizebox{8cm}{!}{
\begin{tikzpicture}
    \begin{axis}[height=0.4\textheight, width=0.8\textwidth, xmin=-3.0, xmax=27, ymin=0.0, ymax=12.4, xtick={5, 10, 15, 20, 25}, xticklabels={5, 10, 15, 20, 25}, ytick={0.5, 1, 1.5, 2.0, 3, 3.5, 4.0, 4.5, 5.5, 6, 6.5, 7.0, 8.0, 8.5, 9.0, 9.5, 10.5, 11.0, 11.5, 12.0}, yticklabels={4,3,2,1,4, 3,2,1,4, 3,2,1, 4, 3,2,1, 4, 3,2,1}, xlabel={Time}, x label style={at={(axis description cs:0.5,-0.1)},anchor=north}]
        \addplot[cmhplot,-,domain=0:9]{2.0};
        \addplot[cmhplot,-,domain=9:25]{1.5};
        \addplot[cmhplot,-,domain=0:25]{4.5};
        \addplot[cmhplot,-,domain=0:21]{7.0};
        \addplot[cmhplot,-,domain=0:7]{9.5};
        \addplot[cmhplot,-,domain=7:11]{9.0};
        \addplot[cmhplot,-,domain=11:16]{8.5};
        \addplot[cmhplot,-,domain=16:25]{8.0};
        \addplot[cmhplot,-,domain=0:2]{12.0};
        \addplot[cmhplot,-,domain=2:5]{11.5};
        \addplot[cmhplot,-,domain=5:13]{11.0};
        \addplot[soldot]coordinates{(9,2.0)};
        \addplot[holdot]coordinates{(25,1.5)};
        \addplot[holdot]coordinates{(25,4.5)};
        \addplot[holdot]coordinates{(21,7.0)};
        \addplot[soldot]coordinates{(7,9.5)};
        \addplot[soldot]coordinates{(11,9.0)};
        \addplot[soldot]coordinates{(16,8.5)};
        \addplot[holdot]coordinates{(25,8.0)};
        \addplot[soldot]coordinates{(2,12)};
        \addplot[soldot]coordinates{(5,11.5)};
        \addplot[holdot]coordinates{(13,11.0)};
        \node [left] at (axis cs: -1.5, 1.25) {E};
        \node [left] at (axis cs: -1.5, 3.75) {D};
        \node [left] at (axis cs: -1.5, 6.25) {C};
        \node [left] at (axis cs: -1.5, 8.75) {B};
        \node [left] at (axis cs: -1.5, 11.25) {A};
    \end{axis}
\end{tikzpicture}}
}
\hfil
    \subfloat[\textbf{Total time}]{
\resizebox{8cm}{!}{
\begin{tikzpicture}
    \begin{axis}[height=0.4\textheight, width=0.8\textwidth, xmin=-3.0, xmax=27, ymin=0.0, ymax=12.4, xtick={5, 10, 15, 20, 25}, xticklabels={5, 10, 15, 20, 25}, ytick={0.5, 1, 1.5, 2.0, 3, 3.5, 4.0, 4.5, 5.5, 6, 6.5, 7.0, 8.0, 8.5, 9.0, 9.5, 10.5, 11.0, 11.5, 12.0}, yticklabels={4,3,2,1,4, 3,2,1,4, 3,2,1, 4, 3,2,1, 4, 3,2,1}, xlabel={Time}, x label style={at={(axis description cs:0.5,-0.1)},anchor=north}]
        \addplot[cmhplot,-,domain=0:9]{2.0};
        \addplot[cmhplot,-,domain=0:25]{1.5};
        \addplot[cmhplot,-,domain=0:25, dashed]{1.0};
        \addplot[cmhplot,-,domain=0:25, dashed]{0.5};
        \addplot[cmhplot,-,domain=0:25]{4.5};
        \addplot[cmhplot,-,domain=0:25, dashed]{4.0};
        \addplot[cmhplot,-,domain=0:25, dashed]{3.5};
        \addplot[cmhplot,-,domain=0:25, dashed]{3.0};
        \addplot[cmhplot,-,domain=0:21]{7.0};
        \addplot[cmhplot,-,domain=0:25, dashed]{6.5};
        \addplot[cmhplot,-,domain=0:25, dashed]{6.0};
        \addplot[cmhplot,-,domain=0:25, dashed]{5.5};
        \addplot[cmhplot,-,domain=0:7]{9.5};
        \addplot[cmhplot,-,domain=0:11]{9.0};
        \addplot[cmhplot,-,domain=0:16]{8.5};
        \addplot[cmhplot,-,domain=0:25]{8.0};
        \addplot[cmhplot,-,domain=0:2]{12.0};
        \addplot[cmhplot,-,domain=0:5]{11.5};
        \addplot[cmhplot,-,domain=0:13]{11.0};
        \addplot[cmhplot,-,domain=0:25, dashed]{10.5};
        \addplot[soldot]coordinates{(9,2.0)};
        \addplot[holdot]coordinates{(25,1.5)};
        \addplot[holdot]coordinates{(25,1.0)};
        \addplot[holdot]coordinates{(25,0.5)};
        \addplot[holdot]coordinates{(25,4.5)};
        \addplot[holdot]coordinates{(25,4.0)};
        \addplot[holdot]coordinates{(25,3.5)};
        \addplot[holdot]coordinates{(25,3.0)};
        \addplot[holdot]coordinates{(21,7.0)};
        \addplot[holdot]coordinates{(25,6.5)};
        \addplot[holdot]coordinates{(25,6.0)};
        \addplot[holdot]coordinates{(25,5.5)};
        \addplot[soldot]coordinates{(7,9.5)};
        \addplot[soldot]coordinates{(11,9.0)};
        \addplot[soldot]coordinates{(16,8.5)};
        \addplot[holdot]coordinates{(25,8.0)};
        \addplot[soldot]coordinates{(2,12)};
        \addplot[soldot]coordinates{(5,11.5)};
        \addplot[holdot]coordinates{(13,11.0)};
        \addplot[holdot]coordinates{(25,10.5)};
        \node [left] at (axis cs: -1.5, 1.25) {E};
        \node [left] at (axis cs: -1.5, 3.75) {D};
        \node [left] at (axis cs: -1.5, 6.25) {C};
        \node [left] at (axis cs: -1.5, 8.75) {B};
        \node [left] at (axis cs: -1.5, 11.25) {A};
    \end{axis}
\end{tikzpicture}}
}
\hfil
    \subfloat[\textbf{Gap time}]{
\resizebox{8cm}{!}{
\begin{tikzpicture}
    \begin{axis}[height=0.4\textheight, width=0.8\textwidth, xmin=-3.0, xmax=27, ymin=0.0, ymax=12.4, xtick={5, 10, 15, 20, 25}, xticklabels={5, 10, 15, 20, 25}, ytick={0.5, 1, 1.5, 2.0, 3, 3.5, 4.0, 4.5, 5.5, 6, 6.5, 7.0, 8.0, 8.5, 9.0, 9.5, 10.5, 11.0, 11.5, 12.0}, yticklabels={4,3,2,1,4, 3,2,1,4, 3,2,1, 4, 3,2,1, 4, 3,2,1}, xlabel={Time}, x label style={at={(axis description cs:0.5,-0.1)},anchor=north}]
        \addplot[cmhplot,-,domain=0:9]{2.0};
        \addplot[cmhplot,-,domain=0:16]{1.5};
        \addplot[cmhplot,-,domain=0:25]{4.5};
        \addplot[cmhplot,-,domain=0:21]{7.0};
        \addplot[cmhplot,-,domain=0:7]{9.5};
        \addplot[cmhplot,-,domain=0:4]{9.0};
        \addplot[cmhplot,-,domain=0:4]{8.5};
        \addplot[cmhplot,-,domain=0:9]{8.0};
        \addplot[cmhplot,-,domain=0:2]{12.0};
        \addplot[cmhplot,-,domain=0:3]{11.5};
        \addplot[cmhplot,-,domain=0:8]{11.0};
        \addplot[soldot]coordinates{(9,2.0)};
        \addplot[holdot]coordinates{(16,1.5)};
        \addplot[holdot]coordinates{(25,4.5)};
        \addplot[holdot]coordinates{(21,7.0)};
        \addplot[soldot]coordinates{(7,9.5)};
        \addplot[soldot]coordinates{(4,9.0)};
        \addplot[soldot]coordinates{(4,8.5)};
        \addplot[holdot]coordinates{(9,8.0)};
        \addplot[soldot]coordinates{(2,12)};
        \addplot[soldot]coordinates{(3,11.5)};
        \addplot[holdot]coordinates{(8,11.0)};
        \node [left] at (axis cs: -1.5, 1.25) {E};
        \node [left] at (axis cs: -1.5, 3.75) {D};
        \node [left] at (axis cs: -1.5, 6.25) {C};
        \node [left] at (axis cs: -1.5, 8.75) {B};
        \node [left] at (axis cs: -1.5, 11.25) {A};
    \end{axis}
\end{tikzpicture}}
}
\caption[Illustration of risk interval formulations (counting process, total time and gap time)]{Illustration of risk interval formulations using the hypothetical example from Figure $\ref{GraphicRE}$ ($\bullet$ = event, $\circ$ = censoring, dashed line = dummy risk interval)}
\label{GraphiT}
\end{figure}

In general, recurrent event methods can be characterized by $5$ components: \textit{time scale}, \textit{risk interval}, \textit{risk set}, \textit{baseline hazard} and \textit{intra-individual correlation}. By means of the hypothetical recurrent event data illustrated in Figure $\ref{GraphicRE}$, these characteristics will be exemplified. 
\newline \newline
\textbf{Time scale} \newline
Recurrent event methods can be specified in two different time scales: \textit{calendar time} and \textit{gap time} \parencite{Cook2007, Kelly2000}. Calendar time corresponds to the time measured from the time origin. In calendar time perspective, time zero may be the onset of disease, start of treatment or randomization in clinical trials. In gap time perspective, time is reset to $0$ after each event and the time scale is based on the time elapsed since the previous event. Gap time is recommended to use when a 'renewal' happens after each event, meaning that the individual under study returns back to a similar state after the recurrence. For instance, when studying the occurrence of repeated bladder infections, health status of women affected by this short-term disease is expected to be completely restored after the infection has been gone. This assumption makes a gap time analysis reasonable. In case the underlying disease evolves over the course of time, such as in PPMS, calendar time is the preferred time scale. In this thesis, the focus lies on calendar time analyses. 
\newline \newline 
\textbf{Risk interval} \newline 
By reference to \citet{Kelly2000}, risk intervals define when an individual is at-risk for experiencing an event along a given time scale (gap time or calendar time). It is distinguished between \textit{total time}, \textit{gap time} and \textit{counting process} formulation. Figure $\ref{GraphiT}$ graphically illustrates the different types of risk intervals for the individuals from the hypothetical example introduced in Figure $\ref{GraphicRE}$. Total time corresponds to the time elapsed from time origin and is depicted in panel (b). For instance, marginal analyses of the time to later events (i.e., time from baseline to the second event) use total time as risk interval formulation. With total time, individual $B$ is at-risk for the first event in the time interval $[0, 7)$, for the second event during $[0,11)$, for the third event during $[0,16)$ and for the fourth event during $[0,25)$. In counting process formulation (panel (a)), individual $B$ is at-risk for the first event between $[0, 7)$ and for the second, third and fourth event during $[7,11)$, $[11,16)$ and $[16,25)$, respectively. Thus, the counting process approach also uses calendar time as time scale but additionally takes truncation schemes (e.g., delayed entry) into account. As seen from panel (a), an individual is not at-risk for a $j^{th}$ event before a $(j-1)^{th}$ event has been observed. With gap time, individual $B$ is assumed to be at risk for the first event during $[0, 7)$ and for the second, third and fourth event during $[0,4)$, $[0,4)$ and $[0,9)$. Comparing the different types of risk intervals, the risk interval for the first event is the same for all three approaches, and the gap time and counting process formulation result in the same length of at-risk periods. \newline 

\begin{table}[H]
\centering
\scalebox{0.45}{
\begin{tabular}{l|ccc}
\multicolumn{1}{c|}{}                       & \multicolumn{3}{c}{\textbf{Risk set / baseline hazard}}                                                                                                                                                                                           \\ \hline
\multicolumn{1}{c|}{\textbf{Risk interval}} & \textit{\begin{tabular}[c]{@{}c@{}}unrestricted /\\ common\end{tabular}} & \textit{\begin{tabular}[c]{@{}c@{}}semi-restricted / \\ event-specific\end{tabular}} & \textit{\begin{tabular}[c]{@{}c@{}}restricted / \\ event-specific\end{tabular}} \\ \hline
\textit{Gap time (GT)}                           & $\surd$                                                                  & X                                                                                    & PWP-GT                                                                          \\
\textit{Total time}                         & LWA                                                                      & WLW                                                                                  & $\surd$                                                                         \\
\textit{Counting process (CP)}                   & AG                                                                       & $\surd$                                                                              & PWP-CP                                                                         
\end{tabular}}
\caption[Classification of recurrent event methods according to model characteristics]{Classification of some recurrent event methods according to different model characteristics ($\surd$ = possible but model without specific name in literature, X = not possible)}
\label{MethodIntroductionOverview}
\end{table}

\textbf{Baseline hazard} \newline 
Another characteristic of recurrent event methods is the type of the baseline hazard. It is differentiated between a \textit{common} and an \textit{event-specific} baseline hazard \parencite{Kelly2000}. A recurrent event model with common baseline hazard has the same underlying hazard for all events, whereas an event-specific baseline hazard is a stratified baseline hazard allowing the baseline hazard to vary with the $j^{th}$ event. 
\newline \newline 
\textbf{Risk set} \newline
The $j^{th}$ risk set at $t-$ includes all individuals who are at-risk for experiencing a $j^{th}$ event at time $t$. It can be differentiated between three different types of risk sets: \textit{unrestricted}, \textit{semi-restricted} and \textit{restricted}. The risk set is said to be unrestricted, if all risk intervals contribute to the risk set for any event, independent on the number of previous events \parencite{Kelly2000}. For instance, an individual's second event time may contribute to the risk set associated with another individual's first event. In counting process formulation, the risk set corresponding to the second event of individual $B$ from the hypothetical example contains information from the third event of $A$, the second event of $B$, the first event of $C$ and $D$, and the second event of $E$. In total time formulation, information on the third event of $A$, the second, third and fourth event of $B$, the first event of individuals $C$ and $D$, and the second event of $E$ is included in the risk set corresponding to $B$'s second event. In gap time formulation, the risk set includes information on the third event of $A$, the first, second, third and fourth event of $B$, the first event of $C$ and $D$, and the first and second event of $E$. An unrestricted risk set has a common baseline hazard for all events. \newline  
A risk set is called restricted, if the $j^{th}$ risk set only includes the $j^{th}$ event risk intervals from individuals who have already experienced $(j-1)$ events. This means that only individuals with $(j-1)$ previous events are considered to be at-risk for a $j^{th}$ event. In all three formulations, the risk set associated with the second event of individual $B$ at time $11$ contains information on the second event of individual $B$ and $E$. Compared to an unrestricted risk set, a restricted risk set has event-specific baseline hazards. \newline
A risk set is semi-restricted, if the risk sets have event-specific baseline hazards but allow individuals who have less than $(j-1)$ events to be at-risk for a $j^{th}$ event by defining so-called dummy risk intervals, as seen in Figure $\ref{GraphiT}$ \parencite{Wei1989}. Using this risk set definition, individuals are considered to be at-risk for all events starting from time origin. For instance, an individual who has already experienced one event is considered to be at-risk for a second, third, ... event simultaneously. However, a semi-restricted risk set does not allow information from the $j^{th}$ risk interval to contribute to the risk for an earlier event. While total time and counting process formulations are compatible with a semi-restricted risk set, gap time in combination with a semi-restricted risk set is not possible. For example, in total time formulation, the risk set associated with the second event of individual $B$ contains information on the second dummy intervals of individuals $C$ and $D$, and the second event of individual $E$. 
Due to event-specific baseline hazards, the third risk intervals are not included.
\newline \newline 
\textbf{Intra-individual correlation} \newline 
For the dependence structure among repeated events on the same individual, it can be accounted for by different approaches: \textit{conditional}, \textit{marginal}, \textit{random effects}.
The conditional approach assumes that dependence between recurrent events is completely explained by time-varying covariates (e.g., previous number of events, time since most recent event). This means that the time increment between events are conditionally uncorrelated given the observed covariates. The marginal approach assumes independence between recurrent events within one individual. The random effect approach incorporates a random effect or a frailty term into the recurrent event model that induces dependence among recurrent event times. 
\newline \newline 
Table $\ref{MethodIntroductionOverview}$ gives an overview of the most famous recurrent event models classified by baseline hazard, risk set and risk interval.
\newpage

\newpage 
\section{Counting processes for recurrent events}
\label{SectionSP}
Counting processes and intensity functions serve as a convenient framework for describing recurrent event data. In the following, an introduction into the theory of counting processes and intensity functions is provided based on \citet{Andersen1993, Cook2007, Aalen2008, Beyersmann2012}. 
\newline 
Suppose $n$ recurrent event processes starting at $t=0$ are under observation. $T_{ij}$ is defined as the $j^{th}$ event time for individual $i$, $j = 1, 2, ... $ and $i = 1, ..., n$, with $T_{ij} \in [0, \infty)$ and $T_{i1} < T_{i2} < T_{i3} < ... \ $. $G_{ij} := T_{ij} - T_{i(j-1)}$ is the $j^{th}$ gap time or interevent time between two consecutive events, with $T_{i0} := 0$ and $G_{i1} := T_{i1} \ \forall \ j = 1, 2, ... \ .$ In other words, $G_{ij}$ is the duration of time between the $(j-1)^{th}$ and $j^{th}$ event for individual $i$. Recurrent event data, as introduced in Section $\ref{CharaRecurrentEvents}$, can now be formulated using counting processes. 

\begin{definition}[Counting process]
A right-continuous stochastic process $\overline{N} = \{ \overline{N}(t): 0 \leq t < \infty\}$ is said to be a counting process, if $\overline{N}(t) \in \{0, 1, 2, ...\}$ is the number of events that have happened up to and including time $t$. A counting process has jumps of size $1$ at the event times and is constant in between, i.e., 
\begin{itemize}
\item $\Delta \overline{N}(t) := \overline{N}(t) - \overline{N}(t-) = \overline{N}(t) - \lim\limits_{{s \nearrow t, s \neq t}} \overline{N}(s) \ \in \{0, 1 \}$, i.e., at most one event can happen in the infinitesimal small time interval $[t, t+dt)$.
\item $\Delta \overline{N}(\cdot) = 1$ at event times, i.e, the counting process jumps at the event times. 
\item $\overline{N}(0) = 0$. 
\end{itemize}
\label{defCP}
\end{definition}
\noindent

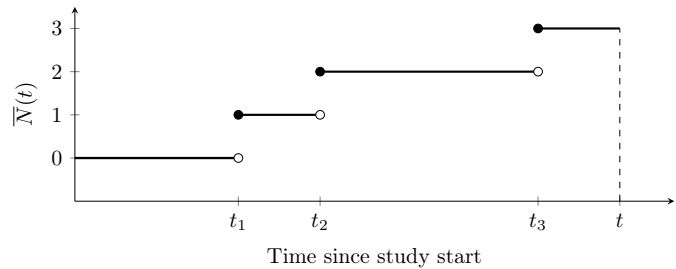
\begin{figure}[H]
\centering
\resizebox{9cm}{!}{
\begin{tikzpicture}
    \begin{axis}[height=0.2\textheight, width=0.75\textwidth, xmin=0, xmax=22, ymin=0.0, ymax=4.5, xtick={6, 9, 17, 20}, xticklabels={$t_{1}$, $t_{2}$, $t_{3}$, $t$}, 
    ytick={1, 2, 3, 4}, yticklabels={0, 1, 2, 3}, xlabel={Time since study start}, ylabel={$\overline{N}(t)$}, y label style={at={(axis description cs:-0.05,.5)},rotate=90,anchor=south},
    x label style={at={(axis description cs:0.5,-0.2)},anchor=north}]
        \addplot[cmhplot,-,domain=0:6]{1};
        \addplot[cmhplot,-,domain=6:9]{2};
        \addplot[cmhplot,-,domain=9:17]{3};
        \addplot[cmhplot,-,domain=17:20]{4};
        \addplot[cmhplot,-,domain=0:6]{1};
        \addplot[holdot]coordinates{(6,1)};
        \addplot[soldot]coordinates{(6,2)};
        \addplot[holdot]coordinates{(9,2)};
        \addplot[soldot]coordinates{(9,3)};
        \addplot[holdot]coordinates{(17,3)};
        \addplot[soldot]coordinates{(17,4)};
        \addplot[dashed]coordinates {(20,0) (20,4)};
    \end{axis}
\end{tikzpicture}
}
\caption[Counting process representation of recurrent event data]{Counting process representation of recurrent event data ($\bullet$ = included, $\circ$ = not included)}
\label{CP}
\end{figure}

Figure $\ref{CP}$ illustrates a realization of an arbitrary recurrent event process in terms of its counting process, where events have been observed at times $t_{1}, t_{2}$ and $t_{3}$. It can be seen that the sample path of the counting process is an increasing step function with jumps at the event times. Initially, as long as the individual is event-free, the counting process is equal to $0$. That is, $\overline{N}(t)=0 \ \forall \ t \in [0, t_{1})$. Exactly at time $t_{1}$, an event can be observed and the counting process jumps from $0$ to $1$, i.e., $\overline{N}(t_{1})=1$. Then, the counting process keeps constant until the second event occurs, i.e., $\overline{N}(t)=1 \ \forall \ t \in [t_{1}, t_{2})$ and $\overline{N}(t_{2})=2$. At time $t_{3}$, the counting process jumps from $2$ to $3$ and stays there as long as no fourth event will happen. The graph also shows that a counting process is continuous from the right. \newline 

Let $\overline{N}_{i}(t)$ denote the fully observed (uncensored) counting process for individual $i$, where $\overline{N}_{i}(t) = \sum_{j=1}^\infty \mathbbm{1}(T_{ij} \leq t)$ counts the number of events experienced by individual $i$ over the time period $[0, t]$. In time-to-first-event settings, $\overline{N}_{i}(t)$ takes values in $\{0, 1 \}$, whereas for recurrent event processes $\overline{N}_{i}(t) \in \mathbb{N}_{0}$. $d\overline{N}_{i}(t) = \overline{N}_{i}(t) - \overline{N}_{i}(t + dt)$ is the increment of $\overline{N}_{i}(t)$ over $[t, t+dt)$ and is defined as the number of events occurring in $[t, t+dt)$. \newline
In general, the process history (= 'past') plays an important role in recurrent event analysis and is crucial to differentiate between conditional and marginal recurrent event methods. Roughly speaking, the past includes all information that has been generated by the counting process from time origin until present. For instance, the past contains previous realizations of the counting process and, thus, provides information on the occurrence and timing of previous events. 
\newpage
More formally, the past is usually formulated as a $\sigma$-algebra generated by the counting processes $\overline{N}_{i}$, $i=1,2, ..., n$. That is, the past up to time t is defined as $\mathcal{\overline{N}}(t) = \sigma \bigl( \bigl( \overline{N}_{i}(s) \bigl)_{s \leq t}: i=1,2,...,n \bigr)$, whereas the past just before time $t$ corresponds to $\mathcal{\overline{N}}(t-) = \sigma \bigl( \bigl( \overline{N}_{i}(s) \bigl)_{s < t} : i=1,2,...,n \bigr)$, respectively. The entire process history can also be seen as an increasing family of $\sigma$-algebras, often known as filtration. 
\begin{definition}[History]
A filtration (or history) $\bigl \{ \mathcal{\overline{N}}(t) \bigr \}_{t \geq 0}$ is an increasing family of sub-$\sigma$-algebras. In other words, $\mathcal{\overline{N}}(t)$ is a $\sigma$-algebra for each $t$ and if $s \leq t$, then $\mathcal{\overline{N}}(s) \subset \mathcal{\overline{N}}(t)$. This means, the amount of knowledge about the past increases, as time passes.
\end{definition}

While the $\sigma$-algebra $\mathcal{\overline{N}}(t)$ represents information available at time $t$, the filtration $\{ \mathcal{\overline{N}}(t)\}_{t \geq 0}$ presents the evolution of information over the course of time. The counting process $\overline{N}_{i}(t)$ is adapted to the history $ \{ \mathcal{\overline{N}}(t) \}_{t \geq 0}$, meaning that at time $t$ the realizations of $\overline{N}_{i}(s)$ are known for all $s \leq t$, i.e., $\overline{N}_{i}(s) \in \mathcal{\overline{N}}(t)$ for all $s \leq t$ and $i=1,2,..., n$. Assuming that only one event can happen in $[t, t+dt)$, the intensity function $\overline{\lambda}_{i}(t)$ of the individual-specific counting process $\overline{N}_{i}(t)$ with regard to the history $\mathcal{\overline{N}}$ is defined via 
\begin{equation}
\overline{\lambda_{i}}(t) dt \ = P(d\overline{N_{i}}(t) = 1 \ | \ \mathcal{\overline{N}}(t-)).
\label{intensityuncensored}
\end{equation}
The intensity function given in Eq. $(\ref{intensityuncensored})$ is the instantaneous probability of an event occurring at time $t$ conditional on the process history up to $t-$. $\mathcal{\overline{N}}$ is the smallest self-exciting filtration to which the counting processes $\overline{N}_{i}$ are adapted.\newline 
When individuals are subject to right-censoring, not all recurrent events in the underlying processes $\overline{N}_{i}$ are observed. Let $C_{i} \in (0, \infty]$ denote the right-censoring or end of follow-up time for individual $i$, $i=1,...,n$. The $n$ individuals under study are then observed over the time interval $[0, C_{i}]$, where $t=0$ corresponds to the start of the recurrent event process. Let $Y_{i}(t) = \mathbbm{1}(C_{i} \geq t) \in \{ 0, 1 \}$ be an indicator function reflecting whether individual $i$ is under observation and at-risk for an event just prior to time $t$. In the following, $Y_{i}(t) $ is often referred to as the at-risk process. The left-continuous at-risk process is assumed to be predictable, i.e., the value of $Y_{i}(t)$ is already known at $t-$. Then, the randomly right-censored (or observable) counting process of observed events over $[0, C_{i}]$ is given by 
\begin{equation}
N_{i}(t) = \int_{0}^{t} Y_{i}(u) d\overline{N}_{i}(u), 
\label{CPobserved}
\end{equation}
with corresponding history $\mathcal{N}(t) =  \sigma \bigl( \bigl( N_{i}(s), Y_{i}(s) \bigl)_{s \leq t}: i=1, 2,...,n \bigr)$, generated by the right-censored counting processes and the at-risk processes. In contrast to the self-exciting filtration $\mathcal{\overline{N}}$, $\mathcal{N}$ contains not only information on past observed events but also on past censoring events. Eq. $(\ref{CPobserved})$ implies that the counting process is only allowed to jump when the at-risk process is equal to $1$ and the individual is under observation. Thus, $N_{i}(t)$ counts the number of observed events in $[0, t]$ and $dN_{i}(t)$ is the number of observed events in $[t, t+dt)$, respectively. The total number of events experienced by individual $i$ over $[0, C_{i}]$ is denoted by $N_{i}(C_{i}) := n_{i}$. The underlying intensity function $\lambda_{i}(t)$ of $N_{i}(t)$ with regard to $\mathcal{N}$ is given via
\begin{equation}
\label{intensityfunction2} \lambda_{i}(t) dt \ = P(dN_{i}(t) = 1 \ | \ \mathcal{N}(t-)).
\end{equation}

It is often of interest to relate the intensity function $(\ref{intensityfunction2})$ to baseline and time-dependent covariates. For example, in RCTs, one is interested in comparing different treatment groups with regard to the occurrence of repeated events. This requires defining a fixed indicator variable representing the treatment group of patients. Or, in MS disease, the risk of disability progression is likely to be increased for patients with high magnetic resonance imaging (MRI) activity which is measured on a regular basis during follow-up and, thus, defines a time-dependent covariate. \newline 
Given individual-specific follow-up periods $[0,C_{i}]$, let $X_{i}(t) = \{ X_{i}(u): 0 \leq u \leq t \}$ be the covariate process for individual $i$ which contains information on $p$ baseline and/or time-dependent covariates up to time $t$, with $X_{i}(t) = \bigl(X_{i1}(t), X_{i2}(t), ..., X_{ip}(t) \bigr) \in \mathbb{R}^{p}$ and $t \leq C_{i}$. Baseline covariates are measured at time origin $(t=0)$ and remain fixed over the course of time, whereas time-dependent (or time-varying) covariates may change their values over time. 
\newpage
Further, it is differentiated between external and internal covariates \parencite{Kalbfleisch2002}: 
\begin{definition}[External and internal covariates] 
External and internal covariates are also often denoted as exogeneous and endogeneous covariates, respectively. 
\begin{itemize}
\item \textbf{External covariate}: \newline 
An external covariate is a covariate that satisfies the following condition:
\begin{equation}
\label{extcov} P(dN_{i}(u)= 1 \ | \ X_{i}(u), \ \mathcal{N}(t-)) \ = \ P(dN_{i}(u)= 1 \ | \ X_{i}(t), \ \mathcal{N}(t-)) \ \textnormal{for} \ u \leq t. 
\end{equation}
This condition implies that external covariates may influence the risk of observing an event but its future path up to time $t > u$ is not affected by the occurrence of an event in $[u, u+du)$. Thus, values of external covariates are determined independently of the recurrent event process, as depicted in Figure $\textit{\ref{external}}$. \newline 
External covariates include both baseline and time-dependent covariates. With regard to time-dependent variables, it is additionally distinguished between 'defined' and 'ancillary' covariates. For a defined covariate, the covariate path can be completely determined in advance. For example, an individual's age is known at any time $t$ or the disease duration can be computed at any time $t$, provided that the time point of diagnosis is given at the outset of study. Baseline covariates can also be assigned to this class, as its constant paths are already known at time origin. An ancillary time-dependent covariate defines an observed path of a stochastic process whose development does not depend on the recurrent event process, e.g., level of air pollution. External covariates may also be observed beyond an individual's censoring time. 
\item \textbf{Internal covariate}: \newline 
Generally speaking, time-dependent covariates are classified as internal when they are not external and their path is influenced by the recurrent event process (cf. Figure $\textit{\ref{internal}}$). For this reason, the before mentioned condition $(\ref{extcov})$ is not fulfilled for internal covariates which can only be measured as long as the individual is under observation and uncensored. Examples of internal time-dependent covariates are measurements of disease indicators recorded at regular follow-up visits such as blood pressure, biomarkers or the volume of lesions in MS patients. In recurrent event analysis, the previous number of observed events $N_{i}(t-)$ or the time since the most recent event constitute internal time-varying covariates. 
\newline
\end{itemize} 
\end{definition}

\begin{figure}[htb]
    \centering
    \begin{minipage}{0.48\textwidth}
        \centering
\begin{displaymath} 
 \resizebox{7cm}{!}{$
    \xymatrix{ & dN(t_1)/N(t_1) \ar[r] & dN(t_2)/N(t_2) \ar[r] & ... \\
               X(t_1)  \ar[ur] \ar[r] & X(t_2) \ar[r] \ar[ur] & ... }$} 
\end{displaymath}
    \caption[Time-dependent external covariates]{Time-dependent external covariate}
    \label{external}
    \end{minipage}
    \begin{minipage}{0.48\linewidth}
        \centering
\begin{displaymath}
        \resizebox{7cm}{!}{$
    \xymatrix{ & dN(t_1)/N(t_1) \ar[r] \ar[d] & dN(t_2)/N(t_2) \ar[r] \ar[d] & ... \\
               X(t_1)  \ar[ur] \ar[r] & X(t_2) \ar[r] \ar[ur] & ... }$}
\end{displaymath}
    \caption[Time-dependent internal covariates]{Time-dependent internal covariate}
    \label{internal}
    \end{minipage}
\end{figure}

Sample paths of both external and internal covariates are included in $X_{i}(t)$. As a result, the covariate process can be decomposed into $X_{i}(t) = \{ (X_{i, ext}(u), X_{i, int}(u)): 0 \leq u \leq t \}$, where $X_{i, ext}(t)$ is a $p_{1}$-dimensional vector expressing external covariates and $X_{i, int}(t)$ is a $p_{2}$-dimensional vector reflecting internal covariates, with $p_{1}+p_{2}=p$. The covariate process is assumed to be left-continuous, meaning that the value of $X_{i}(t)$ is known just before time $t$ and only covariate information before time $t$ affects the intensity function exactly at time $t$. \newline 
In order to incorporate additional covariate information into the past, an extended history must be considered: $\mathcal{F}(t) =  \sigma \bigl( \bigl( N_{i}(s), Y_{i}(s), X_{i}(s) \bigl)_{s \leq t}: i=1,2,...,n \bigr)$. In addition to event and censoring information, $\mathcal{F}(t)$ also contains information on the external and internal covariates up to time $t$. It yields: $\mathcal{N}(t) \subset \mathcal{F}(t)$. The covariate process is also said to be predictable with respect to the filtration $\mathcal{F}$. The intensity function $\lambda_{i}(t)$ of $N_{i}(t)$ with regard to $\mathcal{F}$ is then defined via
\begin{equation}
\nonumber \lambda_{i}(t) dt \ = P(dN_{i}(t) = 1 \ | \ \mathcal{F}(t-)).
\end{equation}


\section{Conditional models}
\label{condmodel}
\begin{figure}[H]
\centering
\resizebox{14cm}{!}{
\begin{tikzpicture}[node distance=1.9cm]
  \tikzset{node style/.style={state, fill=gray!20!white, rectangle}}
        \node[node style]               (I)   {0};
        \node[node style, right=of I]   (II)  {1};
        \node[node style, right=of II]  (III) {2};
        \node[draw=none, right=of III]  (dot)  {$\cdots$};
        \node[node style, right=of dot] (IV)   {$j$};
        \node[node style, right=of IV] (V)   {$j+1$};
        \node[draw=none, right=of V]  (dottwo)  {$\cdots$}; 
    \draw[>=latex,
          auto=left,
          every loop]
         (I)   edge node {$\tiny \alpha_{01}(t)$} (II)
         (II)  edge node {$\alpha_{12}(t)$} (III)
         (III) edge node {$$} (dot)
         (dot) edge node {$$} (IV)
         (IV) edge node {$\tiny \alpha_{j(j+1)}(t)$} (V)
         (V) edge node {$$} (dottwo); 
\end{tikzpicture}}
\caption{Multistate representation of a recurrent event process}
\label{multistatemodel}
\end{figure}
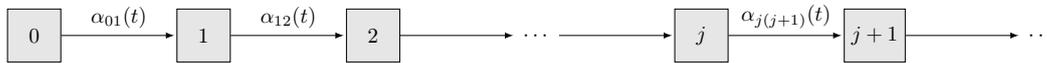

As already mentioned, recurrent event methods can be essentially divided into conditional and marginal approaches. Conditional models aim at providing deep insights into the structure of the recurrent event process and rely on intensity-based modelling, in which the intensity function can depend on arbitrary features of the preceding event history. Specifically, conditional models require full specification of the event process through explicit definitions of the past. In conditional models, the underlying situation can therefore be regarded as a special case of a multistate model based on counting processes \parencite{Andersen2019, Cook2007}. \newline 
Figure $\ref{multistatemodel}$ illustrates this multistate model adopted to the recurrent event setting in the absence of terminal events. States are represented by boxes and possible transitions between the states are depicted by arrows. Let $\bigr(N(t)\bigl)_{t \geq 0}$ be a multistate process in continuous time with right-continuous path (left-hand limits) and state space $\{0, 1, 2, ... \}$. $N(t)$ represents the state occupied by an individual at time $t$, where the index $i$ is dropped for notational convenience. In this specific multistate model, the multistate process $\bigr(N(t)\bigl)_{t \geq 0}$ also defines a counting process, with $N(t)$ representing the cumulative number of events experienced up to and including time $t$. For instance, being in state $0$ is interpreted as being 'event-free'. Individuals who have already experienced one event are in state $1$ and individuals in state $j$ are at-risk for a $(j+1)^{th}$ event. In this setting, $T_{j}$ can be seen as the time of entry into state $j$ and $N(t)=j$ indicates that $T_{j} \leq t < T_{j+1}$, for $j=1,2, ... $ . As seen from Figure $\ref{multistatemodel}$, only transitions from state $j-1$ to state $j$ are possible so that state $j$ can only be reached by individuals who have already experienced $(j-1)$ events. Occurrence of a $j^{th}$ event is modelled as a $(j-1) \longrightarrow j$ transition. Thus, each time an event happens, the individual leaves its current state and moves to the next event state. All individuals are event-free at time origin $t=0$ (e.g., randomization), so that each individual under study starts in state $0$, i.e., $P(N(0)=0)=1$. As a result, there is one common initial state $0$. Since individuals can experience an arbitrary finite number of events, the multistate model does not have a common absorbing state. \newline
If the state space of the multistate process is restricted to $\{ 0, 1 \}$ only, the multistate model considered reduces to a multistate model that describes the conventional time-to-first-event setting (cf. Figure $\ref{MultiStateModelSurvival}$). Generally, the multistate model adapted to the recurrent event setting can be seen as a generalization of the time-to-first-event setting. 

\noindent
Following general multistate model principles, the transition hazard $\alpha_{j(j+1)}(t)$ for the recurrent event process (or multistate process) is defined as follows:
\begin{align}
\nonumber \alpha_{j(j+1)}(t)dt &=  P(N(t+dt) = j+1 \ | \ N(t-)=j, \ \textnormal{past}).   
\end{align}
If the process depends on the past only through the cumulative number of events at $t-$, then the process is Markov and the transition hazard reduces to 
\begin{align}
\nonumber \alpha_{j(j+1)}(t)dt &= P(N(t+dt) = j+1 \ | \ N(t-)=j, \ \textnormal{past})\\
\label{hazard} &\underset{\text{Markov}}{=} P(N(t+dt) = j+1 \ | \ N(t-)=j).
\end{align}
The Markov property states that the risk for a $j \longrightarrow j+1$ transition depends on the current state $j$ and time $t$ since time origin but not on the entry time into state $j$. \newline
Let $N_{j(j+1)}(t)$ be a transition-specific counting process indicating the number of $j \longrightarrow (j+1)$ transitions in the time interval $[0,t]$, for $j = 0, 1, ...$ . Under the assumption of no tied data, the increment $dN_{j(j+1)}(t) := N_{j(j+1)}(t) - N_{j(j+1)}(t-) \in \{0,1\}$ is equal to $1$, if a $j \longrightarrow (j+1)$ transition is observed to happen at time $t$. The at-risk process for transitions out of state $j$ is given by $Y_{j}(t) = \mathbbm{1}(N(t-)=j, C \geq t)$. To be more specific, $Y_{j}(t) = 1$ means that the multistate process is in state $j$ just before time $t$, $t-$, and may be either observed to move out of state $j$ at time $t$ or to remain under observation in state $j$. As already defined, $\mathcal{F}$ corresponds to the history generated by the recurrent event process. 
\newline 
The key-quantities of the multistate model are the transition-specific intensities of the counting processes $N_{j(j+1)}$ which take the following form: 
\begin{align}
\nonumber \lambda_{j(j+1)}(t)dt &:= P(dN_{j(j+1)}(t) = 1 \ | \ \mathcal{F}(t-)) \\
\nonumber &= \ P(N(t+dt) = j+1 \ | \ N(t-)=j, \mathcal{F}(t-)) \\
\label{intensity} &= Y_{j}(t) \cdot \alpha_{j(j+1)}(t)dt, 
\end{align}
with $\alpha_{j(j+1)}(t)$ to be estimated. 
The intensity $\lambda_{j(j+1)}(t)dt$ can be seen as the instantaneous probability of making a $j \longrightarrow (j+1)$ transition or of observing a $(j+1)^{th}$ event in $[t, t+dt)$, given past information. 
The 'overall' counting process $N= \{N(t): t \in [0, \infty) \}$ has the event intensity function 
\begin{align}
\nonumber \lambda(t)dt &:= P(dN(t) = 1 \ | \ \textnormal{past}) \ = \ P(dN(t) = 1 \ | \ N(t-), \mathcal{F}(t-)) \\
\nonumber &= \left\{ \begin{array}{ll} \alpha_{01}(t)dt, & N(t-)=0 , \mathcal{F}(t-) \\
         \alpha_{12}(t)dt, & N(t-)=1, \mathcal{F}(t-)  \\
         ...  & ... \\
         \alpha_{j(j+1)}(t)dt, & N(t-)= j, \mathcal{F}(t-)  \\
         ...  & ... \\
         \end{array} \right. \\
\nonumber &= \sum_{j=0}^{\infty} Y_{j}(t) \cdot \alpha_{j(j+1)}(t)dt. 
\end{align}
The intensity function $\lambda(t)dt$ can be interpreted as the instantaneous probability of observing any event in the small time period $[t, t+dt)$, given the past history. 
\newline \newline 
As intuitively clear from Eq. $(\ref{hazard})$ and Eq. $(\ref{intensity})$, the transition hazard $\alpha_{j(j+1)}(t)$ can depend on the number of events $j$ that have already happened for this patient before time t. This fact further implicates that the order of recurrent events is usually preserved in conditional models. In a more general sense, the transition hazard can depend on any component of the recurrent event process history before time $t$ (e.g., time $t$, previous event times $T_{1}, T_{2},..., T_{j}$, ...). Depending on the specific assumptions on the target quantities $\alpha_{j(j+1)}(t)$, different (conditional) recurrent event models can be formulated: 
\\
\begin{itemize}
\item \textbf{Poisson models}: \ \ $\alpha_{j(j+1)}(t) = \alpha_{0}(t) \cdot r(\beta, X_{ext}(t))$ \ \ or \ \ $\alpha_{j(j+1)}(t) = \alpha_{0} \cdot r(\beta, X_{ext})$,
\hspace*{32.5mm} with $\beta$ regression coefficient and $r(\beta, X_{ext})$ relative risk function
\item \textbf{AG model}: \ \ $\alpha_{j(j+1)}(t) = \alpha_{0}(t) \cdot r(\beta, X(t))$ 
\item \textbf{PWP-CP model}: \ \ $\alpha_{j(j+1)}(t) = \alpha_{0(j+1)}(t) \cdot r(\beta, X(t))$
\item \textbf{PWP-GT model}: \ \ $\alpha_{j(j+1)}(t) = \alpha_{0(j+1)}(t-T_{N(t-)}) \cdot r(\beta, X(t))$
\item \textbf{NB models}: $\alpha_{j(j+1)}(t \ | \ U) = U \cdot \alpha_{0}(t) \cdot r(\beta, X_{ext}(t))$ \ \ or \ \  $\alpha_{j(j+1)}(t \ | \ U) = U \cdot \alpha_{0} \cdot r(\beta, X_{ext})$, \hspace*{20.5mm} U gamma distributed random effect
\newline
\end{itemize}

In the following, the models mentioned above, except for the PWP-GT, will be discussed in more detail. Gap time models are out of the scope of this thesis. 
\newline \newline 
Conditional models can further grouped into classical intensity-based models and random effect (or frailty) models, differing in adjusting for intra-individual correlation. The classical intensity-based model accounts for the dependence structure between recurrent events by regressing the hazard function on information of previous events via internal and external time-varying covariates. Extensions of classical intensity-based models are random effect models, in which an unobserved random effect (or frailty term) additionally induces dependency among repeated events. 

\newpage 
\subsection{Intensity-based models}
The Poisson, AG, and PWP-CP models are general intensity-based models. In this work, Poisson models are introduced as a special case of conditional intensity-based models, as the AG model generalizes the Poisson model. However, Poisson models can also be seen as marginal rate-based models, as described later in Section $\ref{marginalmodel}$. 

\subsubsection{Poisson models}
\label{SectionPoissonModels}
Poisson models are statistical models, where the transition hazard (or intensity function)  does not depend on the prior event history. As defined in Section $\ref{condmodel}$, the transition hazard $\alpha_{j(j+1)}(t)$ of experiencing a $(j+1)^{th}$ event does neither depend on $j$ nor on occurrence or timing of previous events $T_{1}, T_{2}, ..., T_{j}$. Because of its independence on $j$, the transition hazard for any event can be simply denoted by $\alpha(t)$, omitting the indices. That is, the intensity function $\lambda_{j(j+1)}(t) = Y_{j}(t)\alpha_{j(j+1)}(t)$ can be easily written as $\lambda(t) = Y(t)\alpha(t)$. For Poisson models, the underlying counting process is a Poisson process with $\mathbb{E}(N(t))= Var(N(t))$ and the following properties: 
\begin{definition}[Poisson process]
\label{DefPoissonProcess}
A counting process $N(t) = \{ N(t): 0 \leq t < \infty\}$ is a non-homogeneous Poisson process with parameter $\alpha(t)$, if 
 \begin{itemize}
  \item N(0) = 0 
  \item Given $t_{2} < t_{3}$, $N(t_{1}, t_{2})$ is independent of $N(t_{3}, t_{4})$, where $N(t_{1}, t_{2}) := N(t_{2}) - N(t_{1}-)$ is the number of events in $(t_{1}, t_{2}]$, respectively. In other words, if $(t_{1}, t_{2}]$ and $(t_{3}, t_{4}]$ are non-overlapping intervals, then $N(t_{1}, t_{2})$ and $N(t_{3}, t_{4})$ are independent.
  \item For $0 \leq t_{1} < t_{2}$, $N(t_{1}, t_{2})$ is Poisson distributed with mean $\mu(t_{1}, t_{2}) = \mu(t_{2}) - \mu(t_{1}) = \int_{t_{1}}^{t_{2}} \alpha(u) du$ , where $\mu(t_{1}) = \int_{0}^{t_{1}} \alpha(u) du$ and $\mu(t_{2}) = \int_{0}^{t_{2}} \alpha(u) du$, respectively. That is,
  \begin{align}
  \label{PoissonEQ} N(t_{1}, t_{2}) \sim Poisson \biggl( \int_{t_{1}}^{t_{2}} \alpha(u) du \biggr)  = Poisson \biggl( \mu(t_{1},t_{2}) \biggr).
  \end{align}
It follows for $j = 0, 1, ... :$  $P(N(t) = j) = \dfrac{\mu(t)^{j}}{j!} \exp(-\mu(t))$, with $N(t) = N(0,t)$.
\end{itemize}
\end{definition}
A non-homogeneous Poisson process modulated by external covariates can also be described by its intensity function: 
\begin{align}
\lambda_{i}(t)dt &:= P(dN_{i}(t) = 1 \ | \ \underbrace{\mathcal{F}(t-)}_{ = \ \bigl(N_{i}(s), \ Y_{i}(s), \ X_{i, ext}(s), \ X_{i, int}(s)\bigr)_{s < t}}) \\ 
\nonumber &\underset{Poisson}{=} P(dN_{i}(t) = 1 \ | \ (Y_{i}(s), X_{i, ext}(s))_{s < t}). 
\end{align}
This conditionally independent increment property of the Poisson process implies that, given $X_{i,ext}(t)$, the instantaneous probability of an event in $[t, t+dt)$ does not depend on the preceding event history and internal time-varying covariates \parencite{Zhong2019, Cook2007}. If the past includes information on internal covariates (e.g., time since the most recent event or the number of previous events), the recurrent event process is no longer Poisson.\newline 
In particular, Poisson models with 
\begin{align}
\label{PoissonModel} \lambda_{i}(t) = Y_{i}(t) \alpha(t) = Y_{i}(t) \alpha_{0}(t) \exp(\beta^\intercal Z_{i, ext}(t))
\end{align}
ensures multiplicative effects of $Z_{i,ext}(t)$ on the hazard, where $Z_{i,ext}(t)$ is a $q$-dimensional vector of functions of the external covariates $X_{i, ext}(t)$. With regard to the Poisson model $(\ref{PoissonModel})$, $Y_{i}(t) = \mathbbm{1}(C_{i} \geq t)$ is the at-risk process, $\alpha_{0}(t)$ a positive-valued baseline hazard function corresponding to individuals with $Z_{i, ext}(t)=0 \ \forall \ t > 0$ and $\beta \in \mathbb{R}^q$ is a vector of unknown regression coefficients. The baseline hazard $\alpha_{0}(t)$ can be either specified parametrically or non-parametrically. If $\alpha_{0}(t)$ is an arbitrary function, the semiparametric model $(\ref{PoissonModel})$ is a special example of the AG model, and statistical estimation procedures for $\beta$ will be discussed in Section $\ref{IntensityAGmodel}$. In case of a parametric baseline hazard, usual maximum likelihood estimation can be used to estimate $\beta$. 
\newpage
\textbf{Special case: homogeneous Poisson process} \newline
The special case of a homogeneous Poisson process is obtained by assuming constant transition hazards that do not depend on time $t$. That is, with regard to the general multistate setup, $\alpha_{01}(t) = \alpha_{12}(t) = .... = \alpha_{j(j+1)}(t) \equiv \alpha$ with $\alpha > 0$. 
If $\{ N(t): 0 \leq t < \infty \}$ is a time-homogeneous Poisson process with parameter $\alpha > 0$, the gap times $G_{j}$ between successive events are independent and identically (iid) exponential distributed random variables with mean $\alpha^{-1}$ and survival function $P(G_{j} > g) = \exp(- \alpha g)$, for $g > 0$ and $j=1, 2, ...$ . That is, $G_{j} \sim \text{Exp}(\alpha)$. 
\newline 
Under this parametric Poisson model, the intensity function for any event is defined by
\begin{align}
\label{parametricPoisson} \lambda_{i}(t) = Y_{i}(t) \alpha = Y_{i}(t) \alpha_{0} \exp(\beta^\intercal Z_{i, ext}), 
\end{align}
where $\alpha_{0} > 0$ is a constant baseline hazard corresponding to individuals with $Z_{i,ext}=0$ and $Z_{i,ext} \in \mathbb{R}^q$ is a $q$-dimensional vector of baseline covariates. 
\newline \newline 
\textbf{Inference for $\boldsymbol{\beta}$} \newline
Maximum likelihood estimation of $\beta$ in the parametric Poisson model $(\ref{parametricPoisson})$ with constant hazards is based on the following theorem, which yields for any counting process not only for Poisson processes.

\begin{theorem}[Likelihood contribution for individual $i$]
\label{likelihoodcontri}
Conditional on the past \newline $\mathcal{F}(t-)$, the probability density function of the outcome '$n_{i}$ events at times $t_{i1}, ..., t_{in_{i}}$' for a process with intensity $\lambda_{i}(t)dt = P(dN_{i}(t)=1 \ | \ \mathcal{F}(t-))$ over $[0, C_{i}]$ is 
\begin{align}
\prod_{j=1}^{n_{i}} \lambda(t_{ij}) \exp\biggl( - \int_{0}^{C_{i}} \lambda_{i}(u) du \biggr), \ \ \ \ i=1,2,...,n \ . 
\end{align}
Under independent and non-informative censoring, the log-likelihood contribution for individual $i$ having $n_{i}$ events at times $t_{i1} < t_{i2} < ... < t_{in_{i}}$ over the observation period $[0, C_{i}]$ is then  
\begin{align}
\int_{0}^{\infty} Y_{i}(t) \bigl[ \log(\lambda_{i}(t))dN_{i}(t) - \lambda_{i}(t)dt\bigr], \ \ \ \ i=1,2,...,n \ . 
\end{align}
The proof for this statement can be found in \citet{Cook2007}. 
\end{theorem}
According to Theorem $\ref{likelihoodcontri}$, the log-likelihood function for model $(\ref{parametricPoisson})$ is given by 
\begin{align}
\label{LLPoisson} \ell^{Poisson}(\alpha_{0}, \beta) &= \sum_{i=1}^{n} \int_{0}^{\infty} Y_{i}(t) \bigl[ \log(\alpha_{0} \exp(\beta^\intercal Z_{i, ext}))dN_{i}(t) - \alpha_{0} \exp(\beta^\intercal Z_{i, ext})dt \bigr] \\
\nonumber &= \sum_{i=1}^{n} \biggl( n_{i} (\log(\alpha_{0}\exp(\beta^\intercal Z_{i, ext}))) - C_{i}\alpha_{0}\exp(\beta^\intercal Z_{i, ext}) \biggr) \\
\nonumber &=  \sum_{i=1}^{n} \biggl( n_{i} (\log(\alpha_{0}) + \beta^\intercal Z_{i, ext}) - \exp(\log(C_{i}) + \log(\alpha_{0}) + \beta^\intercal Z_{i, ext}) \biggr), 
\end{align}
which is essentially proportional to a Poisson log-likelihood under the assumption of $n_{i} := N_{i}(C_{i}) \sim Poisson(C_{i}\exp(\beta_{0} + \beta^\intercal Z_{i, ext}))$ with $\beta_{0} = \log(\alpha_{0})$. The score functions can be derived as 
\begin{align}
\nonumber U^{Poisson}_{\alpha_{0}} &= \frac{\partial}{\partial \alpha_{0}} \ell^{Poisson}(\alpha_{0}, \beta) = \sum_{i=1}^{n} \biggl(\dfrac{n_{i}}{\alpha_{0}} - C_{i}\exp(\beta^\intercal Z_{i, ext}) \biggr) \\
\nonumber U^{Poisson}_{\beta} &= \frac{\partial}{\partial \beta} \ell^{Poisson}(\alpha_{0}, \beta) = \sum_{i=1}^{n} Z_{i, ext} \bigl( n_{i} - C_{i} \alpha_{0}\exp(\beta^\intercal Z_{i, ext}) \bigr).
\end{align}
The profile likelihood estimate $\tilde{\alpha_{0}}(\beta)$ results from solving $U^{Poisson}_{\alpha_{0}} = 0$. In order to obtain the maximum likelihood estimate $\hat{\beta}$, the profile log-likelihood function $\ell^{Poisson}(\tilde{\alpha_{0}}(\beta), \beta)$ obtained by plugging $\tilde{\alpha_{0}}(\beta)$ into Eq. $(\ref{LLPoisson})$ is maximized with regard to $\beta$. Finally, the maximum likelihood estimate $\hat{\alpha_{0}}$ is obtained by inserting $\hat{\beta}$ into $\tilde{\alpha_{0}}(\beta) \ \Longrightarrow \hat{\alpha_{0}} = \tilde{\alpha_{0}}(\hat{\beta})$. 
\newline
Poisson models assume that all repeated events occur conditionally independent of each other, regardless of whether events have been experienced by the same individual or from different individuals, and that occurrence of an event does not alter the instantaneous probability for a next event. 

\subsubsection{Andersen-Gill model}
\label{IntensityAGmodel}
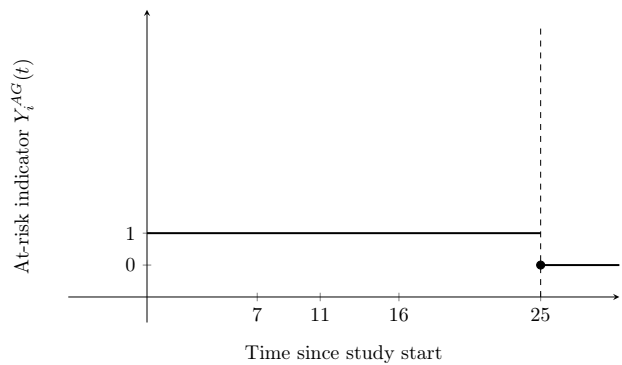
\begin{figure}[H]
\centering
\scalebox{0.75}{ 
\begin{tikzpicture}
    \begin{axis}[height=0.3\textheight, width=0.75\textwidth, xmin=-5.0, xmax=30, ymin=-0.8, ymax=9, xtick={7,11,16,25}, xticklabels={7, 11, 16, 25}, ytick={1, 2}, yticklabels={0,1}, xlabel={Time since study start}, ylabel={At-risk indicator $Y_{i}^{AG}(t)$},y label style={at={(axis description cs:-0.05,.5)},rotate=90,anchor=south},
    x label style={at={(axis description cs:0.5,-0.05)},anchor=north}]
        \addplot[cmhplot,-,domain=25:30]{1};
        \addplot[cmhplot,-,domain=0:25]{2};
        \addplot[soldot]coordinates{(25,1)};
        \addplot[dashed]coordinates {(25,0) (25,8.5)};
    \end{axis}
\end{tikzpicture}}
\caption[At-risk indicator under the Andersen-Gill model]{At-risk indicator under AG model for individual $B$ from hypothetical example}
\label{AtRiskAG}
\end{figure}

The AG model is one of the most famous recurrent event models and can be seen as a generalization of the Cox proportional hazards model proposed by \citet{Cox1972} for time-to-first-event endpoints. \newline
In a Poisson model, the hazard (or intensity) function is independent of the past event history and is only regressed on baseline and external time-dependent covariates. However, in practice, the intensity function often additionally depends on the preceding event history and internal time-dependent covariates in a complex manner. In comparison to Eq. $(\ref{PoissonModel})$, the intensity function for an event is then specified as
\begin{align}
\nonumber \lambda_{i}(t)dt &= P(dN_{i}(t) = 1 \ | \ \mathcal{F}(t-)) \\ 
\nonumber &=  P(dN_{i}(t) = 1 \ | \ (N_{i}(s), Y_{i}^{AG}(s), X_{i, ext}(s), X_{i, int}(s))_{s < t}).
\end{align}
For instance, an individual who experiences an event at time $t$ is more likely to experience the next event than individuals with no occurrences at time $t$. In general, occurrence of a $j^{th}$ event modifies the probability of experiencing a $(j+1)^{th}$ event. The AG model is essentially flexible enough to incorporate aspects of the past event history into the recurrent event analysis, modelled as regression variables in the relative risk function of the model equation \parencite{Andersen1982}. \newline \newline  
In order to allow for dependence among repeated events, \citet{Andersen1982} proposed a semiparametric proportional intensity model for recurrent events of the following form
\begin{align}
\label{AGintensitycondmodel} \lambda_{i}(t) = Y_{i}^{AG}(t) \alpha_{i}(t) &= Y_{i}^{AG}(t) \alpha_{0}(t) \exp( \beta^\intercal Z_{i}(t)), 
\end{align}
where $\alpha_{0}(t)$ is an unspecified baseline hazard common for all events. $Y_{i}^{AG}(t)  = \mathbbm{1}(C_{i} \geq t)$ is the at-risk process taking values in $\{0,1 \}$, $\beta \in \mathbb{R}^{q}$ a vector of regression coefficients and $Z_{i}(t)=(Z_{i1}(t), ..., Z_{iq}(t)) \in \mathbb{R}^{q}$ is comprised of functions of external and internal covariates, $X_{i,ext}(t)$ and $X_{i,int}(t)$, the past event history $N_{i}(t-) = \{ N_{i}(u): 0 \leq u < t \}$ and interaction with time $t$. More specifically, $\alpha_{0}(t)$ can be any integrable, non-negative function with $\int_{0}^{t} \alpha_{0}(u)du < \infty$ and corresponds to individuals for whom $Z_{i}(t)=0 \ \forall \ t$. 
The combination of the non-parametric baseline hazard and the parametric part $\exp( \beta^\intercal Z_{i}(t))$ justifies the semiparametric property of the AG model. For individual $B$ from the hypothetical example used in Figure $\ref{GraphicRE}$, the underlying at-risk process is depicted in Figure $\ref{AtRiskAG}$. In contrast to the conventional Cox model, where individuals are no longer at-risk for the event of interest after occurrence of the first event $(Y_{i}^{Cox}(t) = \mathbbm{1}(\min\{T_{i1}, C_{i} \} \geq t)$, cf. Figure $\ref{AtRiskCox})$, the AG model considers individuals to be at-risk and under observation throughout the whole follow-up period $[0, C_{i}]$. This implies that patients who experience an event remain in the at-risk set for further events. It is illustrated in Figure $\ref{AtRiskAG}$ that the at-risk process continues to be equal to $1$ after each event occurrence (at times $7$, $11$ and $16$) and jumps to $0$ at the right-censoring time $C_{i} = 25$ of individual $B$. 
\newline 
To sum up, the AG model uses counting process formulation, a common baseline hazard and an unrestricted risk set. 

\newpage
\textbf{Specification of $\boldsymbol{Z_{i}(t)}$} \newline 
In the AG model $(\ref{AGintensitycondmodel})$, $Z_{i}(t)$ is a time-dependent component capturing the dependence between recurrent events and can be specified according to the specific structure of the recurrent event process \parencite{Kalbfleisch2002, Cook2007}:

\begin{itemize}
\item If $Z_{i}(t)$ contains only functions of baseline and external time-dependent covariates, the underlying process reduces to a non-homogeneous Poisson process modulated by covariates (cf. Section $\ref{SectionPoissonModels}$).
\item $Z_{i}(t)$ may contain the number of previous events $N_{i}(t-)$. Then, for instance, the AG model is defined via 
  \begin{align}
    \label{specialAGmodel} \lambda_{i}(t) &= Y_{i}^{AG}(t) \alpha_{0}(t) \exp( \beta N_{i}(t-)) \\
    \nonumber &= Y_{i}^{AG}(t) \alpha_{0}(t) \underbrace{\exp(\beta) \exp(\beta) \cdot ... \cdot \exp(\beta)}_{N_{i}(t-) \ \textnormal{times}}.
  \end{align}
Thus, the instantaneous probability of experiencing an event at time $t$ alters by a multiplicative constant factor of $\exp(\beta)$ following each occurrence of an event, as compared to individuals with no prior events at time $t$. If $\beta > 0$, the intensity for an event increases, whereas the intensity decreases for $\beta < 0$. If $\beta=0$, the intensity function is not affected by the number of previous events. \newline
With regard to the general multistate setup, the AG model $(\ref{specialAGmodel})$ results from assuming $\alpha_{j(j+1)}(t) = \alpha_{(j-1)j}(t)\exp(\beta)$ with $\alpha_{01}(t) = \alpha_{0}(t)$. \newline
\citet{Aalen2008} recommend to use $N_{i}(t-)/t$ or $\log(N_{i}(t-)/t)$ to avoid explosion. 
\item Amongst other covariates, $Z_{i}(t)$ may also include $\mathbbm{1}(N_{i}(t-) = 1)\gamma_{1} + \mathbbm{1}(N_{i}(t-) = 2)\gamma_{2} + ...$, in which case the intensity function is allowed to increase or decrease by a multiplicative factor $\exp(\gamma_{j})$ after occurrence of the $j^{th}$ event, as compared to individuals who are event-free.  
  \begin{align}
    \nonumber \lambda_{i}(t) &= Y_{i}^{AG}(t) \alpha_{0}(t) \exp( \beta^\intercal Z_{i}(t) + \mathbbm{1}(N_{i}(t-) = 1)\gamma_{1} + \mathbbm{1}(N_{i}(t-) = 2)\gamma_{2} + ...) \\
    \nonumber &= Y_{i}^{AG}(t) \alpha_{0}(t) \exp( \beta^\intercal Z_{i}(t)) \underbrace{\exp(\mathbbm{1}(N_{i}(t-) = 1)\gamma_{1}) \exp(\mathbbm{1}(N_{i}(t-) = 2)\gamma_{2}) \cdot ...}_{= \left\{ \begin{array}{ll} 1, & N_{i}(t-)=0  \\
         \exp(\gamma_{1})  & N_{i}(t-)=1 \\
         ... & ... \\
         \exp(\gamma_{j}) , & N_{i}(t-)= j  \\
         ...  & ... \\
         \end{array} \right.}
  \end{align}
\item Definition of the covariate part by $\mathbbm{1}(N_{i}(t-) = 1)\beta_{1}^\intercal Z_{i}(t) + \mathbbm{1}(N_{i}(t-) = 2)\beta_{2}^\intercal Z_{i}(t) + ...$ allows the multiplicative effect of both baseline and external time-dependent covariates to arbitrarily depend on the number of previous events. 
\item Adding $\mathbbm{1}(\textnormal{individual i has experienced an event during} \ [t - 12 \ \textnormal{weeks}, t))\gamma$ to $Z_{i}(t)$, it is assumed that the instantaneous probability of an event at time $t$ varies by a multiplicative factor of $\exp(\gamma)$ with presence of an event within the last $3$ months. 
\end{itemize}

\newpage 
\textbf{Inference for $\boldsymbol{\beta}$ and large sample theory} \newline
For the following derivation of the estimation procedure for $\beta$ and the large sample theory, let the true underlying intensity function be given by $\lambda_{i}(t) = Y_{i}^{AG}(t) \alpha_{0}(t) \exp( \beta_{true}^\intercal Z_{i}(t))$. $\beta_{true}$ denotes the true $q$-dimensional regression coefficient vector. As developed by \citet{Cox1972}, partial likelihood functions are useful frameworks for the estimation of $\beta \in \mathbb{R}^{q}$ within the intensity-based AG approach. Under the assumption of no tied recurrent event data (i.e., different counting processes do not jump at the same time), the basis of the partial likelihood function is of the following expression
\setlength\jot{0.3cm}
\begin{align}
\nonumber \pi_{i}(t, \beta) &= P(dN_{i}(t) = 1 \ | \ \mathcal{F}(t-), dN.(t) = 1) \\
\nonumber &= \dfrac{P(dN_{i}(t) = 1, dN.(t) = 1 \ | \ \mathcal{F}(t-))}{P(dN.(t) = 1 \ | \ \mathcal{F}(t-))} \ \ \stackrel{=}{\textnormal{no ties}} \ \ \dfrac{P(dN_{i}(t) = 1 \ | \ \mathcal{F}(t-))}{P(dN.(t) = 1 \ | \ \mathcal{F}(t-))} \\
\nonumber &= \dfrac{Y_{i}^{AG}(t) \exp( \beta^\intercal Z_{i}(t))}{\sum_{l=1}^{n} Y_{l}^{AG}(t) \exp(\beta^\intercal Z_{l}(t))}. 
\end{align}
$\pi_{i}(t, \beta)$ is defined as the conditional probability that it is individual $i$ who has an event at time $t$, conditional on observing an event at time $t$ and the observed history $\mathcal{F}(t-)$. In addition, $N.(t) := \sum_{i=1}^{n} N_{i}(t)$ is the aggregated observed counting process over all individuals $i$, with corresponding increment defined as $dN.(t) := \sum_{i=1}^{n} dN_{i}(t)$. Then, the partial likelihood function is given by 
\begin{align}
\nonumber L^{AG}(\beta) = \prod_{u \leq \tau} \ \prod_{i=1}^{n} \bigl( \pi_{i}(u, \beta) \bigr)^{dN_{i}(u)} = \prod_{u \leq \tau} \ \prod_{i=1}^{n} \biggl(  \dfrac{ \exp( \beta^\intercal Z_{i}(u))}{\sum_{l=1}^{n} Y_{l}^{AG}(u) \exp(\beta^\intercal Z_{l}(u))} \biggr)^{dN_{i}(u)}, 
\end{align}
where $\beta$ is the argument of the likelihood function and the first product is defined over all unique repeated event times $u \leq \tau$, with $\tau < \infty$. The first term reduces to the second one due to the following fact: $dN_{i}(u)=1 \ \Longrightarrow Y_{i}^{AG}(u) = 1$. The resulting log-likelihood function is of the form 
\begin{align}
\nonumber \ell^{AG}(\beta) &= \log \bigl(L^{AG}(\beta) \bigr) = \sum_{u \leq \tau} \ \sum_{i=1}^{n} \underbrace{dN_{i}(u)}_{\in \{ 0,1 \}} \biggl[ \beta^\intercal Z_{i}(u) - \log \biggl( \sum_{l=1}^{n} Y_{l}^{AG}(u) \exp(\beta^\intercal Z_{l}(u)) \biggr) \biggr ] \\
&= \sum_{i=1}^{n} \int_{(0, \tau]} \biggl[ \beta^\intercal Z_{i}(u) - \log \bigl( S^{(0, AG)}(\beta, u) \bigr) \biggr] dN_{i}(u), 
\end{align}
where 
\begin{align}
\nonumber S^{(0), AG}(\beta, t) &= \sum_{i=1}^{n} Y_{i}^{AG}(t) \exp( \beta^\intercal Z_{i}(t)),  \\
\nonumber S^{(1), AG}(\beta, t) &= \sum_{i=1}^{n} Y_{i}^{AG}(t) Z_{i}(t) \exp( \beta^\intercal Z_{i}(t)), \\
\nonumber S^{(2), AG}(\beta, t) &= \sum_{i=1}^{n} Y_{i}^{AG}(t) Z_{i}(t)^{\otimes 2} \exp( \beta^\intercal Z_{i}(t)).
\end{align}
The vectors of the score functions and the observed information matrix are given as 
\begin{align}
\nonumber U^{AG}(\beta) &= \frac{\partial}{\partial \beta} \ell^{AG}(\beta) = \sum_{i=1}^{n} \int_{(0, \tau]} \biggl( Z_{i}(u) - \dfrac{S^{(1), AG}(\beta, u)}{S^{(0), AG}(\beta, u)} \biggr) dN_{i}(u) \\
\nonumber I^{AG}(\beta) &= - \frac{\partial}{\partial \beta^\intercal} U^{AG}(\beta) = \int_{(0, \tau]} \dfrac{S^{(2), AG}(\beta, u)}{S^{(0), AG}(\beta, u)} - \biggl(\dfrac{S^{(1), AG}(\beta, u)}{S^{(0), AG}(\beta, u)}\biggr)^{\otimes 2} dN(u). 
\end{align}
For $\beta_{true}$, it yields: $dN_{i}(t) = \lambda_{i}(t)dt + dM_{i}(t) = Y_{i}^{AG}(t) \alpha_{0}(t) \exp( \beta_{true}^\intercal Z_{i}(t)) dt + dM_{i}(t)$, with $M_{i}(t)$ as a martingale. Using this Doob-Meyer decomposition and properties of vector-valued stochastic integrals, it can be shown that $U^{AG}(\beta_{true})$ is a zero-mean martingale with $\mathbb{E}(U^{AG}(\beta_{true})) = 0$, i.e., 

\begin{align}
\nonumber U^{AG}(\beta_{true}) &= \sum_{i=1}^{n} \int_{(0, \tau]} \biggl( Z_{i}(u) - \dfrac{S^{(1), AG}(\beta_{true}, u)}{S^{(0), AG}(\beta_{true}, u)} \biggr) (Y_{i}^{AG}(u) \alpha_{0}(u) \exp( \beta_{true}^\intercal Z_{i}(u))du + dM_{i}(u)) \\
\nonumber &= \int_{(0, \tau]} \biggl[ \underbrace{\sum_{i=1}^{n} \biggl( Z_{i}(u) - \dfrac{S^{(1), AG}(\beta_{true}, u)}{S^{(0), AG}(\beta_{true}, u)} \biggr) (Y_{i}^{AG}(u) \exp( \beta_{true}^\intercal Z_{i}(u))}_{=0}\biggr] \alpha_{0}(u)du \\
\nonumber &+ \int_{(0, \tau]} \sum_{i=1}^{n} \biggl( Z_{i}(u) - \dfrac{S^{(1), AG}(\beta_{true}, u)}{S^{(0), AG}(\beta_{true}, u)} \biggr) dM_{i}(u) \\
\nonumber &= \sum_{i=1}^{n} \int_{(0, \tau]} \biggl( \underbrace{Z_{i}(u) - \dfrac{S^{(1), AG}(\beta_{true}, u)}{S^{(0), AG}(\beta_{true}, u)}}_{\textnormal{predictable}} \biggr) dM_{i}(u).
\end{align}

Finally, the estimated regression coefficient vector $\hat{\beta}$ is defined as the solution to the equation
\begin{align}
\nonumber U^{AG}(\beta) &\stackrel{!}{=} 0.
\end{align}
Under some regularity  conditions (cf. \citet{Andersen1982}), the probability that $U^{AG}(\beta)$ has an unique solution $\hat{\beta}$ tends to $1$ and $\hat{\beta}$ converges in probability to $\beta_{true}$, i.e., $\hat{\beta} \ \xrightarrow[n \longrightarrow \infty]{P} \ \beta_{true}$. This implies that $\hat{\beta}$ is a consistent estimator for $\beta_{true}$. Since the vector of score functions evaluated at $\beta_{true}$ is a zero-mean martingale, Rebolledo's martingale central limit theorem and martingale-based partial likelihood theory can be used to prove that 
\begin{align}
\label{C} &\dfrac{1}{\sqrt{n}} U^{AG}(\beta_{true}) \ \xrightarrow[n \longrightarrow \infty]{D} \ N \bigl (0, \Sigma \bigr)  \ \ \ \ \textnormal{with} \\
\nonumber &\Sigma =  \int_{(0, \tau]} \biggr( \dfrac{s^{(2), AG}(\beta_{true}, u)}{s^{(0), AG}(\beta_{true}, u)} - \biggl(\dfrac{s^{(1), AG}(\beta_{true}, u)}{s^{(0), AG}(\beta_{true}, u)}\biggr)^{\otimes 2} \biggl) s^{(0), AG}(\beta_{true}, u) \alpha_{0}(u)du \ \ \ \ \textnormal{and}, \\
\nonumber &\dfrac{1}{n} \mathbb{E}(I^{AG}(\beta_{true})) \ \xrightarrow[n \longrightarrow \infty]{P} \  \Sigma, 
\end{align}
where $s^{(m), AG}(\beta_{true}, u)$ is the limit of $S^{(m), AG}(\beta_{true}, u) \ \forall \ m=0, 1, 2$. Therefore, under certain conditions, the vector of score functions $n^{-1/2} U^{AG}(\beta_{true})$ is asymptotically multivariate normal distributed with mean zero and a positive definite covariance matrix function $\Sigma$, given that the AG model is true \parencite{Andersen1982}. Due to the fact that $ n^{-1} I^{AG}(\beta_{true})  \ \ \xrightarrow[n \longrightarrow \infty]{P} \ n^{-1} \mathbb{E}(I^{AG}(\beta_{true}))$ and $\hat{\beta} \ \xrightarrow[n \longrightarrow \infty]{P} \ \beta_{true}$, it follows: 
\begin{align}
\label{D} &\dfrac{1}{n} I^{AG}(\hat{\beta}) \ \xrightarrow[n \longrightarrow \infty]{P} \  \Sigma.  
\end{align}
If $\beta_{true}$ corresponds to the true parameter vector, standard maximum likelihood arguments and Taylor series expansion around $\beta_{true}$ can be used to derive the asymptotic distribution of $\sqrt{n}(\hat{\beta}-\beta_{true})$. That is, 
\begin{align}
\nonumber &0 = U^{AG}(\hat{\beta}) \ \approx \ U^{AG}(\beta_{true}) - I^{AG}(\beta_{true})(\hat{\beta}-\beta_{true}) \\ 
\nonumber &\Longleftrightarrow \sqrt{n}(\hat{\beta}-\beta_{true}) \ \approx \  \biggl( \underbrace{\dfrac{1}{n} I^{AG}(\beta_{true})}_{\xrightarrow[n \longrightarrow \infty]{P} \ \Sigma} \biggr)^{-1} \cdot \underbrace{\dfrac{1}{\sqrt{n}} U^{AG}(\beta_{true})}_{\substack{\xrightarrow[n \longrightarrow \infty]{D} \ N(0, \Sigma) \\ \textnormal{according to} \ Eq. (\ref{C})}} \ \xrightarrow[n \longrightarrow \infty]{D} \ N(0, \Sigma^{-1}). 
\end{align}
Finally, taking Eq. $(\ref{D})$ into account, the asymptotic distribution is given by 
\begin{align}
\label{oui} \sqrt{n}(\hat{\beta}-\beta_{true}) \ \xrightarrow[n \longrightarrow \infty]{D} \ N \biggl (0, \biggl (\dfrac{1}{n}I^{AG}(\hat{\beta}) \biggr)^{-1} \biggr).
\end{align}
Statistical tests of $H_{0}:\beta=0$ and confidence intervals/bands can be constructed based on Eq. $(\ref{oui})$. If the AG model is correctly specified, a robust variance estimator is theoretically not required. Proofs can be found in \citet{Andersen1982}.

\subsubsection{Prentice-Williams-Peterson model}
\label{SectionPWP}
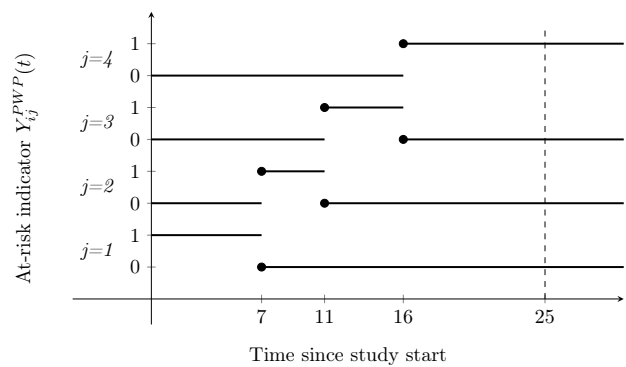
\begin{figure}[H]
\centering
\scalebox{0.75}{
\begin{tikzpicture}
    \begin{axis}[height=0.3\textheight, width=0.75\textwidth, xmin=-5.0, xmax=30, ymin=-0.8, ymax=9, xtick={7,11,16,25}, xticklabels={7, 11, 16, 25}, ytick={1, 2, 3, 4, 5, 6, 7, 8}, yticklabels={0,1,0,1,0,1,0,1}, xlabel={Time since study start}, ylabel={At-risk indicator $Y_{ij}^{PWP}(t)$},y label style={at={(axis description cs:-0.05,.5)},rotate=90,anchor=south},
    x label style={at={(axis description cs:0.5,-0.05)},anchor=north}]
        \addplot[cmhplot,-,domain=7:30]{1};
        \addplot[cmhplot,-,domain=0:7]{2};
        \addplot[soldot]coordinates{(7,1)};
        \addplot[cmhplot,-,domain=7:11]{4};
        \addplot[cmhplot,-,domain=11:30]{3};
        \addplot[cmhplot,-,domain=0:7]{3};
        \addplot[soldot]coordinates{(7,4)};
        \addplot[soldot]coordinates{(11,3)};
        \addplot[cmhplot,-,domain=0:11]{5};
        \addplot[cmhplot,-,domain=11:16]{6};
        \addplot[cmhplot,-,domain=16:30]{5};
        \addplot[soldot]coordinates{(11,6)};
        \addplot[soldot]coordinates{(16,5)};
        \addplot[cmhplot,-,domain=0:16]{7};
        \addplot[cmhplot,-,domain=16:30]{8};
        \addplot[soldot]coordinates{(16,8)};
        \addplot[dashed]coordinates {(25,0) (25,8.5)};
        \node [left] at (axis cs: -2.0, 1.5) {$\textit{j=1}$};
        \node [left] at (axis cs: -2.0, 3.5) {$\textit{j=2}$};
        \node [left] at (axis cs: -2.0, 5.5) {$\textit{j=3}$};
        \node [left] at (axis cs: -2.0, 7.5) {$\textit{j=4}$};
    \end{axis}
\end{tikzpicture}}
\caption[At-risk indicator under the Prentice-Williams-Peterson model]{At-risk indicator under the PWP model for individual $B$ from hypothetical example}
\label{AtRiskPWP}
\end{figure}

In some situations, multiplicative effects of time-dependent covariates reflecting aspects of the preceding event history (e.g., N(t-), ...) may not be reasonable \parencite{Kalbfleisch2002}. Time-dependent stratification provides another method for conditioning on the event history, making the PWP-CP model appealing. \newline \newline 
\textbf{PWP-CP} \newline 
The semiparametric PWP-CP model proposed by \citet{Prentice1981} with common covariate effect estimates is of the following form
\begin{align}
\label{PWPCP} \lambda_{ij}(t) = Y_{ij}^{PWP}(t) \alpha_{ij}(t) &= Y_{ij}^{PWP}(t) \alpha_{0j}(t) \exp( \beta^\intercal Z_{i}(t)), \ j>0, 
\end{align} 
where $\alpha_{0j}(t)$ is an event-specific baseline hazard, $\beta \in \mathbb{R}^{q}$ a vector of regression coefficients and  $Y_{ij}^{PWP}(t) = \mathbbm{1}(N_{i}(t)=j-1, \ C_{i} \geq t) = \mathbbm{1}(T_{i(j-1)} \leq t \leq T_{ij}, \ C_{i} \geq t)$ is the at-risk process for a $j^{th}$ event. $Z_{i}(t)=(Z_{i1}(t), ..., Z_{iq}(t)) \in \mathbb{R}^{q} $ is a vector of functions of external and internal covariates, $X_{i,ext}(t)$ and $X_{i,int}(t)$, the past event history $N_{i}(t-) = \{ N_{i}(u): 0 \leq u < t \}$ and interaction with time $t$. The baseline hazard $\alpha_{0j}(t)$ may be any integrable and non-negative function that depends arbitrarily on the previous number of events, with $\int_{0}^{t} = \alpha_{0j}(u)du$. While in the AG approach the baseline hazard is common for all events, the shape and form of the baseline hazard varies with increasing number of events in the PWP approach. In particular, the at-risk process of the PWP-CP model also differs from the one of the AG model, as illustrated in Figure $\ref{AtRiskPWP}$ for individual B from the hypothetical example. Under the PWP-CP model, individual B is at-risk for its first event in the time interval $[0, 7)$, i.e., $Y_{i1}^{PWP}(t) = 1$ $\forall \ t \in [0, 7)$. After occurrence of the first event at $t=7$, the at-risk process for a second event jumps to $1$ and stays there until the occurrence of the second event at $t=11$. Then, individual B is considered to be at-risk for a third event in $[11, 16)$ and for a fourth event in $[16, 25)$. The PWP-CP model assumes individuals not to be at-risk for a $j^{th}$ event as long as they have not yet experienced a $(j-1)^{th}$ event. This is also reflected in the risk set definition for the PWP-CP model. To sum up, the PWP-CP model uses event-specific baseline hazards and a restricted risk set.  
In general, the PWP model accounts for the dependence between repeated events by stratifying the intensity function on the number of preceding events.

\textbf{Inference for $\boldsymbol{\beta}$ and large sample theory} \newline 
Partial likelihood theory accounting for time-dependent stratification can be used to estimate $\beta \in \mathbb{R}^{q}$. Each individual starts in stratum $j=1$ and moves to stratum $j=2$ upon occurrence of the first event, and so on. At each time $t$, an individual is exactly assigned to one stratum. More counting process notation is required to derive the partial likelihood function. Let $N_{i}(t) = (N_{i1}(t), ..., N_{ij}(t), ...)$ be a multivariate counting process for individual $i$, where $N_{ij}(t)$ counts the number of events that happen in stratum $j$ over the time interval $[0, t)$, for $i=1,2,...,n$ and $j=1,2, ...$ . It is additionally assumed that no more than two counting processes jump simultaneously. The increment $dN_{ij}(t)$ is equal to $1$, if a $j^{th}$ event happens at time $t$. Further, $N_{i.}(t) = \sum_{j=1}^{\infty} N_{ij}(t)$ is defined as the 'overall' counting process for individual $i$ and $N_{.j}(t) = \sum_{i=1}^{n} N_{ij}(t)$ counts the total number of type $j$ events over all individuals. The filtration associated with the underlying stochastic process is given by $\mathcal{F}(t-)$, as introduced in a previous section. Under independent censoring, it follows that 
\begin{align}
\nonumber \lambda_{ij}(t)dt = P(dN_{ij}(t)=1 \ | \ \mathcal{F}(t-)) = Y_{ij}^{PWP}(t) \alpha_{0j}(t)\exp( \beta^\intercal Z_{i}(t))dt.
\end{align}
The conditional probability that it is individual $i$ who has a $j^{th}$ event at time t given that a $j^{th}$ event is observed in stratum $j$ and the past is 
\begin{align}
\nonumber \pi_{ij}(t, \beta) &= P(dN_{ij}(t) = 1 \ | \ \mathcal{F}(t-), dN_{.j}(t)=1) \\
\nonumber &= \dfrac{Y_{ij}^{PWP}(t) \exp( \beta^\intercal Z_{i}(t))}{\sum_{l=1}^{n} Y_{lj}^{PWP}(t) \exp(\beta^\intercal Z_{l}(t))}.
\end{align}
The contribution from the $j^{th}$ stratum to the partial likelihood function is given by 
\begin{align}
\nonumber L_{j}^{PWP}(\beta) = \prod_{u_{j} \leq \tau} \ \prod_{i=1}^{n} \bigl( \pi_{ij}(u_{j}, \beta) \bigr)^{dN_{ij}(u_{j})},
\end{align}
where $u_{j}$ corresponds to the unique event times in stratum $j$ or the unique $j^{th}$ event times.
Then, the partial likelihood function for $\beta$ is defined as the product over $L_{j}^{PWP}(\beta)$ for all $j$: \begin{align}
\nonumber L^{PWP}(\beta) = \prod_{j} L_{j}^{PWP}(\beta) = \prod_{j} \ \prod_{u_{j} \leq \tau} \ \prod_{i=1}^{n} \biggl( \dfrac{\exp( \beta^\intercal Z_{i}(u_{j}))}{\sum_{l=1}^{n} Y_{lj}^{PWP}(u_{j}) \exp(\beta^\intercal Z_{l}(u_{j})} \biggr)^{dN_{ij}(u_{j})}, 
\end{align}
where $\beta$ is the argument of the likelihood function. The first product is defined over all possible strata, while the second product is defined over all unique repeated $j^{th}$ event times $u_{j} \leq \tau$. The resulting log-likelihood function, the likelihood-based score vector and the observed information matrix have the following form: 
\begin{align}
\nonumber \ell^{PWP}(\beta) &= \log \bigl(L^{PWP}(\beta) \bigr) \\
\nonumber &= \sum_{j} \ \sum_{u_{j} \leq \tau} \ \sum_{i=1}^{n} \underbrace{dN_{ij}(u_{j})}_{\in \{ 0,1 \}} \biggl[ \beta^\intercal Z_{i}(u_{j}) - \log \biggl( \sum_{l=1}^{n} Y_{lj}^{PWP}(u_{j}) \exp(\beta^\intercal Z_{l}(u_{j})) \biggr) \biggr ] \\
\nonumber &= \sum_{j} \ \sum_{i=1}^{n} \int_{(0, \tau]} \biggl[ \beta^\intercal Z_{i}(u) - \log \bigl( S_{j}^{(0, PWP)}(\beta, u) \bigr) \biggr] dN_{ij}(u) \\
\nonumber \\
\nonumber U^{PWP}(\beta) &= \frac{\partial}{\partial \beta} \ell^{PWP}(\beta) = \sum_{j} \sum_{i=1}^{n} \int_{(0, \tau]} \biggl( Z_{i}(u) - \dfrac{S_{j}^{(1), PWP}(\beta, u)}{S_{j}^{(0), PWP}(\beta, u)} \biggr) dN_{ij}(u) \\
\nonumber \\
\nonumber I^{PWP}(\beta) &= - \frac{\partial}{\partial \beta^\intercal} U^{PWP}(\beta) = \sum_{j} \int_{(0, \tau]} \dfrac{S_{j}^{(2), PWP}(\beta, u)}{S_{j}^{(0), PWP}(\beta, u)} - \biggl(\dfrac{S_{j}^{(1), PWP}(\beta, u)}{S_{j}^{(0), PWP}(\beta, u)}\biggr)^{\otimes 2} dN_{j.}(u), 
\end{align}
where 
\begin{align}
\nonumber S_{j}^{(0), PWP}(\beta, t) &= \sum_{i=1}^{n} Y_{ij}^{PWP}(t) \exp( \beta^\intercal Z_{i}(t)),  \\
\nonumber S_{j}^{(1), PWP}(\beta, t) &= \sum_{i=1}^{n} Y_{ij}^{PWP}(t) Z_{i}(t) \exp( \beta^\intercal Z_{i}(t)), \\
\nonumber S_{j}^{(2), PWP}(\beta, t) &= \sum_{i=1}^{n} Y_{ij}^{PWP}(t) Z_{i}(t)^{\otimes 2} \exp( \beta^\intercal Z_{i}(t)).
\end{align}
Hence, the observed information matrix arises as the sum of the information matrices obtained from each stratum $j$. The estimated regression coefficient vector $\hat{\beta}$ is obtained by solving $U^{PWP}(\beta) = 0$.   
For $\beta_{true}$, it yields: $dN_{ij}(t) = \lambda_{ij}(t)dt + dM_{ij}(t) = Y_{ij}^{PWP}(t) \alpha_{0j}(t) \exp( \beta_{true}^\intercal Z_{i}(t)) dt + dM_{ij}(t)$, with $M_{ij}(t)$ as orthogonal martingales. Using this Doob-Meyer decomposition and properties of vector-valued stochastic integrals, it can be shown that $U^{PWP}(\beta_{true})$ is a zero-mean martingale. As a consequence, $U^{PWP}(\beta_{true})$ can be written as 
\begin{align}
\nonumber U^{PWP}(\beta_{true}) &= \sum_{j} \sum_{i=1}^{n} \int_{(0, \tau]} \biggl( Z_{i}(u) - \dfrac{S_{j}^{(1), PWP}(\beta_{true}, u)}{S_{j}^{(0), PWP}(\beta_{true}, u)} \biggr) dM_{ij}(u). 
\end{align}
Asymptotic results for $\hat{\beta}$ can be found in \citet{Andersen1984, Andersen1993}. If the PWP-CP model is correctly specified, there is no need for robust variance estimation. 
\newline \newline 
Besides common covariate effects, it is also possible to allow the covariate effects to vary across the events. The PWP-CP model proposed by \citet{Prentice1981} with event-/strata-specific covariate effects is of the following form
\begin{align}
\label{specialPWP} \lambda_{ij}(t) = Y_{ij}^{PWP}(t) \alpha_{ij}(t) &= Y_{ij}^{PWP}(t) \alpha_{0j}(t) \exp( \beta_{j}^\intercal Z_{i}(t)), \ \forall \ j=1,...,K \ , 
\end{align} 
where $\alpha_{0j}(t)$ is an event-specific baseline hazard and $\beta_{j} \in \mathbb{R}^{q}$ is an event-specific vector of regression coefficients. In practice, data may need to be limited to a specific number of recurrent events $K$, if the risk set becomes very small for higher strata. Thus, it is important to choose $K$ such that there is sufficient data to get precise estimates. \newline \newline 
\textbf{Stratification variable} \newline
The PWP-CP models $(\ref{PWPCP})$ and $(\ref{specialPWP})$ can be formulated in a more general way, allowing the baseline hazard to depend on a stratification variable $s=s(N(t), Z(t), t)$, with $s$ as a function of time for a given individual \parencite{Prentice1981, Kalbfleisch2002}.  
\begin{align}
\label{PWPCPgeneral} \lambda_{is}(t) &= Y_{is}^{PWP}(t) \alpha_{0s}(t) \exp( \beta^\intercal Z_{i}(t)), \ \forall \ s=1,2, ... \ ,  \\
\label{PWPCPgeneral1} \lambda_{is}(t) &= Y_{is}^{PWP}(t) \alpha_{0s}(t) \exp( \beta_{s}^\intercal Z_{i}(t))\ \forall \ s=1,2, ... \ . 
\end{align} 
If s=$N(t)+1$, model $(\ref{PWPCPgeneral})$ reduces to the model $(\ref{PWPCP})$. It is also possible to define other stratification variables $s$. However, stratification choices need to be constructed such that each individual has at most a single at-risk interval in each stratum. 
\newpage
\subsection{Random effect models}
General intensity-based models, such as the AG and PWP-CP models, rely on defining explicit expressions for the dependence between repeated events via internal time-varying covariates and time-dependent stratification. Random effects may also be incorporated in conditional models to induce additional dependence on the preceding event history. Such models are often referred to as frailty models in recent literature. Random effects are further useful for reflecting heterogeneity across individuals due to unmeasured and unobserved covariates. \newline
In this work, the concept of random effect models will be explained by means of the NB model. As described in Section $\ref{SectionPoissonModels}$, Poisson models based on homogeneous and non-homogeneous Poisson processes are mainly characterized by the fact that recurrence of events is conditionally independent of the prior event history. However, even after conditioning on external covariates, there may be more variation in event occurrence across individuals than accounted for by a Poisson process, i.e., $\mathbb{E}(N(t)) \neq Var(N(t))$. In this case, NB models attempt to overcome this problem. The idea behind NB models is to formulate recurrent event models through assumptions of conditional independence between events, given a gamma distributed random effect.

\subsubsection{Negative binomial models}
\label{SectionNBmodel}
Poisson models with random effects are based on so-called mixed Poisson processes, in which the conditional ('individual-specific') intensity function is given by 
\begin{align}
\label{intensityPoisson} \lambda_{i}(t \ | \ U_{i})dt &= P(dN_{i}(t) = 1 \ | \ \mathcal{F}(t-), \ U_{i}) \\
\nonumber &= U_{i}  P(dN_{i}(t) = 1 \ | \ (Y_{i}(s), X_{i, ext}(s))_{s < t}) \\
\nonumber &= U_{i} Y_{i}(t) \alpha_{0}(t)dt \exp( \beta^\intercal Z_{i, ext}(t)).  
\end{align}
In model $(\ref{intensityPoisson})$, $Y_{i}(t) = \mathbbm{1}(C_{i} \geq t)$ is the left-continuous at-risk process, $U_{i}$ the non-negative and unobservable random effect and $\alpha_{0}(t)$ is an unspecified baseline hazard. $Z_{i}(t) \in \mathbb{R}^{q}$ is a vector of functions of external covariates and $\beta \in \mathbb{R}^{q}$ denotes a vector of unknown regression coefficients. The baseline hazard $\alpha_{0}(t)$ can be either specified parametrically or non-parametrically. In the following, the covariate process is restricted to functions of baseline covariates only, in which case the intensity function $(\ref{intensityPoisson})$ reduces to
\begin{align}
\lambda_{i}(t \ | \ U_{i}) &= U_{i} Y_{i}(t) \alpha_{0}(t)dt \exp( \beta^\intercal Z_{i}).  
\end{align}
The notation $\lambda_{i}(t \ | \ U_{i})$ has been chosen to clearly emphasize that the intensity function is formulated conditionally on $U_{i}$. Further, the random effects $U_{1}, ..., U_{n}$ are assumed to be independent and identically distributed with finite mean and cumulative distribution function (CDF) $G$. Although many different distribution families can be used for $U$, the gamma distribution with mean $1$ and variance $\phi >0$ is convenient because several process quantities have closed-form expressions (e.g., marginal likelihood function).
\begin{definition}[Gamma distribution]
Let U be an absolutely continuous random variable. $U$ is said to follow a Gamma distribution with scale parameter $\phi^{-1}$ and shape parameter $\phi^{-1}$, i.e., $ U \sim \Gamma(\phi^{-1}, \phi^{-1})$, if its probability density function is given by $g_{U}(u) = \dfrac{exp(-\phi^{-1}u) u^{\phi^{-1}-1}}{\phi^{\phi^{-1}}\Gamma(\phi^{-1})}$, $u >0$. The mean and variance of $U$ are $\mathbb{E}(U)=1$ and $Var(U)=\phi$, respectively. 
\end{definition}
Given the random effect $U_{i}$, the counting process $\{N_{i}(t): 0 \leq t < \infty \}$ follows a non-homogeneous Poisson process with parameter $U_{i}\alpha_{0}(t)\exp( \beta^\intercal Z_{i})$. If $\mu_{i}(t) = \int_{0}^{t} \alpha_{0}(u) \exp( \beta^\intercal Z_{i})du$, the conditional probability of $j$ events in $[0, t]$ given $U_{i}$ is 
\begin{align}
\nonumber P(N(t) = j \ | \ U_{i}) &= \dfrac{(U_{i}\mu_{i}(t))^{j}}{j!} \exp(-U_{i}\mu_{i}(t)).
\end{align} 
\newpage
The marginal probability of $j$ events in $[0, t]$ can be calculated as 
\begin{align}
\label{NegBinProcess} P(N(t) = j) &= \int_{0}^{\infty} \dfrac{(u \mu_{i}(t))^{j}}{j!} \exp(-u\mu_{i}(t)) g(u) du = \dfrac{\Gamma(j + \phi^{-1})}{\Gamma(\phi^{-1})} \dfrac{(\phi \mu_{i}(t))^j}{(1+ \phi \mu_{i}(t))^{j+\phi^{-1}}},
\end{align}
for $j \in \mathbb{N}_{0}$. Eq. $(\ref{NegBinProcess})$ is of negative binomial form so that the observable data $(n_{i}, t_{i1}, ..., t_{in_{i}})$ for individual $i$ arises from a negative binomial process. In other words, the counting process $\{N_{i}(t): 0 \leq t < \infty \}$ is a negative binomial process (i.e., $U \sim \Gamma$) or, more generally, a mixed Poisson process (i.e., $U$ follows arbitrary distribution). If $\phi \longrightarrow 0$, Eq. $(\ref{NegBinProcess})$ gives the Poisson distribution, as introduced in Definition $\ref{DefPoissonProcess}$. 
\newline \newline 
Due to $\mathbb{E}(U)=1$ and $Var(U)=\phi$, the negative binomial process fulfills the following properties:
\begin{align}
\nonumber \mathbb{E}(N_{i}(t)) &= \mathbb{E}(\mathbb{E}(N_{i}(t) \ | \ U_{i})) =  \mathbb{E}(U_{i}\mu_{i}(t)) = \mu_{i}(t), \\ 
\nonumber Var(N_{i}(t)) &= \mathbb{E}(Var(N_{i}(t) \ | \ U_{i})) + Var(\mathbb{E}(N_{i}(t) \ | \ U_{i})) = \mathbb{E}(U_{i}\mu_{i}(t)) + Var(U_{i}\mu_{i}(t)) \\
\nonumber &= \mu_{i}(t) + \phi \mu_{i}(t)^{2} \ \ \ \ \textnormal{and} \\  
\nonumber Cov(N_{i}(t_{1}, t_{2}), \ & N_{i}(t_{3}, t_{4})) = \phi \mu_{i}(t_{1}, t_{2}) \mu_{i}(t_{3}, t_{4}),  
\end{align}
with $t_{1} \leq t_{2} < t_{3} \leq t_{4}$. If $\phi=0$, the marginal expected mean and variance of $N(t)$ reduces to $\mathbb{E}(N(t))=Var(N(t))$, in which case the counting process is Poisson. The third property states that the covariance function for event counts in disjunct time intervals depends on $\phi$, making $N_{i}(t_{1}, t_{2})$ and $N_{i}(t_{3}, t_{4})$ dependent. However, the parameter $\mu_{i}(t)$ of the negative binomial process is independent of $\phi$ and equal to the one under a Poisson process.  
\newline \newline 
The full intensity function of a mixed Poisson process has the form 
\begin{align}
\nonumber \lambda_{i}(t)dt &= P(dN_{i}(t) = 1 \ | \ \mathcal{F}_{i}(t-)) = \int_{0}^{\infty} P(dN_{i}(t) = 1 \ | \ \mathcal{F}_{i}(t-), U_{i}) g(u \ | \ \mathcal{F}_{i}(t-) ) du \\
\nonumber &= Y_{i}(t) \alpha_{0}(t) \exp( \beta^\intercal Z_{i}) \underbrace{\int_{0}^{\infty} u g(u \ | \ \mathcal{F}_{i}(t-)) du}_{= \mathbb{E}(U_{i}\ | \ \mathcal{F}_{i}(t-)) = \mathbb{E}(U_{i}\ | \ N_{i}(t-))} = Y_{i}(t) \alpha_{0}(t) \exp( \beta^\intercal Z_{i}) \mathbb{E}(U_{i} \ | \ N_{i}(t-))
\end{align}
Under independent censoring, it can be shown that $\mathbb{E}(U_{i}\ | \ \mathcal{F}_{i}(t-)) = \mathbb{E}(U_{i}\ | \ N_{i}(t-))$. As a consequence, the full intensity function is a product of the Poisson intensity function and the conditional expectation of the random effect given the number of observed events in $[0, t)$. 
If $U$ arises from a Gamma distribution with mean $1$ and variance $\phi$, the full intensity function is 
\begin{align}
\label{fullintensityNB} \lambda_{i}(t)dt &= Y_{i}(t) \alpha_{0}(t) \exp( \beta^\intercal Z_{i}) \biggl( \dfrac{1+\phi N_{i}(t-)}{1 + \phi \mu_{i}(t)}\biggr). 
\end{align}
More specifically, $\mathbb{E}(U_{i}\ | \ N_{i}(t-)) = \dfrac{1+\phi N_{i}(t-)}{1 + \phi \mu_{i}(t)}$ in Eq. $(\ref{fullintensityNB})$ follows from the fact that $U \sim \Gamma(\phi^{-1}, \phi^{-1})$, $N(t-) \ | \ U \sim$ Poisson($U\mu(t))$ and $N(t-) \sim NB(N(t-), \phi)$, leading to 
\begin{align}
\nonumber P(U \ | \ N(t-)) &= \dfrac{P(U, N(t-))}{P(N(t-))} = \dfrac{P(N(t-) \ | \ U) P(U)}{P(N(t-))} = ... = \\
\nonumber &= \dfrac{u^{(N(t-)+\phi^{-1})-1} \exp(-(\mu(t) + \phi^{-1})u)(\mu(t) + \phi^{-1})^{N(t-)+\phi^{-1}}}{\Gamma(N(t-)+\phi^{-1})}.
\end{align}
Thus, $U_{i} \ | \ N_{i}(t-) \sim \Gamma(N_{i}(t-) + \phi^{-1}, \mu_{i}(t) + \phi^{-1})$. \newline
If $\phi=0$, the intensity function $(\ref{fullintensityNB})$ corresponds to an underlying non-homogeneous Poisson process (cf. Section $\ref{SectionPoissonModels})$. In contrast, when $\phi > 0$, the full intensity function at time $t$ depends on the heterogeneity parameter $\phi$ and on the event history through $N(t-)$. 
\newpage
Additionally, Eq. $(\ref{fullintensityNB})$ demonstrates that the event intensity at time $t$ increases with the number of observed events before time $t$. This justifies the idea of frailty terms: large values of $N(t-)$ are associated with larger realizations of $U$, which in turn are associated with larger event counts beyond $t$. Since the full intensity function depends on the event history only through $N(t-)$, the process is still Markov. 
With regard to the general multistate model, the (unconditional) intensity function of $N_{(j-1)j}(t)$ can be similarly derived: 
\begin{align}
\nonumber \lambda_{ij}(t)dt &= Y_{i}(t) \alpha_{0}(t) \exp( \beta^\intercal Z_{i}) \biggl( \dfrac{1+\phi (j-1)}{1 + \phi \mu_{i}(t)}\biggr) \  \ \ \ j= 1, 2, ... \ . 
\end{align}
\textbf{Special case: mixed homogeneous Poisson process} \newline
Similar to the case of a homogeneous Poisson process, the conditional intensity function for a mixed homogeneous Poisson process is 
\begin{align}
\label{parametricNB} \lambda_{i}(t \ | \ U_{i}) = U_{i} Y_{i}(t) \alpha = U_{i} Y_{i}(t) \alpha_{0} \exp(\beta^\intercal Z_{i}). 
\end{align}
In the following, the focus is on the parameter estimation for $\beta$ when $U$ is gamma distributed with mean $1$ and variance $\phi$. As defined previously, $n_{i}$ is the total number of events experienced by individual $i$ over $[0, C_{i}]$. 
\newline \newline 
\textbf{Inference for $\boldsymbol{\alpha_{0}}, \boldsymbol{\beta}$ and $\boldsymbol{\phi}$} \newline
The likelihood functions for $\alpha_{0}, \beta$ and $\phi$ are constructed from the intensity function $(\ref{parametricNB})$ using Theorem $\ref{likelihoodcontri}$. Since the random effects $U_{i}$ are unobserved, the likelihood contribution for individual $i$ under this specific NB model (= constant hazard and baseline covariates) is 
\begin{align}
\nonumber \int_{0}^{\infty} & \biggl( \biggl[\prod_{j}^{n_{i}} U_{i} \alpha_{0} \exp(\beta^\intercal Z_{i}) \biggr] \exp \bigl( - \int_{0}^{\infty} Y_{i}(s) U_{i} \alpha_{0} \exp(\beta^\intercal Z_{i}) ds  \bigr) \biggr) g(U_{i}) dU_{i} \\
\nonumber &= \biggl[\prod_{j}^{n_{i}} \alpha_{0} \exp(\beta^\intercal Z_{i}) \biggl] \dfrac{\Gamma(n_{i} + \phi^{-1})}{\Gamma(\phi^{-1})} \dfrac{\phi^{n_{i}}}{(1+\phi \mu_{i}(C_{i}))^{n_{i}+\phi^{-1}}}, 
\end{align}
leading to the following log-likelihood function
\begin{align}
\nonumber \ell^{NB}(\alpha_{0}, \beta, \phi) =  & \sum_{i=1}^{n} \biggl( \sum_{j=1}^{n_{i}} \log(\alpha_{0}\exp(\beta^\intercal Z_{i})) + \sum_{j=0}^{n_{i}^{\ast}} \log(1+\phi j) - (n_{i} + \phi^{-1}) \log(1 + \phi \mu_{i}(C_{i})) \biggr),
\end{align}
where $n_{i}^{\ast} = max(0, n_{i}-1)$ and $\mu_{i}(C_{i}) = \int_{0}^{C_{i}} \alpha_{0} \exp(\beta^\intercal Z_{i}) du = C_{i} \alpha_{0} \exp(\beta^\intercal Z_{i})$. In particular, this log-likelihood function is proportional to a NB log-likelihood under the assumption of $n_{i} := N_{i}(C_{i}) \sim NegBin(C_{i} \exp(\beta_{0} + \beta^\intercal Z_{i}), \phi)$ with $\beta_{0} = \log(\alpha_{0})$. The log-likelihood function for the time-homogeneous NB model depends on the event counts and the individual-specific follow-up times, but not on the actual event times. For this reason, the NB model can also be seen as a marginal rate-based model. The corresponding score functions are derived as 
\begin{align}
\nonumber U^{NB}_{\alpha_{0}} &=\frac{\partial}{\partial \alpha_{0}}  \ell^{NB}(\alpha_{0}, \beta, \phi) = \sum_{i=1}^{n} \bigl( \dfrac{n_{i}}{\alpha_{0}} - \dfrac{(1+\phi n_{i}) C_{i}\exp(\beta^\intercal Z_{i})}{(1+\phi \mu_{i}(C_{i}))} \bigr) \\
\nonumber U^{NB}_{\beta} &=\frac{\partial}{\partial \beta}  \ell^{NB}(\alpha_{0}, \beta, \phi) =  \sum_{i=1}^{n} \dfrac{Z_{i}(n_{i} - \mu_{i}(C_{i}))}{(1 + \phi \mu_{i}(C_{i}))} \\
\nonumber U^{NB}_{\phi} &= \frac{\partial}{\partial \phi}  \ell^{NB}(\alpha_{0}, \beta, \phi) = \sum_{i=1}^{n} \biggl( \sum_{j=0}^{n_{i}^{\ast}} \dfrac{j}{1+\phi j} - \dfrac{\mu_{i}(C_{i})(n_{i} + \phi^{-1})}{1+\phi \mu_{i}(C_{i})} + \phi^{-2} \log(1 + \phi \mu_{i}(C_{i})) \biggr). 
\end{align}
Solutions to equations $U^{NB}_{\alpha_{0}}=0, U^{NB}_{\beta}=0$ and $U^{NB}_{\phi}$ yield estimates of $\alpha_{0}$, $\beta$ and $\phi$. 
\newline \newline 
In total, the NB models account for the dependence structure among repeated events by incorporating random effects. The concept of a frailty term seems to be reasonable: the intensity function for an event is increased if a high number of events have already been observed in the past, as this would indicate a high frailty. 

\newpage
\section{Marginal models}
\label{margMoDels}
As an alternative to conditional intensity-based models, there are so-called marginal models that do not intend to give a full specification of the recurrent event process and focus on marginal parameters of the event process. Examples for marginal parameters are the expected number of events $\mathbb{E}(N(t))$ in $[0, t]$, rate functions of events, times from $t=0$ until occupying a certain state (i.e., a certain number of events) or state occupation probabilities $P(N(t) = j)$. In marginal models, it is further differentiated between marginal hazard models which rely on the marginal distribution of event times and marginal rate models. As indicated by the name, marginal rate models rely on rate-based modelling, in which the rate function is either completely unaffected by the past event history of the recurrent event process or may be related to a part of the history only. \newline
After discussing marginal hazard models for recurrent event data, marginal mean and rate models for arbitrary recurrent event processes are introduced. This section also aims at explaining the difference between transition hazards (or transition intensities) and transition rates. 

\subsection{Marginal hazard models}
\label{MarginalHazardsModel}
Marginal hazard models include the WLW model \parencite{Wei1989} and the LWA model \parencite{Lee1992}. The main purpose of marginal hazard approaches is to model the marginal distribution of times to the first, second, third, ... event. The WLW and LWA models are methods that do not apply to the general multistate setup introduced in Section $\ref{condmodel}$.
\subsubsection{Wei-Lin-Weissfeld model}
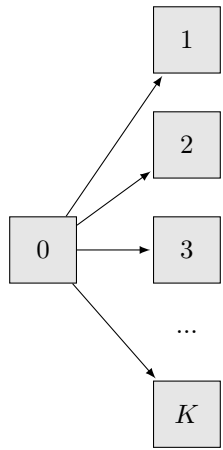
\begin{figure}[H]
\centering
\begin{tikzpicture}
  \tikzset{node style/.style={state, fill=gray!20!white, rectangle}}
  \node[node style]               (I)   {0};
  \node[node style, right=of I]   (III)  {3};
  \node[node style, above=0.5cm of III] (II)  {2};
  \node[node style, above=0.5cm of II] (IV)  {1};
  \node[draw=none, below=0.5cm of III] (dot)  {...};
  \node[node style, below=0.5cm of dot] (K)  {$K$};
  \draw[>=latex, auto=left, every loop]
         (I)   edge node {}   (II)
         (I)  edge node {}  (III)
         (I) edge node {} (IV)
         (I)  edge node {} (K); 
\end{tikzpicture}
\caption[Multistate representation of the Wei-Lin-Weissfeld and Lee-Wei-Amato approaches]{Multistate representation of the WLW and LWA approaches}
\label{WLWmodel}
\end{figure}

The WLW model is an unconditional marginal model based on a total time scale \parencite{Wei1989}. The main idea of the WLW model is to restrict the recurrent event analysis to $K$ events, $K \in \mathbb{N}$, and to apply $K$ distinct Cox proportional hazards model to the time-to-event data. Each individual under study provides information to each marginal Cox model, either an observed $j^{th}$ event time or a censoring time, $j=1, 2, ..., K$. Figure $\ref{WLWmodel}
$ illustrates a multistate model reflecting recurrent event data based on a WLW model formulation. For each event time $T_{j}$, the marginal Cox model with event-specific regression coefficients is formulated by 
\begin{align}
\label{WLWhazard} h_{ij}(t) = h_{0j}(t) \exp( \beta_{j}^\intercal Z_{i}(t)), \ \ j = 1, 2, ..., K, 
\end{align}
where $h_{0j}(t)$ is an unspecified baseline hazard depending on model $j$, $\beta_{j} \in \mathbb{R}^{p}$ a vector of regression coefficients and $Z_{i}(t)=(Z_{i1}(t), ..., Z_{iq}(t)) \in \mathbb{R}^{q}$ is a $q$-dimensional covariate vector. The WLW analysis does not account for the fact that $T_{i1} < T_{i2} < ... < T_{iK}$.
\newline 

\begin{figure}[H]
\centering
\scalebox{0.80}{ 
\begin{tikzpicture}
    \begin{axis}[height=0.3\textheight, width=0.75\textwidth, xmin=-5.0, xmax=30, ymin=-0.8, ymax=9, xtick={7,11,16,25}, xticklabels={7, 11, 16, 25}, ytick={1, 2, 3, 4, 5, 6, 7, 8}, yticklabels={0,1,0,1,0,1,0,1}, xlabel={Time since study start}, ylabel={At-risk indicator $Y_{ij}^{WLW}(t)$},y label style={at={(axis description cs:-0.05,.5)},rotate=90,anchor=south},
    x label style={at={(axis description cs:0.5,-0.05)},anchor=north}]
        \addplot[cmhplot,-,domain=7:30]{1};
        \addplot[cmhplot,-,domain=0:7]{2};
        \addplot[soldot]coordinates{(7,1)};
        \addplot[cmhplot,-,domain=0:11]{4};
        \addplot[cmhplot,-,domain=11:30]{3};
        \addplot[soldot]coordinates{(11,3)};
        \addplot[cmhplot,-,domain=0:16]{6};
        \addplot[cmhplot,-,domain=16:30]{5};
        \addplot[soldot]coordinates{(16,5)};
        \addplot[cmhplot,-,domain=0:25]{8};
        \addplot[cmhplot,-,domain=25:30]{7};
        \addplot[soldot]coordinates{(25,7)};
        \addplot[dashed]coordinates {(25,0) (25,8.5)};
        \node [left] at (axis cs: -2.0, 1.5) {$\textit{j=1}$};
        \node [left] at (axis cs: -2.0, 3.5) {$\textit{j=2}$};
        \node [left] at (axis cs: -2.0, 5.5) {$\textit{j=3}$};
        \node [left] at (axis cs: -2.0, 7.5) {$\textit{j=4}$};
    \end{axis}
\end{tikzpicture}}
\caption[At-risk indicator under the Wei-Lin-Weissfeld and Lee-Wei-Amato models]{At-risk indicator under WLW and LWA model for individual $B$ from hypothetical example, K=4}
\label{AtRiskWLW}
\end{figure}
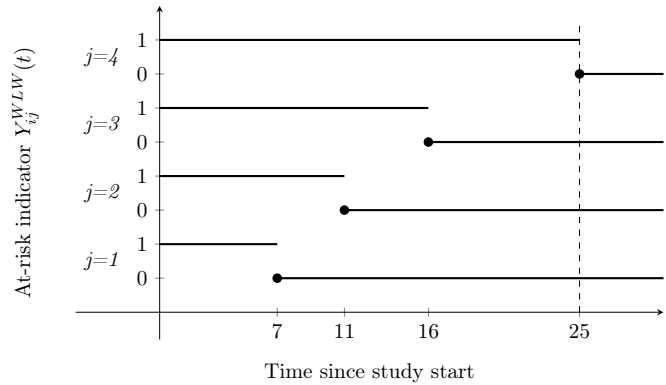

The hazard $h_{ij}(t)$ for event $j$ has the following interpretation:
\begin{align}
\nonumber h_{ij}(t)dt &= P(j^{th} \ \textnormal{event in} \ [t, t+dt) \ \textnormal{for individual} \ i \ | \ \textnormal{censoring and covariate histories}) \\
\nonumber &= P(T_{ij} \in [t, t+dt), \ T_{ij} \leq C_{i} \ | \ T_{ij} \geq t, \ C_{i} \geq t, \ Z_{i}(t)).
\end{align}
In fact, $h_{ij}(t)$ is the instantaneous probability of observing a $j^{th}$ event in the small time interval $[t, t+dt)$, conditional on the covariate history and the fact that neither the event nor censoring have happened before time $t$. Specifically, $h_{ij}(t)$ is defined by disregarding information on past event times $T_{i1}, T_{i2}, ..., T_{i(j-1)}$ in the conditioning set. For this reason, the WLW model is referred to as a marginal approach. 
\newline
Assuming $K=4$, Figure $\ref{AtRiskWLW}$ shows the at-risk indicator $Y_{ij}^{WLW}(t)$ for individual B from the hypothetical example given in Figure $\ref{GraphicRE}$. The WLW model assumes that individuals under study are at-risk for each event starting from time origin. For instance, individual $B$ is at-risk for a first, second, third and fourth event in time interval $[0, 7)$. Between $t=7$ and $t=11-$, individual $B$ is included in the at-risk set for a second, third and fourth event, and so on. Basically, individuals are at-risk for a $j^{th}$ event even before experiencing a $(j-1)^{th}$ event, leading to the following definition of the at-risk indicator: $Y_{ij}^{WLW}(t) = \mathbbm{1}(T_{ij} \geq t, C_{i} \geq t) \in \{0,1 \}$. This concept does not really coincide with the usual recurrent event setting, where events happen successively. As a consequence, the natural order of the repeated events is destroyed and the structure of dependence between the recurrent events remains unspecified. Due to the model properties described above, the WLW model uses event-specific baseline hazards and a semi-restricted risk set.

\textbf{Inference for $\boldsymbol{\beta_{j}}$} \newline 
The regression coefficients $\beta_{1}, \beta_{2}, ..., \beta_{K}$ of the WLW model can be estimated under a working independence assumption. This means that the usual Cox partial likelihood function one would use in a standard time-to-first-event analysis is maximized to get the estimate $\hat{\beta_{j}}$. Specifically, the partial likelihood function is 
\begin{align}
\nonumber L^{WLW}_{j}(\beta_{j}) = \prod_{i=1}^{n} \ \biggl(  \dfrac{ \exp( \beta_{j}^\intercal Z_{i}(\min(T_{ij}, C_{i})))}{\sum_{l=1}^{n} Y_{lj}^{WLW}(\min(T_{ij}, C_{i})) \exp(\beta_{j}^\intercal Z_{l}(\min(T_{ij}, C_{i})))} \biggr)^{\mathbbm{1}(T_{ij} \leq C_{i})}. 
\end{align}
$K$ needs to be chosen such that the event-specific regression coefficients can be reliably estimated. In addition, the $K$ proportinal hazards model given in Eq. $(\ref{WLWhazard})$ can also be redefined by assuming common covariate effects across all marginal models rather than event-specific ones. That is, $h_{ij}(t) = h_{0j}(t) \exp( \beta^\intercal Z_{i}(t))$, in which case the estimated regression coefficient $\hat{\beta} \in \mathbb{R}^{q}$ is obtained by maximizing the likelihood function
\begin{align}
\nonumber L^{WLW}(\beta) = \prod_{j=1}^{K} \prod_{i=1}^{n} \ \biggl(  \dfrac{ \exp( \beta^\intercal Z_{i}(\min(T_{ij}, C_{i})))}{\sum_{l=1}^{n} Y_{lj}^{WLW}(\min(T_{ij}, C_{i})) \exp(\beta^\intercal Z_{l}\min(T_{ij}, C_{i}))} \biggr)^{\mathbbm{1}(T_{ij} \leq C_{i})}. 
\end{align}

Since the proportional hazards assumption is usually not met for each distinct Cox model, \citet{Lin1989} proposed robust variance estimation for both $\hat{\beta}$ and $\hat{\beta_{j}}$ to deal with model misspecification and to ensure control of statistical inference properties (e.g., type I error for tests of null hypotheses). 
\newline \newline 
Apart from the critical risk-set definition, the WLW model has also been criticized by the so-called 'carry-over' effect, meaning that the covariate effect on event $j$ is carried over to subsequent events \parencite{Metcalfe2007}. Suppose a beneficial treatment that only affects the first event but not the subsequent events, i.e., $\beta_{1} < 0$ and $\beta_{2}, ..., \beta_{K} = 0$. Under this setting, the time to the $j^{th}$ event measured from time origin will always be longer for treated patients than for untreated patients, $j = 2, ..., K$. As a consequence, treatment appears to be effective also for the second, third, ... and $K^{th}$ event, although no treatment effect exists on events following the first event. However, the treatment effect diminishes with each successive event, as the relative difference between the control and treatment group becomes smaller with increasing time. Additionally, \citet{Kelly2000} showed in their simulation studies that, in case of $\beta_{1} = \beta_{2} = ... = \beta_{K} < 0$, the event-specific WLW model produces an unbiased estimate for $\beta_{1}$ and leads to overestimation of $\beta_{2}, ...., \beta_{K}$, with the overestimation becoming larger with each successive event. The carry-over effect can also be found in common covariate effect estimates $\beta$ in the WLW model $(\ref{WLWhazard})$. \newline
A disadvantage of applying the WLW model to recurrent event data is to disregard events occurring in individuals who have experienced more than $K$ events, as $K$ must be limited to an adequate value to still get precise estimates for large $j$. It should be noted that the higher $K$, the less patients are included in the analysis and the less precise the covariate effects. 

\subsubsection{Lee-Wei-Amato model}
The Lee-Wei-Amato (LWA) model is a less known recurrent event method and can be classified as an unconditional marginal model based on a total time scale. Proposed by \citet{Lee1992}, the model was originally developed for clustered multivariate failure time data but can also be applied to recurrent event data. In Figure $\ref{WLWmodel}$, a multistate model reflecting recurrent event data based on a LWA model formulation is displayed. As seen from this figure, the analysis of recurrent events is also limited to $K$ events and the time to the $j^{th}$ event measured since study start is modelled via a marginal proportional hazards model of the following form: 
\begin{align}
\label{LWAhazard} h_{ij}(t) = h_{0}(t) \exp( \beta^\intercal Z_{i}(t)), \ \ j = 1, 2, ..., K, \end{align}
where $h_{0}(t)$ is a common unspecified baseline hazard independent of $j$, $\beta \in \mathbb{R}^{q}$ a vector of regression coefficients and $Z_{i}(t)=(Z_{i1}(t), ..., Z_{iq}(t)) \in \mathbb{R}^{q}$ corresponds to a $q$-dimensional covariate vector for individual $i$, $i=1,2,...,n$. As in the WLW model, each individual under study contributes an outcome to the $j^{th}$ model, either an observed $j^{th}$ event time or a censoring time. Eq. $(\ref{LWAhazard})$ states that the LWA model assumes a common baseline hazard across all $K$ distinct Cox models, with an unrestricted risk set and no specific dependence structure among the recurrent events. If $h_{0j}(t)$ in Eq. $(\ref{WLWhazard})$ is constrained to $h_{0}(t)$ and $\beta_{j} = \beta$, the WLW model reduces to the LWA model. Thus, the WLW model can be referred to as an event-specific LWA model. The at-risk process $Y_{ij}^{LWA}(t) \in \{0, 1 \}$ for the LWA model is identical to the one for the WLW approach, i.e., $Y_{ij}^{LWA}(t) = Y_{ij}^{WLW}(t) = \mathbbm{1}(T_{ij} \geq t, C_{i} \geq t)$. As displayed in Figure $\ref{AtRiskWLW}$, the LWA model assumes that each individual is at-risk for a $j^{th}$ event from t=0 onwards, irrespective of whether a $(j-1)^{th}$ event has already been observed. 
\newline \newline 
\textbf{Inference for $\boldsymbol{\beta}$} \newline 
Estimation of $\beta$ in the LWA model $(\ref{LWAhazard})$ is based on the following likelihood function
\begin{align}
\nonumber L^{LWA}(\beta) = \prod_{j=1}^{K} \prod_{i=1}^{n} \ \biggl(  \dfrac{ \exp( \beta^\intercal Z_{i}(\min(T_{ij}, C_{i})))}{\sum_{v=1}^{K} \sum_{l=1}^{n} Y_{lv}^{LWA}(\min(T_{ij}, C_{i})) \exp(\beta^\intercal Z_{l}\min(T_{ij}, C_{i}))} \biggr)^{\mathbbm{1}(T_{ij} \leq C_{i})}. 
\end{align}
 
\newpage 
\subsection{Marginal rate models}
\label{marginalmodel} 
When interest lies in marginal features of recurrent event processes, rate and mean functions are attractive due to their clear interpretation. In order to fit marginal models based on rate and mean functions, unbiased estimating equations and robust (sandwich) variance estimation are generally used. In contrast to intensity-based models, the dependence structure among repeated events in rate-based models must not be exactly specified so that marginal analyses require fewer assumptions on the recurrent event process. However, the censoring process needs to be completely independent of the recurrent event process, which is a very strong condition \parencite{Andersen2019}. Such kind of marginal mean and rate models without terminal events have been extensively studied by \citet{Pepe1993}, \citet{Lawless1997}, \citet{Aalen2008} and \citet{Lin2000}.
\newline \newline 
\textbf{Mean and rate function} \newline
The rate function of an arbitrary counting process $\{ N(t): 0 \leq t < \infty \}$ is defined by 
\begin{equation}
r(t)dt = P(dN(t) = 1), 
\label{rate1}
\end{equation}
which can be interpreted as the marginal (unconditional) instantaneous probability of an event occuring between $t$ and $t+dt$. Since $dN(t)$ takes values in $\{ 0,1\}$, it follows that $r(t)dt = \mathbb{E}(dN(t))$. The rate function can be interpreted as the average intensity function at time $t$ across all possible process histories (cf. Section $\ref{rateversusintensity}$). \newline 
The cumulative mean function (CMF) 
\begin{equation}
\label{CMF} \mu(t)= \mathbb{E}(N(t)) = \int_{0}^{t} r(u) du
\end{equation}
gives the marginal expected number of events in $[0, t]$. \citet{Cook2009} proposed a nonparametric estimator for the marginal mean function $\mu(t)$ based on the heuristic arguments that $d\mu(t) = r(t)dt$ and $\mathbb{E}(dN(t)) = d\mu(t)$ for continuous $\mu(t)$. Correspondingly, the latter argument results in $\mathbb{E}(dN(t) - d\mu(t)) = 0$. If observations of $n$ independent individuals are available, the estimating equation in the absence of covariates is $\sum_{i=1}^n Y_{i}(t) (dN_{i}(t) - d\mu(t)) = 0$, leading to 
\begin{align}
\label{CMFestimator} \hat{\mu}(t) = \int_{0}^{t} d\hat{\mu}(u)du = \int_{0}^{t} \frac{{\sum_{i=1}^n Y_{i}(t)dN_{i}(t)}}{{\sum_{i=1}^n Y_{i}(t)}}.
\end{align}
The estimator $\hat{\mu}(t)$ results from the fact that $\mu(t) = \int_{0}^{t} d\mu(u)$ and is similar to the Nelson Aalen estimator for time-to-first-event data but, in this situation, Eq. $(\ref{CMFestimator})$ is interpreted as the estimated mean function. At time $t$, the CMF shows the estimated number of events experienced by individual $i$ by time $t$. 
\newline \newline
\textbf{Two-sample test for differences in CMFs} \newline 
Let $\{ N_{ki}(t): 0 \leq t < \infty \}$ denote the counting process for individual $i$ in treatment group $k$ and let $Y_{ki}(t)$ be the at-risk process of whether individual $i$ is in group $k$ and at risk at $t-$, with $k=1$ and $k=2$ denoting the treatment group. The aggregated at-risk process in group $k$ is defined as $Y_{k.} = \sum_{i=1}^{n_{k}} Y_{ki}(t)$. The mean and rate functions in group $k$ are given by $\mathbb{E}(dN_{ki}(t)) = r_{k}(t)dt$ and $\mathbb{E}(N_{ki}(t)) = \mu_{k}(t)$ for $i=1,...,n_{k}$. The mean functions $\mu_{1}(t)$ and $\mu_{2}(t)$ are not expected to cross. Then, \citet{Lawless1995} suggested a two-sample test for differences in the CMFs based on the test statistic 
\begin{align}
\label{CFMtest} W(\tau) = \int_{0}^{\tau} \dfrac{Y_{1.}(u) Y_{2.}(u)}{Y_{1.}(u) + Y_{2.}(u)}(d\hat{\mu}_{1}(u) - d\hat{\mu}_{2}(u)), 
\end{align}
where $\tau > 0$ is the maximum follow-up time \parencite{Cook2007}. The null hypothesis of the two-sample pseudo-score test is $H_{0}: \{ r_{2}(t)=r_{1}(t) \}$ versus the alternative hypothesis $H_{1}: \{ r_{2}(t)=r_{1}(t)\exp(\beta) \}$. It can be shown that $W(\tau)/\hat{Var}(W(\tau))$ approaches a $\chi^{2}_{1}$ distribution under the null hypothesis, as $Y_{1.}(t)$ and $Y_{2.}(t)$ become large over $[0, \tau]$. 
\newpage 
\subsubsection{Rate-based versus intensity-based modelling}
\label{rateversusintensity}
Previously, the rate function has been defined as the instantaneous (unconditional) probability of experiencing an event in $[t, t+dt)$, without conditioning on any aspect of the past. Marginal models based on rate functions that completely ignore the dependence on the past can be seen as extreme cases. Models that condition on some information of the past are also classified as rate-based approaches. Therefore, rate functions can differ in the extent to which they condition on the previous event history \parencite{Aalen2008}. In order to clearly emphasize the difference between rate- and intensity-based modelling, the general multistate setup from Section $\ref{condmodel}$ will now be reconsidered.

\begin{figure}[H]
\centering
\resizebox{14cm}{!}{
\begin{tikzpicture}[node distance=1.9cm]
  \tikzset{node style/.style={state, fill=gray!20!white, rectangle}}
        \node[node style]               (I)   {0};
        \node[node style, right=of I]   (II)  {1};
        \node[node style, right=of II]  (III) {2};
        \node[draw=none, right=of III]  (dot)  {$\cdots$};
        \node[node style, right=of dot] (IV)   {$j$};
        \node[node style, right=of IV] (V)   {$j+1$};
        \node[draw=none, right=of V]  (dottwo)  {$\cdots$}; 
    \draw[>=latex,
          auto=left,
          every loop]
         (I)   edge node {$\tiny r_{01}(t)$} (II)
         (II)  edge node {$r_{12}(t)$} (III)
         (III) edge node {$$} (dot)
         (dot) edge node {$$} (IV)
         (IV) edge node {$\tiny r_{j(j+1)}(t)$} (V)
         (V) edge node {$$} (dottwo); 
\end{tikzpicture}}
\caption{Multistate representation of a recurrent event process based on transition rates}
\label{multistateRATEmodel}
\end{figure}
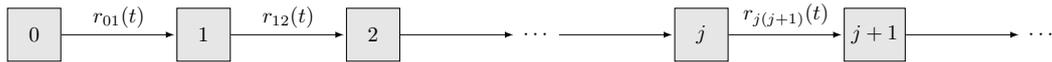

The main difference between intensity and rate functions lies in conditioning on the past. As already mentioned, conditional intensity-based models require full specification of the recurrent event process by mimicking the past through the event history, censoring history and internal/external covariate histories. The transition hazard $\alpha_{i,j(j+1)}(t)$ of the multistate process has been defined as
\begin{align}
\nonumber \alpha_{i, j(j+1)}(t)dt &:= P(j \longrightarrow (j+1) \ \textnormal{transition between} \ t \ \textnormal{and} \ t+dt \ | \ \mathcal{F}(t-)) \\
\nonumber &= P((j+1)^{th} \ \textnormal{event in} \ [t,t+dt) \ | \ \mathcal{F}(t-)) \\
\nonumber &= P(dN_{j(j+1)}(t) = 1 \ | \ \mathcal{F}(t-)) \\
\label{transitionhazard} &= P(N_{i}(t) = j+1 \ | \ N_{i}(t-)=j, \ (\underbrace{N_{i}(s)}_{1.}, \ \underbrace{Y_{i}(s)}_{2.}, \ \underbrace{X_{i, int}(s)}_{3.}, \ \underbrace{X_{i, ext}(s)}_{4.} )_{s < t})
\end{align}

\begin{enumerate}
\item The past event history $(N_{i}(s))_{s<t}$ includes information on occurrences and timing of events up to time $t-$. It is well known whether individual $i$ is event-free at $t-$ or if individual $i$ has already experienced a specific number of repeated events. If $N_{i}(t-) > 0$, then the past event history also provides information on timing of prior events. For instance, for an individual who has already experienced one event at time $t_{1}$, the past event history gives detailed insights into the counting process history: $N_{i}(0)=0, N_{i}(t_{1}-)=0, N_{i}(t_{1})=1, N_{i}(t-)=1$. Further, the time since the most recent event can be derived as: $t-T_{i(N_{i}(t-))}$. 
\item The censoring history $(Y_{i}(s))_{s<t}$ informs about the at-risk status of individual $i$. If $Y_{i}(t-)=0$, then the instantaneous probability for experiencing an event becomes zero. This means, the intensity function can only be $>0$ as long as the individual $i$ is at-risk for an event and under observation. Otherwise, observation of the recurrent event process ceases.  
\item The covariate history $(X_{i, int}(s))_{s <t}$ keeps track on the measurements of internal covariates that may be responsive to event occurrences. 
\item The covariate history $(X_{i, ext}(s))_{s <t}$ contains information on measurements that are 'external' to the recurrent event process. The intensity function $\lambda_{i}(t)$ at time $t$ may depend on the external covariate process up to $t-$ but it is assumed that the complete covariate path of external covariates is already part of $\mathcal{F}(0)$, i.e., $(X_{i, ext}(\infty)) \subset \mathcal{F}(0)$, since realizations of external covariates are not affected by event occurrences. Consequently, the external covariate processes do not contribute to the nested structure of the history/filtration $\mathcal{F}(t)$ \parencite{Cook2007, Kalbfleisch2002}. 
\newline
\end{enumerate}

Intensity-based modelling essentially requires deep knowledge of the true underlying recurrent event process as well as sufficient information by means of observed covariates to adequately capture the past. Conditional intensity-based models are therefore very sensitive to model misspecification. If the past information is 'incomplete', a rate function rather than a intensity function is targeted. 

In contrast to transition hazards (or intensities), the transition rate $r_{i,j(j+1)}(t)$ for a $j \longrightarrow (j+1)$ transition in the multistate setup is defined via:
\begin{align}
\nonumber r_{j(j+1)}(t)dt &:= P(j \longrightarrow (j+1) \ \textnormal{transition between} \ t \ \textnormal{and} \ t+dt \ | \ \mathcal{G}(t-)) \\
\nonumber &= P((j+1)^{th} \ \textnormal{event in} \ [t,t+dt) \ | \ \mathcal{G}(t-)) \\
\nonumber &= P(dN_{j(j+1)}(t) = 1 \ | \ \mathcal{G}(t-)) \\
\label{transitionrate} &= P(N_{i}(t) = j+1 \ | \ N_{i}(t-)=j, \ \mathcal{G}(t-)).
\end{align}

It can be seen from Eq. $(\ref{transitionhazard})$ and Eq. $(\ref{transitionrate})$ that the difference between transition rates and transition hazards is mainly characterized by the conditioning set. While the transition hazard conditions on the complete past, the transition rate conditions only on a part of the underlying process history. In order to give a more precise definition, let $\mathcal{G}(t)$ be the 'incomplete' past up to and including time $t$, with $\mathcal{G}(t) \neq \mathcal{F}(t)$. In doing so, $\mathcal{G}(t-)$ can be defined in several ways. 
\newline \newline
\textbf{Examples of non-nested conditioning sets:}
\begin{itemize}
\item $\mathcal{G}_{1}(t-) = \bigl \{ \bigl(Y(s) \bigr)_{s < t} \bigr \}$
\item $\mathcal{G}_{2}(t-) = \bigl \{ \bigl(Y(s), X_{ext}(s) \bigr)_{s < t} \bigr \}$
\item $\mathcal{G}_{3}(t-) = \bigl \{ \bigl(Y(s), N(s), X_{ext}(s) \bigr)_{s < t}, X_{int}(t) \bigr \}$
\item $\mathcal{G}_{4}(t-) = \bigl \{ N(t-), \bigl(Y(s), X_{ext}(s), X_{int}(s) \bigr)_{s < t} \bigr \}$
\item $\mathcal{G}_{5}(t-) = \bigl \{ N(t-), \bigl(Y(s), X_{ext}(s) \bigr)_{s < t}, X_{int}(t) \bigr \}$ , 
\end{itemize}
A transition rate with $\mathcal{G}_{1}$ as conditioning set is an extreme case, as information on the preceding event and covariate history is completely ignored. In contrast to $\mathcal{G}_{1}$, $\mathcal{G}_{2}$ additionally considers the external covariate processes up to time $t-$ to model the dependence of the past on future transitions. Using $\mathcal{G}_{3}$, the past includes the complete event, censoring and external covariate histories but only current realizations of internal covariates rather than the complete paths up to $t$. The associated transition rate is given as follows:
\begin{align}
\nonumber r_{j(j+1)}(t)dt &= P(N(t) = j+1 \ | \ N(t-)=j, \ \mathcal{G}_{3}(t-)) \\
\nonumber &= P(N(t) = j+1 \ | \ \ N(t-)=j, \ (N(s), Y(s), X_{ext}(s))_{s < t},  \underbrace{X_{int}(t)}_{\substack{\textnormal{Individuals with different} \\ \textnormal{internal covariate paths in [0,t-)} \\ \textnormal{but same value} \\ \textnormal{at t-}}}). 
\end{align}
In fact, $r_{j(j+1)}(t)dt$ is the average transition rate that applies to individuals who are under observation at $t-$, have already experienced $j$ events under a specific event history and whose internal covariate value at $t-$ is $x_{int}(t)$. So, it is only conditioned on the immediate past of the internal covariate history. 
$\mathcal{G}_{4}$ contains information on censoring events and both internal and external covariates up to $t-$ but models the influence of the counting process history on future transitions only through the current state $N(t-)$ (or cumulative number of observed events up to $t-$). Thus, it is only conditioned on the immediate past of the counting process history. The corresponding transition rate is then defined via 
\begin{align}
\nonumber r_{j(j+1)}(t)dt &= P(N(t) = j+1 \ | \ N(t-)=j, \ \mathcal{G}_{4}(t-)) \\
\nonumber &= P(N(t) = j+1 \ | \ \underbrace{N(t-)=j}_{\substack{\textnormal{Individuals who have already} \\ \textnormal{experienced j events at t-,} \\ \textnormal{with different prior event histories} \\ \rightarrow \ \textnormal{previous event/transition times} \\ \textnormal{are disregarded}}}, \bigl(Y(s), X_{ext}(s), X_{int}(s) \bigr)_{s < t}), 
\end{align}
and applies to individuals who are under observation at $t-$, whose external and internal covariate paths are given by $(x_{ext}(s), x_{int}(s))_{s < t}$ and who are in state $j$ at $t-$, irrespective of the preceding event history. Given the censoring and covariate processes up to time $t$, the transition rate can be interpreted as an average instantaneous 'risk' for making a $j \longrightarrow (j+1)$ transition at t, arising from individuals with different event histories but fixed $N(t-)$. $\mathcal{G}_{5}$ is an extension of $\mathcal{G}_{4}$, where only the current value of $X_{int}(t)$ rather than the complete covariate history is taken into account. 
\newline \newline 
Since the conditioning sets $\mathcal{G}$ are not nested in time, there exists no filtration to which $M_{j(j+1)}(t)$ is a martingale. 
While $M_{j(j+1)}(t) = N_{j(j+1)}(t) - \int_{0}^{t} \mathbbm{1}(N(t) = j-1, C \geq t) \alpha_{j(j+1)}(u \ | \ \mathcal{F}(u-))du$ is a zero-mean martingale with regard to $\mathcal{F}$, $M_{j(j+1)}(t) = N_{j(j+1)}(t) - \int_{0}^{t} \mathbbm{1}(N(t) = j-1, C \geq t) r_{j(j+1)}(u \ | \ \mathcal{G}(u-) )du$ does not define a martingale because of the non-nested conditioning sets. In general, 
\begin{equation}
\nonumber \alpha_{j(j+1)}(t \ | \ \mathcal{F}(t-))  \ \neq \ r_{j(j+1)}(t \ | \ \mathcal{G}(t-) ).  
\end{equation} 
\newline \newline 
In summary, it can be differentiated between two different settings:
\begin{itemize}
\item If the information on the past is 'complete' and nested across time, the target quantity is the hazard or intensity function \ $\Longrightarrow$ intensity-based modelling. 
\item If the information on the past is 'incomplete' and the recurrent event process can not be fully specified, the target quantity is the rate function \ $\Longrightarrow$ rate-based modelling.
\newline
\end{itemize}

\subsubsection{Lin-Wei-Yang-Ying model}
Based on the previously explained concept of rate functions, the LWYY model can now be introduced. The LWYY model is one of the most commonly used rate-based models for recurrent event analyses in absence of terminal events \parencite{Lin2000}. It can be seen as analogue to the intensity-based AG model. As described in Section $\ref{IntensityAGmodel}$, the proportional intensity-based AG model assumes the following two major properties
\begin{align}
\label{A} P(dN_{i}(t) = 1 \ | \ \mathcal{F}(t-)) \ &= \ P(dN_{i}(t) = 1 \ | \ Y_{i}^{AG}(t), \ Z_{i}(t)) \\ 
\label{B} P(dN_{i}(t) = 1 \ | \ Y_{i}^{AG}(t), \ Z_{i}(t)) &= Y_{i}^{AG}(t) \alpha_{0}(t) \exp( \beta^\intercal Z_{i}(t)), 
\end{align}
where $Z_{i}(t)$ reflects functions of the covariate process $X_{i}(t) = \{ (X_{i,ext}(u), X_{i,int}(u)): 0 \leq u \leq t \}$, the past event history process $N_{i}(t-) = \{ N_{i}(u): 0 \leq u < t \}$ and interaction with time $t$. Property $(\ref{A})$ assumes that the impact of the process history on further event occurrences is completely explained by measured covariates included in $Z_{i}(t)$. Therefore, the time increments between recurrent events are conditionally uncorrelated given $Z_{i}(t)$. On the other hand, property $(\ref{B})$ specifies a multiplicative effect of $Z_{i}(t)$ on the intensity function \parencite{Lin2000}.\newline 
If there is only insufficient information on the past available to model the complex structure of the recurrent event process, dependence among repeated events may not be appropriately reflected by $Z_{i}(t)$. In this case, property $(\ref{A})$ is not fulfilled anymore and the AG model may be potentially misspecified. Indeed, the LWYY model assumes an arbitrary dependence structure between recurrent events and does not rely on property $(\ref{A})$, but it is mainly defined by property $(\ref{B})$.
\newline 
In contrast to the AG model, the LWYY model is only characterized by the following property: 
\begin{align}
\label{model1234} P(dN_{i}(t) = 1 \ | \ Y_{i}^{LWYY}(t), \ Z_{i}(t)) &= Y_{i}^{LWYY}(t) r_{0}(t) \exp( \beta^\intercal Z_{i}(t)). 
\end{align}
In model $(\ref{model1234})$, $Y_{i}^{LWYY}(t) = \mathbbm{1}(C_{i} \geq t) \in \{0,1\}$ defines the predictable at-risk process for individual $i$, $r_{0}(t)$ is a baseline rate function and $Z_{i}(t)$ is defined as above such that $P(dN_{i}(t) = 1 \ | \ \mathcal{F}_{i}(t-)) \ \neq \ P(dN_{i}(t) = 1 \ | \ Z_{i}(t))$. Thus, the LWYY model targets the rate function rather than the intensity function of the recurrent event process. 
\newpage
More specifically, the proportional rate-based model proposed by \citet{Lin2000} takes the following form
\begin{align}
\label{LWYYmodel} r_{i}(t) &= Y_{i}^{LWYY}(t) r_{0}(t) \exp( \beta^\intercal Z_{i}(t)),  \\
\label{LWYYmodel2} R_{i}(t) &= \int_{0}^{t} r_{i}(u) du = \int_{0}^{t} Y_{i}^{LWYY}(u) r_{0}(u) \exp( \beta^\intercal Z_{i}(u)) du.   
\end{align}
The proposed LWYY model is semiparametric in the sense that the baseline rate function is unspecified, while the form which relates covariates to the rate function is parametrically specified. If $Z_{i}(t)$ excludes internal covariates, model $(\ref{LWYYmodel2})$ can be interpreted as the mean function for recurrent events, i.e., $R(t) = \mathbb{E}(N_{i}(t) \ | \ Z_{i}(t))$. Otherwise, model $(\ref{LWYYmodel2})$ is referred to as the cumulative rate function of recurrent events. If $Z_{i}(t) = Z_{i}$ only includes baseline covariates, model $(\ref{LWYYmodel2})$ is simply given by
\begin{align}
\mathbb{E}(N_{i}(t) \ | \ Z_{i}) = Y_{i}^{LWYY}(t) \exp( \beta^\intercal Z_{i}) R_{0}(t),   \ \ \textnormal{with} \ R_{0}(t) = \int_{0}^{t} r_{0}(u)du.  
\end{align}
It can be seen that the intensity-based AG model implies the LWYY model but not the other way round. The AG model makes much stronger assumptions than the LWYY model. The rate-based LWYY model is less restrict in the sense that it allows arbitrary dependence structure between recurrent events. \newline \newline
\textbf{Inference for $\boldsymbol{\beta}$ and large sample theory} \newline 
In order to describe an estimation procedure for $\beta$ and to apply large sample theory, let $\mathcal{G}(t)$ denote the non-nested conditioning set associated with $Z_{i}(t)$ in the LWYY model. The true regression coefficient vector is denoted by $\beta_{true}$. \citet{Lin2000} further assumed that the censoring mechanism is completely indepedent of the recurrent event process such that $P(dN(t) = 1 \ | \ Z(t), C(t) \geq t) = P(dN(t) = 1 \ | \ Z(t))$. The observable data from individual $i$ $\{N_{i}(\cdot), Y^{LWYY}_{i}(\cdot), Z_{i}(\cdot)\}$ is assumed to be independent and identically distributed. Martingale-based partial likelihood theory for the estimation of $\beta$ can not be applied, as the Doob-Meyer decomposition does generally not follow the LWYY model. Although $dM_{i}(t) = dN_{i}(t) - Y_{i}^{LWYY}(t) \exp( \beta_{true}^\intercal Z_{i}(t)) r_{0}(t)dt$ is no zero-mean martingale, the expectation of $dM_{i}(t)$ given $Z_{i}(t)$ is zero, i.e., 
\begin{align}
\nonumber \mathbb{E}(dM_{i}(t) \ | \ Z_{i}(t)) = \mathbb{E}(dN_{i}(t) - Y_{i}^{LWYY}(t) \exp( \beta_{true}^\intercal Z_{i}(t)) r_{0}(t) dt \ | \ Z_{i}(t)) = 0, 
\end{align}
implying $\mathbb{E}(dM_{i}(t))=0$. An unbiased estimating function can be defined via 
\begin{align}
\nonumber U^{LWYY}(\beta_{true}, t) &= \sum_{i=1}^{n} \int_{0}^{t} \biggl( Z_{i}(u) -  \dfrac{S^{(1), LWYY}(\beta_{true}, u)}{S^{(0), LWYY}(\beta_{true}, u)} \biggr) dM_{i}(u),  \ \ \ \textnormal{with} \\
\nonumber S^{(0), LWYY}(\beta, t) &= \dfrac{1}{n} \sum_{i=1}^{n} Y_{i}^{LWYY}(t) \exp( \beta^\intercal Z_{i}(t))  \\
\nonumber S^{(1), LWYY}(\beta, t) &= \dfrac{1}{n} \sum_{i=1}^{n} Y_{i}^{LWYY}(t) Z_{i}(t) \exp( \beta^\intercal Z_{i}(t)).
\end{align}
The unbiased estimation equation looks identical to the partial likelihood score function of the AG model but, in the LWYY model, the contributions to $U^{LWYY}(\beta_{true}, t)$ at unique event times are correlated because of the non-nested conditioning events. Thus, $U^{LWYY}(\beta_{true}, t)$ should not be associated with partial likelihood theory. Solving $U^{LWYY}(\beta, t) = 0$ for $\beta$ results in the estimated regression coefficient vector $\hat{\beta}_{LWYY}$. For $\hat{\beta}_{LWYY}$, it follows that $\hat{\beta}_{LWYY} \ \xrightarrow[n \longrightarrow \infty]{as} \beta_{true}$, where $\hat{\beta}_{LWYY} = \hat{\beta}$ obtained from the intensity-based AG model. \citet{Lin2000} used empirical process theory to proof the asymptotic distribution of $\hat{\beta}_{LWYY}$, given some regularity conditions. The stochastic process $(1/\sqrt{n}) U^{LWYY}(\beta_{true}, t)$ converges to a continuous Gaussian process with zero mean and covariance matrix $\Pi$. 
\begin{align}
\label{E} &\dfrac{1}{\sqrt{n}} U^{LWYY}(\beta_{true}) \xrightarrow{D} N \bigl (0, \Pi \bigr), \ \ \ \textnormal{with} \\ 
\nonumber &\Pi = \mathbb{E} \biggl[ \int_{(0, \tau]} \biggl( Z_{1}(u) - \dfrac{s^{(1)}(\beta_{true}, u)}{s^{(0)}(\beta_{true}, u)} \biggr) dM_{1}(u) \int_{(0, \tau]}   \biggl( Z_{1}(v) - \dfrac{s^{(1)}(\beta_{true}, v)}{s^{(0)}(\beta_{true}, v)} \biggr)^\intercal dM_{1}(v) \biggr]. 
\end{align}
The covariance matrix function $\Pi$ can be consistently estimated by the empirical estimator 

\begin{align}
\nonumber &\hat{\Pi} = \dfrac{1}{n}  \sum_{i=1}^{n} \int_{(0, \tau]} \biggl( Z_{i}(u) - \dfrac{S^{(1)}(\hat{\beta}, u)}{S^{(0)}(\hat{\beta}, u)} \biggr) d\hat{M}_{i}(u) \int_{(0, \tau]} \biggl( Z_{i}(v) - \dfrac{S^{(1)}(\hat{\beta}, v)}{S^{(0)}(\hat{\beta}, v)} \biggr) d\hat{M}_{i}(v),  \ \ \ \textnormal{with} \\ 
\nonumber &\hat{M}_{i}(t) = N_{i}(t) - \int_{(0, \tau]} Y_{i}^{LWYY}(t) \exp( \hat{\beta}^\intercal Z_{i}(u)) \hat{r}_{0}(u) du \ \ \ \textnormal{and} \\
\nonumber &\hat{R}_{0}(t) = \int_{(0, \tau]} \dfrac{dN.(t)}{n S^{(0)}(\hat{\beta}, u)}, 
\end{align}
where $N.(t) =  \sum_{i=1}^{n} N_{i}(t)$, $t \in [0, \tau]$, is the aggregated counting process and $\hat{R}_{0}(t)$ refers to the Aalen Breslow-type estimator. \citet{Lin2000} also proved that $\hat{R}_{0}(\cdot) \ \xrightarrow[n \longrightarrow \infty]{as} R_{0}(\cdot)$. If $\beta_{true}$ corresponds to the true parameter vector, Taylor series expansion around $\beta_{true}$ can be used to derive the asymptotic distribution of $\sqrt{n}(\hat{\beta}-\beta_{true})$. That is, \begin{align}
\nonumber &0 = U^{LWYY}(\hat{\beta}, t) \ \approx \ U^{LWYY}(\beta_{true}, t) - \frac{\partial}{\partial \beta_{true}} U^{LWYY}(\beta_{true}, t)(\hat{\beta}-\beta_{true}) \\ 
\nonumber &\Longleftrightarrow \sqrt{n}(\hat{\beta}-\beta_{true}) \ \approx \  \biggl( \underbrace{\dfrac{1}{n}  \frac{\partial}{\partial \beta_{true}} U^{LWYY}(\beta_{true}, t)}_{\xrightarrow[n \longrightarrow \infty]{P} \ U } \biggr)^{-1} \cdot \underbrace{\dfrac{1}{\sqrt{n}} U^{LWYY}(\beta_{true}, t)}_{\substack{\xrightarrow[n \longrightarrow \infty]{D} \ N(0, \Pi) \\ \textnormal{according to} \ Eq. (\ref{E})}} \\
\nonumber & \ \ \ \ \ \ \ \ \xrightarrow[n \longrightarrow \infty]{D} \ N\bigl(0, U^{-1} \Pi \bigl(U^{-1}\bigr)^\intercal \bigr), 
\end{align}
where the matrix $U$ is defined as $U := \dfrac{1}{n} \mathbb{E} \bigl(\frac{\partial}{\partial \beta_{true}} U^{LWYY}(\beta_{true}, t) \bigr)$. Further, it yields that
\begin{align}
\label{F} &\hat{U} :=  \dfrac{1}{n} \int_{(0, \tau]} \dfrac{S^{(2), LWYY}(\hat{\beta}, u)}{S^{(0), LWYY}(\hat{\beta}, u)} - \biggl(\dfrac{S^{(1), LWYY}(\hat{\beta}, u)}{S^{(0), LWYY}(\hat{\beta}, u)}\biggr)^{\otimes 2} dN(u) \  \xrightarrow[n \longrightarrow \infty]{P} \ U \\ 
\label{G} &\hat{\Pi} \xrightarrow[n \longrightarrow \infty]{P} \ \Pi
\end{align}
Using Eq. $(\ref{F})$ and $(\ref{G})$, $\sqrt{n}(\hat{\beta}-\beta_{true})$ is asymptotically normal distributed with zero mean and covariance matrix $\hat{U}^{-1} \hat{\Pi} \bigl(\hat{U}^{-1}\bigr)^\intercal$, i.e., 
\begin{align}
\nonumber \sqrt{n}(\hat{\beta}-\beta_{true}) \ \xrightarrow[n \longrightarrow \infty]{D} \ \mathcal{N} \biggl (0, \ \underbrace{\hat{U}^{-1} \hat{\Pi} \bigl(\hat{U}^{-1}\bigr)^\intercal}_{=: W}\biggr), 
\end{align}
\newline
where $W$ corresponds to the robust covariance matrix estimator. It can be concluded that the AG and LWYY models differ in the limiting behaviour of the covariance matrix. However, if the intensity-based AG models holds, then $U=\Sigma$ and $\Pi = \Sigma$, which leads to $W = \Sigma^{-1}$. As a result, the covariance estimators of the AG and LWYY models coincide.

\subsubsection{Partially conditional rate-based model}
The partially conditional rate-based model is obtained by specifying 
\begin{align}
\label{PCRBmodel} r_{ij}(t) &= Y_{ij}^{PCRB}(t) r_{0j}(t) \exp( \beta^\intercal Z_{i}(t)),  
\end{align}
where $r_{0j}(t)$ is an event-specific baseline rate function, $Y_{ij}^{PCRB}(t) = \mathbbm{1}(N_{i}(t)=j-1, \ C_{i} \geq t)$ the at-risk indicator, $\beta \in \mathbb{R}^{q}$ a vector of regression coefficient and $Z_{i}(t)=(Z_{i1}(t), ..., Z_{iq}(t)) \in \mathbb{R}^{q}$ is a vector of functions of external covariates. The partially conditional rate-based model involves time-dependent stratification on the cumulative number of events, similar to the intensity-based PWP-CP model. The term 'partially conditional' is used to reflect that only part of the event history is conditioned upon. 

\section{Recurrent events in randomized clinical trials}
\label{RCTrecurrentevents}
Although a broad range of statistical methods is available for the analysis of recurrent events, only a few methods are qualified for evaluating causal treatment effects on a recurrent event endpoint in RCTs \parencite{Kuramoto2008}. In particular, randomization ensures balance in the distribution of baseline covariates across the treatment groups and mitigates the effect of confounding factors to yield valid causal inference. In RCTs, it is important that causal conclusions on the treatment effect can be drawn based on the random assignment of individuals. If it is conditioned on the prior event history (e.g., N(t-)) of the recurrent event process, this property is not fulfilled anymore because balance in the distribution of other covariates that has been orginally achieved by randomization is lost. This can be shown by the following example: suppose a clinical trial that is conducted to assess the effect of an intervention on repeated disability progression in PPMS patients. The active treatment is assumed to be beneficial in that it reduces the probability of experiencing disability progression. Random assignment of study participants to either the active treatment group or control group ensures that both measured and unmeasured confounders are equally distributed across both treatment groups. When conditioning on the immediate event $N(t-)$, treated patients who have already experienced one event are compared to untreated patients who have already experienced one event. However, these groups are not comparable because treated patients who have already experienced one event are expected to be much worse in their current disease conditions than untreated patients with $N(t-)=1$. This leads to an imbalance across the treatment and control group with respect to other covariates and induces confounding. As a consequence, valid causal conclusion on the treatment effect size can not be drawn. \newline 
Therefore, it is well-known that treatment comparisons in clinical trials should not be carried out by conditioning on post-randomization or intermediate events which may be responsive to the treatment. Conditional intensity-based models require the correct specification of the recurrent event process and may therefore condition on internal time-varying covariates, making treatment comparisons in RCTs difficult. This suggests that treatment effects are better expressed by marginal parameters \parencite{Cook2009}. 

\subsection{Specification of treatment effects}
When interest lies in evaluating treatment effects in clinical trials with recurrent event endpoint, methods based on marginal rate and mean functions are generally recommended to use \parencite{Cook2007}. These quantities are also easy for clinicians and patients to understand, as the implications of using the active treatment rather than the control treatment should also be clear to non-statisticians. For instance, the difference in CMFs is easily interpreted and clearly understandable. In particular, the LWYY model and the standard NB model are appropriate analysis methods for a recurrent event endpoint in RCTs.   
\newpage
\textbf{Lin-Wei-Yang-Ying model} \newline 
The proportional rate model proposed by \citet{Lin2000} is given by 
\begin{align}
\label{EqLWYY1} \mathbb{E}(dN_{i}(t) \ | \ Z_{i}) &= r_{0}(t) \exp( \beta Z_{i}) \\
\label{EqLWYY2} \mathbb{E}(N_{i}(t) \ | \ Z_{i}) &= \mu_{0}(t) \exp( \beta Z_{i}), \ \ \textnormal{with} \ \ \mu_{0}(t) = \int_{0}^{t} r_{0}(u)du,  
\end{align}
where $Z_{i} \in \{0, 1 \}$ is a binary variable indicating the treatment group for individual $i$. In the LWYY model, the estimated treatment effect is expressed as rate ratio (RR) and has therefore simple marginal ('population-average') interpretation (cf. Eq. $(\ref{EqLWYY1})$). The $RR=\exp(\beta)$ is defined as the ratio of relative rates of events for treated versus untreated individuals. The treatment effect also applies to the expected number of events (cf. Eq. $(\ref{EqLWYY2})$). The semiparametric LWYY model assumes that the event rate may depend on time and the baseline rate function $r_{0}(t)$ is unspecified. Another advantage of using model $(\ref{EqLWYY1})$ is that it is not restricted to any specific type of recurrent event process.
\newline \newline
\textbf{Negative binomial model} \newline
The standard time-homogeneous NB approach models the individual-specific rate function as
\begin{align}
\label{EqNB1} \mathbb{E}(dN_{i}(t) \ | \ Z_{i}, U_{i}) &= U_{i} r_{0}(t) \exp( \beta Z_{i}) \\
\label{EqNB2} \mathbb{E}(N_{i}(t) \ | \ Z_{i}, U_{i}) &= U_{i} \mu_{0}(t) \exp( \beta Z_{i}), \ \
\end{align} 
where $U_{i}$ is a gamma distributed random effect reflecting heterogeneity across individuals, with $\mathbb{E}(U_{i})=1$ and $Var(U_{i})=\phi$. The NB model also expresses the treatment effect as RR. As summarized by \citet{Cook2009}, the individual-specific and the population-average relative rates are the same under this model. Since $\mathbb{E}(dN_{i}(t) \ | \ Z_{i}) = r_{0}(t) \exp( \beta Z_{i})$, the treatment effect measure $\exp(\beta)$ represents both individual-specific and population-average effects of the treatment. A further attractive property of this approach is that the random effect $U_{i}$ is independent of the treatment group $Z_{i}$ due to random allocation of treatment. In contrast to the LWYY model, the parametric NB model $(\ref{EqNB1})$ assumes that the event rate is constant over time but may differ across individuals. If the event rate is roughly constant over time, the NB and LWYY analyses have been expected to provide similar treatment effect estimates. 

\subsection{Sample size calculation}
\label{SamplesizeRCT}
Sample size calculation is a crucial point in designing clinical trials to ensure sufficient power to detect treatment effects \parencite{Cook1995}. Sample size calculation for RCTs with recurrent event endpoints have been discussed by \citet{Cook1995, Cook2007, Cook2009, Ingel2014, Bernardo2001, Matsui2005, Rebora2012}. In the following, sample size calculation for the standard time-homogeneous NB model and the LWYY model is described.
\newline \newline 
\textbf{Negative binomial model} \newline 
\citet{Cook2007} proposed a sample size formula for clinical trials with mixed Poisson process data. At the design stage, $n$ individuals are randomly allocated with probability $0.5$ to either the active treatment or control group. Individuals are followed over the time period $[0, \tau]$ but some individuals may withdraw from the study earlier. Let $W_{i}$ denote the withdrawal time for individual $i$ so that $C_{i}=min(W_{i}, \tau)$ corresponds to the right-censoring time. As previously defined, $n_{i}=N_{i}(C_{i})$ is the total number of events experienced by individual $i$ over $[0, C_{i}]$. Recurrent events are assumed to follow a mixed time-homogeneous Poisson process with constant event rates $U_{i}r_{0}\exp(\beta_{1} Z_{i})$, where $U_{i}$ is an individual-specific gamma distributed random effect (with $\mathbb{E}(U_{i})=1$ and $Var(U_{i})=\phi$) and $Z_{i} \in \{0, 1\}$ is the treatment group, $i=1,2,...,n$. Constant event rates are often seen in clinical trials, making the model reasonable. 
\newpage 
Under this assumption, the underlying model conforms to the NB model described in Section $\ref{SectionNBmodel}$ (cf. special case), with $n_{i}=N_{i}(C_{i}) \sim NegBin(C_{i}\exp(\beta_{0} + \beta_{1}Z_{i}), \phi)$ and $\beta_{0}=\log(r_{0})$. It can be shown that 
\begin{align}
\label{Samplesize1} Var(\sqrt{n}(\hat{\beta_{0}}- \beta_{0})) = \biggl[ \dfrac{\exp(\beta_{0})\mathbb{E}(C \ | \ Z = 0)}{1 + \phi \exp(\beta_{0})\mathbb{E}(C \ | \ Z = 0)} \biggr]^{-1} \\
\label{samplesizeNB2} Var(\sqrt{n}(\hat{\beta_{1}}- \beta_{1})) = \sum_{Z=0}^{1} \biggl[ \dfrac{\exp(\beta_{0} + \beta_{1}Z)\mathbb{E}(C \ | \ Z )}{1 + \phi \exp(\beta_{0} +\beta_{1}Z)\mathbb{E}(C \ | \ Z)} \biggr]^{-1}, 
\end{align}
where $C_{1},..., C_{n}$ are iid random variables. It is typically assumed that early withdrawal is independent of treatment and the withdrawal times $W_{1}, ..., W_{n}$ follow an exponential distribution such that calculation of the expectations in Eq. $(\ref{Samplesize1})$ and Eq. $(\ref{samplesizeNB2})$ becomes simpler. \newline 
For this NB model, the minimum number of individuals required to obtain a power of $1-\gamma$ for rejecting the null hypothesis $H_{0}: \{ \beta_{1} = \beta_{1,H_{0}} \}$ against $H_{1}: \{ \beta_{1} = \beta_{1,H_{1}} \}$ at the two-sided significance level $\alpha$, when comparing two treatment groups of the same size, is then given by
\begin{equation}
\label{sampleNB} n \ > \ \dfrac{\bigl( Var_{H_{0}}(\sqrt{n}(\hat{\beta_{1}}-\beta_{1,H_{0}}))z_{1-\alpha/2} + Var_{H_{1}}(\sqrt{n}(\hat{\beta_{1}}-\beta_{1,H_{1}}))z_{1-\gamma}\bigr)^{2}}{(\beta_{1,H_{0}}-\beta_{1,H_{1}})^{2}}.  
\end{equation} 
$Var_{H_{0}}$ and $Var_{H_{1}}$ denote the variances given by Eq.$(\ref{samplesizeNB2})$ under the null and alternative hypotheses. $z_{1-\alpha/2}$ and $z_{1-\gamma}$ correspond to the quantiles of the standard normal distribution such that $P(Z \leq z_{\alpha}) = \alpha$, for $ Z \sim N(0,1)$. Often $\beta_{1, H_{0}}$ is equal to 0, so that the events happen at the same rate in both treatment groups. As seen from the sample size formula $(\ref{sampleNB})$, the following parameters need to be specified in advance: $\beta_{1,H_{0}}$, $\beta_{1,H_{1}}$, $\phi$, $\tau$ and the mean of the exponential distribution. \newline 
Sample size calculation for the NB model has also been discussed by \citet{Matsui2005, Tang2015, Tang2018}. 
\newline \newline 
\textbf{Lin-Wei-Yang-Ying model} \newline
\citet{Tang2019} proposed a procedure for calculating the sample size formula for a LWYY model based on a mixed non-homogeneous Poisson process, while considering two different study designs. The sample size formula is applicable to both study designs. The first design is defined by a fixed treatment duration for all individuals, whereas in the second design individuals are enrolled at different calendar times but administratively right-censored at the same calendar time. Let $Z \in \{ 0, 1 \}$ denote the treatmen group. At the design stage, $n$ individuals are randomly allocated with probability $p_{z}$ to treatment group $z$ and each individual is followed over the time interval $[0, C_{i}]$. As already mentioned, recurrent event data is assumed to follow a mixed non-homogeneous Poisson process such that, given the random effect $U_{i}$, the recurrent event process of individual $i$ follows a Poisson process with mean function $U_{i}R(t)=U_{i}R_{0}(t)\exp(\beta_{1}Z_{i})$, with $R_{0}(t)=\int_{0}^{t} r_{0}(u)du$. Thus, $R_{1}(t)=R_{0}(t)\exp(\beta_{1})$ and $R_{0}(t)$ are the mean event functions for the active and control treatment group, respectively. More generally, it can be formulated as $R_{z}(t)$. For instance, the underlying event rate function $r_{0}(t)$ could be piecewise constant or of Weibull form. The random effect $U_{i}$ arises from an arbitrary distribution with $\mathbb{E}(U_{i})=1$ and $Var(U_{i})=\phi_{z}$, allowing for group-specific heterogeneity parameters. The distribution for the censoring time is denoted by $G_{z}(t) = 1-\pi_{z}(t)$, where $\pi_{z}(t)= P(C > t \ | \ Z = z)$ is the probability that an individual in treatment group $z$ remains in the study and is under observation at time $t$. However, in most cases, it is reasonable to assume the same censoring distribution in both treatment groups, in which case $\pi_{1}(t)=\pi_{0}(t)$. Then, in a superiority trial with common censoring distribution across treatment groups and $\phi=\phi_{1}=\phi_{2}$ independent of $Z$, the number of individuals required to obtain a power of $1-\gamma$ for rejecting the null hypothesis $H_{0}: \{ \exp(\beta_{1}) \geq 1 \}$ against $H_{1}: \{ \exp(\beta_{1}) < 1 \}$ at the significance level $\alpha$, is given by
\begin{align}
\nonumber n = \dfrac{(z_{1-\alpha/2} + z_{1-\gamma})^2 V_{\beta_{1}}}{\beta_{1}^{2}}, \ \ \ \textnormal{where} \\
V_{\beta} = \biggl[ \dfrac{1}{p_{1}\exp(\beta_{1})} + \dfrac{1}{p_{0}} \biggr] \dfrac{1}{E_{0}} +  \biggl[ \dfrac{\phi}{p_{1}} + \dfrac{\phi}{p_{0}} \biggr] \dfrac{2F_{0}}{E_{0}^{2}}, 
\end{align}
with $E_{z} = \int \pi_{z}(t)dR_{z}(t)$, $F_{z}= \int_{t=0}^{\tau} \pi_{z}(t)R_{z}(t)dR_{z}(t)$ and $\tau$ is the maximum follow-up duration. The objective of a superiority trial is to demonstrate that the active treatment is able to lower the event rate, as compared to the control medication. In their publication, \citet{Tang2019} derived analytic expressions for $E_{z}$ and $F_{z}$ for a Weibull and piecewise constant event rate. For more detailed explanations on the factor $V_{\beta_{1}}$, it is referred to \citet{Tang2019}. 
\newline \newline
\textbf{Schoenfeld formula} \newline
The sample size formula proposed by \citet{Schoenfeld1983} for a time-to-first-event endpoint can also be extended to the recurrent event setting. At the design stage, data is assumed to follow a Poisson process with intensity $\lambda(t)=Y(t) \alpha_{0}(t)\exp(\beta_{1}Z)$. The number of events required to achieve a power of $1-\gamma$ for rejecting the null hypothesis $H_{0}: \{ \beta_{1} = 0 \}$ at the two-sided significance level $\alpha$, when comparing two treatment groups of the same size, is given by 
\begin{equation}
\label{Schoenfeldformula} n_{events} = \dfrac{4{(z_{1-\alpha/2} + z_{1-\gamma})}^2}{\beta_{1}^2}, 
\end{equation}
where $\alpha$ is the type I error rate and $\gamma$ is the type II error rate, respectively. Eq. $(\ref{Schoenfeldformula})$ coincides with the formula for the number of required events in a standard survival analysis when the assumption of exponential event times is made or when the logrank test is used to compare treatment groups. Calculation of the sample size based on Eq. $(\ref{Schoenfeldformula})$ can be found in \citet{Bernardo2001, Ingel2014}. \newline \newline 
In all cases, the relation between the sample size, type I error and type II error can be summarized as follows
\begin{equation}
\textnormal{sample size} \ \propto \ {(z_{1-\alpha/2} + z_{1-\gamma})}^2. 
\label{proportionality}
\end{equation}
The sample size is expected to be proportional to the squared sum of the $z_{1-\alpha/2}$ and $z_{1-\gamma}$ quantiles of the standard normal distribution. This proportionality yields exactly, for instance, for all Z test statistics and also for the Schoenfeld sample size formula based on recurrent event data. Assuming equal variances under the null and alternative hypotheses, the proportionality holds approximately for the Wald test statistic within the NB approach as well. The sample size formula for the LWYY model also fulfills this proportionality. 

\chapter{Application}
In this chapter, the time-to-first-event and recurrent event methods described in Chapter $3$ and Chapter $4$ will be applied to both PPMS and RRMS trial data. Analyses are restricted to data collected during the double-blind treatment period. The main objective is to compare time-to-first-event and recurrent event analyses in RCTs but it is also of interest to investigate covariate effects of baseline and time-dependent variables on event occurrences via multivariate intensity-based and rate-based models to get a better understanding of the recurrent event data. While Section $\ref{DAoratorio}$ represents the results from the ORATORIO trial in PPMS, Section $\ref{DAopera}$ reports the results from the two OPERA trials in RRMS. 

\section{ORATORIO trial}
\label{DAoratorio}
The main purpose of the randomized phase III ORATORIO trial was to demonstrate superior efficacy of ocrelizumab (OCR) compared to placebo (PLA) in patients with PPMS \parencite{oratorio}. In this trial, the primary endpoint was the time to the onset of the first 12-week CDP, where confirmed disability progression was defined according to the standard definition introduced in Section $2.2.1$. Statistical methods for analyzing the primary endpoint included a two-sided log-rank test (stratified by region and age) for differences between the OCR and PLA groups and a Cox regression model for the estimation of the treatment effect. The trial involved $732$ patients who were followed for the first occurrence of disability progression. The study participants had to meet certain inclusion criteria, e.g., age between $18$ and $55$ years, diagnosis of PPMS according to the $2005$ revised McDonald criteria, an EDSS score of $3.0$ to $6.5$ at screening, a score on the pyramidal functions component of the FS of at least $2$ and a certain duration of MS symptoms ($10$ or $15$ depending on the EDSS score at screening). The patients were randomized to OCR or PLA in a $2:1$ ratio so that $488$ patients were assigned to active treatment and $244$ patients to the control group. In MS trials involving patients with RRMS and PPMS, the event of interest is not affected by competing risks (e.g., death), precluding a competing risk analysis. The ORATORIO trial was event-driven such that patients under study were exposed to at least $120$ weeks of treatment with OCR or PLA until the occurrence of approximately $253$ CDP12 events. The median follow-up time was $2.9$ years in the OCR group and $2.8$ years in the PLA group. 
\newline \newline
Table $\ref{BCHOratorio}$ shows the baseline demographic and disease characteristics of the ORATORIO patients which are well balanced across the two treatment groups. The mean age at baseline is $44.7$ (range $20-56$) in the OCR group and $44.4$ (range $18-56$) in the PLA group. $51.4 \%$ of the OCR patients are male, compared to $49.2 \%$ male patients in the PLA arm. In both treatment groups, approximately $14 \%$ of the patients come from the United States and the remaining $86\%$ of the patients come from the rest of the world (ROW). Time since onset of MS symptoms ranges from $1.1$ to $32.9$ years in the active treatment arm (mean $6.7$ years) and from $0.9$ to $23.8$ years in the control arm (mean $6.1$), whereas the time since diagnosis of PPMS ranges from $0.1$ to $16.8$ years in the OCR group (mean $2.9$) and from $0.1$ to $23.8$ years in the PLA group (mean $2.7$), respectively. Of $488$ OCR and $244$ PLA patients, $433$ OCR patients and $214$ PLA patients have not taken a disease-modifying therapy before study start. The EDSS score at baseline is also balanced across the two treatment arms, with an average score of $4.7$ (SD $1.2$) in both groups. Presence of gadolinium-enhancing lesions on $T_{1}$-weighted images has been detected in $27.5 \%$ OCR patients and $24.7 \%$ PLA patients, respectively. The average number of lesions on $T_{2}$-weighted images are $48.7$ in the OCR group and $48.2$ in the PLA group, with an average volume of $12.7$ (OCR) and $10.9$ (PLA). While the normalized brain volume in the active treatment group ranges from $1214.3$ to $1711.1$ (mean $1462.9$, SD $84$), a range from $1216.3$ to $1701.7$ (mean $1469.9$, SD $88.7$) can be observed in the control group. 

\begin{table}[t]
\centering
\scalebox{0.75}{
\begin{tabular}{lcc}
\textbf{Characteristic}                                                                                                                                            & \textbf{\begin{tabular}[c]{@{}c@{}}OCR \\ (N = 488)\end{tabular}}                   & \textbf{\begin{tabular}[c]{@{}c@{}}PLA\\ (N = 244)\end{tabular}}                         \\ \hline
\begin{tabular}[c]{@{}l@{}}Age - years\\ $\qquad$ mean\\ $\qquad$ median (range)\end{tabular}                                                                      & \begin{tabular}[c]{@{}c@{}}$$\\ 44.7 $\pm$ 7.9\\ 46.0 (20.0 - 56.0)\end{tabular}          & \begin{tabular}[c]{@{}c@{}}$$\\ 44.4 $\pm$ 8.3\\ 46.0 (18.0 - 56.0)\end{tabular}          \\ \hline
Male sex - no. (\%)                                                                                                                                                & 251 (51.4)                                                                               & 120 (49.2)                                                                               \\ \hline
\begin{tabular}[c]{@{}l@{}}Geographical region - no. (\%)\\ $\qquad$ United States\\ $\qquad$ Rest of the world\end{tabular}                                       & \begin{tabular}[c]{@{}c@{}}$$\\ 67 (13.7)\\ 421 (86.3)\end{tabular}                      & \begin{tabular}[c]{@{}c@{}}$$\\ 34 (13.9)\\ 210 (86.1)\end{tabular}                      \\ \hline
\begin{tabular}[c]{@{}l@{}}Time since onset of MS symptoms - years\\ $\qquad$ mean\\ $\qquad$ median (range)\end{tabular}                                          & \begin{tabular}[c]{@{}c@{}}$$\\ 6.7 $\pm$ 4.0\\ 5.9 (1.1 - 32.9)\end{tabular}             & \begin{tabular}[c]{@{}c@{}}$$\\ 6.1 $\pm$ 3.6\\ 5.5 (0.9 - 23.8)\end{tabular}             \\ \hline
\begin{tabular}[c]{@{}l@{}}Time since diagnosis of PPMS - years\\ $\qquad$ mean\\ $\qquad$ median (range)\end{tabular}                                             & \begin{tabular}[c]{@{}c@{}}$$\\ 2.9 $\pm$ 3.2\\ 1.6 (0.1 - 16.8)\end{tabular}             & \begin{tabular}[c]{@{}c@{}}$$\\ 2.7 $\pm$3.3\\ 1.3 (0.1 - 23.8)\end{tabular}              \\ \hline
\begin{tabular}[c]{@{}l@{}}No previous use of  disease- \\ $\quad$ modifying therapy - no. (\%)\end{tabular}                                                        & 433 (88.7)                                                                               & 214 (87.7)                                                                               \\ \hline
\begin{tabular}[c]{@{}l@{}}EDSS Score\\ $\qquad$ mean\\ $\qquad$ median (range)\end{tabular}                                                                       & \begin{tabular}[c]{@{}c@{}}$$\\ 4.7 $\pm$ 1.2\\ 4.5 (2.5 - 7.0)\end{tabular}              & \begin{tabular}[c]{@{}c@{}}$$\\ 4.7 $\pm$ 1.2\\ 4.5 (2.5 - 6.5)\end{tabular}              \\ \hline
\begin{tabular}[c]{@{}l@{}}Gadolinium-enhancing lesions on $T_{1}$- \\ $\quad$ weighted images - no./total no. (\%)\\ $\qquad$ yes\\ $\qquad$ no\end{tabular} & \begin{tabular}[c]{@{}c@{}}$$\\ $$\\ 133/484 (27.5)\\ 351/484 (72.5)\end{tabular}        & \begin{tabular}[c]{@{}c@{}}$$\\ $$\\ 60/243 (24.7)\\ 182/243 (75.3)\end{tabular}         \\ \hline
\begin{tabular}[c]{@{}l@{}}Number of lesions on $T_{2}$-weighted images\\ $\qquad$ mean\\ $\qquad$ median (range)\end{tabular}                                     & \begin{tabular}[c]{@{}c@{}}$$\\ 48.7 $\pm$ 38.2\\ 42.0 (0.0 - 249.0)\end{tabular}             & \begin{tabular}[c]{@{}c@{}}$$\\ 48.2 $\pm$ 39.3\\ 43.0 (0.0 - 208.0)\end{tabular}             \\ \hline
\begin{tabular}[c]{@{}l@{}}Total volume of lesions on $T_{2}$-weighted \\ $\quad$ images - $\textnormal{cm}^{3}$\\ $\qquad$ mean\\ $\qquad$ median (range)\end{tabular}         & \begin{tabular}[c]{@{}c@{}}$$\\ $$\\ 12.7 $\pm$ 15.1\\ 7.3 (0.0 - 90.3)\end{tabular}      & \begin{tabular}[c]{@{}c@{}}$$\\ $$\\ 10.9 $\pm$ 13.0\\ 6.2 (0.0 - 81.1)\end{tabular}      \\ \hline
\begin{tabular}[c]{@{}l@{}}Normalized brain volume - $\textnormal{cm}^{3}$\\ $\qquad$ mean\\ $\qquad$ median (range)\end{tabular}                                               & \begin{tabular}[c]{@{}c@{}}$$\\ 1462.9 $\pm$ 84.0\\ 1462.2 (1214.3 - 1711.1)\end{tabular} & \begin{tabular}[c]{@{}c@{}}$$\\ 1469.9 $\pm$ 88.7\\ 1464.5 (1216.3 - 1701.7)\end{tabular} \\ \hline
\end{tabular}
}
\caption[ORATORIO - Baseline demographic and disease characteristics]{ORATORIO - Baseline demographic and disease characteristics (plus-minus values are mean $\pm$ SD)}
\label{BCHOratorio}
\end{table}

First, the results obtained from time-to-first-event analyses will be presented. Results presented in this thesis are based on a reanalysis of the original data and might deviate from the pre-specified analyses.

\subsection{Time-to-first-event analysis}
\label{TTFEApplication}
\begin{figure}[H]
\centering
\scalebox{0.72}{
\includegraphics{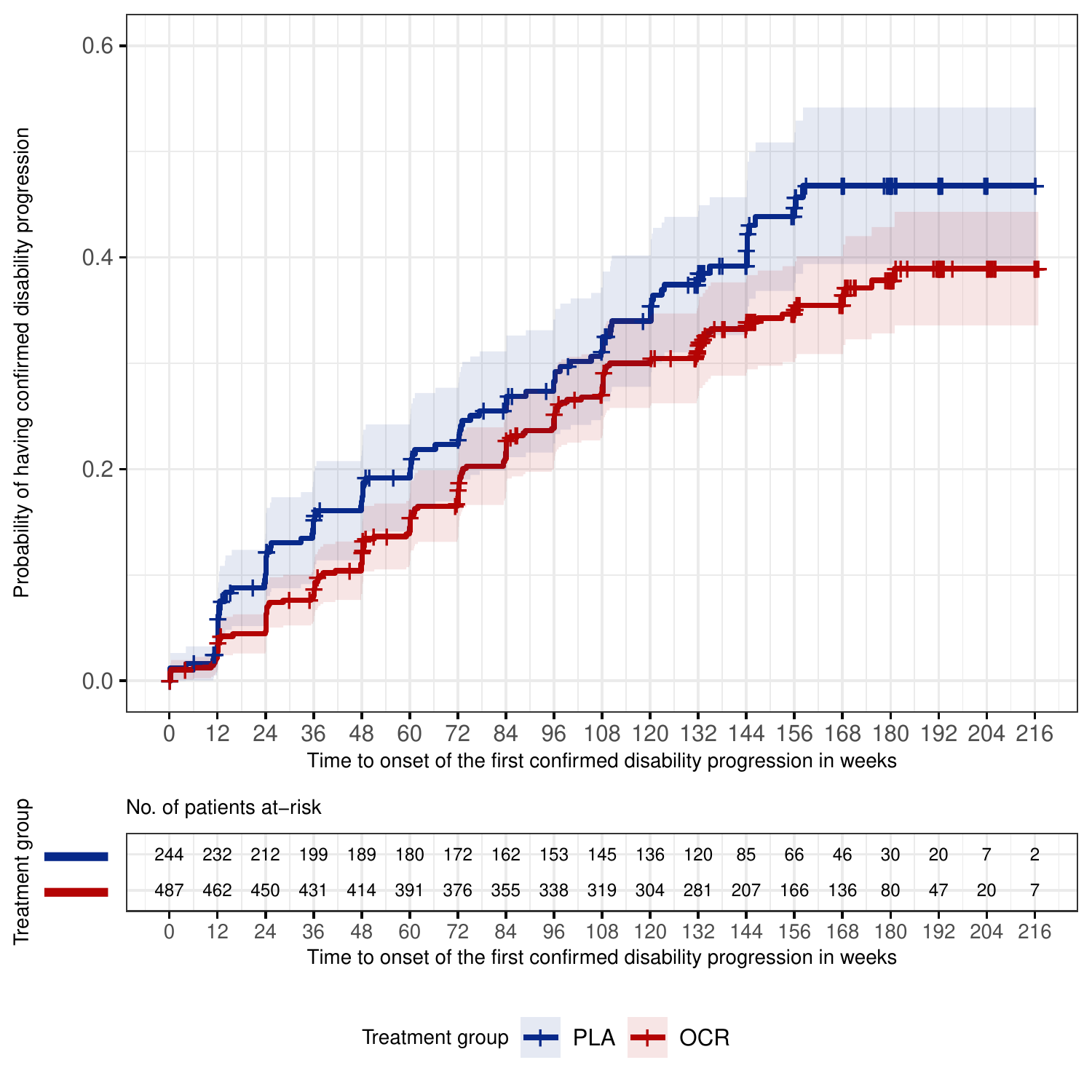}
}
\caption[ORATORIO - One minus Kaplan Meier plot of time-to-onset-of-first-CDP12]{ORATORIO - One minus Kaplan Meier plot of time-to-onset-of-first-CDP12 during double-blind treatment period and $95 \%$ CIs (+ indicates censoring)}
\label{KMoratorio}
\end{figure}

Time to the onset of the first CDP12 is the standard endpoint for primary and key secondary analyses of disability progression in clinical MS trials. The primary endpoint of the ORATORIO trial is analysed using a two-sided log-rank test for differences between the OCR and PLA group, stratified by geographical region (USA versus ROW) and age at baseline ($\leq 45$ versus $> 45$ years). The Cox proportional hazards model is used to estimate the relative treatment effect in terms of a hazard ratio (HR). The results can be found in Table $\ref{ResultsCDP12oratorio}$ and Figure $\ref{KMoratorio}$. \newline 
The time-to-first-event analysis is based on $731$ patients, with $487$ patients in the OCR group and $244$ patients in the PLA group. One patient who was randomized to the active treatment arm was excluded from the analysis because of a missing baseline EDSS value, in which case the derivation of the CDP endpoint is not possible. The time to the onset of the first CDP12 ranges from $0$ (censored) to $217$ (censored) weeks in the OCR group and from $0$ (censored) to $216$ (censored) weeks in the PLA group. The percentage of patients with $12$-week CDP is $32.9 \%$ with OCR versus $39.3 \%$ with PLA. The 1-KM curves for time-to-onset-of-first-CDP12 are shown in Figure $\ref{KMoratorio}$.

\begin{table}[H]
\centering
\scalebox{0.75}{
\begin{tabular}{lcc}
                                                                                                                                                    & \textbf{\begin{tabular}[c]{@{}c@{}}OCR\\ (N=488)\end{tabular}}                                                             & \textbf{\begin{tabular}[c]{@{}c@{}}PLA\\ (N=244\end{tabular}}                                                               \\ \hline
Patients included in analysis                                                                                                                       & 487 (100.0 $\%$)                                                                                                           & 244 (100.0 $\%$)                                                                                                            \\
Patients with event ($\%$)                                                                                                                          & 160 (32.9 $\%$)                                                                                                            & 96 (39.3 $\%$)                                                                                                              \\
                                                                                                                                                    &                                                                                                                            &                                                                                                                             \\
Time-to-first-CDP12 in weeks                                                                                                                        & 0* to 217*                                                                                                                 & 0* to 216*                                                                                                                  \\
                                                                                                                                                    &                                                                                                                            &                                                                                                                             \\
\begin{tabular}[c]{@{}l@{}}Stratified** analysis\\ $\qquad$ p-value (log-rank)\\ $$\\ $\qquad$ HR (95\% CI)\end{tabular}                            & \multicolumn{2}{c}{\begin{tabular}[c]{@{}c@{}}$$\\ 0.0321\\ $$\\ 0.76 {[}0.59, 0.98{]}\end{tabular}}                                                                                                                                                     \\
                                                                                                                                                    &                                                                                                                            &                                                                                                                             \\
\begin{tabular}[c]{@{}l@{}}Time point analysis: 1-KM estimate (95\% CI)\\ $\qquad$ 48 weeks\\ $\qquad$ 96 weeks\\ $\qquad$ 120 weeks\end{tabular} & \begin{tabular}[c]{@{}c@{}}$$\\ 11.05 {[}8.22, 13.88{]}\\ 24.75 {[}20.80, 28.70{]}\\ 30.23 {[}26.00, 34.45{]}\end{tabular} & \begin{tabular}[c]{@{}c@{}}$$\\ 16.96 {[}12.17, 21.76{]}\\ 28.28 {[}22.43, 34.13{]}\\ 33.98 {[}27.77, 40.18{]}\end{tabular} \\ \hline
\end{tabular}}
\caption[ORATORIO - Time-to-onset-of-first-CDP12 analysis]{ORATORIO - Time-to-onset-of-first-CDP12 analysis (* = censored observation, ** = stratified by geographical region (USA versus ROW) and age at baseline ($\leq 45$ versus $> 45$ years))}
\label{ResultsCDP12oratorio}
\end{table}

The graphs show separation from $12$ weeks, with a lower proportion of patients with disability progression in the OCR group throughout the whole double-blind treatment period. The log-rank test provides a p-value of $0.0321$. Since the p-value is less than the significance niveau of $5 \%$, there is a significant difference between the OCR and PLA group. The estimated probabilities of having CDP at week $120$ are $30.23 \%$ ($95 \%$ CI [26.00, 34.45]) with OCR versus $33.98 \%$ ($95 \%$ CI [27.77, 40.18]) with PLA. 

\begin{figure}[H]
\centering
  \subfloat[$-\log(-\log(S(t)))$ versus time $t$]{
  \scalebox{0.4}[0.4]{ 
\includegraphics{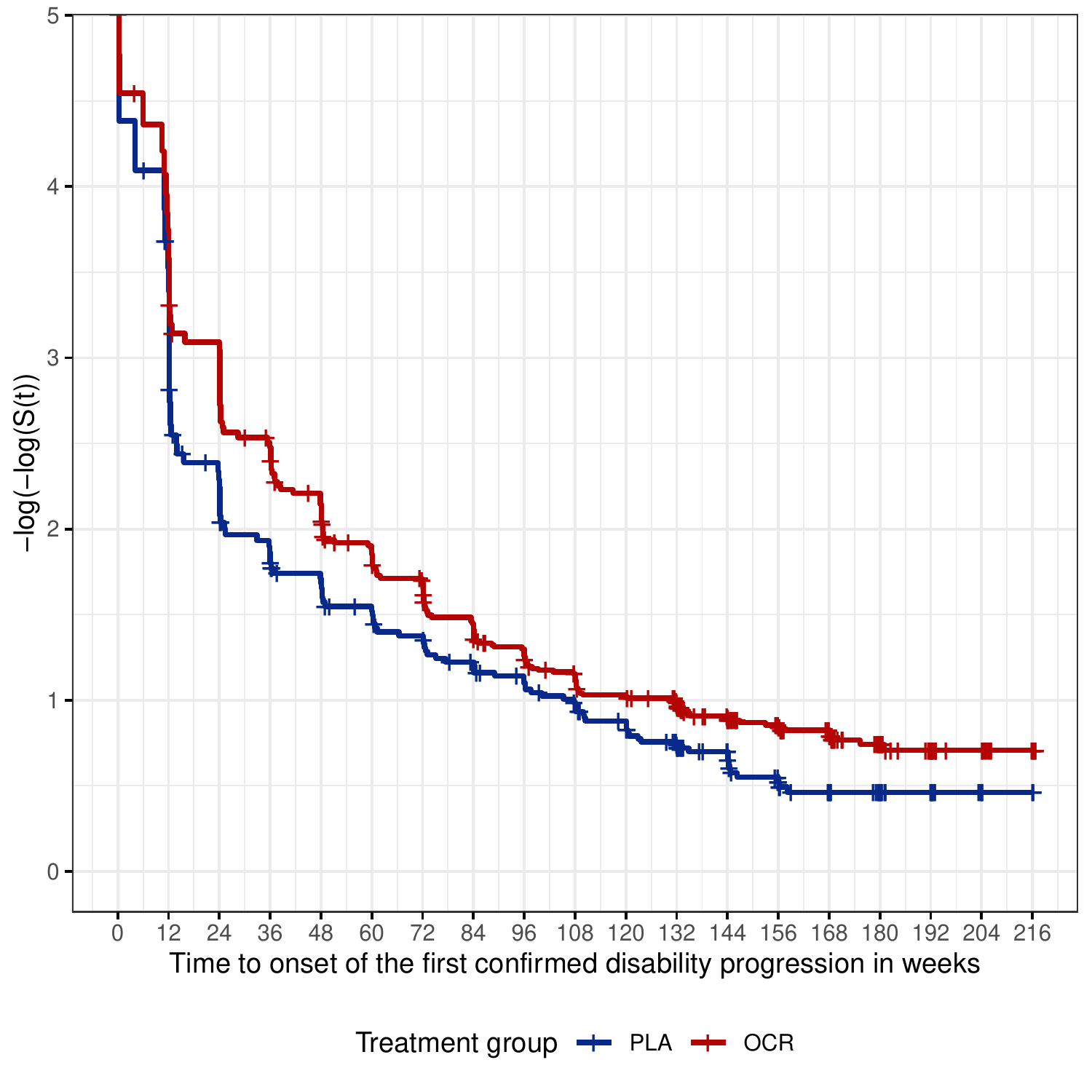}
  }}
  \subfloat[Schoenfeld residuals for treatment group]{
  \scalebox{0.4}[0.4]{ 
\includegraphics{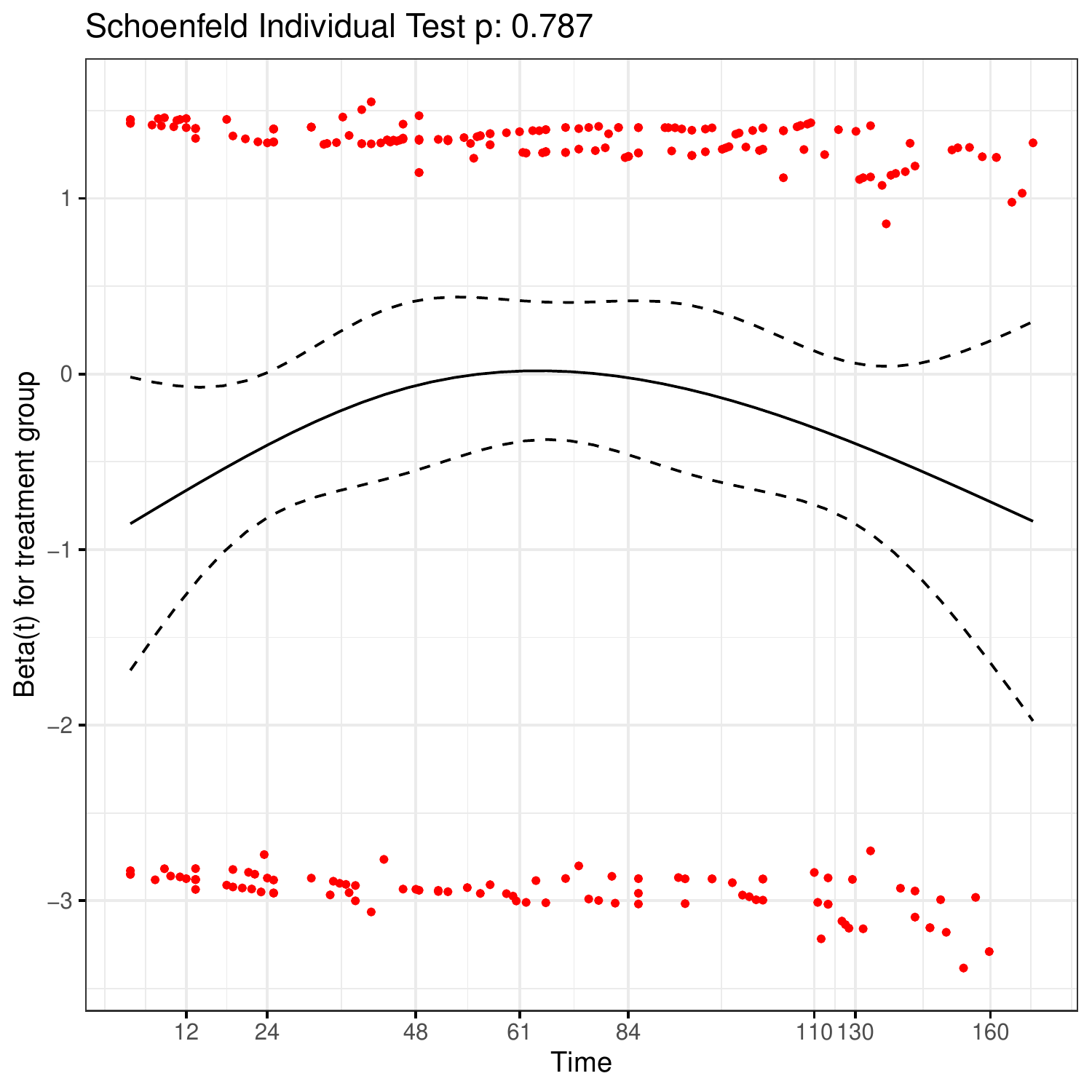}
  }}
\caption{ORATORIO - Proportional hazards assumption}
\label{PropHazardAssumptOratorio}
\end{figure}

The Cox analysis used to estimate the relative treatment effect yields a HR of $0.76$, with corresponding $95 \%$ CI $[0.59, 0.98]$. As a result, treatment with OCR leads to a $24 \%$ reduction in the risk of 12-week CDP compared with PLA. \newline 
When modelling a Cox proportional hazards model, a key property is the proportional hazards assumption. With respect to treatment comparisons in RCTs, this assumption states that the ratio of the hazard for an individual on OCR to the hazard for an individual on PLA remains constant over time, as seen from Eq. $(\ref{ratio})$ in Chapter $3$. That is, the regression coefficient $\beta$ for the treatment group does not vary over time. There are several methods to evaluate the validity of the proportional hazards assumption. As presented in Figure $\ref{PropHazardAssumptOratorio}$, a simple graphical test for categorical covariates is to plot the estimated transformed survival function $-\log(-\log(\hat{S}(t)))$ for both treatment groups against time $t$. Since the survival function under the Cox model complies with $S(t) = \exp(-\exp(\beta X)A_{0}(t))$ and $-\log(-\log(S(t))) = \log(A_{0}(t)) - \beta X$, the transformed survival curves should be approximately parallel, if proportional hazards are met. Figure $\ref{PropHazardAssumptOratorio}$ (a) suggests that the impact of the treatment on the hazard is roughly proportional. The same conclusion can be drawn from the right panel of Figure $\ref{PropHazardAssumptOratorio}$ which plots the scaled Schoenfeld residuals for the treatment indicator against the transformed time, with the solid line representing a smoothing spline fit for $\beta(t)$. The graph (b) depicts that $\hat{\beta}(t)$ is approximately constant around $0$, except for early and later times during follow-up. Due to a p-value of $0.787$, the statistical test based on the Schoenfeld residuals also indicates that the covariate satisfies the proportional hazards assumption. In total, there is no evidence that the Cox model violates the proportional hazards assumption. 

\begin{figure}[H]
\centering
\scalebox{0.65}{
\includegraphics{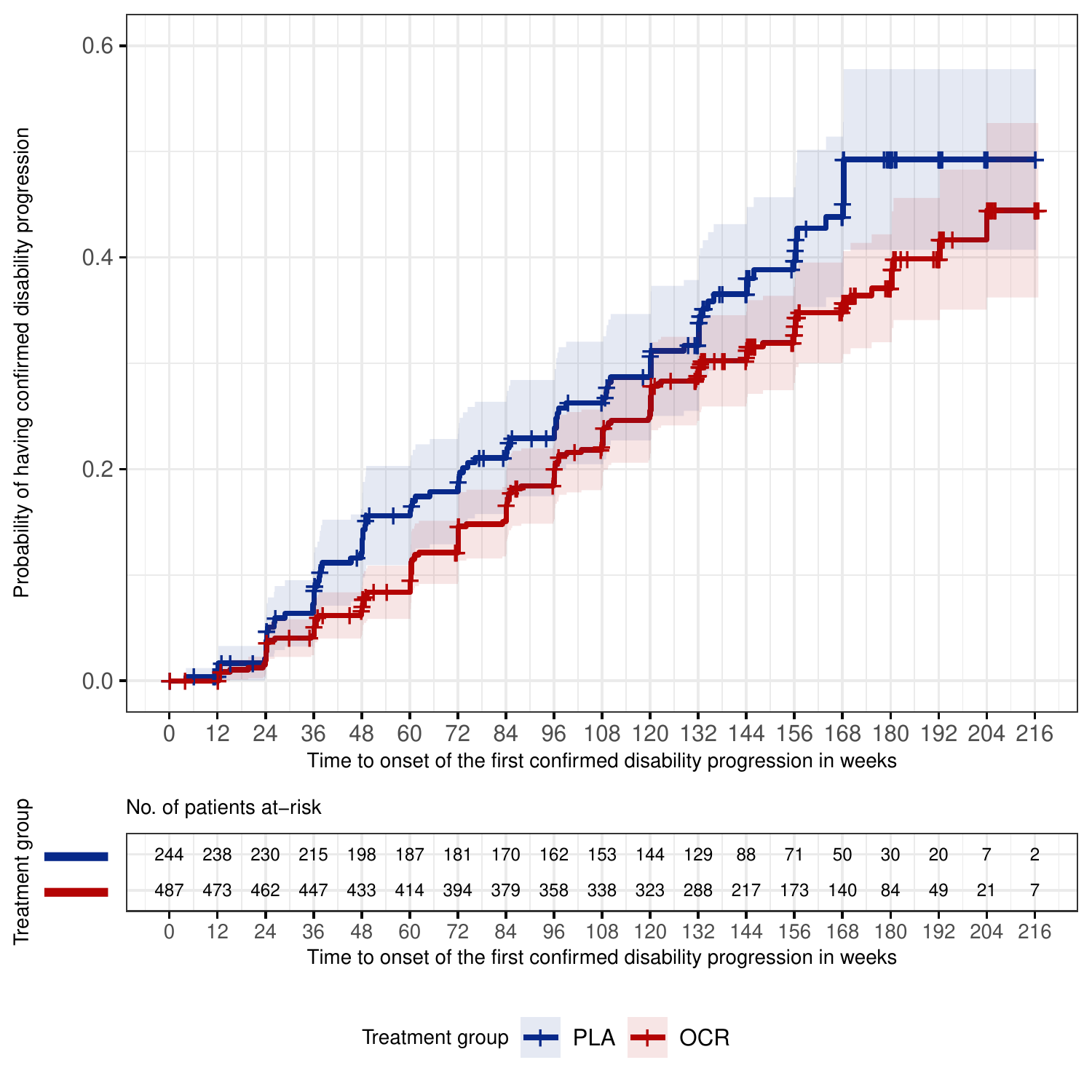}
}
\caption[ORATORIO - One minus Kaplan Meier plot of time-to-confirmation-of-first-CDP12]{ORATORIO - One minus Kaplan Meier plot of time-to-confirmation-of-first-CDP12 during double-blind treatment period and $95 \%$ CIs (+ indicates censoring)}
\label{KMoratorioConfir}
\end{figure}

Table $\ref{ResultsCDP12oratorioConfir}$ and Figure $\ref{KMoratorioConfir}$ represent the results obtained from the time-to-confirmation-of-first-CDP12 analyses. 

\begin{table}[H]
\centering
\scalebox{0.75}{
\begin{tabular}{lcc}
                                                                                                                                                  & \textbf{\begin{tabular}[c]{@{}c@{}}OCR\\ (N=488)\end{tabular}}                                                           & \textbf{\begin{tabular}[c]{@{}c@{}}PLA\\ (N=244)\end{tabular}}                                                             \\ \hline
Patients included in analysis                                                                                                                     & 487 (100.0 $\%$)                                                                                                         & 244 (100.0 $\%$)                                                                                                           \\
Patients with event ($\%$)                                                                                                                        & 155 (31.8 $\%$)                                                                                                          & 90 (36.9 $\%$)                                                                                                             \\
                                                                                                                                                  &                                                                                                                          &                                                                                                                            \\
Time-to-first-CDP12 in weeks                                                                                                                      & 0* to 217*                                                                                                               & 0* to 216*                                                                                                                 \\
                                                                                                                                                  &                                                                                                                          &                                                                                                                            \\
\begin{tabular}[c]{@{}l@{}}Stratified** analysis\\ $\qquad$ p-value (log-rank)\\ $$\\ $\qquad$ HR (95\% CI)\end{tabular}                          & \multicolumn{2}{c}{\begin{tabular}[c]{@{}c@{}}$$\\ 0.0585\\ $$\\ 0.78 {[}0.60, 1.01{]}\end{tabular}}                                                                                                                                                  \\
                                                                                                                                                  &                                                                                                                          &                                                                                                                            \\
\begin{tabular}[c]{@{}l@{}}Time point analysis: 1-KM estimate (95\% CI)\\ $\qquad$ 48 weeks\\ $\qquad$ 96 weeks\\ $\qquad$ 120 weeks\end{tabular} & \begin{tabular}[c]{@{}c@{}}$$\\ 6.62 {[}4.37, 8.87{]}\\ 18.86 {[}15.27, 22.46{]}\\ 25.52 {[}21.49, 29.56{]}\end{tabular} & \begin{tabular}[c]{@{}c@{}}$$\\ 12.06 {[}7.86, 16.25{]}\\ 22.92 {[}17.44, 28.41{]}\\ 28.69 {[}22.72, 34.66{]}\end{tabular} \\ \hline
\end{tabular}
}
\caption[ORATORIO - Time-to-confirmation-of-first-CDP12 analysis]{ORATORIO - Time-to-confirmation-of-first-CDP12 analysis (* = censored observation, ** = stratified by geographical region (USA versus ROW) and age at baseline ($\leq 45$ versus $> 45$ years))}
\label{ResultsCDP12oratorioConfir}
\end{table}

As compared to the standard definition, this alternative endpoint definition results in a loss of $11$ CDP12 events, with $5$ events in the OCR group and $6$ events in the PLA group. The percentage of patients with $12$-week CDP is $31.8 \%$ with OCR versus $36.9 \%$ with PLA. This implies that the reduction in the number of first CDP12 events under the alternative definition appears to be increased in the PLA group, as compared to the OCR group. \newline 
In a perfectly conducted MS trial with complete data (i.e., no right-censoring at the end of study), the number of CDP12 events is not expected to vary across the two endpoint definitions. However, in practice, administrative right-censoring due to study closure is common in clinical trials. The difference in the number of CDP12 events between the endpoint definitions may be explained by the fact that the CDP event must happen within the double-blind treatment period. Under the standard definition (=time-to-onset-of-first-CDP), initial worsening in disability progression must occur during the treatment period but EDSS assessments in the OLE period and safety follow-up can be used for confirmation of IDP. So, although disability progression has been approved at the next confirmatory EDSS assessment following the double-blind treatment period, a CDP event is registered to happen in the double-blind phase. However, such an event is not captured using the alternative time-to-confirmation-of-first-CDP definition because the event would not happen within the double-blind treatment period. This means, a clinical trial based on a time-to-confirmation-of-first-CDP endpoint requires a longer follow-up period to capture all progression events required to assess the treatment effect with adequate statistical power.  \newline
Due to the different timings of CDP12 events, the 1-KM curves and the corresponding KM estimates obviously differ between the two endpoint definitions. In Figure $\ref{KMoratorioConfir}$, the 1-KM curves do not show a clear separation until week $24$ or week $36$, since the event of interest can theoretically not happen until week 24 (expection: imputed events due to early withdrawal from treatment). Only patients who have an unscheduled study visit with IDP shortly after randomization may experience a CDP12 event at their 12-week study visit. As expected, the relative treatment effects estimated by the Cox model are pretty similar across the endpoint definitions.  

Original analyses of the ORATORIO data disregarded all CDPs occurring after the first event. In the following, the ORATORIO trial is reanalysed using information on all recurrent CDP events. 

\newpage
\subsection{Recurrent event analysis}
\begin{figure}[H] 
\centering
  \subfloat[Event plot for PPMS patients with $1$ CDP12 event]{
  \scalebox{0.50}[0.45]{ 
\includegraphics{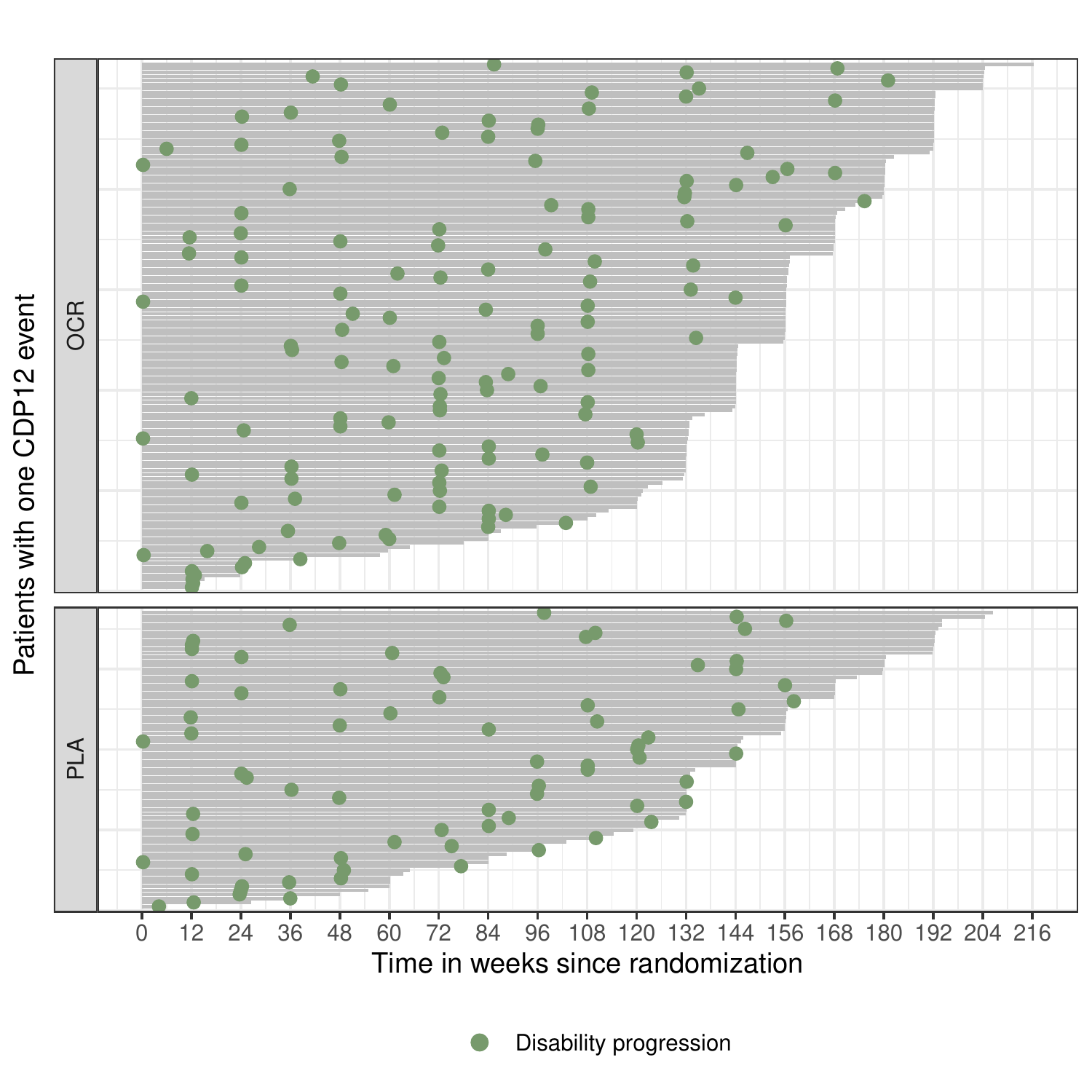}
  }}
  
  \subfloat[Event plot for PPMS patients with $2$ CDP12 events]{
  \scalebox{0.35}{ 
\includegraphics{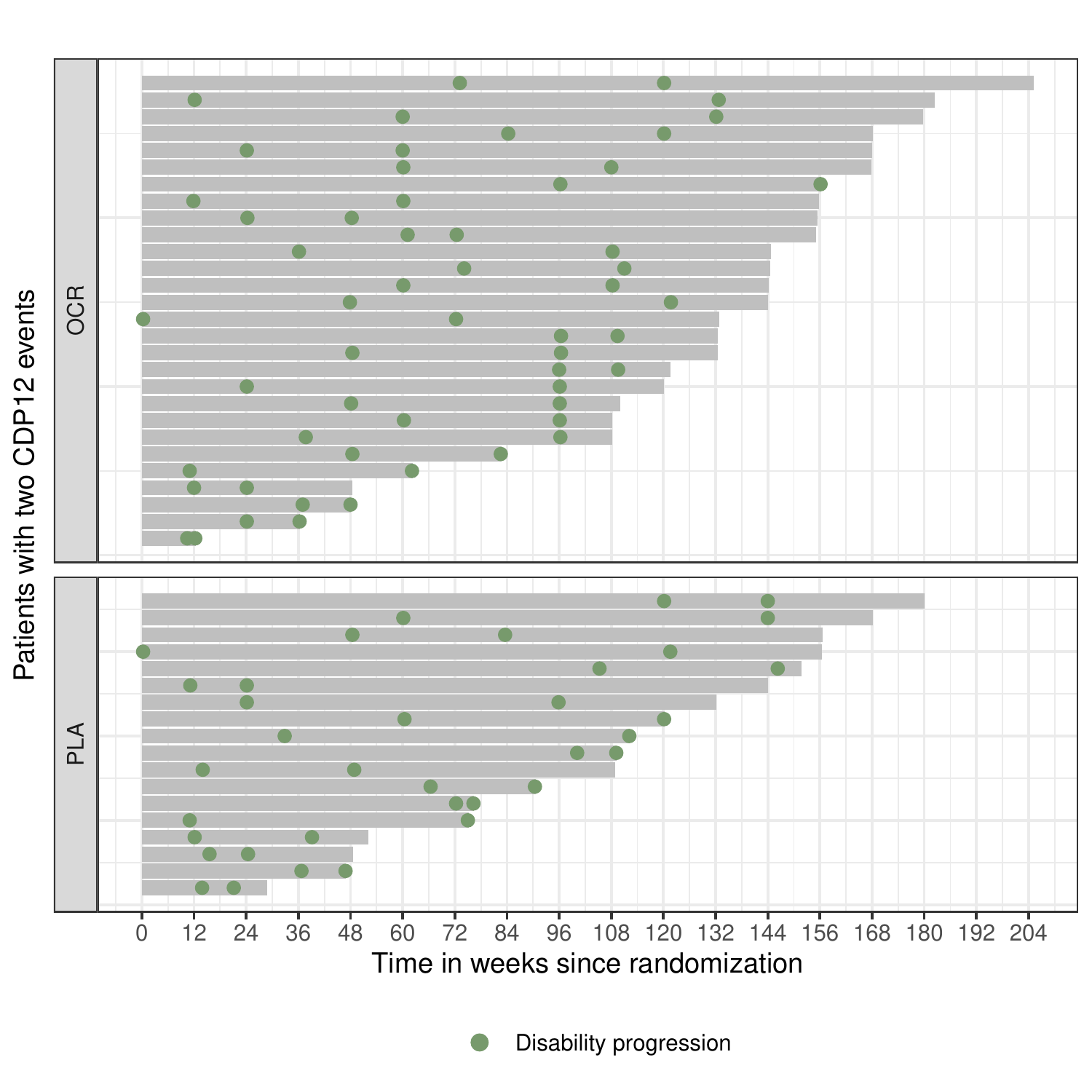}
  }} 
  \subfloat[Event plot for PPMS patients with more than $2$ CDP12 events]{
  \scalebox{0.35}{ 
\includegraphics{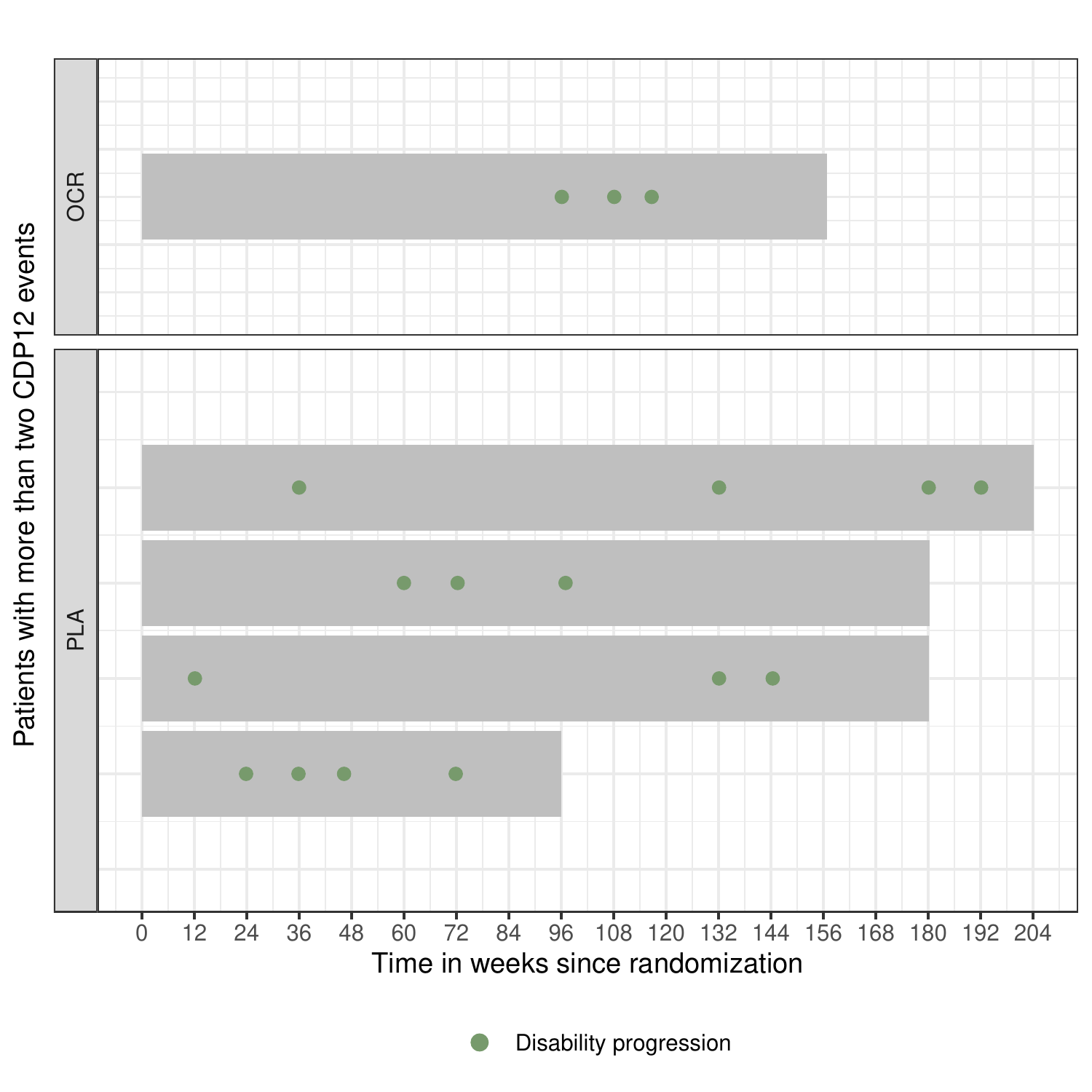}
  }}
  \caption[ORATORIO - Recurrent CDP12 data for PPMS patients]{ORATORIO - Event plot for PPMS patients randomized to OCR or PLA, each horizontal line corresponds to one patient with CDP12 events represented as dots}
  \label{EventPlotsOratorio}
\end{figure}

\begin{landscape}
\begin{figure}[H] 
\centering
\scalebox{1.05}{
\includegraphics{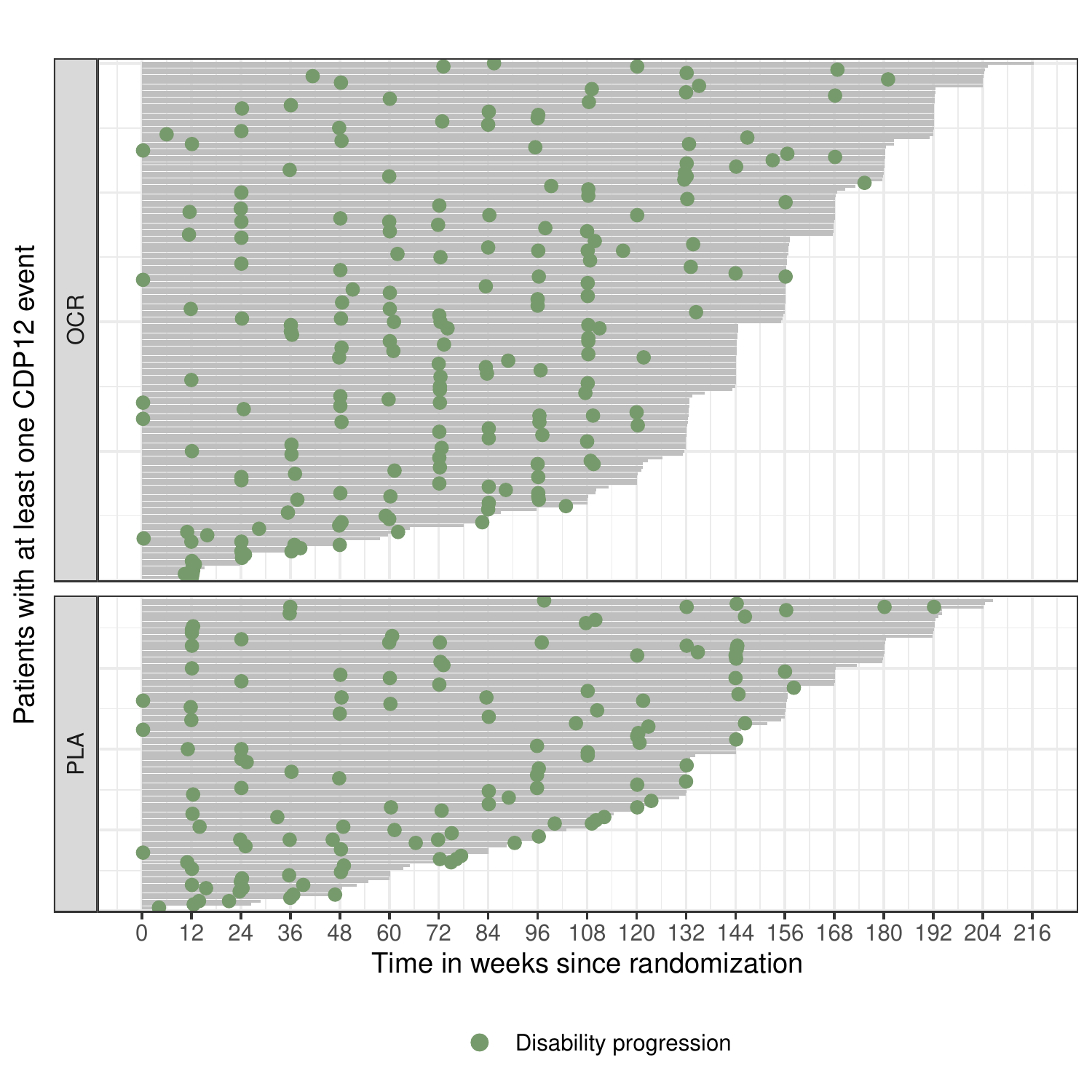}
}
\caption[ORATORIO - Recurrent CDP12 data for PPMS patients with at least one event]{ORATORIO - Event plot for PPMS patients with at least one $1$ CDP12 event, each horizontal line corresponds to one patient with CDP12 events represented as dots}
\label{EventPlotsOratorio1}
\end{figure}
\end{landscape}

Event plots stratified by treatment group, as displayed in Figure $\ref{EventPlotsOratorio}$ and Figure $\ref{EventPlotsOratorio1}$, are useful to get first insights into the recurrent event processes and to identify the frequency and patterns of CDP12 events. An event plot represents the CDP12 events for each individual belonging to the study population (= patient profile), where individuals are displayed on the y-axis (one line per individual) and the time in weeks since randomization is shown on the x-axis. The dots illustrate event occurrences and are placed on the days the events have been registered. Only ORATORIO individuals with at least one progression event are included in the graphs. The plots also show the total follow-up time for each individual, i.e., lengths of the grey lines are associated with the patient-specific follow-up times. While panel (a) includes all individuals with one CDP12 event, panel (b) and panel (c) contain individuals who have experienced two or more than two events, respectively. From all subfigures, it can be summarized that the follow-up times vary considerably across the PPMS patients. Most of the CDP12 events are observed to happen at the regular study visits, causing the clearly visible band patterns. There is also a small proportion of events detected to happen at unscheduled EDSS assessments. At first glance, the band patterns seem to be less evident in the PLA group, which would imply that PLA patients have on average more unscheduled study visits than OCR patients. Further, the event plots give the impression that, in a few PPMS patients treated with OCR or PLA, events often happen immediately at subsequent study visits (see OCR patient in (c)).

\begin{table}[t]
\centering
\begin{tabular}{llcccccc}
\hline
\multicolumn{1}{|c|}{\multirow{2}{*}{\textbf{Definition}}}                                                                                 & \multicolumn{1}{c|}{\multirow{2}{*}{\textbf{\begin{tabular}[c]{@{}c@{}}Treatment \\ group\end{tabular}}}} & \multicolumn{5}{c|}{\textbf{\begin{tabular}[c]{@{}c@{}}No. of CDP12 events \end{tabular}}}                                  & \multicolumn{1}{c|}{\textbf{\begin{tabular}[c]{@{}c@{}}'Unused'\\ events\end{tabular}}} \\ \cline{3-8} 
\multicolumn{1}{|c|}{}                                                                                                                     & \multicolumn{1}{c|}{}                                                                                     & \multicolumn{1}{c|}{0}   & \multicolumn{1}{c|}{1}   & \multicolumn{1}{c|}{2}  & \multicolumn{1}{c|}{3} & \multicolumn{1}{c|}{\textbf{4}} & \multicolumn{1}{l|}{}                                                                   \\ \hline
                                                                                                                                           &                                                                                                           & \multicolumn{1}{l}{}     & \multicolumn{1}{l}{}     & \multicolumn{1}{l}{}    & \multicolumn{1}{l}{}   & \multicolumn{1}{l}{}            & \multicolumn{1}{l}{}                                                                    \\ \hline
\multicolumn{1}{|l|}{\multirow{2}{*}{\textbf{\begin{tabular}[c]{@{}l@{}}Time-to-onset-of-CDP \\ and fixed reference\end{tabular}}}}        & \multicolumn{1}{l|}{PLA}                                                                              & \multicolumn{1}{c|}{148} & \multicolumn{1}{c|}{74}  & \multicolumn{1}{c|}{18} & \multicolumn{1}{c|}{2} & \multicolumn{1}{c|}{2}          & \multicolumn{1}{c|}{\multirow{2}{*}{58}}                                                \\ \cline{2-7}
\multicolumn{1}{|l|}{}                                                                                                                     & \multicolumn{1}{l|}{OCR}                                                                            & \multicolumn{1}{c|}{327} & \multicolumn{1}{c|}{131} & \multicolumn{1}{c|}{28} & \multicolumn{1}{c|}{1} & \multicolumn{1}{c|}{0}          & \multicolumn{1}{c|}{}                                                                   \\ \hline
                                                                                                                                           &                                                                                                           &                          &                          &                         &                        & \multicolumn{1}{l}{}            & \multicolumn{1}{l}{}                                                                    \\ \hline
\multicolumn{1}{|l|}{\multirow{2}{*}{\textbf{\begin{tabular}[c]{@{}l@{}}Time-to-onset-of-CDP \\ and roving reference*\end{tabular}}}}       & \multicolumn{1}{l|}{PLA}                                                                              & \multicolumn{1}{c|}{138} & \multicolumn{1}{c|}{81}  & \multicolumn{1}{c|}{20} & \multicolumn{1}{c|}{3} & \multicolumn{1}{c|}{2}          & \multicolumn{1}{c|}{\multirow{2}{*}{68}}                                                \\ \cline{2-7}
\multicolumn{1}{|l|}{}                                                                                                                     & \multicolumn{1}{l|}{OCR}                                                                            & \multicolumn{1}{c|}{308} & \multicolumn{1}{c|}{145} & \multicolumn{1}{c|}{32} & \multicolumn{1}{c|}{2} & \multicolumn{1}{c|}{0}          & \multicolumn{1}{c|}{}                                                                   \\ \hline
                                                                                                                                           &                                                                                                           & \multicolumn{1}{l}{}     & \multicolumn{1}{l}{}     & \multicolumn{1}{l}{}    & \multicolumn{1}{l}{}   & \multicolumn{1}{l}{}            & \multicolumn{1}{l}{}                                                                    \\ \hline
\multicolumn{1}{|l|}{\multirow{2}{*}{\textbf{\begin{tabular}[c]{@{}l@{}}Time-to-confirmation-of-CDP\\ and fixed reference\end{tabular}}}}  & \multicolumn{1}{l|}{PLA}                                                                              & \multicolumn{1}{c|}{154} & \multicolumn{1}{c|}{74}  & \multicolumn{1}{c|}{12} & \multicolumn{1}{c|}{2} & \multicolumn{1}{c|}{2}          & \multicolumn{1}{c|}{\multirow{2}{*}{47}}                                                \\ \cline{2-7}
\multicolumn{1}{|l|}{}                                                                                                                     & \multicolumn{1}{l|}{OCR}                                                                            & \multicolumn{1}{c|}{332} & \multicolumn{1}{c|}{131} & \multicolumn{1}{c|}{23} & \multicolumn{1}{c|}{1} & \multicolumn{1}{c|}{0}          & \multicolumn{1}{c|}{}                                                                   \\ \hline
                                                                                                                                           &                                                                                                           & \multicolumn{1}{l}{}     & \multicolumn{1}{l}{}     & \multicolumn{1}{l}{}    & \multicolumn{1}{l}{}   & \multicolumn{1}{l|}{}           & \multicolumn{1}{l|}{}                                                                   \\ \hline
\multicolumn{1}{|l|}{\multirow{2}{*}{\textbf{\begin{tabular}[c]{@{}l@{}}Time-to-confirmation-of-CDP\\ and roving reference*\end{tabular}}}} & \multicolumn{1}{l|}{PLA}                                                                              & \multicolumn{1}{c|}{144} & \multicolumn{1}{c|}{81}  & \multicolumn{1}{c|}{14} & \multicolumn{1}{c|}{3} & \multicolumn{1}{c|}{2}          & \multicolumn{1}{c|}{\multirow{2}{*}{57}}                                                \\ \cline{2-7}
\multicolumn{1}{|l|}{}                                                                                                                     & \multicolumn{1}{l|}{OCR}                                                                            & \multicolumn{1}{c|}{313} & \multicolumn{1}{c|}{145} & \multicolumn{1}{c|}{27} & \multicolumn{1}{c|}{2} & \multicolumn{1}{c|}{0}          & \multicolumn{1}{c|}{}                                                                   \\ \hline
\end{tabular}
\caption[ORATORIO - Distribution of the numbers of CDP12 events by treatment group]{ORATORIO - Distribution of the numbers of CDP12 events by treatment group, patients with missing baseline EDSS were excluded from analyses, N=244 (PLA) and N=487 (OCR), column 'unused events' refers to events not used in time-to-first-event analyses only (* = 24-week confirmation period of new reference EDSS score)} 
\label{NOCDPW12oratorio}
\end{table}

\begin{figure}[h]
\centering
\scalebox{0.42}[0.42]{
\includegraphics{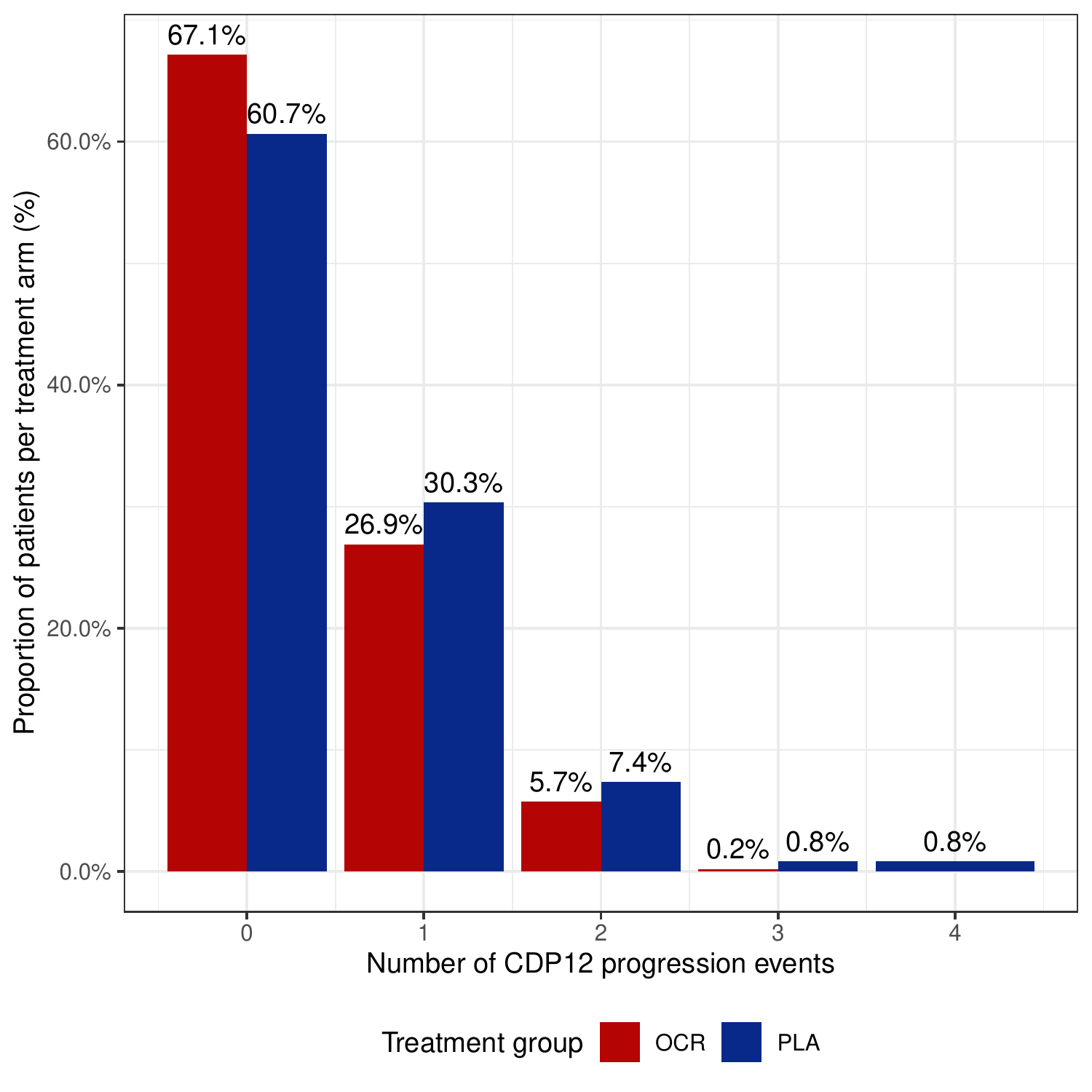}
}
\caption[ORATORIO - Histogram of the number of CDP12 events]{ORATORIO - Histogram of the number CDP12 events during double-blind treatment period, N=244 (PLA) and N=487 (OCR)}
\label{HistCDP12oratorio}
\end{figure}

Summary statistics on the number of CDP12 events for the two treatment groups based on $731$ patients are given in Table $\ref{NOCDPW12oratorio}$ and Figure $\ref{HistCDP12oratorio}$. One patient who was randomly assigned to the active treatment arm was excluded from the recurrent event analysis owing to missing data on the EDSS score at baseline. First, it is focussed on the standard definition using time-to-onset-of-CDP and a fixed reference system. By end of the trial, $96 / 244$  ($ 39.3 \%$) of the PLA patients and  $160 / 487$  ($ 32.9 \%$) of the OCR patients had at least $1$ progression event. In the PLA arm, $74$ patients $(30.3 \%)$ experienced $1$ progression event, $18$ patients $(7.4 \%)$ experienced $2$ progression events, $2$ patients $(0.8 \%)$ experienced $3$ events and $2$ patients $(0.8 \%)$ experienced $4$ progression events during the double-blind treatment period. In the OCR arm, $131$ patients $(26.9 \%)$ progressed once, $28$ patients $(5.7 \%)$ progressed twice and only $1$ patient $(0.2 \%)$ had 3 progression events. No patient was observed to experience more than $3$ events in the OCR group within the treatment period. \newline
As stated in Table $\ref{NOCDPW12oratorio}$ (cf. 'unused' events), time-to-first-event analyses are based on $256$ CDP12 events ($160$ OCR, $96$ PLA), whereas recurrent event analyses incorporate all $314$ CDP12 events ($190$ OCR, $124$ PLA). As a consequence, time-to-first-event analyses disregard $58$ events ($30$ OCR, $28$ PLA) available in the ORATORIO dataset. 

\begin{figure}[h]
\centering
  \subfloat[Time-to-onset-of-CDP and fixed reference]{
  \scalebox{0.38}[0.38]{ 
\includegraphics{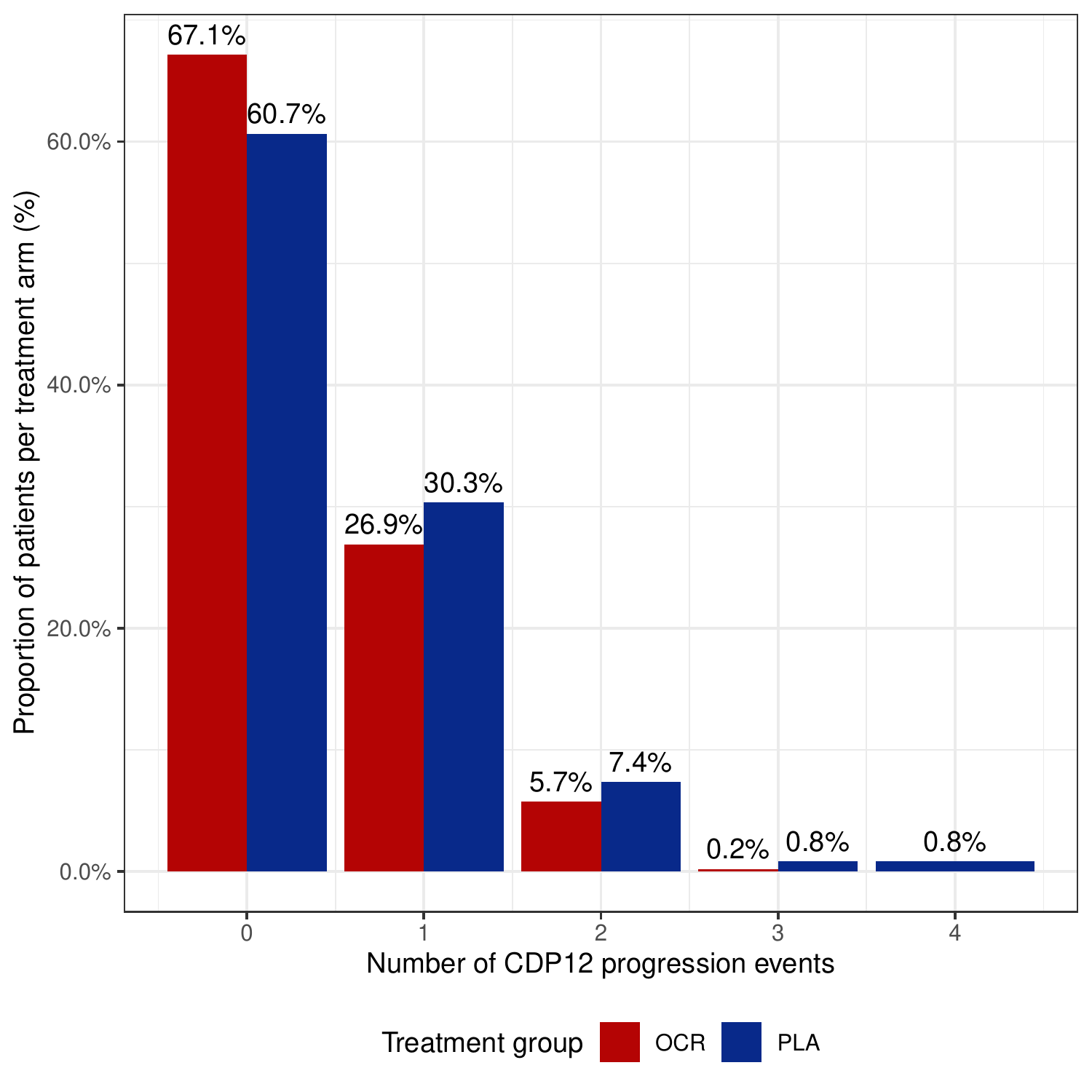}
  }}
  \subfloat[Time-to-onset-of-CDP and roving reference*]{
  \scalebox{0.38}[0.38]{ 
\includegraphics{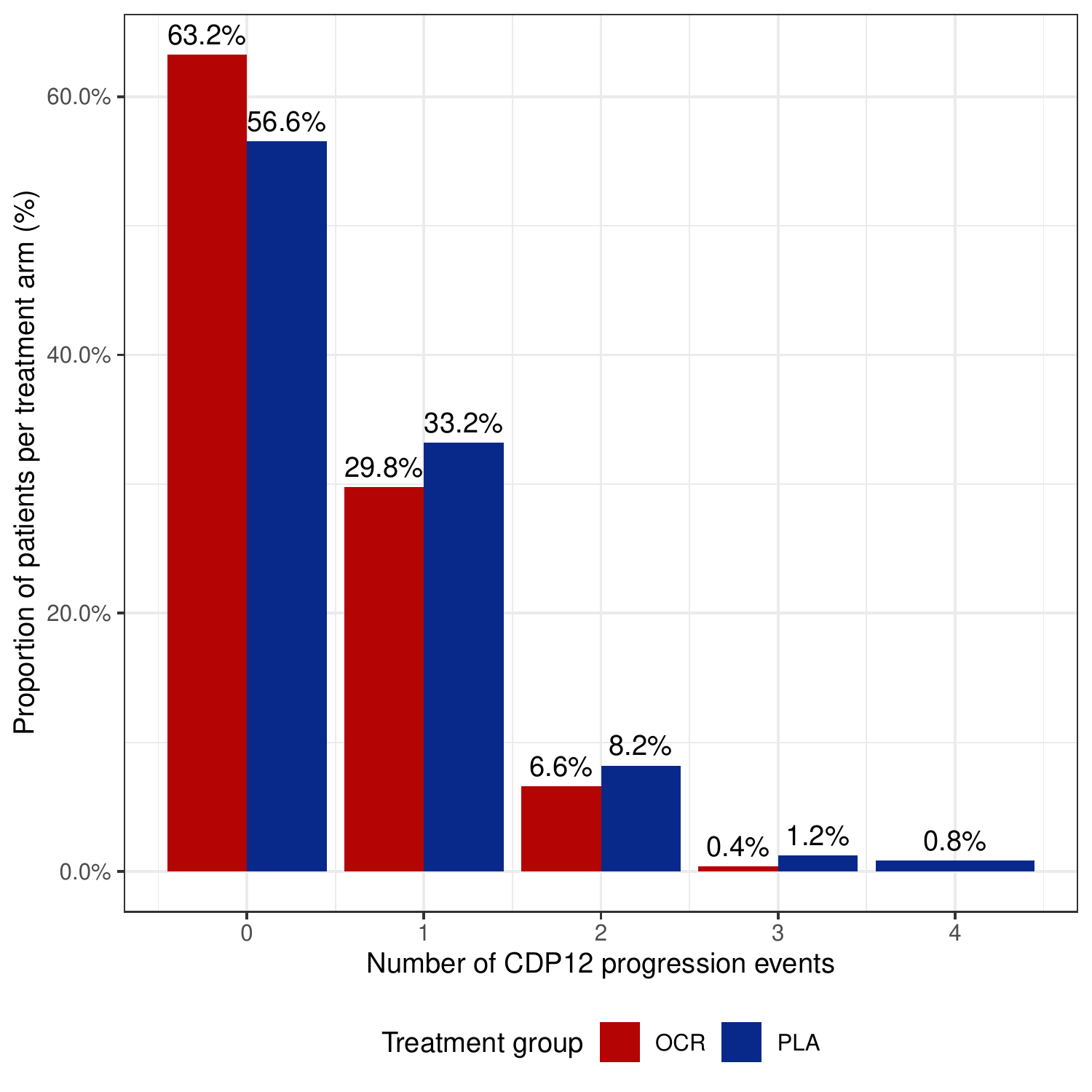}
}}

  \subfloat[Time-to-confirmation-of-CDP and fixed reference]{
  \scalebox{0.38}[0.38]{ 
\includegraphics{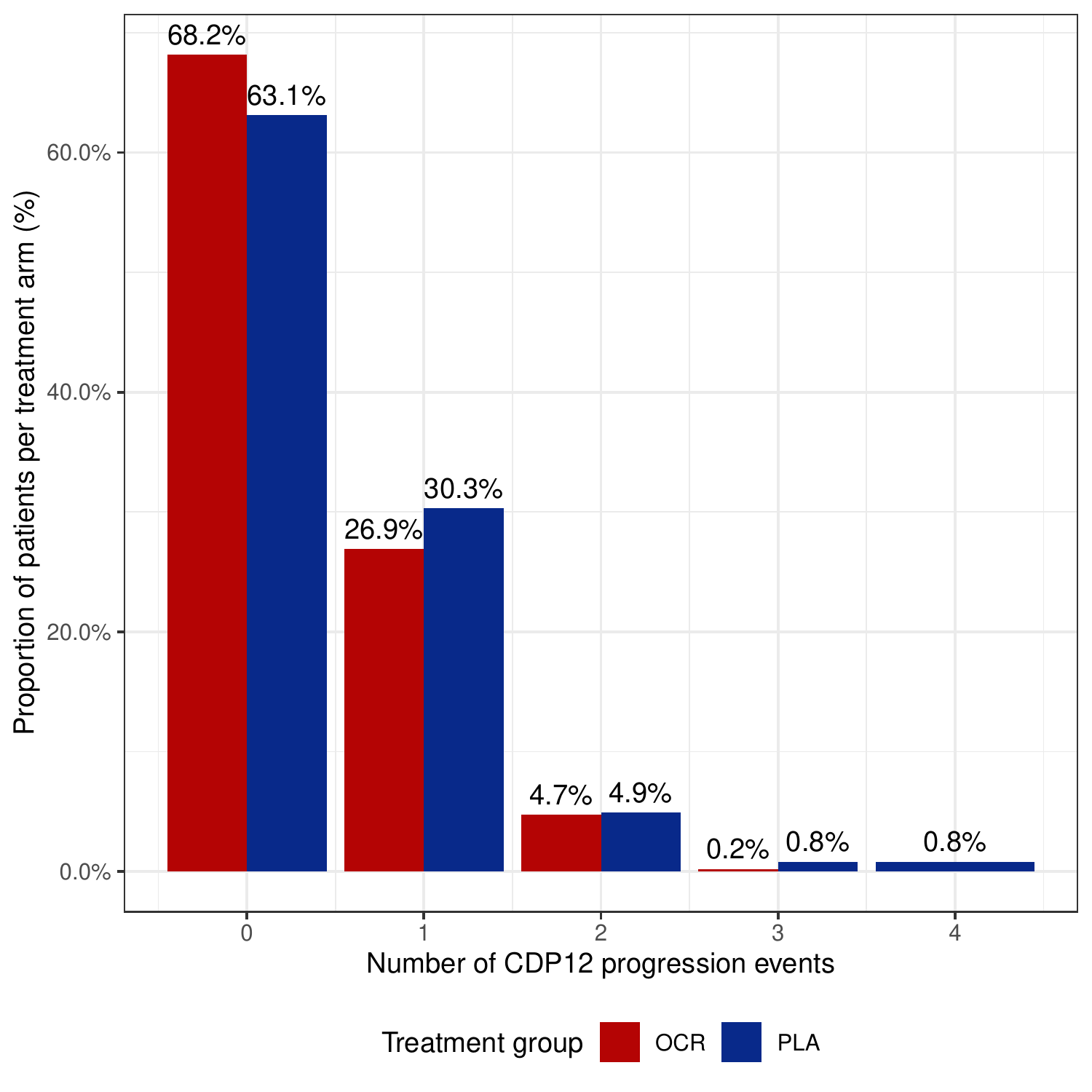}
  }}
  \subfloat[Time-to-confirmation-of-CDP and roving reference*]{
  \scalebox{0.38}[0.38]{ 
\includegraphics{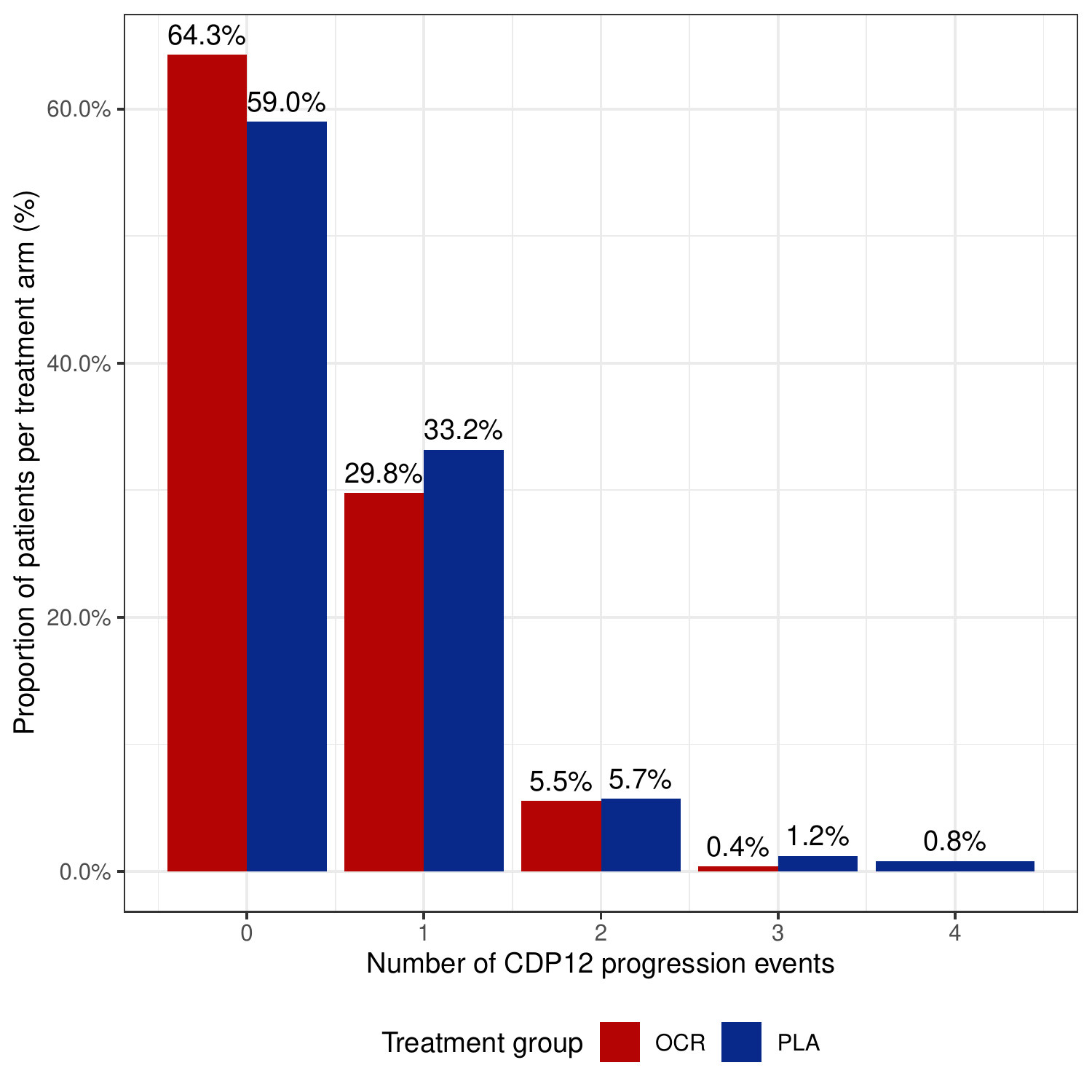}
}}
\caption[ORATORIO - Histogram of the number of CDP12 events according to different endpoint definitions]{ORATORIO - Histogram of the number of CDP12 events during double-blind treatment period according to different endpoint definitions, N=244 (PLA) and N=487 (OCR) (* = 24-week confirmation period of new reference EDSS score)}
\label{HistCDP12oratorioAltDef}
\end{figure}

Table $\ref{NOCDPW12oratorio}$ also contains the frequencies of the numbers of CDP12 events under the alternative endpoint definitions. Regardless of whether time-to-onset-of-CDP or time-to-confirmation-of-CDP is considered, derivation of recurrent CDP events using a roving reference system results in a higher proportion of overall CDP12 events, as compared to the commonly used fixed reference system. In analyses based on time-to-onset-of-CDP and a roving reference system, $285$ PPMS patients ($39.0 \%$) experienced at least one $1$ CDP12 event, as compared to $256$ PPMS patients ($35.0 \%$) under the standard definition. When comparing Figure $\ref{HistCDP12oratorioAltDef}$ (a) with Figure $\ref{HistCDP12oratorioAltDef}$ (b), it seems that both treatment groups are equally affected by disability improvement. Similar findings can also be found from analyses, where time-to-confirmation-of-CDP is kept fixed and the reference system varies (cf. panel (c) versus panel (d)). \newline
While keeping the reference system fixed, time-to-confirmation-of-CDP analyses lead to a reduced number of CDP12 events, as compared to time-to-onset-of-CDP analyses. In each case (cf. panel (a) versus panel (c), panel (b) versus panel (d)), a slightly higher reduction in the number of CDP events is observed in the PLA group. 

Unless otherwise stated, the recurrent endpoint definition based on time-to-onset-of-CPD and a fixed reference system will be used in the following, as the MS-specific simulation study has shown that the look-ahead bias is negligible in such analyses.
\newpage

\begin{figure}[h]
\centering
  \subfloat[CMF of CDP12 events]{
  \scalebox{0.45}[0.45]{ 
\includegraphics{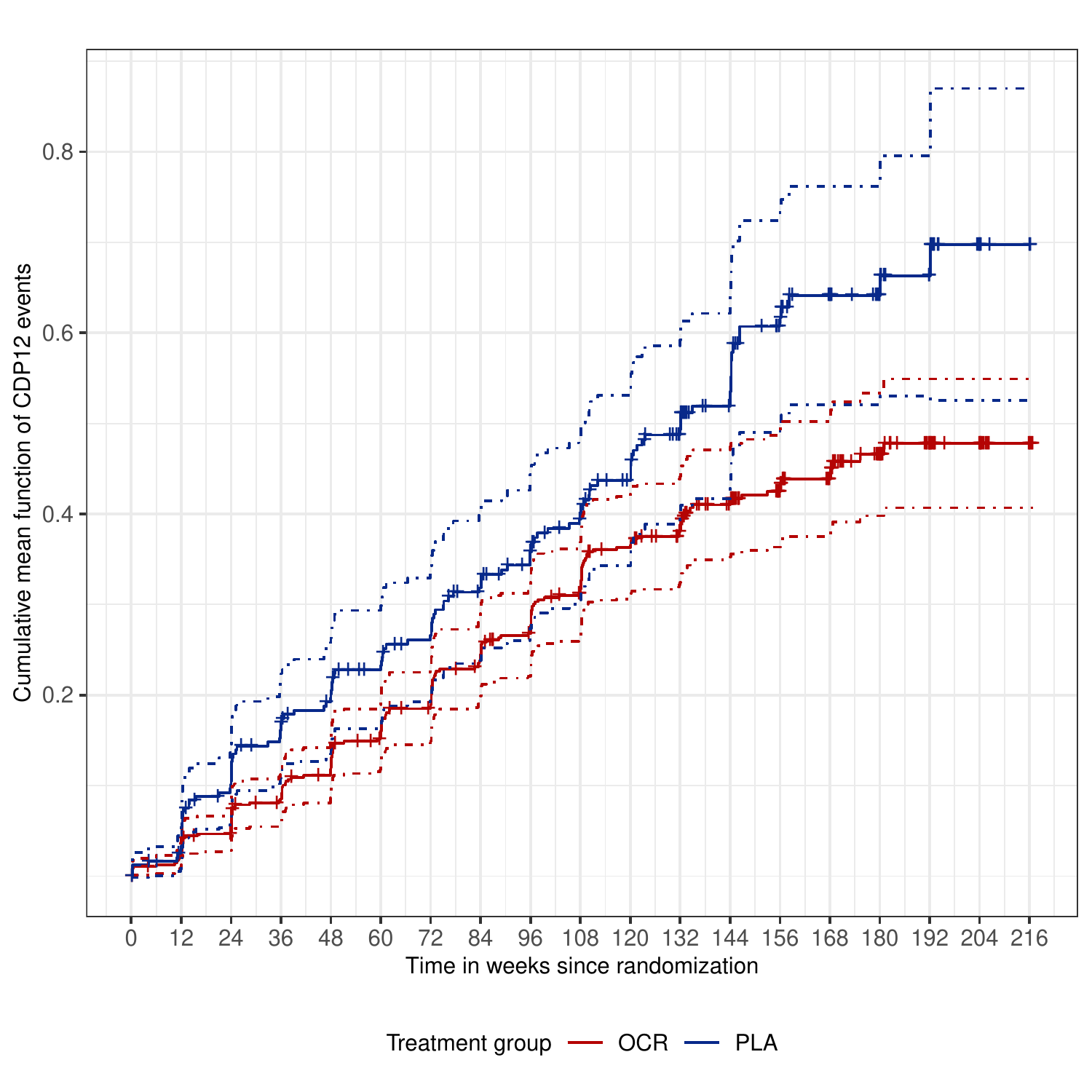}
  }}
  \subfloat[Difference in CMFs between OCR and PLA group]{
  \scalebox{0.45}[0.45]{ 
\includegraphics{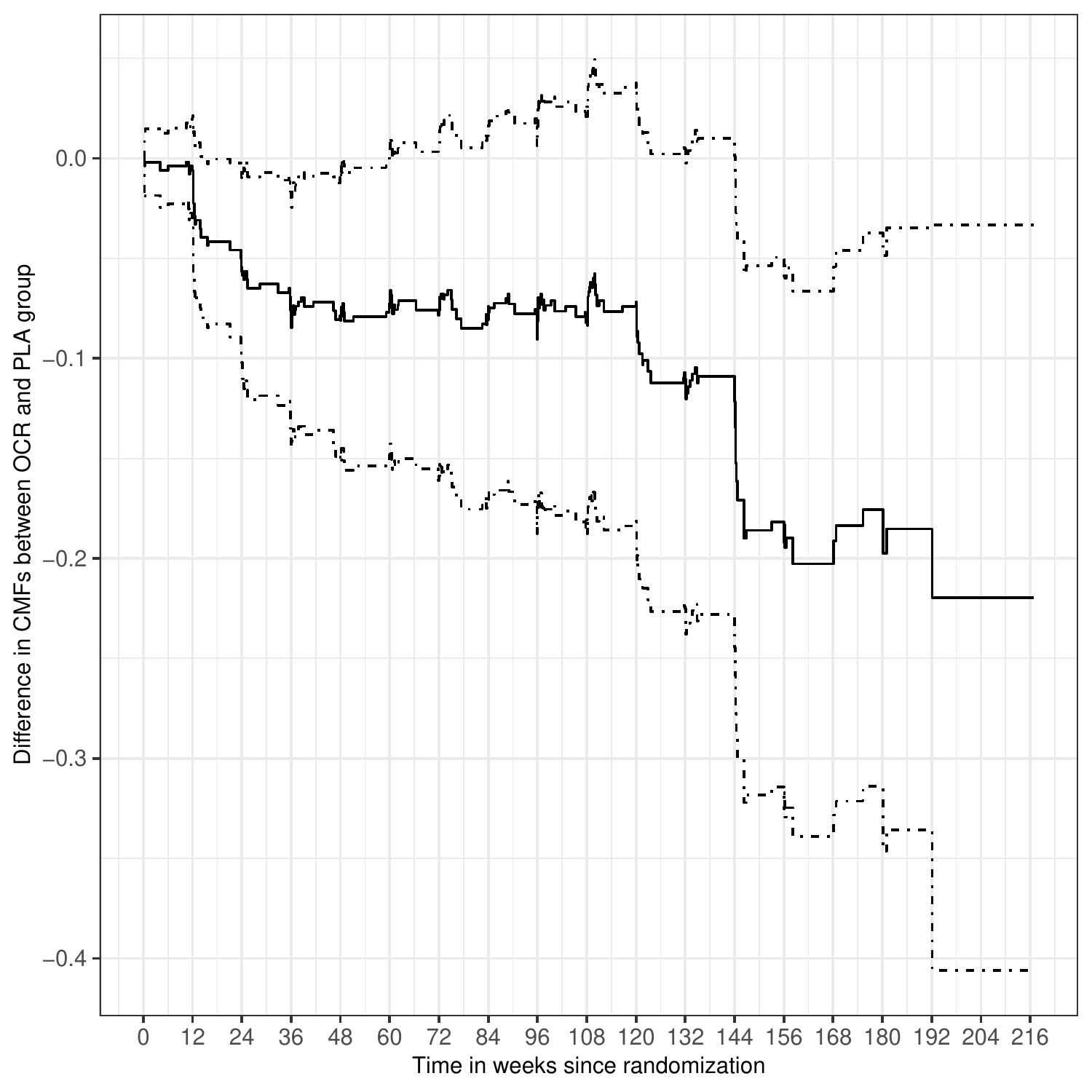}
  }}
\caption[ORATORIO - Cumulative mean function of CDP12 by treatment group]{ORATORIO - Cumulative mean function of CDP12 by treatment group, with $95 \%$ pointwise CIs}
\label{CMForatorio}
\end{figure}

Recurrent event data can also be described by estimating the CMF (cf. Chapter $4$, Section $\ref{marginalmodel}$, Eq. $(\ref{CMFestimator})$), which is the expected mean number of cumulative CDP12 events experienced by an individual at each point in time since time origin. Figure $\ref{CMForatorio}$ (a) illustrates the estimated CMF of CDP12 events for OCR and PLA patients from the ORATORIO trial, suggesting some time trends in the event rate over follow-up. While in the beginning the event rate is roughly constant, it decreases over time. The CMF value for the OCR group is about $0.37$ ($95 \%$ CI $[0.31, 0.43]$) at week $120$, which means that a patient treated with OCR experienced on average $0.37$ 12-week CDPs over the first $120$ weeks of the double-blind treatment period. The corresponding CMF value for the PLA group is $0.46$ ($95 \%$ CI $[0.36, 0.56]$), respectively. Table $\ref{CMFCDP12oratorio}$ gives further CMF estimates at week $48$ and week $96$. In total, there seems to be no early difference in CMFs between the two treatment groups but, from week $12$ onwards, the patients treated with OCR have a lower average number of CDP12 events than patients on PLA, concluding the beneficial effect of OCR. In addition, Figure $\ref{CMForatorio}$ (b) plots the difference in CMFs between the OCR and PLA group. It can be extracted that the difference becomes bigger as time passes. The two-sample pseudo-score test proposed by \citet{Lawless1995} (cf. Eq. $(\ref{CFMtest})$) results in a p-value of $0.0119$, which implies that there is a significant difference between the OCR and PLA group.

\begin{figure}[H] 
\centering
  \subfloat[EDSS category]{
  \scalebox{0.45}[0.45]{ 
\includegraphics{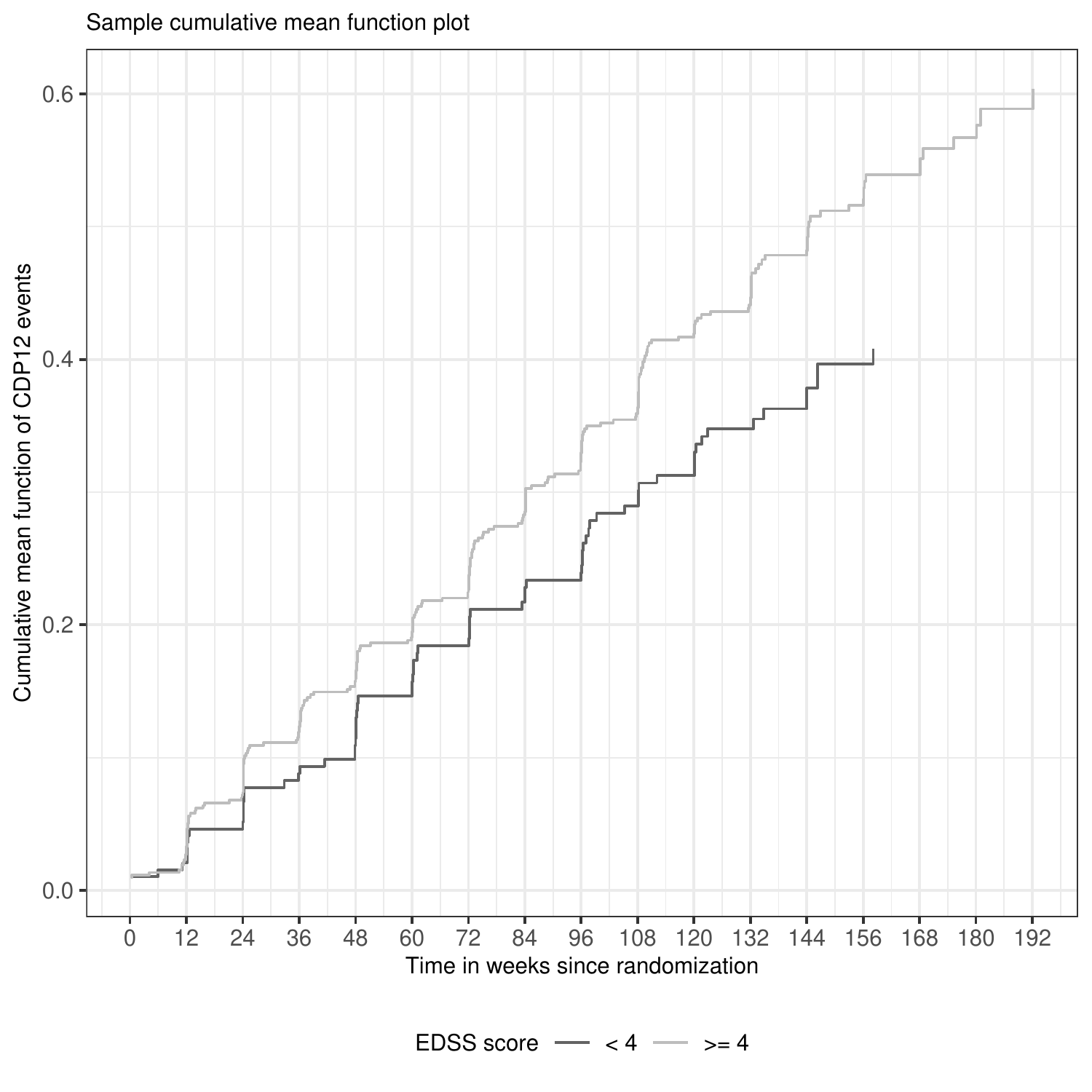}
  }}
  \subfloat[EDSS category and treatment group]{
  \scalebox{0.45}{ 
\includegraphics{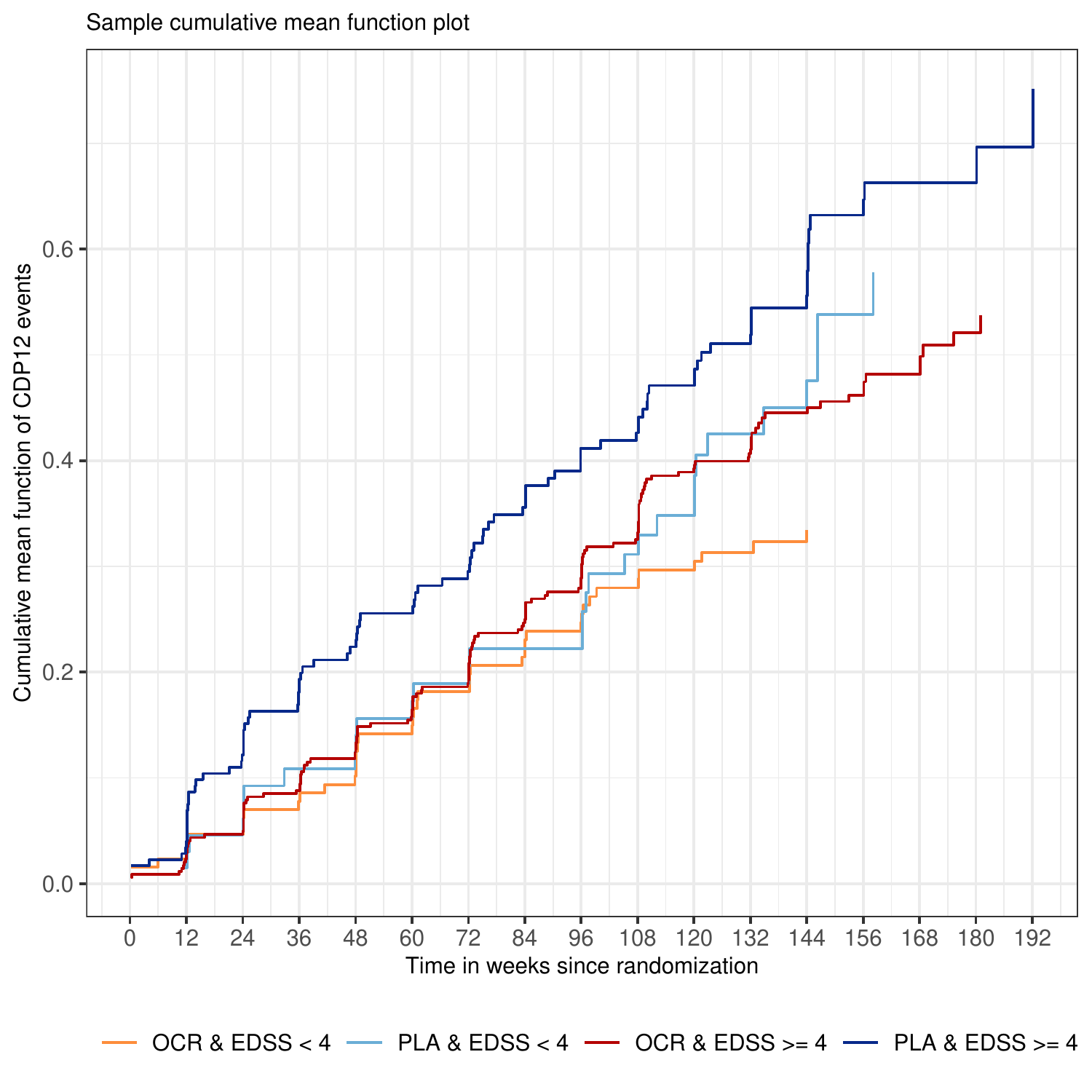}
  }} 
\caption[ORATORIO - Cumulative mean function of CDP12 by treatment group and EDSS]{ORATORIO - Cumulative mean function of CDP12 by treatment group and EDSS category}
\label{CMForatorioEDSS}
\end{figure}

Figure $\ref{CMForatorioEDSS}$ shows the estimated CMFs stratified by EDSS score at baseline ($< 4.0$ versus $\geq 4.0$) and treatment group. PPMS patients whose baseline EDSS score is $< 4.0$ have on average a lower cumulative number of CDP12 events than patients with a baseline EDSS score $\geq 4.0$ (cf. panel (a)). Panel (b) suggests that PLA patients with a baseline EDSS score of $\geq 4.0$ have on average the highest number of cumulative CDP12 events, as compared to the remaining study population. While there is no clear difference in the CMFs between the 'PLA and EDSS $< 4.0$' and 'OCR and EDSS $\geq 4.0$' patients, OCR patients with EDSS score $ < 4.0$ have the lowest number of CDP12 events over time. CMFs stratified by other baseline covariates (e.g., sex, age, $T_{1}$ lesions yes/no, ...) can be found in Appendix $\ref{AppendixCMF}$.

\begin{table}[h]
\centering
\scalebox{0.85}{ 
\begin{tabular}{lcc}
                                                                                                                                                 & \textbf{OCR}                                                                                                        & \textbf{PLA}                                                                                                       \\ \hline
\begin{tabular}[c]{@{}l@{}}Time point analysis: CMF estimate (95\% CI)\\ $\qquad$ 48 weeks\\ $\qquad$ 96 weeks\\ $\qquad$ 120 weeks\end{tabular} & \begin{tabular}[c]{@{}c@{}}$$ \\ 0.13 {[}0.10, 0.16{]}\\ 0.29 {[}0.24, 0.34{]}\\ 0.37 {[}0.31, 0.43{]}\end{tabular} & \begin{tabular}[c]{@{}c@{}}$$\\ 0.21 {[}0.14, 0.27{]}\\ 0.36 {[}0.27, 0.44{]}\\ 0.46 {[}0.36, 0.56{]}\end{tabular} \\ \hline
\end{tabular}}
\caption[ORATORIO - Time point analysis of cumulative mean function for CDP12]{ORATORIO - Time point analysis of CMF}
\label{CMFCDP12oratorio}
\end{table}

\subsubsection{Estimation of treatment effect}
\label{secRCTEstimation}
As seen from Section $\ref{TTFEApplication}$, time-to-first-event analyses of the ORATORIO data showed that treatment with OCR reduces the hazard for a CDP12 event by approximately $24 \%$ (HR $0.76$, $95 \%$ CI $[0.59, 0.98]$). For the time-to-first-CDP12 endpoint, the Cox proportional hazards model is the main analysis method. Now, it is of special interest to investigate the overall treatment effect of OCR on disability progression in PPMS patients, while taking into account all repeated CDP12 events rather than the first CDP12 only. By overall or common treatment effect, the effect of treatment on any event is meant. In clinical trials, where treatments are expected to affect the first as well as subsequent events, the robust semiparametric LWYY model and the parametric NB model are adequate approaches for evaluating the overall treatment effect on a recurrent event endpoint in RCTs. (cf. Chapter $\ref{RETheory}$, Section $\ref{RCTrecurrentevents}$). Both models are classified as marginal rate models that provide the treatment effect estimate (expressed as RR) with a clear causal interpretation. In the following, the LWYY and NB models are applied to the recurrent event data. Unless otherwise stated, the models considered in Section $\ref{secRCTEstimation}$ only control for the treatment group. 
\newline \newline 
\textbf{Negative binomial model} \newline 
The NB analysis is based on $731$ PPMS patients and $314$ CDP12 events, with $190$ CDP12 events in the OCR group and $124$ CDP12 events in the PLA group. Results obtained from fitting a standard time-homogeneous NB model to the ORATORIO data can be found in Table $\ref{NBoratorio}$. By adjusting for age group ($> 45$ versus $\leq 45$ years) and geographical region (USA versus ROW), treatment with OCR results in a statistically significant $28.6 \%$ reduction in the adjusted CDP12 rate compared with PLA (adjusted RR 0.714, $95\%$ CI: [0.565, 0.906], p-value=$0.00489$). The dispersion or heterogeneity parameter $\phi$ is estimated by $0.165$, which implies that there is minimal overdispersion.  

\begin{table}[h]
\centering
\scalebox{0.70}{ 
\begin{tabular}{lcc}
\textbf{Efficacy Variable}                                                                                                                          & \textbf{\begin{tabular}[c]{@{}c@{}}OCR \\ (N = 487)\end{tabular}}  & \textbf{\begin{tabular}[c]{@{}c@{}}PLA \\ (N = 244)\end{tabular}} \\ \hline
\begin{tabular}[c]{@{}l@{}}Total number of progression events \\ Total patient-years followed \end{tabular} & \begin{tabular}[c]{@{}c@{}}190\\ 1340 \end{tabular}               & \begin{tabular}[c]{@{}c@{}}124\\ 627\end{tabular}         \\
                                                                                                                                                    &                                                                          &                                                                   \\
\begin{tabular}[c]{@{}l@{}}Adjusted** RR\\ 95\% CI  of RR\\ p-value\end{tabular}                               & \begin{tabular}[c]{@{}c@{}}0.714\\ {[}0.565, 0.906{]}\\ 0.00489\end{tabular} &                                                                  
\end{tabular}
}
\caption[ORATORIO - Estimates of treatment effect using a negative binomial model]{ORATORIO - Estimates of treatment effect using a NB model, log-transformed exposure time is included as an offset variable (** = adjusted for age (> 45 versus $\leq 45$ years) and geographical region (USA versus ROW)}
\label{NBoratorio}
\end{table}

\newpage 
\textbf{Lin-Wei-Yang-Ying model} \newline 
When controlling for the treatment group, a stratified LWYY analysis yields an estimated RR of $0.723$ ($95\%$ CI: [0.572, 0.915], p-value=$0.00699$), indicating that OCR is significantly effective in reducing the $12$-week CDP rate. More specifically, the expected number of CDP12s per unit time in the OCR group is reduced by $27.3 \%$ compared to the expected number of CDP12s per unit time in the PLA group. Or, in other words, the overall rate of a CDP12 event is $27.3 \%$ lower in the OCR group than in the PLA group. 

\begin{table}[h]
\centering
\begin{tabular}{llccc}
                                                                                              & \textbf{Model} & \textbf{Treatment effect} & \textbf{$\boldsymbol{95 \%}$ CI}                                     & \textbf{p-value}                                        \\ \hline
\textbf{\begin{tabular}[c]{@{}l@{}}Time-to-first-\\ event\\ analysis\end{tabular}}            & Cox model*     & HR 0.759                  & \begin{tabular}[c]{@{}c@{}}$$\\ {[}0.589, 0.978{]}\\ $$\end{tabular} & \begin{tabular}[c]{@{}c@{}}$$\\ 0.033\\ $$\end{tabular} \\ \hline
\multirow{2}{*}{\textbf{\begin{tabular}[c]{@{}l@{}}Recurrent event \\ analyses\end{tabular}}} & NB model**     & RR 0.714                  & {[}0.565, 0.906{]}                                                   & 0.0049                                                 \\
                                                                                              & LWYY model*    & RR 0.723                  & {[}0.572, 0.915{]}                                                   & 0.00699                                                
\end{tabular}
\caption[ORATORIO - Comparisons of treatment effect estimates obtained from time-to-first-event and marginal recurrent event analyses]{ORATORIO - Comparisons of treatment effect estimates obtained from time-to-first-event (Cox) and marginal recurrent event analyses (NB and LWYY), log-transformed exposure time is included as an offset variable in NB model (* = stratified by and ** = adjusted for age group ($> 45$ versus $\leq 45$ years) and geographical region (USA versus ROW))}
\label{TreatmentEffectOverviewOratorioRCTLWYYNB}
\end{table}

\begin{figure}[H]
\centering
\scalebox{0.45}[0.45]{ 
\includegraphics{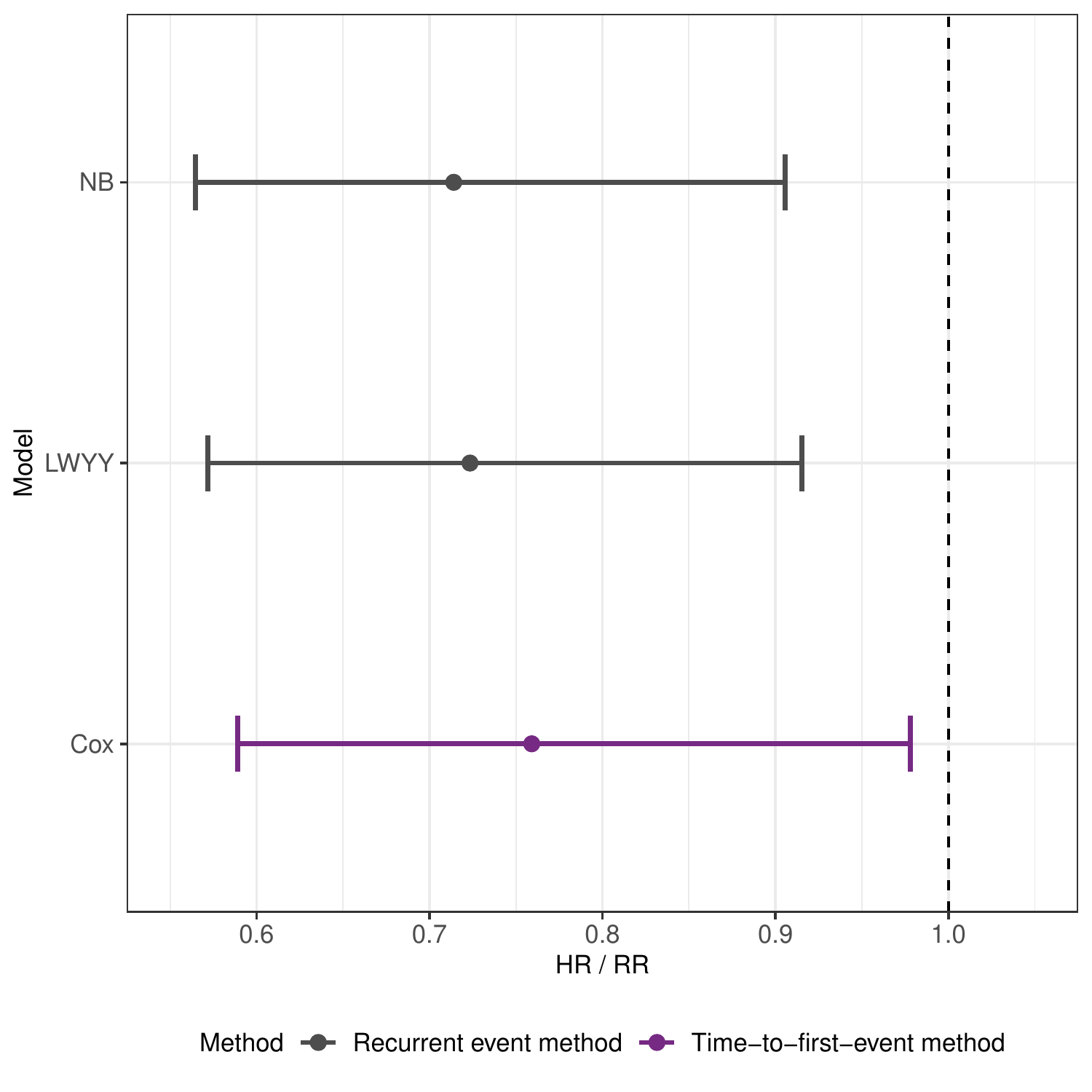}
}
\caption[ORATORIO - Forest plot of treatment effect estimates obtained from time-to-first-event and marginal recurrent event analyses]{ORATORIO - Forest plot of treatment effect estimates obtained from time-to-first-event and marginal recurrent event analyses with $95 \%$ CIs}
\label{forestplotORATORIOonlyRCTmethods}
\end{figure}

Results from the time-to-first-event and the marginal recurrent event analyses of the ORATORIO trial data are summarized in Table $\ref{TreatmentEffectOverviewOratorioRCTLWYYNB}$ and Figure $\ref{forestplotORATORIOonlyRCTmethods}$. Regardless of the statistical method applied, the beneficial effect of OCR on confirmed disability progression in early PPMS patients can be concluded from all analyses. As seen from Table $\ref{TreatmentEffectOverviewOratorioRCTLWYYNB}$, the estimated treatment effect obtained from the time-to-first-event approach is smaller than the treatment effects estimated by the recurrent event methods. In recurrent event analyses, the NB estimate is similar to the estimate obtained from the LWYY model. Compared to the Cox analysis, the NB and LWYY models improve statistical precision, since the widths of the $95 \%$ CIs are smaller with the recurrent event methods (cf. Figure $\ref{forestplotORATORIOonlyRCTmethods}$).
\newline 
In Figure $\ref{CMForatorio}$, the CMFs suggest some minor time trends in the CDP12 event rate over time, which gives preference to the LWYY model over the NB model as analysis method. The robust semiparametric LWYY model is flexible in the sense that neither the baseline rate function nor the heterogeneity induced by the recurrent event processes must be specified by a certain parametric statistical model, as it is the case in the NB approach. If the event rate is roughly constant, both recurrent event methods are appropriate. Due to the semiparametric property of the LWYY model, it is recommended to use the LWYY model as primary analysis for the recurrent CDP12 endpoint and the NB model as sensitivity analysis.

Other recurrent event methods including the partially conditional rate-based model and the WLW model can be considered as supplementary analyses to get a complete picture of the treatment effect and to study treatment effects on the time to later events. For instance, it is also of particular interest to assess the treatment effect on the rate of subsequent CDP12 events among patients who have already experienced prior events. Such a study question can be adressed by a partially conditional rate-based model. The treatment effect on the time to the second or third CDP12 event can be estimated by the WLW model. 
\newline \newline 
\textbf{Supplementary analyses} \newline 
Results from the supplementary analyses are summarized in Table $\ref{OratorioWLWandPWPestimates}$. 

\begin{table}[h]
\begin{tabular}{c|llcccc}
\multicolumn{1}{l|}{}                                                                        & \multicolumn{1}{c|}{\textbf{Model}}                                                                  & \multicolumn{2}{c}{\textbf{Treatment effect}}                                                                                                  & \textbf{SE($\boldsymbol{\beta}$)}                             & \textbf{$\boldsymbol{95 \%}$ CI}                                                                     & \textbf{p-value}                                                 \\ \hline
\multirow{5}{*}{\textbf{\begin{tabular}[c]{@{}c@{}}Recurrent event\\ analyses\end{tabular}}} & \multicolumn{1}{l|}{\multirow{2}{*}{\textbf{\begin{tabular}[c]{@{}l@{}}PCRB\\ model*\end{tabular}}}} & Common                                                                & RR 0.752                                                               & 0.117                                                         & {[}0.599, 0.945{]}                                                                                   & 0.0145                                                           \\ \cline{3-7} 
                                                                                             & \multicolumn{1}{l|}{}                                                                                & \begin{tabular}[c]{@{}l@{}}Event 1\\ Event 2  \\ Event 3\end{tabular} & \begin{tabular}[c]{@{}c@{}}RR 0.759\\ RR 0.775\\ RR 0.156\end{tabular} & \begin{tabular}[c]{@{}c@{}}0.129\\ 0.284\\ 1.085\end{tabular} & \begin{tabular}[c]{@{}c@{}}{[}0.590, 0.977{]}\\ {[}0.444, 1.353{]}\\ {[}0.019, 1.306{]}\end{tabular} & \begin{tabular}[c]{@{}c@{}}0.0322\\ 0.3696\\ 0.0865\end{tabular} \\ \cline{2-7} 
                                                                                             &                                                                                                      &                                                                       &                                                                        &                                                               &                                                                                                      &                                                                  \\ \cline{2-7} 
                                                                                             & \multicolumn{1}{l|}{\multirow{2}{*}{\textbf{\begin{tabular}[c]{@{}l@{}}WLW \\ model*\end{tabular}}}} & \multicolumn{1}{c}{Common}                                            & HR 0.711                                                               & 0.137                                                         & {[}0.544, 0.930{]}                                                                                   & 0.0129                                                           \\ \cline{3-7} 
                                                                                             & \multicolumn{1}{l|}{}                                                                                & \begin{tabular}[c]{@{}l@{}}Event 1\\ Event 2\\ Event 3\end{tabular}   & \begin{tabular}[c]{@{}c@{}}HR 0.759\\ HR 0.615\\ HR 0.105\end{tabular} & \begin{tabular}[c]{@{}c@{}}0.129\\ 0.283\\ 1.126\end{tabular} & \begin{tabular}[c]{@{}c@{}}{[}0.590, 0.977{]}\\ {[}0.353, 1.070{]}\\ {[}0.016, 0.957{]}\end{tabular} & \begin{tabular}[c]{@{}c@{}}0.0322\\ 0.0851\\ 0.0456\end{tabular}
\end{tabular}
\caption[ORATORIO - Common and event-specific estimates of treatment effect using partially conditional rate-based and WLW models]{ORATORIO - Common and event-specific estimates of treatment effect using partially conditional rate-based and WLW models, robust variance estimation in WLW approach (PCRB = partially conditional rate-based, * = stratified by age (> 45 versus $\leq 45$ years) and geographical region (USA versus ROW))}
\label{OratorioWLWandPWPestimates}
\end{table}

\textbf{Partially conditional rate-based model} \newline
In order to investigate whether the treatment effect changes for subsequent events, a partially conditional rate-based model is fitted to the ORATORIO data. Since there are only a few patients with more than $2$ CDP12 events, $3$ time-dependent strata are defined based on no events $(N(t-)=0)$, 1 event $(N(t-)=1)$ and $\geq 2$ events $(N_{i}\geq 2)$. As expected, the estimated RR for the first CDP12 event from the partially conditional rate-based model (RR $0.759$, $95 \%$ CI [$0.589, 0.978$], p-value=$0.0322$) gives the same treatment effect estimate than the Cox model. As seen from Table $\ref{OratorioWLWandPWPestimates}$, the event-specific RR for the second CDP12 event indicates a beneficial effect of OCR in reducing the rate for a second $12$-week CDP among patients who have already experienced one CDP12 event. Provided that the first CDP12 event has already happened, the rate for a second CDP12 event from randomization is $22.5 \%$ lower in the OCR group than in the PLA group. For the first and second CDP12 event, the event-specific RRs remain relatively constant, meaning that the effect of OCR does not considerably vary with increasing number of previous events. The RR for the third CDP12 event appears to be unreliable due to the small number of PPMS patients at-risk in this stratum. In total, this analysis reveals that, conditional on previous CDP12 events, the rates for a first and second event from study start are $24.1 \%$ and $22.5 \%$ lower in the OCR group. There is evidence that OCR does not only affect the first CDP12 event but also the subsequent events. However, treatment comparisons for subsequent CDP12 events are not based on all individuals who had been initially randomized (i.e., randomization is destroyed), making causal inference difficult. 
\newline \newline 
\textbf{Wei-Lin-Weissfeld model} \newline 
For the analysis of recurrent CDP12 data using the WLW model, the maximum number of events is restricted to $K=3$ (cf. Chapter $4$, Section $...$). The WLW analysis for the time to the first CDP12 event is based on $731$ patients ($N=244$ PLA, $N=487$ OCR), with $160$ CDP12 events in the OCR group and $90$ CDP12 events in the PLA group. For the time to the second CDP12, the WLW approach considers $29$ CDP12 events with OCR and $22$ CDP12 events with PLA. The third WLW analysis is based on $1$ CDP12 event in the OCR group and $4$ events in the PLA group. \newpage
Table $\ref{OratorioWLWandPWPestimates}$ illustrates the event-specific HRs obtained from fitting $3$ distinct Cox proportional hazards models to the ORATORIO data. Obviously, the estimated HR for the first CDP12 event is identical to the one resulting from the Cox model. The event-specific HRs decrease from $0.759$ for the first CDP12 event ($95 \%$ CI [$0.590, 0.977$], p-value=$0.0322$) to $0.615$ for the second CDP12 event ($95 \%$ CI [$0.353, 1.070$], p-value=$0.0851$). The marginal effect of OCR on the time to the third CDP12 event may be unreliable due to the fact that only a few ORATORIO patients experienced three CDP12 events during the double-blind treatment period. In summary, the WLW analysis indicates that the marginal hazard for a first and second CDP12 event is reduced by $24.1 \%$ and $38.5 \%$ with OCR, as compared to PLA. \newline 
In contrast to the partially conditional rate-based model, treatment effect estimates under the WLW approach are based on comparisons of the complete randomized treatment groups for each event, since patients are included in the risk set for each distinct event from study start. 
\newline \newline 
\textbf{Andersen-Gill model} \newline
When controlling for the treatment group, the AG intensity model yields a common HR of $0.723$ with a $95 \%$ CI of $[0.577, 0.907]$. The treatment effect estimate obtained from the AG model is the same as the LWYY but estimates differ in interpretation. Treatment with OCR significantly reduces the intensity for a CDP12 event by $27.7 \%$ compared with PLA (p-value = $0.00507$). 
\newline \newline 
\textbf{Estimation of treatment effect using alternative endpoint definitions} \newline 
Table $\ref{TreatmentEffectOverviewOratorioRCTLWYYNBdef2}$ represents the results from the time-to-first-event and marginal recurrent event analyses under the three alternative CDP12 endpoint definitions (cf. Chapter $2$). It can be concluded that, under all alternative endpoint definitions, the recurrent event methods including the LWYY and NB models outperform the Cox model in terms of statistical precision, as the widths of the $95 \%$ CIs are smaller with the recurrent event analyses.  \newline 
Results from the supplementary analyses including partially conditional rate-based and WLW models are not presented. 

\begin{figure}[h]
\centering
  \subfloat[Endpoint: time-to-onset-of-CDP12 and roving reference system]{
  \scalebox{0.8}{ 
\begin{tabular}{llccc}
                                                                                              & \textbf{Model} & \textbf{Treatment effect} & \textbf{$\boldsymbol{95 \%}$ CI}                                     & \textbf{p-value}                                         \\ \hline
\textbf{\begin{tabular}[c]{@{}l@{}}Time-to-first-\\ event\\ analysis\end{tabular}}            & Cox model*     & HR 0.766                  & \begin{tabular}[c]{@{}c@{}}$$\\ {[}0.602, 0.974{]}\\ $$\end{tabular} & \begin{tabular}[c]{@{}c@{}}$$\\ 0.0295\\ $$\end{tabular} \\ \hline
\multirow{2}{*}{\textbf{\begin{tabular}[c]{@{}l@{}}Recurrent event \\ analyses\end{tabular}}} & NB model**     & RR 0.728                    & {[}0.587, 0.905{]}                                                      & 0.00393                                                      \\
                                                                                              & LWYY model*    & RR 0.733                  & {[}0.589, 0.912{]}                                                   & 0.00525                                                 
\end{tabular}
  }}

  \subfloat[Endpoint: time-to-confirmation-of-CDP12 and fixed reference system]{
  \scalebox{0.8}{ 
\begin{tabular}{llccc}
                                                                                              & \textbf{Model} & \textbf{Treatment effect} & \textbf{$\boldsymbol{95 \%}$ CI}                                     & \textbf{p-value}                                         \\ \hline
\textbf{\begin{tabular}[c]{@{}l@{}}Time-to-first-\\ event\\ analysis\end{tabular}}            & Cox model*     & HR 0.779                  & \begin{tabular}[c]{@{}c@{}}$$\\ {[}0.601, 1.010{]}\\ $$\end{tabular} & \begin{tabular}[c]{@{}c@{}}$$\\ 0.0599\\ $$\end{tabular} \\ \hline
\multirow{2}{*}{\textbf{\begin{tabular}[c]{@{}l@{}}Recurrent event \\ analyses\end{tabular}}} & NB model**     & RR 0.752                    & {[}0.594, 0.956{]}                                                      & 0.0188                                                      \\
                                                                                              & LWYY model*    & RR 0.745                  & {[}0.583, 0.951{]}                                                   & 0.0179                                                  
\end{tabular}
  }}
  
  \subfloat[Endpoint: time-to-confirmation-of-CDP12 and roving reference system]{
  \scalebox{0.8}{ 
\begin{tabular}{llccc}
                                                                                              & \textbf{Model} & \textbf{Treatment effect} & \textbf{$\boldsymbol{95 \%}$ CI}                                     & \textbf{p-value}                                         \\ \hline
\textbf{\begin{tabular}[c]{@{}l@{}}Time-to-first-\\ event\\ analysis\end{tabular}}            & Cox model*     & HR 0.778                  & \begin{tabular}[c]{@{}c@{}}$$\\ {[}0.608, 0.995{]}\\ $$\end{tabular} & \begin{tabular}[c]{@{}c@{}}$$\\ 0.0452\\ $$\end{tabular} \\ \hline
\multirow{2}{*}{\textbf{\begin{tabular}[c]{@{}l@{}}Recurrent event \\ analyses\end{tabular}}} & NB model**     & RR 0.762                       & {[}0.611, 0.953{]}                                                      & 0.0162                                                      \\
                                                                                              & LWYY model*    & RR 0.749                  & {[}0.597, 0.939{]}                                                   & 0.0124                                                  
\end{tabular}
  }}
\caption[ORATORIO - Comparisons of treatment effect estimates obtained from time-to-first-event and marginal recurrent event analyses (endpoint: alternative definitions)]{ORATORIO - Comparisons of treatment effect estimates obtained from time-to-first-event (Cox) and marginal recurrent event analyses (NB and LWYY), log-transformed exposure time is included as an offset variable in NB model (* = stratified by and ** = adjusted for age group ($> 45$ versus $\leq 45$ years) and geographical region (USA versus ROW))}
\label{TreatmentEffectOverviewOratorioRCTLWYYNBdef2}
\end{figure}

\subsubsection{Estimation of covariate effects impacting disease progression}
\label{IntensitybasedModelsCovariates}
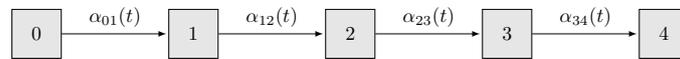
\begin{figure}[H]
\centering
\resizebox{9cm}{!}{
\begin{tikzpicture}[node distance=1.9cm]
  \tikzset{node style/.style={state, fill=gray!20!white, rectangle}}
        \node[node style]               (I)   {0};
        \node[node style, right=of I]   (II)  {$1$};
        \node[node style, right=of II]  (III) {$2$};
        \node[node style, right=of III]  (IV)  {$3$};
        \node[node style, right=of IV]  (V)   {$4$};
    \draw[>=latex,
          auto=left,
          every loop]
         (I)   edge node {$\alpha_{01}(t)$} (II)
         (II)  edge node {$\alpha_{12}(t)$} (III)
         (III) edge node {$\alpha_{23}(t)$} (IV)
         (IV) edge node {$\alpha_{34}(t)$} (V); 
         \end{tikzpicture}}
\caption{Multistate model}
\label{GeneralMultistateModelApplication}
\end{figure}

So far, the main focus of this chapter was to reanalyse the efficacy of OCR from the randomized ORATORIO trial in PPMS patients by using recurrent event methods. Evaluation of treatment effects in RCTs involving recurrent events requires specific statistical methods based on marginal rate functions to ensure valid causal inference. Apart from estimating such marginal parameters of recurrent event processes, intensity models that condition on the past are useful to get deeper insights into the structure of the recurrent event process. The intensity models that will follow aim at identifying potential risk factors associated with CDP12 occurrences and at understanding the event process dynamics. In order to examine process dynamics, a general multistate model for recurrent CDP12 events is a suitable framework. 
\newline \newline
\textbf{General multistate model for recurrent events} \newline
Figure $\ref{GeneralMultistateModelApplication}$ contains the multistate model diagram considered in this analysis, with transition hazards $\alpha_{01}(t), \alpha_{12}(t), \alpha_{23}(t)$, and $\alpha_{34}(t)$. The analysis is restricted to a maximum of $4$ CDP12 events per patient.  

\begin{figure}[H] 
\centering
  \subfloat[Nelson Aalen estimates for PLA group]{
  \scalebox{0.5}[0.5]{ 
\includegraphics{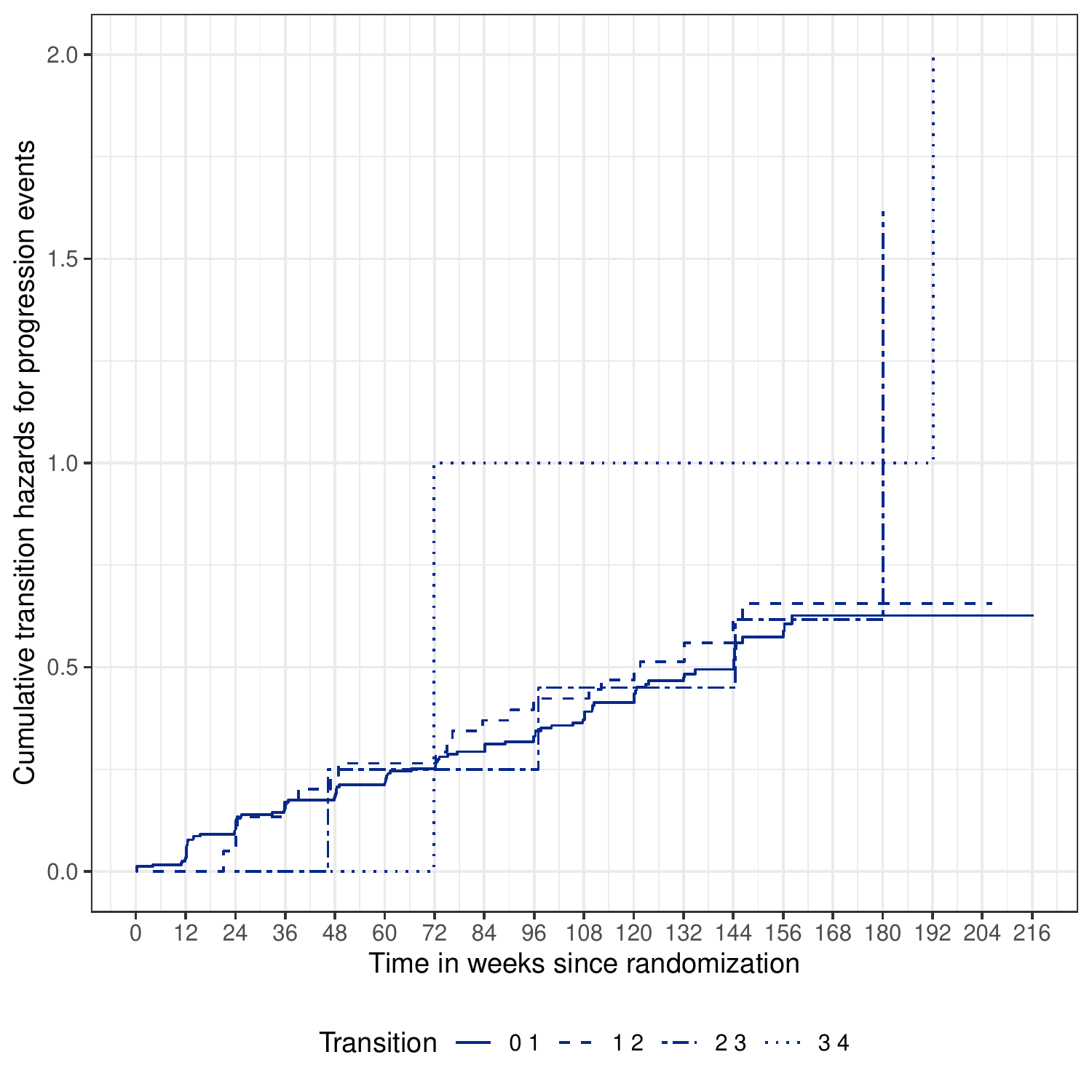}
  }}
  \subfloat[Nelson Aalen estimates for OCR group]{
  \scalebox{0.5}{ 
\includegraphics{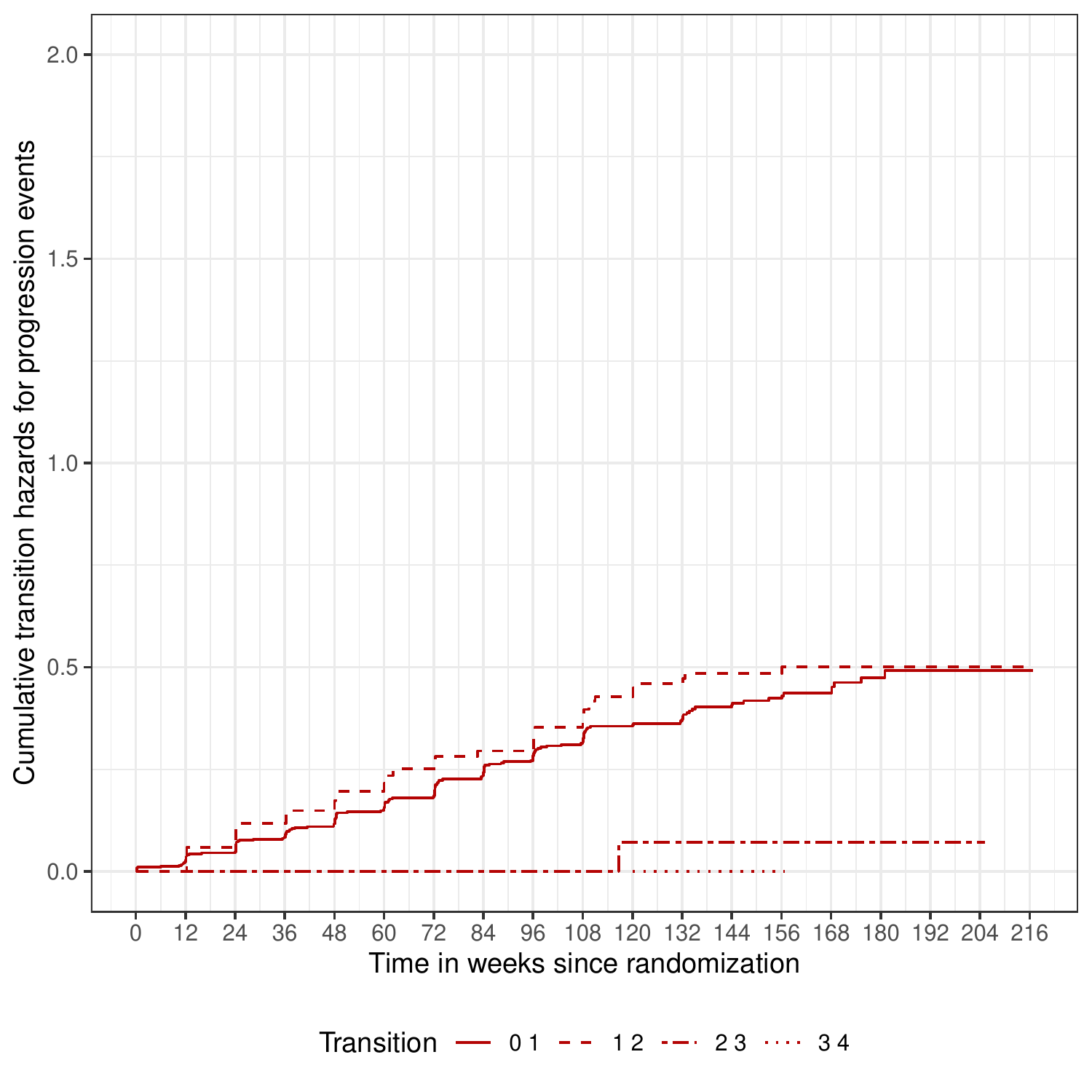}
  }} 
\caption[ORATORIO - Nelson Aalen estimates of cumulative transition hazards]{ORATORIO - Nelson Aalen estimates of cumulative transition hazards for subsequent CDP12 events following $0$, $1$, $2$ or $3$ CDP12 events}
\label{NAoratorio}
\end{figure}

The Nelson Aalen estimators $\hat{A}_{01}(t), \hat{A}_{12}(t), \hat{A}_{23}(t)$ and $\hat{A}_{34}(t)$ for the cumulative transition hazards are displayed in Figure $\ref{NAoratorio}$ for OCR and PLA patients. Among the PLA patients, the similar slopes of $\hat{A}_{01}(t)$ and $\hat{A}_{12}(t)$ reveal that patients who have already experienced $1$ CDP12 event (dashed line) are not at a higher risk for a further event than patients who are still event-free (solid line). Since only a few patients are observed to make $2 \longrightarrow 3$ and $3 \longrightarrow 4$ transitions, the Nelson Aalen estimates $\hat{A}_{23}(t)$ and $\hat{A}_{34}(t)$ are less precise. Similar results can be found for the OCR patients. In both treatment groups, the hazard for a new CDP12 event at time $t$ does not increase with the number of previous CDP12 events. Compared to the PLA group, the Nelson Aalen estimates under active treatment are observed to be reduced, concluding the beneficial effect of OCR. 

\newpage 
\textbf{Analyses of recurrent event data via intensity-based models} \newline 
In Table $\ref{OratorioIntensitybasedModels}$, results obtained from fitting several multiplicative intensity-based models to the ORATORIO data are displayed. The baseline covariates include treatment group (OCR versus PLA), age (years), sex (female versus male), geographical region (USA versus ROW), EDSS score, presence of $T_{1}$ lesions (yes versus no), duration since MS symptoms onset (years) and intake of prior disease-modifying  MS therapies (yes versus no). The regression analyses also involve a patient-specific gamma frailty and/or the time-varying internal covariate $N(t-)$. \newline 
Table $\ref{overviewModelsapplication}$ gives an overview on the specification of the different intensity models. Model 1A is a semiparametric AG model that only controls for the treatment group. The random effect model 1B is an extension of model 1A additionally controlling for unobserved heterogeneity. In addition to treatment group, model 2A adjusts for further important baseline characteristics. In the same manner, model 2B is also a random effect model which extends model 2A by accounting for unobserved between-patient variability. In contrast to model 2A, model 2C is of PWP-type allowing for time-dependent stratification. Model 3A extends model 2A by incorporating the past event history through the cumulative number of prior progression events. Model 3B extends model 3A by including a frailty term. 

\begin{table}[h]
\centering
\scalebox{0.55}{ 
\begin{tabular}{|l|l|l|l|l|l|l|l|}
\hline
\textbf{Components}            & \textbf{1A} & \textbf{1B} & \textbf{2A} & \textbf{2B} & \textbf{2C} & \textbf{3A} & \textbf{3B} \\ \hline
Treatment group                 & $\surd$     & $\surd$     & $\surd$     & $\surd$     & $\surd$     & $\surd$     & $\surd$     \\ \hline
Sex                             &             &             & $\surd$     & $\surd$     & $\surd$     & $\surd$     & $\surd$     \\ \hline
Age                             &             &             & $\surd$     & $\surd$     & $\surd$     & $\surd$     & $\surd$     \\ \hline
Geographical region             &             &             & $\surd$     & $\surd$     & $\surd$     & $\surd$     & $\surd$     \\ \hline
Baseline EDSS                   &             &             & $\surd$     & $\surd$     & $\surd$     & $\surd$     & $\surd$     \\ \hline
$T_{1}$ Gd enhancing lesions    &             &             & $\surd$     & $\surd$     & $\surd$     & $\surd$     & $\surd$     \\ \hline
Previous MS treatment           &             &             & $\surd$     & $\surd$     & $\surd$     & $\surd$     & $\surd$     \\ \hline
Duration since MS symptom onset &             &             & $\surd$     & $\surd$     & $\surd$     & $\surd$     & $\surd$     \\ \hline
Number of prior events $N(t-)$  &             &             &             &             &             & $\surd$     & $\surd$     \\ \hline
                                &             &             &             &             &             &             &             \\ \hline
Time-dependent stratification   &             &             &             &             & $\surd$     &             &             \\ \hline
Patient-specific gamma frailty  &             & $\surd$     &             & $\surd$     &             &             & $\surd$     \\ \hline
\end{tabular}}
\caption{ORATORIO - Recurrent event analyses via intensity-based models}
\label{overviewModelsapplication}
\end{table}

All AG-type and the PWP-type models 1A, 2A, 2C and 3A claim that patients treated with OCR have a reduced intensity for disability progression compared to patients treated with PLA. Based on model 1B, the variance $\phi$ of the gamma distributed random effect is estimated to be equal to $0.1389$. By adjusting for baseline covariates, the frailty term in model 2B reduces to $0.0480$, indicating that the covariates considered may explain some variability in the CDP12 recurrences. In model 2A, the treatment effect estimate represents the relative risk for CDP12 events among patients with the same realization of covariates. After controlling for treatment, sex, age, geographical region, baseline EDSS score, presence of $T_{1}$ Gd-enhancing lesions, intake of prior disease-modifying therapy and duration since MS symptoms onset, OCR significantly reduces the intensity for a CDP12 event by approximately $28.9 \%$ (HR 0.7108 , $95 \%$ CI [0.5617, 0.8994], p-value = $0.00447$). Further, model 2A suggests that patients with a higher baseline EDSS score have a higher risk for CDP12 event recurrence, justifying the findings from Figure $\ref{CMForatorioEDSS}$. Each additional year in duration since MS symptoms onset is associated with a lower risk of disability progression (p-value = $0.074447$). In the stratified model 2C, the estimated regression coefficients and SEs are pretty similar to the ones obtained from model 2A. By comparison, treatment comparison in model 3A is restricted to patients with the same realization of covariates and the same cumulative number of prior CDP12 events. When controlling for the cumulative number of prior CDP12 events (model 3A), the estimated regression coefficients and SEs are similar to the ones obtained from model 2A and the coefficient for $N(t-)$ is not significant. This suggests that the risk of a new CDP12 event is not likely to increase with the number of prior events. This is consistent with the general multistate analysis (cf. Nelson Aalen estimates in Figure $\ref{NAoratorio}$). The variance estimate is almost zero in model 3B, when accounting for $N(t-)$. This implies that the variability in CDP12 occurrences can be adequately explained by fixed effects. 
\newpage 

\begin{landscape}
\begin{table}[]
\centering
\resizebox{23cm}{!}{ 
\begin{tabular}{|l|l|lll|ccc|clc|ccc|}
\hline
\multicolumn{2}{|l|}{\multirow{3}{*}{}}      & \multicolumn{3}{c|}{\textbf{\begin{tabular}[c]{@{}c@{}}unstratified\\ (AG-type)\end{tabular}}} & \multicolumn{3}{c|}{\textbf{\begin{tabular}[c]{@{}c@{}}unstratified\\ (AG-type)\end{tabular}}} & \multicolumn{3}{c|}{\textbf{\begin{tabular}[c]{@{}c@{}}stratified\\ (PWP-CP type)\end{tabular}}} & \multicolumn{3}{c|}{\textbf{\begin{tabular}[c]{@{}c@{}}unstratified\\ (AG-type)\end{tabular}}} \\ \cline{3-14} 
\multicolumn{2}{|l|}{}                       & \multicolumn{3}{c|}{\textbf{Model 1A}}                                                         & \multicolumn{3}{c|}{\textbf{Model 2A}}                                                         & \multicolumn{3}{c|}{\textbf{Model 2C}}                                                           & \multicolumn{3}{c|}{\textbf{Model 3A}}                                                         \\ \cline{3-14} 
\multicolumn{2}{|l|}{}                       & \multicolumn{1}{c}{HR}       & \multicolumn{1}{c}{SE}      & \multicolumn{1}{c|}{p-value}      & HR                            & SE                            & p-value                        & HR                            & \multicolumn{1}{c}{SE}          & p-value                        & HR                            & SE                           & p-value                         \\ \hline
Treatment  group                   & OCR     & 0.7240                       & 0.1155                      & 0.00517                           & 0.7108                        & 0.1201                        & 0.00447                        & 0.7286                        & 0.1215                          & 0.00915                        & 0.7137                        & 0.1204                       & 0.00508                         \\
Sex                                & Male    &                              &                             &                                   & 0.9907                        & 0.0073                        & 0.20510                        & 0.9916                        & 0.0074                          & 0.25232                        & 0.9907                        & 0.0073                       & 0.20392                         \\
Age                                & (years) &                              &                             &                                   & 1.0112                        & 0.1176                        & 0.92459                        & 0.9954                        & 0.1182                          & 0.96871                        & 1.0102                        & 0.1176                       & 0.93105                         \\
Geographical region                & USA     &                              &                             &                                   & 0.8610                        & 0.1858                        & 0.42032                        & 0.8351                        & 0.1883                          & 0.33863                        & 0.8627                        & 0.1858                       & 0.42674                         \\
Baseline EDSS                      &         &                              &                             &                                   & 1.2526                        & 0.0541                        & $<$ 0.001                      & 1.2448                        & 0.0548                          & $<$ 0.001                      & 1.2497                        & 0.0543                       & $<$ 0.001                       \\
$T_{1}$ Gd enhancing lesions       & Yes     &                              &                             &                                   & 1.1073                        & 0.1331                        & 0.44385                        & 1.0947                        & 0.1344                          & 0.50054                        & 1.1048                        & 0.1332                       & 0.45445                         \\
Previous MS treatment              & Yes     &                              &                             &                                   & 1.0382                        & 0.1831                        & 0.83799                        & 1.0077                        & 0.1859                          & 0.96705                        & 1.0379                        & 0.1832                       & 0.83920                         \\
Duration since MS symptom onset  & (years) &                              &                             &                                   & 0.8382                        & 0.0989                        & 0.07447                        & 0.8647                        & 0.1001                          & 0.14610                        & 0.8407                        & 0.0991                       & 0.07993                         \\
Number of prior events $N(t-)$     &         &                              &                             &                                   & \multicolumn{1}{l}{}          & \multicolumn{1}{l}{}          & \multicolumn{1}{l|}{}          & \multicolumn{1}{l}{}          &                                 & \multicolumn{1}{l|}{}          & 1.0664                        & 0.1311                       & 0.62390                         \\
                                   &         &                              &                             &                                   & \multicolumn{1}{l}{}          & \multicolumn{1}{l}{}          & \multicolumn{1}{l|}{}          & \multicolumn{1}{l}{}          &                                 & \multicolumn{1}{l|}{}          &                               &                              &                                 \\
(Penalized) log-likelihood         &         & \multicolumn{3}{c|}{-2007.008}                                                                 & \multicolumn{3}{c|}{-1863.163}                                                                 & \multicolumn{3}{c|}{-1710.252}                                                                   & \multicolumn{3}{c|}{-1863.045}                                                                 \\ \hline
\end{tabular}
}
\caption[ORATORIO - Estimates of covariate effects using multiplicative intensity-based models]{ORATORIO - Estimates of covariate effects using different multiplicative intensity-based models}
\label{OratorioIntensitybasedModels}
\end{table}

\begin{table}[]
\centering
\resizebox{15cm}{!}{ 
\begin{tabular}{|l|l|lll|ccc|ccc|}
\hline
\multicolumn{2}{|l|}{\multirow{3}{*}{}}      & \multicolumn{9}{c|}{\textbf{Random effects models}}                                                                                                                                                                                 \\ \cline{3-11} 
\multicolumn{2}{|l|}{}                       & \multicolumn{3}{c|}{\textbf{Model 1B}}                                                  & \multicolumn{3}{c|}{\textbf{Model 2B}}                              & \multicolumn{3}{c|}{\textbf{Model 3B}}                              \\ \cline{3-11} 
\multicolumn{2}{|l|}{}                       & \multicolumn{1}{c}{HR}     & \multicolumn{1}{c}{SE}     & \multicolumn{1}{c|}{p-value}  & HR                   & SE                   & p-value               & HR                   & SE                   & p-value               \\ \hline
Treatment  group                   & OCR     & \multicolumn{1}{c}{0.7194} & \multicolumn{1}{c}{0.1195} & \multicolumn{1}{c|}{0.005867} & 0.7087               & 0.1216               & 0.0046289             & 0.7118               & 0.1205               & 0.0047956             \\
Sex                                & Male    &                            &                            &                               & 0.9907               & 0.0074               & 0.2078                & 0.9906               & 0.0073               & 0.19796               \\
Age                                & (years) &                            &                            &                               & 1.0115               & 0.1190               & 0.92359               & 1.0117               & 0.1174               & 0.92085               \\
Geographical region                & USA     &                            &                            &                               & 0.8576               & 0.1876               & 0.41279               & 0.8598               & 0.1853               & 0.4151                \\
Baseline EDSS                      &         &                            &                            &                               & 1.2553               & 0.0549               & $< 0.001$             & 1.2511               & 0.0544               & $< 0.001$             \\
$T_{1}$ Gd enhancing lesions       & Yes     &                            &                            &                               & 1.1053               & 0.1348               & 0.4576                & 1.1039               & 0.1330               & 0.45741               \\
Previous MS treatment              & Yes     &                            &                            &                               & 1.0397               & 0.1855               & 0.076451              & 1.1039               & 0.1830               & 0.073510              \\
Duration since MS symptom onset  & (years) &                            &                            &                               & 0.8375               & 0.1001               & 0.83398               & 0.8408               & 0.0969               & 0.83477               \\
Number of prior events $N(t-)$     &         &                            &                            &                               & \multicolumn{1}{l}{} & \multicolumn{1}{l}{} & \multicolumn{1}{l|}{} & 1.0595               & 0.1336               & 0.66544               \\
                                   &         &                            &                            &                               & \multicolumn{1}{l}{} & \multicolumn{1}{l}{} & \multicolumn{1}{l|}{} & \multicolumn{1}{l}{} & \multicolumn{1}{l}{} & \multicolumn{1}{l|}{} \\
Variance                           &         & \multicolumn{3}{c|}{0.1389}                                                             & \multicolumn{3}{c|}{0.0480}                                         & \multicolumn{3}{c|}{0.00006}                                        \\
Penalized marginal log-likelihood  &         & \multicolumn{3}{c|}{-2732.58}                                                           & \multicolumn{3}{c|}{-2563.77}                                       & \multicolumn{3}{c|}{-2563.74}                                       \\ \hline
\end{tabular}
}
\caption[ORATORIO - Estimates of covariate effects using multiplicative intensity-based models and random effect models]{ORATORIO - Estimates of covariate effects using different multiplicative intensity-based models and random effect models}
\label{OratorioIntensitybasedModels2}
\end{table}
\end{landscape}
 
\subsubsection*{Analyses of recurrent event data via rate-based models}
Results obtained from fitting different multiplicative proportional rate models to the ORATORIO data are summarized in Table $\ref{OratorioRatebasedModels}$. While model $5$ only adjusts for the treatment indicator, other baseline covariates are included in model $6$. As described in Chapter $4$, the proportional rate/mean models average the overall intensity over the distribution of the past event history, providing the regression coefficient with a population-averaged interpretation. \newline 
After controlling for treatment, sex, age, geographical region, baseline EDSS score, presence of $T_{1}$ Gd-enhancing lesions, intake of prior disease-modifying therapy and duration since MS symptoms onset, OCR significantly reduces the mean frequency of CDP12 recurrences by approximately $29.9 \%$ (RR 0.7108 , $95 \%$ CI [0.5569, 0.9072], p-value = $0.0061$). Higher EDSS values are associated with a significantly increased mean number of progression events.  

\begin{table}[h]
\centering
\scalebox{0.8}{ 
\begin{tabular}{|l|l|lll|ccc|}
\hline
\multicolumn{2}{|l|}{\multirow{3}{*}{}}      & \multicolumn{6}{c|}{\textbf{Unstratified}}                                                                                                                                                \\ \cline{3-8} 
\multicolumn{2}{|l|}{}                       & \multicolumn{3}{c|}{\textbf{\begin{tabular}[c]{@{}c@{}}Model 5\\ (LWYY-type)\end{tabular}}} & \multicolumn{3}{c|}{\textbf{\begin{tabular}[c]{@{}c@{}}Model 6\\ (LWYY-type)\end{tabular}}} \\ \cline{3-8} 
\multicolumn{2}{|l|}{}                       & \multicolumn{1}{c}{RR}       & \multicolumn{1}{c}{SE}       & \multicolumn{1}{c|}{p-value}  & RR                           & SE                          & p-value                        \\ \hline
Treatment  group                   & OCR     & \multicolumn{1}{c}{0.7240}   & \multicolumn{1}{c}{0.1203}   & \multicolumn{1}{c|}{0.00726}  & 0.7108                       & 0.1201                      & 0.0061                         \\
Sex                                & Male    &                              &                              &                               & 1.0112                       & 0.1176                      & 0.9246                         \\
Age                                & (years) &                              &                              &                               & 0.9907                       & 0.0073                      & 0.1940                         \\
Geographical region                & USA     &                              &                              &                               & 0.8610                       & 0.1858                      & 0.4696                         \\
Baseline EDSS                      &         &                              &                              &                               & 1.2526                       & 0.0541                      & $< 0.001$                      \\
$T_{1}$ Gd enhancing lesions       & Yes     &                              &                              &                               & 1.1073                       & 0.1331                      & 0.4715                         \\
Previous MS treatment              & Yes     &                              &                              &                               & 1.0382                       & 0.1831                      & 0.8381                         \\
Duration since MS symptom onset  & (years) &                              &                              &                               & 0.8382                       & 0.0989                      & 0.1109                         \\ \hline
\end{tabular}
}
\caption[ORATORIO - Estimates of covariate effects using multiplicative rate-based models]{ORATORIO - Estimates of covariate effects using different multiplicative rate-based models}
\label{OratorioRatebasedModels}
\end{table}

\section{OPERA trials}
\label{DAopera}
The OPERA trials, OPERA I and OPERA II, are two identical, double-blinded, randomized phase III trials designed to investigate the efficacy and safety of ocrelizumab (OCR) compared to interferon beta-1a (IFN) in patients with RRMS \parencite{opera}. The two trials used identical protocols but were conducted independently at non-overlapping trial sites. Due to the fact that the disease course in RRMS is typically dominated by relapses and periods of remissions, the primary endpoint in the OPERA trials was the annualized relapse rate (ARR) by $96$ weeks. The ARR is defined as the number of relapses that are observed per person-year of follow-up. The time to the onset of the first 12-week CDP, with the initial event of neurological worsening occurring during the double-blind treatment period, was the important key secondary endpoint. In the secondary analysis, confirmed disability progression was defined according to the standard definition mentioned in Section $2.2.1$. In contrast to the event-driven ORATORIO trial, patients in the OPERA trials were all followed for a fixed duration of $96$ weeks. In total, $1656$ patients underwent $1:1$ randomization, with $821$ patients in the OPERA I trial (N=$410$ OCR and N=$411$ IFN) and $835$ patients in the OPERA II trial (N=$417$ OCR and N=$418$ IFN). The trials involved RRMS patients who were aged between $18$ and $55$ years, diagnosed in accordance with the $2005$ revised McDonald criteria, had an EDSS score of $0.0$ to $5.5$ at screening and who had at least $2$ documented clinical relapses within the previous $2$ years or $1$ clinical relapse within the year before screening. Further eligibility criteria included no neurological worsening for at least $30$ days before screening and baseline and MRI of the brain showing abnormalities consistent with MS. As seen from Figure $\ref{BaselineOPERA}$, baseline demographic and disease characteristics of the RRMS patients are well balanced across the treatment arms in both OPERA trials. 
\newline \newline
In the following, the focus is on the analysis of the secondary endpoint, as this work is motivated by repeated CDP events. Results presented in this work are based on a re-analysis of the original data and might deviate from the prespecified analyses.   

\newpage

\begin{figure}[H] 
\centering
  \subfloat[OPERA I]{
  \scalebox{0.65}{ 
\begin{tabular}{lcc}
\textbf{Characteristic}                                                                                                                                       & \textbf{\begin{tabular}[c]{@{}c@{}}OCR \\ (N = 410)\end{tabular}}                   & \textbf{\begin{tabular}[c]{@{}c@{}}IFN \\ (N = 411)\end{tabular}}                 \\ \hline
\begin{tabular}[c]{@{}l@{}}Age - years\\ $\qquad$ mean\\ $\qquad$ median (range)\end{tabular}                                                                 & \begin{tabular}[c]{@{}c@{}}$$\\ 37.1 $\pm$ 9.3\\ 38.0 (18.0, 56.0)\end{tabular}          & \begin{tabular}[c]{@{}c@{}}$$\\ 36.9 $\pm$ 9.3\\ 37.0 (18.0, 55.0)\end{tabular}          \\ \hline
Male sex - no. (\%)                                                                                                                                           & 140 (34.1)                                                                               & 139 (33.8)                                                                               \\ \hline
\begin{tabular}[c]{@{}l@{}}Geographical region - no. (\%)\\ $\qquad$ United States\\ $\qquad$ Rest of the world\end{tabular}                                  & \begin{tabular}[c]{@{}c@{}}$$\\ 105 (25.6)\\ 305 (74.4)\end{tabular}                     & \begin{tabular}[c]{@{}c@{}}$$\\ 105 (25.5)\\ 306 (74.5)\end{tabular}                     \\ \hline
\begin{tabular}[c]{@{}l@{}}Time since onset of MS symptoms - years\\ $\qquad$ mean\\ $\qquad$ median (range)\end{tabular}                                     & \begin{tabular}[c]{@{}c@{}}$$\\ 6.7 $\pm$ 6.4\\ 4.9 (0.2, 33.6)\end{tabular}             & \begin{tabular}[c]{@{}c@{}}$$\\ 6.2 $\pm$ 6.0\\ 4.6 (0.2, 34.9)\end{tabular}             \\ \hline
\begin{tabular}[c]{@{}l@{}}Time since diagnosis of RRMS - years\\ $\qquad$ mean\\ $\qquad$ median (range)\end{tabular}                                         & \begin{tabular}[c]{@{}c@{}}$$\\ 3.8 $\pm$ 4.8\\ 1.5 (0.0, 28.9)\end{tabular}             & \begin{tabular}[c]{@{}c@{}}$$\\ 3.7 $\pm$ 4.6\\ 1.6 (0.1, 28.0)\end{tabular}             \\ \hline
\begin{tabular}[c]{@{}l@{}}No. of relapses in previous 12 months\\ $\qquad$ mean\\ $\qquad$ median (range)\end{tabular}                                       & \begin{tabular}[c]{@{}c@{}}$$\\ 1.3 $\pm$ 0.7\\ 1.0 (0.0, 5.0)\end{tabular}              & \begin{tabular}[c]{@{}c@{}}$$\\ 1.3 $\pm$ 0.6\\ 1.0 (0.0, 4.0)\end{tabular}              \\ \hline
\begin{tabular}[c]{@{}l@{}}No previous use of disease-modifying\\ $\quad$ therapy no. /total no. (\%)\end{tabular}                                            & \begin{tabular}[c]{@{}c@{}}$$\\ 301/408 (73.8)\end{tabular}                              & \begin{tabular}[c]{@{}c@{}}$$\\ 292/409 (71.4)\end{tabular}                              \\ \hline
\begin{tabular}[c]{@{}l@{}}EDSS Score\\ $\qquad$ mean\\ $\qquad$ median (range)\end{tabular}                                                                  & \begin{tabular}[c]{@{}c@{}}$$\\ 2.8 $\pm$ 1.2\\ 2.5 (0.0, 6.0)\end{tabular}              & \begin{tabular}[c]{@{}c@{}}$$\\ 2.7 $\pm$ 1.3\\ 2.5 (0.0, 6.0)\end{tabular}              \\ \hline
\begin{tabular}[c]{@{}l@{}}Gadolinium-enhancing lesions on $T_{1}$- \\ $\quad$ weighted images - no./total no. (\%)\\ $\qquad$ yes\\ $\qquad$ no\end{tabular} & \begin{tabular}[c]{@{}c@{}}$$\\ $$\\ 172/405 (42.5)\\ 233/405 (57.5)\end{tabular}        & \begin{tabular}[c]{@{}c@{}}$$\\ $$\\ 155/407 (38.1)\\ 252/407 (61.9)\end{tabular}        \\ \hline
\begin{tabular}[c]{@{}l@{}}Number of lesions on $T_{2}$-weighted images\\ $\qquad$ mean\\ $\qquad$ median (range)\end{tabular}                                & \begin{tabular}[c]{@{}c@{}}$$\\ 51.0 $\pm$ (39.0)\\ 40.5 (1.0, 218.0)\end{tabular}       & \begin{tabular}[c]{@{}c@{}}$$\\ 51.1 $\pm$ 39.9\\ 41.0 (1.0, 226.0)\end{tabular}         \\ \hline
\begin{tabular}[c]{@{}l@{}}Total volume of lesions on $T_{2}$-weighted \\ $\quad$ images - $cm^{3}$\\ $\qquad$ mean\\ $\qquad$ median (range)\end{tabular}    & \begin{tabular}[c]{@{}c@{}}$$\\ $$\\ 10.8 $\pm$ 13.9\\ 5.7 (0.0, 83.2)\end{tabular}      & \begin{tabular}[c]{@{}c@{}}$$\\ $$\\ 9.7 $ \pm$ 11.3\\ 6.2 (0.0, 63.5)\end{tabular}      \\ \hline
\begin{tabular}[c]{@{}l@{}}Normalized brain volume - $cm^{3}$\\ $\qquad$ mean\\ $\qquad$ median (range)\end{tabular}                                          & \begin{tabular}[c]{@{}c@{}}$$\\ 1500.9 $\pm$ 84.1\\ 1498.8 (1271.7, 1736.5)\end{tabular} & \begin{tabular}[c]{@{}c@{}}$$\\ 1499.2 $\pm$ 87.7\\ 1503.6 (1251.8, 1729.6)\end{tabular} \\ \hline
\end{tabular}
}}

\subfloat[OPERA II]{
  \scalebox{0.65}{
\begin{tabular}{lcc}
\textbf{Characteristic}                                                                                                                                       & \textbf{\begin{tabular}[c]{@{}c@{}}OCR \\ (N = 417)\end{tabular}}                   & \textbf{\begin{tabular}[c]{@{}c@{}}IFN \\ (N = 418)\end{tabular}}                 \\ \hline
\begin{tabular}[c]{@{}l@{}}Age - years\\ $\qquad$ mean\\ $\qquad$ median (range)\end{tabular}                                                                 & \begin{tabular}[c]{@{}c@{}}$$\\ 37.2 $\pm$ 9.1\\ 37.0 (18.0, 55.0)\end{tabular}          & \begin{tabular}[c]{@{}c@{}}$$\\ 37.4 $\pm$ 9.0\\ 38.0 (18.0, 55.0)\end{tabular}          \\ \hline
Male sex - no. (\%)                                                                                                                                           & 146 (35.0)                                                                               & 138 (33.0)                                                                               \\ \hline
\begin{tabular}[c]{@{}l@{}}Geographical region - no. (\%)\\ $\qquad$ United States\\ $\qquad$ Rest of the world\end{tabular}                                  & \begin{tabular}[c]{@{}c@{}}$$\\ 112 (26.9)\\ 305 (73.1)\end{tabular}                     & \begin{tabular}[c]{@{}c@{}}$$\\ 114 (27.3)\\ 304 (72.7)\end{tabular}                     \\ \hline
\begin{tabular}[c]{@{}l@{}}Time since onset of MS symptoms - years\\ $\qquad$ mean\\ $\qquad$ median (range)\end{tabular}                                     & \begin{tabular}[c]{@{}c@{}}$$\\ 6.7 $\pm$ 6.1\\ 5.2 (0.2, 33.9)\end{tabular}             & \begin{tabular}[c]{@{}c@{}}$$\\ 6.7 $\pm$ 6.1\\ 5.1 (0.2, 31.7)\end{tabular}             \\ \hline
\begin{tabular}[c]{@{}l@{}}Time since diagnosis of RRMS - years\\ $\qquad$ mean\\ $\qquad$ median (range)\end{tabular}                                         & \begin{tabular}[c]{@{}c@{}}$$\\ 4.1 $\pm$ 5.0\\ 2.1 (0.1, 26.9)\end{tabular}             & \begin{tabular}[c]{@{}c@{}}$$\\ 4.1 $\pm$ 5.1\\ 1.8 (0.1, 28.5)\end{tabular}             \\ \hline
\begin{tabular}[c]{@{}l@{}}No. of relapses in previous 12 months\\ $\qquad$ mean\\ $\qquad$ median (range)\end{tabular}                                       & \begin{tabular}[c]{@{}c@{}}$$\\ 1.3 $\pm$ 0.7\\ 1.0 (0.0, 5.0)\end{tabular}              & \begin{tabular}[c]{@{}c@{}}$$\\ 1.3 $\pm$ 0.7\\ 1.0 (0.0, 6.0)\end{tabular}              \\ \hline
\begin{tabular}[c]{@{}l@{}}No previous use of disease-modifying\\ $\quad$ therapy no. /total no. (\%)\end{tabular}                                            & \begin{tabular}[c]{@{}c@{}}$$\\ 303/417 (72.7)\end{tabular}                              & \begin{tabular}[c]{@{}c@{}}$$\\ 313/418 (74.9)\end{tabular}                              \\ \hline
\begin{tabular}[c]{@{}l@{}}EDSS Score\\ $\qquad$ mean\\ $\qquad$ median (range)\end{tabular}                                                                  & \begin{tabular}[c]{@{}c@{}}$$\\ 2.7 $\pm$ 1.3\\ 2.5 (0.0, 6.0)\end{tabular}              & \begin{tabular}[c]{@{}c@{}}$$\\ 2.8 $\pm$ 1.4\\ 2.5 (0.0, 6.0)\end{tabular}              \\ \hline
\begin{tabular}[c]{@{}l@{}}Gadolinium-enhancing lesions on $T_{1}$- \\ $\quad$ weighted images - no./total no. (\%)\\ $\qquad$ yes\\ $\qquad$ no\end{tabular} & \begin{tabular}[c]{@{}c@{}}$$\\ $$\\ 161/413 (39.0)\\ 252/413 (61.0)\end{tabular}        & \begin{tabular}[c]{@{}c@{}}$$\\ $$\\ 172/415 (41.4) \\ 243/415 (58.6)\end{tabular}       \\ \hline
\begin{tabular}[c]{@{}l@{}}Number of lesions on $T_{2}$-weighted images\\ $\qquad$ mean\\ $\qquad$ median (range)\end{tabular}                                & \begin{tabular}[c]{@{}c@{}}$$\\ 49.3 $\pm$ 38.6\\ 39.0 (1.0, 233.0)\end{tabular}         & \begin{tabular}[c]{@{}c@{}}$$\\ 51.0 $\pm$ 35.7\\ 45.0 (0, 218.0)\end{tabular}           \\ \hline
\begin{tabular}[c]{@{}l@{}}Total volume of lesions on $T_{2}$-weighted \\ $\quad$ images - $cm^{3}$\\ $\qquad$ mean\\ $\qquad$ median (range)\end{tabular}    & \begin{tabular}[c]{@{}c@{}}$$\\ $$\\ 10.7 $\pm$ 14.3\\ 5.3 (0.0, 96.0)\end{tabular}      & \begin{tabular}[c]{@{}c@{}}$$\\ $$\\ 10.6$ \pm$ 12.3\\ 6.1 (0.0, 76.1)\end{tabular}      \\ \hline
\begin{tabular}[c]{@{}l@{}}Normalized brain volume - $cm^{3}$\\ $\qquad$ mean\\ $\qquad$ median (range)\end{tabular}                                          & \begin{tabular}[c]{@{}c@{}}$$\\ 1503.9 $\pm$ 92.6\\ 1510.5 (1202.7, 1761.3)\end{tabular} & \begin{tabular}[c]{@{}c@{}}$$\\ 1501.1 $\pm$ 91.0\\ 1506.5 (1245.9, 1751.9)\end{tabular} \\ \hline
\end{tabular}
}}
\caption[OPERA I and OPERA II - Baseline demographic and disease characteristics]{OPERA I and OPERA II - Baseline demographic and disease characteristics (plus-minus values are means $\pm$ SD)}
\label{BaselineOPERA}
\end{figure}
\newpage

\subsection{Time-to-first-event analysis}
\begin{figure}[H] 
\centering
  \subfloat[OPERA I]{
  \scalebox{0.75}{ 
\includegraphics{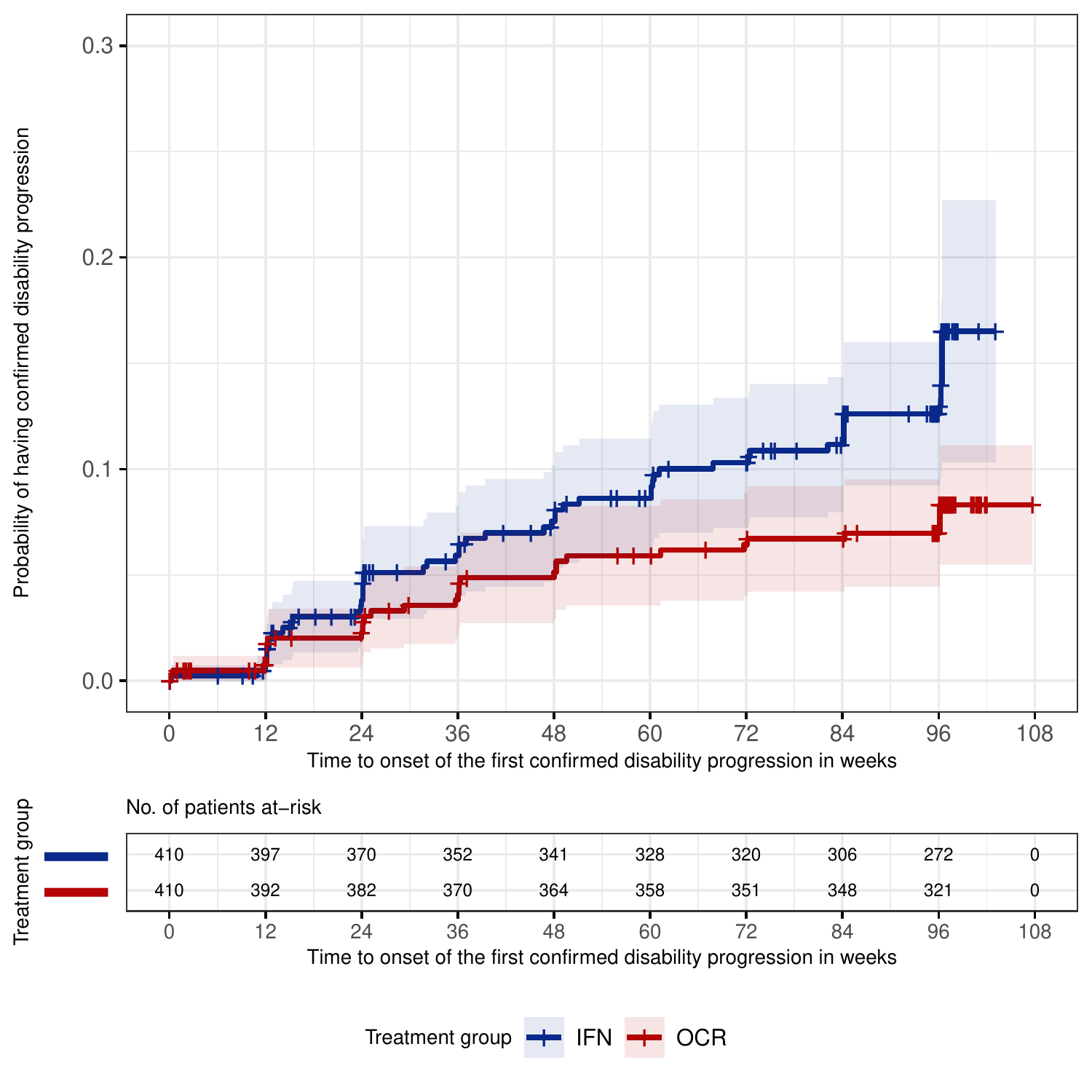}
}}

\subfloat[OPERA II]{
  \scalebox{0.75}{
\includegraphics{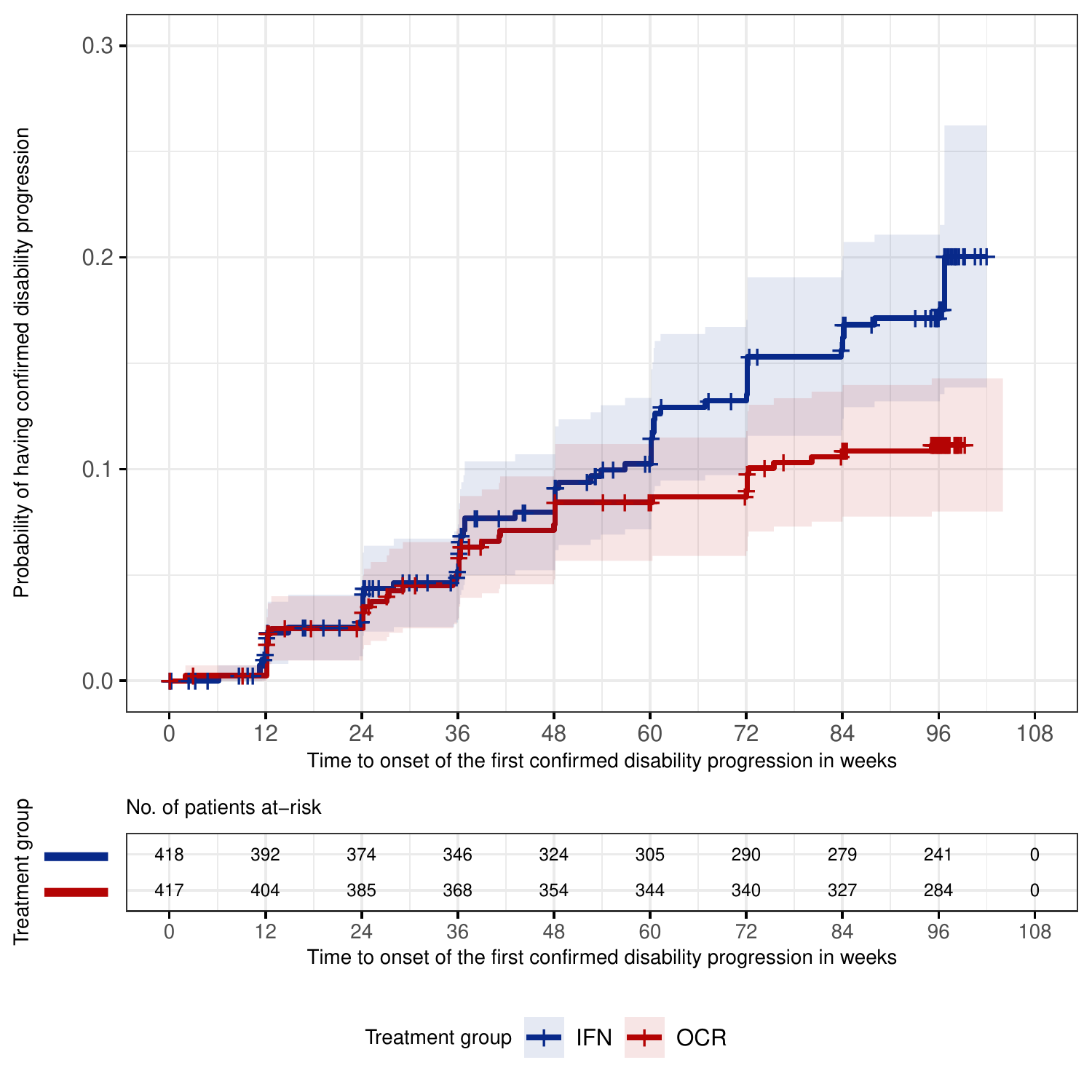}
}}
\caption[OPERA I and OPERA II - One minus Kaplan Meier plot of time-to-onset-of-first-CDP12]{OPERA I and OPERA II - One minus Kaplan Meier plot of time-to-onset-of-first-CDP12 during double-blind treatment period and $95 \%$ CIs (+ indicates censoring)}
\label{KMOPERAs}
\end{figure}

A key secondary endpoint in clinical RRMS trials is the time to the onset of the first $12$-week CDP, where a $12$-week CDP is defined according to the standard definition, as described in Chapter $2$. This endpoint is analysed with the use of a two-sided log-rank test and a Cox proportional hazards model, with stratification according to geographical region (USA versus ROW) and baseline ESSS score ($< 4.0$ versus $\geq 4.0$). 
\newline  \newline
In OPERA I, one patient who was randomly assigned to the IFN group was excluded from the analysis because of a missing EDSS value at baseline. The time-to-first-event analysis is therefore based on $820$ patients, with $410$ patients in the the OCR group and $410$ patients in the PLA group. A total of $31$ of $410$ patients ($7.6 \%$) in the OCR group experienced a CDP12 event during the double-blind treatment period, as compared with $50$ of $410$ patients ($12.2 \%$) in the IFN group. In Figure $\ref{KMOPERAs}$ (a), the 1-KM curves for time to the onset of the first CDP12 are depicted. The 1-KM curves reveal a separation from week $12$ which is also confirmed by the log-rank test (p-value=$0.0139$). Among patients with RRMS, a $43 \%$ reduction in the hazard of a $12$-week CDP on OCR can be seen in OPERA I (HR 0.574, $95\%$ CI: [0.366, 0.899], p-value=$0.0153$). There is no evidence that the proportional hazards assumption for the treatment group is violated (p-value=$0.484$). 

\begin{table}[H]
\centering
\scalebox{0.75}{
\begin{tabular}{lcc}
                                                                                                                              & \textbf{\begin{tabular}[c]{@{}c@{}}OCR\\ (N=410)\end{tabular}}                             & \textbf{\begin{tabular}[c]{@{}c@{}}IFN\\ (N=411)\end{tabular}}                               \\ \hline
Patients included in analysis                                                                                                 & 410 (100.0 $\%$)                                                                           & 410 (100.0 $\%$)                                                                             \\
Patients with event ($\%$)                                                                                                    & 31 (7.6 $\%$)                                                                              & 50 (12.2 $\%$)                                                                               \\
                                                                                                                              &                                                                                            &                                                                                              \\
Time-to-first-CDP12 in weeks                                                                                                  & 0* to 108*                                                                                 & 0* to 103*                                                                                   \\
                                                                                                                              &                                                                                            &                                                                                              \\
\begin{tabular}[c]{@{}l@{}}Stratified** analysis\\ $\qquad$ p-value (log-rank)\\ $$\\ $\qquad$ HR (95\% CI)\end{tabular}      & \multicolumn{2}{c}{\begin{tabular}[c]{@{}c@{}}$$\\ 0.0139\\ $$\\ 0.57 {[}0.37, 0.90{]}\end{tabular}}                                                                                      \\
                                                                                                                              &                                                                                            &                                                                                              \\
\begin{tabular}[c]{@{}l@{}}Time point analysis: 1-KM estimate (95\% CI)\\ $\qquad$ 48 weeks\\ $\qquad$ 96 weeks\end{tabular} & \begin{tabular}[c]{@{}c@{}}$$\\ 5.13 {[}2.94, 7.32{]}\\ 6.98 {[}4.44, 9.52{]}\end{tabular} & \begin{tabular}[c]{@{}c@{}}$$\\ 7.53 {[}4.90, 10.17{]}\\ 12.61 {[}9.23,15.99{]}\end{tabular} \\ \hline
\end{tabular}
}
\caption[OPERA I - Time-to-onset-of-first-CDP12 analysis]{OPERA I - Time-to-onset-of-first-CDP12 analysis (* = censored observation, ** = stratified by geographical region (USA versus ROW) and baseline EDSS score ($< 4.0$ versus $\geq 4.0$))}
\label{ResultsCDP12opera1}
\end{table}

In OPERA II, the time-to-first-event analysis involves $835$ patients, with $417$ patients in the OCR group and $418$ patients in the PLA group. The percentage of patients with CDP12 event is $10.6 \%$ with the OCR group versus $15.1 \%$ with PLA. The 1-KM curves for the time to the onset of the first $12$-week CDP are shown in Figure $\ref{KMOPERAs}$ (b). The treatment effect estimate obtained from fitting a Cox model to the OPERA II data is equal to HR=$0.626$ ($95\%$ CI: [0.425, 0.923], p-value=$0.0182$). Consequently, treatment with OCR reduces the hazard for a CDP12 event by approximately $37 \%$. 

\begin{table}[H]
\centering
\scalebox{0.75}{
\begin{tabular}{lcc}
                                                                                                                              & \textbf{\begin{tabular}[c]{@{}c@{}}OCR\\ (N=417)\end{tabular}}                              & \textbf{\begin{tabular}[c]{@{}c@{}}IFN\\ (N=418)\end{tabular}}                                \\ \hline
Patients included in analysis                                                                                                 & 417 (100.0 $\%$)                                                                            & 418 (100.0 $\%$)                                                                              \\
Patients with event ($\%$)                                                                                                    & 44 (10.6 $\%$)                                                                              & 63 (15.1 $\%$)                                                                                \\
                                                                                                                              &                                                                                             &                                                                                               \\
Time-to-first-CDP12 in weeks                                                                                                  & 0* to 104                                                                                  & 0* to 102*                                                                                    \\
                                                                                                                              &                                                                                             &                                                                                               \\
\begin{tabular}[c]{@{}l@{}}Stratified** analysis\\ $\qquad$ p-value (log-rank)\\ \\ $\qquad$ HR (95\% CI)\end{tabular}      & \multicolumn{2}{c}{\begin{tabular}[c]{@{}c@{}}$$\\ 0.0169\\ $$\\ 0.63 {[}0.42, 0.92{]}\end{tabular}}                                                                                        \\
                                                                                                                              &                                                                                             &                                                                                               \\
\begin{tabular}[c]{@{}l@{}}Time point analysis: 1-KM estimate (95\% CI)\\ $\qquad$ 48 weeks\\ $\qquad$ 96 weeks\end{tabular} & \begin{tabular}[c]{@{}c@{}}$$\\ 7.38 {[}4.79, 9.96{]}\\ 11.14 {[}8.00, 14.29{]}\end{tabular} & \begin{tabular}[c]{@{}c@{}}$$\\ 7.96 {[}5.22, 10.70{]}\\ 17.13 {[}13.2, 21.07{]}\end{tabular} \\ \hline
\end{tabular}
}
\caption[OPERA II - Time-to-onset-of-first-CDP12 analysis]{OPERA II - Time-to-onset-of-first-CDP12 analysis (* = censored observation, ** = stratified by geographical region (USA versus ROW) and baseline EDSS score ($< 4.0$ versus $\geq 4.0$))}
\label{ResultsCDP12opera2}
\end{table}

\newpage

\newpage

\subsection{Recurrent event analysis}
The summary statistics on the number of CDP12 events in Table $\ref{opera1DistNo}$ and Table $\ref{opera2DistNo}$ reveal that repeated CDP12 events are very rare in RRMS patients, due to a shorter study duration in the OPERA trials. 
\newline 
By end of the OPERA I trial, $50$ of $410$ OCR patients ($12.2 \%$) and $31$ of $410$ IFN patients ($7.6 \%$) had experienced at least one CDP12 event. Of the $31$ IFN patients with at least one CDP12 event, $1$ patient experienced $3$ CDP12 events (cf. Table $\ref{opera1DistNo}$). Of the $50$ OCR patients with at least one CDP12 event, $6$ patients had $2$ CDP12 events. No patient was observed to experience more than $3$ CDP12 events. As a result, $8$ events are not used in the time-to-first-event analysis. Similar findings can be found from the alternative endpoint definitions. 

\begin{table}[h]
\centering
\scalebox{0.80}{ 
\begin{tabular}{llccccc}
\hline
\multicolumn{1}{|c|}{\multirow{2}{*}{\textbf{Definition}}}                                                                                 & \multicolumn{1}{c|}{\multirow{2}{*}{\textbf{\begin{tabular}[c]{@{}c@{}}Treatment \\ group\end{tabular}}}} & \multicolumn{4}{c|}{\textbf{\begin{tabular}[c]{@{}c@{}}No. of CDP12 events\end{tabular}}}                               & \multicolumn{1}{c|}{\multirow{2}{*}{\textbf{\begin{tabular}[c]{@{}c@{}}'Unused'\\ events\end{tabular}}}} \\ \cline{3-6}
\multicolumn{1}{|c|}{}                                                                                                                     & \multicolumn{1}{c|}{}                                                                                     & \multicolumn{1}{c|}{\textbf{0}} & \multicolumn{1}{c|}{\textbf{1}} & \multicolumn{1}{c|}{\textbf{2}} & \multicolumn{1}{c|}{\textbf{3}} & \multicolumn{1}{c|}{}                                                                                    \\ \hline
                                                                                                                                           &                                                                                                           & \multicolumn{1}{l}{}            & \multicolumn{1}{l}{}            & \multicolumn{1}{l}{}            & \multicolumn{1}{l}{}            & \multicolumn{1}{l}{}                                                                                     \\ \hline
\multicolumn{1}{|l|}{\multirow{2}{*}{\textbf{\begin{tabular}[c]{@{}l@{}}Time-to-onset-of-CDP \\ and fixed reference\end{tabular}}}}        & \multicolumn{1}{l|}{IFN}                                                                              & \multicolumn{1}{c|}{360}        & \multicolumn{1}{c|}{44}         & \multicolumn{1}{c|}{6}          & \multicolumn{1}{c|}{0}          & \multicolumn{1}{c|}{\multirow{2}{*}{8}}                                                                  \\ \cline{2-6}
\multicolumn{1}{|l|}{}                                                                                                                     & \multicolumn{1}{l|}{OCR}                                                                            & \multicolumn{1}{c|}{379}        & \multicolumn{1}{c|}{30}         & \multicolumn{1}{c|}{0}          & \multicolumn{1}{c|}{1}          & \multicolumn{1}{c|}{}                                                                                    \\ \hline
                                                                                                                                           &                                                                                                           &                                 &                                 &                                 &                                 & \multicolumn{1}{l}{}                                                                                     \\ \hline
\multicolumn{1}{|l|}{\multirow{2}{*}{\textbf{\begin{tabular}[c]{@{}l@{}}Time-to-onset-of-CDP \\ and roving reference*\end{tabular}}}}       & \multicolumn{1}{l|}{IFN}                                                                              & \multicolumn{1}{c|}{358}        & \multicolumn{1}{c|}{46}         & \multicolumn{1}{c|}{6}          & \multicolumn{1}{c|}{0}          & \multicolumn{1}{c|}{\multirow{2}{*}{8}}                                                                  \\ \cline{2-6}
\multicolumn{1}{|l|}{}                                                                                                                     & \multicolumn{1}{l|}{OCR}                                                                            & \multicolumn{1}{c|}{375}        & \multicolumn{1}{c|}{34}         & \multicolumn{1}{c|}{0}          & \multicolumn{1}{c|}{1}          & \multicolumn{1}{c|}{}                                                                                    \\ \hline
                                                                                                                                           &                                                                                                           & \multicolumn{1}{l}{}            & \multicolumn{1}{l}{}            & \multicolumn{1}{l}{}            & \multicolumn{1}{l}{}            & \multicolumn{1}{l}{}                                                                                     \\ \hline
\multicolumn{1}{|l|}{\multirow{2}{*}{\textbf{\begin{tabular}[c]{@{}l@{}}Time-to-confirmation-of-CDP\\ and fixed reference\end{tabular}}}}  & \multicolumn{1}{l|}{IFN}                                                                              & \multicolumn{1}{c|}{368}        & \multicolumn{1}{c|}{39}         & \multicolumn{1}{c|}{3}          & \multicolumn{1}{c|}{0}          & \multicolumn{1}{c|}{\multirow{2}{*}{3}}                                                                  \\ \cline{2-6}
\multicolumn{1}{|l|}{}                                                                                                                     & \multicolumn{1}{l|}{OCR}                                                                            & \multicolumn{1}{c|}{384}        & \multicolumn{1}{c|}{26}         & \multicolumn{1}{c|}{0}          & \multicolumn{1}{c|}{0}          & \multicolumn{1}{c|}{}                                                                                    \\ \hline
                                                                                                                                           &                                                                                                           & \multicolumn{1}{l}{}            & \multicolumn{1}{l}{}            & \multicolumn{1}{l}{}            & \multicolumn{1}{l}{}            & \multicolumn{1}{l}{}                                                                                     \\ \hline
\multicolumn{1}{|l|}{\multirow{2}{*}{\textbf{\begin{tabular}[c]{@{}l@{}}Time-to-confirmation-of-CDP\\ and roving reference*\end{tabular}}}} & \multicolumn{1}{l|}{IFN}                                                                              & \multicolumn{1}{c|}{367}        & \multicolumn{1}{c|}{40}         & \multicolumn{1}{c|}{3}          & \multicolumn{1}{c|}{0}          & \multicolumn{1}{c|}{\multirow{2}{*}{3}}                                                                  \\ \cline{2-6}
\multicolumn{1}{|l|}{}                                                                                                                     & \multicolumn{1}{l|}{OCR}                                                                            & \multicolumn{1}{c|}{381}        & \multicolumn{1}{c|}{29}         & \multicolumn{1}{c|}{0}          & \multicolumn{1}{c|}{0}          & \multicolumn{1}{c|}{}                                                                                    \\ \hline
\end{tabular}}
\caption[OPERA I - Distribution of the numbers of CDP12 events by treatment group]{OPERA I - Distribution of the numbers of CDP12 events by treatment group, patients with missing baseline EDSS were excluded from the analyses, N=410 (IFN) and N=410 (OCR), column 'unused events' refers to events not used in time-to-first-event analyses only (* = 24-week confirmation period of new reference EDSS score)}
\label{opera1DistNo}
\end{table}

By end of the OPERA II trial, $63$ of $418$ OCR patients ($15.1 \%$) and $44$ of $417$ IFN patients ($10.6 \%$) had experienced at least one CDP12 event (cf. Table $\ref{opera2DistNo}$). Only $2$ patients in the OCR group experienced $2$ CDP12 events. In the IFN group, $4$ patients had $2$ CDP12 events and $1$ patient experienced $3$ CDP12 events. Thus, only $8$ CDP12 events are ignored in a time-to-first-event analysis compared with recurrent event analyses. Under the different endpoint definitions, similar findings can be found. 

\begin{table}[h]
\centering
\scalebox{0.80}{ 
\begin{tabular}{llccccc}
\hline
\multicolumn{1}{|c|}{\multirow{2}{*}{\textbf{Definition}}}                                                                                 & \multicolumn{1}{c|}{\multirow{2}{*}{\textbf{\begin{tabular}[c]{@{}c@{}}Treatment \\ group\end{tabular}}}} & \multicolumn{4}{c|}{\textbf{\begin{tabular}[c]{@{}c@{}}No. of CDP12 events\end{tabular}}}                               & \multicolumn{1}{c|}{\multirow{2}{*}{\textbf{\begin{tabular}[c]{@{}c@{}}'Unused'\\ events\end{tabular}}}} \\ \cline{3-6}
\multicolumn{1}{|c|}{}                                                                                                                     & \multicolumn{1}{c|}{}                                                                                     & \multicolumn{1}{c|}{\textbf{0}} & \multicolumn{1}{c|}{\textbf{1}} & \multicolumn{1}{c|}{\textbf{2}} & \multicolumn{1}{c|}{\textbf{3}} & \multicolumn{1}{c|}{}                                                                                    \\ \hline
                                                                                                                                           &                                                                                                           & \multicolumn{1}{l}{}            & \multicolumn{1}{l}{}            & \multicolumn{1}{l}{}            & \multicolumn{1}{l}{}            & \multicolumn{1}{l}{}                                                                                     \\ \hline
\multicolumn{1}{|l|}{\multirow{2}{*}{\textbf{\begin{tabular}[c]{@{}l@{}}Time-to-onset-of-CDP \\ and fixed reference\end{tabular}}}}        & \multicolumn{1}{l|}{IFN}                                                                              & \multicolumn{1}{c|}{355}        & \multicolumn{1}{c|}{58}         & \multicolumn{1}{c|}{4}          & \multicolumn{1}{c|}{1}          & \multicolumn{1}{c|}{\multirow{2}{*}{8}}                                                                  \\ \cline{2-6}
\multicolumn{1}{|l|}{}                                                                                                                     & \multicolumn{1}{l|}{OCR}                                                                            & \multicolumn{1}{c|}{373}        & \multicolumn{1}{c|}{42}         & \multicolumn{1}{c|}{2}          & \multicolumn{1}{c|}{0}          & \multicolumn{1}{c|}{}                                                                                    \\ \hline
                                                                                                                                           &                                                                                                           &                                 &                                 &                                 &                                 & \multicolumn{1}{l}{}                                                                                     \\ \hline
\multicolumn{1}{|l|}{\multirow{2}{*}{\textbf{\begin{tabular}[c]{@{}l@{}}Time-to-onset-of-CDP \\ and roving reference*\end{tabular}}}}       & \multicolumn{1}{l|}{IFN}                                                                              & \multicolumn{1}{c|}{351}        & \multicolumn{1}{c|}{61}         & \multicolumn{1}{c|}{5}          & \multicolumn{1}{c|}{1}          & \multicolumn{1}{c|}{\multirow{2}{*}{9}}                                                                  \\ \cline{2-6}
\multicolumn{1}{|l|}{}                                                                                                                     & \multicolumn{1}{l|}{OCR}                                                                            & \multicolumn{1}{c|}{363}        & \multicolumn{1}{c|}{52}         & \multicolumn{1}{c|}{2}          & \multicolumn{1}{c|}{0}          & \multicolumn{1}{c|}{}                                                                                    \\ \hline
                                                                                                                                           &                                                                                                           & \multicolumn{1}{l}{}            & \multicolumn{1}{l}{}            & \multicolumn{1}{l}{}            & \multicolumn{1}{l}{}           & \multicolumn{1}{l}{}                                                                                    \\ \hline
\multicolumn{1}{|l|}{\multirow{2}{*}{\textbf{\begin{tabular}[c]{@{}l@{}}Time-to-confirmation-of-CDP\\ and fixed reference\end{tabular}}}}  & \multicolumn{1}{l|}{IFN}                                                                              & \multicolumn{1}{c|}{360}        & \multicolumn{1}{c|}{54}         & \multicolumn{1}{c|}{3}          & \multicolumn{1}{c|}{1}          & \multicolumn{1}{c|}{\multirow{2}{*}{7}}                                                                  \\ \cline{2-6}
\multicolumn{1}{|l|}{}                                                                                                                     & \multicolumn{1}{l|}{OCR}                                                                            & \multicolumn{1}{c|}{376}        & \multicolumn{1}{c|}{39}         & \multicolumn{1}{c|}{2}          & \multicolumn{1}{c|}{0}          & \multicolumn{1}{c|}{}                                                                                    \\ \hline
                                                                                                                                           &                                                                                                           & \multicolumn{1}{l}{}            & \multicolumn{1}{l}{}            & \multicolumn{1}{l}{}            & \multicolumn{1}{l}{}            & \multicolumn{1}{l}{}                                                                                     \\ \hline
\multicolumn{1}{|l|}{\multirow{2}{*}{\textbf{\begin{tabular}[c]{@{}l@{}}Time-to-confirmation-of-CDP\\ and roving reference*\end{tabular}}}} & \multicolumn{1}{l|}{IFN}                                                                              & \multicolumn{1}{c|}{359}        & \multicolumn{1}{c|}{54}         & \multicolumn{1}{c|}{4}          & \multicolumn{1}{c|}{1}          & \multicolumn{1}{c|}{\multirow{2}{*}{8}}                                                                  \\ \cline{2-6}
\multicolumn{1}{|l|}{}                                                                                                                     & \multicolumn{1}{l|}{OCR}                                                                            & \multicolumn{1}{c|}{368}        & \multicolumn{1}{c|}{47}         & \multicolumn{1}{c|}{2}          & \multicolumn{1}{c|}{0}          & \multicolumn{1}{c|}{}                                                                                    \\ \hline
\end{tabular}}
\caption[OPERA II - Distribution of the numbers of CDP12 events by treatment group]{OPERA II - Distribution of the numbers of CDP12 events by treatment group, patients with missing baseline EDSS were excluded from the analyses, N=418 (IFN) and N=417 (OCR), column 'unused events' refers to events not used in time-to-first-event analyses only, (* = 24-week confirmation period of new reference EDSS score)}
\label{opera2DistNo}
\end{table}

\subsubsection{Estimation of treatment effect}
\begin{figure}[H] 
\centering
  \subfloat[OPERA I]{
\scalebox{0.45}[0.45]{ 
\includegraphics{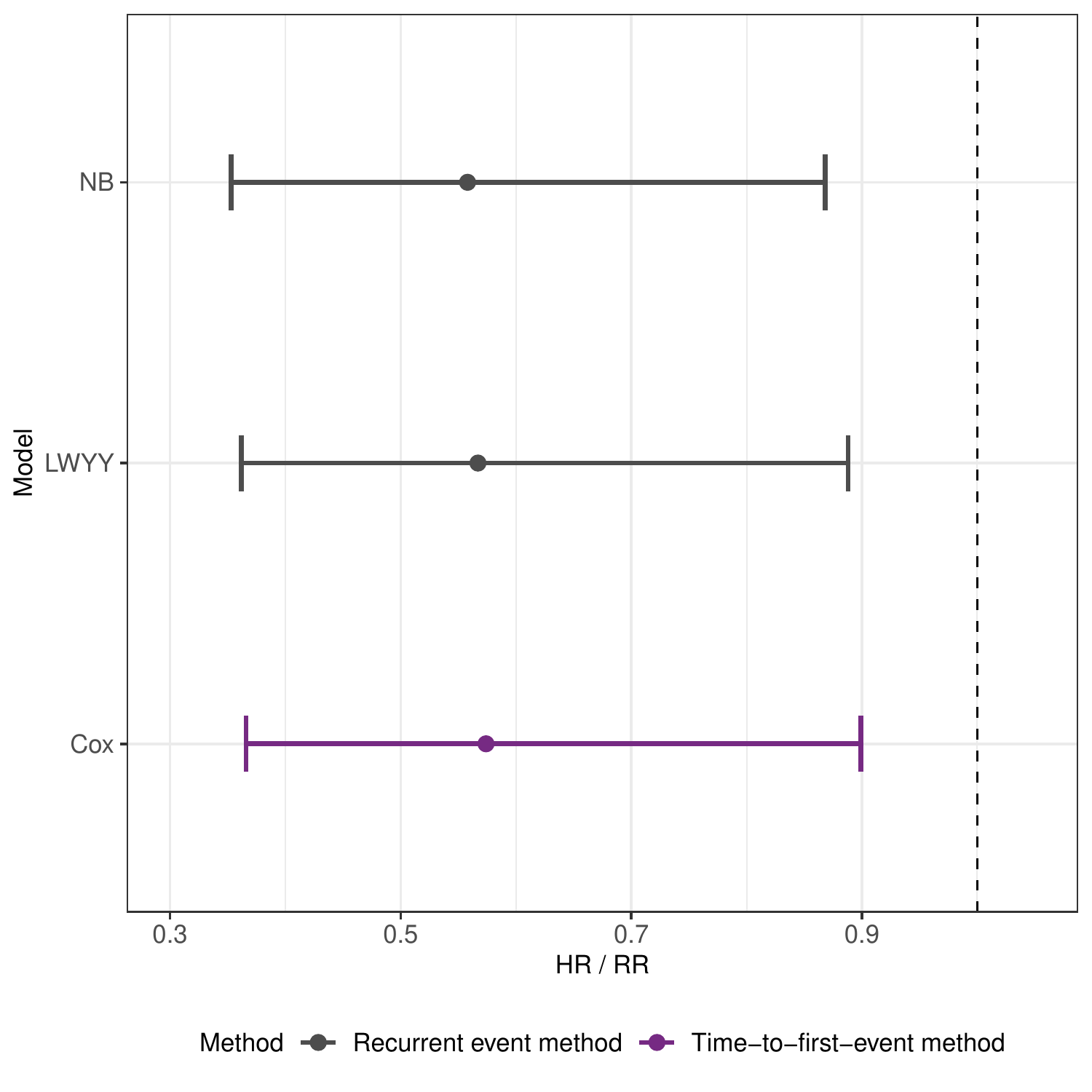}
}}
\subfloat[OPERA II]{
\scalebox{0.45}[0.45]{ 
\includegraphics{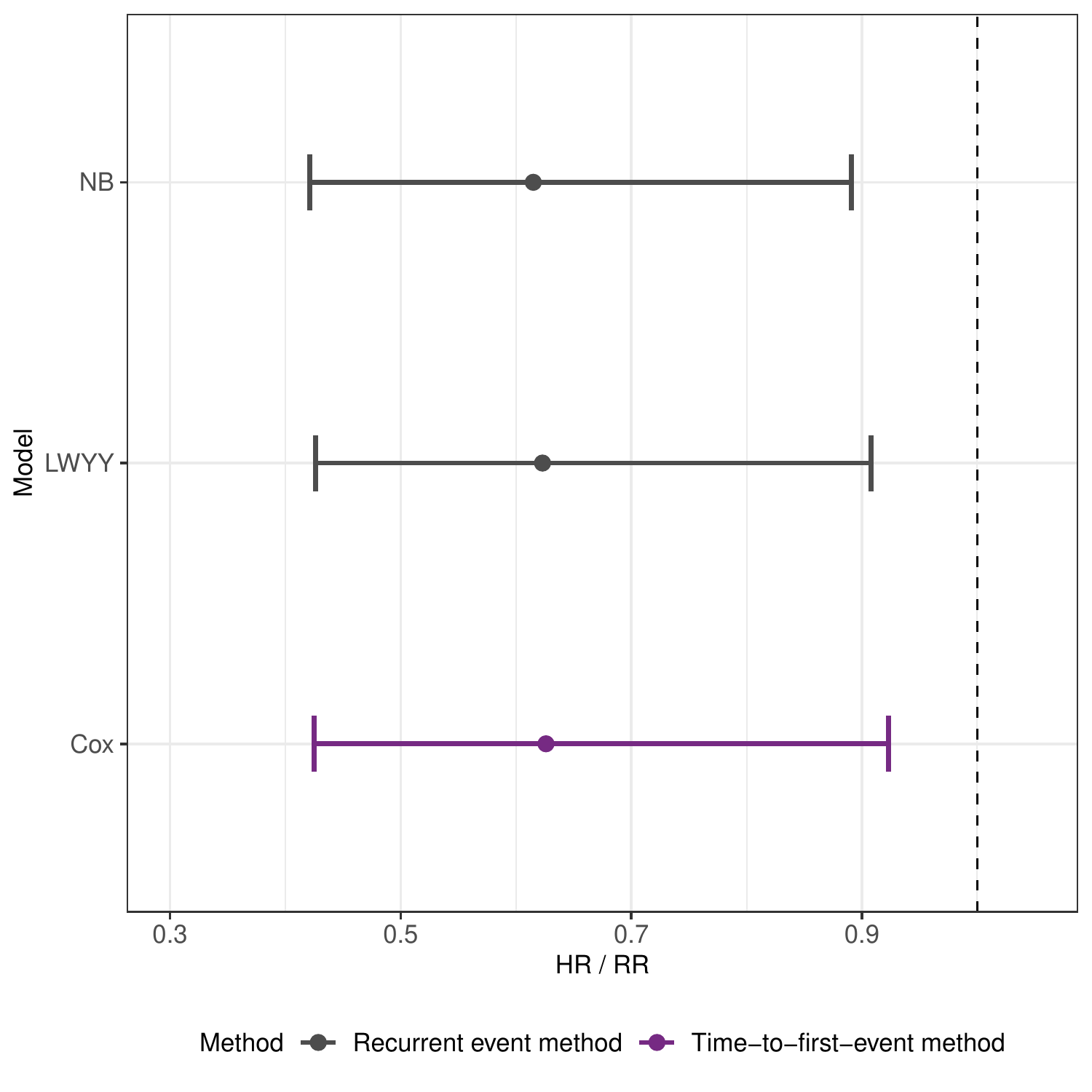}
}}
\caption[OPERA I and OPERA II - Forest plot of treatment effect estimates obtained from time-to-first-event and marginal recurrent event analyses]{OPERA I and OPERA II - Forest plot of treatment effect estimates obtained from time-to-first-event and marginal recurrent event analyses with $95 \%$ CIs}
\label{forestplotOPERA}
\end{figure}

A reanalysis of the OPERA I and OPERA II trials has also been undertaken to estimate the overall treatment effect of OCR on $12$-week CDP using the NB and LWYY models, with results represented in Table $\ref{TreatmentEffectOverviewOPERA1}$ and Table $\ref{TreatmentEffectOverviewOPERA2}$. Due to the small number of repeated CDP12 events in the RRMS population, recurrent event methods do no show a clear benefit over the conventional time-to-first-event approach in both OPERA trials. In particular, there is close agreement between the estimates from the Cox and LWYY models. Clinical interpretation of the treatment effect is therefore comparable across the time-to-first-event and recurrent event analyses. As illustrated in Figure $\ref{forestplotOPERA}$, the Cox and LWYY analyses yield a similar level of statistical precision. 
\begin{table}[h]
\centering
\begin{tabular}{llccc}
                                                                                              & \textbf{Model} & \textbf{Treatment effect} & \textbf{$\boldsymbol{95 \%}$ CI}                                     & \textbf{p-value}                                         \\ \hline
\textbf{\begin{tabular}[c]{@{}l@{}}Time-to-first-\\ event\\ analysis\end{tabular}}            & Cox model*     & HR 0.574                  & \begin{tabular}[c]{@{}c@{}}$$\\ {[}0.366, 0.899{]}\\ $$\end{tabular} & \begin{tabular}[c]{@{}c@{}}$$\\ 0.0153\\ $$\end{tabular} \\ \hline
\multirow{2}{*}{\textbf{\begin{tabular}[c]{@{}l@{}}Recurrent event \\ analyses\end{tabular}}} & NB model**     & RR 0.558                        & {[}0.353, 0.868{]}                                                      & 0.0106                                                      \\
                                                                                              & LWYY model*    & RR 0.567                  & {[}0.362, 0.888{]}                                                   & 0.0133                                                  
\end{tabular}
\caption[OPERA I - Comparisons of treatment effect estimates obtained from time-to-first-event and marginal recurrent event analyses]{OPERA I - Comparisons of treatment effect estimates obtained from time-to-first-event and marginal recurrent event analyses, log-transformed exposure time is included as an offset variable in NB model (* = stratified by and ** = adjusted for EDSS category ($< 4.0$ versus $\geq 4.0$) and geographical region (USA versus ROW))}
\label{TreatmentEffectOverviewOPERA1}
\end{table}

\begin{table}[h]
\centering
\begin{tabular}{llccc}
                                                                                              & \textbf{Model} & \textbf{Treatment effect} & \textbf{$\boldsymbol{95 \%}$ CI}                                     & \textbf{p-value}                                         \\ \hline
\textbf{\begin{tabular}[c]{@{}l@{}}Time-to-first-\\ event\\ analysis\end{tabular}}            & Cox model*     & HR 0.626                  & \begin{tabular}[c]{@{}c@{}}$$\\ {[}0.425, 0.923{]}\\ $$\end{tabular} & \begin{tabular}[c]{@{}c@{}}$$\\ 0.0182\\ $$\end{tabular} \\ \hline
\multirow{2}{*}{\textbf{\begin{tabular}[c]{@{}l@{}}Recurrent event \\ analyses\end{tabular}}} & NB model**     & RR 0.615                  & {[}0.421, 0.891{]}                                                   & 0.0109                                                   \\
                                                                                              & LWYY model*    & RR 0.609                  & {[}0.419, 0.886{]}                                                   & 0.00944                                                 
\end{tabular}
\caption[OPERA II - Comparisons of treatment effect estimates obtained from time-to-first-event and marginal recurrent event analyses]{OPERA II- Comparisons of treatment effect estimates obtained from time-to-first-event and marginal recurrent event analyses, log-transformed exposure time is included as an offset variable in NB model (* = stratified by and ** = adjusted for EDSS category ($< 4.0$ versus $\geq 4.0$) and geographical region (USA versus ROW))}
\label{TreatmentEffectOverviewOPERA2}
\end{table}

\chapter{Simulation setup}
Recurrent event analyses of the randomized ORATORIO trial have shown slightly larger treatment effect sizes, smaller widths of the $95 \%$ CIs for the treatment effect and smaller p-values than for the primary time-to-first-event analysis. This suggests a benefit of recurrent event approaches in terms of statistical precision and power. In order to compare recurrent event methods (including NB, LWYY and AG models) with time-to-first-event methods (Cox model) with respect to treatment evaluation in randomized PPMS trials, two simulation studies are conducted. The first simulation scheme is a general setup for recurrent events, whereas the second MS-specific setup aims at simulating repeated CDP data that closely mimick real clinical MS trial settings. 
\newline \newline 
\textbf{Simulation schemes:}
\begin{enumerate}
\item General recurrent event setup: simulation of recurrent event times according to a mixed non-homogeneous Poisson process (S1).
\item MS-specific setup: simulation of longitudinal EDSS measurements using multistate methodology for panel data and derivation of recurrent CDP events from EDSS data (S2).
\end{enumerate}

This chapter describes two different simulation schemes for generating recurrent CDP data in PPMS. In Section $\ref{SimulationAlgos}$, simulation methods for both studies are described in more detail. Section $\ref{SimulationAdd}
$ considers additional simulation parameters such as recruitment, censoring and generation of covariates and frailty terms. Simulation scenarios, parameter settings and evaluation measures used for both simulation studies are described in Section $\ref{SimulationParameter}$ and Section $\ref{SimulationAnalysis}$. An overview about both simulation algorithms can be finally found in Section $\ref{SimulationOverview}$.

\section{Generation of recurrent event data}
\label{SimulationAlgos}
In general, recurrent event processes can be simulated in several ways. For instance, in cardiovascular diseases, it is realistic to consider that the occurrence of an event may change the instantaneous probability of experiencing a new event. In this case, recurrent events may be generated by a Markov multistate process, where the occurrence of each event increases or decreases the baseline intensity function. Event-dependency may also be modelled by incorporating an internal time-varying covariate (e.g., $N(t-)$) into the simulation model. Simulation of such complex time-to-event data based on multistate models has been extensively studied by \citet{Bluhmki2019, Jenny2018, Beyersmann2012, Allignol2011}. \newline 
Recurrent event analyses of the ORATORIO trial (cf. Section $\ref{IntensitybasedModelsCovariates}$) have not shown a clear indication of event-dependency. Since the disease course of progressive MS forms can strongly vary across patients, heterogeneity between individuals is expected to be present in PPMS populations. Further, the CMF of the expected number of CDP12 events against follow-up time suggests time-varying event rates (cf. Figure $\ref{CMForatorio}$). Under these MS-specific assumptions, the underlying event generation process is likely to be a mixed non-homogeneous Poisson process. The main purpose of the general simulation setup is therefore to provide an algorithm that randomly generates recurrent event processes according to a mixed non-homogeneous Poisson process. 
\subsection{General simulation setup}
\label{GSsimulation}
In the generic simulation setup, recurrent events are generated according to a mixed non-homogeneous Poisson process with 'conditional' intensity function (cf. Section $\ref{SectionNBmodel}$)
\begin{align}
\nonumber \lambda_{i}(t \ | \ U_{i}) &= P(dN_{i}(t) = 1 \ | \ U_{i}, \ \textnormal{past})  \\
\nonumber &= \underbrace{Y_{i}(t)}_{= \mathbbm{1}(C_{i} \geq t)} U_{i} \alpha_{0}(t) \exp(\beta Z_{i}), \ \ \ i=1,2,..., n,  
\end{align}
where $Z_{i}$ defines the treatment arm, $U_{i}$ is an individual-specific random effect (or frailty term), $C_{i}$ is the administrative censoring time and $\alpha_{0}(t)$ is a baseline intensity function. \newline 
As already mentioned, the CMFs in Figure $\ref{CMForatorio}$ suggest a slightly decreasing event intensity over time. Although the event intensity appears to be roughly constant in the beginning of follow-up, it decreases over time. The Weibull distribution with intensity $\eta \nu t^{\nu-1}$ is an appropriate choice for $\alpha_{0}(t)$, as this distribution allows the event intensity to decrease over time ($\nu < 1$). $\nu >0 $ and $\eta >0$ are the shape and scale parameters of the Weibull distribution, respectively. In order to simulate from a realistic baseline intensity function $\alpha_{0}(t)=\eta \nu t^{\nu-1}$, a Weibull regression model is fitted to the ORATORIO placebo data using phreg() in R to get estimates of $\nu$ and $\eta$. Based on this model, $\nu$ is chosen to be $0.9161516$ and $\eta$ is set to $0.0009675564$. Since for $\nu=1$ the Weibull distribution reduces to an exponential distribution, deviations from linearity are only small. 
\newline \newline
\citet{Bender2005} proposed an algorithm to generate non-recurrent event times using the inversion method. This approach can be extended to simulate recurrent event times from a mixed non-homogeneous Poisson process \parencite{JahnEimermacher2015, Penichoux2014}.  \newline 
In order to derive the recursive simulation algorithm, the distribution of the gap time $G_{j}$ conditional on the first $(j-1)$ event times $T_{1}, T_{2}, ..., T_{j-1}$ and the random effect $U$ must be specified. In Chapter $4$, the $j^{th}$ gap time $G_{j}$ has been defined as the duration of time between the $(j-1)^{th}$ and $j^{th}$ event (i.e., $G_{j} := T_{j}-T_{j-1}$). The cumulative distribution function (CDF) of $G_{j} \ | \ T_{1}, ..., T_{j-1}, U$ is given by
\begin{align}
\nonumber F_{G_{j}}(g) &= P(G_{j} \leq g \ | \ T_{j-1} = t_{j-1}, ..., T_{1}=t_{1}, \ U=u) \\
\nonumber &= 1 - P(G_{j} > g \ | \ T_{j-1} = t_{j-1}, ..., T_{1}=t_{1}, \ U=u) \\
\nonumber &= 1 - P(N(t_{j-1} + g) - N(t_{j-1}) = 0 \ |  \ T_{j-1} = t_{j-1}, ..., T_{1}=t_{1}, \ U=u) \\ 
\nonumber &= 1 - P(N(t_{j-1} + g) - N(t_{j-1}) = 0 \ |  \ U=u) \\
\nonumber &= 1- \exp \biggl( - \int_{t_{j-1}}^{t_{j-1}+g} U \alpha_{0}(s) \exp(\beta Z) ds \biggr) = 1- \exp \biggl( - \int_{t_{j-1}}^{t_{j-1}+g} U \eta \nu s^{\nu-1} \exp(\beta Z) ds \biggr), 
\end{align}
where the third equality follows from the fact that, given the random effect $U$, the counting process $\{N(t): 0 \leq t < \infty \}$ is Poisson with mean $U\mu(t)$, with $\mu(t) = \int_{0}^{t} \alpha_{0}(s)\exp(\beta Z)ds$. That is, $N(t) \ | \ U \sim Poisson(U\mu(t))$ is a non-homogeneous Poisson process. The fourth equality follows directly from Eq. $(\ref{PoissonEQ})$. 
\newline \newline 
Using the inversion method, the event times for individual $i$ are then generated according to the following recursive simulation algorithm: 
\begin{enumerate}
\item $T_{i0} := 0.$
\item Generation of uniformly distributed random variable $W_{j} \sim U(0,1)$ for the $j^{th}$ event
\item Generation of gap time $G_{ij} = F_{G_{ij}}^{-1}(W_{j})$ and calculation of time to $j^{th}$ event 
\[ T_{ij} = T_{i(j-1)} + G_{ij} = \biggl( -\dfrac{\log(1-W_{j})}{U_{i} \eta \exp(\beta Z)} \ + T_{i(j-1)}^{\nu} \biggr)^{\dfrac{1}{\nu}}\] 
\item $\begin{cases} \textnormal{Generation of} \ (j+1)^{th} \ \textnormal{event by returning to} \ 2. & , \ \textnormal{if} \ T_{ij} < C_{i} \\ T_{ij} = C_{i} \ \textnormal{and} \ G_{ij}=C_{i}-T_{i(j-1)} & , \  \textnormal{otherwise} \\ \end{cases}$. 
\end{enumerate}

\subsection{MS-specific simulation setup}
\label{MSsetup}
The general simulation setup is mainly characterized by the fact that recurrent events can happen at any continuous time point during the follow-up period. As seen in Section $\ref{GSsimulation}$, survival techniques have been chosen to directly simulate the timing of the $j^{th}$ event. In fact, data resulting from this simulation algorithm does not really mimick 'real' clinical MS trial data, as CDP is expressed on the ordinal EDSS scale and EDSS data is only measured approximately every $3$ months. As a consequence, recurrent CDP12 events are generally separated in time by a minimum of $12$ weeks. In order to closely resemble 'real' CDP data, a second MS-specific simulation study is performed, where multistate model methodology for panel data is used to generate longitudinal EDSS data. Based on these EDSS measurements, CDP events can be easily derived according to the definitions introduced in Chapter $2$. Figure $\ref{IdeaMSsimulation}$ graphically illustrates the idea behind the MS-specific simulation study. 
\newline 

\tikzstyle{block} = [ rectangle, draw, text width=7em, text badly centered, rounded corners, minimum height=4em]  
\tikzstyle{line} = [ -latex', draw]  
\begin{figure}[h]
\centering
\begin{tikzpicture}[node distance=5cm, auto]  
  \node [block]           (puc)  {Time-homogeneous multistate model for EDSS dynamics};  
  \node [block, right of=puc]  (wdt)  {Longitudinal EDSS measurements};  
  \node [block, right of=wdt]  (port) {Recurrent CDP events};  
  \path [line] (puc)  -- (wdt);  
  \path [line] (wdt)  -- (port);  
\end{tikzpicture}
\caption{Idea of MS-specific simulation algorithm for recurrent CDP events}
\label{IdeaMSsimulation}
\end{figure}
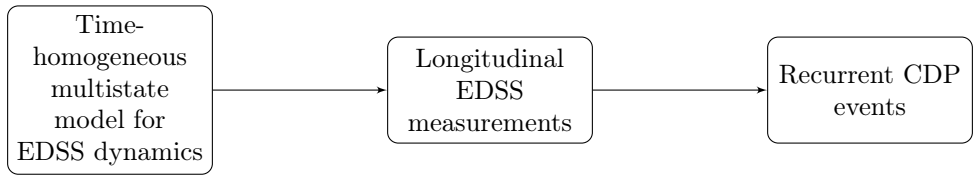

\textbf{Generation of EDSS assessment times} \newline 
In clinical MS trials, study visits are originally scheduled every $3$ months but actual visits deviate slightly from the schedule and can therefore vary between and within individuals. Assuming a maximum length of follow-up of $216$ weeks, the scheduled EDSS assessment times can be described as follows: 
\newline
\begin{table}[h]
\centering
\begin{tabular}{|l|c|c|c|c|c|c|c|c|c|}
\hline
\textbf{Study visit}   & Baseline           & \begin{tabular}[c]{@{}c@{}} Week  \\ $12$ \end{tabular} & \begin{tabular}[c]{@{}c@{}} Week \\ $24$ \end{tabular} & \begin{tabular}[c]{@{}c@{}} Week \\ $36$ \end{tabular} & \begin{tabular}[c]{@{}c@{}} Week \\ $48$\end{tabular} & \begin{tabular}[c]{@{}c@{}}Week \\ $60$\end{tabular} & ... & \begin{tabular}[c]{@{}c@{}}Week  \\ $204$ \end{tabular} & \begin{tabular}[c]{@{}c@{}}Week \\ $216$\end{tabular} \\ \hline
\textbf{Scheduled day} & 1                & 85                                                  & 169                                                 & 253                                                 & 337                                                 & 421                                                 & ... & 1429                                                 & 1513                                                 \\ \hline
\textbf{Notation}      & $\overline{v}_{0}$ & $\overline{v}_{1}$                                  & $\overline{v}_{2}$                                  & $\overline{v}_{3}$                                  & $\overline{v}_{4}$                                  & $\overline{v}_{5}$                                  &     & $\overline{v}_{17}$                                  & $\overline{v}_{18}$                                  \\ \hline
\end{tabular}
\caption{Scheduled EDSS assessment times in clinical MS trials}
\end{table}

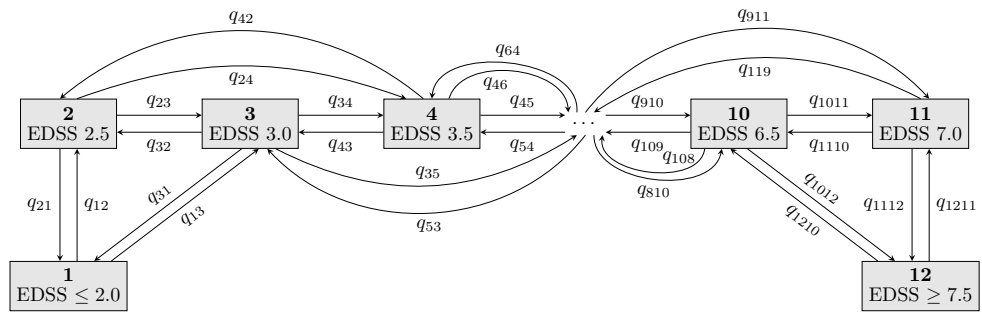
\begin{figure}[h]
\centering
\resizebox{13cm}{!}{
\begin{tikzpicture}
  \tikzset{node style/.style={state, fill=gray!20!white, rectangle}}
        \node[node style]               (II)   {\shortstack{\textbf{2}\\ EDSS $2.5$}};
        \node[node style, below=2cm of II]   (I)  {\shortstack{\textbf{1}\\ EDSS $\leq 2.0 $}};
        \node[node style, right=1.5cm of II]  (III) {\shortstack{\textbf{3}\\ EDSS $3.0$}};
        \node[node style, right=1.5cm of III] (IV)  {\shortstack{\textbf{4}\\ EDSS $3.5$}};
        \node[draw=none, right=1.5cm of IV]  (dot)  {$\cdots$};
        \node[node style, right=1.5cm of dot] (VI)   { \shortstack{\textbf{10}\\ EDSS $6.5$}};
        \node[node style, right=1.5cm of VI] (VII)   {\shortstack{\textbf{11}\\ EDSS $7.0$}};
        \node[node style, below=2cm of VII] (VIII)   {\shortstack{\textbf{12}\\ EDSS $\geq 7.5$}}; 
         \draw[->]([xshift=1ex]I.north) -- ([xshift=1ex]II.south)node[midway, right]{$q_{12}$}; 
         \draw[->]([xshift=-1ex]II.south) -- ([xshift=-1ex]I.north)node[midway, left]{$q_{21}$}; 
         \draw[->]([xshift=5ex]I.north) -- ([xshift=1ex]III.south)node[midway, below, sloped]{$q_{13}$}; 
         \draw[->]([xshift=-1ex]III.south) -- ([xshift=3ex]I.north)node[midway, above, sloped]{$q_{31}$}; 
         \draw[->]([yshift=1ex]II.east) -- ([yshift=1ex]III.west)node[midway, above]{$q_{23}$}; 
         \draw[->]([yshift=-1ex]III.west) -- ([yshift=-1ex]II.east)node[midway, below]{$q_{32}$}; 
         \draw[->, bend left=20] ([xshift=1ex]II.north) to node[below]{$q_{24}$} ([xshift=-3ex]IV.north); 
         \draw[->, bend right=40] ([xshift=-1ex]IV.north) to node[above]{$q_{42}$} ([xshift=-1ex]II.north); 
         \draw[->]([yshift=1ex]III.east) -- ([yshift=1ex]IV.west)node[midway, above]{$q_{34}$}; 
         \draw[->]([yshift=-1ex]IV.west) -- ([yshift=-1ex]III.east)node[midway, below]{$q_{43}$}; 
         \draw[->, bend right=30] ([xshift=3ex]III.south) to node[above]{$q_{35}$} ([xshift=-1ex]dot.south); 
         \draw[->, bend left=50] ([xshift=0ex]dot.south) to node[below]{$q_{53}$} ([xshift=2ex]III.south); 
         \draw[->]([yshift=1ex]IV.east) -- ([yshift=1ex]dot.west)node[midway, above]{$q_{45}$}; 
         \draw[->]([yshift=-1ex]dot.west) -- ([yshift=-1ex]IV.east)node[midway, below]{$q_{54}$};  
         \draw[->]([yshift=1ex]dot.east) -- ([yshift=1ex]VI.west)node[midway, above]{$q_{910}$}; 
         \draw[->]([yshift=-1ex]VI.west) -- ([yshift=-1ex]dot.east)node[midway, below]{$q_{109}$}; 
         \draw[->]([yshift=1ex]VI.east) -- ([yshift=1ex]VII.west)node[midway, above]{$q_{1011}$}; 
         \draw[->]([yshift=-1ex]VII.west) -- ([yshift=-1ex]VI.east)node[midway, below]{$q_{1110}$}; 
         \draw[->]([xshift=1ex]VIII.north) -- ([xshift=1ex]VII.south)node[midway, right]{$q_{1211}$}; 
         \draw[->]([xshift=-1ex]VII.south) -- ([xshift=-1ex]VIII.north)node[midway, left]{$q_{1112}$};
         \draw[->]([xshift=1ex]VI.south) -- ([xshift=-3ex]VIII.north)node[pos=0.45, above, sloped]{$q_{1012}$}; 
         \draw[->]([xshift=-5ex]VIII.north) -- ([xshift=-1ex]VI.south)node[pos=0.45, below, sloped]{$q_{1210}$}; 
         \draw[->, bend right=30] ([xshift=0ex]VII.north) to node[below]{$q_{119}$}([xshift=1ex]dot.north); 
         \draw[->, bend left=50] ([xshift=0ex]dot.north) to node[above]{$q_{911}$}([xshift=1ex]VII.north);
         \draw[->, bend left=90, pos=0.3] ([xshift=-4ex]VI.south) to node[above]{$q_{108}$}([xshift=2ex]dot.south); 
         \draw[->, bend right=85] ([xshift=1ex]dot.south) to node[below]{$q_{810}$}([xshift=-2ex]VI.south);
         \draw[->, bend left=80, pos=0.4] ([xshift=2ex]IV.north) to node[below]{$q_{46}$}([xshift=-2ex]dot.north); 
         \draw[->, bend right=90] ([xshift=-1ex]dot.north) to node[above]{$q_{64}$}([xshift=0ex]IV.north);
\end{tikzpicture}
}
\caption[Multistate model used in the MS-specific simulation study]{Multistate model used in the MS-specific simulation study (excluding $1.5$ step transitions)}
\label{markierung1}
\end{figure}

In order to reflect the variation in actual EDSS assessment times, random noise around the scheduled EDSS assessment times is added. Let $\overline{V}_{r} \in \{ \overline{v}_{1}, \overline{v}_{2}, ..., \overline{v}_{18}\} = \{ 85, 169, ..., 216 \}$ denote the time of the $r^{th}$ scheduled EDSS assessment measured in days since baseline. Accordingly, $V_{ir}$ is the time of the $r^{th}$ actual EDSS assessment for individual $i$, also measured in days since baseline. The corresponding realizations of $\overline{V}_{r}$ and $V_{ir}$ are denoted by $\overline{v}_{r}$ and $v_{ir}$. Inspired by the ORATORIO trial, random noise defined as the deviation between the scheduled time $\overline{V}_{r}$ and the actual time $V_{ir}$ is assumed to be iid $t$-distributed with $3.54$ degrees of freedom and a non-centrality parameter of $0.25$. Then, the actual time for the $r^{th}$ EDSS assessment for individual $i$ is generated by  
\begin{equation}
V_{ir} = \overline{V}_{r} + \epsilon_{ir}, \ \ \ r = 1, 2, ... 18 \ \ \text{and} \ \ i=1,2,..., n, 
\label{markierung9}
\end{equation}
where $\epsilon_{ir} \sim t_{3.54, 0.25}$ and $v_{i0} := \overline{v}_{0} = 1$ (= baseline study visit). \newline
Since study visits can only be made for individuals who are still involved into the trial and under observation, censoring at $C_{i}$ terminates the assessment process. If the assessment times for individual $i$ are denoted by $v_{i}=(v_{i0}, v_{i1}, ..., v_{ir_{i}})$, study visits are generated according to Eq. $(\ref{markierung9})$ only until $v_{ir} > C_{i}$. The assessment at $v_{i(r-1)}$ is then determined as the last study visit observed for individual $i$ and $r_{i}$ is equal to $r-1$. That is, patients are censored at the date of their last EDSS assessment, as requested from the classical CDP endpoint definition (cf. Chapter $2$). 
\newpage
\textbf{Generation of baseline EDSS score} \newline
The baseline EDSS score $E_{i}(v_{i0}) \in \{1, ..., J \}$ is generated from a multinomial distribution with the following $J=12$ potential outcomes: $\leq 2.0, \ 2.5, \ 3.0,\ 3.5,\ 4.0, \ 4.5, \ 5.0,\ 5.5,\ 6.0, \ 6.5, \ 7.0, \ \geq 7.5 $. For instance, $E_{i}(v_{i0}) = 1$ indicates that the EDSS score at baseline is less or equal to $2.0$, while $E_{i}(v_{i0})=2$ corresponds to an EDSS score of 2.5, and so on. Since only a few ORATORIO patients have EDSS scores less than $2.0$ or greater than $7.5$, EDSS values ranging from $0.0$ to $2.0$ and from $7.5$ to $10.0$ are summarized into 2 categories. If $\pi_{j} = P(E_{i}(v_{i0})=j) \in (0,1)$ denotes the probability of observing outcome $j$ and $\sum_{j=1}^{J} \pi_{j} = 1$ for $j=1,...,J$, then
\[ E_{i}(v_{i0}) \sim Mult(1, (\pi_{1}, \pi_{2}, ..., \pi_{J})). \]
To obtain realistic probabilities $\pi_{j}$, maximum likelihood estimation has been applied to the ORATORIO data, leading to the following choices: 
\begin{align} 
\nonumber \pi = ( & 0.00000, 0.00274, 0.08208, 0.18331, 0.17921, 0.09439, \\
        & 0.05746, 0.09986, 0.18057, 0.11902, 0.00137, 0.00000). \label{piPPMS}  
\end{align}
Based on the initial EDSS score and the subsequent EDSS assessment times, post-baseline EDSS values can be generated by making use of a time-homogeneous multistate model. \newline \newline 
\textbf{Post-baseline EDSS scores} \newline 
Longitudinal measurements of the EDSS scale at post-baseline study visits are simulated using a time-homogeneous multistate model. \newline
Figure $\ref{markierung1}$ portrays the multistate model used in this PPMS simulation study, with $J=12$ different states defined according to an individual's EDSS score. States are represented by boxes and possible transitions by arrows. For simulation purposes, a multistate process $\bigl (E(t) \bigr)_{t \geq 0}$ with finite state space $\{ 1, 2, 3, ..., J \}$ is considered, where $E(t)$ denotes the state occupied by an individual at time $t$, $t \geq 0$. For example, $E(t)=6$ means that the EDSS score at time $t$ is $4.5$. Since baseline EDSS scores recorded in the ORATORIO trial range from $\leq 2$ to $\geq 7.5$, there is no initial state and individuals can start in each state, i.e., $E(v_{0}) \in \{ 1, 2, 3, ..., J \}$. Occurrence of a 'new' (higher or lower) EDSS score at a subsequent study visit is modelled by transitions into the state defined by the corresponding EDSS value. For instance, an increase in EDSS score from $3.0$ to $4.0$ between two subsequent study visits is modelled by a transition from state $3$ to state $5$. \newline 
Transition models for EDSS dynamics have also been discussed by \citet{Mandel2013}. 

This multistate model in continuous time can be specified in terms of transition intensity functions
\begin{equation} 
q_{hj}(t) dt := P(E(t+dt)=j \ | \ E(t-)=h, \textnormal{past}) \stackrel{\textnormal{Markov}} {=} P(E(t+dt)=j \ | \ E(t-)=h),  \label{equation1} 
\end{equation}
for $h \neq j$ and $h,j \in \{ 1, 2, ..., J \}$. Past denotes the history up to just prior time $t$ of the multistate process and relevant covariates. The transition intensity $q_{hj}(t)dt$ defines the conditional probability of moving from state $h$ to state $j$ in the next very small time interval $[t, t+dt)$, provided that the state at $t-$ is $h$. The Markov property states that the momentary risk of a $h \longrightarrow j$ transition depends on the current state $h$ and time $t$ since time origin but not on the entry time into state $h$. 
\newpage
EDSS data on MS patients are intermittently collected at fixed study visits $v_{i0} < v_{i1} < ...$, so that individuals' current states are only known at the assessment times, although transitions from one state to another can generally happen at any continuous time point. As a result, EDSS trajectories and states occupied between observation times $v_{i(r-1)}$ and $v_{ir}$ are unknown. The only available data from the multistate process $\bigl (E(t) \bigr)_{t \geq 0}$ are the observed states $E_{i}(v_{i0}), E_{i}(v_{i1}), ..., E_{i}(v_{ir_{i}})$ at assessment times $v_{i0}, ...., v_{ir_{i}}$. Due to this specific panel data, it is simulated from a time-homogeneous multistate process with a $(J \ \textnormal{x} \ J)$ transition intensity matrix $Q = \bigl (q_{hj} \bigr)_{h,j}$, where  
\begin{align}
\nonumber q_{hj} &= \left\{ \begin{array}{ll} q_{hj,0} \exp(\beta_{hj} Z) &, h \leq j \ \text{and} \ j-h \leq 3  \\
          q_{hj,0} &, h > j \ \text{and} \ h-j \leq 3   \\
          0  &, \text{otherwise}
         \end{array} \right. .
\end{align}
The resulting transition intensity matrix $Q(Z)$ is depicted in Figure $\ref{labelmatrix}$, where entries $q_{hj}$ of transitions that are not explicitely modelled are equal to zero. $Q(Z)$ has off-diagonal entries $q_{hj}, \ h \neq j$, and diagonal entries $q_{hh} = - \sum_{j, j\neq h }{} q_{hj}, \ h \in \{ 1,2,...,J \}$, which implies that the off-diagonal entries must be non-negative and the rows of $Q(Z)$ sum up to $0$. 

\begin{figure}[h]
\centering
\resizebox{10cm}{!}{
$ Q(Z) = \begin{pmatrix}
$$ & $\boldsymbol{$\leq 2$}$ & $\boldsymbol{$2.5$}$ & $\boldsymbol{$3.0$}$ & $\boldsymbol{$3.5$}$ & $\boldsymbol{$4.0$}$ & $\boldsymbol{$4.5$}$ & $\boldsymbol{$5.0$}$ & $\boldsymbol{$5.5$}$ & $\boldsymbol{$6.0$}$ & $\boldsymbol{$6.5$}$ & $\boldsymbol{$7.0$}$ & $\boldsymbol{$\geq 7.5$}$ \\
$\boldsymbol{$\leq 2$}$ & q_{11} & q_{12} & q_{13} & q_{14} & 0 & 0 & 0 & 0 & 0 & 0 & 0 & 0 \\
$\boldsymbol{$2.5$}$ & q_{21} & q_{22} & q_{23} & q_{24} & q_{25} & 0 & 0 & 0 & 0 & 0 & 0 & 0 \\
$\boldsymbol{$3.0$}$ & q_{31} & q_{32} & q_{33} & q_{34} & q_{35} & q_{36} & 0 & 0 & 0 & 0 & 0 & 0 \\
$\boldsymbol{$3.5$}$ & q_{41} & q_{42} & q_{43} & q_{44} & q_{45} & q_{46} & q_{47} & 0 & 0 & 0 & 0 & 0  \\
$\boldsymbol{$4.0$}$ & 0 & q_{52}  & q_{53} & q_{54} & q_{55} & q_{56} & q_{57} & q_{58} & 0 & 0 & 0 & 0    \\
$\boldsymbol{$4.5$}$ & 0 & 0 & q_{63} & q_{64} & q_{65} & q_{66} & q_{67} & q_{68} & q_{69} & 0 & 0 & 0 \\
$\boldsymbol{$5.0$}$ & 0 & 0 & 0 & q_{74} & q_{75} & q_{76} & q_{77} & q_{78} & q_{79} & q_{710} & 0 & 0  \\
$\boldsymbol{$5.5$}$ & 0 & 0 & 0 & 0 & q_{85} & q_{86} & q_{87} & q_{88} & q_{89}& q_{810} &  q_{811} & 0  \\
$\boldsymbol{$6.0$}$ & 0 & 0 & 0 & 0 & 0 & q_{96} & q_{97} & q_{98} & q_{99} & q_{910} & q_{911} & q_{912}\\
$\boldsymbol{$6.5$}$ & 0 & 0 & 0 & 0 & 0 & 0 & q_{107}  & q_{108} & q_{109} & q_{1010} & q_{1011} & q_{1012}  \\
$\boldsymbol{$7.0$}$ & 0 & 0 & 0 & 0 & 0 & 0 & 0 & q_{118} & q_{119} & q_{1110} & q_{1111} & q_{1112} \\
$\boldsymbol{$\geq 7.5$}$ & 0 & 0 & 0 & 0 & 0 & 0 & 0 & 0 & q_{129} & q_{1210} & q_{1211} & q_{1212}
\end{pmatrix}, $
}
\caption{Transition intensity matrix for MS-specific simulation study} 
\label{labelmatrix}
\end{figure}

The specific structure of $Q(Z)$ is justified by the following facts: 
\begin{itemize}
\item  PPMS patients are not free to move among the total of $J$ possible states because they are usually not expected to experience an increase or decrease in EDSS score by $>1.5$ points within $3$ months. A similar conclusion can be found from the ORATORIO analyses, where only a very small proportion of patients is observed to make upward or downward transitions of $> 1.5$ points. 
\item PPMS is a progressive disease meaning that disease conditions of patients usually never improve. EDSS scores of PPMS patients should theoretically tend to be continuously increasing or to be at least stable over time. However, in practice, decreasing EDSS trajectories are still common. In order to mimick real clinical trial data in PPMS patients as closely as possible, decreases in EDSS scores by $\leq 1.5$ points are allowed. 
\item The ORATORIO trial showed a significant reduction in disability progression in PPMS patients treated with OCR, as compared to PLA patients. Since disability progression in MS patients is measured on the discrete EDSS scale, transition intensities of the multistate model obviously depend on the treatment group. Ocrelizumab is a recombinant humanized monoclonal antibody designed primarily to stave off disability progression, preserve neurological functions such as coordination and cognitivity and to suppress ongoing disease activity. Nevertheless, ocrelizumab does not seem to be able to reverse the damage that has already been caused by the disease. Transferred to multistate modelling, ocrelizumab works by preventing transitions into higher EDSS categories or by keeping current EDSS scores stable but the molecule is not assumed to directly affect transitions into lower EDSS categories. As a result, the effect of treatment $Z$ on EDSS transitions is constrained to the non-zero transition intensities $q_{hj}$ with $h \leq j$ (i.e., entries above diagonal). Due to the specific properties of a transition intensity matrix (i.e., rows sum up to $0$), it does not matter whether the treatment effect is simulated on the diagonal elements or not.  \newline
For simplicity, the treatment effect is constrained to be equal for all transitions, i.e., $\beta_{hj}$ does not depend on score $h$ or $j$, for $h \leq j$ and $j-h \leq 3$.  
\end{itemize}

In order to get a realistic choice for the baseline transition intensity matrix $Q_{0} = \bigl ( q_{hj,0} \bigr)_{h,j=1,...,J}$, the multistate model depicted in Figure $\ref{markierung1}$ has been applied to the placebo arm of the ORATORIO data using the msm package in R. The resulting baseline transition intensities $q_{hj,0}$ used in this simulation study are summarized in Figure $\ref{Q0PPMS}$. Given $Q_{0}$, $Z$ and $\beta_{hj}$, the transition intensity matrix $Q(Z)$ can be generated. \newline
For a time-homogeneous multistate process, the ($J \ \textnormal{x} \ J$) transition probability matrix $P(t; Z)$ with entries $p_{hj}(t) =  P(E(s + t) = j \ | \ E(s) = h, Z) $ is given by the Chapman-Kolmogorov equation \parencite{Cox1965}: 
\begin{equation} P(t; Z) = \textnormal{Exp}\bigl(tQ(Z)\bigr). \label{markierung10} \end{equation}
According to Eq. $(\ref{markierung10})$, the transition probability matrix $P(t; Z)$ can be calculated for each individual from the generated $Q(Z)$. The transition probability matrix for $Z=0$ and $t=12$ weeks is illustrated in Figure $\ref{PControlHomo}$.  
\newline \newline 
Given the transition probability matrix $P(t; Z_{i})$, the baseline EDSS score $E_{i}(v_{i0})$ and the subsequent EDSS assessment times $v_{i}=(v_{i0}, v_{i1}, ..., v_{ir_{i}})$ for individual $i$, the EDSS score $E_{i}(v_{ir})$ can be simulated from a multinomial distribution with
\begin{align}
\nonumber E_{i}(v_{ir}) \  &| \  E_{i}(v_{i(r-1)}), v_{i(r-1)}, v_{ir} \ \sim \\ \nonumber
&Mult(1, (p_{E_{i}(v_{i(r-1)})1}(v_{ir}-v_{i(r-1)}; Z_{i}), ..., p_{E_{i}(v_{i(r-1)})J}(v_{ir}-v_{i(r-1)}; Z_{i}))), 
\end{align}
where $r=1,...,r_{i}$ and $i=1,..., n$. 
\newline \newline 
\textbf{Specification of heterogeneity} \newline 
The multistate approach incorporates frailty terms to differentiate between patients who are more prone to move through the states ('movers') and patients who prefer to stay in the same EDSS state ('stayers') \parencite{Hout2016}. However, in the MS-specific simulation study, the concept of how frailties affect a patient's risk for disability progression is more complex and interpretation of the frailty term depends on the specification of the heterogeneity matrix. 
\newline 
In general, the transition intensity matrix of the multistate model can be graphically displayed as follows: 
\begin{figure}[H]
\centering
\resizebox{5cm}{!}{
\begin{tikzpicture}
\matrix[matrix of math nodes, column sep=3mm, row sep=3mm, left delimiter=(, right delimiter=),nodes in empty cells] (m)
{
\pgfmatrixnextcell \pgfmatrixnextcell \pgfmatrixnextcell \pgfmatrixnextcell \pgfmatrixnextcell \\
\pgfmatrixnextcell  \pgfmatrixnextcell  \pgfmatrixnextcell \Huge{\textbf{\textnormal{worsening}}} \pgfmatrixnextcell  \pgfmatrixnextcell \\
\pgfmatrixnextcell \pgfmatrixnextcell  \rotatebox{-25}{\Huge{\textbf{\textnormal{stability}}}} \pgfmatrixnextcell  \pgfmatrixnextcell \pgfmatrixnextcell \\
\pgfmatrixnextcell \pgfmatrixnextcell  \pgfmatrixnextcell  \pgfmatrixnextcell \pgfmatrixnextcell \\
\pgfmatrixnextcell \Huge{\textbf{\textnormal{improvement}}} \pgfmatrixnextcell \pgfmatrixnextcell  \pgfmatrixnextcell  \pgfmatrixnextcell \\
\pgfmatrixnextcell \pgfmatrixnextcell  \pgfmatrixnextcell \pgfmatrixnextcell  \pgfmatrixnextcell  \\
};
\fill[gray!30, opacity=0.4] (m-1-2.south west) -| (m-5-6.north east);
\fill[gray!30, opacity=0.4] (m-2-1.north east) |- (m-6-5.south west);
\end{tikzpicture}
}
\caption{Simplified illustration of transition intensity matrix}
\label{SimplyTM}
\end{figure}
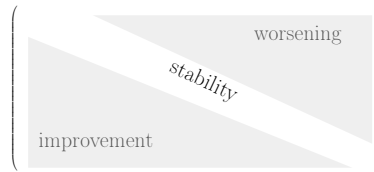

In Figure $\ref{SimplyTM}$, the upper diagonal of the transition matrix represents upward transitions into higher EDSS scores (= worsening), the lower diagonal corresponds to downward transitions into lower EDSS values (= improvement), and the main diagonal symbolizes stability of the disease process (= stability). In this MS-specific multistate model reflecting EDSS dynamics, there are two possibilities of how frailties can be defined. 

\begin{figure}[H]
\centering
\subfloat[$U_{1}$]{
\resizebox{0.5\linewidth}{!}{
$
\begin{pmatrix}{}
$$ & $\boldsymbol{$\leq 2$}$ & $\boldsymbol{$2.5$}$ & $\boldsymbol{$3.0$}$ & $\boldsymbol{$3.5$}$ & $\boldsymbol{$4.0$}$ & $\boldsymbol{$4.5$}$ & $\boldsymbol{$5.0$}$ & $\boldsymbol{$5.5$}$ & $\boldsymbol{$6.0$}$ & $\boldsymbol{$6.5$}$ & $\boldsymbol{$7.0$}$ & $\boldsymbol{$\geq 7.5$}$ \\
$\boldsymbol{$\leq 2$}$ & 1 & \tikzmark{top1}{U} & \tikzmark{top2}{U} & \tikzmark{top3}{U} & 1 & 1 & 1 & 1 & 1 & 1 & 1 & 1 \\ 
$\boldsymbol{$2.5$}$ &  1 & 1 & U & U & U & 1 & 1 & 1 & 1 & 1 & 1 & 1 \\ 
$\boldsymbol{$3.0$}$ &  1 & 1 & 1 & U & U & U & 1 & 1 & 1 & 1 & 1 & 1 \\ 
$\boldsymbol{$3.5$}$ &  1 & 1 & 1 & 1 & U & U & U & 1 & 1 & 1 & 1 & 1 \\ 
$\boldsymbol{$4.0$}$ &  1 & 1 & 1 & 1 & 1 & U & U & U & 1 & 1 & 1 & 1 \\ 
$\boldsymbol{$4.5$}$ &  1 & 1 & 1 & 1 & 1 & 1 & U & U & U & 1 & 1 & 1 \\ 
$\boldsymbol{$5.0$}$ &  1 & 1 & 1 & 1 & 1 & 1 & 1 & U & U & U & 1 & 1 \\ 
$\boldsymbol{$5.5$}$ &  1 & 1 & 1 & 1 & 1 & 1 & 1 & 1 & U & U & U & 1 \\ 
$\boldsymbol{$6.0$}$ &  1 & 1 & 1 & 1 & 1 & 1 & 1 & 1 & 1 & U & U & \tikzmark{bottom3}{U} \\ 
$\boldsymbol{$6.5$}$ &  1 & 1 & 1 & 1 & 1 & 1 & 1 & 1 & 1 & 1 & U & \tikzmark{bottom2}{U} \\ 
$\boldsymbol{$7.0$}$ &  1 & 1 & 1 & 1 & 1 & 1 & 1 & 1 & 1 & 1 & 1 & \tikzmark{bottom1}{U} \\ 
$\boldsymbol{$\geq 7.5$}$ &  1 & 1 & 1 & 1 & 1 & 1 & 1 & 1 & 1 & 1 & 1 & 1 \\ 
\end{pmatrix}
$
\begin{tikzpicture}[overlay,remember picture]
     \draw[opacity=.05,line width=3mm,line cap=round] (top1.center) -- (bottom1.center);
     \draw[opacity=.05,line width=3mm,line cap=round] (top2.center) -- (bottom2.center);
     \draw[opacity=.05,line width=3mm,line cap=round] (top3.center) -- (bottom3.center);
\end{tikzpicture}
}}
\subfloat[$U_{2}$]{
\resizebox{0.5\linewidth}{!}{
$ 
\begin{pmatrix}{}
$$ & $\boldsymbol{$\leq 2$}$ & $\boldsymbol{$2.5$}$ & $\boldsymbol{$3.0$}$ & $\boldsymbol{$3.5$}$ & $\boldsymbol{$4.0$}$ & $\boldsymbol{$4.5$}$ & $\boldsymbol{$5.0$}$ & $\boldsymbol{$5.5$}$ & $\boldsymbol{$6.0$}$ & $\boldsymbol{$6.5$}$ & $\boldsymbol{$7.0$}$ & $\boldsymbol{$\geq 7.5$}$ \\
$\boldsymbol{$\leq 2$}$ & 1 & \tikzmark{top1}{U} & \tikzmark{top2}{U} & \tikzmark{top3}{U} & 1 & 1 & 1 & 1 & 1 & 1 & 1 & 1 \\ 
$\boldsymbol{$2.5$}$ &  \tikzmark{top6}{U} & 1 & U & U & U & 1 & 1 & 1 & 1 & 1 & 1 & 1 \\ 
$\boldsymbol{$3.0$}$ &  \tikzmark{top5}{U} & U & 1 & U & U & U & 1 & 1 & 1 & 1 & 1 & 1 \\ 
$\boldsymbol{$3.5$}$ &  \tikzmark{top4}{U} & U & U & 1 & U & U & U & 1 & 1 & 1 & 1 & 1 \\ 
$\boldsymbol{$4.0$}$ &  1 & U & U & U & 1 & U & U & U & 1 & 1 & 1 & 1 \\ 
$\boldsymbol{$4.5$}$ &  1 & 1 & U & U & U & 1 & U & U & U & 1 & 1 & 1 \\ 
$\boldsymbol{$5.0$}$ &  1 & 1 & 1 & U & U & U & 1 & U & U & U & 1 & 1 \\ 
$\boldsymbol{$5.5$}$ &  1 & 1 & 1 & 1 & U & U & U & 1 & U & U & U & 1 \\ 
$\boldsymbol{$6.0$}$ &  1 & 1 & 1 & 1 & 1 & U & U & U & 1 & U & U & \tikzmark{bottom3}{U} \\ 
$\boldsymbol{$6.5$}$ &  1 & 1 & 1 & 1 & 1 & 1 & U & U & U & 1 & U & \tikzmark{bottom2}{U} \\ 
$\boldsymbol{$7.0$}$ &  1 & 1 & 1 & 1 & 1 & 1 & 1 & U & U & U & 1 & \tikzmark{bottom1}{U} \\ 
$\boldsymbol{$\geq 7.5$}$ &  1 & 1 & 1 & 1 & 1 & 1 & 1 & 1 & \tikzmark{bottom4}{U} & \tikzmark{bottom5}{U} & \tikzmark{bottom6}{U} & 1 \\ 
\end{pmatrix}
$
\begin{tikzpicture}[overlay,remember picture]
     \draw[opacity=.05,line width=3mm,line cap=round] (top1.center) -- (bottom1.center);
     \draw[opacity=.05,line width=3mm,line cap=round] (top2.center) -- (bottom2.center);
     \draw[opacity=.05,line width=3mm,line cap=round] (top3.center) -- (bottom3.center);
     \draw[opacity=.05,line width=3mm,line cap=round] (top4.center) -- (bottom4.center);
     \draw[opacity=.05,line width=3mm,line cap=round] (top5.center) -- (bottom5.center);
     \draw[opacity=.05,line width=3mm,line cap=round] (top6.center) -- (bottom6.center);
\end{tikzpicture}
}}
\caption{Heterogeneity matrices $U_{1}$ and $U_{2}$}
\label{HeterogeneityMatrices}
\end{figure}

\begin{landscape}
\begin{figure}
\centering
\scalebox{0.70}{
$ Q_{0} = 
\begin{pmatrix}
$$ & $\boldsymbol{$\leq 2$}$ & $\boldsymbol{$2.5$}$ & $\boldsymbol{$3.0$}$ & $\boldsymbol{$3.5$}$ & $\boldsymbol{$4.0$}$ & $\boldsymbol{$4.5$}$ & $\boldsymbol{$5.0$}$ & $\boldsymbol{$5.5$}$ & $\boldsymbol{$6.0$}$ & $\boldsymbol{$6.5$}$ & $\boldsymbol{$7.0$}$ & $\boldsymbol{$\geq 7.5$}$ \\
$\boldsymbol{$\leq 2$}$ & -0.00571457 & 0.00344927 & 0.00213957 & 0.00012573 & 0 & 0 & 0 & 0 & 0 & 0 & 0 & 0 \\
$\boldsymbol{$2.5$}$ & 0.00278110 & -0.00979778 & 0.00410048 & 0.00113692 & 0.00177928 & 0 & 0 & 0 & 0 & 0 & 0 & 0 \\
$\boldsymbol{$3.0$}$ & 0.00097883 & 0.00173411 & -0.00907440 & 0.00436886 & 0.00124637 & 0.00074624  & 0 & 0 & 0 & 0 & 0 & 0 \\
$\boldsymbol{$3.5$}$ & 0.00016261 & 0.00036194 & 0.00197995 & -0.00531696 & 0.00179672 & 0.00079744 & 0.00021830 & 0 & 0 & 0 & 0 & 0  \\
$\boldsymbol{$4.0$}$ & 0 & 0.00038329 &  0.00084327 & 0.00212851 & -0.00556463 & 0.00161647 & 0.00012261 & 0.00047048 & 0 & 0 & 0 & 0    \\
$\boldsymbol{$4.5$}$ & 0 & 0 & 0.00058258 & 0.00107844 & 0.00164630 & -0.00716624 & 0.00204277 & 0.00052749 & 0.00128865 & 0 & 0 & 0 \\
$\boldsymbol{$5.0$}$ & 0 & 0 & 0 & 0.00067075 & 0.00071686 & 0.00431262 & -0.01315725 & 0.00525261 & 0.00195771 & 0.00024671 & 0 & 0 \\
$\boldsymbol{$5.5$}$ & 0 & 0 & 0 & 0 & 0.00065346 & 0.00159607 & 0.00321215 & -0.01148634 & 0.00588884  & 0.00013581 & 0 & 0  \\
$\boldsymbol{$6.0$}$ & 0 & 0 & 0 & 0 & 0 &  0.00046211 & 0.00022775 & 0.00054228 & -0.00288351  & 0.00158707 & 0.00006375 & 0.00000055 \\
$\boldsymbol{$6.5$}$ & 0 & 0 & 0 & 0 & 0 & 0 & 0  & 0.00000346 & 0.00135177 & -0.00263882 & 0.00120201  & 0.00008158  \\
$\boldsymbol{$7.0$}$ & 0 & 0 & 0 & 0 & 0 & 0 & 0 & 0 & 0.00016236 & 0.00481588 & -0.01026036 & 0.00528211 \\
$\boldsymbol{$\geq 7.5$}$ & 0 & 0 & 0 & 0 & 0 & 0 & 0 & 0 & 0 & 0.00000891 & 0.00219949 & -0.00220840  \\
\end{pmatrix} $
}
\caption[Baseline transition intensity matrix in MS-specific simulation study]{Baseline transition intensity matrix $Q_{0}$ in MS-specific simulation study, time-constant transition intensities $q_{hj,0}$ are rounded off to $8$ digits for illustration purposes}
\label{Q0PPMS}
\end{figure} 

\begin{figure}
\centering
\scalebox{0.70}{
$ P_{{PPMS, Z=0, U=1}} = 
\begin{pmatrix}{}
$$ & $\boldsymbol{$\leq 2$}$ & $\boldsymbol{$2.5$}$ & $\boldsymbol{$3.0$}$ & $\boldsymbol{$3.5$}$ & $\boldsymbol{$4.0$}$ & $\boldsymbol{$4.5$}$ & $\boldsymbol{$5.0$}$ & $\boldsymbol{$5.5$}$ & $\boldsymbol{$6.0$}$ & $\boldsymbol{$6.5$}$ & $\boldsymbol{$7.0$}$ & $\boldsymbol{$\geq 7.5$}$ \\
$\boldsymbol{$2.0$}$& 0.6435 & 0.1625 & 0.1272 & 0.0388 & 0.0213 & 0.0055 & 0.0006 & 0.0004 & 0.0003 & 0.0000 & 0.0000 & 0.0000 \\ 
$\boldsymbol{$2.5$}$& 0.1334 & 0.4716 & 0.1805 & 0.0970 & 0.0976 & 0.0149 & 0.0019 & 0.0020 & 0.0010 & 0.0001 & 0.0000 & 0.0000 \\ 
$\boldsymbol{$3.0$}$ & 0.0564 & 0.0789 & 0.5041 & 0.2194 & 0.0842 & 0.0454 & 0.0052 & 0.0029 & 0.0033 & 0.0002 & 0.0000 & 0.0000 \\ 
$\boldsymbol{$3.5$}$ & 0.0157 & 0.0263 & 0.1021 & 0.6718 & 0.1081 & 0.0526 & 0.0131 & 0.0051 & 0.0048 & 0.0004 & 0.0000 & 0.0000 \\ 
$\boldsymbol{$4.0$}$ & 0.0054 & 0.0226 & 0.0538 & 0.1291 & 0.6472 & 0.0897 & 0.0146 & 0.0239 & 0.0126 & 0.0009 & 0.0000 & 0.0000 \\ 
$\boldsymbol{$4.5$}$ & 0.0021 & 0.0045 & 0.0337 & 0.0713 & 0.0930 & 0.5759 & 0.0809 & 0.0404 & 0.0904 & 0.0072 & 0.0004 & 0.0001 \\ 
$\boldsymbol{$5.0$}$ & 0.0006 & 0.0017 & 0.0089 & 0.0414 & 0.0509 & 0.1805 & 0.3678 & 0.1693 & 0.1548 & 0.0227 & 0.0012 & 0.0003 \\ 
$\boldsymbol{$5.5$}$ & 0.0002 & 0.0008 & 0.0039 & 0.0115 & 0.0385 & 0.0929 & 0.1075 & 0.4124 & 0.3002 & 0.0301 & 0.0017 & 0.0003 \\ 
$\boldsymbol{$6.0$}$ & 0.0000 & 0.0001 & 0.0008 & 0.0020 & 0.0033 & 0.0303 & 0.0153 & 0.0289 & 0.8015 & 0.1082 & 0.0077 & 0.0019 \\ 
$\boldsymbol{$6.5$}$ & 0.0000 & 0.0000 & 0.0000 & 0.0001 & 0.0001 & 0.0018 & 0.0009 & 0.0019 & 0.0917 & 0.8209 & 0.0618 & 0.0207 \\ 
$\boldsymbol{$7.0$}$ & 0.0000 & 0.0000 & 0.0000 & 0.0000 & 0.0000 & 0.0004 & 0.0002 & 0.0004 & 0.0233 & 0.2457 & 0.4556 & 0.2743 \\ 
$\boldsymbol{$\geq 7.5$}$ & 0.0000 & 0.0000 & 0.0000 & 0.0000 & 0.0000 & 0.0000 & 0.0000 & 0.0000 & 0.0019 & 0.0257 & 0.1138 & 0.8585 \\ 
\end{pmatrix}
$}
\caption[Baseline transition probability matrix in MS-specific simulation study]{Baseline transition probability matrix $P_{PPMS, Z=0, U=1}$ in MS-specific simulation study, t=12 weeks, entries are rounded off to $4$ digits for illustration purposes}
\label{PControlHomo}
\end{figure}
\end{landscape}

\begin{itemize}
\item \textbf{Option $\boldsymbol{U_{1}}$}:  \newline 
The patient-specific frailties are only added to transition intensities that correspond to upward transitions, i.e., worsening. In this case, the transition intensities $q_{hj}$ are specified as follows: 
\begin{align}
\nonumber q_{hj} &= \left\{ \begin{array}{ll} U q_{hj,0} \exp(\beta_{hj} Z) &, h \leq j \ \text{and} \ j-h \leq 3  \\
          q_{hj,0} &, h > j \ \text{and} \ h-j \leq 3   \\
          0  &, \text{otherwise}
         \end{array} \right.,
\end{align}
where $U$ is a random effect with $\mathbb{E}(U)=1$ and $Var(U)=\phi$, with $\phi>0$. The resulting heterogeneity matrix $\boldsymbol{U_{1}}$ is specified in Figure $\ref{HeterogeneityMatrices}$ (a). Patients with a large frailty term $U \gg 1$ are so-called 'upward movers', as they are very frail to move through the different EDSS states but only in one direction, namely towards higher EDSS scores. Given the multistate process, upward movers are patients whose disease conditions tend to be much worse than average and whose chance for improvement is consequently considerably reduced. Thus, the EDSS score at study visit $k+1$ is on average at least as high as the EDSS score at visit $k$, in which case the EDSS curves tend to be monotonically increasing. 
By comparison, patients with a very small frailty term, i.e., $U < 1$ and close to $0$, are most likely to stay in the current disease state, followed by certain probabilities for improvement. Upward transitions occur with zero probability. So, if $U \longrightarrow 0$, the transition probability matrix becomes a lower triangular matrix, leading to constant or decreasing EDSS curves. This implies that the EDSS score at study visit $k+1$ is at least as low as the EDSS score at visit $k$, in which case the EDSS curves tend to be monotonically decreasing. In this context, patients who are less frail are referred to as 'stayers / downward movers'.
Figure $\ref{TransProbMatrixZ1}$ summarizes how the transition probability matrix $P_{Z=0, U=1}$ is affected by frailties under the assumption of $U_{1}$.
\item \textbf{Option $\boldsymbol{U_{2}}$}: \newline
The patient-specific frailties are added to transitions that correspond to up- and downward transitions, i.e., worsening and improvement. In this case, the transition intensities $q_{hj}$ are specified as follows: 
\begin{align}
\nonumber q_{hj} &= \left\{ \begin{array}{ll} U q_{hj,0} \exp(\beta_{hj} Z) &, h \leq j \ \text{and} \ j-h \leq 3  \\
          U q_{hj,0} &, h > j \ \text{and} \ h-j \leq 3   \\
          0  &, \text{otherwise}
         \end{array} \right. , 
\end{align}
where $U$ is a random effect with $\mathbb{E}(U)=1$ and $Var(U)=\phi$, with $\phi>0$. Figure $\ref{HeterogeneityMatrices}$ (b) illustrates the resulting heterogeneity matrix $\boldsymbol{U_{2}}$. Patients with a large realization of $U $, $U > 1$, are both 'up- and downward movers' in the sense that they are very likely to transition into higher and lower states but they are unlikely to stay. Those patients are assumed to be pretty unstable in their current disease condition, making the underlying EDSS trajectory over time more variable. In contrast to corresponding patients under $U_{1}$, the EDSS curves do not generally show a clear increasing trend but are characterized by the occurrence of both worsening and improvement transitions. 
Patients with very small frailty term, i.e., U close to $0$, are the so-called 'stayers' who show neurological and physical stability. EDSS curves of such patients look like a horizontal line, with almost no variability. As depicted in Figure $\ref{TransProbMatrixZ2}$, the corresponding transition probability matrix approaches the identity matrix, if $U \longrightarrow 0$.
\end{itemize}
Assuming $U_{1}$, heterogeneity reflects suceptibility for disease progression, while under $U_{2}$ heterogeneity can be interpreted as suceptibility for being in unstable disease conditions. \newline 
In scenarios with heterogeneity, the transition intensity matrix $Q(Z, U)$ can be calculated by means of $Q_{0}$, $Z$, $U$ and $\beta_{hj}$, using either specification $U_{1}$ or $U_{2}$. The transition probability matrix with entries $p_{hj}(t) = P(E(s + t) = j \ | \ E(s) = h, Z, U)$ is determined by $P(t; Z, U) = \textnormal{Exp}\bigl(tQ(Z, U)\bigr)$. Based on $P(t; Z_{i}, U_{i})$, $E_{i}(v_{i0})$ and $v_{i}=(v_{i0}, v_{i1}, ..., v_{ir_{i}})$, the EDSS score $E_{i}(v_{ir})$ at study visit $v_{ir}$ is generated from a multinomial distribution with 
\begin{align}
\nonumber E_{i}(v_{ir}) \  &| \  E_{i}(v_{i(r-1)}), v_{i(r-1)}, v_{ir} \ \sim \\ \nonumber
&Mult(1, (p_{E_{i}(v_{i(r-1)})1}(v_{ir}-v_{i(r-1)}; Z_{i}, U_{i}), ..., p_{E_{i}(v_{i(r-1)})J}(v_{ir}-v_{i(r-1)}; Z_{i}, U_{i}))), 
\end{align}
where $r=1,...,r_{i}$ and $i=1,..., n$. \newline 
The time to the $j^{th}$ CDP12 can be finally derived from the longitudinal EDSS measurements. 

\newpage 
\begin{figure}[H]
\newcommand\bigzero{\makebox(0,0){\text{\huge0}}}
\centering
\begin{minipage}[t]{0.475\textwidth}
\centering
\subfloat[$U \longrightarrow 0$]{
\resizebox{0.75\textwidth}{!}{
\begin{tikzpicture}
\matrix[matrix of math nodes, column sep=3mm, row sep=3mm, left delimiter=(, right delimiter=),nodes in empty cells] (m)
{
 1 \pgfmatrixnextcell \pgfmatrixnextcell \pgfmatrixnextcell \pgfmatrixnextcell \pgfmatrixnextcell \\
 > \pgfmatrixnextcell \gg \pgfmatrixnextcell \pgfmatrixnextcell \pgfmatrixnextcell \bigzero \pgfmatrixnextcell \\
 \pgfmatrixnextcell > \pgfmatrixnextcell \gg  \pgfmatrixnextcell \pgfmatrixnextcell \pgfmatrixnextcell \\
 \pgfmatrixnextcell \pgfmatrixnextcell \cdots \pgfmatrixnextcell \cdots \pgfmatrixnextcell \pgfmatrixnextcell \\
 \pgfmatrixnextcell \bigzero \pgfmatrixnextcell \pgfmatrixnextcell > \pgfmatrixnextcell \gg \pgfmatrixnextcell \\
 \pgfmatrixnextcell \pgfmatrixnextcell  \pgfmatrixnextcell \pgfmatrixnextcell > \pgfmatrixnextcell \ll  \\
};
\fill[gray!30, opacity=0.4] (m-1-1.north east) -| (m-6-6.north east);
\fill[gray!30, opacity=0.4] (m-1-1.south west) |- (m-6-6.south west);
\end{tikzpicture}}} 

\subfloat[$U < 1.0$]{
\resizebox{0.75\textwidth}{!}{
  \begin{tikzpicture}
\matrix[matrix of math nodes, column sep=3mm, row sep=3mm, left delimiter=(, right delimiter=),nodes in empty cells] (m)
{
 > \pgfmatrixnextcell \ll \pgfmatrixnextcell \pgfmatrixnextcell \pgfmatrixnextcell \pgfmatrixnextcell \\
 > \pgfmatrixnextcell > \pgfmatrixnextcell \ll \pgfmatrixnextcell \pgfmatrixnextcell \bigzero \pgfmatrixnextcell \\
 \pgfmatrixnextcell > \pgfmatrixnextcell >  \pgfmatrixnextcell \ll  \pgfmatrixnextcell \pgfmatrixnextcell \\
 \pgfmatrixnextcell \pgfmatrixnextcell \cdots \pgfmatrixnextcell \cdots \pgfmatrixnextcell \cdots \pgfmatrixnextcell \\
 \pgfmatrixnextcell \bigzero \pgfmatrixnextcell \pgfmatrixnextcell > \pgfmatrixnextcell > \pgfmatrixnextcell \ll \\
 \pgfmatrixnextcell \pgfmatrixnextcell  \pgfmatrixnextcell \pgfmatrixnextcell > \pgfmatrixnextcell < \\
};
\fill[gray!30, opacity=0.4] (m-1-1.north east) -| (m-6-6.north east);
\fill[gray!30, opacity=0.4] (m-1-1.south west) |- (m-6-6.south west);
\end{tikzpicture}}}

\subfloat[$U > 1.0$]{
\resizebox{0.75\textwidth}{!}{
\begin{tikzpicture}
\matrix[matrix of math nodes, column sep=3mm, row sep=3mm, left delimiter=(, right delimiter=),nodes in empty cells] (m)
{
 < \pgfmatrixnextcell > \pgfmatrixnextcell \pgfmatrixnextcell \pgfmatrixnextcell \pgfmatrixnextcell \\
 < \pgfmatrixnextcell < \pgfmatrixnextcell > \pgfmatrixnextcell \pgfmatrixnextcell \bigzero \pgfmatrixnextcell \\
 \pgfmatrixnextcell < \pgfmatrixnextcell < \pgfmatrixnextcell > \pgfmatrixnextcell \pgfmatrixnextcell \\
 \pgfmatrixnextcell \pgfmatrixnextcell \cdots  \pgfmatrixnextcell \cdots  \pgfmatrixnextcell \cdots \pgfmatrixnextcell \\
 \pgfmatrixnextcell \bigzero \pgfmatrixnextcell \pgfmatrixnextcell < \pgfmatrixnextcell < \pgfmatrixnextcell > \\
 \pgfmatrixnextcell \pgfmatrixnextcell \pgfmatrixnextcell \pgfmatrixnextcell < \pgfmatrixnextcell >  \\
};
\fill[gray!30, opacity=0.4] (m-1-1.north east) -| (m-6-6.north east);
\fill[gray!30, opacity=0.4] (m-1-1.south west) |- (m-6-6.south west);
\end{tikzpicture}}}

\subfloat[$U \gg 1.0$]{
\resizebox{0.75\textwidth}{!}{
\begin{tikzpicture}
\matrix[matrix of math nodes, column sep=3mm, row sep=3mm, left delimiter=(, right delimiter=),nodes in empty cells] (m)
{
 \ll \pgfmatrixnextcell >/< \pgfmatrixnextcell \pgfmatrixnextcell \gg \pgfmatrixnextcell \pgfmatrixnextcell  0 \\
  \approx 0 \pgfmatrixnextcell \ll \pgfmatrixnextcell >/< \pgfmatrixnextcell \pgfmatrixnextcell \gg \pgfmatrixnextcell \\
 \pgfmatrixnextcell  \approx 0 \pgfmatrixnextcell \ll \pgfmatrixnextcell > \pgfmatrixnextcell \pgfmatrixnextcell \gg \\
 \pgfmatrixnextcell \pgfmatrixnextcell \cdots \pgfmatrixnextcell \cdots \pgfmatrixnextcell \cdots \pgfmatrixnextcell \\
 \pgfmatrixnextcell \bigzero \pgfmatrixnextcell \pgfmatrixnextcell \approx 0 \pgfmatrixnextcell \ll \pgfmatrixnextcell > \\
 \pgfmatrixnextcell \pgfmatrixnextcell \pgfmatrixnextcell \pgfmatrixnextcell \approx 0 \pgfmatrixnextcell \gg \\
};
\fill[gray!30, opacity=0.4] (m-1-1.north east) -| (m-6-6.north east);
\fill[gray!30, opacity=0.4] (m-1-1.south west) |- (m-6-6.south west);
\end{tikzpicture}}}
\caption[Association of $U$ and $P$ using heterogeneity matrix $U_{1}$]{Association of $U$ and $P$ using heterogeneity matrix $U_{1}$ \newline 
The symbols '>' and '$\gg$' stand for an increase in transition probabilities, whereas the symbols '<' and '$\ll$' represent a decrease in transition probabilities, as compared to the corresponding probabilities of $P_{Z=0, U=1}$ (= reference probability matrix). The figure shall be interpreted in the following way: e.g., patients with $U \gg 1.0$ have greatly reduced probabilities on the diagonal, strongly increased probabilities on the upper diagonal and almost 0 probabilities on the lower diagonal, as compared to patients with $U=1.0$.}
\label{TransProbMatrixZ1}
\end{minipage} 
\begin{minipage}[t]{0.475\textwidth}
\centering
\subfloat[$U \longrightarrow 0$ ]{
\resizebox{0.75\textwidth}{!}{
\begin{tikzpicture}
\matrix[matrix of math nodes, column sep=3mm, row sep=3mm, left delimiter=(, right delimiter=), nodes in empty cells] (m)
{
 1 \pgfmatrixnextcell \pgfmatrixnextcell \pgfmatrixnextcell \pgfmatrixnextcell \pgfmatrixnextcell \\
 \pgfmatrixnextcell 1 \pgfmatrixnextcell \pgfmatrixnextcell \pgfmatrixnextcell \bigzero \pgfmatrixnextcell \\
 \pgfmatrixnextcell \pgfmatrixnextcell 1 \pgfmatrixnextcell \pgfmatrixnextcell \pgfmatrixnextcell \\
 \pgfmatrixnextcell \pgfmatrixnextcell \pgfmatrixnextcell \cdots \pgfmatrixnextcell \pgfmatrixnextcell \\
 \pgfmatrixnextcell \bigzero \pgfmatrixnextcell \pgfmatrixnextcell \pgfmatrixnextcell 1 \pgfmatrixnextcell \\
 \pgfmatrixnextcell \pgfmatrixnextcell \pgfmatrixnextcell \pgfmatrixnextcell \pgfmatrixnextcell 1 \\
};
\fill[gray!30, opacity=0.4] (m-1-1.north east) -| (m-6-6.north east);
\fill[gray!30, opacity=0.4] (m-1-1.south west) |- (m-6-6.south west);
\end{tikzpicture}}} 

\subfloat[$U < 1.0 $]{
\resizebox{0.75\textwidth}{!}{
  \begin{tikzpicture}
\matrix[matrix of math nodes, column sep=3mm, row sep=3mm, left delimiter=(, right delimiter=),nodes in empty cells] (m)
{
 > \pgfmatrixnextcell <\pgfmatrixnextcell \pgfmatrixnextcell \pgfmatrixnextcell \pgfmatrixnextcell \\
 < \pgfmatrixnextcell > \pgfmatrixnextcell < \pgfmatrixnextcell \pgfmatrixnextcell \bigzero \pgfmatrixnextcell \\
 \pgfmatrixnextcell < \pgfmatrixnextcell > \pgfmatrixnextcell < \pgfmatrixnextcell \pgfmatrixnextcell \\
 \pgfmatrixnextcell \pgfmatrixnextcell \cdots \pgfmatrixnextcell \cdots \pgfmatrixnextcell \cdots \pgfmatrixnextcell \\
 \pgfmatrixnextcell \pgfmatrixnextcell \pgfmatrixnextcell < \pgfmatrixnextcell > \pgfmatrixnextcell < \\
 \pgfmatrixnextcell \pgfmatrixnextcell \pgfmatrixnextcell \pgfmatrixnextcell < \pgfmatrixnextcell > \\
};
\fill[gray!30, opacity=0.4] (m-1-1.north east) -| (m-6-6.north east);
\fill[gray!30, opacity=0.4] (m-1-1.south west) |- (m-6-6.south west);
\end{tikzpicture}}}

\subfloat[$U > 1.0$]{
\resizebox{0.75\textwidth}{!}{
  \begin{tikzpicture}
\matrix[matrix of math nodes, column sep=3mm, row sep=3mm, left delimiter=(, right delimiter=),nodes in empty cells] (m)
{
 < \pgfmatrixnextcell > \pgfmatrixnextcell \pgfmatrixnextcell \pgfmatrixnextcell \pgfmatrixnextcell \\
 > \pgfmatrixnextcell < \pgfmatrixnextcell > \pgfmatrixnextcell \pgfmatrixnextcell \bigzero \pgfmatrixnextcell \\
 \pgfmatrixnextcell > \pgfmatrixnextcell < \pgfmatrixnextcell > \pgfmatrixnextcell \pgfmatrixnextcell \\
 \pgfmatrixnextcell \pgfmatrixnextcell \cdots \pgfmatrixnextcell \cdots \pgfmatrixnextcell \cdots \pgfmatrixnextcell \\
 \pgfmatrixnextcell \pgfmatrixnextcell \pgfmatrixnextcell > \pgfmatrixnextcell < \pgfmatrixnextcell > \\
 \pgfmatrixnextcell \pgfmatrixnextcell \pgfmatrixnextcell \pgfmatrixnextcell > \pgfmatrixnextcell <  \\
};
\fill[gray!30, opacity=0.4] (m-1-1.north east) -| (m-6-6.north east);
\fill[gray!30, opacity=0.4] (m-1-1.south west) |- (m-6-6.south west);
\end{tikzpicture}}} 

\subfloat[$U \gg 1.0$]{
\resizebox{0.75\textwidth}{!}{
  \begin{tikzpicture}
\matrix[matrix of math nodes, column sep=3mm, row sep=3mm, left delimiter=(, right delimiter=),nodes in empty cells] (m)
{
 \ll \pgfmatrixnextcell >/< \pgfmatrixnextcell \pgfmatrixnextcell \gg  \pgfmatrixnextcell \pgfmatrixnextcell 0 \\
 >/<\pgfmatrixnextcell \ll \pgfmatrixnextcell  >/< \pgfmatrixnextcell \pgfmatrixnextcell \gg \pgfmatrixnextcell \\
 \pgfmatrixnextcell >/< \pgfmatrixnextcell \ll \pgfmatrixnextcell > \pgfmatrixnextcell \pgfmatrixnextcell \gg \\
 \gg \pgfmatrixnextcell \pgfmatrixnextcell \cdots \pgfmatrixnextcell \cdots \pgfmatrixnextcell \cdots \pgfmatrixnextcell \\
 \pgfmatrixnextcell \gg \pgfmatrixnextcell \pgfmatrixnextcell > \pgfmatrixnextcell \ll \pgfmatrixnextcell > \\
 0 \pgfmatrixnextcell \pgfmatrixnextcell \gg \pgfmatrixnextcell \pgfmatrixnextcell > \pgfmatrixnextcell \ll \\
};
\fill[gray!30, opacity=0.4] (m-1-1.north east) -| (m-6-6.north east);
\fill[gray!30, opacity=0.4] (m-1-1.south west) |- (m-6-6.south west);
\end{tikzpicture}}} 
\caption{Association of $U$ and $P$ using heterogeneity matrix $U_{2}$}
\label{TransProbMatrixZ2}
\end{minipage} 
\end{figure}

\newpage
\section{Generation of recruitment, censoring, covariates and frailties}
\label{SimulationAdd}
This section describes simulation mechanisms for the treatment arm, frailty term, entry times (recruitment) and administrative/non-administrative censoring times. The following settings apply to both simulation studies (S1 and S2). \newline \newline 
\textbf{Simulation of treatment group} \newline 
The treatment group $Z \in \{0, 1\}$ is a binary covariate taking the values $1$ (treatment) and $0$ (control). Block randomization with a fixed block length of $4$ is used to ensure equal sample sizes in the treatment and control group (i.e., $n_{trt} = n_{control}$). Patients are randomized to receive either treatment or control in a $1:1$ ratio.  
\newline \newline 
\textbf{Simulation of frailty term} \newline
The frailty term or random effect $U$ is generated from a gamma distribution $\Gamma(\phi^{-1}, \phi^{-1})$ with $\mathbb{E}(U)=1$ and $Var(U)=\phi$, $\phi >0$. In the ORATORIO trial, the value for the heterogeneity parameter estimated from a frailty model is approximately $0.15$. In order to evaluate the extent of heterogeneity on statistical properties, an additional value of $\phi=1.0$ is employed.
\newline \newline 
\textbf{Simulation of recruitment} \newline 
Assuming a time-constant recruitment rate, the entry time $T_{recruit}$ is simulated from an uniform distribution $U(0, end.recruit)$ over the time interval $(0, end.recruit)$, where $end.recruit$ denotes the maximum length of the recruitment period. In both simulation studies, the duration of the recruitment period is restricted to $1$ year (i.e., $end.recruit=365$ days), following the common guidelines for clinical MS trials. 
\newline \newline 
\textbf{Simulation of non-administrative censoring} \newline 
The non-administrative censoring time $\overline{C} \in (0, \infty)$ follows an exponential distribution Exp$(\lambda)$, with $\lambda > 0$. Inspired by the ORATORIO trial, $\lambda$ is chosen to be equal to $0.00025$. The censoring time $\overline{C}$ is a random time at which an individual may withdraw from study early or is lost to follow-up. 
\newline \newline 
\textbf{Simulation of administrative censoring} \newline 
ORATORIO was an event-driven trial, where patients were treated with OCR or PLA for at least $120$ weeks and until approximately $253$ events had been accrued. In order to evaluate the benefit from recurrent event analyses over time-to-first-event analyses, length of follow-up should be consistent in both approaches. Motivated by the event-driven ORATORIO trial, patients in the simulated PPMS trials are followed for the recurrence of CDP events until a prespecified number of first CDP events have been observed (type II censoring). Patients who are still at-risk at study closure are right-censored. \newline 
As before, $T_{i1}$ is defined as the time to the first event measured since baseline and $\overline{C}_{i}$ is the non-administrative censoring time due to early withdrawal from study. Let $\delta = \mathbbm{1}(T_{i1} \leq \overline{C}_{i})$ denote the censoring indicator taking level $0$, if an individual has been censored and level $1$, if a first CDP event has been observed. 
\begin{figure}[h]
\centering
\begin{tikzpicture}[x=.5cm, y=.5cm,domain=-9:9,smooth,decoration=brace]
\draw[->] (-4,0) -- (18,0) node[right] {calendar time};
\draw[thick] (-4,-0.5) -- (-4,0.5) node[below=12pt] {};
\draw[thick] (2,-0.5) -- (2,0.5) node[below=12pt] {};
\draw[thick] (4,-0.5) -- (4,0.5) node[below=12pt] {};
\draw[thick] (14,-0.5) -- (14,0.5) node[below=12pt] {};
\coordinate[label=right: \small{study start}](A) at (-5.4,-0.9);
\coordinate[label=right: \small{end of recruitment}](A) at (2.2,-0.9);
\coordinate[label=right: \small{period}](A) at (2.9,-1.5);
\coordinate[label=right: $T_{calendar}$](A) at (13.4,-0.9);
\coordinate[label=right: $T_{recruit}$](A) at (1.0,0.9);
\draw[dashed] (-4,0.8) -- (-4, 2); 
\draw[dashed] (2,0.8) -- (2, 4); 
\draw[dashed] (4,0.8) -- (4, 2); 
\draw[dashed] (14,0.8) -- (14, 4); 
\draw[decorate, yshift=7ex] (-4,0) -- node[above=0.4ex] {\small{recruitment period}} (4,0);
\draw[decorate, yshift=7ex] (2,2) -- node[above=0.4ex] {{$\min \{ T_{1}, \overline{C} \}$}} (14,2);
\end{tikzpicture}
\caption{Follow-up of study in calendar time}
\label{markierung8}
\end{figure}
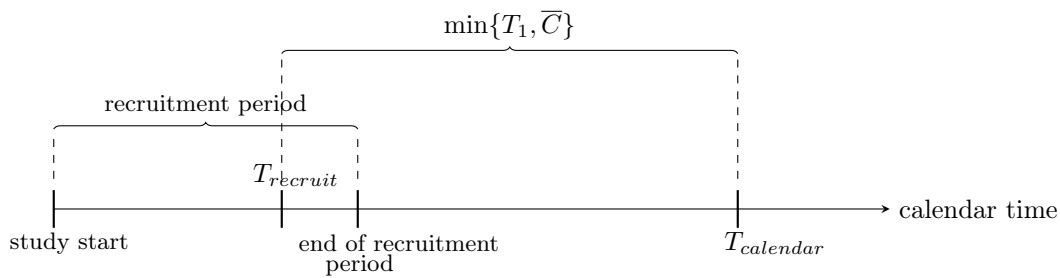

\newpage
As illustrated in Figure $\ref{markierung8}$, the calendar time point for either the first event or non-administrative censoring is then given by 
\[ T_{i, calendar} := T_{i, recruit} + \min \{ T_{i1}, \overline{C}_{i} \} = \begin{cases} T_{i, recruit} +  T_{i1} & , \ \textnormal{if} \ \delta = 1 \\  T_{i, recruit} +  \overline{C}_{i}& ,  \  \textnormal{if} \ \delta = 0  \\ \end{cases}. \]
In order to find the calendar time point for study closure, the calendar event times $T_{calendar, \delta=1}$ must be ordered increasingly: 
\[ T_{(calendar, \delta=1)_{1}} < T_{(calendar, \delta=1)_{2}} < T_{(calendar, \delta=1)_{3}}  < ... < T_{(calendar, \delta=1)_{n_{first.events}}}. \]
The administrative censoring time $C_{A} := T_{(calendar, \delta=1)_{n_{first.events}}}$ defines the calendar time point for study closure. Consequently, the individual-specific administrative censoring time $C_{i}$ (measured in time since baseline) can be derived as follows: 
\[ C_{i} = \min \{ T_{i, recruit} + \overline{C}_{i}, \ C_{A} \} - T_{i, recruit}, \ \textnormal{for} \ i=1,2,...,n. \]
\textbf{Number of first events} \newline
For comparing two equally-sized treatment groups, \citet{Schoenfeld1983} derived the following formula for time-to-event endpoints: 
\begin{equation}
n_{first.events} = \dfrac{4 {(z_{1-\alpha/2} + z_{1-\beta})}^2}{{\log(HR)}^2}, 
\label{label.nevents}
\end{equation}
where $n_{first.events}$ is the total number of events, $\alpha$ the type I error rate, $\beta$ the type II error rate and $HR$ is the hazard ratio the randomized trial wants to detect with power $1-\beta$. $z_{1-\alpha/2}$ and $z_{1-\beta}$ are quantiles of the standard normal distribution, i.e., $P(Z \leq z_{\alpha}) = \alpha$, for $ Z \sim N(0,1)$. Common choices for the type I and II errors are $\alpha=5\%$ and $\beta=20\%$. The number of first events is set to $n_{first.events}=246$, which gives approximately $80 \%$ power to detect a HR of $0.7$ in a time-to-first-event analysis at the $5\%$ significance level ($2$-sided test). \\

\section{Simulation scenarios and parameter settings}
\label{SimulationParameter}
In order to cover the most common scenarios in time-to-first-event and recurrent event analyses, recurrent event datasets are simulated based on the characteristics of $6$ different scenarios, with varying treatment effect $\beta \in \{\log(1.0), \log(0.7)\}$ (S1) or $\beta_{hj} \in \{\log(1.0), \log(0.7)\}$ (S2), and heterogeneity parameter $\phi \in \{ 0.0, 0.15, 1.0 \}$. Table $\ref{ScenariosS1}$ and Table $\ref{ScenariosS2}$ represent the different simulation scenarios labelled according to the following scheme: 
\[ \underbrace{\textnormal{setup}}_{= \ \textnormal{\{S1, S2\}}} / \underbrace{\textnormal{MS type}}_{\in \ \{ \textnormal{PPMS}\}}/ \underbrace{\textnormal{effect of treatment}}_{\in \ \{\textnormal{noeffect, effect}\}} / \underbrace{\textnormal{homo- or heterogeneity}}_{\in \ \{\textnormal{homo, hetero1, hetero2} \}}. \]  

\begin{table}[h]
\centering
\scalebox{0.85}{ 
\begin{tabular}{|l|c|c|c|c|c|}
\hline
                                                                                                                    & \multicolumn{2}{c|}{\textbf{Treatment effect}}                                                                                            & \multicolumn{3}{c|}{\textbf{Heterogeneity parameter}}                                                                                                                            \\ \hline
\textbf{Scenario}                                                                                                   & \multicolumn{1}{l|}{\textbf{$\boldsymbol{\exp(\beta)=1}$}}          & \multicolumn{1}{l|}{\textbf{$\boldsymbol{\exp(\beta)=0.7}$}}         & \multicolumn{1}{l|}{\textbf{$\boldsymbol{\phi} = 0$}}     & \multicolumn{1}{l|}{\textbf{$\boldsymbol{\phi}= 0.15$}}   & \multicolumn{1}{l|}{\textbf{$\boldsymbol{\phi}=1.0$}}     \\ \hline
\begin{tabular}[c]{@{}l@{}}S1/PPMS/noeffect/homo\\ S1/PPMS/noeffect/hetero1\\ S1/PPMS/noeffect/hetero2\end{tabular} & \begin{tabular}[c]{@{}c@{}}$\surd$\\ $\surd$\\ $\surd$\end{tabular} &                                                                     & \begin{tabular}[c]{@{}c@{}}$\surd$\\ $$\\ $$\end{tabular} & \begin{tabular}[c]{@{}c@{}}$$\\ $\surd$\\ $$\end{tabular} & \begin{tabular}[c]{@{}c@{}}$$\\ $$\\ $\surd$\end{tabular} \\ \hline
\begin{tabular}[c]{@{}l@{}}S1/PPMS/effect/homo\\ S1/PPMS/effect/hetero1\\ S1/PPMS/effect/hetero2\end{tabular}       &                                                                     & \begin{tabular}[c]{@{}c@{}}$\surd$\\ $\surd$\\ $\surd$\end{tabular} & \begin{tabular}[c]{@{}c@{}}$\surd$\\ $$\\ $$\end{tabular} & \begin{tabular}[c]{@{}c@{}}$$\\ $\surd$\\ $$\end{tabular} & \begin{tabular}[c]{@{}c@{}}$$\\ $$\\ $\surd$\end{tabular} \\ \hline
\end{tabular}}
\caption[Scenarios in general simulation study]{Scenarios in general simulation study (S1)}
\label{ScenariosS1}
\end{table}

The MS-specific simulation setup is also used to determine the look-ahead bias present in time-to-onset-of-CDP analyses by evaluating the impact of different CDP endpoint definitions (time-to-confirmation-of-CDP versus time-to-onset-of-CPD) on statistical properties. 
\newpage
\begin{table}[h]
\centering
\scalebox{0.85}{ 
\begin{tabular}{|l|c|c|c|c|c|}
\hline
                                                                                                                    & \multicolumn{2}{c|}{\textbf{Treatment effect}}                                                                                            & \multicolumn{3}{c|}{\textbf{Heterogeneity parameter}}                                                                                                                             \\ \hline
\textbf{Scenario}                                                                                                   & \multicolumn{1}{l|}{\textbf{$\boldsymbol{\exp(\beta_{hj})=1}$}}          & \multicolumn{1}{l|}{\textbf{$\boldsymbol{\exp(\beta_{hj})=0.7}$}}         & \multicolumn{1}{l|}{\textbf{$\boldsymbol{\phi} = 0$}}     & \multicolumn{1}{l|}{\textbf{$\boldsymbol{\phi}= 0.15$}}   & \multicolumn{1}{l|}{\textbf{$\boldsymbol{\phi}=1.0$}}     \\ \hline
\begin{tabular}[c]{@{}l@{}}S2/PPMS/noeffect/homo\\ S2/PPMS/noeffect/hetero1\\ S2/PPMS/noeffect/hetero2\end{tabular} & \begin{tabular}[c]{@{}c@{}}$\surd$\\ $\surd$\\ $\surd$\end{tabular} &                                                                     & \begin{tabular}[c]{@{}c@{}}$\surd$\\ $$\\ $$\end{tabular} & \begin{tabular}[c]{@{}c@{}}$$\\ $\surd$\\ $$\end{tabular} & \begin{tabular}[c]{@{}c@{}}$$\\ $$\\ $\surd$\end{tabular} \\ \hline
\begin{tabular}[c]{@{}l@{}}S2/PPMS/effect/homo\\ S2/PPMS/effect/hetero1\\ S2/PPMS/effect/hetero2\end{tabular}       &                                                                     & \begin{tabular}[c]{@{}c@{}}$\surd$\\ $\surd$\\ $\surd$\end{tabular} & \begin{tabular}[c]{@{}c@{}}$\surd$\\ $$\\ $$\end{tabular} & \begin{tabular}[c]{@{}c@{}}$$\\ $\surd$\\ $$\end{tabular} & \begin{tabular}[c]{@{}c@{}}$$\\ $$\\ $\surd$\end{tabular} \\ \hline
\end{tabular}}
\caption[Scenarios in MS-specific simulation study]{Scenarios in MS-specific simulation study (S2)}
\label{ScenariosS2}
\end{table}

While the approaches for generating recurrent event data are described in Section $\ref{GSsimulation}$ and Section $\ref{MSsetup}$, Table $\ref{markierung4}$ gives an overview of the fixed parameter settings in both simulation studies. $N=10000$ datasets of $n=1000$ patients ($n_{trt}=n_{control}=500$ patients in each treatment group) are simulated for each scenario. Each simulated dataset emulates a simplified clinical PPMS trial comparing two treatment arms. The treatment effect is estimated using the Cox proportional hazards model (= time-to-first-event method), the NB model, LWYY model and the AG model (= recurrent event methods). 

\begin{table}[h]
\centering
\scalebox{0.75}{
\begin{tabular}{lll}
\hline
Parameter                                                                          & Notation           & Settings   \\ \hline
Number of simulation                                                               & $N$                & $10000$     \\
Sample size                                                                        & $n$                & $1000$      \\
Non-administrative censoring rate                                                  & $\lambda$          & $0.00025$      \\ 
Type I error (used in Eq. $(\ref{label.nevents})$)                                 & $\alpha$           & $0.05$ \\
Power for time-to-first-event analysis (used in Eq. $(\ref{label.nevents})$)                                & $1-\beta$            & $0.8$ \\
HR (used in Eq. $(\ref{label.nevents})$)                                           & $HR$               & $0.70$ \\
Number of first CDP12 events                                                 & $n_{first.events}$ & $246$    \\
Duration of recruitment period (in days)                                           & $end.recruit$      & $365$      \\ 
\end{tabular}}
\caption[Parameter settings in the general and MS-specific simulation study]{Parameter settings in the general (S1) and MS-specific simulation study (S2)}
\label{markierung4}
\end{table}

\section{Evaluation measures}
\label{SimulationAnalysis}
Let $\beta$ be the true regression coefficient. The estimate of $\beta$ from the $l^{th}$ simulation is denoted by ${\widehat{\beta}}_{l}$, for $l=1,2,..., N$. Further, $\widehat{\beta}_{lower,l}$ and $\widehat{\beta}_{upper,l}$ are the upper and lower confidence limit from the $l^{th}$ simulation. The p-value returned by the $l^{th}$ simulation is denoted by $\text{pval}_{l}$. 
For each statistical method in each scenario, the following performance measures are reported:
\newline 
\begin{table}[h]
\centering
\scalebox{0.65}{
\begin{tabular}{lll}
\hline
\textbf{Evaluation measure}                                                                 & \textbf{Definition}                                                                                                                                        & \textbf{Explanation}                                                                               \\ \hline
\begin{tabular}[c]{@{}l@{}}Average\\ treatment effect \\ $$\end{tabular}                    & \begin{tabular}[c]{@{}l@{}}$$\\ $\exp \bigl (\overline{\beta} \bigr) = \exp \bigl(N^{-1} \sum_{l=1}^{N} \widehat{\beta}_{l} \bigr)$\\ $$\end{tabular}      & \begin{tabular}[c]{@{}l@{}}Estimated mean \\ treatment effect \\ (on hazard scale)\end{tabular}    \\ \hline
\begin{tabular}[c]{@{}l@{}}$$\\ Bias\\ $$\end{tabular}                                      & \begin{tabular}[c]{@{}l@{}}$$\\ $N^{-1} \sum_{l=1}^{N} \widehat{\beta}_{l} - \beta$\\ $$\end{tabular}                                                      & \begin{tabular}[c]{@{}l@{}}$$\\ -\\ $$\end{tabular}                                                \\ \hline
\begin{tabular}[c]{@{}l@{}}$$\\ MSE\\ $$\end{tabular}                                       & \begin{tabular}[c]{@{}l@{}}$$\\ $N^{-1} \sum_{l=1}^{N} (\widehat{\beta}_{l} - \beta)^{2}$\\ $$\end{tabular}                                                & \begin{tabular}[c]{@{}l@{}}$$\\ -\\ $$\end{tabular}                                                \\ \hline
\begin{tabular}[c]{@{}l@{}}$$\\ SE\\ $$\end{tabular}                                        & \begin{tabular}[c]{@{}l@{}}$$\\ $\sqrt{(N-1)^{-1} \sum_{l=1}^{N} (\widehat{\beta}_{l} - \overline{\beta})^{2}}$\\ $$\end{tabular}                          & \begin{tabular}[c]{@{}l@{}}Standard deviation \\ of estimators\\ across simulations\end{tabular}   \\ \hline
\begin{tabular}[c]{@{}l@{}}$$\\ SEE\\ $$\end{tabular}                                       & \begin{tabular}[c]{@{}l@{}}$$\\ $N^{-1} \sum_{l=1}^{N} SE(\widehat{\beta}_{l})$\\ $$\end{tabular}                                                          & \begin{tabular}[c]{@{}l@{}}Mean standard error \\ of estimators \\ across simulations\end{tabular} \\ \hline
\begin{tabular}[c]{@{}l@{}}$$\\ Coverage probability\\ $$\end{tabular}                      & \begin{tabular}[c]{@{}l@{}}$$\\ $N^{-1} \sum_{l=1}^{N} \mathbbm{1}(\widehat{\beta}_{lower,l} \leq \beta \leq \widehat{\beta}_{upper,l})$\\ $$\end{tabular} & \begin{tabular}[c]{@{}l@{}}$$\\ -\\ $$\end{tabular}                                                \\ \hline
\begin{tabular}[c]{@{}l@{}}Power \\ Type I error\\ ($H_{0}: \{ \beta = 0  \}$)\end{tabular} & \begin{tabular}[c]{@{}l@{}}$$\\ $N^{-1} \sum_{l=1}^{N} \mathbbm{1}(\text{pval}_{r} \leq \alpha)$\\ $$\end{tabular}                                         & \begin{tabular}[c]{@{}l@{}}$$\\ -\\ $$\end{tabular}                                               
\end{tabular}
}
\caption{Evaluation measures}
\end{table}

\newpage
\section{Overview}
\label{SimulationOverview}
In order to generate recurrent event data using the general or MS-specific setup, it is proceeded as follows: 
\begin{enumerate}
\item Generation of binary treatment covariate $Z_{i} \in \{ 0, 1 \}$ using block randomization with a fixed block length of $4 \ \forall i=1,2,...,n$. 
\item Generation of individual-specific random effect $U_{i} \sim \Gamma\biggl(\dfrac{1}{\phi},\dfrac{1}{\phi}\biggr) $ with mean $1$ and variance $\phi$. 
\item Generation of non-administrative censoring time $\overline{C_{i}} \sim \text{Exp}(\lambda), \ \lambda > 0$. 
\item Generation of entry time $T_{i, recruit} \sim U(0, end.recruit)$ and calculation of follow-up time $C_{i} =  \min \{ T_{i, recruit} + \overline{C}_{i}, C_{A} \} - T_{i, recurit}$, where $C_{A}$ is administrative censoring time. 
\item 
\begin{itemize}
\item S1: Generation of the $j^{th}$ event time using the recursive simulation algorithm described in Section $\ref{GSsimulation}$.
\item S2: \begin{enumerate}
          \item Generation of individual-specific EDSS assessment times $v_{i} = (v_{i0}, v_{i1}, ..., v_{ir_{i}}) $ with $v_{i0}:=1, v_{ir} = \overline{v_{ir}} + \epsilon_{ir}, \epsilon_{ir} \sim t_{df, \xi}$ and $r_{i} \in \mathbb{N}$, where $\overline{v_{ir}}$ is the time of the $r^{th}$ scheduled assessment time. 
          \item Generation of baseline EDSS score: $E_{i}(v_{i0}) \sim Mult(1, (\pi_{1}, \pi_{2}, ..., \pi_{J})),$ where $\pi_{j}=P(E_{i}(v_{i0}) = j)$ \ for $j=1, ..., J$ and $\sum_{j=1}^{J} \pi_{j} = 1$.
          \item Generation of post-baseline EDSS scores: given $E_{i}(v_{i(r-1)}), v_{i(r-1)}$ and $v_{ir}$, 
\begin{align}
\nonumber E_{i}(v_{ir}) \  &| \  E_{i}(v_{i(r-1)}), v_{i(r-1)}, v_{ir} \ \sim \\
\nonumber & Mult(1, (p_{E_{i}(v_{i(r-1)})1}(v_{ir}-v_{i(r-1)}; Z_{i}, U_{i}), ..., p_{E_{i}(v_{i(r-1)})J}(v_{ir}-v_{i(r-1)}; Z_{i}, U_{i})))  \\
\nonumber & \textnormal{using either heterogeneity matrix} \  U_{1} \ \text{or} \ U_{2}.
\end{align}
          \item Derivation of CDP events according to the endpoint definitions introduced in Chapter $2$. 
          \end{enumerate}
\end{itemize}
\item Repeat step 2 - step 5 for each individual $i$, $i = 1,2,...,n$.  
\end{enumerate}

\chapter{Simulation results}
This chapter represents the results from the simulation studies. The first simulation study is generic and recurrent event data is simulated according to a mixed non-homogeneous Poisson process. The second simulation study is MS-specific: longitudinal measurements of the ordinal EDSS scale are simulated using a time-homogeneous multistate model and recurrent event data is derived based on the resulting EDSS scores. As described in Chapter $6$, simulation parameters are chosen to mimic typical MS trial populations in PPMS and the simulation studies include scenarios with frailties. Recurrent event methods including LWYY, NB and AG models are compared to the conventional Cox proportional hazards model in terms of unbiasedness of treatment effect estimates, statistical power, type I error and clinical interpretation. \newline 
Section $7.1$ represents the results from the general simulation study (S1). In Section $7.2$, findings obtained from the second MS-specific simulation (S2) are reported. 

\section{General simulation study}
\label{GeneralStudy}
\subsection{Characteristics of simulated data}
\label{SimulatedDataS1}
\subsubsection*{Frailty distribution} 
\begin{figure}[h]
\centering
\begin{minipage}{\textwidth}
  \begin{minipage}[b]{0.5\textwidth}
    \centering
\scalebox{0.45}{
\includegraphics{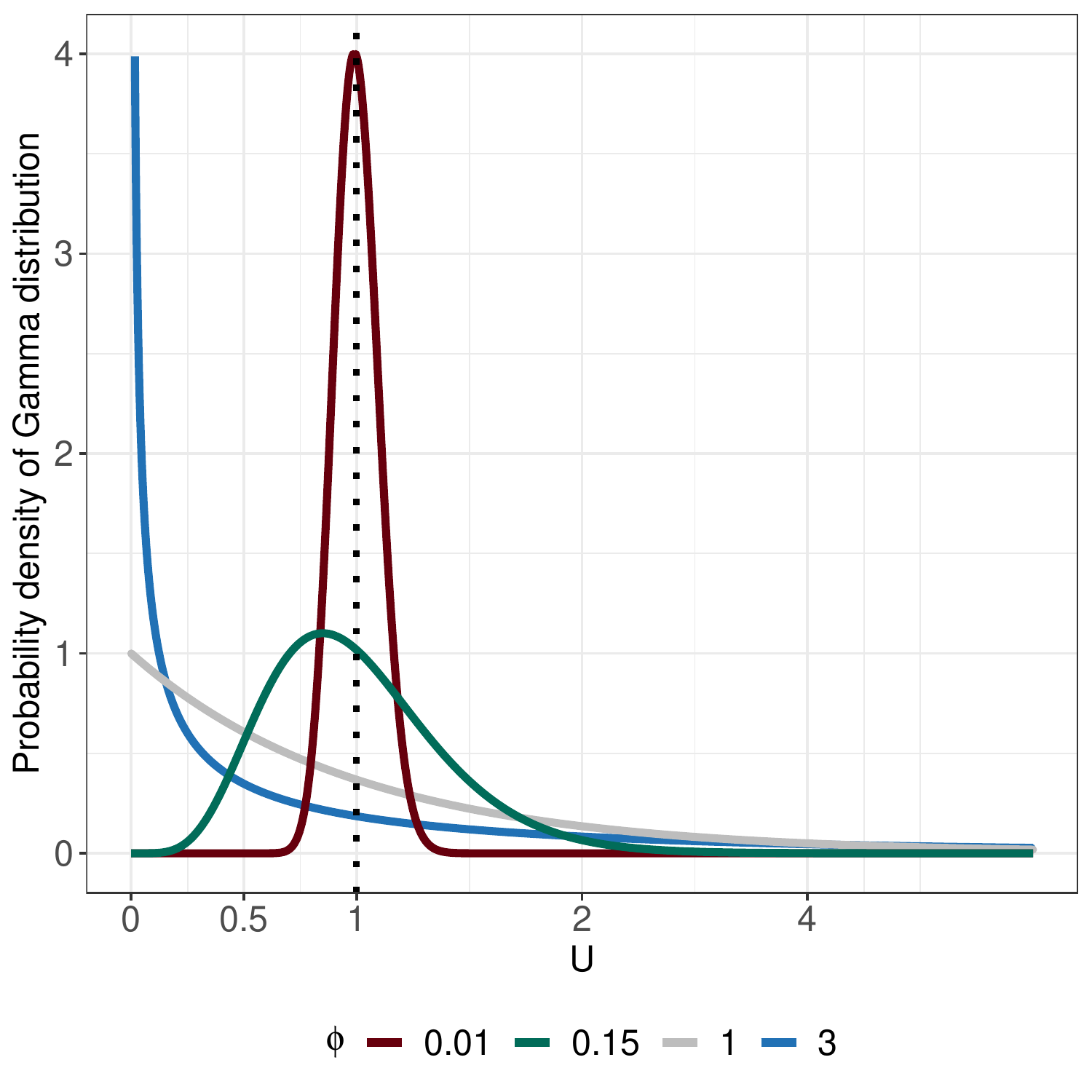}
}
\end{minipage}
\hfill
\begin{minipage}[b]{0.5\textwidth}
\centering 
\scalebox{.75}{
\begin{tabular}{l|cccc}
\textbf{$\boldsymbol{\phi}$} & \multicolumn{1}{l}{\textbf{$\boldsymbol{Q_{10}}$}} & \multicolumn{1}{l}{\textbf{Mean}} & \multicolumn{1}{l}{\textbf{Median}} & \multicolumn{1}{l}{\textbf{$\boldsymbol{Q_{90}}$}} \\ \hline
\textbf{0.0}                 & 1.000                            & 1.000                             & 1.000                               & 1.000                            \\
\textbf{0.15}                & 0.547                            & 1.000                             & 0.950                               & 1.517                            \\
\textbf{1.0}                 & 0.105                            & 1.000                             & 0.693                               & 2.303                           
\end{tabular}
}
$$
$$
$$
$$
$$
$$
$$
$$
\end{minipage}
\caption[Distribution and summary statistics of the gamma distributed frailty term]{Distribution and summary statistics of the gamma distributed random effect $U$ (based on exact probability calculations) according to different heterogeneity parameters, $Q_{10}$ and $Q_{90}$ are the $10 \%$ and $90 \%$ quantiles}
\label{GammaDist}
\end{minipage}
\end{figure}

The mixed non-homogeneous Poisson process incorporates frailty terms to distinguish PPMS patients who are more frail to progress from those who are very unlikely to progress. 
As described in Chapter $6$, the patient-specific random effect $U$ is generated from a gamma distribution $\Gamma \bigl(\phi^{-1}, \phi^{-1} \bigr)$ with mean $\mathbb{E}(U) = 1$, variance $Var(U) = \phi > 0$ and $\phi \in \bigl\{ 0.0, 0.15, 1.0 \bigr \}$. The case $\phi=0.0$ corresponds to a homogeneous study population, where recurrent events are generated from a non-homogeneous Poisson process without frailty terms (i.e., $U=1$). Patients with a high realization of $U$ (i.e., $U > 1$) tend to experience events earlier than patients who are less frail to progress (i.e., $U < 1$). In a recurrent event perspective, frail patients are likely to experience more repeated events than less frail patients. In this work, patients with large $U$ are referred to as 'high-risk' patients or 'fast progressors' (i.e., $U >1$) and patients with small $U$ correspond to 'low-risk' patients or 'slow progressors' (i.e., $U < 1$). Patients with $0.7 \leq U \leq 1.3$ are called 'moderate-risk' patients or 'normal progressors'. Note that this categorization is somewhat arbitrary but it is useful for illustration purposes. 
\newline
The distribution of $U$ is determined by two parameters: the mean $1$ giving the average frailty and the coefficient of variation $\phi$ reflecting the spread of frailties. The corresponding probability density functions for the gamma distribution with mean $1$ and different values of the variance $\phi$ are shown in Figure $\ref{GammaDist}$. As illustrated in this figure, the shape of the gamma density strongly depends on $\phi$. For $\phi=0.01$, it approximately approaches the bell-shape of the normal distribution centered at $1$.  With increasing variance, the distribution becomes unsymmetric, heavily right-skewed and more spread. The variance parameter $\phi$ of the gamma distribution therefore determines the composition of the study population: the higher the variance, the more different the individual-specific frailties and the more frequent some specific frailties. In other words, as $\phi$ increases, the degree of susceptibility varies. \newpage 

\begin{figure}[H]
\centering
\vspace*{-29mm}
\scalebox{0.85}{ 
\includegraphics{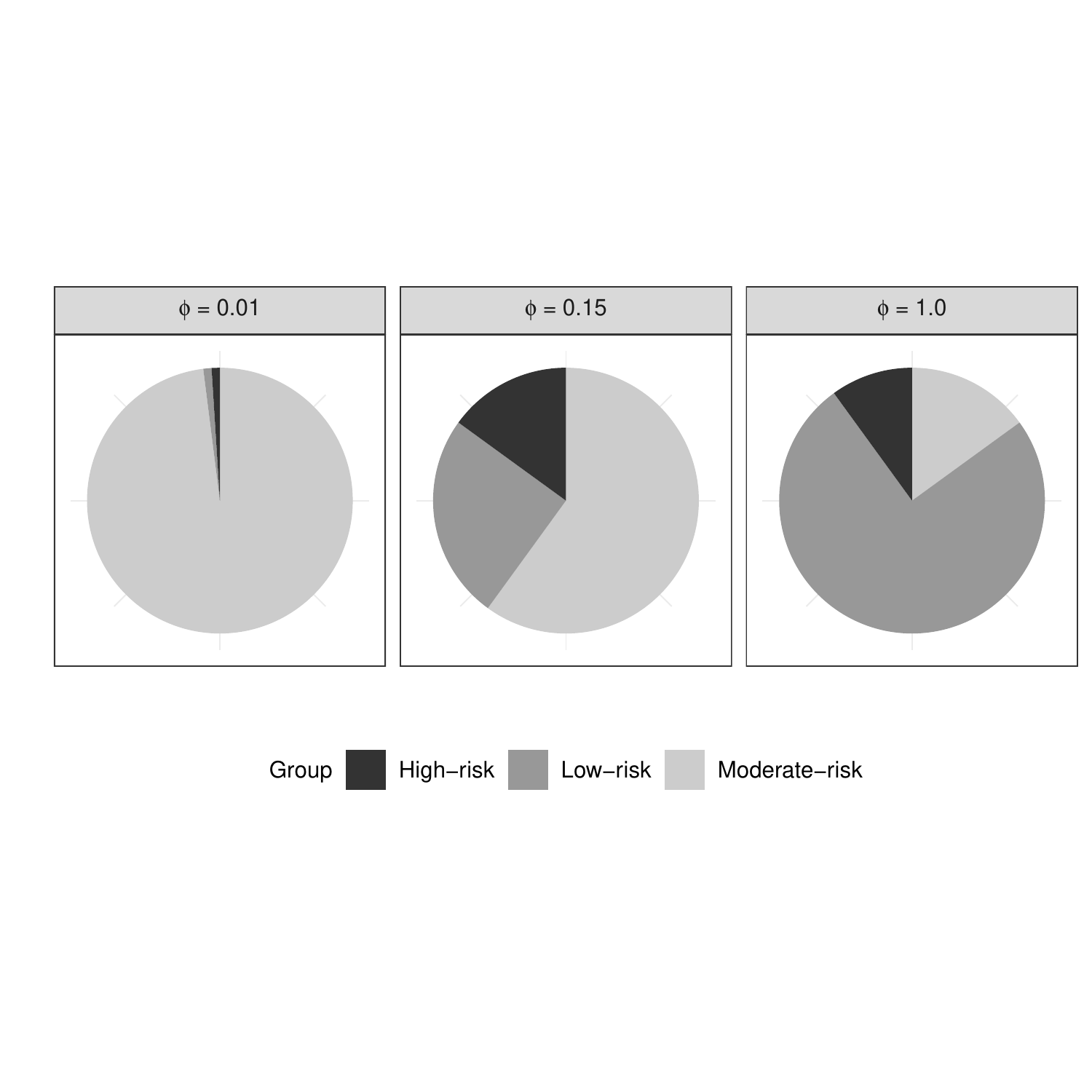}
}
\vspace*{-26mm}
\caption[Schematic distribution of high-, moderate- and low-risk patients according to different heterogeneity parameters]{Schematic distribution of high-, moderate- and low-risk patients according to different heterogeneity parameters (for illustration purposes only), high-risk patients: $U > 1$, moderate-risk patients: $0.7 \leq U \leq 1.3$, low-risk patients: $U < 1$}
\label{DistHighLowAverageGamma}
\end{figure}

In case of homogeneity ($\phi=0.0$), all patients share a common risk for disability progression and the study population simply consists of normal progressors. This can also be seen from the summary statistics added to Figure $\ref{GammaDist}$. For $\phi=0.01$, the frailty is still quite similar across the patients with small variations. If $U$ is simulated from a $\Gamma \bigl({0.15}^{-1}, {0.15}^{-1} \bigr)$ distribution, most patients have a relatively similar moderate frailty but there are also patients who deviate from the majority. Specifically, a group of patients have a high or low frailty. Thus, the study population under $\phi=0.15$ is mainly represented by moderate-risk patients, followed by low- and high-risk patients. Given $\phi=1.0$, many patients share a frailty close to $0$ and a small number of patients have moderate and high frailties. Compared to $\phi=0.15$, the study population is characterized by a higher proportion of slow progressors, a decreased number of moderate progressors and a relatively small number of fast progressors. Figure $\ref{DistHighLowAverageGamma}$ schematically summarizes the composition of the study populations according to different heterogeneity parameters.

\newpage 
\subsubsection*{Study duration and distribution of CDP12 events}
Table $\ref{S1PPMSStudyDuration}$ and Table $\ref{S1PPMSNOCDP}$ show summary statistics of the study duration and average numbers of overall CDP12 events under different scenarios. Study duration is defined as the time from the first patient randomized to the time the target number of events is reached ($n_{first.events}=246$). As the heterogeneity parameter $\phi$ increases, study duration becomes longer and higher numbers of recurrent CDP12 events can be observed. In this PPMS simulation, trials continue until $n_{first.events}=246$ first CDP12 events are observed. Under the scenarios S1/PPMS/noeffect/hetero2 and \newline S1/PPMS/effect/hetero2, study populations are dominated by low-risk patients who are less prone to progress and whose event times tend to be prolonged compared to moderate-risk patients. As a consequence, it takes much longer to reach the prespecified number of first CDP12 events, leading to an increased study duration. However, with longer follow-up, high-risk patients are capable to progress twice or even several times, which explains the higher numbers of recurrent CDP12 events, as the extent of heterogeneity increases.

\begin{table}[H]
\centering
\begin{tabular}{|l|c|c|cccc|}
\hline
                                        &                                                                                    &                              & \multicolumn{4}{c|}{\textbf{Study duration (in days)}}                                              \\ \cline{4-7} 
\multicolumn{1}{|c|}{\textbf{Scenario}} & \textbf{\begin{tabular}[c]{@{}c@{}}\\ $\boldsymbol{\exp(\beta)}$\end{tabular}} & \textbf{$\boldsymbol{\phi}$} & \textbf{$\boldsymbol{Q_{10}}$} & \textbf{Median} & \textbf{Mean} & \textbf{$\boldsymbol{Q_{90}}$} \\ \hline
\textbf{S1/PPMS/noeffect/homo}          & \multirow{3}{*}{\textbf{1.0}}                                                      & \textbf{0.0}                 & 664.86                          & 714.37          & 715.43        & 768.23                          \\
\textbf{S1/PPMS/noeffect/hetero1}       &                                                                                    & \textbf{0.15}                & 678.00                          & 728.92          & 730.64        & 785.66                          \\
\textbf{S1/PPMS/noeffect/hetero2}       &                                                                                    & \textbf{1.0}                 & 750.96                          & 820.48          & 822.68        & 897.95                          \\ \hline
\textbf{S1/PPMS/effect/homo}            & \multirow{3}{*}{\textbf{0.70}}                                                     & \textbf{0.0}                 & 769.24                          & 830.66          & 832.31        & 897.29                          \\
\textbf{S1/PPMS/effect/hetero1}         &                                                                                    & \textbf{0.15}                & 784.11                          & 848.36          & 849.98        & 918.02                          \\
\textbf{S1/PPMS/effect/hetero2}         &                                                                                    & \textbf{1.0}                 & 879.36                          & 967.22          & 969.79        & 1064.81                         \\ \hline
\end{tabular}
\caption[S1.PPMS - Summary statistics of study duration]{S1.PPMS - Summary statistics of study duration according to different treatment effect sizes and heterogeneity parameters, study duration is defined as time from first patient randomized to target number of events reached  ($n_{first.events}=246$), recurrent CDP12 events are generated from a (mixed) non-homogeneous Poisson process, $Q_{10}$ and $Q_{90}$ are the $10 \%$ and $90 \%$ quantiles, N=10000 simulations, n=1000 patients}
\label{S1PPMSStudyDuration}
\end{table}

\begin{table}[H]
\centering
\begin{tabular}{|l|c|c|cccc|}
\hline
                                        &                                                                                    &                              & \multicolumn{4}{c|}{\textbf{Total number of CDP12 events}}                                          \\ \cline{4-7} 
\multicolumn{1}{|c|}{\textbf{Scenario}} & \textbf{\begin{tabular}[c]{@{}c@{}}\\ $\boldsymbol{\exp(\beta)}$\end{tabular}} & \textbf{$\boldsymbol{\phi}$} & \textbf{$\boldsymbol{Q_{10}}$} & \textbf{Median} & \textbf{Mean} & \textbf{$\boldsymbol{Q_{90}}$} \\ \hline
\textbf{S1/PPMS/noeffect/homo}          & \multirow{3}{*}{\textbf{1.0}}                                                      & \textbf{0.0}                 & 276                             & 285             & 285           & 294                             \\
\textbf{S1/PPMS/noeffect/hetero1}       &                                                                                    & \textbf{0.15}                & 282                             & 292             & 292           & 302                             \\
\textbf{S1/PPMS/noeffect/hetero2}       &                                                                                    & \textbf{1.0}                 & 317                             & 332             & 333           & 349                             \\ \hline
\textbf{S1/PPMS/effect/homo}            & \multirow{3}{*}{\textbf{0.70}}                                                     & \textbf{0.0}                 & 278                             & 286             & 287           & 296                             \\
\textbf{S1/PPMS/effect/hetero1}         &                                                                                    & \textbf{0.15}                & 283                             & 293             & 293           & 304                             \\
\textbf{S1/PPMS/effect/hetero2}         &                                                                                    & \textbf{1.0}                 & 319                             & 335             & 336           & 353                             \\ \hline
\end{tabular}
\caption[S1.PPMS - Summary statistics of number of CDP12 events]{S1.PPMS - Summary statistics of number of CDP12 events according to different treatment effect sizes and heterogeneity parameters, recurrent CDP12 events are generated from a (mixed) non-homogeneous Poisson process, $Q_{10}$ and $Q_{90}$ are the $10 \%$ and $90 \%$ quantiles, N=10000 simulations, n=1000 patients}
\label{S1PPMSNOCDP}
\end{table}

\begin{figure}[H]
\vspace*{-35mm}
\resizebox{\textwidth}{!}{
\includegraphics{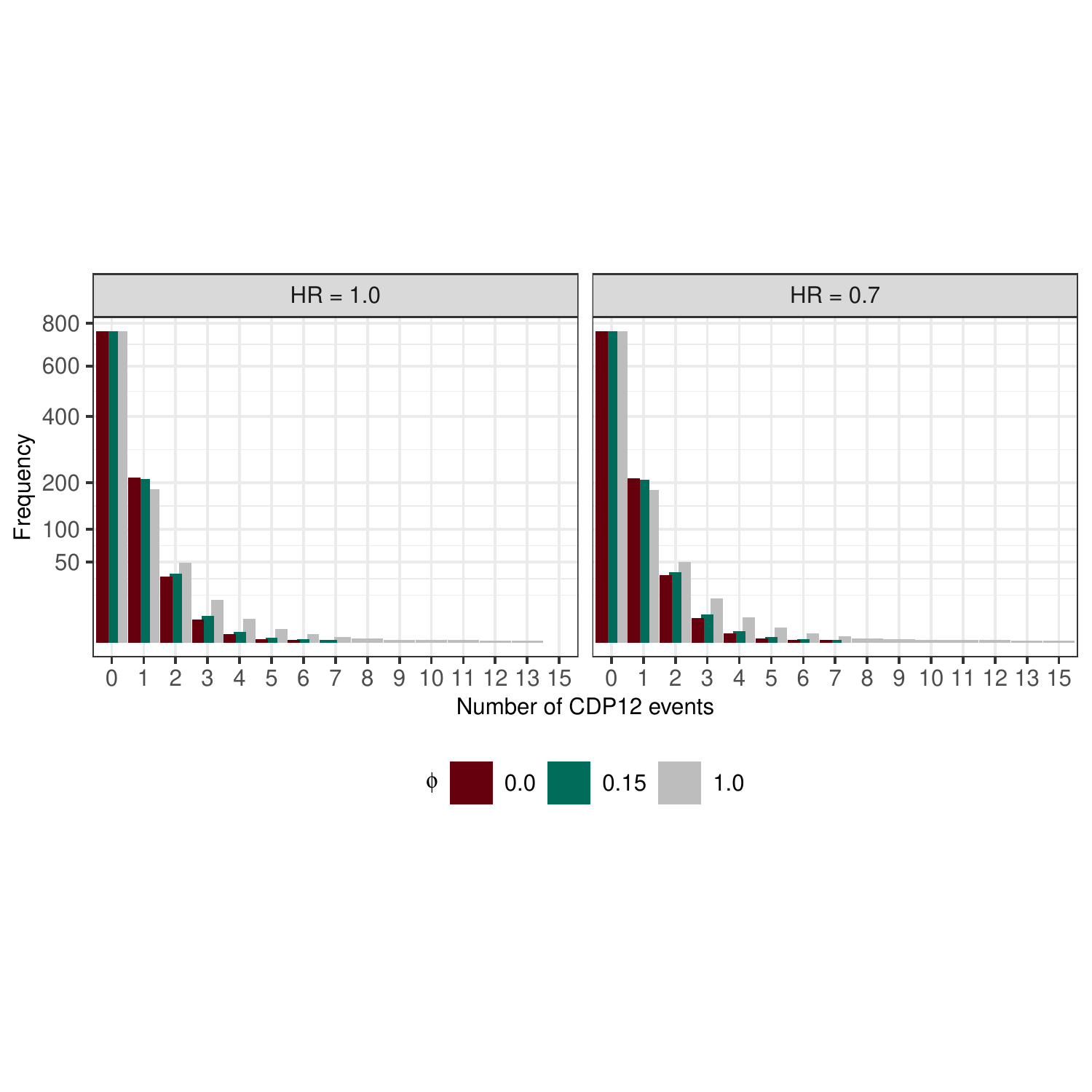}
}
\vspace*{-40mm}
\caption[S1.PPMS - Distribution of number of CDP12 events]{S1.PPMS - Distribution of number of CDP12 events according to different treatment effect sizes and heterogeneity parameters, recurrent CDP12 events are generated from a (mixed) non-homogeneous Poisson process, N=10000 simulations, n=1000 patients}
\label{S1PPMSHist}
\end{figure}

Figure $\ref{S1PPMSHist}$ represents the right-skewed distribution of the number of CDP12 events in dependence of the heterogeneity parameter. Due to the event-driven trial design, the proportion of patients without disability progression is constant across all scenarios and the distributions do not differ in the level '0 event'. While the maximum number of CDP12 events per patient is $6$ or $7$ in scenarios defined by $\phi=0.0$ and $\phi=0.15$, a small proportion of patients under $\phi=1.0$ are observed to experience even up to $13$ or $15$ CDP12 events. Caused by the increasing presence of high-risk patients, the range of the distribution becomes bigger when the between-patient variability increases. Thus, the right tail of the distribution is mainly driven by a relatively small number of high-risk patients. \newline

\subsection{Comparison of time-to-first-event and recurrent event methods}
\subsubsection*{Negative binomial model versus Poisson regression}
Table $\ref{S1NBversusPoisson}$ reports information on non-convergence of the NB model. When data was generated from a non-homogeneous Poisson process (i.e., $\phi=0.0$), the NB model did not converge in approximately $58 \%$ of all simulation runs. In case of non-convergence, the Poisson model rather than the NB model was used. For moderate heterogeneity ($\phi=0.15$), the Poisson model was still applied in $24 \%$ of all cases.  When data was generated from a mixed non-homogeneous Poisson process with $\phi=1.0$, no convergence issues were reported for the NB model. 

\begin{table}[H]
\centering
\begin{tabular}{c|cc}
                & \multicolumn{2}{c}{\textbf{\begin{tabular}[c]{@{}c@{}}Percentage of using Poisson regression\\ rather than NB model (in \%)\end{tabular}}} \\ \hline
\textbf{$\phi$} & \multicolumn{1}{c|}{$\boldsymbol{\exp(\beta)=1.0}$}                                         & $\boldsymbol{\exp(\beta)=0.7}$                                   \\ \hline
\textbf{0.0}    & \multicolumn{1}{c|}{58.08}                                                     & 58.26                                                     \\
\textbf{0.15}   & \multicolumn{1}{c|}{23.86}                                                     & 23.51                                                     \\
\textbf{1.0}    & \multicolumn{1}{c|}{0.00}                                                      & 0.00                                                     
\end{tabular}
\caption[S1.PPMS - Convergence issues of negative binomial model]{S1.PPMS - Convergence issues of NB model, N=10000 simulations, n=1000 patients}
\label{S1NBversusPoisson}
\end{table}

\newpage
\subsubsection*{Treatment effect estimation}
\begin{figure}[H]
\centering
\scalebox{0.40}{
\includegraphics{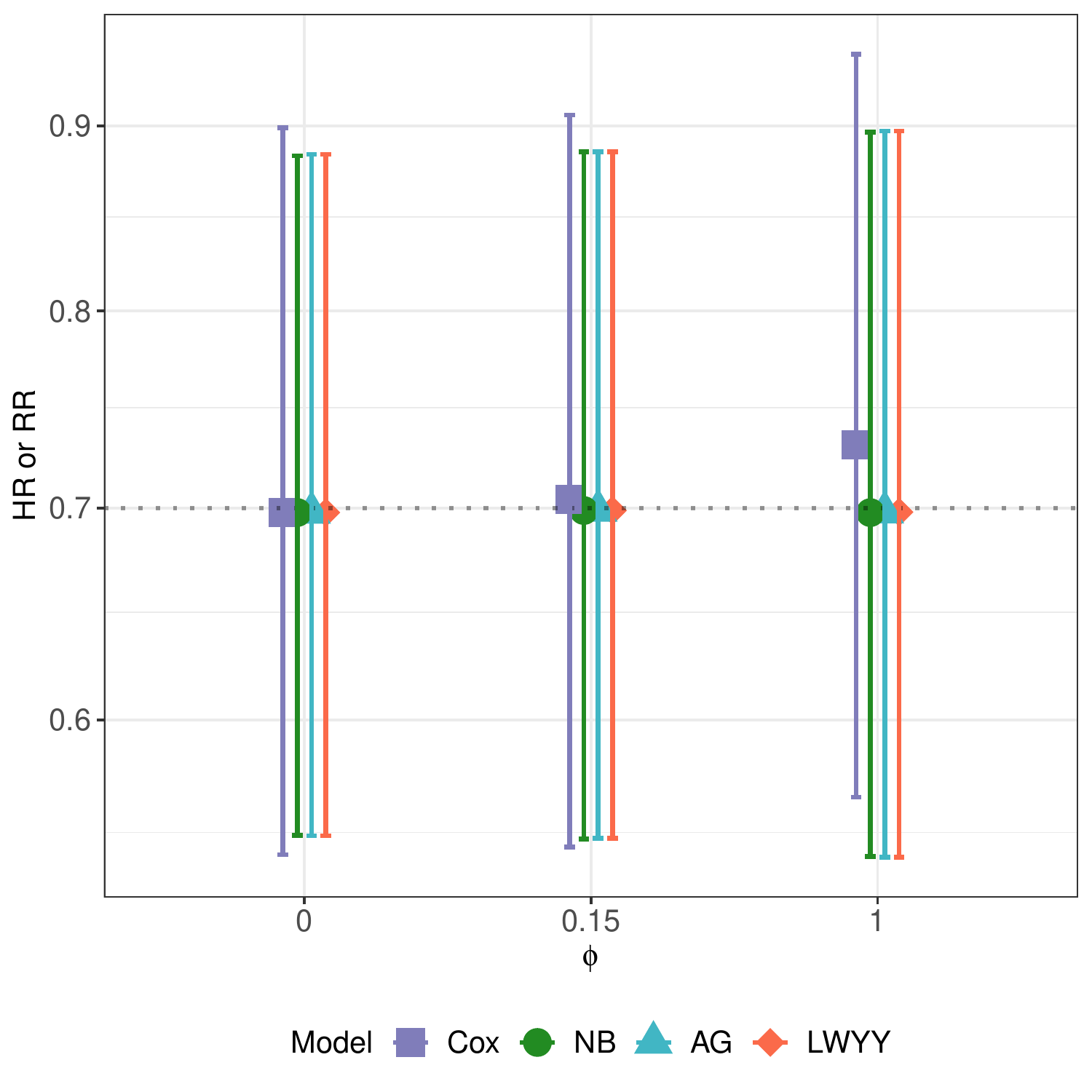}
}
\caption[S1.PPMS - Treatment effect estimates obtained from time-to-first-event and recurrent event methods in generic simulation study]{S1.PPMS - Treatment effect estimates obtained from time-to-first-event and recurrent event methods according to different heterogeneity parameters, error bars indicate $2.5\%$ and $97.5\%$ quantiles, N=10000 simulations, n=1000 patients}
\label{S1.PPMS.precision}
\end{figure}

\begin{figure}[H]
\centering
\scalebox{0.60}{
\includegraphics{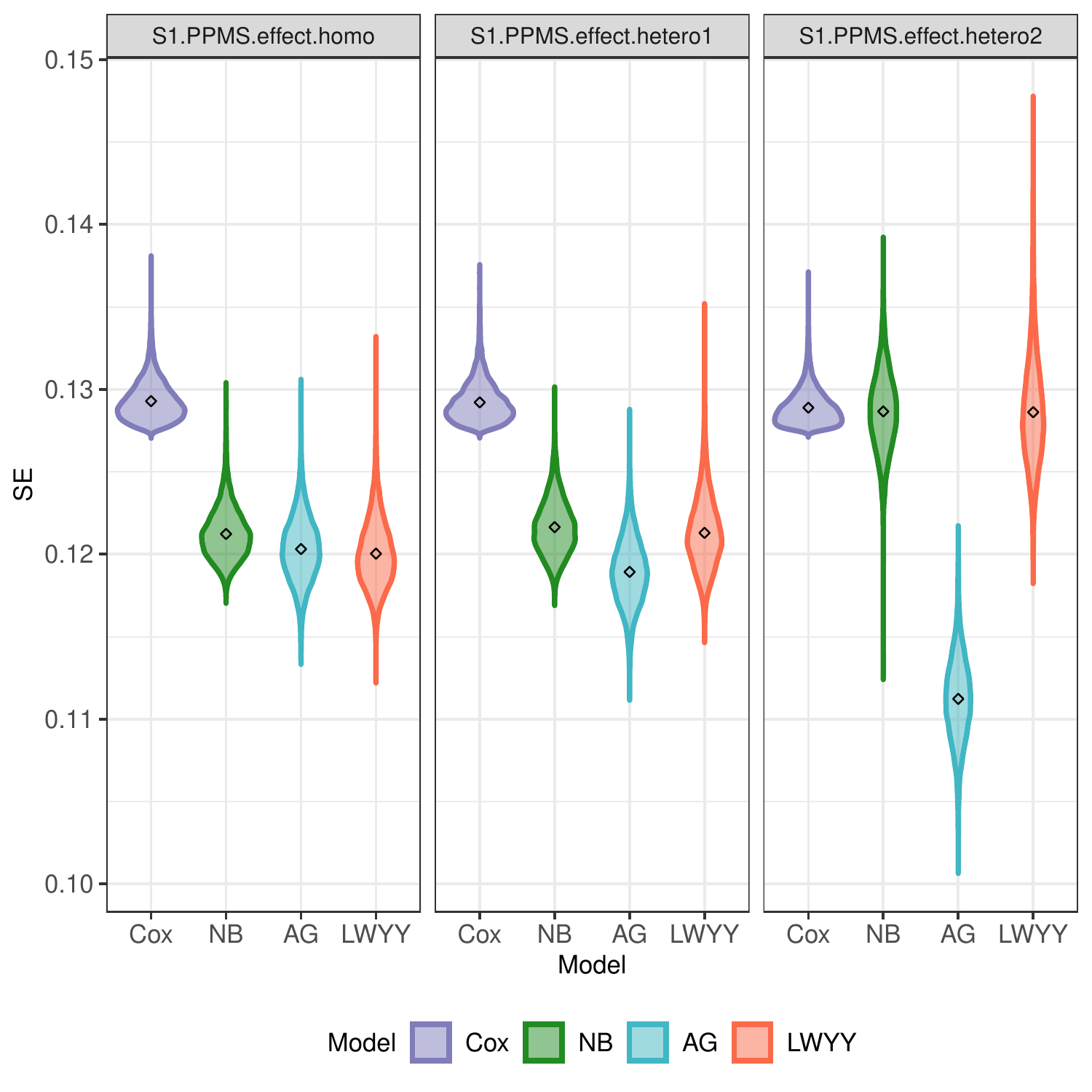}
}
\caption[S1.PPMS - Distribution of standard error estimates in the generic simulation study]{S1.PPMS - Distribution of estimated standard errors under $H_{1}: \{ \beta=\log(0.7) \}$, N=10000 simulations, n=1000 patients} 
\label{S1.PPMS.violin.se}
\end{figure}

\textbf{Case: $\boldsymbol{\beta=\log(0.7)}$}  \newline 
Results of the PPMS simulation study under $H_{1}: \{ \beta=\log(0.7) \}$ are presented in Table $\ref{SimulationResultsS1H1}$. 
When data is simulated from a non-homogeneous Poisson process $(\phi=0.0)$, treatment effect estimates resulting from the Cox, NB, AG and LWYY models are approximately unbiased. However, all recurrent event methods (NB, AG and LWYY models) provide lower MSEs compared to the Cox proportional hazards model. Since the rate-based LWYY model and the intensity-based AG model yield the same treatment effect estimate, the resulting bias, MSE and Monte Carlo SD are exactly the same under both approaches but interpretation differs. While the AG model assumes a naive variance estimator and estimates a HR, the LWYY model makes use of robust variance estimation and gives an estimated RR. The SE estimates are pretty close to the empirically determined Monte Carlo SD, and this applies especially for the Cox and the NB model. The coverage probability is around $95 \%$ with all methods. Although all methods yield valid inferences, recurrent event methods achieve greater precision in the treatment effect estimate than the conventional time-to-first-event method. As expected, results from the AG and LWYY analyses do not show major differences when data is generated from a non-homogeneous Poisson process. \newline
When $\phi \neq 0$, the estimated regression coefficients obtained from the Cox model are biased towards $0$ by an amount that depends on the variability of the frailty term. In case of $\phi=0.15$, the average ${\textnormal{HR}}_{Cox}$ of $0.7042$ is still very close to the true HR and the bias is therefore negligible. However, when going from $\phi=0.15$ to $\phi=1.0$, the bias increases notably. For instance, under $\phi=1.0$, the Cox model estimates an average ${\textnormal{HR}}_{Cox}$ of 0.7314, leading to an underestimation of the treatment effect of around $3 \%$ (see Figure $\ref{S1.PPMS.precision}$). In contrast, recurrent event methods continue to yield approximately unbiased estimates of the treatment effect as frailty variance increases, but with a loss of precision compared to $\phi=0.0$. As seen from Figure $\ref{S1.PPMS.precision}$, the treatment effect estimates across the simulation replicates become more variable leading to higher Monte Carlo SDs and increased MSEs, when $\phi$ increases. The accuracy of average SE estimates to the Monte Carlo SD decreases, as the variance of the frailty increases. The LWYY model gives identical estimates to the AG model but with an appropriately larger estimate of the SE. Due to the fact that the AG approach does not account for within-patient correlation induced by the frailty term, the corresponding SE estimates of the AG model become too small with increasing heterogeneity and the naive variance estimator underestimates the true variance. This leads to a reduced coverage probability of $90.7 \%$ for $\phi=1.0$. 
Moreover, Figure $\ref{S1.PPMS.violin.se}$ illustrates the distribution of the estimated SEs under $H_{1}: \{ \beta=\log(0.7) \}$ for each model. The NB and LWYY approaches result in a similar mean SE (cf. Table $\ref{SimulationResultsS1H1}$) but the variance is considerably larger for LWYY in all scenarios. This may be due to the robust sandwich estimator which is sometimes rather imprecise.

\begin{landscape}
\begin{table}[]
\centering
\scalebox{0.8}{
\begin{tabular}{|c|cccccccc|cccccccc|}
\hline
\multirow{2}{*}{\textbf{$\boldsymbol{\phi}$}} & \multicolumn{8}{c|}{\textbf{Time-to-first-event method}}                                                                                                                                                                                                                                                                                                                                                                                                                                                 & \multicolumn{8}{c|}{\textbf{Recurrent event methods}}                                                                                                                                                                                                                                                                                                                                                                                                                                                                                                     \\ \cline{2-17} 
                                 & \textbf{\begin{tabular}[c]{@{}c@{}}Model\\ $$\end{tabular}} & \textbf{\begin{tabular}[c]{@{}c@{}}HR\\ $$\end{tabular}} & \textbf{\begin{tabular}[c]{@{}c@{}}Bias\\ $$\end{tabular}} & \textbf{\begin{tabular}[c]{@{}c@{}}MSE\\ $$\end{tabular}} & \textbf{\begin{tabular}[c]{@{}c@{}}SE\\ $$\end{tabular}} & \textbf{\begin{tabular}[c]{@{}c@{}}SEE\\ $$\end{tabular}} & \textbf{\begin{tabular}[c]{@{}c@{}}CP\\ $$\end{tabular}} & \textbf{\begin{tabular}[c]{@{}c@{}}Type I \\ $$error$$\end{tabular}} & \textbf{\begin{tabular}[c]{@{}c@{}}Model\\ $$\end{tabular}}      & \textbf{\begin{tabular}[c]{@{}c@{}}HR/RR\\ $$\end{tabular}}         & \textbf{\begin{tabular}[c]{@{}c@{}}Bias\\ $$\end{tabular}}          & \textbf{\begin{tabular}[c]{@{}c@{}}MSE \\ $$\end{tabular}}       & \textbf{\begin{tabular}[c]{@{}c@{}}SE\\ $$\end{tabular}}         & \textbf{\begin{tabular}[c]{@{}c@{}}SEE\\ $$\end{tabular}}        & \textbf{\begin{tabular}[c]{@{}c@{}}CP\\ $$\end{tabular}}      & \textbf{\begin{tabular}[c]{@{}c@{}}Type I \\ $$error$$\end{tabular}} \\ \hline
\textbf{0.0}                     & \textbf{Cox}                                                & \begin{tabular}[c]{@{}c@{}}$$\\ 0.9997\\ $$\end{tabular} & \begin{tabular}[c]{@{}c@{}}$$\\ -0.0003\\ $$\end{tabular}  & \begin{tabular}[c]{@{}c@{}}$$\\ 0.0161\\ $$\end{tabular}  & \begin{tabular}[c]{@{}c@{}}$$\\ 0.1268\\ $$\end{tabular} & \begin{tabular}[c]{@{}c@{}}$$\\ 0.1277\\ $$\end{tabular}  & \begin{tabular}[c]{@{}c@{}}$$\\ 0.953\\ $$\end{tabular}  & \begin{tabular}[c]{@{}c@{}}$$\\ 0.047\\ $$\end{tabular}              & \textbf{\begin{tabular}[c]{@{}c@{}}NB\\ AG\\ LWYY\end{tabular}}  & \begin{tabular}[c]{@{}c@{}}0.9993\\ 0.9993\\ 0.9993\end{tabular} & \begin{tabular}[c]{@{}c@{}}-0.0007\\ -0.0007\\ -0.0007\end{tabular} & \begin{tabular}[c]{@{}c@{}}0.0139\\ 0.0139\\ 0.0139\end{tabular} & \begin{tabular}[c]{@{}c@{}}0.1178\\ 0.1178\\ 0.1178\end{tabular} & \begin{tabular}[c]{@{}c@{}}0.1196\\ 0.1187\\ 0.1184\end{tabular} & \begin{tabular}[c]{@{}c@{}}0.952\\ 0.952\\ 0.951\end{tabular} & \begin{tabular}[c]{@{}c@{}}0.047\\ 0.048\\ 0.049\end{tabular}        \\ \hline
\textbf{0.15}                    & \textbf{Cox}                                                & \begin{tabular}[c]{@{}c@{}}$$\\ 0.9991\\ $$\end{tabular} & \begin{tabular}[c]{@{}c@{}}$$\\ -0.0009\\ $$\end{tabular}  & \begin{tabular}[c]{@{}c@{}}$$\\ 0.0161\\ $$\end{tabular}  & \begin{tabular}[c]{@{}c@{}}$$\\ 0.1270\\ $$\end{tabular} & \begin{tabular}[c]{@{}c@{}}$$\\ 0.1277\\ $$\end{tabular}  & \begin{tabular}[c]{@{}c@{}}$$\\ 0.952\\ $$\end{tabular}  & \begin{tabular}[c]{@{}c@{}}$$\\ 0.048\\ $$\end{tabular}              & \textbf{\begin{tabular}[c]{@{}c@{}}NB\\ AG \\ LWYY\end{tabular}} & \begin{tabular}[c]{@{}c@{}}0.9994\\ 0.9994\\ 0.9994\end{tabular} & \begin{tabular}[c]{@{}c@{}}-0.0006\\ -0.0006\\ -0.0006\end{tabular} & \begin{tabular}[c]{@{}c@{}}0.0144\\ 0.0143\\ 0.0143\end{tabular} & \begin{tabular}[c]{@{}c@{}}0.1198\\ 0.1198\\ 0.1198\end{tabular} & \begin{tabular}[c]{@{}c@{}}0.1201\\ 0.1173\\ 0.1197\end{tabular} & \begin{tabular}[c]{@{}c@{}}0.948\\ 0.945\\ 0.948\end{tabular} & \begin{tabular}[c]{@{}c@{}}0.051\\ 0.055\\ 0.052\end{tabular}        \\ \hline
\textbf{1.0}                     & \textbf{Cox}                                                & \begin{tabular}[c]{@{}c@{}}$$\\ 1.0036\\ $$\end{tabular} & \begin{tabular}[c]{@{}c@{}}$$\\ 0.0036\\ $$\end{tabular}   & \begin{tabular}[c]{@{}c@{}}$$\\ 0.0164\\ $$\end{tabular}  & \begin{tabular}[c]{@{}c@{}}$$\\ 0.1278\\ $$\end{tabular} & \begin{tabular}[c]{@{}c@{}}$$\\ 0.1277\\ $$\end{tabular}  & \begin{tabular}[c]{@{}c@{}}$$\\ 0.952\\ $$\end{tabular}  & \begin{tabular}[c]{@{}c@{}}$$ \\ 0.048\\ $$\end{tabular}             & \textbf{\begin{tabular}[c]{@{}c@{}}NB\\ AG\\ LWYY\end{tabular}}  & \begin{tabular}[c]{@{}c@{}}1.0042\\ 1.0042\\ 1.0042\end{tabular} & \begin{tabular}[c]{@{}c@{}}0.0042\\ 0.0041\\ 0.0041\end{tabular}    & \begin{tabular}[c]{@{}c@{}}0.0165\\ 0.0164\\ 0.0164\end{tabular} & \begin{tabular}[c]{@{}c@{}}0.1283\\ 0.1282\\ 0.1282\end{tabular} & \begin{tabular}[c]{@{}c@{}}0.1275\\ 0.1100\\ 0.1275\end{tabular} & \begin{tabular}[c]{@{}c@{}}0.947\\ 0.909\\ 0.948\end{tabular} & \begin{tabular}[c]{@{}c@{}}0.053\\ 0.091\\ 0.052\end{tabular}        \\ \hline
\end{tabular}}
\caption[S1.PPMS - Results of the general simulation study ($\beta=\log(1.0)$) ]{S1.PPMS - Results of the general simulation study when the true treatment effect is $\beta=\log(1.0)$, recurrent CDP12 events are generated from a (mixed) non-homogeneous Poisson process, N=10000 simulations, n=1000 patients (1:1 randomization) \\
Evaluation measures: HR/RR = mean treatment effect across simulations, Bias = mean of the estimators of $\beta$ minus $\beta$, MSE = mean squared error, SE = standard deviation of estimators across simulations, $SEE$ = mean standard error across simulations, CP = coverage probabilities of the corresponding $95 \%$ CIs}
\label{SimulationResultsS1H0}
\end{table}

\begin{table}[]
\centering
\scalebox{0.80}{ 
\begin{tabular}{|c|cccccccc|cccccccc|}
\hline
\multirow{2}{*}{\textbf{$\boldsymbol{\phi}$}} & \multicolumn{8}{c|}{\textbf{Time-to-first-event method}}                                                                                                                                                                                                                                                                                                                                                                                                                                         & \multicolumn{8}{c|}{\textbf{Recurrent event methods}}                                                                                                                                                                                                                                                                                                                                                                                                                                                                                              \\ \cline{2-17} 
                                              & \textbf{\begin{tabular}[c]{@{}c@{}}Model\\ $$\end{tabular}} & \textbf{\begin{tabular}[c]{@{}c@{}}HR\\ $$\end{tabular}} & \textbf{\begin{tabular}[c]{@{}c@{}}Bias\\ $$\end{tabular}} & \textbf{\begin{tabular}[c]{@{}c@{}}MSE\\ $$\end{tabular}} & \textbf{\begin{tabular}[c]{@{}c@{}}SE\\ $$\end{tabular}} & \textbf{\begin{tabular}[c]{@{}c@{}}SEE\\ $$\end{tabular}} & \textbf{\begin{tabular}[c]{@{}c@{}}CP\\ $$\end{tabular}} & \textbf{\begin{tabular}[c]{@{}c@{}}Power \\ $$\end{tabular}} & \textbf{\begin{tabular}[c]{@{}c@{}}Model\\ $$\end{tabular}}      & \textbf{\begin{tabular}[c]{@{}c@{}}HR/RR\\ $$\end{tabular}}         & \textbf{\begin{tabular}[c]{@{}c@{}}Bias\\ $$\end{tabular}}          & \textbf{\begin{tabular}[c]{@{}c@{}}MSE \\ $$\end{tabular}}       & \textbf{\begin{tabular}[c]{@{}c@{}}SE\\ $$\end{tabular}}         & \textbf{\begin{tabular}[c]{@{}c@{}}SEE\\ $$\end{tabular}}        & \textbf{\begin{tabular}[c]{@{}c@{}}CP\\ $$\end{tabular}}      & \textbf{\begin{tabular}[c]{@{}c@{}}Power\\ $$\end{tabular}}   \\ \hline
\textbf{0.0}                                  & \textbf{Cox}                                                & \begin{tabular}[c]{@{}c@{}}$$\\ 0.6978\\ $$\end{tabular} & \begin{tabular}[c]{@{}c@{}}$$\\ -0.0032\\ $$\end{tabular}  & \begin{tabular}[c]{@{}c@{}}$$\\ 0.0168\\ $$\end{tabular}  & \begin{tabular}[c]{@{}c@{}}$$\\ 0.1294\\ $$\end{tabular} & \begin{tabular}[c]{@{}c@{}}$$\\ 0.1293\\ $$\end{tabular}  & \begin{tabular}[c]{@{}c@{}}$$\\ 0.950\\ $$\end{tabular}  & \begin{tabular}[c]{@{}c@{}}$$\\ 0.800\\ $$\end{tabular}      & \textbf{\begin{tabular}[c]{@{}c@{}}NB\\ AG\\ LWYY\end{tabular}}  & \begin{tabular}[c]{@{}c@{}}0.6978\\ 0.6978\\ 0.6978\end{tabular} & \begin{tabular}[c]{@{}c@{}}-0.0031\\ -0.0031\\ -0.0031\end{tabular} & \begin{tabular}[c]{@{}c@{}}0.0147\\ 0.0147\\ 0.0147\end{tabular} & \begin{tabular}[c]{@{}c@{}}0.1213\\ 0.1214\\ 0.1214\end{tabular} & \begin{tabular}[c]{@{}c@{}}0.1212\\ 0.1203\\ 0.1200\end{tabular} & \begin{tabular}[c]{@{}c@{}}0.949\\ 0.948\\ 0.948\end{tabular} & \begin{tabular}[c]{@{}c@{}}0.848\\ 0.850\\ 0.850\end{tabular} \\ \hline
\textbf{0.15}                                 & \textbf{Cox}                                                & \begin{tabular}[c]{@{}c@{}}$$\\ 0.7042\\ $$\end{tabular} & \begin{tabular}[c]{@{}c@{}}$$\\ 0.0060\\ $$\end{tabular}   & \begin{tabular}[c]{@{}c@{}}$$\\ 0.0170\\ $$\end{tabular}  & \begin{tabular}[c]{@{}c@{}}$$\\ 0.1301\\ $$\end{tabular} & \begin{tabular}[c]{@{}c@{}}$$\\ 0.1292\\ $$\end{tabular}  & \begin{tabular}[c]{@{}c@{}}$$\\ 0.950\\ $$\end{tabular}  & \begin{tabular}[c]{@{}c@{}}$$\\ 0.779\\ $$\end{tabular}      & \textbf{\begin{tabular}[c]{@{}c@{}}NB\\ AG \\ LWYY\end{tabular}} & \begin{tabular}[c]{@{}c@{}}0.6988\\ 0.6989\\ 0.6989\end{tabular} & \begin{tabular}[c]{@{}c@{}}-0.0017\\ -0.0016\\ -0.0016\end{tabular} & \begin{tabular}[c]{@{}c@{}}0.0150\\ 0.0150\\ 0.0150\end{tabular} & \begin{tabular}[c]{@{}c@{}}0.1225\\ 0.1224\\ 0.1224\end{tabular} & \begin{tabular}[c]{@{}c@{}}0.1216\\ 0.1189\\ 0.1213\end{tabular} & \begin{tabular}[c]{@{}c@{}}0.947\\ 0.944\\ 0.948\end{tabular} & \begin{tabular}[c]{@{}c@{}}0.840\\ 0.849\\ 0.842\end{tabular} \\ \hline
\textbf{1.0}                                  & \textbf{Cox}                                                & \begin{tabular}[c]{@{}c@{}}$$\\ 0.7314\\ $$\end{tabular} & \begin{tabular}[c]{@{}c@{}}$$\\ 0.0439\\ $$\end{tabular}   & \begin{tabular}[c]{@{}c@{}}$$\\ 0.0188\\ $$\end{tabular}  & \begin{tabular}[c]{@{}c@{}}$$\\ 0.1299\\ $$\end{tabular} & \begin{tabular}[c]{@{}c@{}}$$\\ 0.1289\\ $$\end{tabular}  & \begin{tabular}[c]{@{}c@{}}$$\\ 0.934\\ $$\end{tabular}  & \begin{tabular}[c]{@{}c@{}}$$\\ 0.680\\ $$\end{tabular}      & \textbf{\begin{tabular}[c]{@{}c@{}}NB\\ AG\\ LWYY\end{tabular}}  & \begin{tabular}[c]{@{}c@{}}0.6978\\ 0.6981\\ 0.6981\end{tabular} & \begin{tabular}[c]{@{}c@{}}-0.0031\\ -0.0028\\ -0.0028\end{tabular} & \begin{tabular}[c]{@{}c@{}}0.0169\\ 0.0169\\ 0.0169\end{tabular} & \begin{tabular}[c]{@{}c@{}}0.1300\\ 0.1300\\ 0.1300\end{tabular} & \begin{tabular}[c]{@{}c@{}}0.1287\\ 0.1112\\ 0.1286\end{tabular} & \begin{tabular}[c]{@{}c@{}}0.948\\ 0.907\\ 0.949\end{tabular} & \begin{tabular}[c]{@{}c@{}}0.801\\ 0.865\\ 0.801\end{tabular} \\ \hline
\end{tabular}}
\caption[S1.PPMS - Results of the general simulation study ($\beta=\log(0.7)$) ]{S1.PPMS - Results of the general simulation study when the true treatment effect is $\beta=\log(0.7)$, recurrent CDP12 events are generated from a (mixed) non-homogeneous Poisson process, N=10000 simulations, n=1000 patients (1:1 randomization) \\
Evaluation measures: HR/RR = mean treatment effect across simulations, Bias = mean of the estimators of $\beta$ minus $\beta$, MSE = mean squared error, SE = standard deviation of estimators across simulations, $SEE$ = mean standard error across simulations, CP = coverage probabilities of the corresponding $95 \%$ CIs}
\label{SimulationResultsS1H1}
\end{table}
\end{landscape}

\textbf{Selection effects} \newline
The commonly used Cox model as a time-to-first-event approach is misspecified in presence of heterogeneity due to selection effects. In time-to-first-event analyses, high-risk patients tend to experience their first event earlier and they consequently leave the risk set sooner than low-risk patients who are less frail to progress. Hence, the at-risk population undergoes a compositional change over time: the number of high-risk patients rapidly decreases with larger $t$, leaving the less frail patients to dominate in the at-risk set at later follow-up times. Such a depletion of susceptible patients induces continuously decreasing event rates over time and happens more quickly in the control group. Therefore, heterogeneity leads to an attenuation of the treatment difference over time, which concurrently implies a violation of the proportional hazards assumption. As a result of non-proportional hazards, the effect of treatment is underestimated in the Cox model. \newline
In recurrent event settings, patients remain at risk after experiencing the first event and high-risk patients still contribute to later follow-up. Thus, selection effects do not take place when considering recurrent events and the NB, AG and LWYY models can provide unbiased treatment effect estimates even in the presence of heterogeneity. \newline
There exists substantial evidence in biostatistical literature that confirms this finding from the generic simulation study: if between-patient variability is present but not accounted for in a Cox proportional hazards model, underestimation of covariate effects can be observed \citep{Aalen1994, Aalen2014, Aalen2015, McNamee2017}. Selection bias in the Cox model has also been demonstrated in several simulation studies with data following a homogeneous or non-homogeneous mixed Poisson process \parencite{Metcalfe2006, Cheung2010, Hengelbrock2016, JahnEimermacher2017}. 
\newline \newline 
\textbf{Case: $\boldsymbol{\beta=\log(1.0)}$} \newline 
Table $\ref{SimulationResultsS1H0}$ makes the same comparisons between different statistical methods for data generated under $H_{0}: \{ \beta=\log(1.0) \}$. In this case, both time-to-first-event and recurrent event methods are able to provide unbiased treatment effect estimates, even in presence of heterogeneity. Specifically, the performance of the Cox proportional hazards model is not affected by between-patient variability. Apart from that, similar results as under $H_{1}: \{ \beta=\log(0.7) \}$ can be observed. In particular, when $\phi = 1.0$, the mean SE estimate for the AG model has little bias and, thus, the confidence intervals do not have proper coverage probabilities. 

\newpage 
\subsubsection*{Power and type I error} 
\begin{figure}[H]
\centering
\begin{minipage}{\textwidth}
  \begin{minipage}[c]{0.49\textwidth}
    \centering
\scalebox{0.50}{
\includegraphics{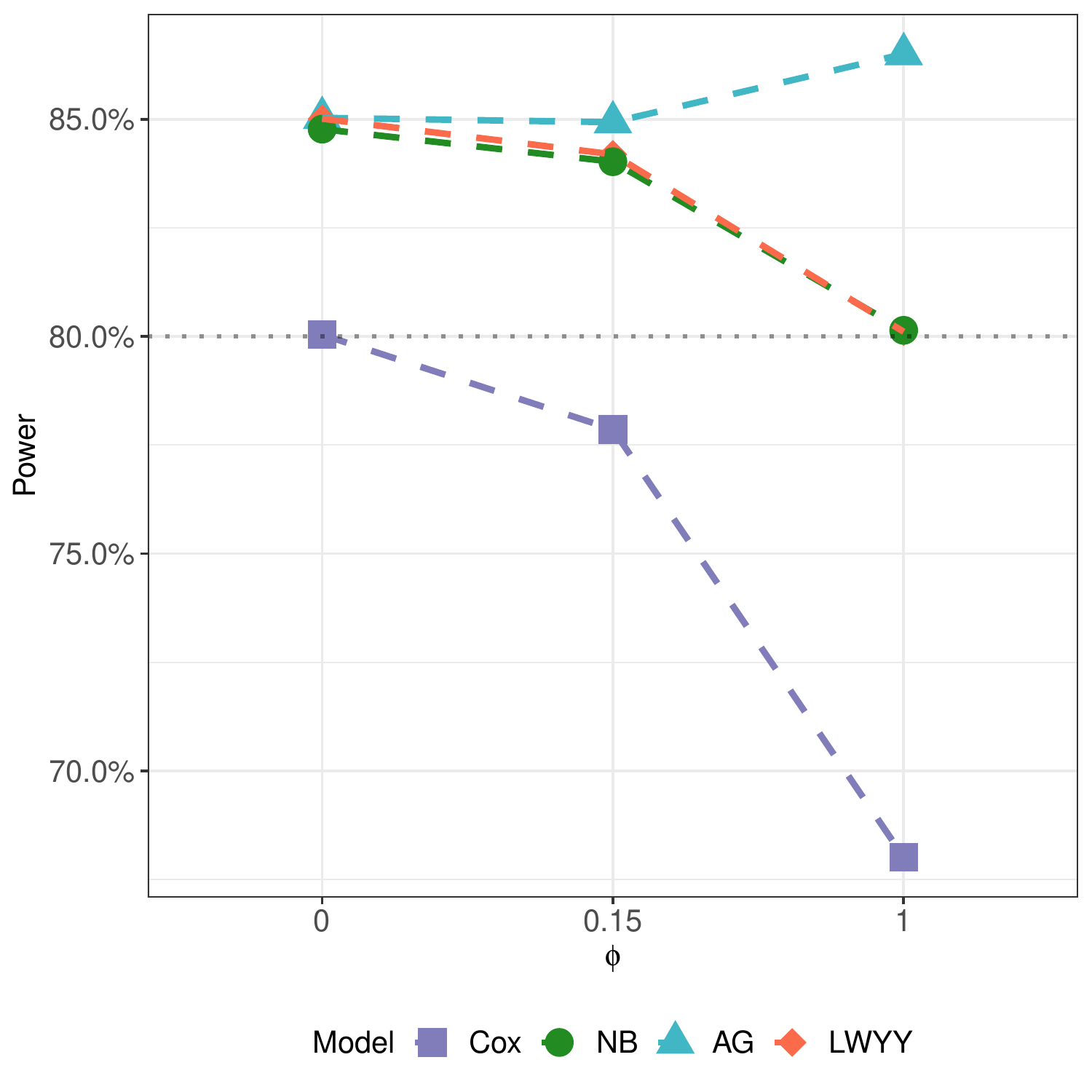}
}
\end{minipage}
\hfill
\begin{minipage}[c]{0.49\textwidth}
\centering 
\scalebox{0.50}{
\includegraphics{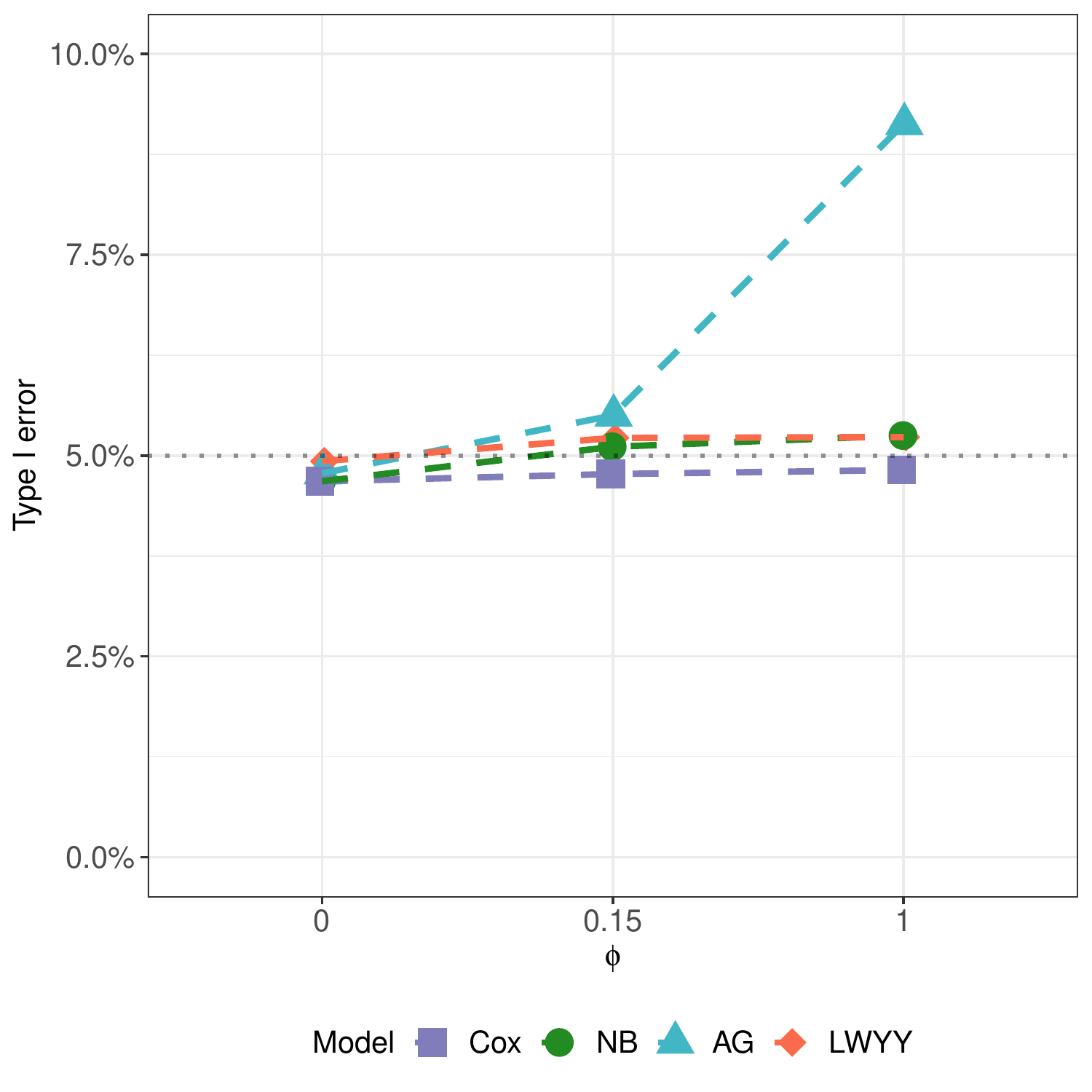}
}
\end{minipage}
$$
$$
\begin{minipage}[c]{\textwidth}
\centering
\scalebox{0.82}{
\begin{tabular}{|c|c|ccc|c|c|ccc|}
\hline
                       & \multicolumn{4}{c|}{\textbf{\begin{tabular}[c]{@{}c@{}}Power = $\boldsymbol{1 - \beta}$\\ (in \%)\end{tabular}}}                                                                                 &           & \multicolumn{4}{c|}{\textbf{\begin{tabular}[c]{@{}c@{}}Type I error = $\boldsymbol{\alpha}$\\ (in \%)\end{tabular}}}                                                                              \\ \cline{1-5} \cline{7-10} 
\multicolumn{1}{|l|}{} & \textbf{\begin{tabular}[c]{@{}c@{}}Time-to-$1^{st}$-event \\ method\end{tabular}} & \multicolumn{3}{c|}{\textbf{\begin{tabular}[c]{@{}c@{}}Recurrent event \\ methods\end{tabular}}} & \textbf{} & \textbf{\begin{tabular}[c]{@{}c@{}}Time-to-$1^{st}$-event \\ method\end{tabular}} & \multicolumn{3}{c|}{\textbf{\begin{tabular}[c]{@{}c@{}}Recurrent event \\ methods\end{tabular}}} \\ \cline{1-5} \cline{7-10} 
\textbf{$\boldsymbol{\phi}$}        & \textbf{Cox}                                                                      & \textbf{NB}                    & \textbf{AG}                   & \textbf{LWYY}                   &           & \textbf{Cox}                                                                      & \textbf{NB}                    & \textbf{AG}                   & \textbf{LWYY}                   \\ \cline{1-5} \cline{7-10} 
\textbf{0.0}           & 80.0                                                                             & 84.8                          & 85.0                         & 85.0                           &           & 4.7                                                                              & 4.7                           & 4.8                          & 4.9                            \\
\textbf{0.15}          & 77.9                                                                             & 84.0                          & 84.9                         & 84.2                           &           & 4.8                                                                              & 5.1                           & 5.5                          & 5.2                            \\
\textbf{1.0}           & 68.0                                                                             & 80.1                          & 86.5                         & 80.1                           &           & 4.8                                                                              & 5.3                           & 9.1                          & 5.2                            \\ \hline
\end{tabular}
}
\end{minipage}
\caption[S1.PPMS - Power and type I error]{S1.PPMS - Power and type I error (based on Table $\ref{SimulationResultsS1H0}$ and Table $\ref{SimulationResultsS1H1}$), N=10000 simulations, n=1000 patients}
\label{S1.PPMS.power}
\end{minipage}
\end{figure}

Figure $\ref{S1.PPMS.power}$ contains the empirical type I error rates of the statistical tests when $\exp(\beta)=1.0$ and the power when $\exp(\beta)=0.7$. Under the non-homogeneous Poisson process ($\phi=0.0$), there is an adequate control of the type I error based on the Cox, NB, AG and LWYY analyses, with probabilities less than $5 \%$. In the presence of heterogeneity, the tests based on the Cox, NB and LWYY models satisfy the nominal type I error rate, whereas the AG model fails to control it. The low standard error estimates resulting from the AG approach when data is generated from a mixed non-homogeneous Poisson process cause the inflation of the type I error. It turns out that the higher the frailty variance, the more extreme the inflation of the type I error rate. \newline 
The left panel of Figure $\ref{S1.PPMS.power}$ plots the trend of empirical power as a function of $\phi$. Typically, it can be concluded that recurrent event analyses generally outperform the time-to-first-event approach in terms of statistical power, provided that treatment does not only affect the timing of the first event but also continues to affect subsequent events as well. Under homogeneity, the power of tests based on recurrent event methods is on average increased by $5\%$, as compared to the Cox model. According to the study design, the Cox time-to-first-event analysis gives a power of $80.0 \%$, as the simulation study was originally powered for $80\%$ for the time-to-first-event endpoint to detect the true hazard ratio of $0.70$. The power obtained from the tests based on NB, AG and LWYY are equal to $84.8\%$, $85.0\%$ and $85.0\%$. As the extent of heterogeneity increases, a decreasing trend of the empirical power can be observed for all approaches, except for the AG model, where the increased power is driven by the higher type I error rate under $H_{0}$. However, the loss of power is of greater magnitude in Cox than in NB or LWYY analyses. When going from $\phi=0.0$ to $\phi=1.0$, the power of the statistical test based on the Cox model reduces from $80\%$ to $68 \%$, caused by the underestimation of the treatment effect in presence of heterogeneity. In contrast, the power of the NB and LWYY analyses remains at approximately $80 \%$ in scenarios with $\phi=1.0$. 

\subsubsection*{Implication for sample size} 
It is of major interest to roughly assess the potential savings of the study design based on recurrent events relative to the study design based on first events only in terms of sample size. To illustrate the sample size implications for the generic simulation, arguments rely on the Schoenfeld sample size formula and homogeneity ($\phi=0.0$).
\begin{enumerate}
\item As seen from Figure $\ref{S1.PPMS.power}$ and Table $\ref{SimulationResultsS1H1}$, all recurrent event methods provide greater statistical power compared to the time-to-first-event method. In simulations with $\phi=0.0$, power increased from $80 \%$ for the time-to-first-event method to approximately $85 \%$ for the recurrent event analyses. 
\item A recurrent event analysis with $80 \%$ rather than $85 \%$ power would require $ \sim 12.6 \%$ less recurrent events. This is due to the fact that the number of events required to detect the treatment effect is proportional to $(z_{1-\alpha/2} + z_{1-\gamma})^2$ (cf. Section $\ref{SamplesizeRCT}$). \newline 
$\bigl[$ Calculation: $1 - \dfrac{{(z_{1-0.05/2} + z_{1-0.2})}^2}{{(z_{1-0.05/2} + z_{1-0.15})}^2} = 0.1258 \ \bigr]$
\item A $12.6 \%$ reduction in the number of recurrent events can be translated into a $12.6 \%$ lower sample size (= number of patients) to achieve $80 \%$ power for a study powered for the recurrent event endpoint compared to a study powered for the time-to-first-event endpoint, assuming the same recruitment period and study duration for both studies. \newline Moreover, one would expect that the lower sample size for the recurrent event analysis could be recruited quicker which would lead to additional gains in study duration. \item In simulations with $\phi \neq 0.0$, power gains of recurrent event analyses were even larger (cf. Figure $\ref{S1.PPMS.power}$ and Table $\ref{SimulationResultsS1H1}$). The value of $12.6 \%$ is therefore a conservative estimate of the sample size savings and can be seen as a lower bound of expected gains.
\end{enumerate}

\newpage
\section{MS-specific simulation study}
\subsection{Characteristics of simulated data}
\subsubsection*{Frailty distribution and heterogeneity matrix}
\begin{figure}[h]
\centering 
\vspace*{-10mm}
\scalebox{0.78}{
\includegraphics{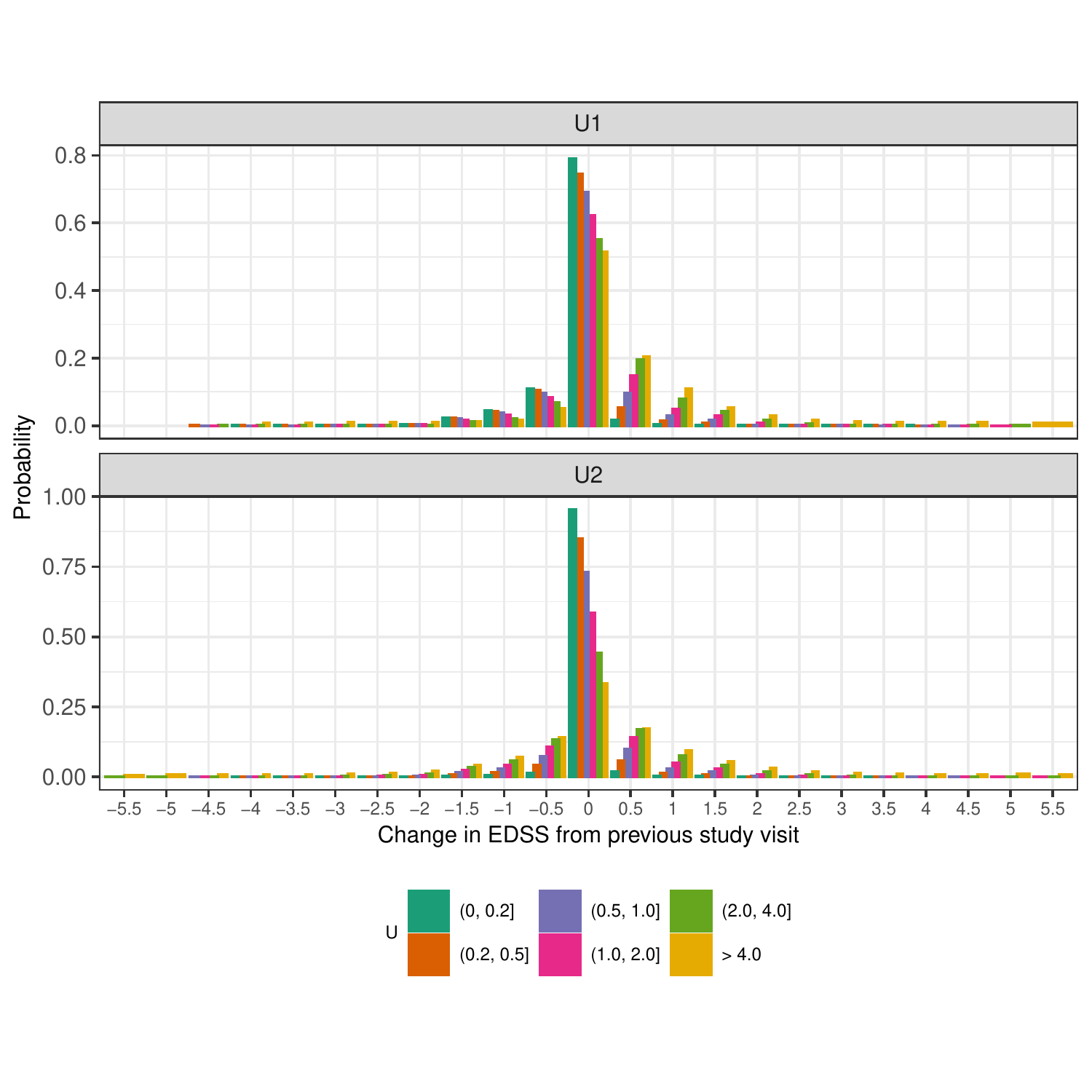}
}
\vspace*{-10mm}
\caption[S2.PPMS - Change in EDSS from previous visit stratified by frailty term]{S2.PPMS - Change in EDSS from previous visit stratified by frailty term $U$, a positive change $> 0$ indicates worsening and negative values $< 0$ correspond to improvement, e.g., the level '0.5' refers to a $0.5$-step transition into a higher EDSS score and the level '-1.0' indicates an improvement in EDSS score of 1.0 point, N=10000 simulations, n=1000 patients}
\label{S2.PPMS.MeanCHG.EDSS.Z}
\end{figure}

The multistate model used to describe EDSS dynamics of PPMS patients includes frailty terms to distinguish between patients who are frail to move through the different EDSS states ('movers') from those who are most likely to stay in the same EDSS state ('stayers'). The patient-specific random effect is generated from a gamma distribution $\Gamma(\phi^{-1}, \phi^{-1})$ with mean $\mathbb{E}(U)=1$, variance $Var(U) = \phi > 0$ and $\phi \in \{0.0, 0.15, 1.0 \}$. The case $\phi=0.0$ corresponds to a homogeneous study population with $U=1$ for all patients. \newline
As explained in Chapter $6$, there are two different ways in which heterogeneity can be defined on the transition intensities. Option $U_{1}$ is defined by adding a frailty term to upward transitions only, whereas option $U_{2}$ is specified by adding frailty terms to upward and downward transitions. In option $U_{1}$, patients with large $U$ are so-called upward movers, while patients with small $U$ are stayers or downward movers (cf. Figure $\ref{TransProbMatrixZ1}$). In contrast, using $U_{2}$, patients with large $U$ are referred to as upward and downward movers, while patients with small $U$ are most likely to stay (cf. Figure $\ref{TransProbMatrixZ2}$). In particular, characteristics of simulated EDSS measurements vary across the two options. In order to describe the difference between the heterogeneity patterns $U_{1}$ and $U_{2}$, Figure $\ref{S2.PPMS.MeanCHG.EDSS.Z}$ plots the distribution of the change in EDSS from the previous study visit, stratified by categories of the frailty term and specification of the heterogeneity matrix.

\begin{figure}[h]
\centering
\scalebox{0.68}{
\includegraphics{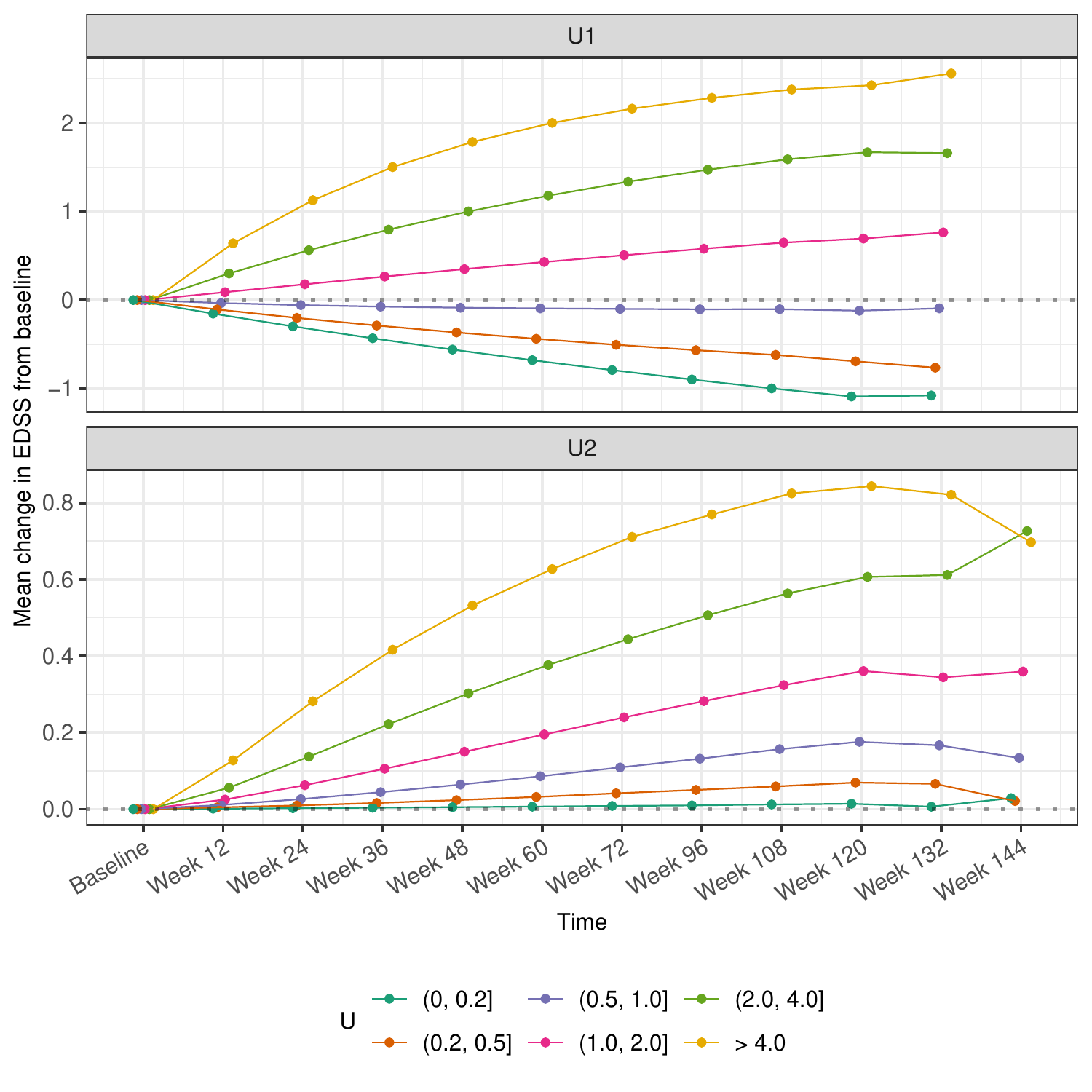}
}
\caption[S2.PPMS - Mean change in EDSS from baseline stratified by frailty term]{S2.PPMS - Mean change in EDSS from baseline stratified by frailty term $U$, N=10000 simulations, n=1000 patients}
\label{S2.PPMS.meanchange.Z}
\end{figure}

For both patterns $U_{1}$ and $U_{2}$, the probability of keeping the same EDSS score within $3$ months decreases, as $U$ increases. Under $U_{1}$, patients who have high probabilities of making upward transitions (i.e., large U) have small probabilities for making downward transitions. Consequently, the mean change in EDSS score from baseline is continuously increasing and positive for patients with large $U$, as depicted in Figure $\ref{S2.PPMS.meanchange.Z}$. 
\newpage 
On the other hand, patients with high probabilities of moving into a lower EDSS state (i.e., small U) have small probabilities of moving into a higher EDSS score. This explains the continuously decreasing negative mean change in EDSS from baseline for patients with small $U$ (cf. Figure $\ref{S2.PPMS.meanchange.Z}$). Hence, the probability to transition into a higher score (= positive change) increases as U increases, whereas the probability to transition into a lower score (= negative change) increases as U decreases. The distribution of the mean change in EDSS from the previous study visit is almost symmetric for $U \in (0.5, 1]$, approximately right-skewed for $U > 1$ and approximately left-skewed for $U\leq 0.5$. In fact, all patients are most likely to stay in the same disease state, followed by either upward or downward transitions of a certain magnitude.
Under $U_{2}$, the distribution is approximately symmetric for all realizations of $U$, meaning that patients who have certain probabilities of making upward transitions also have a similar (slightly reduced) chance for making downward transitions. In contrast to $U_{1}$, the probability of making transitions into both higher or lower EDSS scores increases as $U$ increases. The lower panel of Figure $\ref{S2.PPMS.meanchange.Z}$ indicates that upward transitions are still more likely to occur than downward transitions, as the mean changes in EDSS from baseline are greater than 0. For patients with $U$ close to $0$, the mean change in EDSS from baseline is almost constant and close to $0$ over all study visits.

\begin{figure}[H]
\centering
\scalebox{0.50}{
\includegraphics{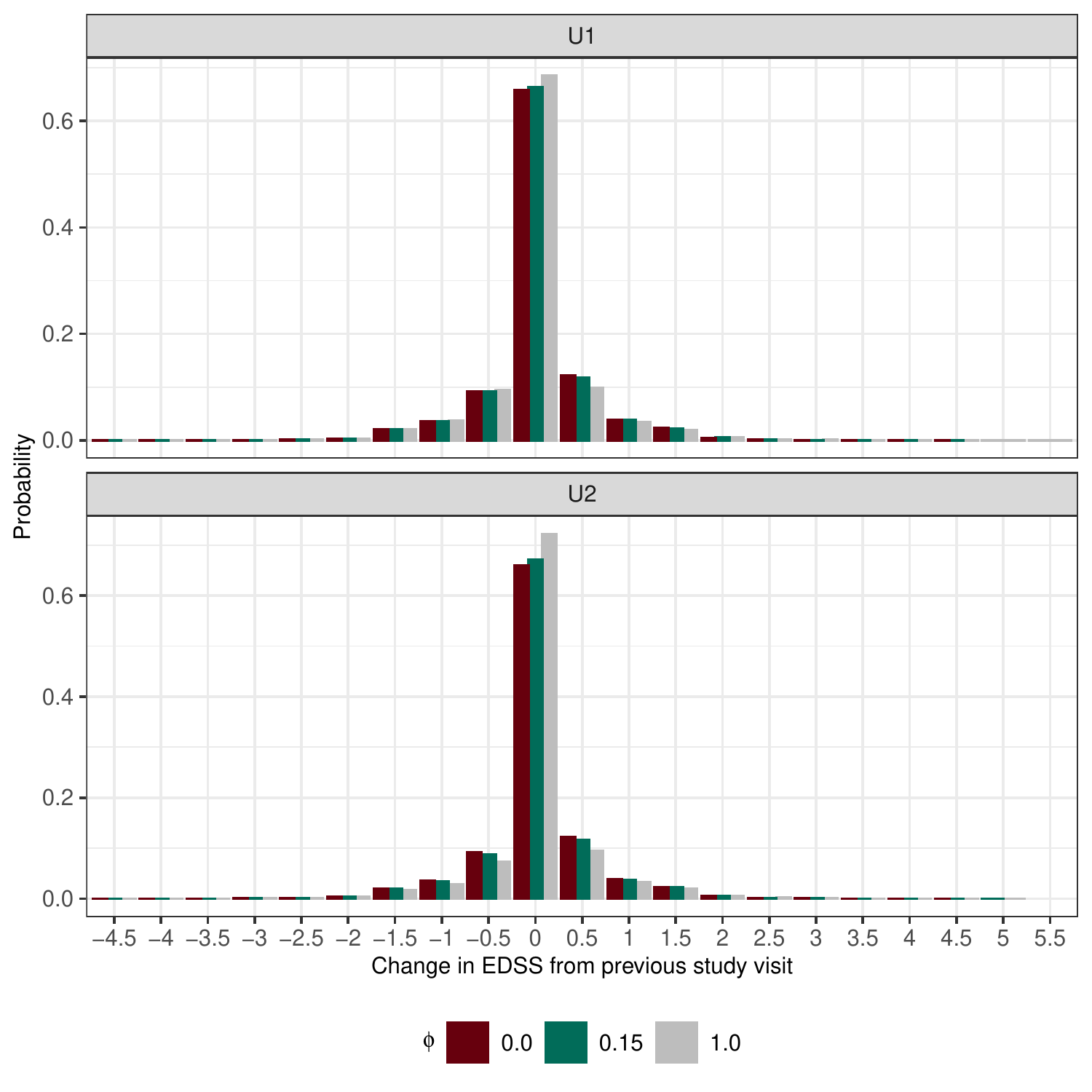}
}
\caption[S2.PPMS - Change in EDSS from previous visit according to different heterogeneity parameters]{S2.PPMS - Change in EDSS from previous visit according to different heterogeneity parameters $\phi$, N=10000 simulations, n=1000 patients}
\label{S2.PPMS.CHG.EDSS.Z}
\end{figure}

\begin{figure}[H]
\centering
\scalebox{0.50}{
\includegraphics{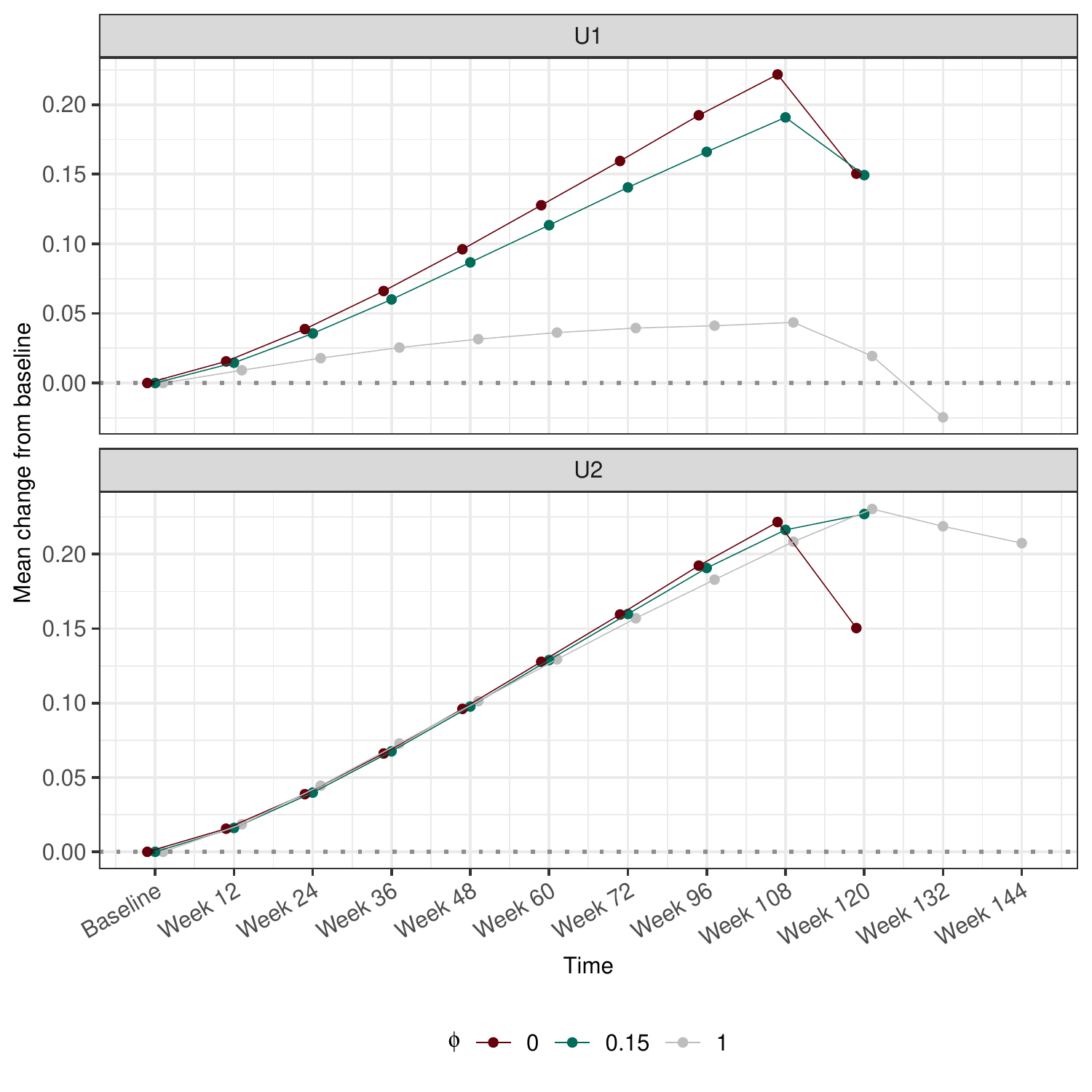}
}
\caption[S2.PPMS - Mean change in EDSS from baseline according to different heterogeneity parameters]{S2.PPMS - Mean change in EDSS from baseline according to different heterogeneity parameters $\phi$, N=10000 simulations, n=1000 patients}
\label{S2.PPMS.meanchange}
\end{figure}

According to Section $\ref{SimulatedDataS1}$ and Figure $\ref{GammaDist}$, the variance parameter $\phi$ specifies the shape of the gamma distribution and defines the composition of the study population. Figure $\ref{S2.PPMS.CHG.EDSS.Z}$ and Figure $\ref{S2.PPMS.meanchange}$ plot the distribution of the change in EDSS from the previous study visit and the mean change in EDSS from baseline to the end of follow-up according to different heterogeneity parameters. \newline
For $\phi=0.0$, patients share the same transition probabilities and the EDSS scores are generated from $P_{{PPMS, Z=\cdot, U=1}}$. With increasing $\phi$, the study population is mainly dominated by patients with very small $U$ and only a few patients share a high frailty. Under $U_{1}$, the study population is therefore represented by a high proportion of stayers / downward movers and a relatively small number of upward movers. In accordance with Figure $\ref{S2.PPMS.MeanCHG.EDSS.Z}$, the high proportion of stayers / downward movers explains the increased probability for making 0-step transitions and reduced probabilities for making upward transitions, as compared to $\phi=0.0$ (cf. Figure $\ref{S2.PPMS.CHG.EDSS.Z}$). On average, the probabilities for making downward transitions do not vary, as $\phi$ increases. Due to the reduced number of upward transitions, the mean change in EDSS from baseline is decreasing over time (cf. upper panel of Figure $\ref{S2.PPMS.meanchange}$). Under $U_{2}$, the study population for $\phi=1.0$ is represented by a high proportion of stayers and a relatively small number of upward and downward movers. Similarily, the high proportion of stayers implies the increased probability for making 0-step transitions and reduced probabilities for making upward and downward transitions, as compared to $0.0$. As seen in the lower panel of Figure $\ref{S2.PPMS.CHG.EDSS.Z}$, the ratio between downward and upward transitions is independent of $\phi$, when heterogeneity is specified via $U_{2}$. Therefore, the mean change from baseline in EDSS does, on average, not change with increasing $\phi$. 

\newpage 

\subsubsection*{Study duration and distribution of CDP12 events}
In Table $\ref{S2PPMSDEF1.FU}$ and Table $\ref{S2PPMSDEF1.NOCDP12}$, summary statistics of the study duration and average number of CDP12 events under different scenarios are presented. Study duration is defined as the time from the first patient randomized to the time the target number of events is reached  ($n_{first.events}=246$). For both specifications of heterogeneity matrices $U_{1}$ and $U_{2}$, it generally yields: as the extent of heterogeneity $\phi$ increases, study duration becomes longer and higher numbers of overall CDP12 events can on average be observed. 
Comparing $U_{1}$ with $U_{2}$, it turns out that study durations are longer under $U_{2}$, whereas the numbers of CDP12 events are higher under $U_{1}$. Thus, given $U_{1}$, a higher number of CDP12 events occur in a shorter period of time. This is due to the following fact: when the frailty term $U$ is simulated from a gamma distribution with variance $\phi=1.0$, the population at-risk for $U_{1}$ is mainly led by a large proportion of downward movers / stayers whose EDSS values are constant or even continuously decreasing over time, and a small number of upward movers whose EDSS values are constant or increasing over time, respectively. EDSS curves resulting from patients with either a very high or a very small frailty term tend to be monotonically increasing or decreasing, with less variability. Because of these extreme characteristics of the EDSS curves, the latters tend to never progress as an IDP occurs with zero probability and the formers are expected to progress several times as the requirements for a CDP12 event (i.e., IDP and confirmation) are very likely to be fulfilled. The higher proportion of CDP12 events is experienced by a relatively large number of frail upward movers in a certain time period. In general, it follows: the less variable the EDSS curve, the easier to detect a CDP12 event. Compared to $U_{1}$, the population at-risk for $U_{2}$ is mainly led by a large proportion of stayers whose EDSS values are constant over time, and a small number of upward and downward movers whose EDSS curves are variable in the sense that the curve includes worsening, improvement and stability. For instance, a PPMS patient who initially improves needs to first progress back to the reference EDSS score, then needs to worsen again to obtain an IDP and thereafter the patient needs to be at least stable to confirm the disability progression. Due to high variability in the EDSS curves, it takes much longer to obtain the target number of CDP12 events and study duration is prolonged using $U_{2}$.  

\begin{table}[H]
\resizebox{\textwidth}{!}{ 
\begin{tabular}{|l|c|c|cccc|l|cccc|}
\hline
\multicolumn{3}{|l|}{} & \multicolumn{4}{c|}{\textbf{$\boldsymbol{U_{1}}$}} & \multirow{9}{*}{} & \multicolumn{4}{c|}{\textbf{$\boldsymbol{U_{2}}$}} \\ \cline{1-7} \cline{9-12} 
 &  &  & \multicolumn{4}{c|}{\textbf{Study duration (in days)}} &  & \multicolumn{4}{c|}{\textbf{Study duration (in days)}} \\ \cline{4-7} \cline{9-12} 
\multicolumn{1}{|c|}{\textbf{Scenario}} & \textbf{\begin{tabular}[c]{@{}c@{}}\\ $\boldsymbol{\exp(\beta_{hj})}$\end{tabular}} & \textbf{$\phi$} & \textbf{$\boldsymbol{Q_{10}}$} & \textbf{Median} & \textbf{Mean} & \textbf{$\boldsymbol{Q_{90}}$} &  & \textbf{$\boldsymbol{Q_{10}}$} & \textbf{Median} & \textbf{Mean} &\textbf{$\boldsymbol{Q_{90}}$} \\ \cline{1-7} \cline{9-12} 
\textbf{S2/PPMS/noeffect/homo} & \multirow{3}{*}{\textbf{1.0}} & \textbf{0.0} & 664.00 & 694.00 & 695.46 & 730.00 &  & 664.00 & 694.00 & 695.46 & 730.00 \\
\textbf{S2/PPMS/noeffect/hetero1} &  & \textbf{0.15} & 669.00 & 699.00 & 701.84 & 739.00 &  & 679.00 & 710.00 & 712.48 & 751.00 \\
\textbf{S2/PPMS/noeffect/hetero2} &  & \textbf{1.0} & 701.00 & 751.00 & 751.31 & 801.00 &  & 768.00 & 816.00 & 817.27 & 867.00 \\ \cline{1-7} \cline{9-12} 
\textbf{S2/PPMS/effect/homo} & \multirow{3}{*}{\textbf{0.70}} & \textbf{0.0} & 745.00 & 784.00 & 787.82 & 836.00 &  & 745.00 & 784.00 & 787.82 & 836.00 \\
\textbf{S2/PPMS/effect/hetero1} &  & \textbf{0.15} & 751.00 & 793.00 & 796.66 & 847.00 &  & 767.00 & 813.00 & 814.10 & 862.00 \\
\textbf{S2/PPMS/effect/hetero2} &  & \textbf{1.0} & 799.00 & 862.00 & 866.20 & 937.00 &  & 895.00 & 961.00 & 965.52 & 1037.00 \\ \hline
\end{tabular}
}
\caption[S2.PPMS - Summary statistics of study duration]{S2.PPMS - Summary statistics of study duration according to different treatment effect sizes and heterogeneity parameters, study duration is defined as time from first patient randomized to target number of events reached ($n_{first.events}=246$), CDP12 events are derived according to a time-to-confirmation-of-CDP endpoint with fixed reference system, $Q_{10}$ and $Q_{90}$ are the $10 \%$ and $90 \%$ quantiles, N=10000 simulations, n=1000 patients}
\label{S2PPMSDEF1.FU}
\end{table}

\begin{table}[H]
\scalebox{0.75}{
\begin{tabular}{|l|c|c|cccc|l|cccc|}
\hline
\multicolumn{3}{|l|}{} & \multicolumn{4}{c|}{\textbf{$\boldsymbol{U_{1}}$}} & \multirow{9}{*}{} & \multicolumn{4}{c|}{\textbf{$\boldsymbol{U_{2}}$}} \\ \cline{1-7} \cline{9-12} 
 &  &  & \multicolumn{4}{c|}{\textbf{\begin{tabular}[c]{@{}c@{}}Total number of \\ CDP12 events\end{tabular}}} &  & \multicolumn{4}{c|}{\textbf{\begin{tabular}[c]{@{}c@{}}Total number of \\ CDP12 events\end{tabular}}} \\ \cline{4-7} \cline{9-12} 
\multicolumn{1}{|c|}{\textbf{Scenario}} & \textbf{\begin{tabular}[c]{@{}c@{}}\\ $\boldsymbol{\exp(\beta_{hj})}$\end{tabular}} & \textbf{$\boldsymbol{\phi}$} & \textbf{$\boldsymbol{Q_{10}}$} & \textbf{Median} & \textbf{Mean} & \textbf{$\boldsymbol{Q_{90}}$} &  &\textbf{$\boldsymbol{Q_{10}}$} & \textbf{Median} & \textbf{Mean} & \textbf{$\boldsymbol{Q_{90}}$} \\ \cline{1-7} \cline{9-12} 
\textbf{S2/PPMS/noeffect/homo} & \multirow{3}{*}{\textbf{1.0}} & \textbf{0.0} & 279 & 288 & 288 & 297 &  & 279 & 288 & 288 & 297 \\
\textbf{S2/PPMS/noeffect/hetero1} &  & \textbf{0.15} & 284 & 294 & 294 & 305 &  & 282 & 291 & 291 & 301 \\
\textbf{S2/PPMS/noeffect/hetero2} &  & \textbf{1.0} & 313 & 328 & 328 & 344 &  & 297 & 309 & 309 & 321 \\ \cline{1-7} \cline{9-12} 
\textbf{S2/PPMS/effect/homo} & \multirow{3}{*}{\textbf{0.70}} & \textbf{0.0} & 283 & 293 & 293 & 303 &  & 283 & 293 & 293 & 303 \\
\textbf{S2/PPMS/effect/hetero1} &  & \textbf{0.15} & 290 & 301 & 301 & 312 &  & 286 & 296 & 297 & 307 \\
\textbf{S2/PPMS/effect/hetero2} &  & \textbf{1.0} & 323 & 339 & 340 & 357 &  & 303 & 315 & 316 & 328 \\ \hline
\end{tabular}
}
\caption[S2.PPMS - Summary statistics of number of CDP12 events]{S2.PPMS - Summary statistics of number of CDP12 events according to different treatment effect sizes and heterogeneity parameters, CDP12 events are derived according to a time-to-confirmation-of-CDP endpoint with fixed reference system, $Q_{10}$ and $Q_{90}$ are the $10 \%$ and $90 \%$ quantiles, N=10000 simulations, n=1000 patients}
\label{S2PPMSDEF1.NOCDP12}
\end{table}

\subsection{Comparison of time-to-first-event and recurrent event methods}
\subsubsection*{Negative binomial model versus Poisson regression}
Table $\ref{S2PPMSNegversusPoisson}$ summarizes non-convergence proportions of the NB model. Under homogeneity, the NB model failed to converge in $75 \%$ or $53 \%$ of all simulation runs in the scenarios S2.PPMS.noeffect.homo and S2.PPMS.effect.homo, respectively. In case of non-convergence, the Poisson model rather than the NB model was used to fix this issue. For moderate heterogeneity ($\phi=0.15$), the non-convergence proportions of the NB model differ across the two heterogeneity options. Higher proportions of non-convergence were reported with option $U_{2}$. Under high heterogeneity ($\phi=1.0$), the NB model converged across all simulation runs. 

\begin{figure}[H]
\centering
\subfloat[$U_{1}$]{
\resizebox{0.5\textwidth}{!}{
$
\begin{tabular}{c|cc}
                & \multicolumn{2}{c}{\textbf{\begin{tabular}[c]{@{}c@{}}Percentage of using Poisson regression\\ rather than NB model (in \%)\end{tabular}}} \\ \hline
\textbf{$\phi$} & \multicolumn{1}{c|}{$\boldsymbol{\exp(\beta) = 1.0}$}                                         & $\boldsymbol{\exp(\beta) = 0.7}$                                         \\ \hline
\textbf{0.0}    & \multicolumn{1}{c|}{75.03}                                                     & 52.86                                                     \\
\textbf{0.15}   & \multicolumn{1}{c|}{41.80}                                                     & 19.24                                                     \\
\textbf{1.0}    & \multicolumn{1}{c|}{0.00}                                                      & 0.00                                                     
\end{tabular}
$
}}
\hfil
\subfloat[$U_{2}$]{
\resizebox{0.5\textwidth}{!}{
$ 
\begin{tabular}{c|cc}
                & \multicolumn{2}{c}{\textbf{\begin{tabular}[c]{@{}c@{}}Percentage of using Poisson regression\\ rather than NB model (in \%)\end{tabular}}} \\ \hline
\textbf{$\phi$} & \multicolumn{1}{c|}{$\boldsymbol{\exp(\beta) = 1.0}$}                                         & $\boldsymbol{\exp(\beta) = 0.7}$                                      \\ \hline
\textbf{0.0}    & \multicolumn{1}{c|}{75.03}                                                     & 52.86                                                     \\
\textbf{0.15}   & \multicolumn{1}{c|}{56.61}                                                     & 34.16                                                     \\
\textbf{1.0}    & \multicolumn{1}{c|}{0.02}                                                      & 0.01                                                     
\end{tabular}
$
}}
\caption[S2.PPMS - Convergence issues of negative binomial model]{S2.PPMS - Convergence issues of NB model, N=10000 simulations, n=1000 patients}
\label{S2PPMSNegversusPoisson}
\end{figure}

\subsubsection*{Treatment effect estimation}
Properties of statistical methods are usually evaluated based on the assumption that the true model is known. In the MS-specific simulation study, data generation is a two-step procedure and the true underlying model assumptions of the time-to-first-event and recurrent event approaches are unknown. In a first step, EDSS scores are simulated using a time-homogeneous multistate model, where the treatment effect is assumed to act multiplicatively on the transition hazards. Hence, the proportional transition-specific hazard models assume each transition hazard to follow a Cox model \parencite{Cox1972}, while a proportional effect of the treatment on the transition hazards is claimed. In a second step, the recurrent event endpoint is derived based on the simulated longitudinal measurements of the ordinal EDSS scale, according to the rules described in Chapter $2$. However, a proportional treatment effect on the transition hazards does not generally imply a proportional treatment effect on the recurrent event intensity. Based on the recurrent event data, the proportional hazards assumption may not hold and associated time-to-first-event and recurrent event methods may be misspecified. \citet{Hjort1992} claimed that a misspecified model still provides a consistent effect estimate, although not of the regression coefficient of the misspecified model but of the so-called least-false parameter. Following the definition of \citet{Beyersmann2012}, the least-false parameter is 'least-false' in the sense that it gives the best approximation of the misspecified model towards the true model that generated the data. The approximation is optimal with regard to an appropriate distance (e.g., Kullback-Leibler) between the misspecified and the true model \parencite{Hjort1992, Claeskens2008}. The least-false parameter yields a time-average hazard ratio. 
\newline 
In the MS-specific simulation study, comparisons of time-to-first-event and recurrent event methods follow a slightly different evaluation concept compared to the evaluation strategy used in Section $\ref{GeneralStudy}$. Note that the simulated treatment effect sizes on transition intensities do not translate 1:1 to effect sizes for recurrent events. Since the true treatment effect size obtained from the Cox, NB, AG and LWYY models is unknown, statements on unbiasedness of treatment effect estimates can not be made. One is rather interested in how the potentially misspecified models behave under different scenarios with regard to precision, power and type I error. \newpage 

\begin{figure}[H]
\centering
\scalebox{0.60}{
\includegraphics{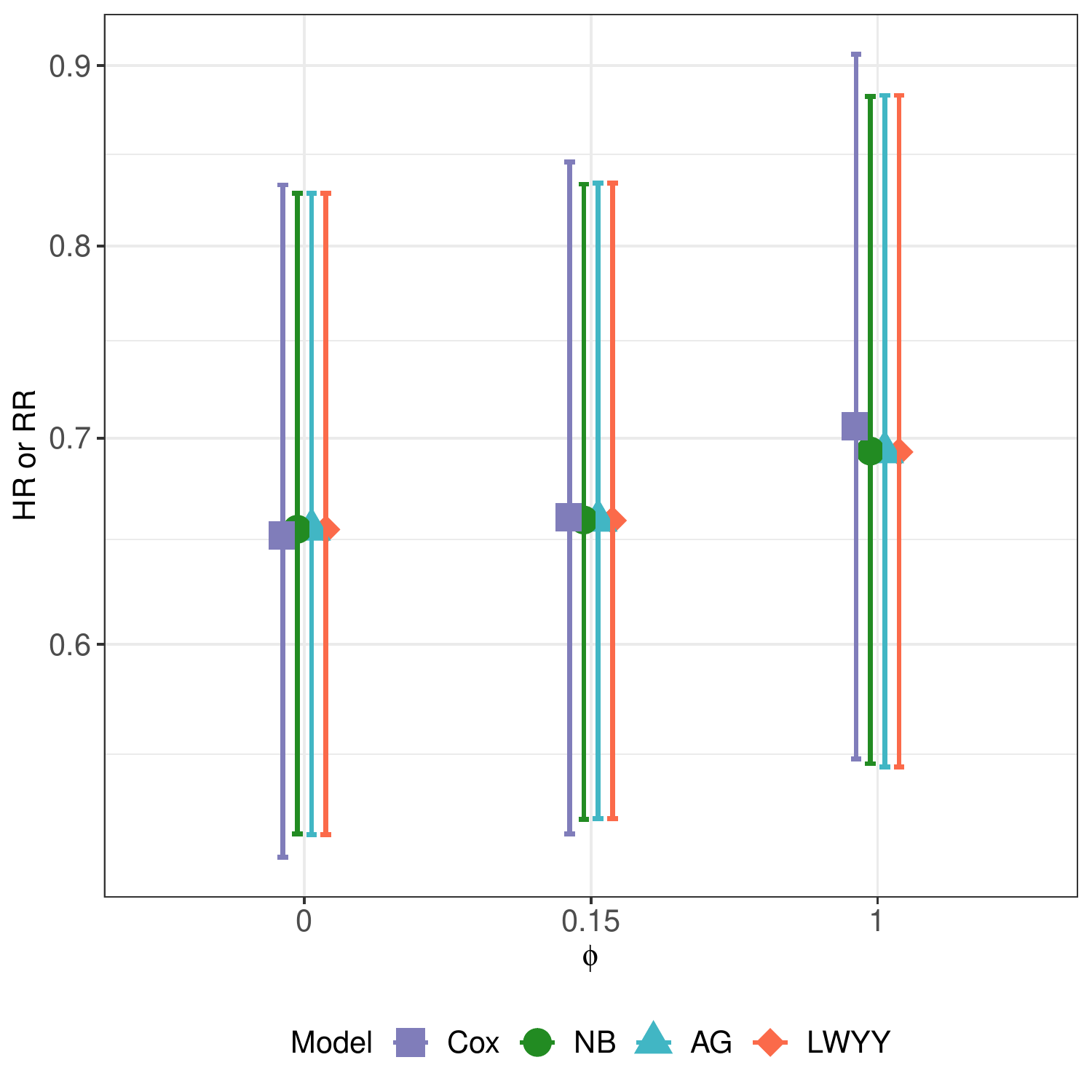}
}
\caption[S2.PPMS - Treatment effect estimates obtained from time-to-first-event and recurrent event methods in MS-specific simulation study ($U_{1}$)]{S2.PPMS - Treatment effect estimates obtained from time-to-first-event and recurrent event methods according to different heterogeneity parameters and based on option $U_{1}$, CDP12 events are derived according to a time-to-confirmation-of-CDP endpoint with fixed reference system, error bars indicate $2.5\%$ and $97.5\%$ quantiles, N=10000 simulations, n=1000 patients}
\label{S2.PPMS.precision.Z1}
\end{figure}

\textbf{Option $\boldsymbol{U_{1}}$} \newline \newline 
\textbf{Case: $\boldsymbol{\beta_{hj}=\log(0.7)}$}  \newline 
Results of the MS-specific simulation study under $H_{1}: \{ \beta_{hj}=\log(0.7) \}$ are summarized in Table $\ref{SimulationResultsS2H1Z1}$. Given $\phi=0.0$, the Cox, NB, AG and LWYY analyses result in estimated treatment effects of $0.6519$, $0.6549$, $0.6548$ and $0.6548$. In a homogeneous study population, time-to-first-event and recurrent event methods provide similar treatment effect estimates. Evaluation measures like bias, MSE and coverage probability can not be applied to the simulation results, as the true treatment effect size for CDP12 events is not clearly known. In terms of precision of treatment effect estimation, recurrent event methods outperform the time-to-first-event approach because the variability in treatment effect estimators across the simulations is larger for the Cox model. Among the recurrent event approaches, all methods provide an equally precise treatment effect estimate. The mean SE estimate is close to the empirically determined Monte Carlo SD with all approaches. \newline 
As frailty variance increases, the estimated treatment effects obtained from the time-to-first-event and the recurrent event methods become smaller. In case of $\phi=1.0$, the average treatment effects resulting from the Cox, NB, AG and LWYY analyses are equal to $0.7059$, $0.6935$, $0.6930$ and $0.6930$. In heterogeneous study populations, time-to-first-event and recurrent event methods still provide similar treatment effect estimates (cf. Figure $\ref{S2.PPMS.precision.Z1}$) but precision is higher with the NB, AG and LWYY models. Further, all methods yield an accurate approximation of the mean SE estimate to the Monte Carlo SD, except for the AG model. With increasing heterogeneity, the naive average SE estimate underestimates the variability of $\hat{\beta}$ for the AG approach. \newline \newline 
\textbf{Case: $\boldsymbol{\beta_{hj}=\log(1.0)}$}  \newline 
Table $\ref{SimulationResultsS2H0Z1}$ summarizes the results of the MS-specific simulation study under $H_{0}: \{ \beta_{hj}=\log(1.0) \}$. Under $H_{0}$, similar results can be observed. Both time-to-first-event and recurrent event methods yield similar estimates of the treatment effect, even in presence of heterogeneity. Recurrent event methods are also seen to outperform the conventional Cox proportional hazards model with regard to precision. 

\begin{landscape}
\begin{table}[]
\centering
\scalebox{0.85}{ 
\begin{tabular}{|c|cccccc|cccccc|}
\hline
\multirow{2}{*}{$\boldsymbol{\phi}$} & \multicolumn{6}{c|}{\textbf{Time-to-first-event method}} & \multicolumn{6}{c|}{\textbf{Recurrent event methods}} \\ \cline{2-13} 
 & \textbf{\begin{tabular}[c]{@{}c@{}}Model\\ $$\end{tabular}} & \textbf{\begin{tabular}[c]{@{}c@{}}$\boldsymbol{\log{(\text{HR})}}$\\ $$\end{tabular}} & \textbf{\begin{tabular}[c]{@{}c@{}}HR \\ $$\end{tabular}} & \textbf{\begin{tabular}[c]{@{}c@{}}SE\\ $$\end{tabular}} & \textbf{\begin{tabular}[c]{@{}c@{}}SEE\\ $$\end{tabular}} & \textbf{\begin{tabular}[c]{@{}c@{}}Power\\ $$\end{tabular}} & \textbf{\begin{tabular}[c]{@{}c@{}}Model\\ $$\end{tabular}} & \textbf{\begin{tabular}[c]{@{}c@{}}$\boldsymbol{\log{(\text{HR/RR})}}$\\ $$\end{tabular}} & \textbf{\begin{tabular}[c]{@{}c@{}}HR/RR\\ $$\end{tabular}} & \textbf{\begin{tabular}[c]{@{}c@{}}SE\\ $$\end{tabular}} & \textbf{\begin{tabular}[c]{@{}c@{}}SEE\\ $$\end{tabular}} & \textbf{\begin{tabular}[c]{@{}c@{}}Power\\ $$\end{tabular}} \\ \hline
\textbf{0.0} & \textbf{Cox} & \begin{tabular}[c]{@{}c@{}}$$\\ 0.0012\\ $$\end{tabular} & \begin{tabular}[c]{@{}c@{}}$$\\ 1.0012\\ $$\end{tabular} & \begin{tabular}[c]{@{}c@{}}$$\\ 0.1265\\ $$\end{tabular} & \begin{tabular}[c]{@{}c@{}}$$\\ 0.1277\\ $$\end{tabular} & \begin{tabular}[c]{@{}c@{}}$$\\ 0.047\\ $$\end{tabular} & \textbf{\begin{tabular}[c]{@{}c@{}}NB\\ AG\\ LWYY\end{tabular}} & \begin{tabular}[c]{@{}c@{}}-0.0003\\ -0.0001\\ -0.0001\end{tabular} & \begin{tabular}[c]{@{}c@{}}0.9997\\ 0.9999\\ 0.9999\end{tabular} & \begin{tabular}[c]{@{}c@{}}0.1178\\ 0.1178\\ 0.1178\end{tabular} & \begin{tabular}[c]{@{}c@{}}0.1186\\ 0.1182\\ 0.1168\end{tabular} & \begin{tabular}[c]{@{}c@{}}0.049\\ 0.051\\ 0.054\end{tabular} \\ \hline
\textbf{0.15} & \textbf{Cox} & \begin{tabular}[c]{@{}c@{}}$$\\ 0.0013\\ $$\end{tabular} & \begin{tabular}[c]{@{}c@{}}$$\\ 1.0013\\ $$\end{tabular} & \begin{tabular}[c]{@{}c@{}}$$\\ 0.1273\\ $$\end{tabular} & \begin{tabular}[c]{@{}c@{}}$$\\ 0.1277\\ $$\end{tabular} & \begin{tabular}[c]{@{}c@{}}$$\\ 0.049\\ $$\end{tabular} & \textbf{\begin{tabular}[c]{@{}c@{}}NB\\ AG \\ LWYY\end{tabular}} & \begin{tabular}[c]{@{}c@{}}0.0012\\ 0.0013\\ 0.0013\end{tabular} & \begin{tabular}[c]{@{}c@{}}1.0012\\ 1.0013\\ 1.0013\end{tabular} & \begin{tabular}[c]{@{}c@{}}0.1187\\ 0.1191\\ 0.1191\end{tabular} & \begin{tabular}[c]{@{}c@{}}0.1185\\ 0.1169\\ 0.1178\end{tabular} & \begin{tabular}[c]{@{}c@{}}0.051\\ 0.056\\ 0.053\end{tabular} \\ \hline
\textbf{1.0} & \textbf{Cox} & \begin{tabular}[c]{@{}c@{}}$$\\ -0.0005\\ $$\end{tabular} & \begin{tabular}[c]{@{}c@{}}$$\\ 0.9995\\ $$\end{tabular} & \begin{tabular}[c]{@{}c@{}}$$\\ 0.1274\\ $$\end{tabular} & \begin{tabular}[c]{@{}c@{}}$$\\ 0.1277\\ $$\end{tabular} & \begin{tabular}[c]{@{}c@{}}$$\\ 0.048\\ $$\end{tabular} & \textbf{\begin{tabular}[c]{@{}c@{}}NB\\ AG\\ LWYY\end{tabular}} & \begin{tabular}[c]{@{}c@{}}-0.0001\\ -0.0001\\ -0.0001\end{tabular} & \begin{tabular}[c]{@{}c@{}}0.9999\\ 0.9999\\ 0.9999\end{tabular} & \begin{tabular}[c]{@{}c@{}}0.1208\\ 0.1214\\ 0.1214\end{tabular} & \begin{tabular}[c]{@{}c@{}}0.1234\\ 0.1108\\ 0.1216\end{tabular} & \begin{tabular}[c]{@{}c@{}}0.046\\ 0.075\\ 0.049\end{tabular} \\ \hline
\end{tabular}}
\caption[S2.PPMS - Results of the MS-specific simulation study ($\beta_{hj}=\log(1.0)$ and $U_{1}$)]{S2.PPMS - Results of the MS-specific simulation study when the true treatment effect is $\beta_{hj}=\log(1.0)$ and heterogeneity is specified via $U_{1}$, CDP12 events are derived according to a time-to-confirmation-of-CDP endpoint with fixed reference system, N=10000 simulations, n=1000 patients (1:1 randomization) \\
Evaluation measures: HR/RR = mean treatment effect across simulations, SE = standard deviation of estimators across simulations, $SEE$ = mean standard error across simulations}
\label{SimulationResultsS2H0Z1}
\end{table}

\begin{table}[]
\centering
\scalebox{0.85}{ 
\begin{tabular}{|c|cccccc|cclccc|}
\hline
\multirow{2}{*}{$\boldsymbol{\phi}$} & \multicolumn{6}{c|}{\textbf{Time-to-first-event method}}                                                                                                                                                                                                                                                                                                                                              & \multicolumn{6}{c|}{\textbf{Recurrent event methods}}                                                                                                                                                                                                                                                                                                                                                                                           \\ \cline{2-13} 
                                 & \textbf{\begin{tabular}[c]{@{}c@{}}Model\\ $$\end{tabular}} & \textbf{\begin{tabular}[c]{@{}c@{}}$\boldsymbol{\log{(\text{HR})}}$\\ $$\end{tabular}} & \textbf{\begin{tabular}[c]{@{}c@{}}HR \\ $$\end{tabular}} & \textbf{\begin{tabular}[c]{@{}c@{}}SE\\ $$\end{tabular}} & \textbf{\begin{tabular}[c]{@{}c@{}}SEE\\ $$\end{tabular}} & \textbf{\begin{tabular}[c]{@{}c@{}}Power \\ $$\end{tabular}} & \textbf{\begin{tabular}[c]{@{}c@{}}Model\\ $$\end{tabular}}      & \textbf{\begin{tabular}[c]{@{}c@{}}$\boldsymbol{\log{(\text{HR/RR})}}$\\ $$\end{tabular}} & \multicolumn{1}{c}{\textbf{\begin{tabular}[c]{@{}c@{}}HR/RR\\ $$\end{tabular}}} & \textbf{\begin{tabular}[c]{@{}c@{}}SE\\ $$\end{tabular}}         & \textbf{\begin{tabular}[c]{@{}c@{}}SEE\\ $$\end{tabular}}        & \textbf{\begin{tabular}[c]{@{}c@{}}Power\\ $$\end{tabular}}    \\ \hline
\textbf{0.0}                     & \textbf{Cox}                                                & \begin{tabular}[c]{@{}c@{}}$$\\ -0.4279\\ $$\end{tabular}                              & \begin{tabular}[c]{@{}c@{}}$$\\ 0.6519\\ $$\end{tabular} & \begin{tabular}[c]{@{}c@{}}$$\\ 0.1281\\ $$\end{tabular} & \begin{tabular}[c]{@{}c@{}}$$\\ 0.1300\\ $$\end{tabular}  & \begin{tabular}[c]{@{}c@{}}$$\\ 0.919\\ $$\end{tabular}      & \textbf{\begin{tabular}[c]{@{}c@{}}NB\\ AG\\ LWYY\end{tabular}}  & \begin{tabular}[c]{@{}c@{}}-0.4233\\ -0.4235\\ -0.4235\end{tabular}                    & \begin{tabular}[c]{@{}l@{}}0.6549\\ 0.6548\\ 0.6548\end{tabular}             & \begin{tabular}[c]{@{}c@{}}0.1204\\ 0.1207\\ 0.1207\end{tabular} & \begin{tabular}[c]{@{}c@{}}0.1208\\ 0.1197\\ 0.1202\end{tabular} & \begin{tabular}[c]{@{}c@{}}0.945\\ 0.946\\ 0.945\end{tabular} \\ \hline
\textbf{0.15}                    & \textbf{Cox}                                                & \begin{tabular}[c]{@{}c@{}}$$\\ -0.4145\\ $$\end{tabular}                              & \begin{tabular}[c]{@{}c@{}}$$\\ 0.6607\\ $$\end{tabular} & \begin{tabular}[c]{@{}c@{}}$$\\ 0.1277\\ $$\end{tabular} & \begin{tabular}[c]{@{}c@{}}$$\\ 0.1298\\ $$\end{tabular}  & \begin{tabular}[c]{@{}c@{}}$$\\ 0.896\\ $$\end{tabular}      & \textbf{\begin{tabular}[c]{@{}c@{}}NB\\ AG \\ LWYY\end{tabular}} & \begin{tabular}[c]{@{}c@{}}-0.4164\\ -0.4168\\ -0.4168\end{tabular}                    & \begin{tabular}[c]{@{}l@{}}0.6594\\ 0.6592\\ 0.6592\end{tabular}             & \begin{tabular}[c]{@{}c@{}}0.1204\\ 0.1203\\ 0.1203\end{tabular} & \begin{tabular}[c]{@{}c@{}}0.1212\\ 0.1182\\ 0.1211\end{tabular} & \begin{tabular}[c]{@{}c@{}}0.937\\ 0.942\\ 0.937\end{tabular}  \\ \hline
\textbf{1.0}                     & \textbf{Cox}                                                & \begin{tabular}[c]{@{}c@{}}$$\\ -0.3483\\ $$\end{tabular}                              & \begin{tabular}[c]{@{}c@{}}$$\\ 0.7059\\ $$\end{tabular} & \begin{tabular}[c]{@{}c@{}}$$\\ 0.1273\\ $$\end{tabular} & \begin{tabular}[c]{@{}c@{}}$$\\ 0.1292\\ $$\end{tabular}  & \begin{tabular}[c]{@{}c@{}}$$\\ 0.777\\ $$\end{tabular}      & \textbf{\begin{tabular}[c]{@{}c@{}}NB\\ AG\\ LWYY\end{tabular}}  & \begin{tabular}[c]{@{}c@{}}-0.3660\\ -0.3667\\ -0.3667\end{tabular}                    & \begin{tabular}[c]{@{}l@{}}0.6935\\ 0.6930\\ 0.6930\end{tabular}             & \begin{tabular}[c]{@{}c@{}}0.1229\\ 0.1234\\ 0.1234\end{tabular} & \begin{tabular}[c]{@{}c@{}}0.1263\\ 0.1107\\ 0.1245\end{tabular} & \begin{tabular}[c]{@{}c@{}}0.835\\ 0.891\\ 0.844\end{tabular}  \\ \hline
\end{tabular}}
\caption[S2.PPMS - Results of the MS-specific simulation study ($\beta_{hj}=\log(0.7)$ and $U_{1}$)]{S2.PPMS - Results of the MS-specific simulation study when the true treatment effect is $\beta_{hj}=\log(0.7)$ and heterogeneity is specified via option $U_{1}$, CDP12 events are derived according to a time-to-confirmation-of-CDP endpoint with fixed reference system, N=10000 simulations, n=1000 patients (1:1 randomization) \\
Evaluation measures: HR/RR = mean treatment effect across simulations, SE = standard deviation of estimators across simulations, $SEE$ = mean standard error across simulations}
\label{SimulationResultsS2H1Z1}
\end{table}
\end{landscape}

\begin{landscape}
\begin{table}[]
\centering
\resizebox{\textwidth}{!}{ 
\begin{tabular}{|c|cccccc|cccccc|}
\hline
\multirow{2}{*}{$\boldsymbol{\phi}$} & \multicolumn{6}{c|}{\textbf{Time-to-first-event method}} & \multicolumn{6}{c|}{\textbf{Recurrent event methods}} \\ \cline{2-13} 
 & \textbf{\begin{tabular}[c]{@{}c@{}}Model\\ $$\end{tabular}} & \textbf{\begin{tabular}[c]{@{}c@{}}$\boldsymbol{\log{(\text{HR})}}$\\ $$\end{tabular}} & \textbf{\begin{tabular}[c]{@{}c@{}}HR\\ $$\end{tabular}} & \textbf{\begin{tabular}[c]{@{}c@{}}SE\\ $$\end{tabular}} & \textbf{\begin{tabular}[c]{@{}c@{}}SEE\\ $$\end{tabular}} & \textbf{\begin{tabular}[c]{@{}c@{}}Power\\ $$\end{tabular}} & \textbf{\begin{tabular}[c]{@{}c@{}}Model\\ $$\end{tabular}} & \textbf{\begin{tabular}[c]{@{}c@{}}$\boldsymbol{\log{(\text{HR/RR})}}$\\ $$\end{tabular}} & \textbf{\begin{tabular}[c]{@{}c@{}}HR/RR\\ $$\end{tabular}} & \textbf{\begin{tabular}[c]{@{}c@{}}SE\\ $$\end{tabular}} & \textbf{\begin{tabular}[c]{@{}c@{}}SEE\\ $$\end{tabular}} & \textbf{\begin{tabular}[c]{@{}c@{}}Power\\ $$\end{tabular}} \\ \hline
\textbf{0.0} & \textbf{Cox} & \begin{tabular}[c]{@{}c@{}}$$\\ 0.0012\\ $$\end{tabular} & \begin{tabular}[c]{@{}c@{}}$$\\ 1.0012\\ $$\end{tabular} & \begin{tabular}[c]{@{}c@{}}$$\\ 0.1265\\ $$\end{tabular} & \begin{tabular}[c]{@{}c@{}}$$\\ 0.1277\\ $$\end{tabular} & \begin{tabular}[c]{@{}c@{}}$$\\ 0.047\\ $$\end{tabular} & \textbf{\begin{tabular}[c]{@{}c@{}}NB\\ AG\\ LWYY\end{tabular}} & \begin{tabular}[c]{@{}c@{}}-0.0003\\ -0.0001\\ -0.0001\end{tabular} & \begin{tabular}[c]{@{}c@{}}0.9997\\ 0.9999\\ 0.9999\end{tabular} & \begin{tabular}[c]{@{}c@{}}0.1178\\ 0.1178\\ 0.1178\end{tabular} & \begin{tabular}[c]{@{}c@{}}0.1186\\ 0.1182\\ 0.1168\end{tabular} & \begin{tabular}[c]{@{}c@{}}0.049\\ 0.051\\ 0.054\end{tabular} \\ \hline
\textbf{0.15} & \textbf{Cox} & \begin{tabular}[c]{@{}c@{}}$$\\ 0.0000\\ $$\end{tabular} & \begin{tabular}[c]{@{}c@{}}$$\\ 1.0000\\ $$\end{tabular} & \begin{tabular}[c]{@{}c@{}}$$\\ 0.1291\\ $$\end{tabular} & \begin{tabular}[c]{@{}c@{}}$$\\ 0.1278\\ $$\end{tabular} & \begin{tabular}[c]{@{}c@{}}$$\\ 0.053\\ $$\end{tabular} & \textbf{\begin{tabular}[c]{@{}c@{}}NB\\ AG \\ LWYY\end{tabular}} & \begin{tabular}[c]{@{}c@{}}-0.0002\\ -0.0001\\ -0.0001\end{tabular} & \begin{tabular}[c]{@{}c@{}}0.9998\\ 0.9999\\ 0.9999\end{tabular} & \begin{tabular}[c]{@{}c@{}}0.1198\\ 0.1204\\ 0.1204\end{tabular} & \begin{tabular}[c]{@{}c@{}}0.1185\\ 0.1175\\ 0.1174\end{tabular} & \begin{tabular}[c]{@{}c@{}}0.053\\ 0.056\\ 0.056\end{tabular} \\ \hline
\textbf{1.0} & \textbf{Cox} & \begin{tabular}[c]{@{}c@{}}$$\\ 0.0000\\ $$\end{tabular} & \begin{tabular}[c]{@{}c@{}}$$\\ 1.0000\\ $$\end{tabular} & \begin{tabular}[c]{@{}c@{}}$$\\ 0.1265\\ $$\end{tabular} & \begin{tabular}[c]{@{}c@{}}$$\\ 0.1277\\ $$\end{tabular} & \begin{tabular}[c]{@{}c@{}}$$\\ 0.047\\ $$\end{tabular} & \textbf{\begin{tabular}[c]{@{}c@{}}NB\\ AG\\ LWYY\end{tabular}} & \begin{tabular}[c]{@{}c@{}}0.0004\\ 0.0004\\ 0.0004\end{tabular} & \begin{tabular}[c]{@{}c@{}}1.0004\\ 1.0004\\ 1.0004\end{tabular} & \begin{tabular}[c]{@{}c@{}}0.1190\\ 0.1194\\ 0.1194\end{tabular} & \begin{tabular}[c]{@{}c@{}}0.1208\\ 0.1141\\ 0.1201\end{tabular} & \begin{tabular}[c]{@{}c@{}}0.046\\ 0.060\\ 0.048\end{tabular} \\ \hline
\end{tabular}}
\caption[S2.PPMS - Results of the MS-specific simulation study ($\beta_{hj}=\log(1.0)$ and $U_{2}$)]{S2.PPMS - Results of the MS-specific simulation study when the true treatment effect is $\beta_{hj}=\log(1.0)$ and heterogeneity is specified via $U_{2}$, CDP12 events are derived according to a time-to-confirmation-of-CDP endpoint with fixed reference system, N=10000 simulations, n=1000 patients (1:1 randomization) \\
Evaluation measures: HR/RR = mean treatment effect across simulations, SE = standard deviation of estimators across simulations, $SEE$ = mean standard error across simulations}
\label{SimulationResultsS2H0Z2}
\end{table}

\begin{table}[]
\centering
\scalebox{0.75}{ 
\begin{tabular}{|c|cccccc|cccccc|}
\hline
\multirow{2}{*}{$\boldsymbol{\phi}$} & \multicolumn{6}{c|}{\textbf{Time-to-first-event method}}                                                                                                                                                                                                                                                                                                                                             & \multicolumn{6}{c|}{\textbf{Recurrent event methods}}                                                                                                                                                                                                                                                                                                                                                                              \\ \cline{2-13} 
                                             & \textbf{\begin{tabular}[c]{@{}c@{}}Model\\ $$\end{tabular}} & \textbf{\begin{tabular}[c]{@{}c@{}}$\boldsymbol{\log{(\text{HR})}}$\\ $$\end{tabular}} & \textbf{\begin{tabular}[c]{@{}c@{}}HR\\ $$\end{tabular}} & \textbf{\begin{tabular}[c]{@{}c@{}}SE\\ $$\end{tabular}} & \textbf{\begin{tabular}[c]{@{}c@{}}SEE\\ $$\end{tabular}} & \textbf{\begin{tabular}[c]{@{}c@{}}Power\\ $$\end{tabular}} & \textbf{\begin{tabular}[c]{@{}c@{}}Model\\ $$\end{tabular}}      & \textbf{\begin{tabular}[c]{@{}c@{}}$\boldsymbol{\log{(\text{HR/RR})}}$\\ $$\end{tabular}} & \textbf{\begin{tabular}[c]{@{}c@{}}HR/RR\\ $$\end{tabular}}         & \textbf{\begin{tabular}[c]{@{}c@{}}SE\\ $$\end{tabular}}         & \textbf{\begin{tabular}[c]{@{}c@{}}SEE\\ $$\end{tabular}}        & \textbf{\begin{tabular}[c]{@{}c@{}}Power\\ $$\end{tabular}}   \\ \hline
\textbf{0.0}                                 & \textbf{Cox}                                                & \begin{tabular}[c]{@{}c@{}}$$\\ -0.4279\\ $$\end{tabular}                              & \begin{tabular}[c]{@{}c@{}}$$\\ 0.6519\\ $$\end{tabular} & \begin{tabular}[c]{@{}c@{}}$$\\ 0.1281\\ $$\end{tabular} & \begin{tabular}[c]{@{}c@{}}$$\\ 0.1300\\ $$\end{tabular}  & \begin{tabular}[c]{@{}c@{}}$$\\ 0.919\\ $$\end{tabular}     & \textbf{\begin{tabular}[c]{@{}c@{}}NB\\ AG\\ LWYY\end{tabular}}  & \begin{tabular}[c]{@{}c@{}}-0.4233\\ -0.4235\\ -0.4235\end{tabular}                    & \begin{tabular}[c]{@{}c@{}}0.6549\\ 0.6548\\ 0.6548\end{tabular} & \begin{tabular}[c]{@{}c@{}}0.1204\\ 0.1207\\ 0.1207\end{tabular} & \begin{tabular}[c]{@{}c@{}}0.1208\\ 0.1197\\ 0.1202\end{tabular} & \begin{tabular}[c]{@{}c@{}}0.945\\ 0.946\\ 0.945\end{tabular} \\ \hline
\textbf{0.15}                                & \textbf{Cox}                                                & \begin{tabular}[c]{@{}c@{}}$$\\ -0.4283\\ $$\end{tabular}                              & \begin{tabular}[c]{@{}c@{}}$$\\ 0.6516\\ $$\end{tabular} & \begin{tabular}[c]{@{}c@{}}$$\\ 0.1293\\ $$\end{tabular} & \begin{tabular}[c]{@{}c@{}}$$\\ 0.1299\\ $$\end{tabular}  & \begin{tabular}[c]{@{}c@{}}$$\\ 0.917\\ $$\end{tabular}     & \textbf{\begin{tabular}[c]{@{}c@{}}NB\\ AG \\ LWYY\end{tabular}} & \begin{tabular}[c]{@{}c@{}}-0.4276\\ -0.4278\\ -0.4278\end{tabular}                    & \begin{tabular}[c]{@{}c@{}}0.6521\\ 0.6520\\ 0.6520\end{tabular} & \begin{tabular}[c]{@{}c@{}}0.1208\\ 0.1210\\ 0.1210\end{tabular} & \begin{tabular}[c]{@{}c@{}}0.1210\\ 0.1191\\ 0.1208\end{tabular} & \begin{tabular}[c]{@{}c@{}}0.948\\ 0.952\\ 0.949\end{tabular} \\ \hline
\textbf{1.0}                                 & \textbf{Cox}                                                & \begin{tabular}[c]{@{}c@{}}$$\\ -0.4309\\ $$\end{tabular}                              & \begin{tabular}[c]{@{}c@{}}$$\\ 0.6499\\ $$\end{tabular} & \begin{tabular}[c]{@{}c@{}}$$\\ 0.1300\\ $$\end{tabular} & \begin{tabular}[c]{@{}c@{}}$$\\ 0.1299\\ $$\end{tabular}  & \begin{tabular}[c]{@{}c@{}}$$\\ 0.920\\ $$\end{tabular}     & \textbf{\begin{tabular}[c]{@{}c@{}}NB\\ AG\\ LWYY\end{tabular}}  & \begin{tabular}[c]{@{}c@{}}-0.4486\\ -0.4491\\ -0.4491\end{tabular}                    & \begin{tabular}[c]{@{}c@{}}0.6385\\ 0.6382\\ 0.6382\end{tabular} & \begin{tabular}[c]{@{}c@{}}0.1238\\ 0.1240\\ 0.1240\end{tabular} & \begin{tabular}[c]{@{}c@{}}0.1236\\ 0.1157\\ 0.1233\end{tabular} & \begin{tabular}[c]{@{}c@{}}0.956\\ 0.965\\ 0.957\end{tabular} \\ \hline
\end{tabular}}
\caption[S2.PPMS - Results of the MS-specific simulation study ($\beta_{hj}=\log(0.7)$ and $U_{2}$)]{S2.PPMS - Results of the MS-specific simulation study when the true treatment effect is $\beta_{hj}=\log(0.7)$ and heterogeneity is specified via $U_{2}$, CDP12 events are derived according to a time-to-confirmation-of-CDP endpoint with fixed reference system, N=10000 simulations, n=1000 patients (1:1 randomization) \\
Evaluation measures: HR/RR = mean treatment effect across simulations, SE = standard deviation of estimators across simulations, $SEE$ = mean standard error across simulations}
\label{SimulationResultsS2H1Z2}
\end{table}
\end{landscape}
\newpage 
\textbf{Option $\boldsymbol{U_{2}}$} \newline \newline 
\textbf{Case: $\boldsymbol{\beta_{hj}=\log(0.7)}$}  \newline 
Results of the MS-specific simulation study under $H_{1}: \{ \beta_{hj}=\log(0.7) \}$ are represented in Table $\ref{SimulationResultsS2H1Z2}$. The scenario S2.PPMS.effect.homo has been already described previously. When $\phi \neq 0$, the simulation suggests that the presence of between-patient variability does not really affect the treatment effect sizes. As before, the time-to-first-event and recurrent event methods provide similar treatment effect estimates in scenarios with frailties. Variability of the treatment effect estimators is smaller for NB, AG and LWYY analyses than for the conventional time-to-first-event method. The mean SE estimate is close to the empirically determined Monte Carlo SD with all approaches.
\newline \newline 
\textbf{Case: $\boldsymbol{\beta_{hj}=\log(1.0)}$}  \newline 
Table $\ref{SimulationResultsS2H0Z1}$ summarizes the results of the MS-specific simulation study under $H_{0}: \{ \beta_{hj}=\log(1.0) \}$. Similar results as under the alternative hypothesis can be observed. Both time-to-first-event and recurrent event methods result in similar treatment effect sizes, even in presence of heterogeneity. With regard to precision, the variance of the treatment effect estimators across the simulations is increased for Cox. 

\begin{figure}[H]
\centering
\scalebox{0.60}{
\includegraphics{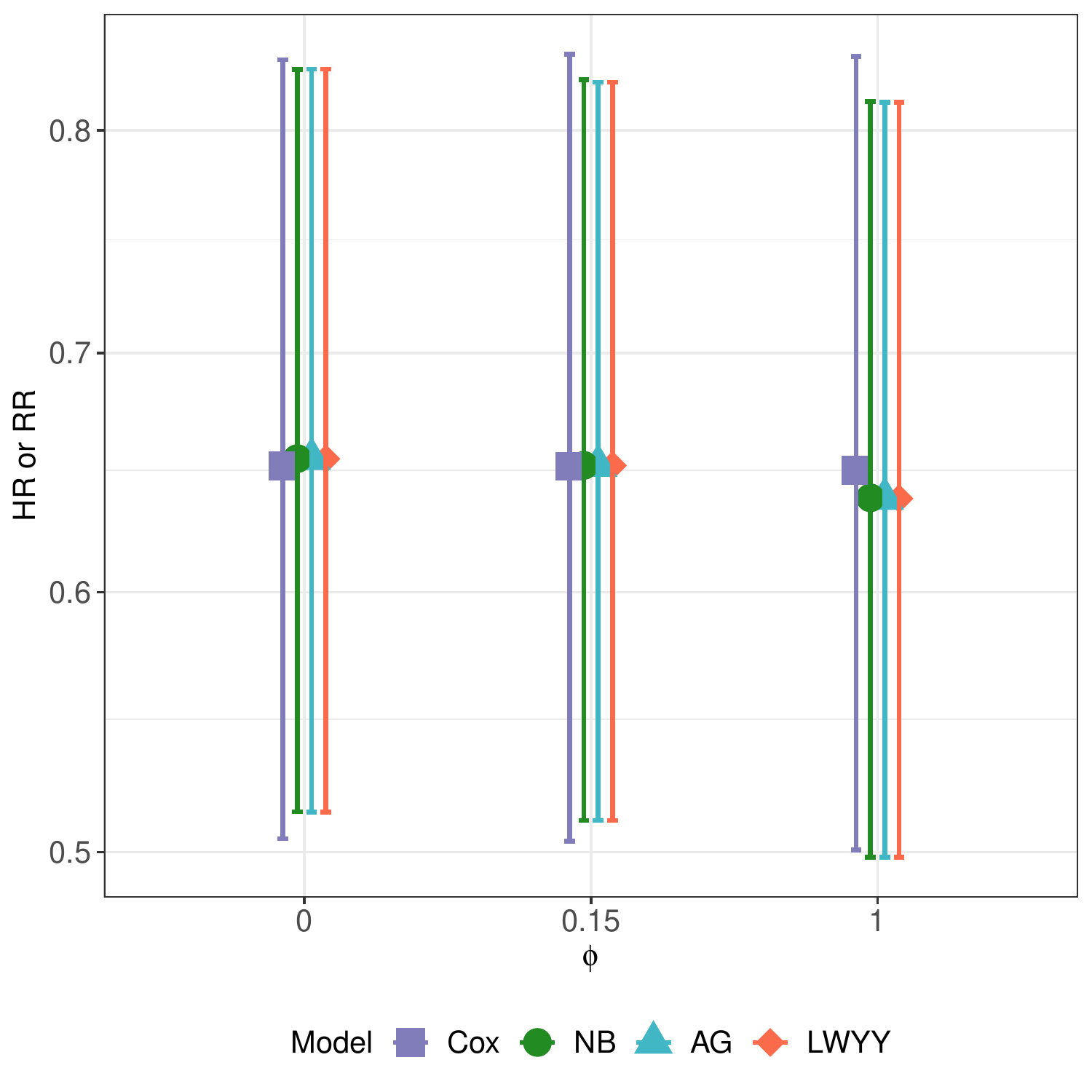}
}
\caption[S2.PPMS - Treatment effect estimates obtained from time-to-first-event and recurrent event methods in MS-specific simulation study ($U_{2}$)]{S2.PPMS - Treatment effect estimates obtained from time-to-first-event and recurrent event methods according to different heterogeneity parameters and based on option $U_{2}$, CDP12 events are derived according to a time-to-confirmation-of-CDP endpoint with fixed reference system, error bars indicate $2.5\%$ and $97.5\%$ quantiles, N=10000 simulations, n=1000 patients}
\label{S2.PPMS.precision.Z2}
\end{figure}

\subsubsection*{Power and type I error}
\begin{figure}[H]
\centering
\begin{minipage}{\textwidth}
  \begin{minipage}[c]{0.49\textwidth}
    \centering
\scalebox{0.50}{
\includegraphics{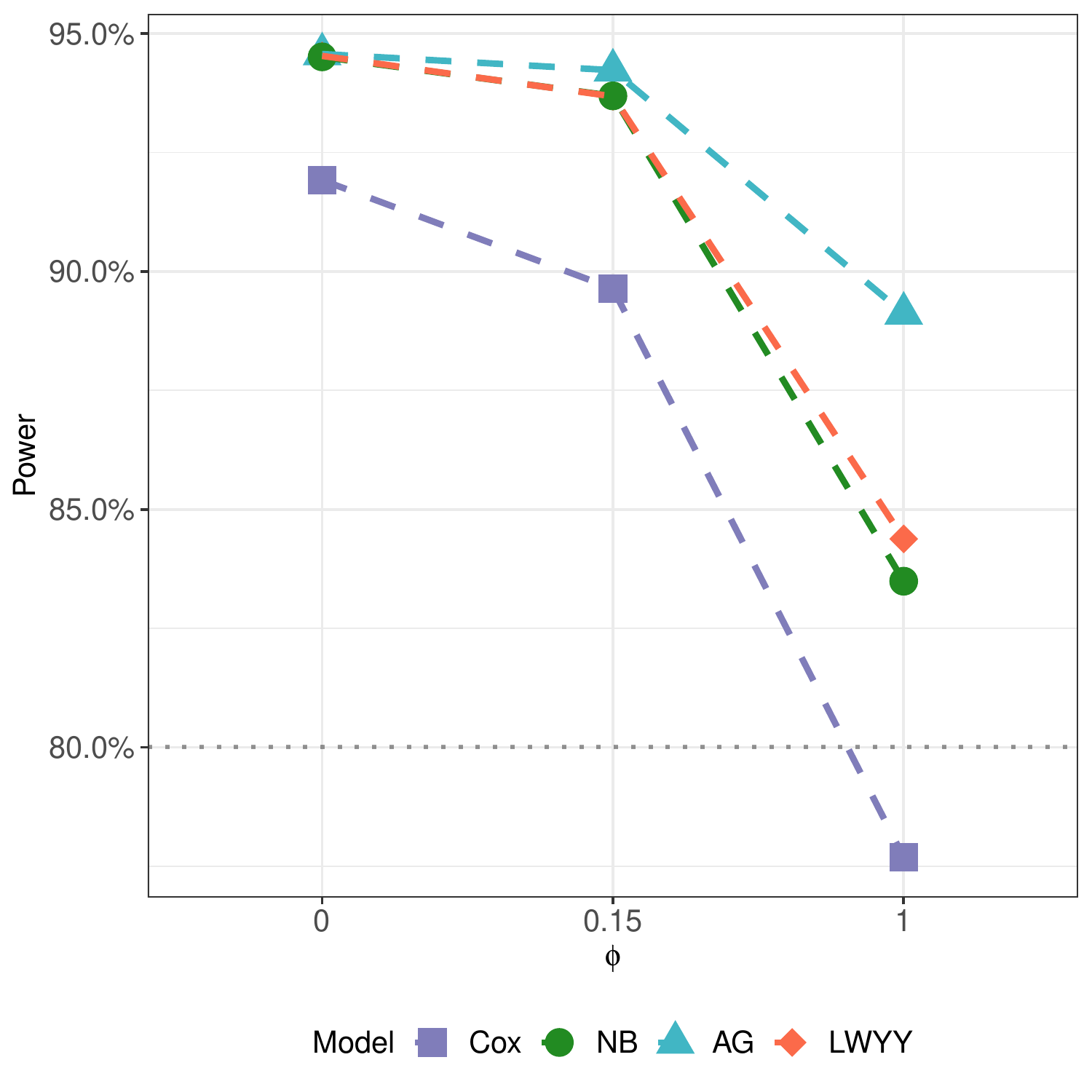}
}
\end{minipage}
\hfill
\begin{minipage}[c]{0.49\textwidth}
\centering 
\scalebox{0.50}{
\includegraphics{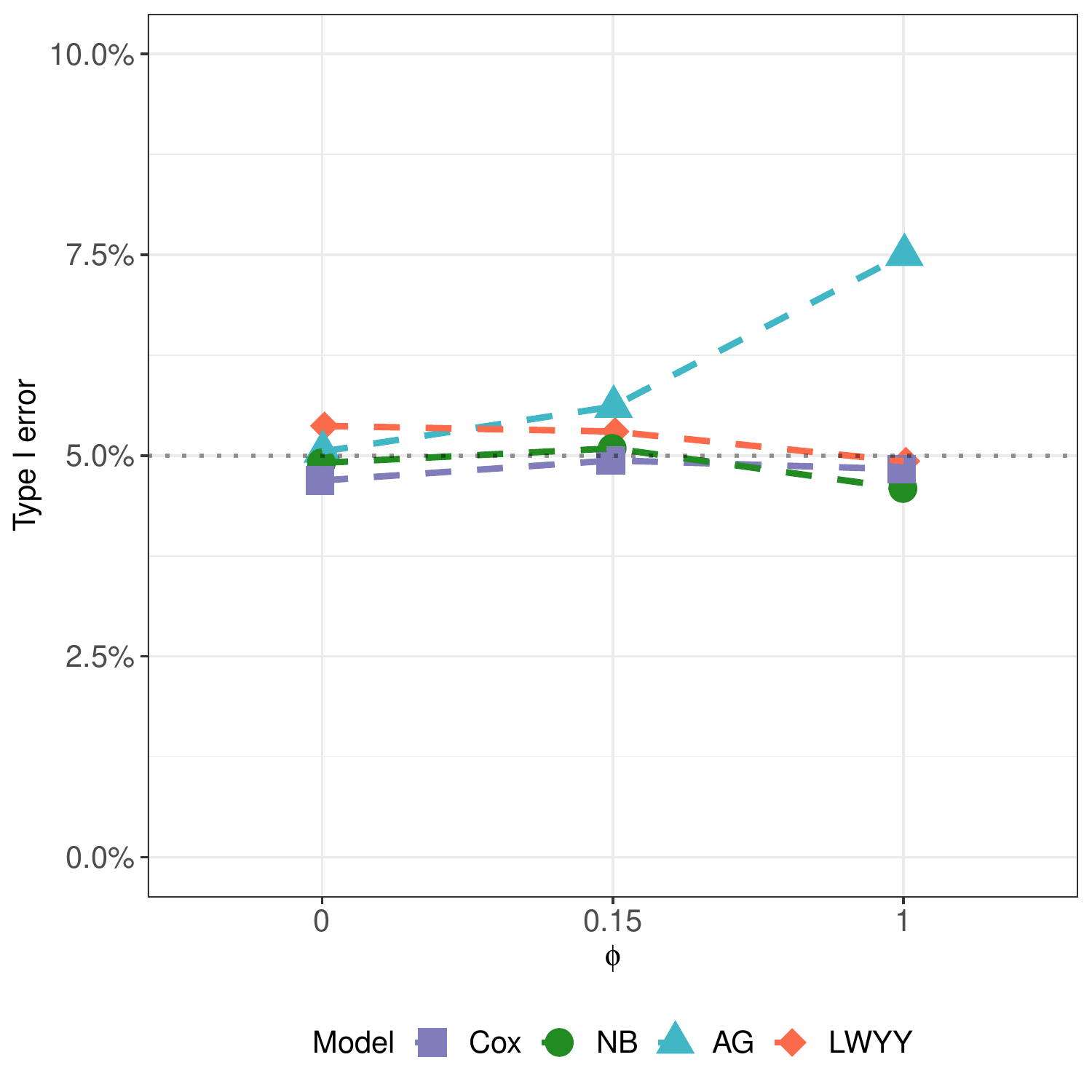}
}
\end{minipage}
$$
$$
\begin{minipage}[c]{\textwidth}
\centering
\scalebox{0.82}{
\begin{tabular}{|c|c|ccc|c|c|ccc|}
\hline
                       & \multicolumn{4}{c|}{\textbf{\begin{tabular}[c]{@{}c@{}}Power = $ 1 - \boldsymbol{\beta}$\\ (in \%)\end{tabular}}}                                                                                 &           & \multicolumn{4}{c|}{\textbf{\begin{tabular}[c]{@{}c@{}}Type I error = $\boldsymbol{\alpha}$\\ (in \%)\end{tabular}}}                                                                                 \\ \cline{1-5} \cline{7-10} 
\multicolumn{1}{|l|}{} & \textbf{\begin{tabular}[c]{@{}c@{}}Time-to-$1^{st}$-event \\ method\end{tabular}} & \multicolumn{3}{c|}{\textbf{\begin{tabular}[c]{@{}c@{}}Recurrent event \\ methods\end{tabular}}} & \textbf{} & \textbf{\begin{tabular}[c]{@{}c@{}}Time-to-$1^{st}$-event \\ method\end{tabular}} & \multicolumn{3}{c|}{\textbf{\begin{tabular}[c]{@{}c@{}}Recurrent event \\ methods\end{tabular}}} \\ \cline{1-5} \cline{7-10} 
\textbf{$\boldsymbol{\phi}$}        & \textbf{Cox}                                                                      & \textbf{NB}                    & \textbf{AG}                   & \textbf{LWYY}                   &           & \textbf{Cox}                                                                         & \textbf{NB}                    & \textbf{AG}                   & \textbf{LWYY}                   \\ \cline{1-5} \cline{7-10} 
\textbf{0.0}           & 91.9                                                                             & 94.5                          & 94.6                         & 94.5                          &           & 4.7                                                                                 & 4.9                           & 5.1                          & 5.4                            \\
\textbf{0.15}          & 89.6                                                                            & 93.7                          & 94.2                         & 93.7                           &           & 4.9                                                                                 & 5.1                           & 5.6                          & 5.3                            \\
\textbf{1.0}           & 77.7                                                                             & 83.5                          & 89.1                         & 84.4                           &           & 4.8                                                                                 & 4.6                           & 7.5                          & 4.9                            \\ \hline
\end{tabular}
}
\end{minipage}
\caption[S2.PPMS - Power and type I error ($U_{1}$)]{S2.PPMS - Power and type I error using heterogeneity specification $U_{1}$ (based on Table $\ref{SimulationResultsS2H0Z1}$ and Table $\ref{SimulationResultsS2H1Z1}$), N=10000 simulations, n=1000 patients}
\label{S2.PPMS.DEF1.Z1.power}
\end{minipage}
\end{figure}

\textbf{Option $\boldsymbol{U_{1}}$} \newline 
In Figure $\ref{S2.PPMS.DEF1.Z1.power}$, type I error rates under $H_{0}: \{ \beta_{hj}=\log(1.0) \} $ and the power under $H_{1}: \{ \beta_{hj}=\log(0.7) \}$ are reported. When $\phi=0.0$, the type I error is well controlled with all methods. As in the general simulation study, the tests based on Cox, NB and LWYY models maintain the type I errors around the nominal level in the presence of heterogeneity, whereas the AG model is not able to control it. The inflation of the type I error for the AG model can be explained by the low SE estimates. Apparently, the inflation is more extreme when recurrent event data is generated from a mixed non-homogeneous Poisson process (S1.PPMS.noeffect.hetero2: $\alpha = 9.1 \%$ versus S2.PPMS.noeffect.hetero2: $\alpha=7.5 \%$ using heterogeneity matrix $U_{1}$). \newpage 

\begin{figure}[H]
\centering
\begin{minipage}{\textwidth}
  \begin{minipage}[c]{0.5\textwidth}
    \centering
\scalebox{0.50}{
\includegraphics{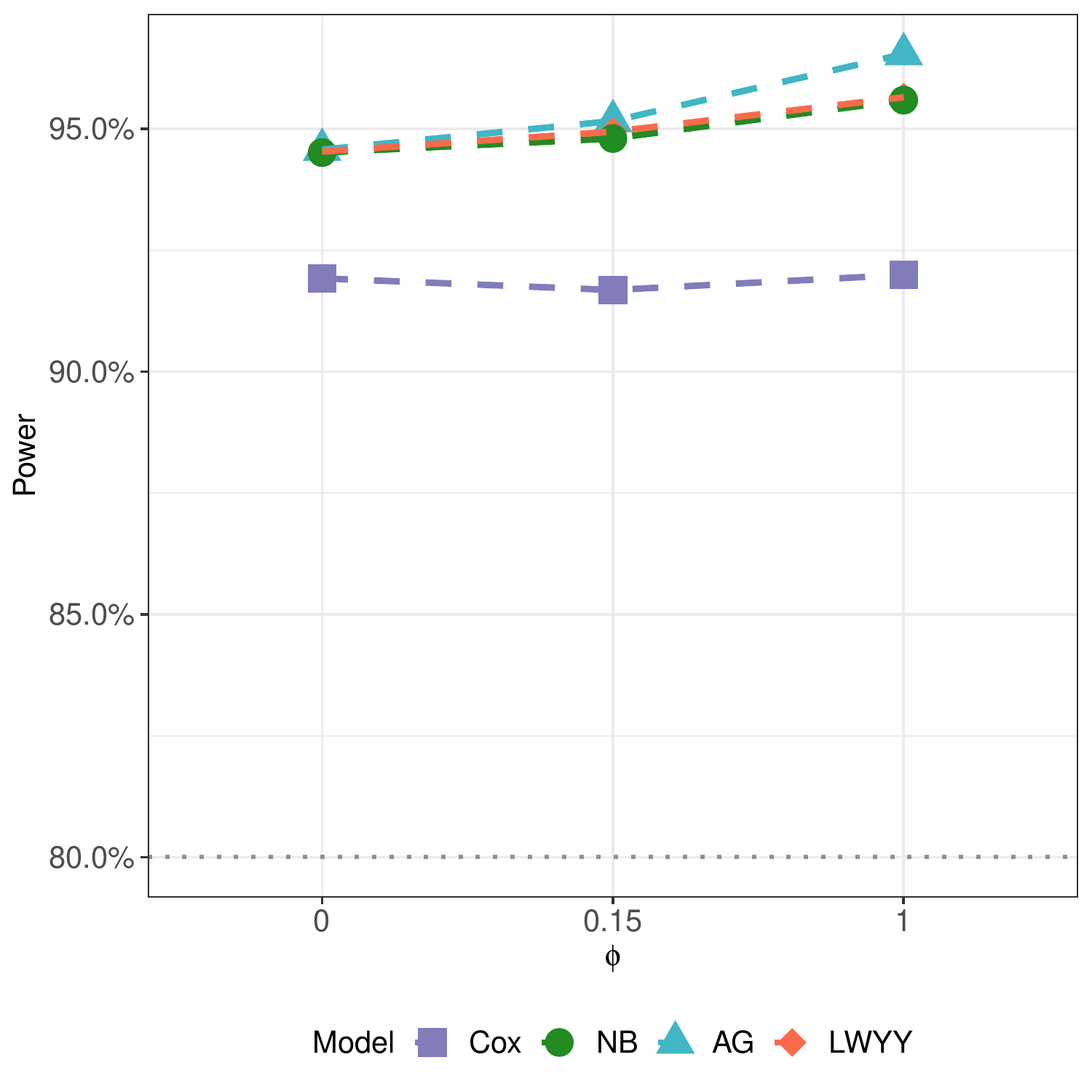}
}
\end{minipage}
\hfill
\begin{minipage}[c]{0.5\textwidth}
\centering 
\scalebox{0.50}{
\includegraphics{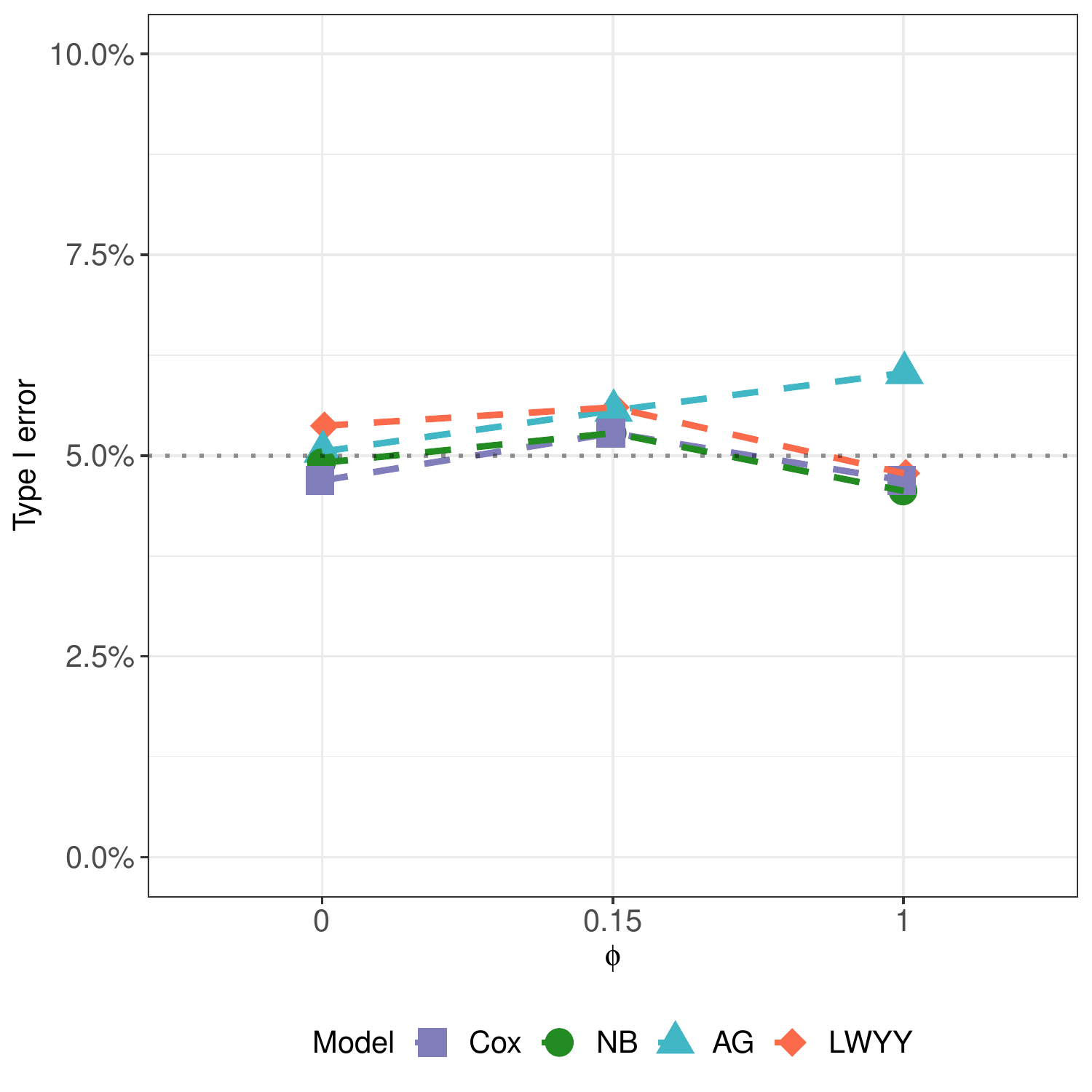}
}
\end{minipage}
$$
$$
\begin{minipage}[c]{\textwidth}
\centering
\scalebox{0.82}{
\begin{tabular}{|c|c|ccc|c|c|ccc|}
\hline
                       & \multicolumn{4}{c|}{\textbf{\begin{tabular}[c]{@{}c@{}}Power = $ 1 - \boldsymbol{\beta}$\\ (in \%)\end{tabular}}}                                                                                 &           & \multicolumn{4}{c|}{\textbf{\begin{tabular}[c]{@{}c@{}}Type I error = $\boldsymbol{\alpha}$\\ (in \%)\end{tabular}}}                                                                              \\ \cline{1-5} \cline{7-10} 
\multicolumn{1}{|l|}{} & \textbf{\begin{tabular}[c]{@{}c@{}}Time-to-$1^{st}$-event \\ method\end{tabular}} & \multicolumn{3}{c|}{\textbf{\begin{tabular}[c]{@{}c@{}}Recurrent event \\ methods\end{tabular}}} & \textbf{} & \textbf{\begin{tabular}[c]{@{}c@{}}Time-to-$1^{st}$-event \\ method\end{tabular}} & \multicolumn{3}{c|}{\textbf{\begin{tabular}[c]{@{}c@{}}Recurrent event \\ methods\end{tabular}}} \\ \cline{1-5} \cline{7-10} 
\textbf{$\boldsymbol{\phi}$}        & \textbf{Cox}                                                                      & \textbf{NB}                    & \textbf{AG}                   & \textbf{LWYY}                   &           & \textbf{Cox}                                                                      & \textbf{NB}                    & \textbf{AG}                   & \textbf{LWYY}                   \\ \cline{1-5} \cline{7-10} 
\textbf{0.0}           & 91.9                                                                             & 94.5                         & 94.6                         & 94.5                           &           & 4.7                                                                             & 4.9                           & 5.1                          & 5.4                            \\
\textbf{0.15}          & 91.7                                                                             & 94.8                          & 95.2                         & 94.9                           &           & 5.3                                                                              & 5.3                          & 5.6                          & 5.6                            \\
\textbf{1.0}           & 92.0                                                                             & 95.6                          & 96.5                         & 95.7                           &           & 4.7                                                                              & 4.6                           & 6.0                          & 4.8                            \\ \hline
\end{tabular}
}
\end{minipage}
\caption[S2.PPMS - Power and type I error ($U_{2}$)]{S2.PPMS - Power and type I error using heterogeneity specification $U_{2}$ (based on Table $\ref{SimulationResultsS2H0Z2}$ and Table $\ref{SimulationResultsS2H1Z2}$), N=10000 simulations, n=1000 patients}
\label{S2.PPMS.DEF1.Z2.power}
\end{minipage}
\end{figure}

The left panel of Figure $\ref{S2.PPMS.DEF1.Z1.power}$ plots the power as a function of $\phi$ and illustrates the potential gain of study designs based on recurrent CDP events. It can be observed that recurrent event methods including NB, AG and LWYY provide greater statistical power than the Cox model over all scenarios considered. In order to keep the study designs in both simulation setups as similar as possible, the MS-specific study design was also designed considering that the chance of detecting a true overall HR of $0.7$ is $80 \%$. As seen from the previous section, a treatment effect of $\exp(\beta_{hj})=0.7$ on the transition hazards is associated with an estimated overall treatment effect of $0.65$ obtained from Cox, NB, AG and LWYY analyses, for $\phi=0.0$. Consequently, the study tends to be 'overpowered'. In case of a homogeneous study population, power increased from $92 \%$ for the time-to-first-event method to approximately $95 \%$ for the recurrent event analyses. Similar results have been found in the general simulation study, where power increased from $80 \%$ for the time-to-first-event method to approximately $85 \%$ for the recurrent event analyses. Further, the MS-specific simulation also suggests that between-patient variability has a big impact on statistical power. With increasing heterogeneity, a decline in the empirical power can be observed for all approaches, which is of greater magnitude within the time-to-first-event method. When going from $\phi=0.0$ to $\phi=1.0$, the power of the statistical test based on the Cox model reduces from $92 \%$ to $78 \%$, leading to a $14 \%$ reduction due to the presence of heterogeneity. By comparison, a power loss of approximately $11 \%$, $5 \%$ and $10 \%$ can be deduced for the NB, AG and LWYY models. Overall, NB and LWYY models are comparable in terms of type I error control and power. 
\newline \newline 
\textbf{Option $\boldsymbol{U_{2}}$} \newline 
Figure $\ref{S2.PPMS.DEF1.Z2.power}$ includes the empirical type I error rates when $\exp(\beta_{hj})=1.0$ and the power when $\exp(\beta_{hj})=0.7$. When $\phi=0.0$, there is a good control of the type I error based on the Cox, NB, AG and LWYY analyses. When $\phi \neq 0$, the tests based on Cox, NB and LWYY maintain the type I error rate. For the AG model, the type I error is increased . 
\newpage 
From the left panel of Figure $\ref{S2.PPMS.DEF1.Z2.power}$, the empirical power rates resulting from the Cox, NB, AG and LWYY analyses can be extracted. The case $\phi=0.0$ is identical to the one under $U_{1}$. In contrast to $U_{1}$, between-patient variability specified via the heterogeneity matrix $U_{2}$ does not seem to have a meaningful impact on statistical power. While the test based on the time-to-first-event method yields constant power approximations even with increasing heterogeneity, a slight increase in power can be observed with the recurrent event methods, as $\phi$ varies. 

\subsection{Summary}
For homogeneous study populations, the MS-specific simulation study clearly demonstrates the benefit of recurrent event over time-to-first-event methods in terms of statistical power. The power increased from $92 \%$ for the time-to-first-event method to approximately $95 \%$ for the recurrent event analyses. Consequently, the sample size of a clinical PPMS trial with recurrent CDP endpoint could be $ \sim 12.9 \%$ lower compared to the time-to-first-CDP endpoint.
For moderate and high between-patient variability, simulation results differ across the heterogeneity options $U_{1}$ and $U_{2}$. Using option $U_{1}$, heterogeneity is only simulated on the upper diagonal of the EDSS transition intensity matrix (= worsening), while heterogeneity is generated on the upper and lower diagonal (= worsening and improvement) under $U_{2}$. With regard to power and type I error, the MS-specific simulation study under $U_{1}$ leads to conclusions that are in accordance with those obtained from the generic simulation study. Specifically, statistical power of both recurrent event and time-to-first-event methods is reduced by increasing heterogeneity and the AG model fails to control the type I error for high heterogeneity. LWYY and NB analyses are comparable in terms of power and type I error. In contrast, when heterogeneity is specified via $U_{2}$, the treatment effect estimates and statistical power are not considerably affected by increasing variance of the frailty term. This can potentially be explained by the constant frailty term being applied to the whole transition intensity matrix, keeping the ratio between upward and downward transitions equal. 
\newpage
In previous MS studies, disability improvement was observed for some MS patients treated with anti-inflammatory treatments, leading to a negative change from baseline in mean EDSS \parencite{Coles2016, Panitch2008}. As illustrated in the upper panel of Figure $\ref{S2.PPMS.MeanCHG.EDSS.Z}$, patients with negative mean change in EDSS from baseline can be captured when heterogeneity is generated via option $U_{1}$. Therefore, heterogeneity specified via $U_{1}$ is of higher clinical relevance. 

\chapter{Discussion}
This thesis examined whether recurrent event analyses are more efficient in RCTs as compared to conventional time-to-first-event analyses. Simulation studies based on a PPMS population clearly demonstrate that there is a potential for recurrent CDP analyses of progressive MS trials. 
\section{Recurrent event analyses in RCTs}
Over the last decades, many recurrent event methods have been developed \parencite{Cook2007}. As an extension of the Cox proportional hazards model, \citet{Andersen1982} proposed a semiparametric multiplicative intensity model based on a (conditionally) independent increment assumption. \citet{Prentice1981} discussed a semiparametric model in counting process formulation that involves time-dependent stratification and permits the baseline intensity function to vary with increasing number of events. \citet{Wei1989} developed semiparametric methods based on marginal proportional hazards analyses for each distinct event. Recurrent event analyses based on semiparametric rate function models have been studied by \citet{Lin2000}. Parametric negative binomial models are also popular approaches for the analysis of recurrent events.  \newline 
However, only specific models are appropriate for the analysis of a recurrent event endpoint in RCTs.
Treatment comparisons in clinical trials should primarily rely on statistical approaches that provide easily interpretable effect measures and do not condition on post-baseline data such as an individual's past event history during follow-up to retain the beneficial aspects of randomization \parencite{Zhong2019, Cook2009}. Conditional intensity-based models \parencite{Andersen1982, Prentice1981} necessitate full specification of the recurrent event process by modelling the past through previous events and/or internal time-varying covariates and do therefore not fulfill this fundamental requirement. As a result, intensity-based models are not recommended for the analysis of recurrent events in RCTs. 
\newline 
Since marginal rate-based models yield a treatment effect estimate with a simple causal interpretation, semiparametric LWYY models are suitable for efficacy analyses in RCTs, in which treatment is expected to impact the first as well as subsequent events \parencite{Zhong2019}. The rate ratio as resulting effect measure of the LWYY model targets the overall treatment effect, provided that random censoring and multiplicative assumptions are reasonable. To investigate whether treatment effects also persist for events subsequent to the first event, partially conditional rate-based models are recommended to use as supporting analysis. Particularly, the LWYY model assumes that the recurrent event process is independent of the censoring process. As discussed by \citet{Cook2009}, inverse probability of censoring weighting (IPCW) may be used to account for event-dependent censoring. This IPCW approach requires to fully model the underlying censoring mechanism. Recently, \citet{Zhong2019} examined the consequences of model misspecification in the LWYY model through omission of covariates. \citet{Lee2019} proposed semiparametric rate-based models for recurrent episodes and risk-free periods in clinical trials. Recent work by \citet{Tang2019} and \citet{Muetze2019} discusses sample size calculation and group sequential designs with robust semiparametric LWYY models, showing an increasing interest in recurrent event endpoints in future RCTs. 
\newpage 
An alternative method to the LWYY model is the parametric NB model commonly endorsed for recurrent relapses in RRMS. Due to the fact that the LWYY model is semiparametric and does not require to specify the heterogeneity parametrically, the marginal LWYY model is recommended as the primary analysis in RCTs. 

\section{Recurrent CDP events in MS trials}
Traditional endpoints used in clinical MS trials do not generally incorporate all relevant information on disease progression. Progressive forms of MS are characterized by repeated CDP events but only the first CDP is usually considered in the primary or secondary analyses of RCTs, evaluated as time-to-first-event endpoint. Analyses based on the time to the first CDP ignore meaningful information occurring after the first disability progression and utilize available data inefficiently. Specifically, $18\%$ of observed CDP12 events are not used in a time-to-first-event analysis of the ORATORIO trial. In contrast, recurrent event analyses use all clinically relevant disability progression data. 
\newline \newline 
Due to the fact that only a few RRMS patients experience repeated CDP12 events, no major difference between recurrent event and time-to-first-event methods can be seen from reanalyses of the OPERA trials. In RRMS trials, recurrent CDP analyses do not confer any advantages and a time-to-first-CDP analysis using survival methods appears to be the most appropriate endpoint. \newline 
In contrast, reanalyses of the ORATORIO trial in early PPMS show that recurrent event methods including all CDP12 events can demonstrate a larger treatment benefit and increased statistical precision (LWYY: RR $0.723$, $95\%$ CI $[0.572, 0.915]$, p-value = $0.00699$, $314$ CDP12 events) than the original time-to-first-event analysis (Cox: HR $0.759$, $95\%$ CI $[0.589, 0.978]$, p-value = $0.033$, $256$ events). For the ORATORIO trial, NB analyses result in similar findings (RR $0.714$, $95\%$ CI $[0.565, 0.906]$, p-value = $0.0049$, $314$ CDP12 events). 
\newline \newline 
Benefits of recurrent event methods including the LWYY, NB and AG models over the time-to-first-event method in terms of fundamental statistical properties can also be deduced from both PPMS simulation studies. Inclusion of recurrent events leads to considerable gains in statistical power and improved precision compared with analyses that incorporate the first event only.
\newline
In the generic simulation study (S1), power increased from $80 \%$ for the time-to-first-event method to $85 \%$ for the recurrent event analyses in case of a homogeneous study population. This indicates that a trial with $80 \%$ power for a recurrent event CDP endpoint in PPMS could have a $10-15 \%$ reduced sample size compared to a trial powered for the conventional time-to-first-event CDP endpoint. 
The simulation results further reveal that the power of all approaches is negatively affected by increasing heterogeneity, and this applies especially to the Cox model. Due to selection effects and a violation of the proportional hazards assumption, the Cox model is known to result in biased treatment effect estimates in presence of heterogeneity, which explains the diminished power \parencite{Struthers1986}. In contrast, recurrent event methods provide unbiased treatment effect estimates, even with increasing heterogeneity. This is because patients continue to contribute follow-up information after their first event so that high-risk patients are not systematically removed from the recurrent event analysis in later follow-up. In heterogeneous study populations, increases in statistical power with recurrent event methods are even larger so that the approximate gain in sample size becomes larger as well. In terms of power, LWYY and NB analyses yield comparable results. While the Cox, LWYY and NB models provide an adequate type I error control, the AG model is associated with an inflated type I error.   
\newline
The MS-specific simulation study (S2) is more complex in the sense that simulated effect sizes on EDSS transitions do not translate $1:1$ to effect sizes for recurrent CDP12 events. Unbiasedness of treatment effect estimation could therefore not be evaluated but precision is higher with LWYY and NB analyses, as compared to Cox. With regard to power and type I error, the MS-specific simulation study (S2) with heterogeneity option $U_{1}$ leads to conclusions which are in accordance with those from the generic simulation study. Specifically, a recurrent event analysis with a $10-15 \%$ lower sample size would result in the same precision obtained from a time-to-first-event analysis, assuming same recruitment period and study duration.
\newline 
In summary, simulation results are comparable across the generic and MS-specific simulations. Sample size of a trial with a recurrent CDP endpoint could be $10 - 20\%$ lower compared to a time-to-first-CDP endpoint in the PPMS setting. A clinical trial with reduced sample size usually involves faster recruitment of study participants and a shorter study duration. As a consequence, both simulation studies indicate that recurrent event analyses are more efficient than time-to-first-event analyses in PPMS trials.  
\newline \newline 
Several questions regarding the design of future clinical trials with recurrent CDP events may be considered as extensions of the simulation results provided in this work. First of all, it is important to determine the most appropriate definition of recurrent CDP events that are derived from longitudinal EDSS measurements. Since recurrent CDP endpoints have not been considered in clinical MS trials so far, there is no definition accepted by regulators yet. The repeated CDP definitions proposed in this work were discussed with a clinician who considered them to be clinically meaningful. However, it may be still possible to improve it. For instance, other summary measures than the EDSS value at IDP could be used as reference EDSS score for subsequent CDP events. Besides that, a weighted event definition could account for different step sizes (e.g., increase of $1.0$ or $2.0$ points). According to the current definition, an increase of $2.0$ points (e.g., EDSS $4.0 \longrightarrow$ EDSS $6.0$) is considered as $1$ CDP, while two increases of $1.0$ point at consecutive study visits  (e.g., EDSS $4.0 \longrightarrow$ EDSS $5.0$ and EDSS $5.0 \longrightarrow$ EDSS $6.0$) are counted as $2$ CDPs. This is a point which could also be reassessed. 
\newline \newline 
In future clinical MS trials, it may also be of interest to study different types of recurrent progression events simultaneously. Besides EDSS progression, impaired manual dexterity is a frequently reported disability in advanced progressive MS and is measured using the 9-Hole Peg Test (9HPT). Many PPMS and SPMS patients who are more advanced in their disease state may potentially transition into wheelchair so that maintaining upper extremity functions is of major importance. A further outcome measure for disability progression in ambulatory functions is the timed 25-Foot Walk (T25FW). In accordance to CDP progression, time-to-event endpoints based on 9HPT and T25FW are defined as the time to a $20\%$ increase in the 9HPT or the T25FW that is confirmed for at least 12 weeks, respectively. This clearly shows that disease progression in MS patients can be expressed in several ways, making multitype recurrent event methods appealing \parencite{Cook2007}. As an analogue to the single-type LWYY model, \citet{Cai2004} proposed semiparametric marginal rate models for multitype recurrent event data. Frailty modelling for multitype recurrent events in clinical trials has been recently discussed by \citet{Brown2019}.  
 \newline 
\citet{Cadavid2017} proposed the so-called composite 'EDSS-Plus' endpoint combining the CDP, 9HPT and T25FW events to increase the expected number of overall events and power in progressive MS trials. This composite endpoint is analyzed using standard survival  methods (including log-rank test and Cox proportional hazards model) by evaluating the time to the first occurring event (either CDP, 9HPT or T25FW whichever occurs first). However, this approach ignores the fact that MS patients may experience more than one event of any type, leading to loss of information. Based on the findings of this work, recurrent event methods for composite endpoints or multitype recurrent event analyses are expected to be more efficient in this setting. 
\newline \newline 
Results from this thesis demonstrate that innovative study designs based on recurrent endpoints can advance clinical PPMS research. This has the potential to accelerate drug developement and quicker access of new drugs to MS patients. 

\appendix
\chapter{Additional outputs}
\section{Recurrent event analysis}
\subsubsection*{Cumulative mean functions}
\label{AppendixCMF}
\begin{figure}[H] 
\centering
  \subfloat[Sex]{
  \scalebox{0.4}[0.4]{ 
\includegraphics{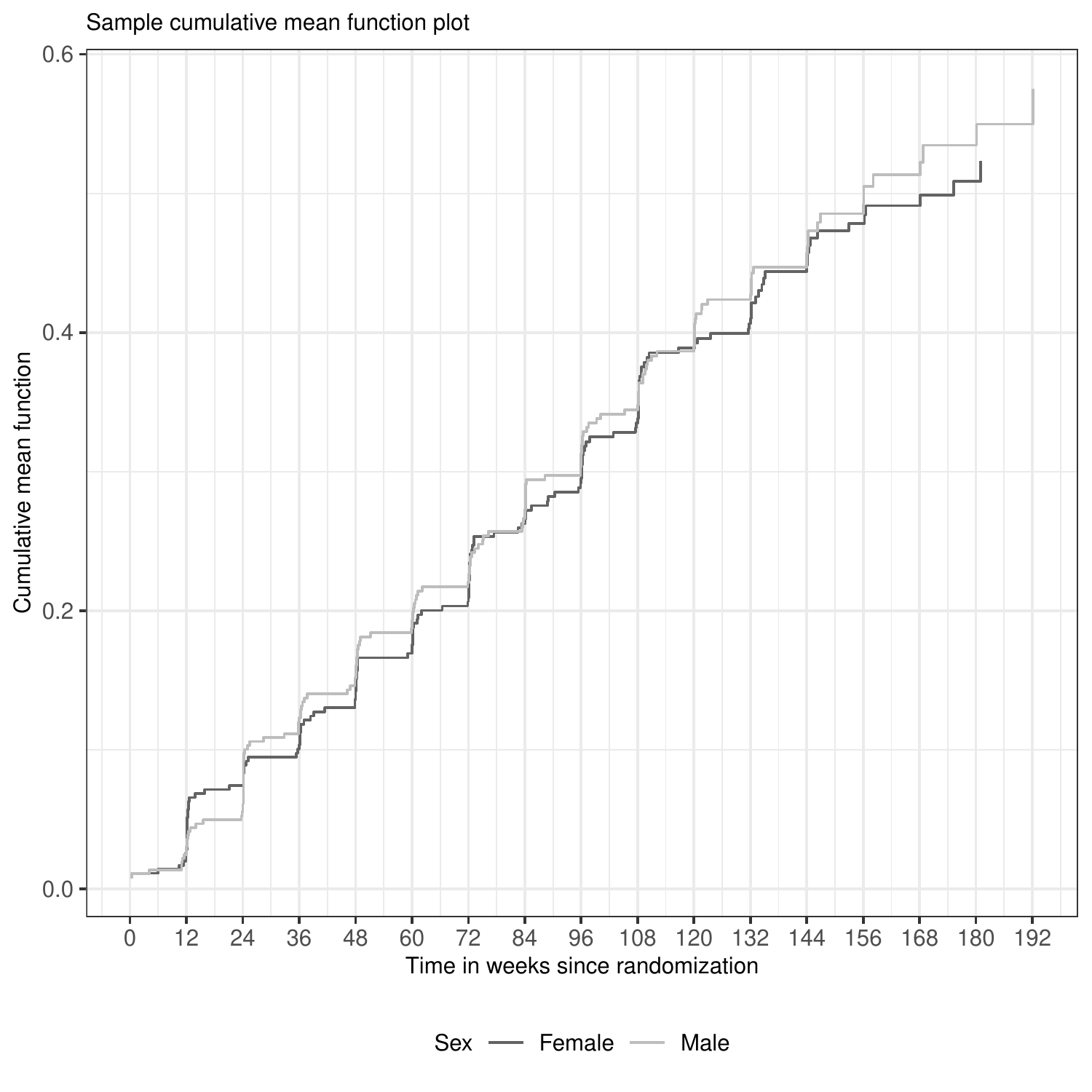}
  }}
  \subfloat[Sex and treatment group]{
  \scalebox{0.4}{ 
\includegraphics{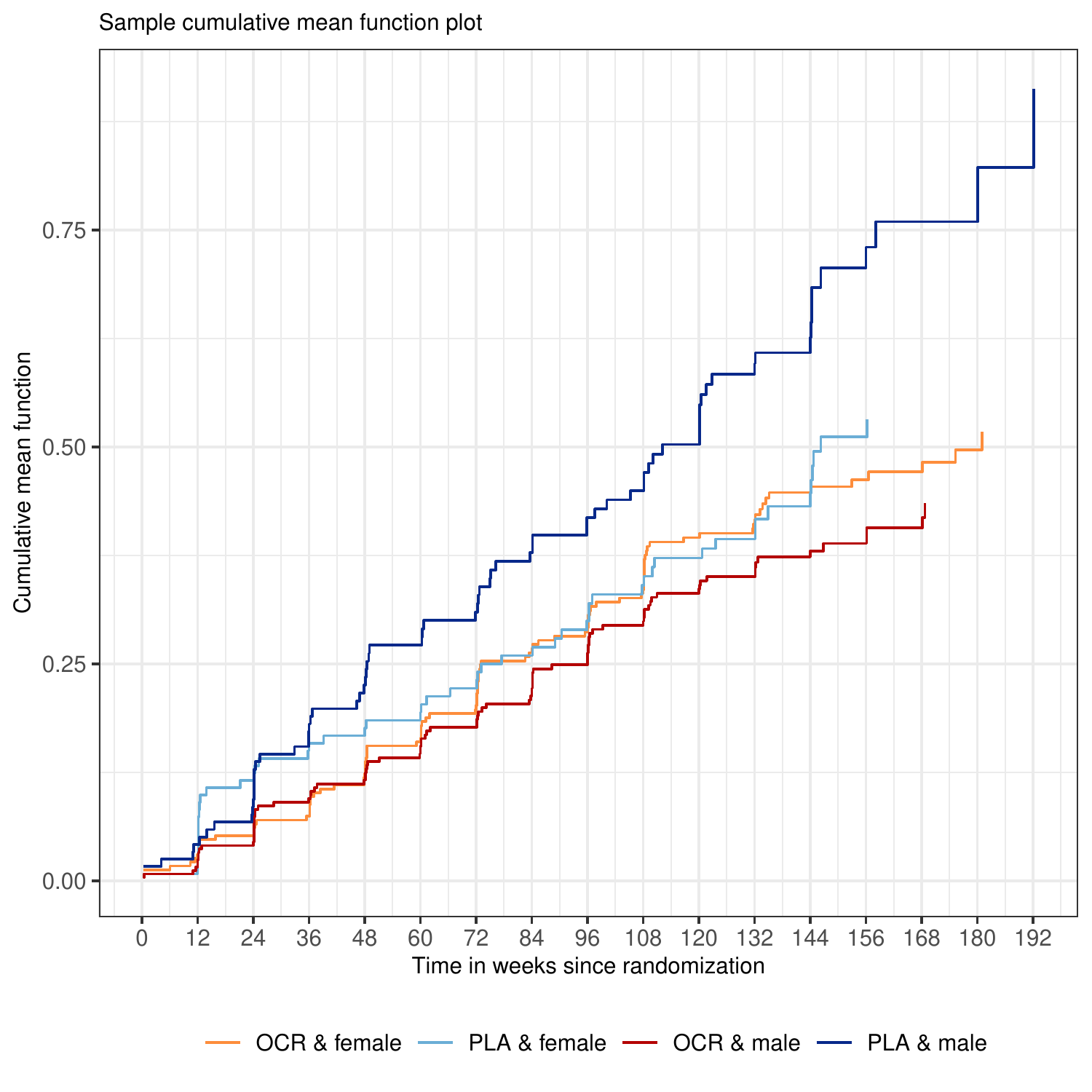}
  }} 
  
  \subfloat[Age]{
  \scalebox{0.4}[0.4]{ 
\includegraphics{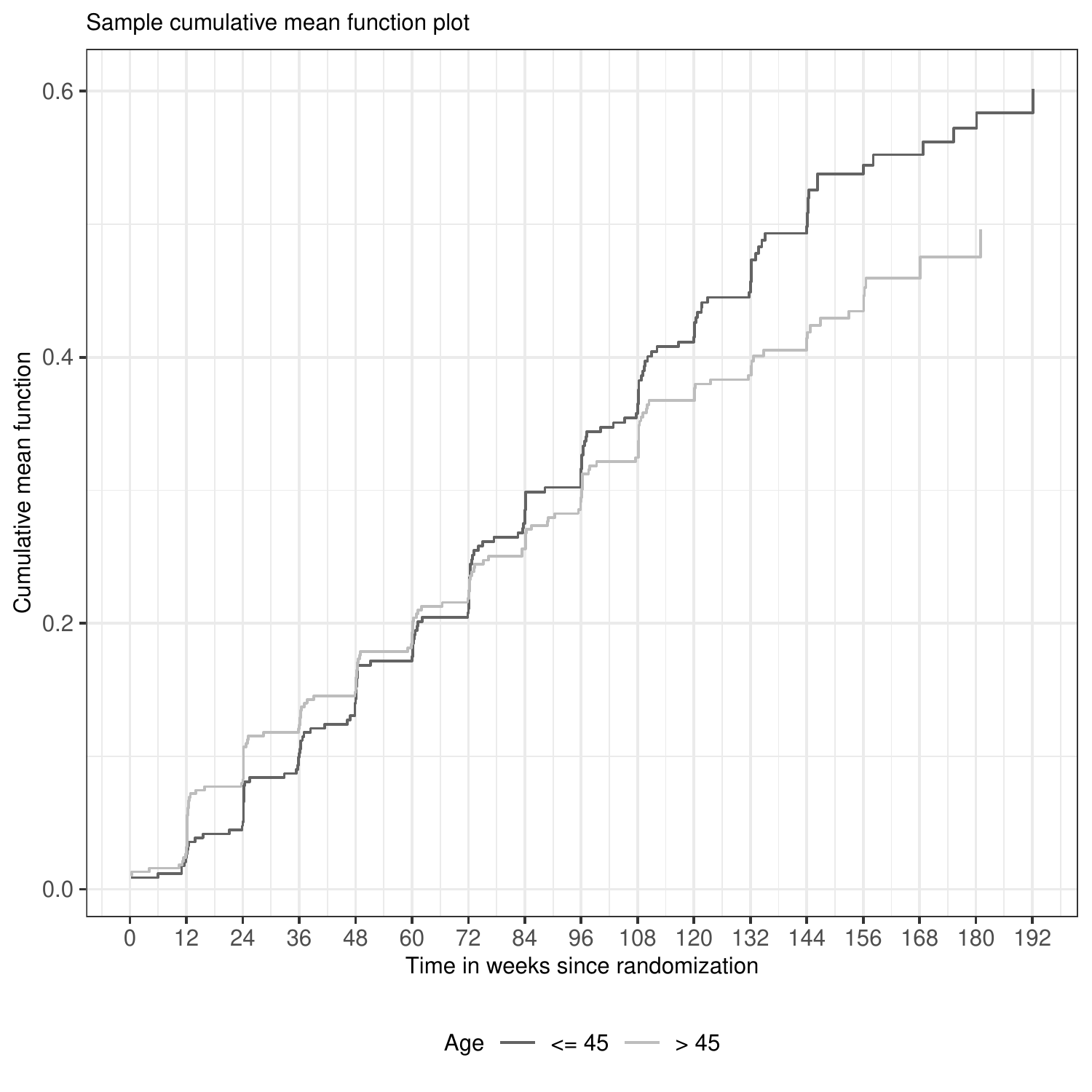}
  }}
  \subfloat[Age and treatment group]{
  \scalebox{0.4}{ 
\includegraphics{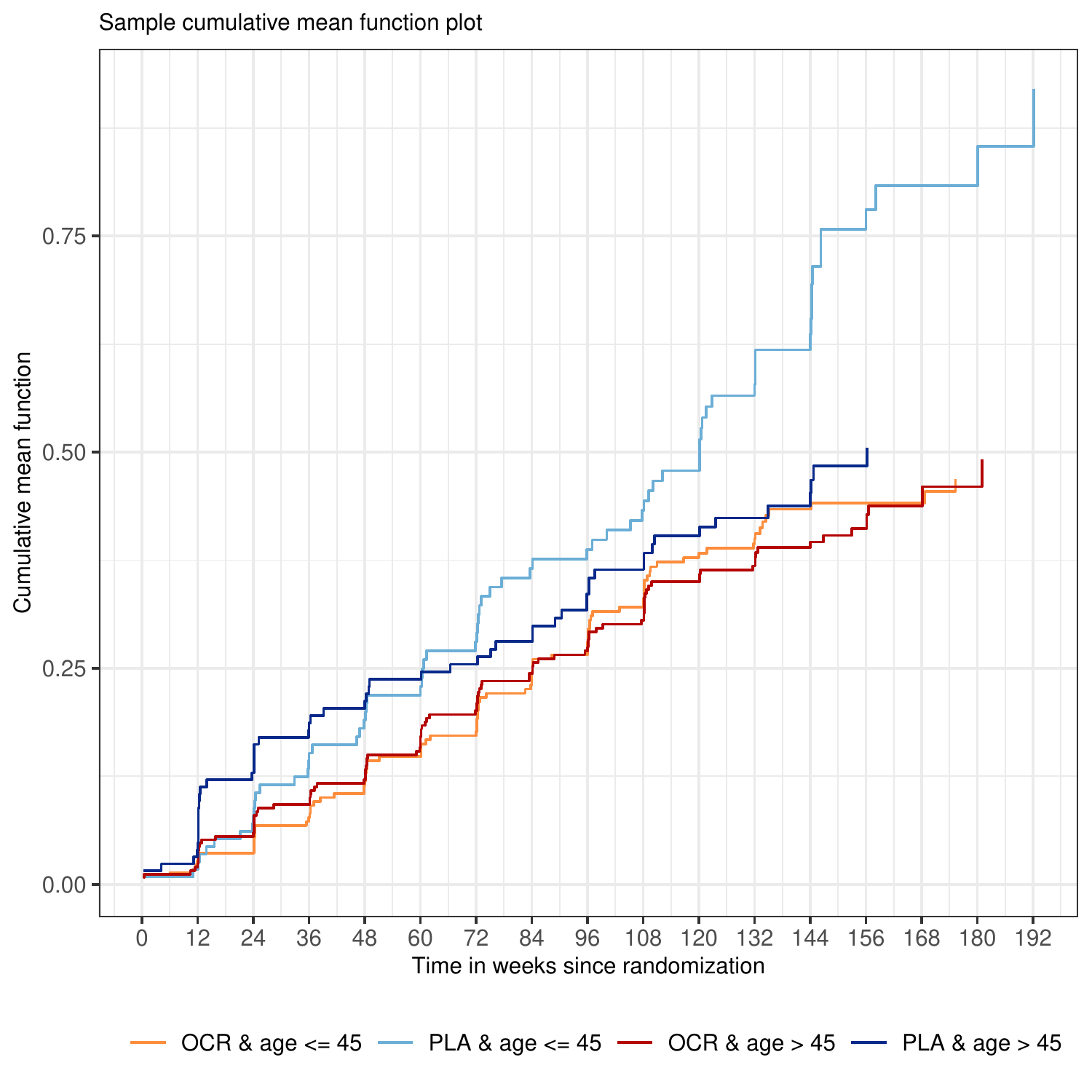}
  }} 
\caption{ORATORIO - Cumulative mean functions of CDP12 by age and sex}
\end{figure}

\begin{figure}[H] 
\centering
  \subfloat[Region]{
  \scalebox{0.4}[0.4]{ 
\includegraphics{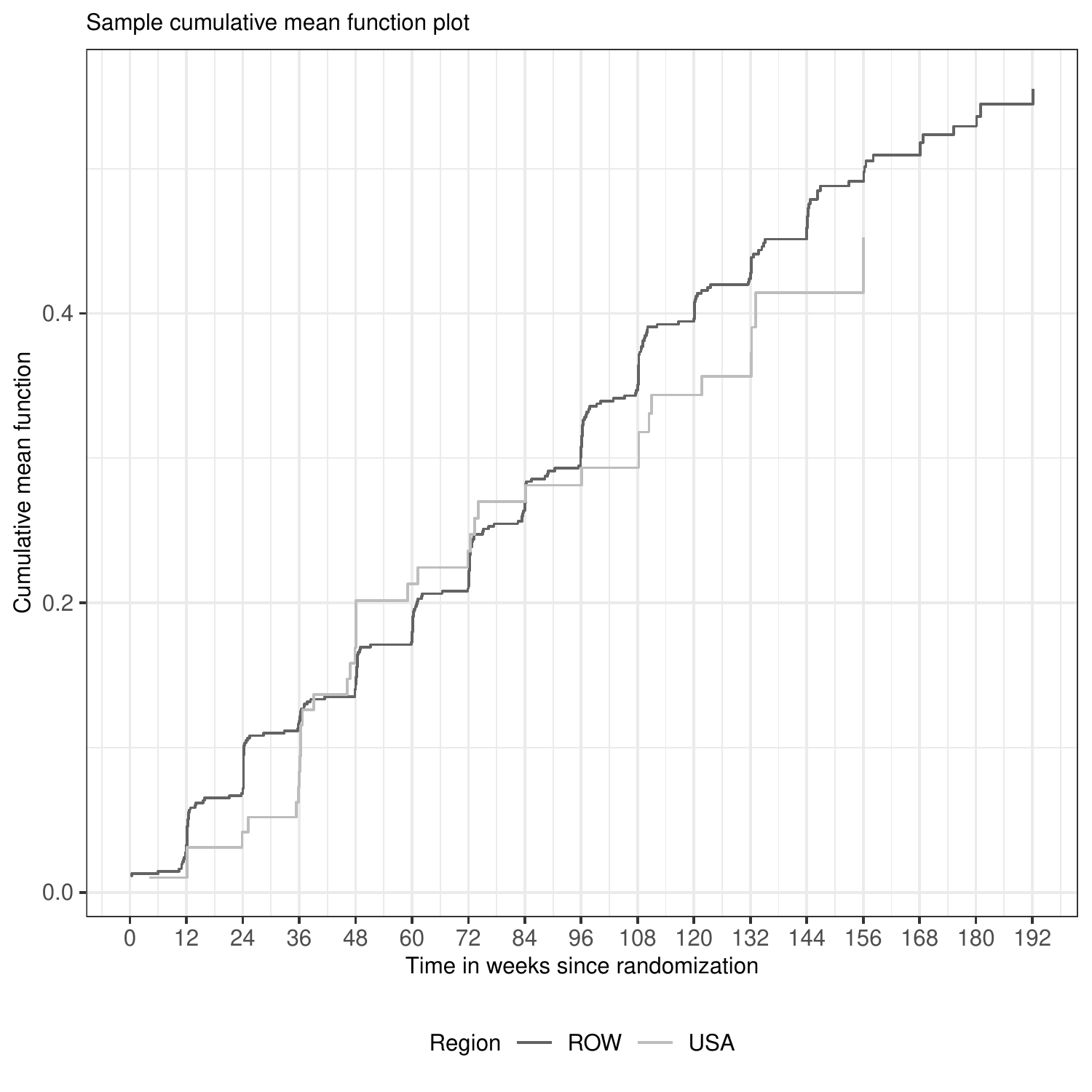}
  }}
  \subfloat[Region and treatment group]{
  \scalebox{0.4}{ 
\includegraphics{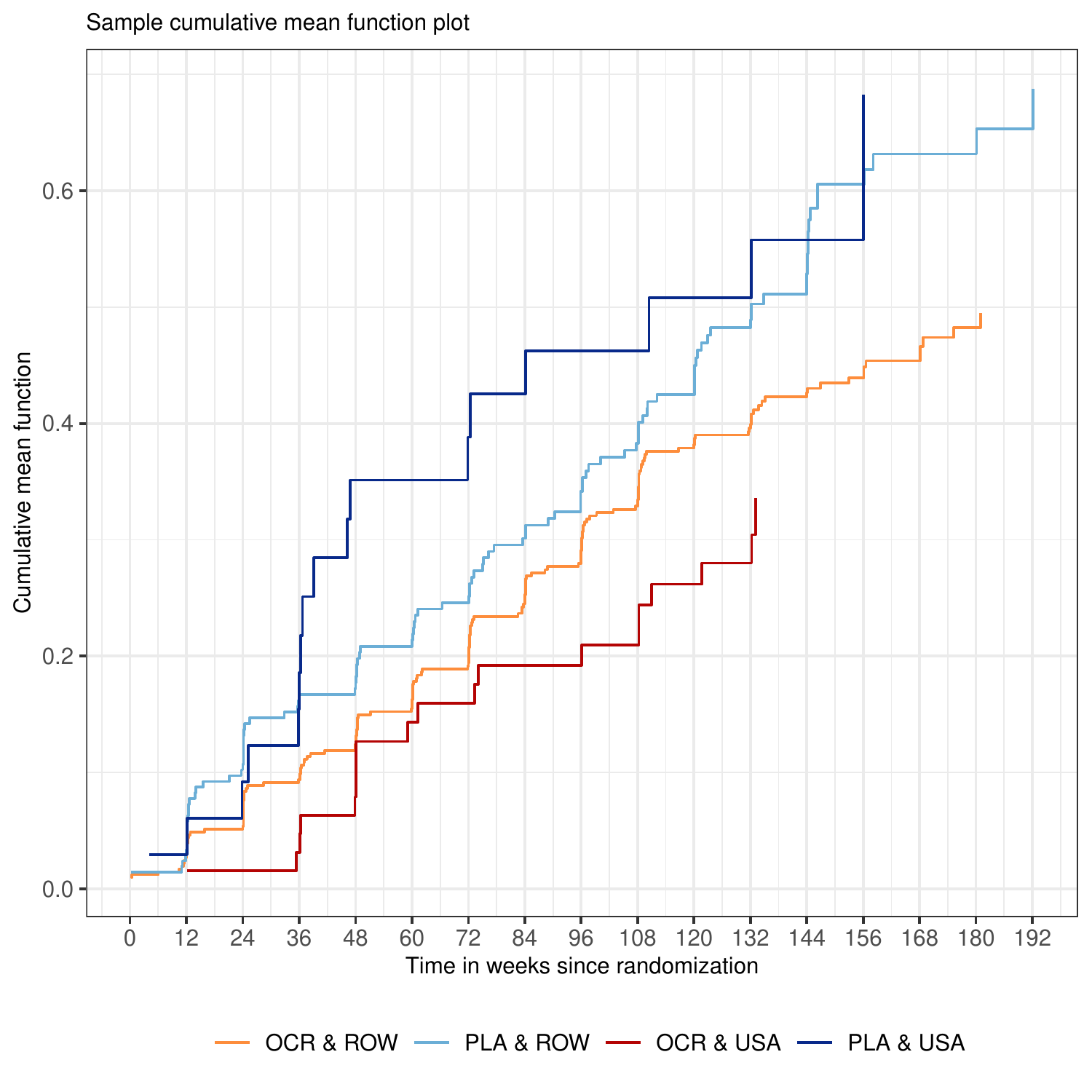}
  }} 

  \subfloat[BMI]{
  \scalebox{0.4}[0.4]{ 
\includegraphics{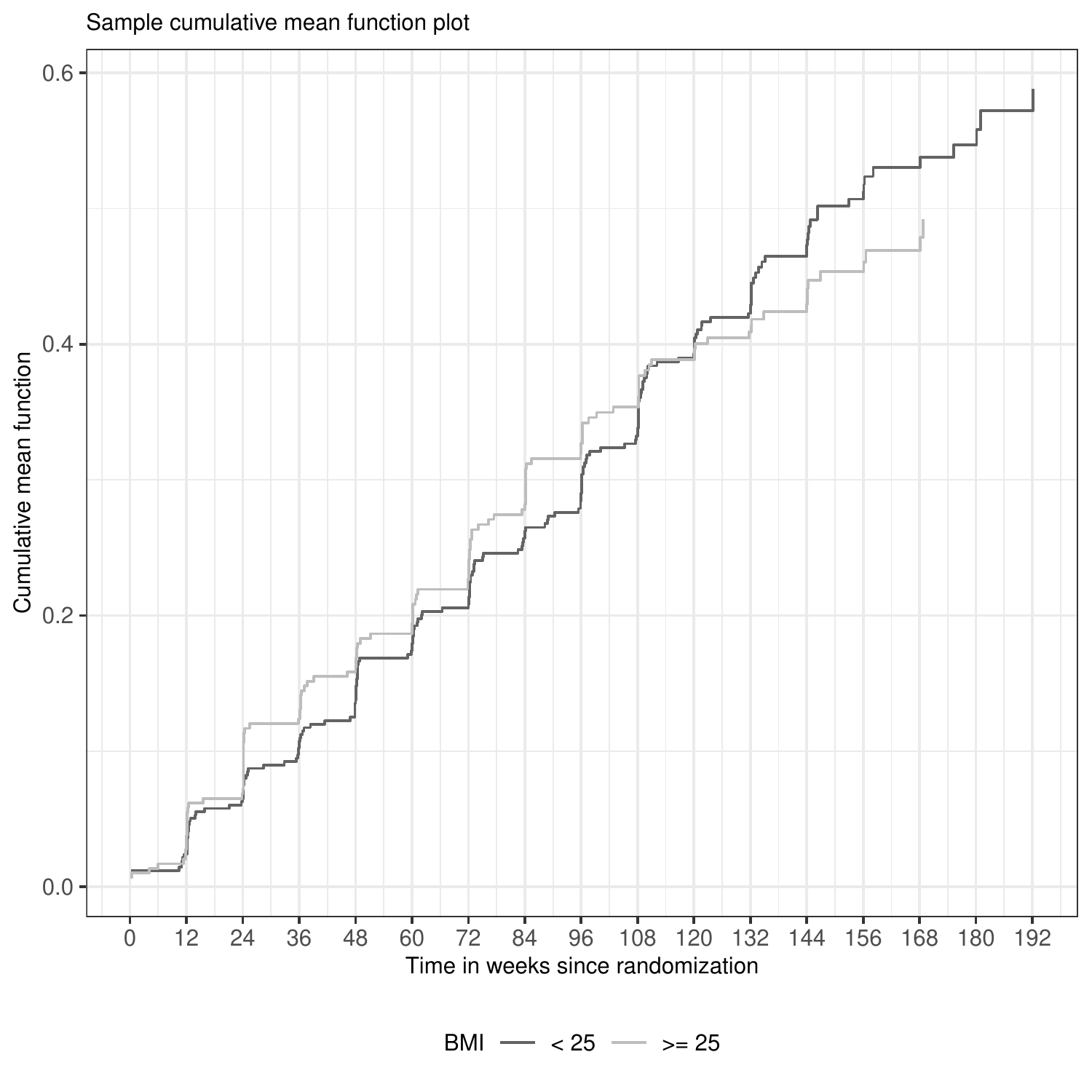}
  }}
  \subfloat[BMI and treatment group]{
  \scalebox{0.4}{ 
\includegraphics{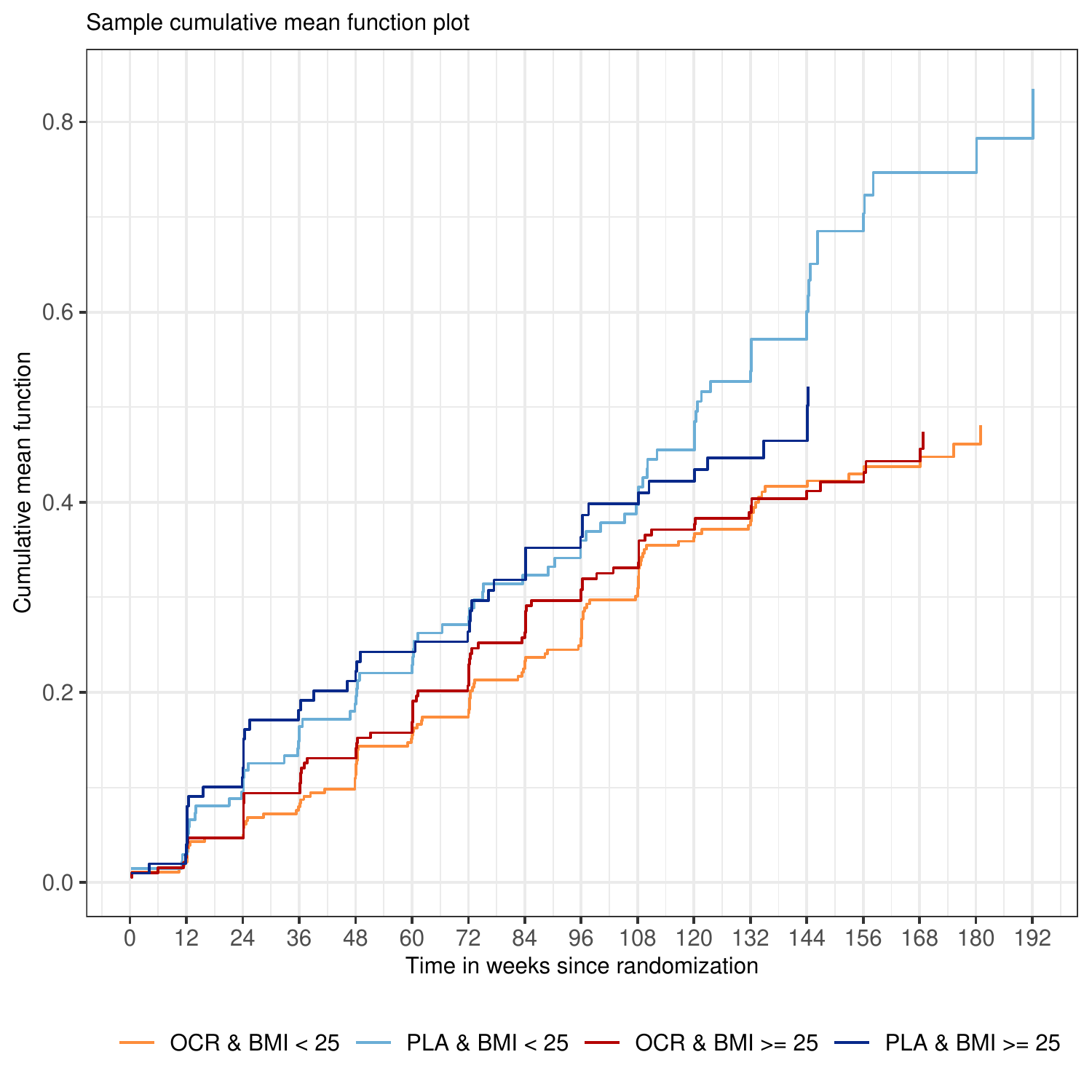}
  }}   

  \subfloat[T1 Gd-enhancing lesions at baseline]{
  \scalebox{0.4}[0.4]{ 
\includegraphics{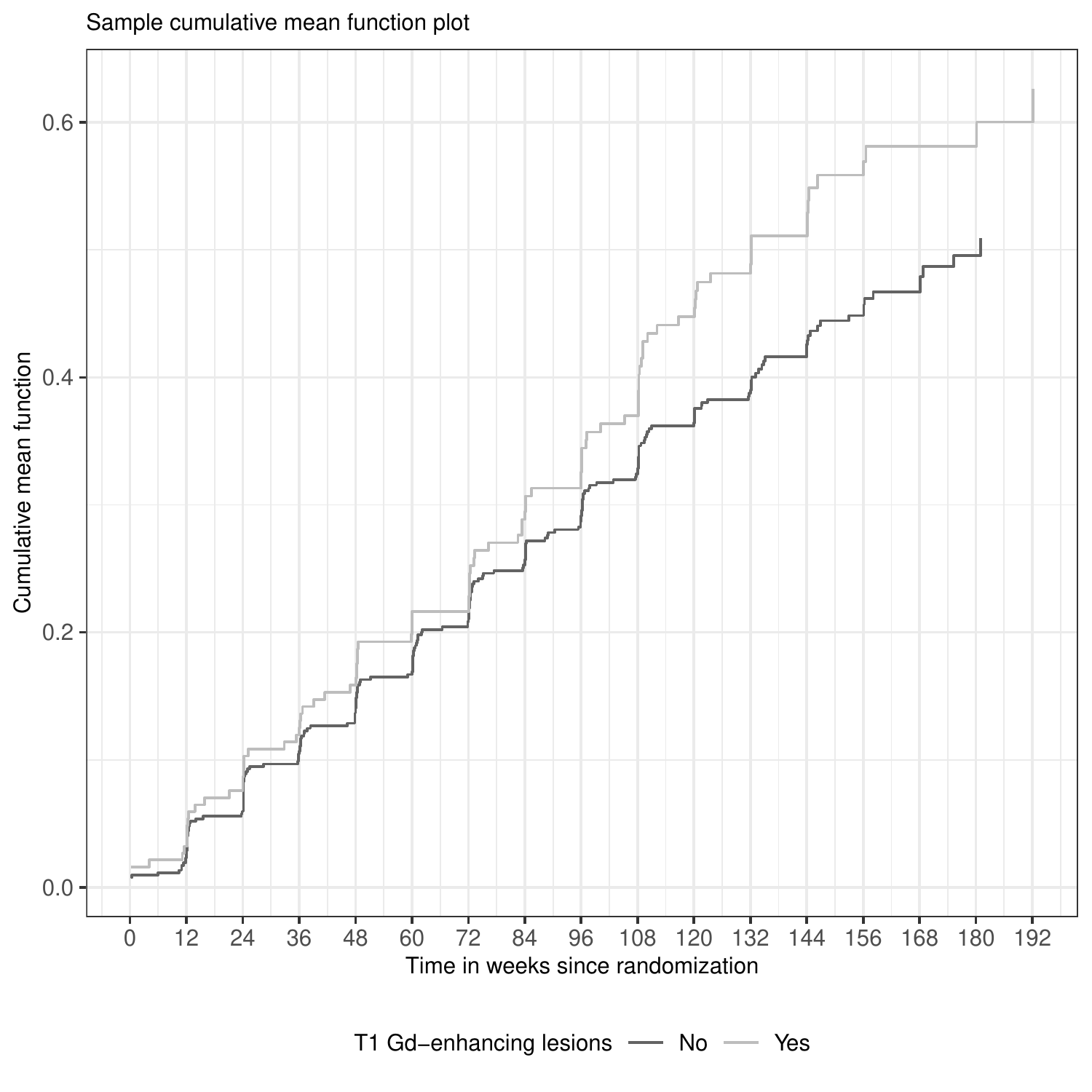}
  }}
  \subfloat[T1 Gd-enhancing lesions at baseline and treatment group]{
  \scalebox{0.4}{ 
\includegraphics{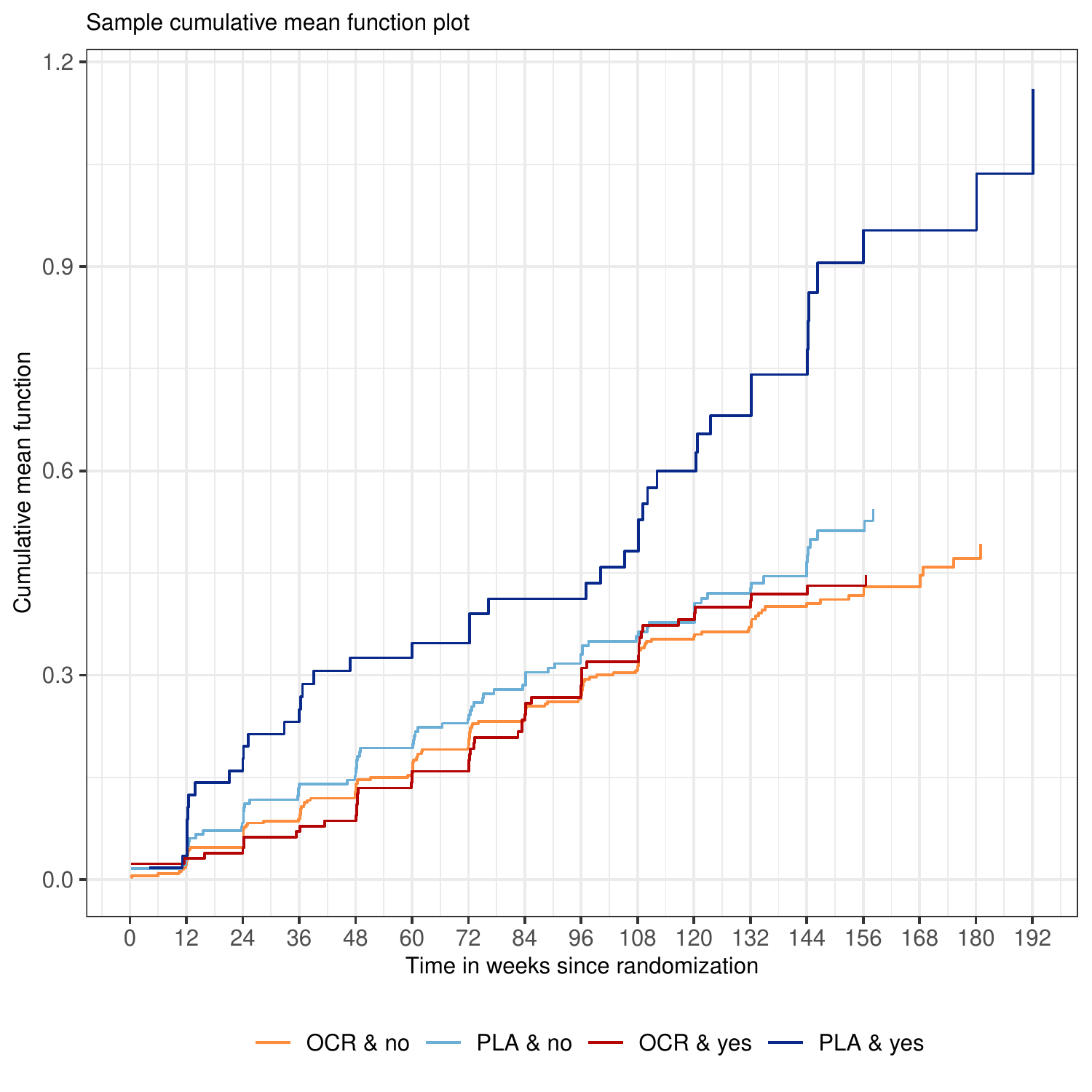}
  }} 
\caption{ORATORIO - Cumulative mean functions of CDP12 by region, BMI and T1 lesions}
\end{figure}

\begin{figure}[H] 
\centering
  \subfloat[Prior MS disease-modifying therapies]{
  \scalebox{0.4}[0.4]{ 
\includegraphics{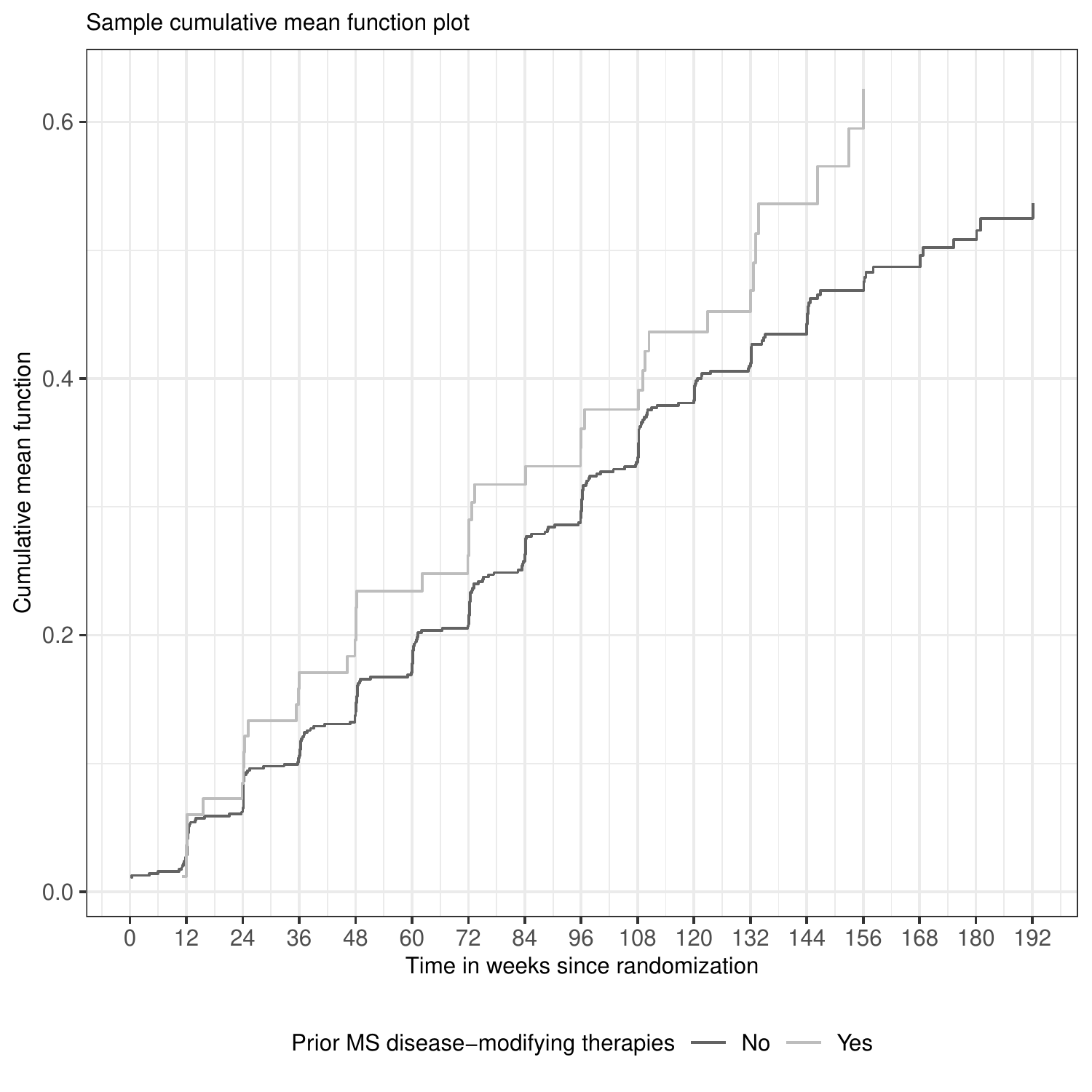}
  }}
  \subfloat[Prior MS disease-modifying therapies and treatment group]{
  \scalebox{0.4}{ 
\includegraphics{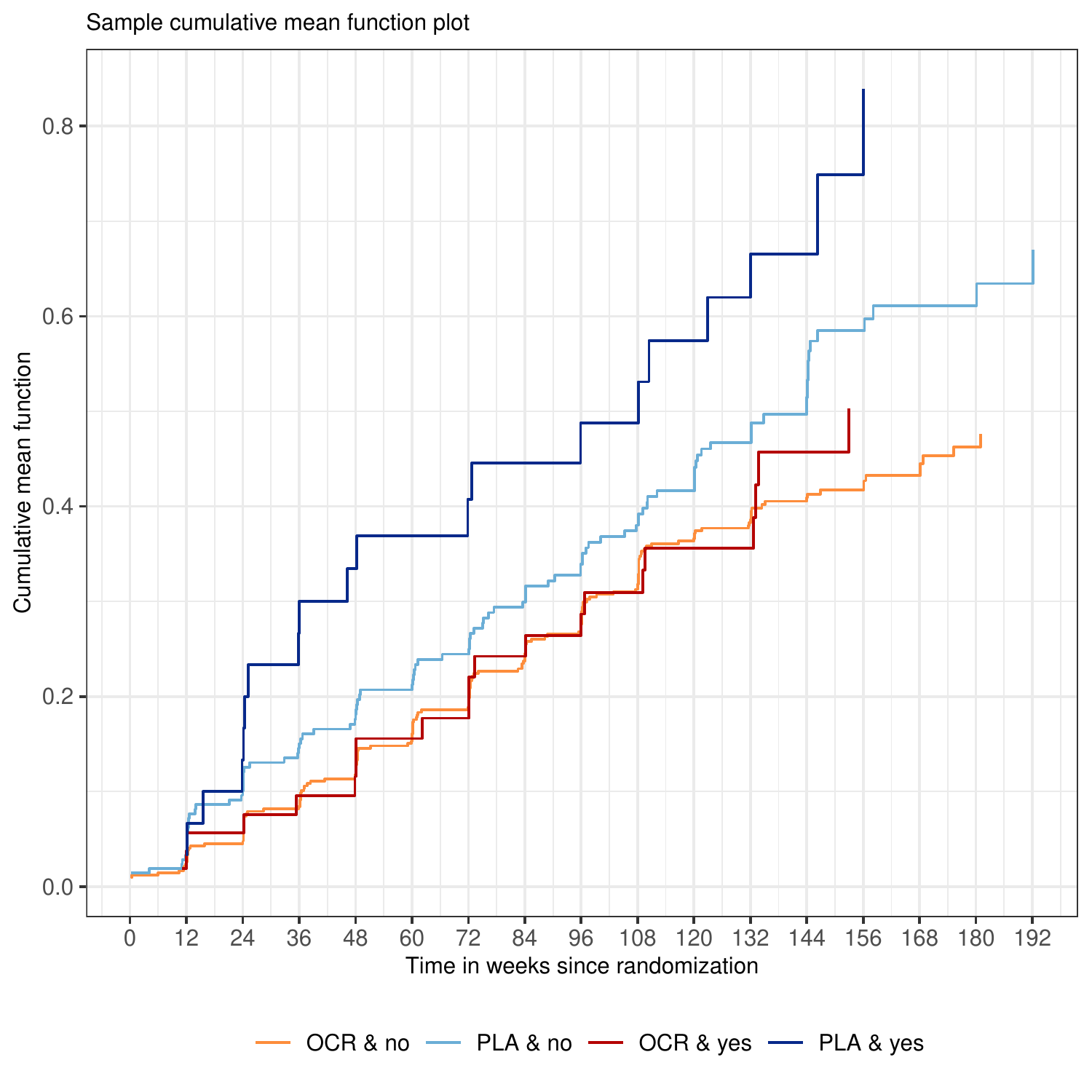}
  }} 
\caption{ORATORIO - Cumulative mean functions of CDP12 by prior MS therapy}
\end{figure}

\chapter{R code}
Statistical analyses are performed using the computing environment R, version 3.5.2 (R Foundation for Statistical Computing).
\section{Time-to-first-event and recurrent event methods}
\subsubsection*{Description of datasets}
A dataset for recurrent event analyses without terminal event should be structured as follows: 
\begin{figure}[h]
    \centering
    \scalebox{0.58}{
    \includegraphics{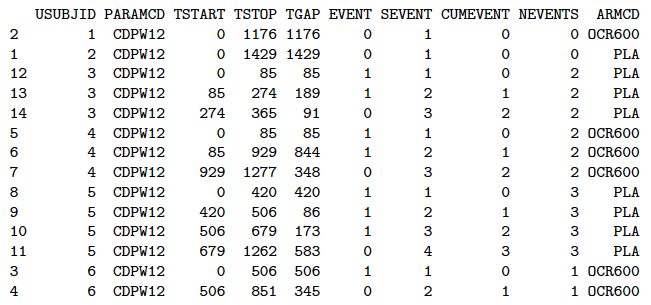}}
\end{figure}
\newline 
As shown above, this extract of the ORATORIO dataset (\textit{USUBJID} changed) includes the variables \textit{USUBJID}, \textit{PARAMCD}, \textit{TSTART}, \textit{TSTOP}, \textit{TGAP}, \textit{EVENT}, \textit{SEVENT}, \textit{CUMEVENT}, \textit{NEVENTS} and \textit{ARMCD}. While the variable \textit{USUBJID} is an unique patient identifier, \textit{PARAMCD} defines the endpoint considered in the analysis. The variable \textit{TSTART} contains the time $t_{i0}=0$ and the previous event times $t_{ij}$. The $\textit{TSTOP}$ variable contains the event times $t_{ij}$ and the right-censoring time $C_{i}$, $i=1,...,n$ and $j=1,...,n_{i}$. In counting process formulation, intervals are assumed to be open on the left and closed on the right, i.e., (TSTART, TSTOP]. The variable \textit{EVENT} $\in \{ 0, 1 \}$ indicates whether \textit{TSTOP} is an observed event time or a right-censoring time. It yields that \textit{EVENT} is equal to $1$, if an event has been observed at time $\textit{TSTOP}$ and $0$, if $\textit{TSTOP}$ is a right-censoring time. The $\textit{TGAP}$ variable is defined as $\textit{TSTOP}- \textit{TSTART}$ and specifies the number of days between two successive events. For use in stratified analyses, $\textit{SEVENT}$ simply records the cumulative number of lines for each patient. If data is restricted to \textit{SEVENT}=1, the recurrent event dataset reduces to a time-to-first-event dataset. The variable $\textit{CUMEVENT}$ corresponds to the number of previous events $N_{i}(t-)$ experienced by the patient at time $\textit{TSTART}$ and $\textit{NEVENTS}$ summarizes the total number of events experienced by a patient during follow-up. \textit{ARMCD} states a patient's treatment group such that patients treated with OCR have 'OCR600' in \textit{ARMCD} and patients on PLA have 'PLA'. 
\newpage
In this dataset, patients without disability progression (= 0 events) have only 1 line (e.g., patients $1$ and $2$), whereas patients with at least one progression event have $(n_{i} + 1$) lines, with the last line corresponding to the time of right-censoring. 
\newline \newline
The WLW model requires a specific dataset structure:
\begin{figure}[h]
    \centering
    \scalebox{0.56}{
    \includegraphics{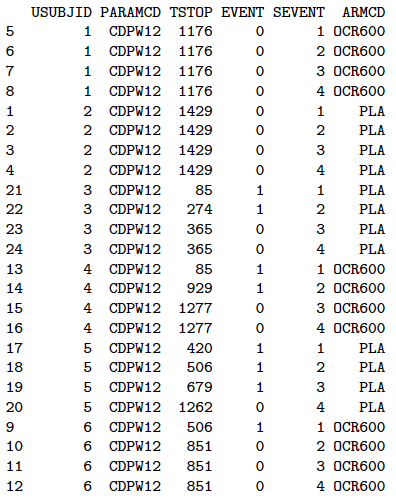}}
\end{figure}
\newline 
For WLW models, the dataset includes the variables \textit{USUBJID}, \textit{PARAMCD}, \textit{TSTOP}, \textit{EVENT}, \textit{SEVENT} and \textit{ARMCD}. As defined previously, the variable \textit{USUBJID} is an unique patient identifier and \textit{PARAMCD} defines the endpoint considered in the analysis. $\textit{TSTOP}$ is the time variable containing the event time $t_{ij}$ or the right-censoring time $C_{i}$, $i=1,...,n$ and $j=1,...,n_{i}$. The variable \textit{EVENT} $\in \{ 0, 1 \}$ indicates whether \textit{TSTOP} is an observed event time or a right-censoring time. The maximum number of events is $K=4$ in the data and each individual appears in all strata $\textit{SEVENT}$. $\textit{ARMCD}$ is the treatment arm such that patients treated with OCR have 'OCR600' in \textit{ARMCD} and patients on PLA have 'PLA'. In this dataset, each patient has $K$ lines.

\subsubsection*{Cumulative mean function}
\begin{lstlisting}[caption={Cumulative mean function}]
library(reda)
MCF <- mcf(Survr(USUBJID, TSTOP/7, EVENT) ~ ARMCD, data = oratorio.data)
mcfDiff.test(MCF)
\end{lstlisting}

\subsubsection*{Cox proportional hazards model}
\begin{lstlisting}[caption={Cox proportional hazards model}]
Cox.model <- function(data) {
  fit <- coxph(Surv(TSTOP, EVENT) ~ ARMCD, ties="breslow", data=data, 
               subset=(SEVENT==1))
  return(fit)
}
Cox.model(data=oratorio.data)
\end{lstlisting}

\subsubsection*{Poisson regression model}
\begin{lstlisting}[caption={Poisson regression model}]
Poisson.model <- function(data) {
  data <- data %>% group_by(USUBJID) %>% 
    summarise(ARMCD=first(ARMCD), COUNT=as.numeric(sum(EVENT)), 
              EXPTIME=last(TSTOP))
  fit <- glm(COUNT ~ offset(log(EXPTIME)) + ARMCD, 
             family=poisson(link=log), data=data)
  return(fit)
}
Poisson.model(data=oratorio.data)
\end{lstlisting}

\subsubsection*{Negative binomial model}
\begin{lstlisting}[caption={NB model}]
NB.model <- function(data) {
  library(dplyr)
  library(MASS)
  data <- data %>% group_by(USUBJID) %>% 
    summarise(ARMCD=first(ARMCD), COUNT=as.numeric(sum(EVENT)), 
              EXPTIME=last(TSTOP))
  fit <- glm.nb(COUNT ~ offset(log(EXPTIME)) + ARMCD, data=data)
  return(fit)
}
NB.model(data=oratorio.data)
\end{lstlisting}

\subsubsection*{Andersen-Gill model}
\begin{lstlisting}[caption={AG model}]
AG.model <- function(data) {
  fit <- coxph(Surv(TSTART, TSTOP, EVENT) ~ ARMCD, data=data)
  return(fit)
}
AG.model(data=oratorio.data)
\end{lstlisting}

\subsubsection*{Prentice-Williams-Peterson CP model}
\begin{lstlisting}[caption={PWP-CP model}]
PWP.model.common <- function(data) {
  fit <- coxph(Surv(TSTART, TSTOP, EVENT) ~ ARMCD + strata(SEVENT), 
               data=data)
  return(fit)
}
PWP.model.common(data=oratorio.data)

PWP.model.specific <- function(data) {
  library(data.table)
  data <- data[, ARMCD.BIN := ifelse(ARMCD=="PLA" | ARMCD=="REBIF", 0, 1)]
  data <- data[, ARMCD1 := ARMCD.BIN * (SEVENT==1)]
  data <- data[, ARMCD2 := ARMCD.BIN * (SEVENT==2)]
  data <- data[, ARMCD3 := ARMCD.BIN * (SEVENT>=3)]
  fit <- coxph(Surv(TSTART, TSTOP, EVENT) ~ ARMCD1 + ARMCD2 + ARMCD3 + 
               strata(SEVENT), data=data)
  return(fit)
}
PWP.model.specific(data=oratorio.data)
\end{lstlisting}

\subsubsection*{Wei-Lin-Weissfeld model}
\begin{lstlisting}[caption={WLW model}]
WLW.model.common <- function(data) {
  fit <- coxph(Surv(TSTOP, EVENT) ~ ARMCD + strata(SEVENT) + cluster(USUBJID), 
               data=data)
  return(fit)
}
WLW.model.common(data=oratorio.data.WLW)

WLW.model.specific <- function(data) {
  library(data.table)
  data <- data[, ARMCD.BIN := ifelse(ARMCD=="PLA" | ARMCD=="REBIF", 0, 1)]
  data <- data[, ARMCD1 := ARMCD.BIN * (SEVENT==1)]
  data <- data[, ARMCD2 := ARMCD.BIN * (SEVENT==2)]
  data <- data[, ARMCD3 := ARMCD.BIN * (SEVENT==3)]
  data <- data[, ARMCD4 := ARMCD.BIN * (SEVENT==4)]
  fit <- coxph(Surv(TSTOP, EVENT) ~ ARMCD1 + ARMCD2 + ARMCD3 + ARMCD4 + 
               strata(SEVENT) + cluster(USUBJID), data=data)
  return(fit)
}
WLW.model.specific(data=oratorio.data.WLW)
\end{lstlisting}

\subsubsection*{Lin-Wei-Yang-Ying model}
\begin{lstlisting}[caption={LWYY model}]
LWYY.model <- function(data) {
  fit <- coxph(Surv(TSTART, TSTOP, EVENT) ~ ARMCD + cluster(USUBJID), 
               data=data)
  return(fit)
}
LWYY.model(data=oratorio.data)
\end{lstlisting}

\subsubsection*{Partially conditional rate-based model}
\begin{lstlisting}[caption={Partially conditional rate-based model}]
PCRB.model.specific <- function(data) {
  library(data.table)
  data <- data[, ARMCD.BIN := ifelse(ARMCD=="PLA" | ARMCD=="REBIF", 0, 1)]
  data <- data[, ARMCD1 := ARMCD.BIN * (SEVENT==1)]
  data <- data[, ARMCD2 := ARMCD.BIN * (SEVENT==2)]
  data <- data[, ARMCD3 := ARMCD.BIN * (SEVENT>=3)]
  fit <- coxph(Surv(TSTART, TSTOP, EVENT) ~ ARMCD1 + ARMCD2 + ARMCD3 + 
               strata(SEVENT) + cluster(USUBJID), data=data)
  return(fit)
}
PCRB.model.specific(data=oratorio.data)
\end{lstlisting}

\newpage 
\section{Simulation}
\subsection{General simulation study}
General input parameters: 
\begin{lstlisting}[caption={Parameter settings}]
require(parallel)
require(dplyr)
require(data.table)
require(MASS)
require(survival)

n <- 1000
trial.design <- "event-driven"
end.recruit <- 365
n.first.events <- 246
lambda <- 0.00025
scale.weibull <- 0.0009675564
shape.weibull <- 0.9161516
\end{lstlisting}

\begin{itemize}
\item \textbf{n} Total number of individuals under study 
\item \textbf{trial.design} Design of simulated study (either 'event-driven' or 'time-fixed')
\item \textbf{n.first.events} Number of first events to define study closure
\item \textbf{lambda} Parameter of exponentially distributed censoring times (cf. Section $\ref{SimulationAdd}
$)
\item \textbf{scale.weibull} Scale parameter of weibull distribution (cf. Section $\ref{GSsimulation}$)
\item \textbf{shape.weibull} Shape parameter of weibull distribution (cf. Section $\ref{GSsimulation}$)
\newline 
\end{itemize}

Input parameter of $\boldsymbol{\textit{simulate}\_\textit{setting1}()}$:
\begin{itemize}
\item \textbf{seed} Random number 
\item \textbf{phi} Variance of frailty ($\phi \in \{0.0, 0.15, 1.0 \}$)
\item \textbf{HR} Simulated treatment effect size on hazard scale ($\exp(\beta) \in \{1.0, 0.70 \}$)
\item \textbf{scenario} Unique scenario identifier 
\end{itemize}

\begin{lstlisting}
simulate_setting1 <- function(seed, phi, HR, scenario) {
  
  set.seed(82*seed)
  
  generate.event.times <- function(C, T1, Z, X, scale.weibull, 
                                   shape.weibull, HR) {
    t <- T1 
    uncens <- t < C
    times <- ifelse(uncens==TRUE, t, C)
    event <- ifelse(uncens==TRUE, 1, 0)
    while(uncens==TRUE) {
      t <- ((-log(1-runif(n=1, min=0, max=1))/(Z* scale.weibull*exp(log(HR)*X))) 
            + t^shape.weibull)^(1/shape.weibull)
      uncens <- t < C
      if(uncens==TRUE) {
        times <- c(times, t)
      } else {
        times <- c(times, C)
      }
    }
    return(times)  
  }
  
  #Cox proportional hazards model
  Cox.model <- function(data) {
    fit <- coxph(Surv(TSTART, TSTOP, EVENT) ~ X, ties="breslow", data=data, 
                 subset=(SEVENT==1))
    #Output: beta, exp(beta), se(beta), p-value, lower.CI, upper.CI
    output <- c(summary(fit)$coef[1, c(1, 2, 3, 5)], 
                summary(fit)$conf.int[1, c(3,4)])
    return(output)
  }
  
  #Negative binomial model 
  NB.model <- function(data) {
    data.NB <- as.data.table(data %>% group_by(ID) %>% 
                               summarise(X=first(X), 
                                         COUNT=as.numeric(sum(EVENT)), 
                                         EXPTIME=last(TSTOP)))
    fit <- glm.nb(COUNT ~ offset(log(EXPTIME)) + X, data=data.NB)
    output <- c(summary(fit)$coef[2, c(1, 2, 4)], confint(fit)[2,])
    #Output: beta, exp(beta), se(beta), p-value
    output <- c(output[1], exp(output[1]), output[2], output[3], 
                exp(output[4]), exp(output[5])) 
    return(output)
  }
  
  #Poisson model 
  Poisson.model <- function(data) {
    data <- as.data.table(data %>% group_by(ID) %>% 
                            summarise(X=first(X), COUNT=as.numeric(sum(EVENT)), 
                                      EXPTIME=last(TSTOP)))
    fit <- glm(COUNT ~ offset(log(EXPTIME)) + X, 
               family=poisson(link=log), data=data)
    output <- c(summary(fit)$coef[2, c(1, 2, 4)], confint(fit)[2,])
    #output: beta, exp(beta), se(beta), p-value
    output <- c(output[1], exp(output[1]), output[2], output[3], 
                exp(output[4]), exp(output[5])) 
    return(output)
  }
  
  #Lin-Wei-Yang-Ying model
  LWYY.model <- function(data) {
    fit <- coxph(Surv(TSTART, TSTOP, EVENT) ~ X + cluster(ID), 
                 ties="breslow", data=data)
    #output: beta, exp(beta), robust se(beta), p-value, lower.CI, upper.CI
    output <- c(summary(fit)$coef[1, c(1,2,4,6)], 
                summary(fit)$conf.int[1, c(3,4)])
    return(output)
  }
  
  #Andersen-Gill model 
  AG.model <- function(data) {
    fit <- coxph(Surv(TSTART, TSTOP, EVENT) ~ X, ties="breslow", data=data)
    #output: beta, exp(beta), se(beta), p-value, lower.CI, upper.CI
    output <- c(summary(fit)$coef[1, c(1, 2, 3, 5)], 
                summary(fit)$conf.int[1, c(3,4)])
    return(output)
  }
  
  create.dataset <- function(trial.design, n, phi, lambda, end.recruit, 
                             n.first.events, scale.weibull, shape.weibull, HR) {
    ID <- 1:n
    #Generation of treatment group X using block randomization of block length 4
    X <- NULL 
    for(i in 1:ceiling(n/4)) {
      X <- c(X, sample(c(0,0,1,1), replace=FALSE))
    }
    X <- X[1:n]
    #Generation of gamma-distributed patient-specific random effect Z with 
    #mean 1 and variance phi
    if(phi != 0) {
      Z <- rgamma(n=n, shape=1/phi, scale=phi)
    } else {
      Z <- rep(1, n)
    }
    #Generation of exponentially distributed non-administrative censoring
    #time C.bar with rate lambda
    C.bar <- rexp(n=n, rate=lambda)
    #Generation of time to the first event 
    T1 <- (-log(1-runif(n=n, min=0, max=1))/
             (Z* scale.weibull*exp(log(HR)*X)))^(1/shape.weibull)
    if(trial.design=="event-driven") {
      #Generation of uniformly distributed entry time T.recruit with 
      #min=0 and max=end.recruit
      T.recruit <- runif(n=n, min=0, max=end.recruit)
      #Calculation of administrative censoring time C
      delta <- ifelse(T1 <= C.bar, 1, 0)
      T.calendar <- T.recruit + pmin(T1, C.bar)
      data <- data.table(ID, X, Z, C.bar, T.recruit, T1, delta, T.calendar)
      CA <- rep(data[delta==1][order(T.calendar)][n.first.events,]$T.calendar,n) 
            + 0.0001
      if (CA[[1]] < end.recruit) stop("Target number of first events reached 
                                      before last patient recruited. Please 
                                      modify settings!")
      C <- pmin(T.recruit + C.bar, CA) - T.recruit
      data <- data.table(ID, X, Z, C, T1)
    } else {
      #Calculation of administrative censoring time 
      CA <- 1513
      C <- pmin(C.bar, CA)
      data <- data.table(ID, X, Z, C, T1)
    }
    #Generation of recurrent events
    data <- data[, TSTOP := .(list(generate.event.times(C, T1, Z, X,
                                    scale.weibull, shape.weibull, HR))), by=ID]
    #Creation of TSTART, TSTOP and TGAP notation                                                   
    data <- data[, .(TSTOP = unlist(TSTOP)), by = setdiff(names(data), 'TSTOP')]
    data <- data[, EVENT := ifelse(TSTOP != C, 1, 0)]
    data <- data[, TSTART := dplyr::lag(TSTOP, 1, default=0), by=ID]
    data <- data[, TGAP := TSTOP-TSTART]
    if((sum(data$TSTART) < 0) > 0) stop("TSTART < 0")
    if((sum(data$TSTOP) < 0) > 0) stop("TSTOP < 0")
    #NEVENTS = total number of progression events
    data <- data[, NEVENTS := sum(EVENT), by=ID]
    #SEVENT: strata 
    data <- data[, SEVENT := 1]
    data <- data[, SEVENT := cumsum(SEVENT) , by=ID]
    data <- data[, .(ID, X, Z, TSTART, TSTOP, TGAP, EVENT, NEVENTS, SEVENT)]
    return(data)
  }
  
  data <- create.dataset(trial.design=trial.design, n=n, phi=phi, lambda=lambda, 
                         end.recruit=end.recruit, n.first.events=n.first.events, 
                         scale.weibull=scale.weibull, shape.weibull=shape.weibull, HR=HR)
  cox.fit <- tryCatch({ c(scenario, 1, HR, Cox.model(data=data), 0)}, 
                      error=function(e) {
                        return(c(scenario, 1, HR, rep(NA, 7)))})
  NB.fit <- tryCatch({ c(scenario, 2, HR, NB.model(data=data), 0) }, 
                     warning=function(w) {
                       return(c(scenario, 2, HR, Poisson.model(data=data), 1))}, 
                     error=function(e) { 
                       return(c(scenario, 2, HR, Poisson.model(data=data), 1))})
  LWYY.fit <- tryCatch({ c(scenario, 3, HR, LWYY.model(data=data), 0)},
                       error=function(e) {
                         return(c(scenario, 3, HR, rep(NA, 7)))})
  AG.fit <- tryCatch({ c(scenario, 4, HR, AG.model(data=data), 0)}, 
                     error=function(e) {
                       return(c(scenario, 4, HR, rep(NA, 7)))})
  return(list(data, cox.fit, NB.fit, LWYY.fit, AG.fit))
} 
\end{lstlisting}

\newpage
\subsection{MS-specific simulation study}

General input parameters: 
\begin{lstlisting}[caption={Parameter settings}]
require(parallel)
require(dplyr)
require(data.table)
require(MASS)
require(survival)

n <- 1000
trial <- "PPMS"
trial.design <- "event-driven"
end.recruit <- 365
n.first.events <- 246
lambda <- 0.00025
n.states <- 12 
Q.init <- PPMS.QPLA
frailty.matrix <- "Z1"
probs.baseline.EDSS <- c(0.00000, 0.00274, 0.08208, 0.18331, 0.17921, 0.09439, 
                         0.05746, 0.09986, 0.18057, 0.11902, 0.00137, 0.00000) 
type <- "confirmation"
weeks <- 12 
reference.method <- "fixed"
roving.period <- NA
\end{lstlisting}

\begin{itemize}
\item \textbf{n} Total number of individuals under study  
\item \textbf{trial} PPMS or RRMS trial ('PPMS' or 'RRMS')
\item \textbf{trial.design} Design of simulated study (either 'event-driven' or 'time-fixed')
\item \textbf{n.first.events} Number of first events to define study closure   
\item \textbf{lambda} Parameter of exponentially distributed censoring times (cf. Section $\ref{SimulationAdd}
$)
\item \textbf{n.states} Total number of states in time-homogeneous EDSS multistate model (cf. Figure $\ref{markierung1}$)
\item \textbf{Q.init} Baseline transition intensity function (cf. Figure $\ref{Q0PPMS}$)
\item \textbf{frailty.matrix} Specification of heterogeneity matrix (either 'Z1' (=$U_{1}$) or 'Z2' (=$U_{2}$))
\item \textbf{probs.baseline.EDSS} Probabilities of baseline EDSS scores (cf. Eq. $(\ref{piPPMS}$))
\item \textbf{type} Time-to-onset-of-CDP or time-to-confirmation-of-CDP (either 'onset' or 'confirmation')
\item \textbf{weeks} Confirmation period of IDP (12-week CDP or 24-week CDP)
\item \textbf{reference.method} Use of fixed or roving reference system (either 'fixed' or 'roving')
\item \textbf{roving.period} Confirmation period of the new reference score (cf. Section $\ref{sectionRoving}$)
\newline 
\end{itemize}

Input parameter of $\textit{simulate}\_\textit{setting2}()$:
\begin{itemize}
\item \textbf{seed} Random number 
\item \textbf{phi} Variance of frailty ($\phi \in \{0.0, 0.15, 1.0 \}$)
\item \textbf{HR.transition} Simulated treatment effect size on EDSS transitions ($\exp(\beta_{hj}) \in \{1.0, 0.70 \}$)
\item \textbf{scenario} Unique scenario identifier 
\end{itemize}

\newpage 
\begin{lstlisting}
simulate_setting2 <- function(seed, phi, HR.transition, scenario) {
  
  set.seed(28*seed)
  
  #Generation of patient-specific actual EDSS assessment times 
  generate.visits <- function(C) {
    visits <- seq(from=1, to=C, by=84)
    #Add random noise
    noise <- c(0, round(rt(n=length(visits)-1, df=3.54, ncp=0.25), digit=0))
    noise[noise < -10] <- -10
    noise[noise > 10] <- 10
    visits.noise <- visits + noise
    return(visits.noise) 
  }
  
  upper.matrix <- function(v, k) {
    n <- length(v)
    m <- n + abs(k)
    y <- matrix(0, nrow = m, ncol = m)
    y[col(y) == row(y) + k] <- v
    return(y)
  } 
  
  #Generation of transition intensity matrix Q
  qmatrix <- function(X, Z) { 
    n.states <- length(probs.baseline.EDSS)
    #Creation of treatment matrix 
    if(X[1]==0) {
      Q.trt <- matrix(1, ncol=n.states, nrow=n.states)
    } else {
      Q.trt <- diag(HR.transition, n.states) + 
               upper.matrix(rep(HR.transition, n.states-1), 1) + 
               upper.matrix(rep(HR.transition, n.states-2), 2) + 
               upper.matrix(rep(HR.transition, n.states-3), 3) 
      Q.trt[Q.trt==0] <- 1
    }
    #Heterogeneity matrix 
    if(frailty.matrix == "Z1") { 
      #Z1
      Q.Z <- diag(Z[1], n.states) + upper.matrix(rep(Z[1], n.states-1), 1) + 
             upper.matrix(rep(Z[1], n.states-2), 2) + 
             upper.matrix(rep(Z[1], n.states-3), 3) 
    } else {
      #Z2
      Q.Z <- diag(Z[1], n.states) + upper.matrix(rep(Z[1], n.states-1), 1) + 
             upper.matrix(rep(Z[1], n.states-2), 2) + 
             upper.matrix(rep(Z[1], n.states-3), 3) +
             upper.matrix(rep(Z[1], n.states-1), -1) + 
             upper.matrix(rep(Z[1], n.states-2), -2) + 
             upper.matrix(rep(Z[1], n.states-3), -3) 
    }
    Q.Z[Q.Z==0] <- 1
    #Calculation of patient-specific transition intensity matrix
    Q <- Q.init * Q.Z * Q.trt 
    #Check properties of Q
    diag(Q) <- 0
    diag(Q) <- 0 - apply(Q, 1, sum)
    return(Q)
  }
  
  #Generation of EDSS scores at baseline and post-baseline study visits
  generate.EDSS <- function(ID, DY, DIFF, X, Z) {
    n.states <- length(probs.baseline.EDSS)
    #Generation of baseline EDSS score using multinomial distribution
    which.state <- rmultinom(n=1, size=1, prob=probs.baseline.EDSS)
    values <- c(1:n.states)[which.state==1]
    #Generation of patient-specific transition intensity matrix Q
    Q <- qmatrix(X, Z)
    #EDSS scores at subsequent study visits
    if(length(DY) > 1) {
      for(j in 2:length(DY)) {
        #Calculation of transition probability matrix P
        PP <- MatrixExp(Q, t=DIFF[j], method=NULL)
        which.state <- rmultinom(n=1, size=1, prob=PP[values[j-1], ])
        values <- c(values, c(1:n.states)[which.state==1])
      }
    }
    return(values)
  } 
  
  #Function to calculate the required EDSS increase (1.0 point or 0.5 point) 
  #depending on the current reference EDSS score
  EDSS.increase <- function(reference) {
    increase <- ifelse((reference >= 0 & reference <= 5.5), 1.0, 0.5)
    return(increase)
  }
  
  #Function to update change from current reference EDSS score
  change.from.reference <- function(index, reference, AVAL, DY) {
    change <- c(rep(NA, index-1), AVAL[(index):length(DY)]) 
              - rep(reference, length(DY))
    return(change)
  }
  
  #Function to derive time-to-CDP-endpoint from longitudinal EDSS scores 
  recurrent.events <- function(type, ID, BASE, AVAL, CHG, DY, weeks) {
    event <- rep(0, length(DY))
    prog <- 0 
    days <- ifelse(weeks==12, 84, 161)
    increase <- EDSS.increase(BASE[1])
    i <- 1 
    
    while(i <= length(DY)) {
      if(prog==0 & CHG[i]>=increase) {
        idp <- DY[i]
        index.idp <- i
        prog <- 1
        i <- i + 1
      } else if (CHG[i] >= increase) {
        #Requirements fulfilled: confirmation period 12 or 24 weeks
        if(DY[i] >= idp + days) {
          if(type=="onset") {
            #Time-to-onset-of-CDP
            event[DY == idp] <- 1
          } else {
            #Time-to-confirmation-of-CDP
            event[i] <- 1
          }
          reference <- AVAL[index.idp]
          increase <- EDSS.increase(reference)
          CHG <- change.from.reference(index.idp, reference, AVAL, DY)
          prog <- 0
          i <- ifelse(type=="onset", index.idp, i)
          #Increase in EDSS yes but other requirements not fulfilled
        } else {
          i <- i + 1 
        }
        #Roving reference system: confirmation period = 24 weeks
        #If previous EDSS score is missing, use previous previous EDSS
      } else if(((reference.method=="roving") & 
                 (roving.period==24) & (i>5) & (CHG[i] < 0) & 
                 (CHG[i]==ifelse(is.na(lag(CHG,1)[i])==TRUE, lag(CHG,2)[i], 
                                 lag(CHG,1)[i])) & 
                 (CHG[i]==lag(CHG,2)[i]) & (DY[i]-lag(DY,2)[i] >= 161)) |
                ((reference.method=="roving") & (roving.period==24) & (i>5) & 
                 (CHG[i] < 0) & (CHG[i]==ifelse(is.na(lag(CHG,1)[i])==TRUE, 
                                                lag(CHG,2)[i], lag(CHG,1)[i])) & 
                 (CHG[i]==lag(CHG,2)[i]) & (CHG[i]==lag(CHG,3)[i]) & 
                 (DY[i]-lag(DY,3)[i] >= 161))) {
        reference <- AVAL[i]
        increase <- EDSS.increase(reference)
        CHG <- change.from.reference(i, reference, AVAL, DY)
        i <- i + 1
      } else {
        i <- i + 1 
        prog <- 0
      }
    }
    return(event)
  }
  
  create.dataset <- function(trial, trial.design, n, phi, lambda, end.recruit, 
                             n.first.events, probs.baseline.EDSS, Q.init, 
                             HR.transition, type, weeks) {
    ID <- as.factor(1:n) 
    #Generation of treatment group X using block randomization of block length 4
    X <- NULL 
    for(i in 1:ceiling(n/4)) {
      X <- c(X, sample(c(0,0,1,1), replace=FALSE))
    }
    X <- X[1:n] 
    #Generation of gamma-distributed patient-specific random effect Z with 
    #mean 1 and variance phi
    if(phi > 0) { 
      Z <- rgamma(n=n, shape=1/phi, scale=phi)
    } else {
      Z <- 1 
    }
    #Generation of exponentially distributed non-administrative censoring 
    #time C.bar with rate lambda
    C.bar <- pmin(rexp(n=n, rate=lambda), 2000)
    
    if(trial.design=="event-driven") {
      #Generation of uniformly distributed entry time T.recruit with 
      #min=0 and max=end.recruit
      T.recruit <- round(runif(n=n, min=0, max=end.recruit), digit=0)
      #Calculation of administrative censoring time C.bar
      C.bar <- pmax(1, round(C.bar, digit=0)) 
      data <- data.table(ID, X, Z, C.bar, T.recruit)
      #Generation of EDSS assessment times
      data <- data[, DY := .(list(generate.visits(C.bar))), by=ID]
      data <- data[, .(DY = unlist(DY)), by=setdiff(names(data), 'DY')]
      data <- data[, DIFF := lag(DY, 1), by=ID]
      data <- data[, DIFF := DY - DIFF]
      #Generation of EDSS measurements at study visits
      data <- data[, STATE := generate.EDSS(ID, DY, DIFF, X, Z), by=ID]
      if(trial=="PPMS") { 
        data <- data[, AVAL := ifelse(STATE > 1, (STATE + 3)/2, 
                                      ifelse(STATE==1, 2, 7.5))]
      } else {
        data <- data[, AVAL := (STATE / 2) + 0.5]
      } 
      #Baseline EDSS score
      data <- data[, BASE := AVAL[1], by=ID]
      #Change from baseline in EDSS score
      data <- data[, CHG := AVAL - BASE, by=ID]
      #Derivation of recurrent CDP events from longitudinal EDSS measurements
      data <- data[, EVENT := recurrent.events(type, ID, BASE, AVAL, 
                                               CHG, DY, weeks), by=ID]
      #Generation of non-administrative censoring time 
      data <- data[, DY.calendar := T.recruit + DY, by=ID]
      tmp <- data %>% filter(EVENT==1) %>% group_by(ID) %>% 
        summarise(T.calendar=first(DY.calendar)) %>% arrange(T.calendar)
      data <- data[, CA := tmp[n.first.events, ]$T.calendar]
      if (data$CA[[1]] < end.recruit) stop("Target number of first events 
                                           reached before last patient 
                                           recruited. Please modify settings!")
      data <- data[DY.calendar <= CA]
    } else {
      #Calculation of administrative censoring time 
      CA <- 673
      C <- pmax(1, round(pmin(C.bar, CA), digit=0))
      data <- data.table(ID, X, Z, C)
      data <- as.data.table(data)
      #Generation of EDSS assessment times
      data <- data[, DY := .(list(generate.visits(C))), by=ID]
      data <- data[, .(DY = unlist(DY)), by=setdiff(names(data), 'DY')]
      data <- data[, DIFF := DY - lag(DY, 1, default=NA), by=ID]
      #Generation of EDSS scores
      data <- data[, STATE := generate.EDSS(ID, DY, DIFF, X, Z), by=ID]
      if(trial=="PPMS") { 
        data <- data[, AVAL := ifelse(STATE > 1, (STATE + 3)/2, 
                                      ifelse(STATE==1, 2, 7.5))]
      } else {
        data <- data[, AVAL := (STATE / 2) + 0.5]
      }      
      data <- data[, BASE := AVAL[1], by=ID]
      data <- data[, CHG := AVAL - BASE, by=ID]
      #Derivation of recurrent CDP events from longitudinal EDSS measurements
      data <- data[, EVENT := recurrent.events(type, ID, BASE, AVAL, CHG, DY, weeks), 
                   by=ID]
    }
    data <- data[, .(ID, X, Z, DY, DIFF, STATE, AVAL, BASE, CHG, EVENT)]
    #Creation of recurrent event dataset 
    data1 <- data[data[, .(select = (DY == max(DY) | EVENT == 1)), 
                       by = ID]$select]
    #Creation of TSTART, TSTOP and TGAP notation
    data1 <- data1[, TSTART := dplyr::lag(DY, 1, default=0), by=ID]
    data1 <- data1[, TSTOP := DY]
    data1 <- data1[, TGAP := TSTOP - TSTART]
    #NEVENTS = total number of progression events
    data1 <- data1[, NEVENTS := sum(EVENT), by=ID]
    #SEVENT = cumulative number of lines in the data frame for each patient
    data1 <- data1[, SEVENT := 1 , by=ID]
    data1 <- data1[, SEVENT := cumsum(SEVENT), by=ID]
    data1 <- data1[, .(ID, X, Z, TSTART, TSTOP, TGAP, EVENT, NEVENTS, SEVENT)]
    return(list(data, data1)) 
  }
  
  Cox.model <- function(data) {
    fit <- coxph(Surv(TSTART, TSTOP, EVENT) ~ X, ties="breslow", data=data, 
                 subset=(SEVENT==1))
    #Output: beta, exp(beta), se(beta), p-value, lower.CI, upper.CI
    output <- c(summary(fit)$coef[1, c(1, 2, 3, 5)], 
                summary(fit)$conf.int[1, c(3,4)])
    return(output)
  }
  
  #Negative binomial model
  NB.model <- function(data) {
    data <- as.data.table(data %>% group_by(ID) %>% 
                            summarise(X=first(X), COUNT=as.numeric(sum(EVENT)), 
                                      EXPTIME=last(TSTOP)))
    fit <- glm.nb(COUNT ~ offset(log(EXPTIME)) + X, data=data)
    output <- c(summary(fit)$coef[2, c(1, 2, 4)], confint(fit)[2,])
    #Output: beta, exp(beta), se(beta), p-value
    output <- c(output[1], exp(output[1]), output[2], output[3], 
                exp(output[4]), exp(output[5])) 
    return(output)
  }
  
  #Poisson model 
  Poisson.model <- function(data) {
    data <- as.data.table(data %>% group_by(ID) %>% 
                            summarise(X=first(X), COUNT=as.numeric(sum(EVENT)), 
                                      EXPTIME=last(TSTOP)))
    fit <- glm(COUNT ~ offset(log(EXPTIME)) + X, 
               family=poisson(link=log), data=data)
    output <- c(summary(fit)$coef[2, c(1, 2, 4)], confint(fit)[2,])
    #Output: beta, exp(beta), se(beta), p-value
    output <- c(output[1], exp(output[1]), output[2], output[3], 
                exp(output[4]), exp(output[5])) 
    return(output)
  }
  
  #Lin-Wei-Yang-Ying model
  LWYY.model <- function(data) {
    fit <- coxph(Surv(TSTART, TSTOP, EVENT) ~ X + cluster(ID), 
                 ties="breslow", data=data)
    #Output: beta, exp(beta), robust se(beta), p-value, lower.CI, upper.CI
    output <- c(summary(fit)$coef[1, c(1,2,4,6)], 
                summary(fit)$conf.int[1, c(3,4)])
    return(output)
  }
  
  #Andersen-Gill model 
  AG.model <- function(data) {
    fit <- coxph(Surv(TSTART, TSTOP, EVENT) ~ X, ties="breslow", 
                 data=data)
    #Output: beta, exp(beta), se(beta), p-value, lower.CI, upper.CI
    output <- c(summary(fit)$coef[1, c(1, 2, 3, 5)], 
                summary(fit)$conf.int[1, c(3,4)])
    return(output)
  }
  
  out <- create.dataset(trial=trial, trial.design=trial.design, n=n, phi=phi, 
                        lambda=lambda, end.recruit=end.recruit, 
                        n.first.events=n.first.events, 
                        probs.baseline.EDSS=probs.baseline.EDSS, Q.init=Q.init, 
                        HR.transition=HR.transition, type=type, weeks=weeks)
  data <- out[[2]]
  data.edss <- out[[1]]
  
  cox.fit <- tryCatch({ c(scenario, 1, HR.transition, Cox.model(data=data), 0)}, 
                      error=function(e) {
                        return(c(scenario, 1, HR.transition, rep(NA, 7)))}) 
  NB.fit <- tryCatch({ c(scenario, 2, HR.transition, NB.model(data=data), 0) }, 
                     warning=function(w) {
                       return(c(scenario, 2, HR.transition, 
                                Poisson.model(data=data), 1))}, 
                     error=function(e) { 
                       return(c(scenario, 2, HR.transition, 
                                Poisson.model(data=data), 1))})
  LWYY.fit <- tryCatch({ c(scenario, 3, HR.transition, LWYY.model(data=data), 0)},
                       error=function(e) {
                         return(c(scenario, 3, HR.transition, rep(NA, 7)))})
  AG.fit <- tryCatch({ c(scenario, 4, HR.transition, AG.model(data=data), 0)}, 
                     error=function(e) {
                       return(c(scenario, 4, HR.transition, rep(NA, 7)))})
  return(list(data.edss, data, cox.fit, NB.fit, LWYY.fit, AG.fit)) 
}


is.qmatrix <- function(Q) {
  Q2 <- Q
  diag(Q2) <- 0
  isTRUE(all.equal(-diag(Q), rowSums(Q2))) && 
    isTRUE(all(diag(Q)<=0)) && isTRUE(all(Q2>=0))
}


MatrixExp <- function(mat, t = 1, method=NULL, ...){
  if (!is.matrix(mat) || (nrow(mat)!= ncol(mat)))
    stop("\"mat\" must be a square matrix")
  qmodel <- if (is.qmatrix(mat) && !is.null(method) && method=="analytic") 
    msm.form.qmodel(mat) else list(iso=0, perm=0, qperm=0)
  if (!is.null(method) && method=="analytic") {
    if (!is.qmatrix(mat))
      warning("Analytic method not available since matrix is not a 
              Markov model intensity matrix. Using \"pade\".")
    else if (qmodel$iso==0) warning("Analytic method not available for this 
                                    Markov model structure. Using \"pade\".")
    
  }
  if (length(t) > 1) res <- array(dim=c(dim(mat), length(t)))
  for (i in seq_along(t)) {
    if (is.null(method) || !(method %in% c("pade","series","analytic"))) {
      if (is.null(method)) method <- eval(formals(expm::expm)$method)
      resi <- expm::expm(t[i]*mat, method=method, ...)
    } else {
      ccall <- .C("MatrixExpR", as.double(mat), as.integer(nrow(mat)), 
                  res=double(length(mat)), as.double(t[i]),
                  as.integer(match(method, c("pade","series"))), 
                  as.integer(qmodel$iso), as.integer(qmodel$perm), 
                  as.integer(qmodel$qperm),
                  as.integer(0), NAOK=TRUE)
      resi <- matrix(ccall$res, nrow=nrow(mat))
    }
    if (length(t)==1) res <- resi
    else res[,,i] <- resi
  }
  res
}
\end{lstlisting}

\newpage 
\addcontentsline{toc}{chapter}{Bibliography}
\printbibliography

\newpage
\pagestyle{empty}
\section*{Statutory declaration}
\begin{verbatim}
\end{verbatim}
\begin{verbatim}
 
\end{verbatim}
I herewith declare that I have composed the present thesis myself and without use of any other than the cited sources and aids. Sentences or parts of sentences quoted literally are marked as such; other references with regard to the statement and scope are indicated by full details of the publications concerned. The thesis in the same or similar form has not been submitted to any examination body and has not been published. This thesis was not yet, even in part, used in another examination or as a course performance. Furthermore I declare that the submitted written (bound) copies  of the  presentthesis  and  the version submitted on a data carrier are consistent with each other in contents.
\newline 
 
\vspace{2cm}
\begin{tabularx}{\textwidth}[b]{p{5cm} p{2cm} p{5cm}} \cline{1-1} \cline{3-3}
\textit{Place and date} & & \textit{Signature} 
\end{tabularx}

\end{document}